%% file: nnpdf40.tex
\documentclass[11pt,a4paper]{article}

\usepackage[colorlinks=true, linkcolor=black!50!blue, urlcolor=blue, citecolor=blue, anchorcolor=blue]{hyperref}
\usepackage[font=small,labelfont=bf,margin=0mm,labelsep=period,tableposition=top]{caption}
\usepackage[a4paper,top=3cm,bottom=2.5cm,left=2.5cm,right=2.5cm,bindingoffset=0mm]{geometry}

\usepackage{graphicx,placeins}
\usepackage{float}
\usepackage{afterpage}
\usepackage{epsfig,cite}
\usepackage{amssymb}
\usepackage{amsmath}
\usepackage{dsfont}
\usepackage{multirow}
\usepackage{url}
\usepackage{xcolor,colortbl}
\usepackage{float}
\usepackage{afterpage}
\usepackage{url}
\usepackage{hyperref}
\usepackage{booktabs}

\usepackage{tikz}
\usepackage{tikz-3dplot}

\usepackage{enumitem}
\usepackage{hyperref}
\usepackage{cite}

\usepackage{pifont}
\newcommand{\cmark}{\cellcolor{blue!20}\textcolor{blue}{\ding{51}}}%
\newcommand{\xmark}{\cellcolor{red!20}\textcolor{red}{\ding{55}}}%
\newcommand{\ymark}{\cellcolor{yellow!30}\textcolor{blue}{(\ding{51})}}%

\usetikzlibrary{shapes, arrows}
\usetikzlibrary{decorations.pathreplacing}
\usetikzlibrary{positioning, calc}
\tikzstyle{fitted} = [rectangle, minimum width=5cm, minimum height=1cm, text centered, draw=black, fill=red!30]
\tikzstyle{operations} = [rectangle, rounded corners, minimum width=2cm,text centered, draw=black, fill=red!30]
\tikzstyle{roundtext} = [rectangle, rounded corners, minimum width=2cm, minimum height=0.8cm, text centered, draw=black, fill=red!30]
\tikzstyle{n3py} = [rectangle, rounded corners, minimum width=3cm, minimum height=1cm, text centered, draw=black, fill=green!30]
\tikzstyle{myarrow} = [thick,->,>=stealth]
\tikzstyle{line} =[draw, -latex']
\tikzstyle{decision} = [diamond, draw, fill=red!20, text width=7.5em, text centered,  inner sep=0pt, minimum height=2em, aspect=4]
\tikzstyle{cloud} = [draw, ellipse,fill=green!20, minimum height=2em]
\tikzstyle{inout} = [rectangle, draw, fill=green!20, text width=9.5em, text centered, rounded corners, minimum height=2em, minimum width=10em]
\tikzstyle{block}=[rectangle, draw, fill=blue!20, text width=9.5em, 
                   text centered, rounded corners, minimum height=2em, 
                   minimum width=10em]

\definecolor{darkgreen}{rgb}{0.0, 0.5, 0.13}

\newcommand{\be}{\begin{equation}}
\newcommand{\ee}{\end{equation}}
\newcommand{\bea}{\begin{eqnarray}}
\newcommand{\eea}{\end{eqnarray}}
\newcommand{\bi}{\begin{itemize}}
\newcommand{\ei}{\end{itemize}}
\newcommand{\ben}{\begin{enumerate}}
\newcommand{\een}{\end{enumerate}}
\newcommand{\la}{\left\langle}
\newcommand{\ra}{\right\rangle}
\newcommand{\lc}{\left[}
\newcommand{\rc}{\right]}
\newcommand{\lp}{\left(}
\newcommand{\rp}{\right)}

\def\frac#1#2{{{#1}\over {#2}}}
\def\gsim{\mathrel{\rlap{\lower4pt\hbox{\hskip1pt$\sim$}}
    \raise1pt\hbox{$>$}}}         
\def\lsim{\mathrel{\rlap{\lower4pt\hbox{\hskip1pt$\sim$}}
    \raise1pt\hbox{$<$}}}         

\newcommand{\rep}{\mathrm{rep}}

\newcommand{\tr}{\mathrm{tr}}

\newcommand{\draft}[1]{}

\def\beq{\begin{equation}}
\def\eeq{\end{equation}}

\def \y{{\bf y}} 
\def\lapprox{\lower .7ex\hbox{$\;\stackrel{\textstyle <}{\sim}\;$}}
\def\gapprox{\lower .7ex\hbox{$\;\stackrel{\textstyle >}{\sim}\;$}}

\numberwithin{equation}{section}
\numberwithin{figure}{section}
\numberwithin{table}{section}

\usepackage{tabularx}
\newcolumntype{C}[1]{>{\centering\arraybackslash}p{#1}}

\usepackage{amsmath}
\usepackage{amsfonts}
\usepackage{amssymb}
\usepackage{dsfont}
\usepackage{pifont}
\usepackage{booktabs}
\usepackage{graphicx}
\usepackage{epstopdf}
\usepackage{epsfig}
\usepackage{framed}
\usepackage{makeidx}
\usepackage{hyperref}
\usepackage{placeins}
\usepackage[font=small,labelfont=bf]{caption}

\newcommand{\levone}{z}
\newcommand{\levtwo}{y}
\newcommand{\law}{f}
\newcommand{\model}{g}
\newcommand{\shift}{\eta}
\newcommand{\noise}{\epsilon}
\newcommand{\covmat}{C}
\newcommand{\biascov}{\Sigma^{\rm bias}}
\newcommand{\varcov}{\Sigma^{\rm var}}

\newcommand{\biasvarratio}{\mathcal{R}_{bv}}

\newcommand{\ndata}{N_{\rm dat}}
\newcommand{\nfits}{n_{\rm fit}}
\newcommand{\nreps}{n_{\rm rep}}
\newcommand{\nx}{n_x}
\newcommand{\nflav}{n_{\rm flav}}

\newcommand{\erf}{{\rm erf}}
\newcommand{\repind}{k}
\newcommand{\datind}{i}
\newcommand{\datindj}{j}
\newcommand{\invcov}[1]{\covmat^{-1}_{#1}}
\newcommand{\erep}[1]{\mathbf{E}_{\noise}\left[ #1 \right]}
\newcommand{\eshift}[1]{\mathbf{E}_{\shift}\left[ #1 \right]}

\newcommand{\repchis}{{\chi^2}^{(\repind)}}
\newcommand{\ie}{{\it i.e.}}

\newcommand{\diffcentunder}{\left( \erep{\model} - \law \right)}
\newcommand{\diffcentrep}{\left( \erep{\model} - \model^{(\repind)}\right)}

\newcommand{\bias}{{\rm bias}}
\newcommand{\var}{{\rm variance}}

\newcommand{\secref}[1]{Sect.~\ref{#1}}
\newcommand{\tableref}[1]{Table~\ref{#1}}

\begin{document}
\newgeometry{top=1.5cm,bottom=1.5cm,left=1.5cm,right=1.5cm,bindingoffset=0mm}
\begin{figure}[h]
  \includegraphics[width=0.32\textwidth]{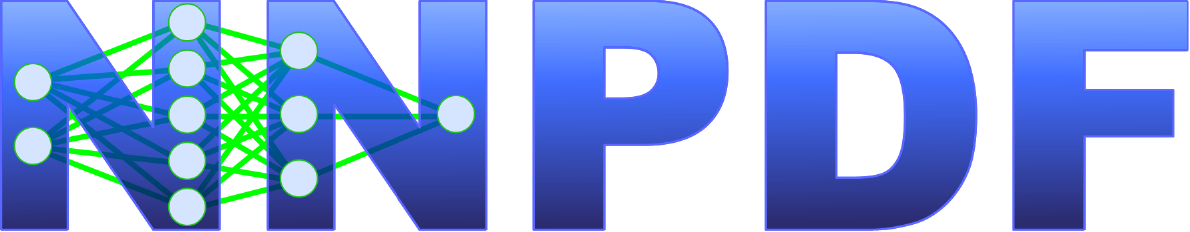}
\end{figure}
\vspace{-2.0cm}
\begin{flushright}
Edinburgh 2021/12 \\
Nikhef-2021-013\\
TIF-UNIMI-2021-11\\
\end{flushright}
\vspace{0.3cm}

\begin{center}
  {\Large \bf The Path to Proton Structure at  One-Percent Accuracy}
\vspace{1.1cm}

  {\small
  {\bf  The NNPDF Collaboration:} \\[0.2cm]
Richard D. Ball,$^{1}$
Stefano Carrazza,$^{2}$
Juan Cruz-Martinez,$^{2}$
Luigi Del Debbio,$^{1}$
Stefano Forte,$^{2}$\\[0.1cm]
Tommaso Giani,$^{1,8}$
Shayan Iranipour,$^{3}$
Zahari Kassabov,$^{3}$
Jose I. Latorre,$^{4,5,6}$
Emanuele R. Nocera,$^{1,8}$\\[0.1cm]
Rosalyn L. Pearson,$^{1}$
Juan Rojo,$^{7,8}$
Roy Stegeman,$^{2}$
Christopher Schwan,$^{2}$
Maria Ubiali,$^{3}$\\[0.1cm]
Cameron Voisey,$^{9}$ and
Michael Wilson$^{1}$
}\\

 \vspace{0.7cm}
 
 {\it \small

 ~$^1$The Higgs Centre for Theoretical Physics, University of Edinburgh,\\
   JCMB, KB, Mayfield Rd, Edinburgh EH9 3JZ, Scotland\\[0.1cm]
    ~$^2$Tif Lab, Dipartimento di Fisica, Universit\`a di Milano and\\
   INFN, Sezione di Milano, Via Celoria 16, I-20133 Milano, Italy\\[0.1cm]
   ~$^3$DAMTP, University of Cambridge, Wilberforce Road, \\ Cambridge, CB3 0WA, United Kingdom\\[0.1cm]
   ~$^4$  Quantum Research Centre, Technology Innovation Institute, Abu Dhabi, UAE\\[0.1cm]
~$^5$ Center for Quantum Technologies, National University of Singapore, Singapore\\[0.1cm]
~$^6$ Qilimanjaro Quantum Tech, Barcelona, Spain\\[0.1cm]
   ~$^7$Department of Physics and Astronomy, Vrije Universiteit, NL-1081 HV Amsterdam\\[0.1cm]
~$^8$Nikhef Theory Group, Science Park 105, 1098 XG Amsterdam, The Netherlands\\[0.1cm]
~$^9$Cavendish Laboratory, University of Cambridge, Cambridge, CB3
0HE, United Kingdom\\[0.1cm]
}

\vspace{1.0cm}

{\bf \large Abstract}

\end{center}

We present a new set of parton distribution functions (PDFs) based on a fully
global dataset and machine learning techniques: NNPDF4.0.
We expand the NNPDF3.1 determination with 44 new datasets, mostly from the LHC.
We derive a novel methodology through hyperparameter optimization, leading to
an efficient fitting algorithm built upon stochastic gradient descent.
We use NNLO QCD calculations and account for NLO electroweak corrections
and nuclear uncertainties. Theoretical improvements in the PDF description
include a systematic implementation of positivity constraints and integrability
of sum rules. We validate  our methodology by means of closure tests and
``future tests'' (i.e. tests of backward and forward data compatibility), and
assess its stability, specifically upon changes of PDF
parametrization basis. We study the internal compatibility of our dataset, and
investigate the dependence of results both upon the choice of input
dataset and of fitting methodology. We perform a  first study of the
phenomenological implications of NNPDF4.0 on representative LHC processes.
The software framework used to produce NNPDF4.0 is made available as an
open-source package together with documentation and examples.

\clearpage

\tableofcontents

\clearpage

\input{sec-introduction}

\input{sec-expdata}

\input{sec-methodology}

\input{sec-dataselection}

\input{sec-results}

\input{sec-closuretests}

\input{sec-dataset}

\input{sec-tests}

\input{sec-pheno}

\input{sec-summary}

\appendix

\input{app-codedoc}

\input{app-datacomp}

\clearpage
\bibliography{nnpdf40}

\end{document}

%% file: sec-introduction.tex
\section{Introduction}
\label{sec:introduction}

It is now an accepted fact that frontier high-energy physics at
colliders requires percent-level accuracy both in theory and
experiment~\cite{Salam:2018rwo}. On the theoretical side, the two
main obstacles to achieving this are missing higher order
corrections in perturbative computations~\cite{Heinrich:2020ybq}, and
uncertainties in parton distribution functions (PDFs)~\cite{Gao:2017yyd,Ethier:2020way}.
The main aim of this paper is to show how percent-level accuracy might be
achieved for PDFs.

The most recent set of PDFs determined by NNPDF, 
NNPDF3.1~\cite{Ball:2017nwa}, was the first to extensively include LHC data,
and was able to reach 3-5\% precision in the PDF uncertainties.
It was based on NNPDF3.x fitting methodology, the first to
be validated by means of closure tests, thereby ensuring
that this precision was matched by a comparable accuracy.

The NNPDF4.0 PDF set presented here is a major step forward
in three significant aspects:
{\it i)} the systematic inclusion of an extensive set of run I LHC at
7 and 8~TeV
data and, for the first time, of LHC Run II data
at $\sqrt{s}=13$~TeV and of several new processes not considered before for
PDF determinations;
{\it ii)} the deployment of state-of-the-art machine learning
algorithms which result in a methodology that is considerably faster and leads to more precise PDFs;
{\it iii)} the validation of these PDF uncertainties both in the data and in
the extrapolation regions using closure and future tests.

All in all, the main accomplishment of this new PDF
set is to go one step further in achieving the main goal that
motivated the NNPDF methodology in the first
place~\cite{Forte:2020yip}, namely, to reduce sources of bias in PDF
determination. The use of a wider dataset reduces sources of bias that
might be related to the dominance of a particular process. The use of
a machine learned methodology reduces sources of bias related to
methodological choices, that are now mostly made through an automated
procedure. Finally, the 
extensive set of validation tools explicitly
checks the absence of bias: in fact, ``future tests'', to be
discussed below, can expose the historical bias that was present in previous PDF
determinations.

The NNPDF4.0 global analysis includes 44 new datasets in comparison with
NNPDF3.1. These involve a number of new LHC measurements of processes already
present in NNPDF3.1, but also data from several new processes, whose impact on
PDFs has been the object of dedicated studies. Specifically, direct photon
production (studied in
Ref.~\cite{Campbell:2018wfu}),  single-top production (studied in
Ref.~\cite{Nocera:2019wyk}), dijets (studied in
Ref.~\cite{AbdulKhalek:2020jut}), $W$+jet  (studied in
Ref.~\cite{Faura:2020oom}), and deep-inelastic
jet production. A significant consequence of this extension of the  dataset
is that now the PDFs are largely controlled by LHC data:
unlike in the past, a DIS-only PDF determination leads to much larger
uncertainties and visibly different results.

NNPDF4.0 is the first PDF determination based on a methodology that is selected
automatically rather than through manual iterations and human experience.
All aspects of the neural
network PDF parametrization and optimization (such as neural net
architecture, learning rates or minimization algorithm) are selected
through a hyperparameter
optimization procedure~\cite{Carrazza:2019mzf}, an automated
scan of the space of models that selects the optimal methodology.
A quality control method is used in
order to make sure that the optimization does not produce a methodology
that leads to overfitted PDFs. This is done through
$K$-folding~\cite{Forte:2020yip}, checking iteratively the
effectiveness of any given methodology on sets of data
excluded in turn from the fit. All this is made possible by
a speedup of the NNPDF fitting code, which is now able to fit an individual replica 
about twenty times faster,
thanks mostly to the use of stochastic gradient descent methods provided by
the {\tt TensorFlow} library, rather than through the genetic algorithm
minimization used previously, along with various technical improvements to be
discussed below~\cite{Carrazza:2019mzf,Carrazza:2019agm,Cruz-Martinez:2020tte}.

The widening of the dataset (with fixed methodology), and especially the
methodological improvements (with fixed dataset) lead to a reduction of
PDF uncertainties, so their combination brings us close to percent
precision. This demands a careful  validation  of these uncertainties, which is
achieved by means of two classes  of tests.

The first is closure tests, already introduced in
NNPDF3.0~\cite{Ball:2014uwa}, which here are  considerably extended and
systematized, thanks to the much greater fitting
efficiency. These consist of fitting PDFs to pseudo-data generated
assuming a certain underlying  true PDF, and comparing the result
of the fit to the known true PDF by means of suitable 
statistical estimators. The closure test verifies that PDF
uncertainties are faithful, specifically in comparison to the data
used to fit them.
The second is future tests~\cite{Cruz-Martinez:2021rgy}: these compare the
results obtained fitting PDFs to a subset of the data, which covers a small
kinematic region compared to the full dataset.  For example, PDFs are fitted to
a pre-HERA dataset, and the result is compared to LHC data. The future test
verifies that PDF uncertainties are faithful when extrapolated outside the
region of the data used to fit them.

As a further test of methodological reliability, we study the
robustness of results upon methodological variations, and in particular
we show that
PDFs are stable upon changes of the parametrization basis (i.e.
the particular linear
combination of PDFs that is parametrized by neural nets), thereby confirming 
that results are parametrization-independent. 

NNPDF4.0 PDFs also include a number of improvements at all stages of
the PDF determination procedure. The most relevant ones are the following:

\begin{itemize}

\item While the main PDF determination is performed with NNLO QCD
  (with further sets provided at NLO and LO), NLO electroweak (EW) and
  mixed QCD-EW processes are implemented for all LHC processes using recent
  dedicated tools~\cite{Carrazza:2020gss} and assessed both for phenomenology
  and in the determination of the input dataset to be used for PDF fitting.

\item Whenever heavy nuclear or deuteron targets are involved, nuclear
  effects are accounted for as theoretical uncertainties using the 
  methodology of Refs.~\cite{Ball:2018lag,Ball:2018twp,Ball:2020xqw}, and the 
  results of the nNNPDF2.0 nuclear PDF determination~\cite{AbdulKhalek:2020yuc}.
  
\item Strict positivity of $\overline{\rm MS}$ PDFs is implemented
  following the results of Ref.~\cite{Candido:2020yat}.
  
\item Finiteness of non-singlet baryon number, i.e.,  integrability
  of all non-singlet PDF first moments is 
  enforced. This specifically implies finiteness of the Gottfried
  sum~\cite{Forte:1992df} $U-D$ and of the strangeness sum $U+D-2 S$,
  where $U$, $D$ and $S$ denote respectively the first moment of the
  sum of quark and antiquark PDFs for up, down and strange quarks.
  
\item  The selection of a consistent dataset is based on an objective two-stage 
  procedure. Potentially problematic datasets are identified on the 
  basis of either poor compatibility with the global dataset, or indications 
  of instability of their experimental covariance matrix. These datasets are 
  then subjected in turn to a dedicated fit in which the failed dataset 
  is given a large weight, and then accepted or rejected depending on the
  outcome. 
 
\end{itemize}

The main missing features of the current PDF determination, which are left
for future work, are the inclusion of theory uncertainties
(specifically missing higher order corrections), which
could be done using the methods of
Refs.~\cite{AbdulKhalek:2019bux,AbdulKhalek:2019ihb}, and  the full inclusion
of EW and mixed QCD-EW corrections directly at the fitting level,
which will be possible using the tools of
Ref.~\cite{Carrazza:2020gss}.

\begin{figure}[!t]
  \centering
  \includegraphics[width=0.49\linewidth]{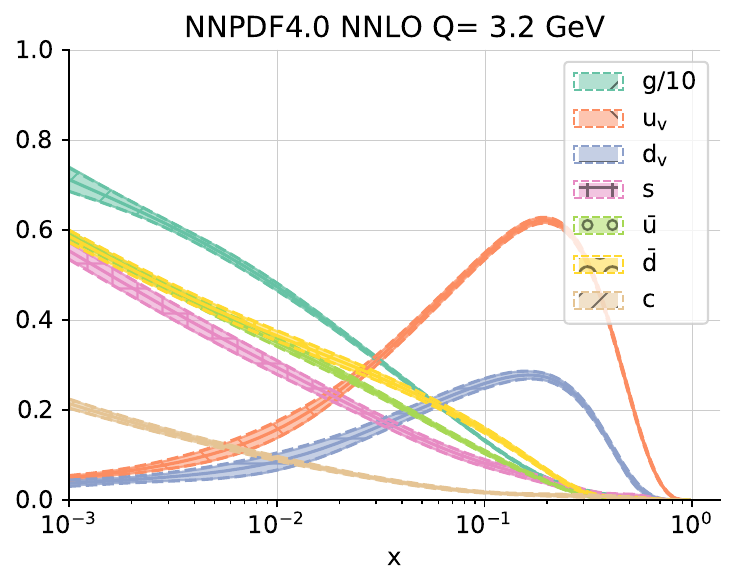}
  \includegraphics[width=0.49\linewidth]{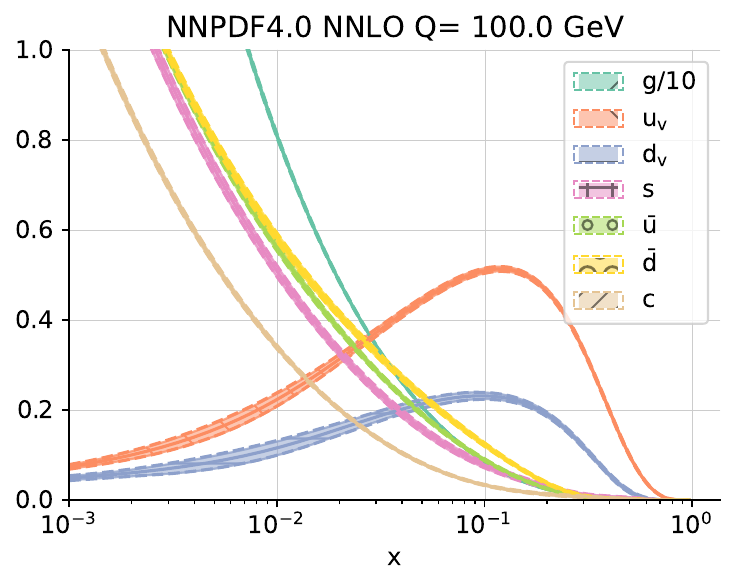}\\
  \caption{The NNPDF4.0 NNLO PDFs at $Q=3.2$~GeV (left) and
    $Q=10^2$~GeV (right).}
   \label{fig:pdg}
\end{figure}

The NNPDF4.0 PDF set is released at LO, NLO and NNLO QCD, for a
variety of values of $\alpha_s$. The default PDF sets are
provided in the FONLL variable-flavor number
scheme~\cite{Forte:2010ta} with maximum number of flavors $n_f=5$, and
an independently parametrized charm PDF. PDF sets with different maximum
number of flavors and with a perturbatively generated charm PDF are
also made available, along with PDF sets determined using  reduced
datasets, which may be useful for specific
applications. The main sets are delivered 
in the following formats: a Monte Carlo representation with 1000 replicas; a
Hessian set with 50 eigenvectors obtained from the Monte Carlo set via the
{\tt MC2Hessian} algorithm~\cite{Carrazza:2015aoa,Carrazza:2016htc}; and 
a compressed set of 100 Monte Carlo replicas, obtained from the original 1000
through the {\tt Compressor} algorithm~\cite{Carrazza:2015hva}  as implemented
in the new {\tt Python} code of Ref.~\cite{Carrazza:2021hny}. The final
NNPDF4.0 NNLO PDFs are shown in Fig.~\ref{fig:pdg} both at a low ($Q=3.2$ GeV)
and a high ($Q=100$ GeV) scale.

More importantly, the full NNPDF software framework is released as an open
source package~\cite{nnpdfcode}. This includes the full dataset; the methodology
hyperoptimization; the PDF parametrization and optimization; the computation of
physical processes; the set of validation tools; and the suite of
visualization tools. The code and the
corresponding  documentation are discussed in a companion
paper~\cite{NNPDF:2021uiq}.

The structure of this paper is the following.
First, in Sect.~\ref{sec:datatheory} we present the input experimental
data and the associated theoretical calculations that will be used in our
analysis, with emphasis on the new datasets added in comparison to NNPDF3.1.
Then in Sect.~\ref{sec:methodology} we discuss the fitting
methodology, in particular the parametrization of
PDFs in terms of neural networks, their training,
and the algorithmic determination of their  hyperparameters.
The procedure adopted to select the NNPDF4.0 baseline dataset is described
in Sect.~\ref{sec:dataselection}.
The main result of this work, the NNPDF4.0
determination of parton distributions, is presented
in Sect.~\ref{sec:results}, where we also compare with
previous NNPDF releases and with other PDF sets.
The closure test and future test used to validate the methodology are
described in Sect.~\ref{sec:closure}.

Subsequently, we assess the dependence of our PDFs on the 
dataset in Sect.~\ref{sec:dataset}, where
we study the impact of new data in comparison with NNPDF3.1,
and verify the impact of individual processes by studying PDF
determinations in which data corresponding to individual classes of
processes are removed in turn. Also, we present PDFs determined by
adding
specific datasets, such 
as the EMC charm structure function, the NOMAD neutrino
dimuon structure functions, and the HERA DIS jet data.
Then in Sect.~\ref{sec:tests}
we assess the dependence of PDFs on the methodology and verify the
robustness of our results, by comparing with
PDFs obtained using the previous NNPDF3.1 methodology and 
by studying the impact of
new positivity and integrability
constraints, checking the
independence of results of the choice of PDF parametrization, discussing the
impact of independently parametrizing the charm PDF, and studying
the role of nuclear corrections.
We finally present a first  assessment of the implications of NNPDF4.0 for LHC
phenomenology in Sect.~\ref{sec:pheno}, by computing PDF luminosities,
fiducial cross-sections, and differential distributions for representative
processes. In Sect.~\ref{sec:summary} we summarize and list the
NNPDF4.0 grid files that are made available through the {\tt LHAPDF}
interface~\cite{Buckley:2014ana} and provide a  summary and outlook.

A brief overview of the NNPDF fitting code is presented in
App.~\ref{sec:nnpdffitter}, while a more extensive description is
provided by the companion publication~\cite{NNPDF:2021uiq}.
In App.~\ref{app:datacomp} we 
compare the NNPDF4.0 dataset to that adopted in
other PDF determinations.

%% file: sec-expdata.tex
\section{Experimental and theoretical input}
\label{sec:datatheory}

We present the NNPDF4.0 dataset in detail. After a general overview, we examine
each of the processes for which new measurements are considered in NNPDF4.0,
we present the details of the measurements, and, for each dataset, we describe
how the corresponding theoretical predictions are obtained. In
NNPDF4.0, theoretical predictions for data taken on nuclear targets are
supplemented by nuclear corrections, which are specifically discussed
in a dedicated section. Experimental statistical and systematic
uncertainties are treated as in previous NNPDF determinations: see in
particular Sect.~2.4.2 of Ref.~\cite{Ball:2014uwa} for a detailed discussion.

The global dataset presented in this section
is the basis for the final NNPDF4.0 dataset, which
will be selected from it by applying criteria based on
testing for dataset consistency and compatibility, and for
perturbative stability upon the inclusion of electroweak
corrections. The selection of the final dataset will be discussed in
Sect.~\ref{sec:dataselection} below.

\subsection{Overview of the NNPDF4.0 dataset}
\label{subsec:dataset_overview}

The NNPDF4.0 dataset includes essentially all the data already
included in NNPDF3.1, the only exceptions being a few datasets that
are replaced by a more recent final version, and single-inclusive jet
datasets which are 
now partly replaced by dijet data, as we discuss below.
All the new datasets that were not included in NNPDF3.1
are more extensively discussed in
Sect.~\ref{subsec:dataset_extension}. For all those already included
in NNPDF3.1 we refer to Sect.~2 of Ref.~\cite{Ball:2017nwa} for a
detailed discussion. Nevertheless we give a summary below.

The NNPDF3.1 dataset included data for lepton-nucleon,
neutrino-nucleus, proton-nucleus 
and proton-(anti)proton scattering processes. The bulk of it consisted
of deep inelastic scattering (DIS) measurements: these included fixed-target
neutral current (NC) structure function data from
NMC~\cite{Arneodo:1996kd,Arneodo:1996qe},
SLAC~\cite{Whitlow:1991uw} and BCDMS~\cite{Benvenuti:1989rh}, fixed-target
inclusive and dimuon charged current (CC) cross-section data from
CHORUS~\cite{Onengut:2005kv} and NuTeV~\cite{Goncharov:2001qe,MasonPhD}, and
collider NC and CC cross-section data from the HERA legacy
combination~\cite{Abramowicz:2015mha}. The combined H1 and ZEUS measurement
of the charm cross-section~\cite{Abramowicz:1900rp} and the separate
H1~\cite{Aaron:2009af} and ZEUS~\cite{Abramowicz:2014zub} measurements of the
bottom cross-section were also included, both to be replaced by more
recent data as we discuss below. The charm structure function
measured by the EMC experiment~\cite{Aubert:1982tt} was also
studied in a variant
fit, in which its constraining power on the intrinsic component of the charm
PDF was explicitly assessed, and the same will be done here.

In addition to the DIS measurements, the NNPDF3.1 dataset
included fixed-target DY data from the Fermilab E605~\cite{Moreno:1990sf}
and E866~\cite{Webb:2003ps,Towell:2001nh} experiments, inclusive gauge boson
production~\cite{Aaltonen:2010zza,Abazov:2007jy,Abazov:2013rja,D0:2014kma}
and single-inclusive jet production~\cite{Abulencia:2007ez}
cross-section data from the Tevatron.
A sizable amount of LHC data were also
included, 
specifically: inclusive gauge boson production data from
ATLAS~\cite{Aad:2011dm,Aaboud:2016btc,Aad:2014qja,Aad:2013iua},
CMS~\cite{Chatrchyan:2012xt,Chatrchyan:2013mza,Chatrchyan:2013tia,
  Khachatryan:2016pev}
and LHCb~\cite{Aaij:2012mda,Aaij:2015gna,Aaij:2015vua,Aaij:2015zlq};
$Z$-boson transverse momentum production data from ATLAS~\cite{Aad:2015auj} and
CMS~\cite{Khachatryan:2015oaa}; 
and top pair production total and differential cross-section data
from ATLAS~\cite{Aad:2014kva,Aaboud:2016pbd,Aad:2015mbv}
and
CMS~\cite{Khachatryan:2016mqs,Khachatryan:2015uqb,Khachatryan:2015oqa}. 
Single-inclusive jet production data from
ATLAS~\cite{Aad:2011fc,Aad:2013lpa,Aad:2014vwa}
and CMS~\cite{Chatrchyan:2012bja,Khachatryan:2015luy} were also
included. These will be partly replaced by dijet data as we discuss
below.  
For the determination of NLO PDFs, $W$ production measurements in
association with a charm jet from CMS~\cite{Chatrchyan:2013uja} were
also included.
Most of these LHC measurements were performed at
$\sqrt{s}=7$~TeV~\cite{Aad:2011dm,Aaboud:2016btc,Aad:2014qja,Aad:2013iua,
  Chatrchyan:2012xt,Chatrchyan:2013mza,Chatrchyan:2013tia,
  Aaij:2012mda,Aaij:2015vua,Aaij:2015gna,
  Aad:2011fc,Aad:2014vwa,Chatrchyan:2012bja,
  Aad:2014kva,Khachatryan:2016mqs,
  Chatrchyan:2013uja};
two single-inclusive jet measurements were performed at
$\sqrt{s}=2.76$~TeV~\cite{Aad:2013lpa,Khachatryan:2015luy};
two gauge boson production
measurements~\cite{Khachatryan:2016pev,Aaij:2015zlq}, the $Z$-boson transverse
momentum measurements~\cite{Aad:2015auj,Khachatryan:2015oaa}
and some top pair production measurements~\cite{Aad:2015mbv,Aad:2014kva,
  Khachatryan:2015oqa,Khachatryan:2016mqs}
were performed at $\sqrt{s}=8$~TeV; and two top pair total cross-section
measurements~\cite{Aaboud:2016pbd,Khachatryan:2015uqb}
were performed at $\sqrt{s}=13$~TeV.

The NNPDF4.0 dataset builds upon NNPDF3.1, by adding
various new datasets to it. On the one hand,
a variety of new LHC measurements for processes
already present in NNPDF3.1, on the other hand
 data corresponding to new processes.
New datasets for existing LHC processes are added for electroweak boson
production, both inclusive and in association with charm, single-inclusive jet
production, and top pair production. The new processes are gauge boson with
jets, single top production, inclusive isolated photon production, and dijet
production. 

For inclusive electroweak boson production we consider: at
$\sqrt{s}=7$~TeV, the ATLAS $W$ and $Z$ distributions~\cite{Aaboud:2016btc} in
the central and forward rapidity regions (only the subset corresponding to the
central region was included in NNPDF3.1); at $\sqrt{s}=8$~TeV, the ATLAS $Z$
double- and triple-differential distributions~\cite{Aad:2016zzw,Aaboud:2017ffb},
the ATLAS $W$ differential distribution~\cite{Aad:2019rou}
and the LHCb $W$ differential distribution~\cite{Aaij:2016qqz};
at $\sqrt{s}=13$~TeV, the ATLAS $W$ and $Z$ total
cross-section~\cite{Aad:2016naf} and the LHCb $Z$ differential
distributions~\cite{Aaij:2016mgv}. For electroweak gauge boson production
with charm, we consider the ATLAS~\cite{Aad:2014xca} and
CMS~\cite{Sirunyan:2018hde} differential distributions at $\sqrt{s}=7$~TeV and
$\sqrt{s}=8$~TeV, respectively. Given that the corresponding NNLO QCD
corrections are not available in a format suitable for inclusion in a
fit~\cite{Czakon:2020coa}, these two
datasets are included only in the determination of NLO PDFs.

For single-inclusive
jet production we consider the ATLAS~\cite{Aaboud:2017dvo} and
CMS~\cite{Khachatryan:2016mlc} double differential cross-sections at
$\sqrt{s}=8$~TeV. For top  pair production we consider:
at $\sqrt{s}=5.02$~TeV, the CMS total cross-section~\cite{Sirunyan:2017ule}; 
at $\sqrt{s}=8$~TeV, the ATLAS differential distributions~\cite{Aaboud:2016iot}
and the CMS double differential distributions~\cite{Sirunyan:2017azo}, both of
which are measured in the dilepton final state; at $\sqrt{s}=13$~TeV,
the CMS differential distributions measured in the
lepton+jets~\cite{Sirunyan:2018wem} and in the dilepton~\cite{Sirunyan:2018ucr}
final states. For $W$-boson production with  jets
we consider the ATLAS differential distributions at
$\sqrt{s}=8$~TeV~\cite{Aaboud:2017soa}. For single top production,
we consider only measurements in the $t$-channel, specifically: at
$\sqrt{s}=7$~TeV, the ATLAS top to antitop total cross-section
ratio, with the corresponding differential distributions~\cite{Aad:2014fwa} and
the CMS combined top and antitop total cross-sections~\cite{Chatrchyan:2012ep}; at $\sqrt{s}=8$~TeV, the
ATLAS~\cite{Aaboud:2017pdi} and CMS~\cite{Khachatryan:2014iya} top
to antitop total cross-section ratios and the ATLAS differential
distributions~\cite{Aaboud:2017pdi}; at $\sqrt{s}=13$~TeV the
ATLAS~\cite{Aaboud:2016ymp} and CMS~\cite{Sirunyan:2016cdg} top
to antitop cross-section ratios. For inclusive isolated photon
production we consider the ATLAS differential cross-sections at
$\sqrt{s}=8$~TeV~\cite{Aad:2016xcr} and at
$\sqrt{s}=13$~TeV~\cite{ATLAS:2017nah}. For  dijet production we
consider, at $\sqrt{s}=7$~TeV, the ATLAS~\cite{Aaboud:2017dvo}
and CMS~\cite{Chatrchyan:2012bja} double differential distributions
and, at $\sqrt{s}=8$~TeV, the CMS triple differential
distributions~\cite{Khachatryan:2016mlc}.

Additional LHC measurements at
$\sqrt{s}=13$~TeV  for processes relevant to PDF
determination are in principle available: specifically,
the ATLAS~\cite{Aad:2019wmn} and CMS~\cite{Sirunyan:2018owv}
$Z$ transverse momentum distributions; the CMS $W$+jets
distributions~\cite{Sirunyan:2017wgx}; the ATLAS~\cite{Aaboud:2017wsi}
and CMS~\cite{Khachatryan:2016wdh} single-inclusive jet distributions;
and the ATLAS~\cite{Aad:2019ntk} and LHCb~\cite{Aaij:2018imy} top pair
distributions. We do not include these measurements because either
they are first analyses based on a still reduced
luminosity sample, or because they do not come
with complete information on experimental uncertainties, or
because NNLO QCD corrections are not yet available.

The non-LHC dataset is also expanded in NNPDF4.0. For DIS,
we now also consider the dimuon to inclusive cross-section ratio measured by
the NOMAD experiment~\cite{Samoylov:2013xoa}, though only in a variant
determination, see Sect.~\ref{subsubsec:NOMAD}. We also consider a selection of
differential cross-sections for single-inclusive and dijet production in DIS
measured by ZEUS~\cite{ZEUS:2002nms,ZEUS:2006xvn,ZEUS:2010vyw} and
H1-HeraII~\cite{H1:2016goa,H1:2014cbm}, again only in a variant determination
that will be discussed in Sect.~\ref{subsubsec:HERAjets}.
For fixed-target DY, we include the recent measurement
for the proton-deuteron to proton-proton differential cross-section ratio
performed by the E906/SeaQuest experiment~\cite{Dove:2021ejl}.

The theoretical treatment of the data already included in NNPDF3.1 is
the same in all respects as in that analysis, to which we refer for details. The
general NNPDF3.1 settings will in fact be adopted throughout, with
specific aspects relevant for the new data to be discussed in
Sect.~\ref{subsec:dataset_extension} below.  Fast interpolation grids, 
accurate to NLO in perturbative QCD, are produced in the {\tt APFELgrid}
format~\cite{Bertone:2016lga}; {\tt APFEL}~\cite{Bertone:2013vaa}
and various fixed-order Monte Carlo event
generators~\cite{Campbell:1999ah,Campbell:2011bn,Campbell:2015qma,
  Bothmann:2019yzt,Frederix:2018nkq,Alwall:2014hca,Nagy:2001fj}
(possibly interfaced to {\tt APPLgrid}~\cite{Carli:2010rw}
or {\tt FastNLO}~\cite{Kluge:2006xs,Wobisch:2011ij,Britzger:2012bs}
with {\tt MCgrid}~\cite{DelDebbio:2013kxa,Bothmann:2015dba} or
{\tt aMCfast}~\cite{Bertone:2014zva})
are utilized for the computation of DIS and non-DIS observables, respectively.
The charm PDF is parametrized by default and the FONLL general-mass variable
flavor number scheme~\cite{Forte:2010ta,Ball:2015tna,
  Ball:2015dpa} is utilized to compute DIS structure functions.

Except for DIS and for DIS jets, for which we also make use of
NNLO fast interpolation
grids, NNLO QCD corrections to matrix elements are 
implemented  by multiplying the NLO predictions by a $K$-factor. This is
defined as the bin-by-bin ratio of the NNLO to NLO prediction computed with a
pre-defined NNLO PDF set (see Sect.~2.3 in~\cite{Ball:2014uwa} for details).
For all of the fixed-target DY data and for all of the new LHC datasets
considered in NNPDF4.0, this PDF set is
{\sc NNPDF3.1\_nnlo\_as\_0118}~\cite{Ball:2017nwa}; for the
Tevatron and LHC datasets already included in NNPDF3.1, we used the same PDF
sets specified in Sect.~2.1 of~\cite{Ball:2017nwa}. For these datasets the PDF
dependence of the $K$-factors is generally smaller than all the other relevant
uncertainties, as explicitly shown in~\cite{Ball:2017nwa}. We have
checked this explicitly by recomputing the $K$-factors for all of the inclusive
gauge boson production measurements, for both fixed-target and collider
experiments, and for all of the top-quark pair production measurements with
the baseline NNPDF4.0 set, and then repeating the NNLO  PDF
determination. The ensuing PDFs turn out to be statistically
equivalent to the  NNPDF4.0 baseline.
The values of all  physical parameters are the same as
in NNPDF3.1.

The NNPDF4.0 dataset is thus a superset of  NNPDF3.1 with
the following exceptions. First, in the NNPDF4.0 baseline the
single-inclusive jet data are replaced by their dijet counterparts
(though the single-inclusive jet data will be considered in a variant
NNPDF4.0 determination, see
Sect.~\ref{subsubsection:singleinclusivejets_7TeV}
below). Furthermore, a number of small alterations is made to the original
set of NNPDF3.1 data, or to their theoretical treatment, as we now
discuss.

In terms of data, the
total cross-section results from Ref.~\cite{Aaboud:2016pbd} are no
longer used, as they are  replaced by the  more recent
measurement~\cite{Aad:2020tmz} based on the full Run II luminosity, to
be discussed in Sect.~\ref{subsubsec:toppair} below. For the differential 
distributions
measured by ATLAS at $\sqrt{s}=8$~TeV in the lepton+jets final
state~\cite{Aad:2015mbv} only one distribution out of the
four available was included in  NNPDF3.1 while all of them  are
included in NNPDF4.0, because the correlations between distributions
have become available meanwhile.
The single-inclusive jet
measurements from ATLAS~\cite{Aad:2013lpa} and
CMS~\cite{Khachatryan:2015luy} at $\sqrt{s}=2.76$~TeV and from
ATLAS~\cite{Aad:2011dm} at $\sqrt{s}=7$~TeV are no longer included
in NNPDF4.0 because 
NNLO QCD corrections, which are provided with the optimal scale choice of
Ref.~\cite{Currie:2018xkj}, 
are not
available for these measurements. For the  same reason the
CDF single-inclusive jet data~\cite{Abulencia:2007ez}
are also not included. These datasets were already removed in
intermediate updates of the NNPDF3.1
determination~\cite{Nocera:2019wyk,Faura:2020oom}
or in subsequent studies~\cite{Ball:2018iqk,AbdulKhalek:2019ihb,
  AbdulKhalek:2019bux,Ball:2020xqw}.

In terms of theoretical  treatment the changes are the following. For  DIS we correct a bug in the
{\sc APFEL} computation of the NLO CC structure functions, that
mostly affects the large-$x$ region; and we re-analyze the NuTeV dimuon
cross-section data by including the NNLO charm-quark massive
corrections~\cite{Gao:2017kkx,Berger:2016inr}, as explained
in~\cite{Faura:2020oom}, and by updating the value of the branching ratio of
charmed hadrons into muons to the PDG value~\cite{Zyla:2020zbs},
as explained in~\cite{Ball:2018twp}. For fixed-target DY, we include the
NNLO QCD corrections for the E866 measurement~\cite{Towell:2001nh} of the
proton-deuteron to proton-proton cross-section ratio: these corrections had
been inadvertently overlooked in NNPDF3.1.  For gauge boson production
at the Tevatron, we correct a small bug affecting the CDF $Z$ rapidity
distribution~\cite{Aaltonen:2010zza}, whereby the last two bins had not been
merged consistently with the updated measurement.
For jets, we update the 
theoretical treatment of the single-inclusive jet measurements at
$\sqrt{s}=7$~TeV~\cite{Aad:2014vwa,Chatrchyan:2012bja},
in that NLO and NNLO theoretical predictions are now computed with
factorization and renormalization scales equal to the optimal scale choice
advocated in Ref.~\cite{Currie:2018xkj}, namely, the scalar sum of
the transverse momenta of all partons in the event, 
see Ref.~\cite{AbdulKhalek:2020jut}.

To assess the impact of these changes in dataset and theoretical treatment,
we will consider a variant of  NNPDF3.1 in which all of these changes,
but not the replacement of single-inclusive jets with dijets, are
taken into account.
This determination will be referred to as
NNPDF3.1-like henceforth. It will be used to carry out various
methodological tests in Sects.~\ref{sec:methodology} and~\ref{sec:closure}.
The NNPDF3.1-like determination contains 4092 data points for a NNLO fit.

The data included in NNPDF4.0 are summarized
in Tables~\ref{tab:DIS_dataset}, \ref{tab:DISJETS_dataset},
\ref{tab:FTDY_dataset}, \ref{tab:collider_dataset_1}
and~\ref{tab:collider_dataset_2},
respectively for DIS, DIS jets, fixed-target DY, collider inclusive gauge boson
production and other LHC processes. For each process we indicate the
name of the dataset used throughout this paper, the corresponding reference,
the number of data points in the NLO/NNLO fits before (and after) 
kinematic cuts (see Sect.~\ref{sec:dataselection}), the kinematic coverage in
the relevant variables after cuts, and the codes used to compute the
corresponding predictions. Datasets not previously considered in NNPDF3.1 are
indicated with an asterisk. Datasets not included in the baseline determination
are indicated in brackets.

The total number of data points included in the default PDF determination is
4426 at NLO and 4618 at NNLO, to be compared to 4295 at NLO 4285 at NNLO in
NNPDF3.1 and to 4092 (at NNLO) in NNPDF3.1-like fits presented here. 
A comparison between the datasets considered in
NNPDF4.0 and the datasets included in NNPDF3.1 and in other recent PDF
determinations, namely ABMP16~\cite{Alekhin:2017kpj}, CT18~\cite{Hou:2019efy}
and MSHT20~\cite{Bailey:2020ooq}, is presented in App.~\ref{app:datacomp},
see Tables~\ref{tab:dataset_noLHC}-\ref{tab:dataset_LHCb}.

\begin{table}[!t]
  \scriptsize
  \centering
  \renewcommand{\arraystretch}{1.4}
  \input{tables/tab-DIS_dataset}
  \caption{The DIS datasets analyzed in the NNPDF4.0 PDF determination.
    For each of them we indicate the name of the dataset used throughout this
    paper, the corresponding reference, the number of data points in the
    NLO/NNLO fits before (and after) kinematic cuts
    (see Sect.~\ref{sec:dataselection}), the kinematic coverage in the relevant
    variables after cuts, and the codes used to compute the corresponding
    predictions. Datasets not previously considered in NNPDF3.1 are indicated
    with an asterisk. Datasets not included in the baseline determination
    are indicated in square brackets. The $Q$ coverage indicated for NOMAD is
    to be interpreted as an integration range (see text).}
  \label{tab:DIS_dataset}
\end{table}

\begin{table}[!t]
  \scriptsize
  \centering
  \renewcommand{\arraystretch}{1.4}
  \input{tables/tab-DISJETS_dataset}
  \caption{Same as Table~\ref{tab:DIS_dataset} for DIS jet data.}
  \label{tab:DISJETS_dataset}
\end{table} 

\begin{table}[!t]
  \scriptsize
  \centering
  \renewcommand{\arraystretch}{1.4}
  \input{tables/tab-FTDY_dataset}
  \caption{Same as Table~\ref{tab:DIS_dataset} for fixed-target DY data.}
  \label{tab:FTDY_dataset}
\end{table} 

\begin{table}[!t]
  \scriptsize
  \centering
  \renewcommand{\arraystretch}{1.4}
  \input{tables/tab-GAUGEBOSON_dataset}
  \caption{Same as Table~\ref{tab:DIS_dataset} for collider (Tevatron, top, and
    LHC, bottom) inclusive gauge boson production data.}
  \label{tab:collider_dataset_1}
\end{table} 

\begin{table}[!t]
  \scriptsize
  \centering
  \renewcommand{\arraystretch}{1.4}
  \input{tables/tab-OTHERLHCPROCESSES_dataset}
  \caption{Same as Table~\ref{tab:DIS_dataset} for other LHC processes.
    From top to bottom we list: $W$-boson production in association with a jet
    of charm or of light quarks; $Z$-boson transverse momentum production;
    total and differential top pair production; single-inclusive and
    dijet production; inclusive isolated photon production; and single
    top $t$-channel total and differential production.}
  \label{tab:collider_dataset_2}
\end{table} 

\begin{figure}[!p]
  \centering
  \includegraphics[width=0.9\textwidth]{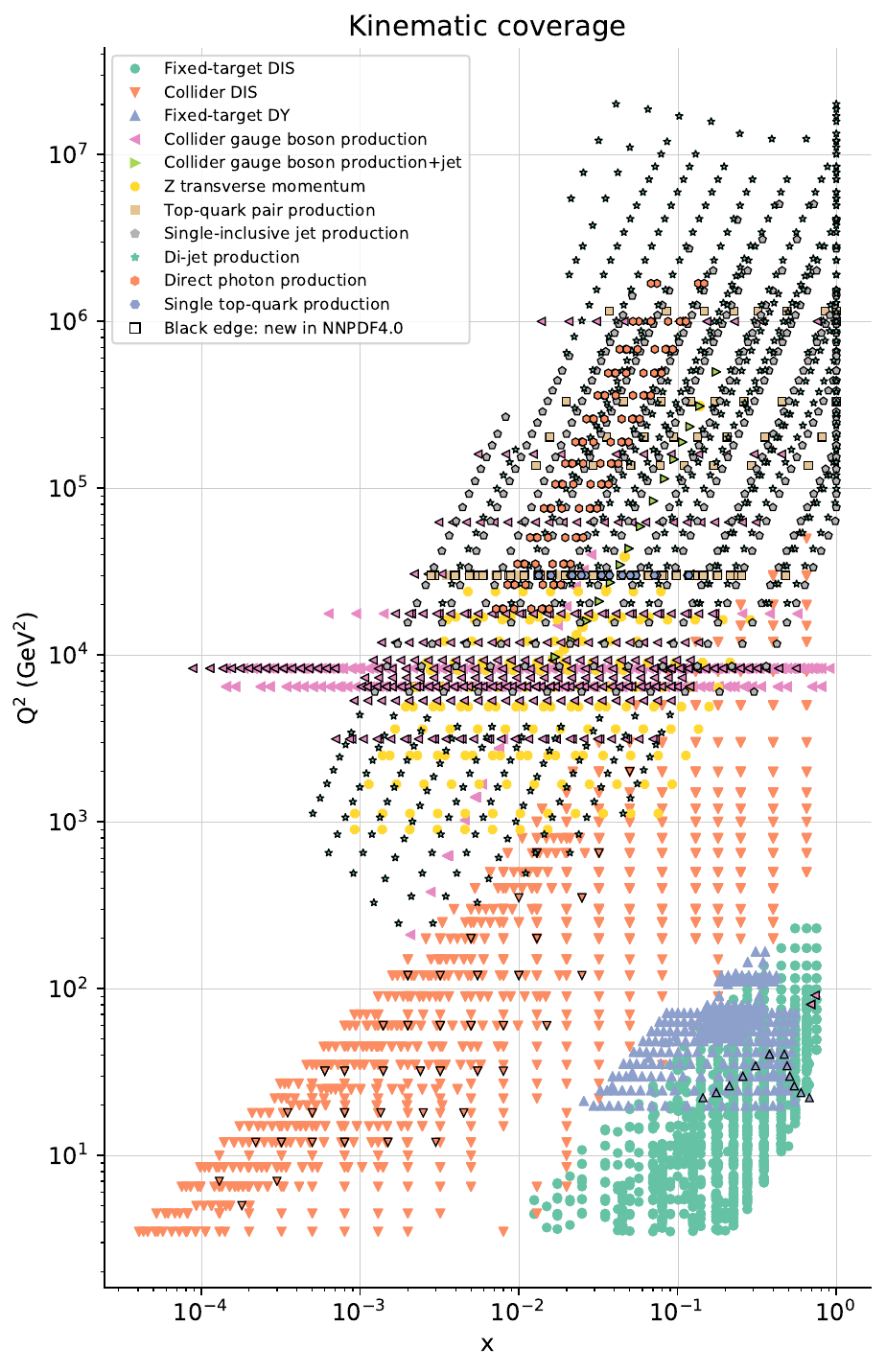}\\
  \caption{The kinematic coverage of the NNPDF4.0 dataset in the $(x,Q^2)$
    plane.}
   \label{fig:kinplot}
\end{figure}

The kinematic coverage in the $(x,Q^2)$ plane of the NNPDF4.0 dataset 
entering the default NNLO fit is displayed in Fig.~\ref{fig:kinplot}. 
For hadronic
data, kinematic variables are determined using LO kinematics. Whenever
an observable is integrated over rapidity, the center of the integration range
is used to compute the values of $x$. The data points
corresponding to datasets that are new in NNPDF4.0 are indicated with a black
edge.

The complete information on experimental uncertainties, including the breakdown
into different sources of systematic uncertainties and their correlations, is
taken into account whenever available from the corresponding publications or
from the {\sc HEPData} repository~\cite{Maguire:2017ypu}. No decorrelation
models are used, except when explicitly recommended by the collaboration.  This is the case of
the single-inclusive jet cross-section measurement performed by ATLAS at
$\sqrt{s}=8$~TeV~\cite{Aaboud:2017dvo}. Decorrelation
models~\cite{Harland-Lang:2017ytb,AbdulKhalek:2020jut,
  Bailey:2019yze,ATLAS:2018owm,Amoroso:2020lgh} were
studied for the ATLAS jet measurements at $\sqrt{s}=7$~TeV~\cite{Aad:2014vwa}
and for the ATLAS top pair measurements
at $\sqrt{s}=8$~TeV~\cite{Aad:2015mbv}. However these are not considered in
our default determination, but only in variant fits, see Sect.~\ref{sec:regcovmat}.

\subsection{New data in NNPDF4.0}
\label{subsec:dataset_extension}

We now discuss in detail the new datasets considered in NNPDF4.0.
These are indicated with an asterisk in
Tables~\ref{tab:DIS_dataset}-\ref{tab:collider_dataset_2}. The data are 
presented by process, with the processes already considered in NNPDF3.1
addressed first.

\subsubsection{Deep-inelastic scattering}
\label{subsubsection:DIS}
We include the combined H1 and ZEUS measurements of reduced
electron-proton NC DIS cross-sections for the production of open charm and
bottom quarks~\cite{H1:2018flt}. These measurements extend the previous
combination of open charm production cross-sections~\cite{Abramowicz:1900rp}
and supersede the separate H1~\cite{Aaron:2009af}
and ZEUS~\cite{Abramowicz:2014zub} datasets
for the structure function $F_2^b$ that were included in NNPDF3.1.
As for the other DIS measurements included in the NNPDF4.0
dataset, they are analyzed in the FONLL scheme~\cite{Forte:2010ta,Ball:2015tna,
  Ball:2015dpa} within fixed order perturbative accuracy (i.e. not
including resummation).

We also consider the measurements of the ratio $\mathcal{R}_{\mu\mu}$ of dimuon
to inclusive neutrino-nucleus CC DIS cross-sections performed by the NOMAD
experiment~\cite{Samoylov:2013xoa}. These measurements are presented
alternatively as a function of the neutrino beam energy $E_\nu$, of the momentum
fraction $x$, or of the final state invariant mass $W$. Because experimental
correlations are not provided among the three distributions, they cannot be
included in the fit at the same time. We therefore select only one of them,
namely the measurement as a function of the neutrino beam energy,
the only variable among the three that is directly measured by the experiment.
This choice is based on the  previous study~\cite{Faura:2020oom}, carried out in
the context of a variant of the NNPDF3.1 determination, in which it was shown
that the three distributions have a similar impact in the fit. 

The treatment of this
dataset in NNPDF4.0 closely follows Ref.~\cite{Faura:2020oom}. Specifically we
incorporate the recently computed NNLO charm-quark massive
corrections~\cite{Berger:2016inr,Gao:2017kkx}  by means of a $K$-factor
(see Sect.~2.2.2 in~\cite{Faura:2020oom}).
The NOMAD data are not included in our default determination, however
we assess its impact on the NNLO PDFs by means of
Bayesian reweighting~\cite{Ball:2010gb,Ball:2011gg}. The reason for this choice
is dictated by the fact that the observable is integrated over $Q$ and $x$
(see {\it e.g.} Eq.~(2.1) in Ref.~\cite{Faura:2020oom}), which complicates
the generation of fast interpolation tables in the {\tt APFELgrid} format.

\subsubsection{Jet production in deep-inelastic scattering}
\label{subsubsection:DISjet}
We consider a selection of DIS single-inclusive jet (1j) and dijet production
(2j) cross-sections measured by
ZEUS~\cite{ZEUS:2002nms,ZEUS:2006xvn,ZEUS:2010vyw}
in the high-$Q$ (HQ) region and by H1-HeraII~\cite{H1:2016goa,H1:2014cbm}
in the HQ and low-$Q$ (LQ) regions. Specifically we consider
cross-sections double differential in $Q^2$ and in the transverse momentum
of the jet or of the jet pair, listed in Table~\ref{tab:DISJETS_dataset}.
Experimental correlations between single-inclusive jet and dijet measurements,
which are available only for H1, are taken into account. These allow us to
include single-inclusive jet and dijet datasets simultaneously. Additional
available measurements, in particular from
H1-HeraI~\cite{H1:2007xjj,H1:2010mgp}, are left for future studies. Likewise, variants of the
H1-HeraII measurements~\cite{H1:2016goa,H1:2014cbm}, in which  cross-sections are
normalized to the inclusive NC cross-section integrated over the width of each
$Q^2$ bin, are not yet considered. These normalized cross-sections might
benefit from cancellations of systematic uncertainties and uncertainty
correlation with HERA inclusive DIS measurements.

Theoretical predictions for the ZEUS and H1-HeraII datasets are obtained 
using fast interpolation grids precomputed with {\tt NNLOjet}. These
incorporate the recently determined NNLO QCD
corrections~\cite{Britzger:2019kkb}. Multiplicative hadronization correction
factors, as provided in the experimental analyses, are included throughout.
Because this theoretical input has become available only very recently, the
ZEUS and H1-HeraII datasets are not included in our default determination,
but only in a variant NNLO set by means of Bayesian reweighting, see
Sect.~\ref{subsubsec:HERAjets}.

\subsubsection{Fixed-target Drell-Yan production}
\label{subsubsec:FTDY}
We consider the new measurement recently performed by the SeaQuest experiment
at Fermilab~\cite{Dove:2021ejl} for
production of a $Z$ boson decaying into muon pairs. Like
the previous NuSea measurement~\cite{Towell:2001nh}, which was
included in the 
NNPDF3.1 dataset, the SeaQuest experiment measures the ratio of the scattering
cross-section of a proton beam
off a deuterium target to  the  cross-section off
a proton target. The measurement is double differential in the partonic
momentum fractions of the struck partons. The
SeaQuest data extend the NuSea data to larger values of $x$,
$0.15\lesssim x \lesssim 0.40$, with the aim of constraining
the antiquark asymmetry in this region~\cite{Towell:2001nh}.
Theoretical predictions are computed by taking into account acceptance
corrections, according to Eq.~(10) in Ref.~\cite{Dove:2021ejl}.
Fast interpolation tables accurate to NLO are generated with {\tt APFEL}; these
are then supplemented with a NNLO $K$-factor computed with a version
of {\tt Vrap}~\cite{Anastasiou:2003ds} that we modified to account for the
isoscalarity of the deuteron target. Nuclear effects are taken into account
by means of the procedure discussed in Ref.~\cite{Ball:2020xqw} and further
summarized in Sect.~\ref{subsec:nuclear}.

\subsubsection{Inclusive collider electroweak gauge boson production}
\label{subsubsec:colliderEW}
The new datasets we consider for inclusive $W$ and $Z$ boson production
and decay are from the ATLAS and LHCb experiments.

We include the ATLAS measurements of the $W$ and $Z$  differential cross-section at
$\sqrt{s}=7$~TeV~\cite{Aaboud:2016btc} in the central and forward rapidity
regions. As mentioned above, these data were already included in
NNPDF3.1, but only the  subset corresponding to the central region.
The measurements cover, respectively, the pseudo-rapidity
range $|\eta_\ell|<2.5$ (for $W$ bosons) and the rapidity range of the
lepton pair $|y_{\ell\ell}|<3.6$ (for the $Z$ boson). In the latter case,
the invariant mass of the lepton pair is $46\le m_{\ell\ell}\le 150$~GeV.
The measurements correspond to an integrated luminosity of 4.6~fb$^{-1}$. We
consider the combination of measurements in the electron and muon decays.

We consider  the ATLAS measurements of the double and
triple differential DY lepton pair
production cross-section  at $\sqrt{s}=8$~TeV~\cite{Aad:2016zzw,Aaboud:2017ffb}.
The differential variables are the invariant mass and rapidity of the lepton
pair, $m_{\ell\ell}$ and $y_{\ell\ell}$, and, in addition to these for the latter
case, the cosine of the Collins-Soper angle $\cos\theta^*$. The measurements
cover  two separate invariant mass ranges, respectively
$116\le m_{\ell\ell}\le 1500$~GeV and $46\le m_{\ell\ell}\le 200$~GeV,
in the same central rapidity range $|y_{\ell\ell}|<2.4$. The
same data sample corresponding to an integrated luminosity of 20.2~fb$^{-1}$
is  used in the two cases, which therefore overlap in the interval
$116\le m_{\ell\ell}\le 200$~GeV. The two analyses are consistent in this region,
however because the one in~\cite{Aad:2016zzw} is optimized to high
invariant masses, we remove the overlapping bins from the dataset
in~\cite{Aaboud:2017ffb}. In both cases we consider the measurements in which
the electron and muon decay channels have been combined; for the
triple differential distribution, we consider the measurement
integrated over $\cos\theta^*$ in order to reduce  
sensitivity to the value of the Weinberg angle $\sin^2\theta_W$.

We include  the ATLAS measurement of the $W$ production cross-section
and decay at $\sqrt{s}=8$~TeV~\cite{Aad:2019rou}.
The data are differential in the pseudo-rapidity of the decay muon
$\eta_\mu$, which is accessed in the central pseudo-rapidity
range $|\eta_\mu|<2.4$ by analyzing a data sample corresponding to an integrated
luminosity of 20.2~fb$^{-1}$. As for the companion ATLAS measurement at
$\sqrt{s}=7$~TeV~\cite{Aaboud:2016btc}, we consider the separate $W^+$
and $W^-$ differential distributions rather than their asymmetry.

We consider  the ATLAS measurement of the total  $W$ and $Z$  cross-section
and decay into leptons at
$\sqrt{s}=13$~TeV~\cite{Aad:2016naf}. The measurement corresponds to an
integrated luminosity of 81~pb$^{-1}$.

We include  the LHCb measurement of the  $W$  cross-section at  $\sqrt{s}=8$~TeV~\cite{Aaij:2016qqz}. The data are
differential in the pseudo-rapidity of the decay electron $\eta_e$, which is
accessed in the forward range $2.00<|\eta_e|<4.25$. The data sample
corresponds to an integrated luminosity of 2~fb$^{-1}$. In this case,
we cannot consider the separate $W^+$
and $W^-$ differential distributions, because we find that
the correlated experimental uncertainties lead to a  covariance matrix
that is not positive definite. Therefore, in this case
we make use of the  asymmetry measurement,
which is not affected by this problem since  most of the correlations
cancel out.

Finally, we include the  LHCb measurement of the $Z$  cross-section at
$\sqrt{s}=13$~TeV~\cite{Aaij:2016mgv}. The data are differential in the
$Z$ boson rapidity $y_Z$~\cite{Aaij:2016mgv}, with $2.00<|y_Z|<4.50$, and
it covers the $Z$-peak lepton pair invariant mass range
$60\le m_{\ell\ell}\le 120$~GeV. The data sample corresponds to
an integrated luminosity of 294~pb$^{-1}$. We include separately the 
datasets in the dimuon and dielectron decay channels.

These datasets, specifically from ATLAS, are
particularly precise, with systematic uncertainties of the order of percent or
less and even smaller statistical uncertainties. They are dominated
by the luminosity uncertainty, which is of the order of 1.9-2.1\% (1.2-3.9\%) 
for ATLAS (LHCb) respectively at $\sqrt{s}=8$~TeV and at $\sqrt{s}=13$~TeV.

Theoretical predictions are computed at NLO with {\tt MCFM}
(v6.8)~\cite{Campbell:1999ah,Campbell:2011bn,Campbell:2015qma} and are
 benchmarked against those obtained with
{\tt mg5\_aMC} (v3.1)~\cite{Frederix:2018nkq,Alwall:2014hca}.
The NNLO $K$-factor is computed with
{\tt FEWZ} (v3.1)~\cite{Gavin:2010az,Gavin:2012sy,Li:2012wna}
for all the datasets excepting those of~\cite{Aaboud:2017ffb,Aad:2019rou}, for which
{\tt DYNNLO}~\cite{Catani:2007vq,Catani:2009sm} is used instead.
We benchmarked these calculations  against {\tt MCFM} (v9.0)~\cite{Campbell:2019dru},
and found the relative difference between different computations to
be negligible in comparison to the data uncertainties. The renormalization and
factorization scales are set equal to the mass of the gauge boson, for total
cross-sections and for cross-sections differential in rapidity or pseudorapidity
variables, or to the central value of the corresponding invariant mass bin, for
cross-sections that are also differential in the invariant mass of the lepton
pair.

\subsubsection{Gauge boson production with additional jets}
\label{subsubsec:EW_collider_rad}
On top of inclusive gauge boson production, we consider more exclusive
measurements in which a $W$ boson is produced in association with $N_{\rm jets}$
jets of light quarks, or with a single jet of charm quarks.

Specifically, we include the ATLAS data for  $W$ production 
with $N_{\rm jets}\geq 1$~\cite{Aaboud:2017soa} at
$\sqrt{s}=8$~TeV. The measurement corresponds to an integrated luminosity of
20.2~fb$^{-1}$. We select the distribution differential in the transverse
momentum of the $W$ boson, $p_T^W$, which covers
the range $0\leq p_T^W\leq 800$~GeV. Theoretical predictions are determined
as in the ATLAS study of~\cite{ATLAS:2021qnl}: at NLO, fast interpolation grids
are generated with {\tt MCFM}; at NNLO, QCD corrections are implemented by
means of $K$-factors determined with the {\tt $N_{\rm jetti}$}
program~\cite{Boughezal:2015dva,Ridder:2015dxa}. The factorization and
renormalization scales are set equal to the mass of the $W$ boson.

We further include the ATLAS~\cite{Aad:2014xca}
and CMS~\cite{Sirunyan:2018hde} data for production of  $W$ 
with a charm jet, at
$\sqrt{s}=7$~TeV and $\sqrt{s}=13$~TeV, respectively. The two measurements
correspond to integrated luminosities of 4.6~fb$^{-1}$ and 35.7~fb$^{-1}$.
In both cases, we utilize the cross-sections differential in the pseudo-rapidity
of the decay lepton $\eta_\ell$, which is accessed in the range $|\eta_\ell|<2.5$
for ATLAS and $|\eta_\ell|<2.4$ for CMS. In the case of ATLAS, separate
distributions for the production of positively and negatively charged
bosons are provided; in the case of CMS, only the distribution for the sum of
the two is available. Theoretical predictions are computed at NLO with
{\tt MCFM}; NNLO QCD corrections have been computed very
recently~\cite{Czakon:2020coa}, although in a format that does not allow for
their ready implementation. These datasets are therefore not included
in the determination of NNLO PDFs. The factorization and renormalization
scales are set equal to the mass of the $W$ boson.

All the measurements discussed in this section have been included in a PDF
determination, in a specific study based on NNPDF3.1~\cite{Faura:2020oom}.

\subsubsection{Top pair production}
\label{subsubsec:toppair}

We consider several new datasets for top pair production at the LHC. At
$\sqrt{s}=8$~TeV, we include  the ATLAS normalized differential cross-section~\cite{Aaboud:2016iot}
and the CMS normalized double differential
cross-section~\cite{Sirunyan:2017azo}, both of which are measured in the dilepton
channel. Companion measurements in the lepton+jets
channel~\cite{Aad:2015mbv,Khachatryan:2015oqa} were already part of
NNPDF3.1. These measurements correspond respectively to an integrated 
luminosity of
20.2~fb$^{-1}$ and 19.7~fb$^{-1}$. At $\sqrt{s}=8$~TeV, we include
the ATLAS total cross-section~\cite{Aad:2020tmz} and
the CMS absolute differential distributions in the
lepton+jets channel~\cite{Sirunyan:2018wem} and in the
dilepton channel~\cite{Sirunyan:2018ucr}. The ATLAS measurement
is based on the full Run~II sample, corresponding to an integrated luminosity
of 139~fb$^{-1}$ and replaces the corresponding measurement, determined from a
partial luminosity~\cite{Aaboud:2016pbd}, included in NNPDF3.1; the CMS
measurements are for an integrated luminosity of 35.8~fb$^{-1}$.

Various differential distributions are available for each of these
measurements. Because correlations between different distributions are
not available,
only one distribution at a time can be included. Rapidity distributions
are generally affected by small higher order
corrections~\cite{Czakon:2016olj}, hence we chose the rapidity of the
top quark, when available,  as our preferred observable, and otherwise, 
the rapidity of the top pair.  Specifically,
we select the distributions differential in the
rapidity of the top pair in the case of~\cite{Aaboud:2016iot}, the
double-differential distribution 
in the rapidity of the top quark and the invariant mass of the top pair
in the case of~\cite{Sirunyan:2017azo} and in the rapidity of the top quark in
the case of~\cite{Sirunyan:2018wem,Sirunyan:2018ucr}. 
We have explicitly verified that the choice of any other distributions does not
alter the results. The kinematic coverage of the
distributions that we included is shown in Table~\ref{tab:collider_dataset_2}.

Theoretical predictions are computed at NLO with {\tt mg5\_aMC}
(v2.6.6)~\cite{Alwall:2014hca}; NNLO QCD
corrections are determined from publicly available {\tt FastNLO}
tables~\cite{Czakon:2017dip,Czakon:2019yrx} for differential distributions and
from {\tt top++}~\cite{Czakon:2011xx} for the total cross-section. The
renormalization and factorization scales are set as in NNPDF3.1, see
Sect.~2.7 in~\cite{Ball:2017nwa} for details.

\subsubsection{Single-inclusive and dijet production}
\label{subsubsec:jets}
In NNPDF4.0, following the study of Ref.~\cite{AbdulKhalek:2020jut},
we consider both single-inclusive jets (as in previous NNPDF
determinations) and dijets, that have several desirable theoretical features~\cite{Currie:2018xkj}.

For  single-inclusive jet production, we include the
ATLAS~\cite{Aaboud:2017dvo} and CMS~\cite{Khachatryan:2016mlc}
measurements at $\sqrt{s}=8$~TeV. They correspond
to integrated luminosities of 20.2~fb$^{-1}$ and 19.7~$^{-1}$, respectively.
In both cases the measurements are provided for the cross-section differential
in the transverse momentum, $p_T^{\rm jet}$, and of the rapidity, $y^{\rm jet}$,
of the jet. The data cover the range $70\le p_T^{\rm jet}\le 2.5$~TeV
and $|y^{\rm jet}|\le 3.0$. Theoretical
predictions are computed at NLO with {\tt NLOJet++}
(v4.1.3)~\cite{Nagy:2001fj} and benchmarked against the independent computation
presented in~\cite{Gehrmann-DeRidder:2019ibf}. NNLO QCD corrections are
incorporated by means of the $K$-factor computed in the same publication.
The factorization and renormalization scales are set equal to the
optimal scale choice recommended in Ref.~\cite{Currie:2018xkj},
namely, the scalar sum
of the transverse momenta of all partons in the event.

For  dijet production we consider the ATLAS~\cite{Aad:2013tea}
and CMS~\cite{Chatrchyan:2012bja} measurements at
$\sqrt{s}=7$~TeV and the CMS measurement~\cite{Sirunyan:2017skj} at
$\sqrt{s}=8$~TeV. They correspond to integrated
luminosities of 4.5~fb$^{-1}$ (at 7~TeV) and of 19.7~fb$^{-1}$ (at 8~TeV).
For ATLAS, the cross-section is  double
differential in the dijet invariant mass $m_{jj}$ and in the absolute
difference of the rapidities of the two jets $y^*$. The corresponding ranges
are $260\le m_{jj}\le 4.27$~TeV and $0.0\le y^* \le 3.0$. For CMS, the
cross-section is double differential in
$m_{jj}$ and in the maximum absolute rapidity of the two jets $|y_{\rm max}|$
(at 7~TeV) and triple differential in the average transverse momentum of the
jet pair $p_{T,{\rm avg}}$, the dijet boost $y_b$, and $y^*$ (at 8~TeV).
The corresponding ranges are $133\le p_{T,{\rm avg}}\le 1.78$~TeV and
$0.0\le y_b,y^*\le 3.0$. As in the case of single-inclusive jets, theoretical
predictions are computed at NLO with {\tt NLOJet++} and are benchmarked
against the independent computation of Ref.~\cite{Gehrmann-DeRidder:2019ibf}.
This computation is also used to determine the NNLO QCD corrections,
implemented as $K$-factors. The renormalization and
factorization scale used in the computation are set to the invariant mass of the
dijet system, again following the recommendation of Ref.~\cite{Currie:2018xkj}.

Single-inclusive jet and dijet observables cannot be simultaneously
included  because full knowledge of the experimental correlations between 
them is not available. The selection of the
optimal set of jet observables will be performed and discussed  in
Sect.~\ref{sec:dataselection}, in the context of the final dataset selection.

\subsubsection{Inclusive isolated-photon production}
\label{subsusbsec:photon}
Isolated photon production was not included in previous NNPDF
releases  and is included in NNPDF4.0 for the first time. We specifically
consider the ATLAS measurements at 
$\sqrt{s}=8$~TeV~\cite{Aad:2016xcr} and $\sqrt{s}=13$~TeV~\cite{Aaboud:2017cbm}.
They correspond to integrated luminosities of 20.2~fb$^{-1}$ and 3.2~fb$^{-1}$,
respectively. The measurements are provided for the cross-section differential
in the photon transverse energy $E_T^\gamma$ in different bins of the photon
pseudorapidity $\eta_\gamma$. The accessed ranges are, in both cases,
$E_T^\gamma<1500$~GeV and $|\eta_\gamma|<2.37$. Theoretical predictions are
computed at NLO with {\tt MCFM} and benchmarked against the independent
computation presented in~\cite{Campbell:2016lzl}. The smooth cone isolation
criterion~\cite{Frixione:1998jh} is adopted accordingly, with the parameter
values determined in~\cite{Campbell:2016yrh}. NNLO QCD corrections are
incorporated by means of the $K$-factors computed in~\cite{Campbell:2016lzl};
$K$-factors are also used to incorporate corrections due to resummation of the
electroweak Sudakov logarithms at leading-logarithmic accuracy, according to
the procedure presented in~\cite{Becher:2013zua,Becher:2015yea}. The
factorization and renormalization scales are set equal to the central value
of $E_T^\gamma$ for each bin.
The impact of the measurements presented above on a PDF determination was
studied in~\cite{Campbell:2018wfu} in the context of a variant of the
NNPDF3.1 fit. These data were found to be generally well described, except 
in the most forward rapidity region, and to have a mild impact on the gluon PDF
at intermediate values of $x$. 

\subsubsection{Single top production}
\label{subsubsec:singletop}
Another process included for the first time in an NNPDF release is
$t$-channel single top
production. We consider the
ATLAS~\cite{Aad:2014fwa,Aaboud:2017pdi,Aaboud:2016ymp} and
CMS~\cite{Chatrchyan:2012ep,Khachatryan:2014iya,Sirunyan:2016cdg} measurements
at $\sqrt{s}=7$, 8 and 13~TeV that correspond, for ATLAS (CMS),
to integrated luminosities of 4.59, 20.2 and 3.2~fb$^{-1}$
(2.73, 19.7 and 2.2~fb$^{-1}$), respectively. In the case of ATLAS,
we consider the ratio of the top to antitop inclusive cross-sections at 7 and
13~TeV and the distributions differential in the top or antitop rapidity
$y_{t,\bar{t}}$ at 7 and 8~TeV normalized to the corresponding total cross-section.
The rapidity ranges are $|y_{t,\bar t}|<3.0$ and
$|y_{t,\bar t}|<2.2$ at $\sqrt{s}=7$ and 8~TeV, respectively.
In the case of CMS, we consider the sum of the
top and antitop inclusive cross-sections at 7~TeV and the ratio of the top to
antitop inclusive cross-sections at 8 and 13~TeV. Theoretical predictions are
computed in the five-flavor scheme. At NLO the calculation is performed with
{\tt mg5\_aMC} (v2.6.6)~\cite{Alwall:2014hca}; NNLO corrections are
incorporated by means of the $K$-factors determined
in~\cite{Berger:2016oht,Berger:2017zof}. The renormalization and factorization
scales are set equal to the top mass.

The measurements presented above were extensively studied in the context of a
variant of the NNPDF3.1 fit in~\cite{Nocera:2019wyk}.
The choice of observables included for PDF
determinations is based on the results of that reference. In particular,
distributions differential in the transverse momentum of the
top quark or antiquark are also provided by the experimental collaborations.
However, their inclusion would result in a double counting, given that
experimental correlations across uncertainties for different distributions are
not provided. In~\cite{Nocera:2019wyk} these measurements were found to have a
mild impact on the up and down PDFs at $x\gtrsim 0.1$.

Single top $t$-channel
production is in principle also sensitive to the theoretical details of the
matching schemes and, in the five-flavor scheme, to the bottom PDF.
Here we determine the bottom PDF using perturbative
matching conditions, but it could in principle be parametrized 
independently, like the charm PDF. However, while this may become relevant in
the future, it does not seem necessary at present given the 
precision and kinematic coverage of the existing data.

\subsection{Treatment of nuclear corrections}
\label{subsec:nuclear}

The NNPDF4.0 dataset, like its predecessors, includes a significant
amount of data involving  deuterium or heavy nuclear targets, both
for deep inelastic and hadronic processes. These are
summarized in Table~\ref{tab:nuclear_dataset}, where we also report the
corresponding reference, the number of data points in the NLO and NNLO baseline
fits, and the species of the nuclear target. Overall, 1416 and 1417 data points
come from nuclear measurements in the NLO and NNLO fits respectively, which
amount to about 30\% of the full dataset. All of these datasets but
SeaQuest~\cite{Dove:2021ejl} were already included in the previous NNPDF3.1
determination~\cite{Ball:2017nwa}.

\begin{table}[!t]
  \scriptsize
  \centering
  \renewcommand{\arraystretch}{1.4}
  \input{tables/tab-nucleardata}
  \caption{The nuclear datasets in NNPDF4.0 involving deuterium
    targets (left) or heavier nuclear targets (right) and
    corresponding targets; $N_{\rm dat}$ denotes the
    number of data points included in the NLO/NNLO fits.
    Note that the EMC $F_2^c$ dataset is not included in the
    default NNPDF4.0 PDF set.}
  \label{tab:nuclear_dataset}
\end{table} 

The inclusion of nuclear data in a fit of proton PDFs requires accounting for
changes in the PDFs induced by the nuclear medium. The impact of such
changes was studied by us in~\cite{Ball:2013gsa, Ball:2014uwa} and
found to be subdominant in comparison to the PDF uncertainty at that time.
Specifically, it was shown  (see Sect.~4.11 in~\cite{Ball:2017nwa})
that, upon removing data with nuclear targets from the dataset, the 
precision of up, down and
strange quark and anti-quark PDFs deteriorated by an amount larger
than the size of the effect of the nuclear corrections estimated on the
basis of models. Nuclear corrections were  consequently not included in  the NNPDF3.1 determination.

In NNPDF4.0 we revisit this state of affairs, motivated by the significant
reduction of the PDF uncertainty  in comparison to
NNPDF3.1, which suggests that nuclear effects can no longer be neglected.
We now account for nuclear effects by viewing them as a
theoretical uncertainty. The way this is determined and included
follows the methodology developed
in~\cite{Ball:2018twp,Ball:2020xqw}, to which we refer for
details. The basic idea is to determine the uncertainty from
the difference between the values of  observables computed with the proton
and nuclear PDFs, with each different determination of the nuclear PDF
taken as an independent nuisance parameter. This can then be used to
compute a theoretical covariance matrix, that can 
be added to the experimental covariance matrix.

In order to apply this methodology an underlying set of nuclear PDFs is
needed for the computation of the shifts. Heavy nuclear and
deuteron corrections are treated separately because of the 
substantial difference in the observed size and expected origin of the
nuclear effects. 
For heavier nuclei (Fe, Cu and Pb targets) we
will use the nNNPDF2.0 nuclear PDFs~\cite{AbdulKhalek:2020yuc}. For
deuterium, we use the self-consistent
procedure described in~\cite{Ball:2020xqw}, whereby the proton and deuteron PDFs
are determined simultaneously, each including the uncertainties on the
other. This procedure thus requires in turn the use of a PDF
determination without deuterium corrections in order to initiate the
self-consistent iteration. Here we will apply it by starting with the
NNPDF4.0 determination itself. The deuterium PDF determined by this procedure 
will be described in Sect.~\ref{subsec:nuclearimpact} below.

While nuclear effects will  be included as an extra
uncertainty in the default NNPDF4.0 determination, we will also
discuss for comparison PDFs obtained by neglecting nuclear
effects altogether, or by using the nuclear corrections computed as discussed
above as a correction to the data and not just as an additional uncertainty, 
again following the methodology of
Refs.~\cite{Ball:2018twp,Ball:2020xqw}. These alternative treatments of
nuclear effects will be compared and discussed in
Sect.~\ref{subsec:nuclear_impact} below and provide the motivation for
including nuclear uncertainties without a correction in the default 
PDF determination.

%% file: tables/tab-DIS_dataset.tex
\begin{tabularx}{\textwidth}{Xccccc}
  \toprule
  Dataset
  & Ref.
  & $N_{\rm dat}$
  & $x$
  & $Q$~[GeV]
  & Theory\\
  \midrule
  NMC $F_2^d/F_2^p$
  & \cite{Arneodo:1996kd}
  & 260 (121/121)
  & [0.012, 0.680]
  & [2.1, 10.]
  & {\tt APFEL}\\
  NMC $\sigma^{{\rm NC},p}$
  & \cite{Arneodo:1996qe}
  & 292 (204/204)
  & [0.012, 0.500]
  & [1.8, 7.9]
  & {\tt APFEL} \\
  SLAC $F_2^p$
  & \cite{Whitlow:1991uw}
  & 211 (33/33)
  & [0.140, 0.550]
  & [1.9, 4.4]
  & {\tt APFEL}\\
  SLAC $F_2^d$
  & \cite{Whitlow:1991uw}
  & 211 (34/34)
  & [0.140, 0.550]
  & [1.9, 4.4]
  & {\tt APFEL}\\
  BCDMS $F_2^p$
  & \cite{Benvenuti:1989rh}
  & 351 (333/333)
  & [0.070, 0.750]
  & [2.7, 15.]
  & {\tt APFEL}\\
  BCDMS $F_2^d$
  & \cite{Benvenuti:1989rh}
  & 254 (248/248)
  & [0.070, 0.750]
  & [2.7, 15.]
  & {\tt APFEL}\\
  CHORUS $\sigma_{CC}^{\nu}$
  & \cite{Onengut:2005kv}
  & 607 (416/416)
  & [0.045, 0.650]
  & [1.9, 9.8]
  & {\tt APFEL}\\
  CHORUS $\sigma_{CC}^{\bar{\nu}}$
  & \cite{Onengut:2005kv}
  & 607 (416/416)
  & [0.045, 0.650]
  & [1.9, 9.8]
  & {\tt APFEL}\\
  NuTeV $\sigma_{CC}^{\nu}$ (dimuon)
  & \cite{Goncharov:2001qe,MasonPhD}
  & 45 (39/39)
  & [0.020, 0.330]
  & [2.0, 11.] 
  & {\tt APFEL+NNLO}\\
  NuTeV $\sigma_{CC}^{\bar{\nu}}$ (dimuon)
  & \cite{Goncharov:2001qe,MasonPhD}
  & 45 (36/37)
  & [0.020, 0.210]
  & [1.9, 8.3] 
  & {\tt APFEL+NNLO}\\
  {[NOMAD $\mathcal{R}_{\mu\mu}(E_\nu)$]}~{\bf (*)}
  & \cite{Samoylov:2013xoa}
  & 15 (---/15)
  & [0.030, 0.640]
  & [1.0, 28.]
  & {\tt APFEL+NNLO}\\
  {[EMC $F_2^c$]}
  & \cite{Aubert:1982tt}
  & 21 (---/16)
  & [0.014, 0.440]
  & [2.1, 8.8]
  & {\tt APFEL}\\
  \midrule
  HERA I+II
  $\sigma_{\rm NC,CC}^{p}$
  & \cite{Abramowicz:2015mha}
  & 1306 (1011/1145)
  & [4$\cdot 10^{-5}$, 0.65]
  & [1.87, 223]
  & {\tt APFEL}\\
  HERA I+II $\sigma_{\rm NC}^{c}$~{\bf (*)}
  & \cite{H1:2018flt}
  & 52 (---/37)
  & [7$\cdot 10^{-5}$, 0.05]
  & [2.2, 45]
  & {\tt APFEL}\\
  HERA I+II $\sigma_{\rm NC}^{b}$~{\bf (*)}
  & \cite{H1:2018flt}
  & 27 (26/26)
  & [2$\cdot 10^{-4}$, 0.50]
  & [2.2, 45]
  & {\tt APFEL}\\
  \bottomrule
\end{tabularx}

%% file: tables/tab-DISJETS_dataset.tex
\begin{tabularx}{\textwidth}{Xccccc}
  \toprule
  Dataset
  & Ref.
  & $N_{\rm dat}$
  & $Q^2$ [GeV$^2$]
  & $p_T$~[GeV]
  & Theory\\
  \midrule
  {[ZEUS 820 (HQ) (1j)]}~{\bf (*)}
  & \cite{ZEUS:2002nms}
  & 30 (---/30)
  & [125,10000]
  & [8,100]
  & {\tt NNLOjet}\\
  {[ZEUS 920 (HQ) (1j)]}~{\bf (*)}
  & \cite{ZEUS:2006xvn}
  & 30 (---/30)
  & [125,10000]
  & [8,100]
  & {\tt NNLOjet}\\
  {[H1 (LQ) (1j)]}~{\bf (*)}
  & \cite{H1:2016goa}
  & 48 (---/48)
  & [5.5,80]
  & [4.5,50]
  & {\tt NNLOjet}\\
  {[H1 (HQ) (1j)]}~{\bf (*)}
  & \cite{H1:2014cbm}
  & 24 (---/24)
  & [150,15000]
  & [5,50]
  & {\tt NNLOjet}\\
  {[ZEUS 920 (HQ) (2j)]}~{\bf (*)}
  & \cite{ZEUS:2010vyw}
  & 22 (---/22)
  & [125,20000]
  & [8,60]
  & {\tt NNLOjet}\\
  {[H1 (LQ) (2j)]}~{\bf (*)}
  & \cite{H1:2016goa}
  & 48 (---/48)
  & [5.5,80]
  & [5,50]
  & {\tt NNLOjet}\\
  {[H1 (HQ) (2j)]}~{\bf (*)}
  & \cite{H1:2014cbm}
  & 24 (---/24)
  & [150,15000]
  & [7,50]
  & {\tt NNLOjet}\\  
  \bottomrule
\end{tabularx}

%% file: tables/tab-FTDY_dataset.tex
\begin{tabularx}{\textwidth}{Xccccc}
  \toprule
  Dataset
  & Ref.
  & $N_{\rm dat}$
  & $y_{\ell\ell}$
  & $m_{\ell\ell}$~[GeV]
  & Theory
  \\
  \midrule
  E866 $\sigma^d/2\sigma^p$ (NuSea)
  & \cite{Towell:2001nh}
  &  \ 15 (15/15)
  & [0.07, 1.53]
  & [4.60, 12.9]
  & {\tt APFEL+Vrap}
  \\
  E866 $\sigma^p$ (NuSea)
  & \cite{Webb:2003ps}
  & 184 (89/89)
  & [0.00, 1.36]
  & [4.50, 8.50]
  & {\tt APFEL+Vrap}
  \\
  E605 $\sigma^p$
  & \cite{Moreno:1990sf}
  & 119 (85/85)
  & [-0.20, 0.40]
  & [7.10, 10.9]
  & {\tt APFEL+Vrap}
  \\
  E906 $\sigma^d/2\sigma^p$ (SeaQuest)~{\bf (*)}
  & \cite{Dove:2021ejl}
  & 6 (6/6)
  & [0.11, 0.77]
  & [4.71, 6.36]
  & {\tt APFEL+Vrap}
  \\
  \bottomrule
\end{tabularx}

%% file: tables/tab-GAUGEBOSON_dataset.tex
\begin{tabularx}{\textwidth}{Xccccc}
  \toprule
  Dataset
  & Ref.
  & $N_{\rm dat}$
  & Kin$_1$
  & Kin$_2$~[GeV]
  & Theory
  \\
  \midrule
  CDF $Z$ differential
  & \cite{Aaltonen:2010zza}
  & 29 (29/29)
  & $0.0\le y_{\ell\ell}\le 2.9$
  & $66\le m_{\ell\ell}\le 116$
  & {\tt Sherpa+Vrap}
  \\
  D0 $Z$ differential
  & \cite{Abazov:2007jy}
  & 28 (28/28)
  & $0.0\le y_{\ell\ell}\le 2.8$
  & $66\le m_{\ell\ell}\le 116$
  & {\tt Sherpa+Vrap}
  \\
  {[D0 $W$ electron asymmetry]}
  & \cite{Abazov:2013rja}
  & 13 (13/8)
  & $0.0\le y_{e}\le 2.9$
  & $Q=m_W$
  & {\tt MCFM+FEWZ}
  \\
  D0 $W$ muon asymmetry
  & \cite{D0:2014kma}
  & 10 (10/9)
  & $0.0\le y_{\mu}\le 1.9$
  & $Q=m_W$
  & {\tt MCFM+FEWZ}
  \\  
  \midrule
  ATLAS low-mass DY 7 TeV
  & \cite{Aad:2014qja}
  & 6 (4/6)
  & $|\eta_\ell|\leq 2.1$
  & $14\le m_{\ell\ell}\le 56$
  & {\tt MCFM+FEWZ}
  \\
  ATLAS high-mass DY 7 TeV
  & \cite{Aad:2013iua}
  & 13 (5/5)
  & $|\eta_\ell|\leq 2.1$
  & $116\le m_{\ell\ell}\le 1500$
  & {\tt MCFM+FEWZ}
  \\
  ATLAS $W,Z$ 7 TeV ($\mathcal{L}=35$~pb$^{-1}$)
  & \cite{Aad:2011dm}
  & 30 (30/30)
  & $|\eta_\ell,y_Z|\leq 3.2$
  & $Q=m_W,m_Z$
  & {\tt MCFM+FEWZ}
  \\
  ATLAS $W,Z$ 7 TeV ($\mathcal{L}=4.6$~fb$^{-1}$)~{\bf (*)}
  & \cite{Aaboud:2016btc}
  & 61 (53/61)
  & $|\eta_\ell,y_Z|\leq 2.5,3.6$
  & $Q=m_W,m_Z$
  & {\tt MCFM+FEWZ}
  \\
  CMS $W$ electron asymmetry 7 TeV
  & \cite{Chatrchyan:2012xt}
  & 11 (11/11)
  & $|\eta_e|\leq 2.4$
  & $Q=m_W$
  & {\tt MCFM+FEWZ}
  \\
  CMS $W$ muon asymmetry 7 TeV
  & \cite{Chatrchyan:2013mza}
  & 11 (11/11)
  & $|\eta_\mu|\leq 2.4$
  & $Q=m_W$
  & {\tt MCFM+FEWZ}
  \\
  CMS DY 2D 7 TeV
  & \cite{Chatrchyan:2013tia}
  & 132 (88/110)
  & $|\eta_{\ell\ell}|\leq 2.2$
  & $20.0\leq m_{\ell\ell}\leq 200$
  & {\tt MCFM+FEWZ}
  \\
  LHCb $Z\to ee$ 7 TeV 
  & \cite{Aaij:2012mda}
  & 9 (9/9)
  & $2.0\leq \eta_\ell\leq 4.5$
  & $Q=m_Z$
  & {\tt MCFM+FEWZ}
  \\
  LHCb $W,Z \to \mu$ 7 TeV
  & \cite{Aaij:2015gna}
  & 33 (29/29)
  & $2.0\leq \eta_\ell\leq 4.5$
  & $Q=m_W$
  & {\tt MCFM+FEWZ}  
  \\
  {[ATLAS $W$ 8 TeV]}~{\bf (*)}
  & \cite{Aad:2019rou}
  & 22 (---/22)
  & $|\eta_\ell|<2.4$
  & $Q=m_W$
  & {\tt MCFM+DYNNLO}
  \\
  ATLAS low-mass DY 2D 8 TeV~{\bf (*)}
  & \cite{Aaboud:2017ffb}
  & 84 (47/60)
  & $|y_{\ell\ell}|<2.4$
  & $46\le m_{\ell\ell}\le 200$
  & {\tt MCFM+DYNNLO}
  \\
  ATLAS high-mass DY 2D 8 TeV~{\bf (*)}
  & \cite{Aad:2016zzw}
  & 48 (48/48)
  & $|y_{\ell\ell}|<2.4$
  & $116\le m_{\ell\ell}\le 1500$
  & {\tt MCFM+FEWZ}
  \\
  CMS $W$ rapidity 8 TeV
  & \cite{Khachatryan:2015oaa}
  & 22 (22/22)
  & $|\eta_\ell|\leq 2.3$
  &$Q=m_W$
  & {\tt MCFM+FEWZ}
  \\
  LHCb $Z\to ee$ 8 TeV
  & \cite{Aaij:2015vua}
  & 17 (17/17)
  & $2.00<|\eta_e|<4.25$
  & $Q=m_Z$
  & {\tt MCFM+FEWZ}
  \\
  LHCb $W,Z\to \mu$ 8 TeV
  & \cite{Aaij:2015zlq}
  & 34 (29/30)
  & $2.00<|\eta_\mu|<4.25$
  & $Q=m_Z$
  & {\tt MCFM+FEWZ}
  \\
  {[LHCb $W \to e$ 8 TeV]}~{\bf (*)}
  & \cite{Aaij:2016qqz}
  & 8 (---/8)
  & $2.00<|\eta_e|<4.25$
  & $Q=m_W$
  & {\tt MCFM+FEWZ}
  \\
  ATLAS $\sigma_{W,Z}^{\rm tot}$ 13 TeV~{\bf (*)}
  & \cite{Aad:2016naf}
  & 3 (3/3)
  & ---
  & $Q=m_W,m_Z$
  & {\tt MCFM+FEWZ}
  \\
  LHCb $Z\to ee$ 13 TeV~{\bf (*)}
  & \cite{Aaij:2016mgv}
  & 17 (15/15)
  & $2.00<|y_Z|<4.25$
  & $Q=m_Z$
  & {\tt MCFM+FEWZ}
  \\
  LHCb $Z\to \mu\mu$ 13 TeV~{\bf (*)}
  & \cite{Aaij:2016mgv}
  & 18 (16/16)
  & $2.00<|y_Z|<4.50$
  & $Q=m_Z$
  & {\tt MCFM+FEWZ}
  \\
  \bottomrule
\end{tabularx}

%% file: tables/tab-OTHERLHCPROCESSES_dataset.tex
\begin{tabularx}{\textwidth}{Xccccc}
  \toprule
  Dataset
  & Ref.
  & $N_{\rm dat}$
  & Kin$_1$
  & Kin$_2$~[GeV]
  & Theory
  \\
  \midrule
  ATLAS $W^\pm+c$ 7 TeV~{\bf (*)}
  & \cite{Aad:2014xca}
  & 22 (22/---)
  & $|\eta_\ell|<2.5$
  & $Q=m_W$
  & {\tt MCFM}
  \\
  CMS $W^\pm+c$ 7 TeV
  & \cite{Chatrchyan:2013uja}
  & 10 (10/---)
  & $|\eta_\ell|<2.1$
  & $Q=m_W$
  & {\tt MCFM}
  \\
  CMS $W^\pm+c$ 13 TeV~{\bf (*)}
  & \cite{Sirunyan:2018hde}
  & 5 (5/---)
  & $|\eta_\ell|<2.4$
  & $Q=m_W$
  & {\tt MCFM}
  \\
  ATLAS $W^\pm$+jet 8 TeV~{\bf (*)}
  & \cite{Aaboud:2017soa}
  & 32 (30/30)
  & $0\le p_T^W\le 800$~GeV
  & $Q=m_W$
  & {\tt MCFM+N$_{\rm jetti}$}
  \\
  \midrule
  ATLAS $Z$ $p_T$ 8 TeV ($p_T,m_{\ell\ell}$)
  & \cite{Aad:2015auj}
  & 64 (40/44)
  & $12\le m_{\ell\ell}\le 150$~GeV
  & $30\le p_T^Z\le 900$
  & {\tt MCFM+N$_{\rm jetti}$}
  \\
  ATLAS $Z$ $p_T$ 8 TeV ($p_T,y_Z$)
  & \cite{Aad:2015auj}
  & 120 (18/48)
  & $|y_Z|<2.4$
  & $30\le p_T^Z\le 150$
  & {\tt MCFM+N$_{\rm jetti}$}
  \\
  CMS $Z$ $p_T$ 8 TeV
  & \cite{Khachatryan:2015oaa}
  & 50 (28/28))
  & $|y_Z|<1.6$
  & $30\le p_T^Z\le 170$
  & {\tt MCFM+N$_{\rm jetti}$}
  \\
  \midrule
  CMS $\sigma_{tt}^{\rm tot}$ 5 TeV~{\bf (*)}
  & \cite{Sirunyan:2017ule}
  & 1 (1/1)
  & ---
  & $Q=m_t$
  & {\tt MCFM+top++}
  \\
  ATLAS $\sigma_{tt}^{\rm tot}$ 7, 8 TeV
  & \cite{Aad:2014kva}
  & 2 (2/2)
  & ---
  & $Q=m_t$
  & {\tt MCFM+top++}
  \\
  CMS $\sigma_{tt}^{\rm tot}$ 7, 8 TeV
  & \cite{Spannagel:2016cqt}
  & 2 (2/2)
  & ---
  & $Q=m_t$
  & {\tt MCFM+top++}
  \\
  ATLAS $\sigma_{tt}^{\rm tot}$ 13 TeV
  ($\mathcal{L}$=139~fb$^{-1}$)~{\bf (*)}
  & \cite{Aad:2020tmz}
  & 1 (1/1)
  & ---
  & $Q=m_t$
  & {\tt MCFM+top++}
  \\
  CMS $\sigma_{tt}^{\rm tot}$ 13 TeV
  & \cite{Khachatryan:2015uqb}
  & 1 (1/1)
  & ---
  & $Q=m_t$
  & {\tt MCFM+top++}
  \\
  {[ATLAS $t\bar{t}~\ell$+jets 8 TeV ($1/\sigma d\sigma/dp_T^t$)]}
  & \cite{Aad:2015mbv}
  & 8 (---/8)
  & $0\le p_T^t\le 500$~GeV
  & $Q=m_t$
  & {\tt Sherpa+NNLO}
  \\
  ATLAS $t\bar{t}~\ell$+jets 8 TeV ($1/\sigma d\sigma/dy_t$)
  & \cite{Aad:2015mbv}
  & 5 (4/4)
  & $|y_t|<2.5$
  & $Q=m_t$
  & {\tt Sherpa+NNLO}
  \\
  ATLAS $t\bar{t}~\ell$+jets 8 TeV ($1/\sigma d\sigma/dy_{t\bar t}$)
  & \cite{Aad:2015mbv}
  & 5 (4/4)
  & $|y_{t\bar t}|<2.5$
  & $Q=m_t$
  & {\tt Sherpa+NNLO}
  \\
  {[ATLAS $t\bar{t}~\ell$+jets 8 TeV ($1/\sigma d\sigma/dm_{t\bar t}$)]}
  & \cite{Aad:2015mbv}
  & 7 (---/7)
  & $345\le m_{t\bar t}\le 1600$~GeV
  & $Q=m_t$
  & {\tt Sherpa+NNLO}
  \\
  ATLAS $t\bar{t}~2\ell$ 8 TeV ($1/\sigma d\sigma/dy_{t\bar t}$)~{\bf (*)}
  & \cite{Aaboud:2016iot}
  & 5 (5/5)
  & $|y_{t\bar t}|<2.8$
  & $Q=m_t$
  & {\tt mg5\_aMC+NNLO}
  \\
  CMS $t\bar{t}~\ell$+jets 8 TeV ($1/\sigma d\sigma/dy_{t\bar t}$)
  & \cite{Khachatryan:2015oqa}
  & 10 (9/9)
  & $-2.5<y_{t\bar t}<2.5$
  & $Q=m_t$
  & {\tt Sherpa+NNLO}
  \\
  CMS $t\bar{t}$ 2D $2\ell$ 8 TeV ($1/\sigma d\sigma/dy_tdm_{t\bar t}$)~{\bf (*)}
  & \cite{Sirunyan:2017azo}
  & 16 (16/16)
  & $|y_t|<2.5$
  & $340\le m_t\le 1500$
  & {\tt mg5\_aMC+NNLO}
  \\
  CMS $t\bar{t}~\ell$+jet 13 TeV ($d\sigma/dy_t$)~{\bf (*)}
  & \cite{Sirunyan:2018wem}
  & 10 (10/10)
  & $|y_t|<2.5$
  & $Q=m_t$
  & {\tt mg5\_aMC+NNLO}
  \\
  CMS $t\bar{t}~2\ell$ 13 TeV ($d\sigma/dy_t$)~{\bf (*)}
  & \cite{Sirunyan:2018ucr}
  & 11 (11/11)
  & $|y_t|<2.5$
  & $Q=m_t$
  & {\tt mg5\_aMC+NNLO}
  \\
  \midrule
  {[ATLAS incl. jets 7 TeV, R=0.6]}
  & \cite{Aad:2014vwa}
  & 90 (---/90)
  & $|y^{\rm jet}|<3.0$
  & $100\le p_T^{\rm jet}\le 1992$
  & {\tt NNLOjet}
  \\
  {[CMS incl. jets 7 TeV]}
  & \cite{Chatrchyan:2014gia}
  & 133 (---/133)
  & $|y^{\rm jet}|<2.5$
  & $100\le p_T^{\rm jet}\le 2000$
  & {\tt NNLOjet}
  \\
  ATLAS incl. jets 8~TeV, R=0.6~{\bf(*)}
  & \cite{Aaboud:2017dvo}
  & 171 (171/171)
  & $|y^{\rm jet}|<3.0$
  & $70\le p_T^{\rm jet}\le 2500$
  & {\tt NNLOjet}  
  \\
  CMS incl. jets 8 TeV~{\bf (*)}
  & \cite{Khachatryan:2016mlc}
  & 185 (185/185)
  & $|y^{\rm jet}|<3.0$
  & $74\le p_T^{\rm jet}\le 2500$
  & {\tt NNLOjet}
  \\
  ATLAS dijets 7 TeV, R=0.6~{\bf (*)}
  & \cite{Aad:2013tea}
  & 90 (90/90)
  & $0.0\le y^*\le 3.0$
  & $260\le m_{jj} \le 4270$
  & {\tt NNLOjet}
  \\
  CMS dijets 7 TeV~{\bf (*)}
  & \cite{Chatrchyan:2012bja}
  & 54 (54/54)
  & $|y_{\rm max}|<2.5$
  & $200\le m_{jj}\le 5000$
  & {\tt NNLOjet}
  \\
  {[CMS 3D dijets 8 TeV]}~{\bf (*)}
  & \cite{Sirunyan:2017skj}
  & 122 (122/122)
  & $0.0<y_b,y^*<3.0$
  & $133\le p_{T,{\rm avg}}\le 1780$
  & {\tt NNLOjet}
  \\
  \midrule
  {[ATLAS isolated $\gamma$ prod. 8 TeV]}~{\bf (*)}
  & \cite{Aad:2016xcr}
  & 49 (---/---)
  & $|\eta_\gamma|<2.37$
  & $E_T^\gamma<1500$
  & {\tt MCFM+NNLO}
  \\
  ATLAS isolated $\gamma$ prod. 13 TeV~{\bf (*)}
  & \cite{ATLAS:2017nah}
  & 53 (53/53)
  & $|\eta_\gamma|<2.37$
  & $E_T^\gamma<1500$
  & {\tt MCFM+NNLO}
  \\
  \midrule
  ATLAS single~$t$ $R_{t}$ 7 TeV~{\bf (*)}
  & \cite{Aad:2014fwa}
  & 1 (1/1)
  & ---
  & $Q=m_t$
  & {\tt mg5\_aMC+NNLO}
  \\
  CMS single~$t$ $\sigma_{t}+\sigma_{\bar{t}}$ 7 TeV~{\bf (*)}
  & \cite{Chatrchyan:2012ep}
  & 1 (1/1)
  & ---
  & $Q=m_t$
  & {\tt mg5\_aMC+NNLO}
  \\
  ATLAS single~$t$ $R_{t}$ 8 TeV~{\bf (*)}
  & \cite{Aaboud:2017pdi}
  & 1 (1/1)
  & ---
  & $Q=m_t$
  & {\tt mg5\_aMC+NNLO}
  \\
  CMS single~$t$ $R_{t}$ 8 TeV~{\bf (*)}
  & \cite{Khachatryan:2014iya}
  & 1 (1/1)
  & ---
  & $Q=m_t$
  & {\tt mg5\_aMC+NNLO}
  \\
  ATLAS single~$t$ $R_{t}$ 13 TeV~{\bf (*)}
  & \cite{Aaboud:2016ymp}
  & 1 (1/1)
  & ---
  & $Q=m_t$
  & {\tt mg5\_aMC+NNLO}
  \\
  CMS single~$t$ $R_{t}$ 13 TeV~{\bf (*)}
  & \cite{Sirunyan:2016cdg}
  & 1 (1/1)
  & ---
  & $Q=m_t$
  & {\tt mg5\_aMC+NNLO}
  \\
  ATLAS single~$t$ 7 TeV ($1/\sigma d\sigma/dy_t$)~{\bf (*)}
  & \cite{Aad:2014fwa}
  & 4 (3/3)
  & $|y_t|<3.0$
  & $Q=m_t$
  & {\tt mg5\_aMC+NNLO}
  \\
  ATLAS single~$t$ 7 TeV ($1/\sigma d\sigma/dy_{\bar t}$)~{\bf (*)}
  & \cite{Aad:2014fwa}
  & 4 (3/3)
  & $|y_{\bar t}|<3.0$
  & $Q=m_t$
  & {\tt mg5\_aMC+NNLO}
  \\
  ATLAS single~$t$ 8 TeV ($1/\sigma d\sigma/dy_t$)~{\bf (*)}
  & \cite{Aaboud:2017pdi}
  & 4 (3/3)
  & $|y_t|<2.2$
  & $Q=m_t$
  & {\tt mg5\_aMC+NNLO}
  \\
  ATLAS single~$t$ 8 TeV ($1/\sigma d\sigma/dy_{\bar t}$)~{\bf (*)}
  & \cite{Aaboud:2017pdi}
  & 4 (3/3)
  & $|y_{\bar t}|<2.2$
  & $Q=m_t$
  & {\tt mg5\_aMC+NNLO}
  \\
  \bottomrule
\end{tabularx}

%% file: tables/tab-nucleardata.tex
\begin{tabularx}{\textwidth}{XcccXccc}
  \toprule
  Dataset
  & Ref.
  & $N_{\rm dat}$
  & Target
  &
  Dataset
  & Ref.
  & $N_{\rm dat}$
  & Target
  \\
  \midrule
  NMC $F_2^d/F_2^p$
  & \cite{Arneodo:1996kd}
  & 121/121
  & $p$, $d$
  & CHORUS $\sigma_{CC}^{\nu}$
  & \cite{Onengut:2005kv}
  & 416/416
  & $^{208}_{\ 82}$Pb
  \\
  SLAC $F_2^d$
  & \cite{Whitlow:1991uw}
  & 34/34
  & $d$
  & CHORUS $\sigma_{CC}^{\bar{\nu}}$
  & \cite{Onengut:2005kv}
  & 416/416
  & $^{208}_{\ 82}$Pb
  \\
  BCDMS $F_2^d$
  & \cite{Benvenuti:1989rh}
  & 248/248
  & $d$
  & NuTeV $\sigma_{CC}^{\nu}$ (dimuon)
  & \cite{Goncharov:2001qe,MasonPhD}
  & 39/39
  & $^{56}_{26}$Fe
  \\
  E866 $\sigma^d/2\sigma^p$ (NuSea)
  & \cite{Towell:2001nh}
  & 15/15
  & $p$, $d$
  & NuTeV $\sigma_{CC}^{\bar{\nu}}$ (dimuon)
  & \cite{Goncharov:2001qe,MasonPhD}
  & 36/37
  & $^{56}_{26}$Fe
  \\
  E906 $\sigma^d/2\sigma^p$ (seaQuest)
  & \cite{Dove:2021ejl}
  & 6/6
  & $p$, $d$
  & E605 $\sigma^p$
  & \cite{Moreno:1990sf}
  & 85/85
  & $^{64}_{32}$Cu
  \\
  \midrule
  &
  &
  &
  & EMC $F_2^c$
  & \cite{Aubert:1982tt}
  & ---/16
  & $^{56}_{26}$Fe
  \\
  \bottomrule
\end{tabularx}

%% file: sec-methodology.tex
\section{Fitting methodology}
\label{sec:methodology}

As discussed in the introduction,
NNPDF4.0 is the first PDF set to be based on a methodology fully selected
through a machine learning algorithm.
This means that, whereas the basic structure of the NNPDF4.0
methodology is the same as in previous NNPDF releases, specifically the
use of a Monte Carlo representation of PDF uncertainties and
correlations, and the use of neural networks as basic interpolating
functions~\cite{Ball:2014uwa,Ball:2017nwa}, all the details of the
implementation, such as the choice 
of neural network architecture and the minimization algorithm, are now
selected through an automated hyperoptimization procedure.
This is
 possible thanks to an extensive rewriting and reorganization of
the NNPDF  framework.
Furthermore, some theory constraints built into the PDF parametrization are
implemented for the first time in NNPDF4.0. Also for the first time we
consider PDF determinations performed with different choices of
parametrization basis.

In \secref{sec:methparametrisation} we start by discussing the PDF
parametrization and choice of basis and the way they implement
theoretical constraints. In \secref{sec:methimplementationdetails}
we then present the new NNPDF fitting framework, which is the basis of
the hyperoptimization procedure. The hyperoptimization in turn is
discussed   in \secref{sec:hyperparam}, along with its
output, which defines the baseline NNPDF4.0 methodology. We conclude in
\secref{sec:benchmark} with quantitative benchmarks assessing
both the efficiency and speed of this final methodology
compared to the methodology used for NNPDF3.1.

\input{subsec-parametrisation.tex}
\input{subsec-fittingframework.tex}
\input{subsec-hyperopt.tex}

\subsection{Performance and quality benchmarks}
\label{sec:benchmark}

The new NNPDF fitting framework features a significantly improved computational performance compared to 
previous NNPDF. This improvement
is mostly driven by the availability of the gradient-based optimizers
provided by the {\tt TensorFlow} library, combined with the dedicated
hyperparameter optimization
and other technical improvements in key parts of the code.
Furthermore, the new fitting framework is able
to take  advantage of  Graphical Processing
Units (GPUs), which, when available, can further improve speed
(although currently setting the same training and validation split for all
replicas is needed for optimal performance).

\begin{table}[!t]
  \scriptsize
  \centering
  \renewcommand{\arraystretch}{1.4}
  \input{tables/tab-performance}
  \vspace{0.2cm}
  \caption{The average fitting time per replica, speed up factor (as compared
    to the NNPDF3.1 performance), and the RAM requirements in global PDF fits based on the
    NNPDF3.1 and NNPDF4.0 frameworks for the same input dataset.
    In the NNPDF4.0 case, we compare the performance obtained
    on CPUs with that on GPUs.}
  \label{table:performance}
\end{table}

To quantify the performance of the new fitting code, in
Table~\ref{table:performance} we show the average fitting time per replica
in PDF fits based on the NNPDF3.1 and
NNPDF4.0 fitting frameworks.
The same global input dataset is used in both cases, in order to ensure
a consistent comparison.
In the case of NNPDF4.0, we compare the performances of running the code
either in CPUs or in GPUs.
These benchmark tests have been carried out
on an Intel(R)
Core(TM) i7-4770 at 3.40GHz CPU and on a NVIDIA Titan V GPU.

The comparisons in Table~\ref{table:performance} show that, while in NNPDF3.1 the typical
fitting time per Monte Carlo replica was around 15 hours, in NNPDF4.0 this has been reduced
on average
by a factor 24 (down to around 40 minutes) when running on CPUs, and by a factor
of 140 (down to 7 minutes) when running on GPUs.
This implies that, in the same time that it takes to run 100 replicas of NNPDF3.1,
one can now run 2400 replicas of NNPDF4.0 or, alternatively, 24 variations
(with different datasets or theory settings)
of the same 100 NNPDF4.0 replicas.
The enhanced performance of NNPDF4.0 is essential for the 
implementation of the hyperoptimization program: one can explore
thousands of different hyperparameter configurations if the fits are fast enough.
Furthermore, we note that this significant increase in speed greatly facilitates several physics
applications, from the
$\alpha_s(m_Z)$ determination~\cite{Ball:2018iqk} to the simultaneous fits of PDFs and EFT Wilson
coefficients~\cite{Carrazza:2019sec,Greljo:2021kvv},
which rely on producing a sufficiently large sample of replicas.

From Table~\ref{table:performance} one can also observe that this 
increase in speed has as a trade-off a greater RAM memory consumption by 
around a factor of four.
These demanding requirements arise because the code needs to hold in memory not only
the FK-tables (as was already the case in NNPDF3.1) but also the $\chi^2$
gradients used for the minimization, which were not stored before.
While this increase in memory may appear limiting, we note that the FK-tables and the 
functional form of the gradient can be shared between Monte Carlo replicas 
running simultaneously on the same processor.
This makes it possible to run a large number of replicas
in parallel on a GPU, and is the main reason for
the reduction of the average fit time per replica
reported in Table~\ref{table:performance}.

\begin{figure}[!t]
  \centering
  \includegraphics[width=0.49\textwidth]{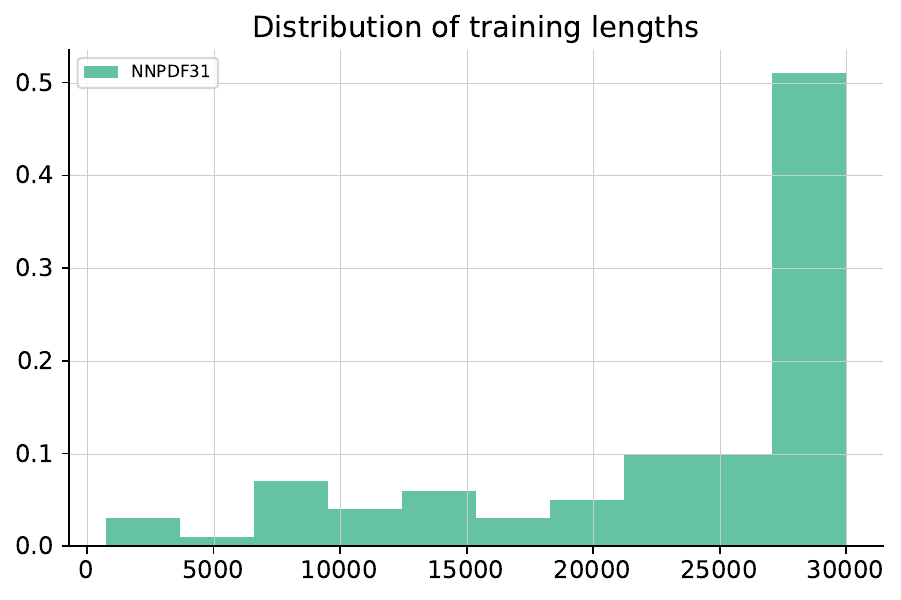}
  \includegraphics[width=0.49\textwidth]{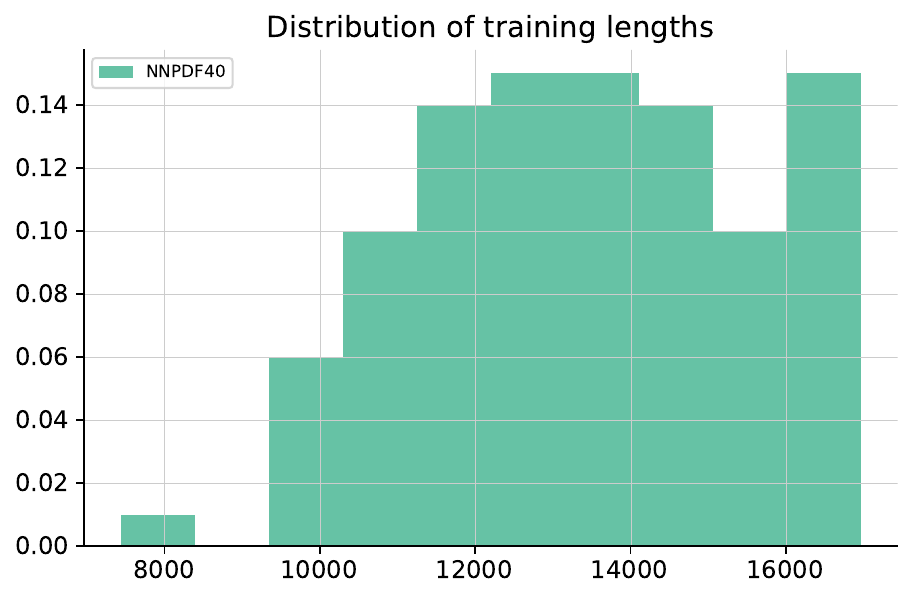}\\
  \caption{Distribution of training lengths, defined by the optimal stopping
    point of each replica, in fits to a common global dataset based on the
    NNPDF3.1 (left) and NNPDF4.0 (right panel) methodologies.}
  \label{fig:benchtlength}
\end{figure}

In addition to the improved computational performance, the new 
framework underlying the NNPDF4.0 fits
exhibits other benefits that impact in a positive manner the actual outcome of
the global fit.
To illustrate these, Fig.~\ref{fig:benchtlength} compares the
distribution over replicas of the
training lengths, defined as the optimal stopping point of each replica,
between fits  based on the NNPDF3.1 
and NNPDF4.0  methodologies for a common dataset. While the number of iterations
of the two different optimization algorithms are incomparable, it is interesting
to note that the rightmost bin of the distribution is populated by the replicas
whose stopping point is determined by the maximum number of iterations, rather
than by satisfying the look-back cross-validation stopping condition. These are
thus replicas for which full convergence has not been reached. The 
fact that replica training does stop through  cross-validation is what
guarantees that the $\chi^2$ minimzation is sufficiently accurate to
actually determine the optimal fit.

From this comparison one finds that in NNPDF3.1, based on nodal genetic algorithms, around half of the replicas
stop at the maximum number of generations, while for the SGD-based NNPDF4.0 fit
this fraction is much smaller, around 15\%.
This observation implies that while in NNPDF3.1 many replicas might stop before proper training
has been achieved, and may be affected by underlearning,
this issue is much less severe
in NNPDF4.0.
Indeed, now 85\% of the replicas stop when
the optimal stopping point has been identified by the look-back cross-validation
algorithm.
One can therefore expect a reduction in the PDF uncertainties thanks
to the new methodology, given that the fraction of replicas
with potential underlearning is markedly reduced, leading to overall smoother and more similar replicas.
We will study in more detail in Sect.~\ref{sec:tests} the impact at the PDF
level of the new methodology.

\begin{figure}[!t]
  \centering
  \includegraphics[width=0.49\textwidth]{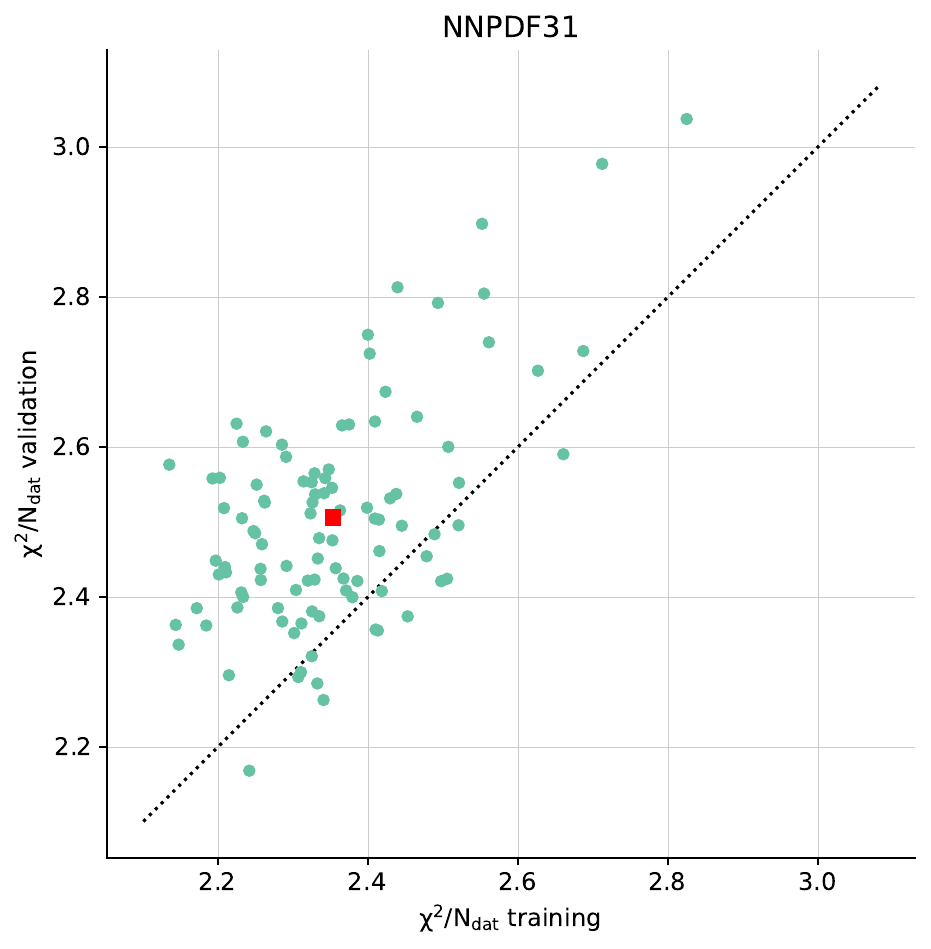}
  \includegraphics[width=0.49\textwidth]{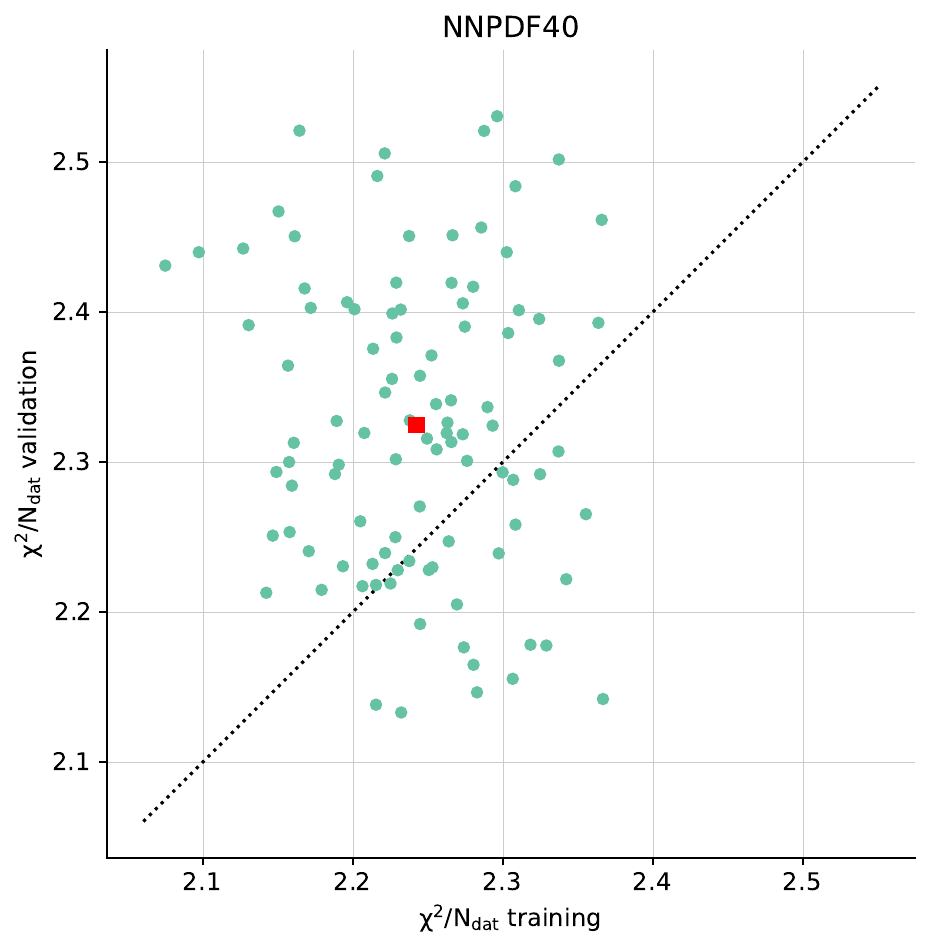}\\
  \caption{Comparison of
    the values of the training and validation $\chi^2$ for each replica between
    the NNPDF3.1 and NNPDF4.0 methodoligies, when fitting a common dataset.  The
    red square indicates the mean value over the replicas.}
  \label{fig:benchtrval}
\end{figure}

Similar considerations can be drawn from Fig.~\ref{fig:benchtrval}, which
compares scatter plots with the values of $\chi^2_{\rm tr}$ and $\chi^2_{\rm val}$
for the $N_{\rm rep}=100$ replicas between fits based on the NNPDF3.1 and NNPDF4.0
methodologies and the same global dataset.
In these plots, the  red square
indicates the position of the mean value over the replicas,
and a dashed line with unit slope is added in order to facilitate visualization.
Note that $\chi^2_{\rm val}$ is expected to be  (on average) somewhat higher than
$\chi^2_{\rm tr}$ given that validation data are not used for the optimization.

From this comparison, one can see that the spread in the values of
$\chi^2_{\rm tr}$ and $\chi^2_{\rm val}$ is reduced when going from NNPDF3.1
to NNPDF4.0.
Furthermore, in the latter case there are no outliers, while this is not the
case in the NNPDF3.1-like fits.
Also, for NNPDF4.0 around one quarter of the replicas have $\chi^2_{\rm val}<\chi^2_{\rm tr} $,
which is another indicator of proper training and stopping.
This fraction
is smaller in NNPDF3.1, again possibly signaling underlearning in some replicas.

All in all, the results presented in here indicate that the methodological
improvements introduced in NNPDF4.0 not only lead to a significant
improvement in terms of computational performance, but also to a more robust procedure
where proper training is achieved for the majority of 
neural network replicas.

%% file: subsec-parametrisation.tex
\subsection{PDF parametrization and theoretical constraints}
\label{sec:methparametrisation}

We now turn to  the general structure of the PDF
parametrization, and the theory constraints that are imposed upon it:
specifically sum rules, positivity and integrability.

\subsubsection{Parametrization bases}
\label{sec:flavev}
A PDF analysis requires a choice of basis, namely
a set of linearly independent PDF flavor combinations that are parametrized
at the input evolution scale $Q_0$. In the NNPDF approach, this
corresponds to choosing which are the PDF combinations whose value is
the output of a neural network. 
Optimal results should in principle be independent of this specific choice of basis.
Previous NNPDF releases adopted the  so-called evolution basis, in
which the basis PDFs are chosen as the singlet quark $\Sigma$ and
gluon $g$ that mix upon QCD evolution,  and
valence $V_i$ and nonsinglet sea $T_i$ combinations that are
eigenstates of evolution, namely 
\begin{align}
  \Sigma  &=  u+\bar{u} + d+\bar{d} + s+\bar{s} + 2c  \, ,  \nonumber \\
  T_3     &=  \lp u+\bar{u}\rp - \lp  d+\bar{d}  \rp \, ,  \nonumber\\
  T_8     &=  \lp u+\bar{u} +  d+\bar{d}  \rp - 2\lp s+\bar{s} \rp \, \label{eq:evol_basis} \\
  V     &= \lp u-\bar{u}\rp + \lp d-\bar{d}\rp + \lp s-\bar{s}\rp   \, ,\nonumber \\
  V_3     &=  \lp u-\bar{u}\rp - \lp  d-\bar{d}  \rp \, , \nonumber\\
  V_8     &=  \lp u-\bar{u} +  d-\bar{d}  \rp - 2\lp s-\bar{s} \rp \nonumber \, .
\end{align}

In NNPDF3.1, this set of linearly independent
flavor combinations was supplemented by an independently parametrized
total charm PDF $c+\bar c$, with the charm asymmetry $c-\bar c$
assumed to vanish at scale $Q_0$.
Here we will instead supplement the basis
Eq.~(\ref{eq:evol_basis}) with a further nonsinglet combination,
namely
\begin{equation}\label{eq:t15}
 T_{15}     =  \lp u+\bar{u} +  d+\bar{d} +  s+\bar{s} \rp - 3\lp
 c+\bar{c}\rp
\end{equation}
still assuming $c-\bar c=0$ at the parametrization scale. At
NNLO a small charm asymmetry is then generated by perturbative
evolution. 
The union of
Eqs.~(\ref{eq:evol_basis}-\ref{eq:t15}) will be referred to as the
evolution basis henceforth.

We will also consider PDF determnations carried out in the flavor basis, in which the PDFs that are
parametrized are 
\begin{align}
	\label{eq:flav_basis}
\tilde{f}_{k} =\{ u,\,\bar{u},\,d,\bar{d},\,s,\,\bar{s},\, c,\, g\},
\end{align}
related to their evolution basis counterparts
\begin{align}
	\label{eq:evolution_basis}
{f}_{k}=\{V,\, V_3,\, V_8,\, T_3,\, T_8,\, T_{15},\, \Sigma,\, g\},
\end{align}
by means of Eqns.~(\ref{eq:evol_basis}-\ref{eq:t15}).

The evolution and flavor bases each have advantages and disadvantages.
For instance, if one chooses a factorization scheme in which PDFs are
non-negative~\cite{Candido:2020yat}, positivity is easier to implement
in the flavor basis.
On the other hand, the integrability of the valence
  distributions $V,V_3,V_8$, as required by the valence sum rules, is
  simpler in the evolution basis.
  In this work, we take the evolution basis as our standard choice,
  however we will  explicitly check basis independence, by verifying
  that equivalent results
  are obtained in the data region if the flavor basis is adopted
  instead, see Sect.~\ref{subsec:flavbasis} below.

  The output of the neural network is supplemented by a preprocessing  factor
  and by normalization constants.
The relation between the PDFs and the neural network
output is 
\begin{align}
	\label{eq:evolution_basis_param}
	xf_k\left( x,Q_0; {\boldsymbol \theta} \right) = A_k\,x^{1-\alpha_k}(1-x)^{\beta_k}{\rm NN}_k(x;
        {\boldsymbol\theta}) \, ,\quad k=1,\ldots,8\,,
\end{align}
where $k$ runs over the elements of the PDF basis,  ${\rm NN}_k(x;{\boldsymbol\theta})$
is the $k$-th output of a neural network, and 
${\boldsymbol\theta}$ collectively indicates the full set of neural
network parameters.
The preprocessing function has the purpose of speeding up the training
of the neural net. In order to make sure that it does not bias the
result, the  exponents $\alpha_k$ and $\beta_k$ are varied in a range
that is determined  iteratively
in a self-consistent manner as described in~\cite{Ball:2014uwa},
supplemented by
a further integrability constraint, to be discussed in
Sec.~\ref{sec:pdfint}. The independence of result of the choice of
preprocessing ranges has been recently validated in
Ref.~\cite{Carrazza:2021yrg}, where it is shown that results otained
here can be obtained by a suitable rescaling on the neural network
input that avoids preprocessing altogether.
The normalization constants $A_k$ are constrained by the valence and
momentum sum rules, also to be discussed below, in
Sec.~\ref{subsec:sumrules}.

When using the flavor basis, the small-$x$ preprocessing is removed
from Eq.~(\ref{eq:evolution_basis_param}), i.e. $\alpha_k=1$ for all
$k$.
This is because standard Regge theory arguments (see
e.g.~\cite{Roberts:1990ww}) imply that the singlet and nonsinglet
have a different small $x$ behavior, and in particular the nonsinglet
has a finite first moment, while the singlet first moment
diverges. This means that the small-$x$ behavior of flavor-basis PDFs
is the linear combination of a leading singlet small-$x$ growth and a
subleading nonsinglet power behavior characterized by a different
exponent.
Hence, factoring out a common preprocessing exponent is not
advantageous in this case.

\subsubsection{Sum rules}
\label{subsec:sumrules}

Irrespectively of the choice of fitting basis, PDFs should satisfy both the momentum sum rule
\be
\label{eq:mom_sum_rule}
\int_0^1 dx\,x\lp g\left(x, Q\right) + \Sigma\left(x, Q\right)\rp = 1 \, ,
\ee
and the three valence sum rules,
\bea
\int_0^1 dx\,\lp u(x,Q)-\bar{u}(x,Q)\rp &=& 2 \, , \nonumber \\
\int_0^1 dx\,\lp d(x,Q)-\bar{d}(x,Q)\rp &=& 1 \, , \label{eq:valence_sum_rules_f}\\
\int_0^1 dx\,\lp s(x,Q)-\bar{s}(x,Q)\rp &=& 0 \, , \nonumber 
\eea
for all values of $Q$.
Provided these sum rules are imposed at the initial parametrization scale, $Q_0$,
perturbative QCD ensures that they will hold for any other value $Q\ne Q_0$.
When transformed to the evolution basis, Eq.~(\ref{eq:valence_sum_rules}), the valence sum rules read
\begin{align}\label{eq:valence_sum_rules}
\int_0^1 dx\, V\left(x, Q\right) = \int_0^1 dx\, V_8\left(x, Q\right) = 3\,,   \quad
    \int_0^1 dx\, V_3\left(x, Q\right) = 1\,.
\end{align}
We have then four equations that fix four of the normalization
constants $A_k$, namely $A_V$, $A_{V_8}$,$A_{V_3}$ and $A_g$.

In the present analysis we always impose the sum rules in the
evolution basis.
This means that when performing a fit in the flavor basis, we express
the evolution basis PDFs $f_k$   Eq.~(\ref{eq:evolution_basis}) in terms of the flavor basis PDFs
$\tilde{f}_{k}$  Eq.~(\ref{eq:flav_basis}) through a transformation matrix  $R_{kk'}$:
\begin{align}
	\label{eq:flavour_basis_param_b}
	xf_k\left(x,Q_0; {\boldsymbol \theta}\right) =
        A_k \sum_{k'} R_{kk'} \,x\tilde{f}_{k'}\left(x,Q_0; {\boldsymbol \theta}\right),
\end{align}
and then impose Eqs.~(\ref{eq:mom_sum_rule},\ref{eq:valence_sum_rules}).

The integrals in Eqs.(\ref{eq:mom_sum_rule},\ref{eq:valence_sum_rules}) are evaluated between $x_{\rm min}=10^{-9}$
and $x_{\rm max}=1$. Each time the neural network parameters ${\boldsymbol \theta}$
are modified by the minimization algorithm, using an adaptative strategy that achieves
a relative precision of $\mathcal{O}\lp 10^{-5}\rp$ across the whole range of $x$.

\subsubsection{Positivity of PDFs and physical observables}
\label{sec:positivity}

Hadron-level cross-sections are non-negative  quantities, because they are
probability distributions. However, PDFs beyond LO are not
probabilities, and thus they may be negative. 
The reason is that, beyond LO, PDFs include a collinear subtraction
which is necessary in order for the partonic  cross-sections to be finite. Whether they remain
positive or not then depends on the form of the subtraction, i.e. on
the factorization scheme.
Consequently, in previous NNPDF determinations, in order to exclude
unphysical PDFs, we imposed 
positivity of a number of  cross-sections,  by means of  Lagrange
multipliers which penalize PDF configurations leading to negative
physical observables.
Specifically, we imposed positivity of the  $F_2^u$, $F_2^d$, $F_2^s$, and $F_{L}$ structure functions
and of the  flavor-diagonal Drell-Yan rapidity
distributions $\sigma_{{\rm DY},u\bar{u}}$, $\sigma_{{\rm DY},d\bar{d}}$,
$\sigma_{{\rm DY},s\bar{s}}$.
However, since this set of positivity observables is not exhaustive, in some extreme kinematic
regions physical observables (e.g. very high-mass $W'$ production)
could still become negative within uncertainties.

It was recently shown in Ref.~\cite{Candido:2020yat} that
PDFs for individual quark flavors and the gluon in the $\overline{\rm MS}$
factorization scheme are non-negative.\footnote{It has been
  recently~\cite{Collins:2021vke} argued that the positivity argument
  of Ref.~\cite{Candido:2020yat} only holds if the ultraviolet
  renormalization
  scale used to define PDFs is chosen to be high enough, and that PDFs
  renormalized at low enough scale can become negative. This is
  relevant when comparing PDFs extracted from 
  high-energy processes with those computed as lattice matrix
  elements~\cite{Lin:2017snn,Constantinou:2020hdm}, as well as when
  extending factorization as low scales, as emphasized in
  Ref.~\cite{Collins:2021vke}. However, here we focus on
  PDFs extracted from and
  relevant for the computation of high-scale hard processes. The
  independence of NNPDF results on the cutoff used to remove
  low-scale data was
  studied in Ref.~\cite{Ball:2013gsa} in the framework of NNPDF2.3,
  and holds with stronger 
  arguments for more recent NNPDF sets, based on a dataset dominated
  by hadron collider data, see also Sect.~\ref{subsec:reduced} below.} 
We thus now also impose this positivity condition along with the
constraint of positivity of physical
cross-sections discussed above. Indeed, note that the positivity of
$\overline{\rm MS}$ PDFs is neither necessary nor sufficient in order
to ensure cross-section positivity~\cite{Candido:2020yat}: they are
independent (though of course related) constraints that limit the
space of acceptable PDFs.

We  impose positivity of the
gluon and of the
up, down and strange quark and antiquark PDFs. The charm PDF is also
positive in the $n_f=3$ scheme, but it needs not be positive in the
$n_f=4$ scheme because perturbative matching conditions neglect the
quark mass and this generally spoils positivity for a massive quark
PDF~\cite{Candido:2020yat}. We do, however, add a positivity constraint
for the charm structure function $F_2^c$, similar to the ones for other
structure functions of individual flavors. Note that this constraint
was not included in NNPDF3.1, though it was included in a more recent
study based on NNPDF3.1 dataset and methodology~\cite{Faura:2020oom},
where it was found to
have a significant impact on the strange PDF.

In the same manner as for the cross-sections, PDF positivity is implemented 
by means of Lagrange multipliers.
Specifically, for each flavor basis PDF $\tilde{f}_{k}$  Eq.~(\ref{eq:flav_basis}),
one adds a contribution to the total cost function used for the neural network training given by
\begin{align}
	\label{eq:chi2pos_k}
	\chi^2_{\rm tot} \to \chi^2_{\rm tot}+\sum_{k=1}^8  \Lambda_k \,\sum_{i=1}^{n_i} \,\text{Elu}_{\alpha}\left(-\tilde{f}_k\left(x_i,Q^2\right)\right)\,,
\end{align}
with  $Q^2 = 5\, \text{GeV}^2$ and with the $n_i$ values $x_i$ given
by 10 points logarithmically spaced between $5\cdot10^{-7}$ and $10^{-1}$ and 10 points
linearly spaced between $0.1$ and $0.9$. The Elu function is given by 
\begin{align}
	\label{eq:Elu}
	\text{Elu}_{\alpha}\left(t\right) = 
	\begin{cases}
		t \,\,\,\,\,\,\,\,\,\,\,\,\,\,\,\,\,\,\,\,\,\,\,\,\,\,\,\,\,\,\,\text{if}\,\,\,\, t>0 \\
		\alpha\left(e^t-1\right)\,\,\,\,\,\,\,\text{if}\,\,\,\, t<0
	\end{cases}\,,
\end{align} 
with the parameter $\alpha=10^{-7}$.
Eq.~\eqref{eq:chi2pos_k} indicates that negative PDFs receive a penalty which is proportional
both to the corresponding Lagrange multipliers $\Lambda_k$ and to the absolute magnitude of the
PDF itself, and therefore these configurations will be strongly
disfavored during the minimization.
The Lagrange multiplier increases exponentially during the
minimization, with a maximum value  $\Lambda_k^{\rm max}$  attained
when the maximum training  length is reached. We choose
$\Lambda_k^{\rm max}=10^{10} $ for the three Drell-Yan observables,
and  $\Lambda_k^{\rm max}=10^6 $ for all the other positivity
observables.  These values are chosen in
such a way that the constraint is enforced with sufficient accuracy in
all cases.
The starting values of the  Lagrange multipliers and the maximum training
length instead are
determined as part of the hyperoptimization procedure
described in Sect.~\ref{sec:hyperparam} below.

When performing fits in the evolution basis, this
PDF positivity constraint is applied
after performing the inverse transformation to
Eq.~(\ref{eq:flavour_basis_param_b}) in order to express the flavor
basis PDFs  $\tilde{f}_{k}$  Eq.~(\ref{eq:flav_basis}) in terms of their
evolution basis counterparts  $f_{k}$.

\subsubsection{PDF integrability}
\label{sec:pdfint}

The small-$x$ behavior of the PDFs is constrained
by integrability requirements.
First, the gluon and singlet PDFs must satisfy the momentum sum rule,
Eq.~(\ref{eq:mom_sum_rule}), which implies that
\be
\label{eq:cond_integ_1}
\lim_{x\rightarrow 0} \, x^2f_k(x,Q)= 0 \, ,\quad \forall~Q \, ,\qquad f_k=g,\,\Sigma \, ,
\ee
while the valence sum rules, Eq.~(\ref{eq:valence_sum_rules}), constrain the small-$x$
behavior of the valence distributions,
\be
\label{eq:cond_integ_2}
\lim_{x\rightarrow 0}\,  xf_k(x,Q)= 0 \, ,\quad \forall~Q \, ,\qquad f_k=V,\,V_3\,,V_8 \, .
\ee
Furthermore, as mentioned, standard Regge theory arguments suggest that 
the first moments of  the non-singlet combinations $T_3$ and $T_8$
are also finite, so for instance the  Gottfried sum (which is
proportional to the first moment of $T_3$) is finite.
This implies that also for these two combinations one has
\be
\label{eq:cond_integ_3}
\lim_{x\rightarrow 0}\,  xf_k(x,Q)= 0 \, ,\quad \forall~Q \, ,\qquad f_k=T_3,\,T_8 \, .
\ee

To ensure that these integrability requirements are satisfied,
first of all we constrain the range of the small-$x$ preprocessing
exponents $\alpha_i$ Eq.~(\ref{eq:evolution_basis_param}).
We supplement the iterative determination of the exponents described in
Ref.~\cite{Ball:2014uwa} with the constraints  $\alpha_k <2$ for the singlet and
gluon and $\alpha_k <1$ for  the nonsinglet combinations 
$xV,\,xV_3,\, xV_8,\, xT_3$ and $xT_8$. Indeed if the preprocessing 
exponent were to violate these bounds, the neural net ${\rm NN}(x;
{\boldsymbol\theta})$ in Eq.~(\ref{eq:evolution_basis_param})
would have to compensate this behavior in order for
integrability to hold. Preprocessing would then be slowing
the minimization rather than speeding it up. Note that, in
the flavor basis, the small-$x$ preprocessing exponents
are absent, so this requirement only applies to the evolution basis.

We observe that while Eq.~(\ref{eq:cond_integ_1}) always 
turns out to be satisfied automatically when fitting to the experimental
data, the additional constraints Eq.~(\ref{eq:cond_integ_2})
and~(\ref{eq:cond_integ_3}) can sometimes be violated by the fit, and
thus must be imposed. This is also achieved
through Lagrange multipliers.
We include in the total cost function  additional
contributions of the form
\begin{align}
	\label{eq:chi2int_k}
	\chi^2_{\rm tot} \to \chi^2_{\rm tot}+ \sum_k \Lambda_k^{\rm (int)} \sum_{i=1}^{n_i}\,\left[xf_k\left(x_{\rm int}^{(i)},Q^2_i\right)\right]^2\,,
\end{align}
where $f_k= T_3, T_8$ in the evolution basis while $f_k=V,V_3,V_8,T_3,T_8$ in the flavor basis. The points
$\{ x_{\rm int}^{(i)}\}$ are a set of  values in the small $x$
region, $Q^2_i$ is a suitable reference scale, and, like in the case
of positivity, the Lagrange multipliers $\Lambda_k^{(\rm int)}$
grow exponentially during the minimization, with a maximum value
$\Lambda_k^{(\rm int)}=100$   attained at maximum
training length. We choose $Q_i^2=5$ GeV$^2$ and in the evolution
basis $n_i=1$ and 
$x_{\rm int}^{(1)} = 10^{-9}$, while in the flavor basis $n_i=3$  and
$x_{\rm  int}^{(i)}=10^{-9},\,10^{-8},\,10^{-7}$.  As for
the positivity multiplier, the starting values of the Lagrange
multipliers
(as well as the maximum
training length) are hyperoptimization parameters.

Finally, we introduce a post-selection criterion, in order to discard
replicas that fail to satisfy the integrability and retain a large
value at small $x$ despite the Lagrange multiplier. It turns out that imposing
\be
\label{eq:integ_def}
\sum_{i=1}^{n_{i}} \left|  x_{\rm int}^{(i)} f_k\lp x_{\rm int}^{(i)}\rp \right|<\frac{1}{2} \, , \qquad f_k=V,V_3,V_8,T_3,T_8 \, ,
\ee
is enough to preserve integrability for all replicas. This is due to
the fact that the function $xf(x)$ at its maximum is of order one, so
the condition Eq.~(\ref{eq:integ_def}) ensures that at small $x$ it is
decreasing. When determining PDF replicas, we have
explicitly checked a posteriori that the numerical computation of the
first moment yields a finite result for all PDF replicas.

%% file: subsec-fittingframework.tex
\subsection{Fitting framework}
\label{sec:methimplementationdetails}

The machine learning approach to PDF determination that we will
discuss shortly has been made possible by a
complete restructuring of the NNPDF fitting framework.
Further motivations for this are the need to deal with a particularly
large dataset, and the  goal of releasing the NNPDF code as  open
source, which imposes stringent requirements of  quality and
accessibility. The code was written in the {\tt Python} programming language and
has been documented and tested thoroughly.
The original developments of our new fitting framework were presented in
Ref.~\cite{Carrazza:2019mzf}. The main differences between the
NNPDF3.1 and NNPDF4.0 codes are summarized in Tab.~\ref{tab:codebase}.

\begin{table}[!t]
  \scriptsize
  \centering
  \renewcommand{\arraystretch}{1.4}
  \input{tables/tab-31vs40.tex}
  \vspace{0.2cm}
  \caption{Summary of the main differences between the NNPDF3.1 and
    the NNPDF4.0 code.}
  \label{tab:codebase}
\end{table}

\subsubsection{General structure}
\begin{figure}[!t]
  \centering
  \begin{tikzpicture}[node distance=1.0cm]\scriptsize
    \definecolor{vp1}{RGB}{102,194,165}
    \definecolor{vp2}{RGB}{252,141,98}
    \definecolor{vp3}{RGB}{117,112,179}
    \tikzstyle{startingNode} = [rectangle, rounded corners, minimum width=3cm, minimum height=0.5cm, text centered, draw=black, fill=vp1!30];
    
    \node[startingNode] (hyperopt) {Hyperopt};
    \node[startingNode, above = 1.0cm of hyperopt] (fktables) {FK Tables};
    \node[startingNode, below = 1.0cm of hyperopt] (expdata) {Experimental Data};
    
    \tikzstyle{n3fitNode} = [rectangle, rounded corners, minimum width=2cm, minimum height=0.5cm, text centered, draw=black, fill=vp2!30];
    \node[n3fitNode, right = 2.2cm of hyperopt] (stopping) {Stop?};
    \node[n3fitNode, above = 0.65cm of stopping] (chisq) {Compute $\chi^{2}$};
    \node[n3fitNode, below = 0.65cm of stopping] (optimize) {Optimization};
    \draw[myarrow] (chisq) -- (stopping);
    \draw[myarrow] (stopping) -- (optimize);
    
    \tikzstyle{outputNode} = [rectangle, rounded corners, text width=1.4cm, minimum height=1.0cm, text centered, draw=black, fill=vp1!30];
    \node[outputNode, right = 1.8cm of stopping] (evolution) {APFEL evolution};
    \node[outputNode, right = 0.7cm of evolution] (postfit) {post-fit selection};
    \draw[myarrow] (evolution) -- (postfit);
    \node[outputNode, right = 0.7cm of postfit] (lhapdf) {LHAPDF grid};
    \draw[myarrow] (postfit) -- (lhapdf);
    
    \coordinate [left = 0.3cm of optimize] (lopt);
    \coordinate [left = 0.3cm of chisq] (lchi);
    \draw[myarrow] (optimize) -- (lopt) -- (lchi) -- (chisq);
    
    \draw[draw=vp3, rounded corners] ($(lchi)+(-0.4, 0.60)$) rectangle ($(optimize.south east)+(0.5, -0.45)$) coordinate (myr);
    \node at ($(myr) + (-0.4, 0.2)$) {\color{vp3}~} ;
    \draw[myarrow] (fktables) -- ($(fktables.east)+(1.5, 0.0)$);
    \draw[myarrow] (expdata) -- ($(expdata.east)+(1.5, 0.0)$);
    \draw[myarrow] (hyperopt) -- ($(hyperopt.east)+(1.5, 0.0)$);
    \draw[myarrow] (stopping)--(evolution);
  \end{tikzpicture}
  \caption{Diagrammatic representation of the NNPDF fitting
    framework. The blue box contains the minimization of the $\chi^2$
    figure of merit, whose computation is
    illustrated in Fig.~\ref{fig:n3fit}.
  }
  \label{fig:fullfitfram}
\end{figure}
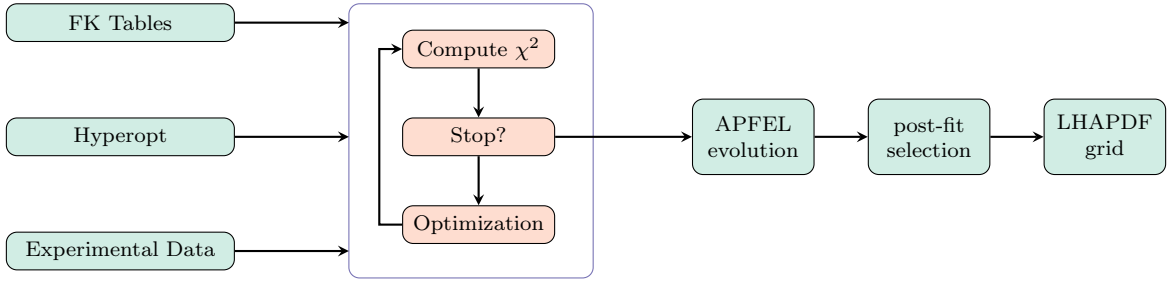

A schematic representation of
the NNPDF4.0 fitting framework is displayed in Fig.~\ref{fig:fullfitfram}.
The fit requires three main inputs, which are managed by the NNPDF framework as
discussed in Ref.~\cite{NNPDF:2021uiq}: first, theoretical calculations of physical
processes, which are encoded in precomputed tables (FK-tables, see below)
possibly supplemented by QCD and EW $K$-factors.
Second, experimental data provided in a common format, including fully
correlated uncertainties encoded in a covariance matrix (possibly also
including theoretical uncertainties).
Third,  hyperparameter settings that determine the particular fitting
methodology adopted,  determined
through a hyperoptimization procedure as discussed below.
The neural network optimization algorithm, with settings determined by the
hyperparameters, finds the best fit of predictions to data by minimizing a
figure of merit whose computation is illustrated in Fig.~\ref{fig:n3fit}.
Following a post-fit selection, where outliers with insufficient quality
are discarded, the  final  PDFs
are stored in {\tt LHAPDF} grid format so that they are readily available 
for use.      

\subsubsection{Evaluation of cross-sections and cost function}
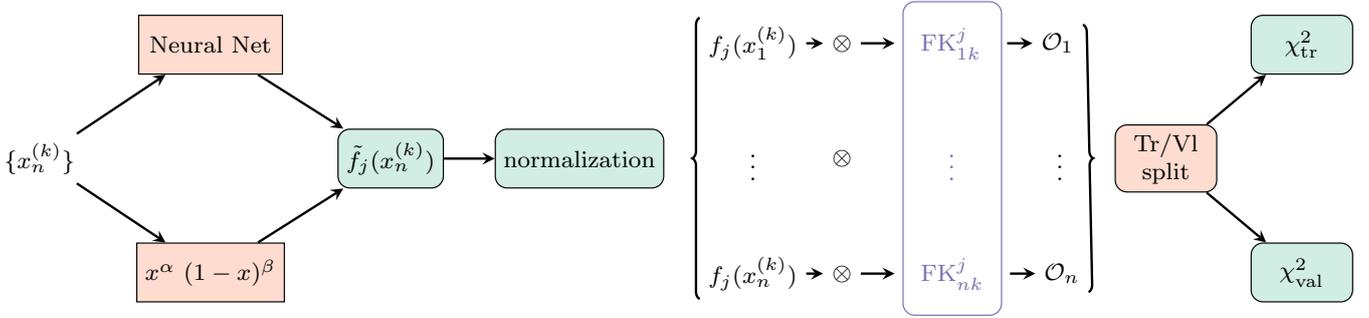
\begin{figure}[!t]
  \centering
  \resizebox{1.0\textwidth}{!}{%
    \begin{tikzpicture}[node distance = 1.0cm]\scriptsize
    \definecolor{vp1}{RGB}{102,194,165}
    \definecolor{vp2}{RGB}{252,141,98}
    \definecolor{vp3}{RGB}{117,112,179}
      \node (xinput) {$\{x_{n}^{(k)}\}$};
      
      \coordinate [right = 1.5cm of xinput] (NNghost) {};
      \node[fitted, fill=vp2!30, above = 1.0cm of NNghost, minimum width=1.7cm, minimum height=0.7cm]
      (pdf) { Neural Net};
      \node[fitted, fill=vp2!30, below = 1.0cm of NNghost, minimum width=1.7cm, minimum height=0.7cm]
      (preproc) { $x^{\alpha}$ $(1-x)^{\beta}$};
      
      \node[operations, fill=vp1!30, minimum width=1.2cm, minimum height=0.7cm, right = 1.5cm of NNghost]
      (fitbasis) {$\tilde{f}_{j}(x^{(k)}_n)$};
      \node[operations, fill=vp1!30, minimum width=1.2cm, minimum height=0.7cm, right = 0.6cm of fitbasis]
      (normalizer) {normalization};
      
      \node[right = 0.9cm of normalizer] (pdfdots) {\vdots};
      \node[above = 0.7cm of pdfdots]
      (pdf1) {${f}_{j}(x^{(k)}_1)$};
      \node[below = 0.7cm of pdfdots]
      (pdfn) {${f}_{j}(x^{(k)}_n)$};
      
      \node[right = 0.2cm of pdf1] (conv1) {$\otimes$};
      \node[right = 0.2cm of pdfn] (convn) {$\otimes$};
      \node at ($(conv1)!0.5!(convn)$) (convdots) {$\otimes$};
      
      \node[vp3, right = 0.6cm of conv1] (f1) {FK$_{1k}^{j}$};
      \node[vp3, right = 0.6cm of convn] (fn) {FK$_{nk}^{j}$};
      \node[vp3] at ($(f1)!0.5!(fn)$) (fd) {\vdots};
      \draw[draw=vp3, rounded corners] ($(f1.north west)+(-0.1, 0.2)$) rectangle ($(fn.south east)+(0.1,-0.2)$);
      
      \node[right = 0.5 cm of f1] (o1) {$\mathcal{O}_{1}$};
      \node[right = 0.5 cm of fn] (on) {$\mathcal{O}_{n}$};
      \node at ($(o1)!0.5!(on)$) (od) {\vdots};
      
      \node[operations, fill=vp2!30, right = 0.5cm of od, minimum width = 1.2cm, text width=1cm, minimum height=0.7cm]
      (trvl) {Tr/Vl split};
      \coordinate [right = 1.0cm of trvl] (ending) {};
      \path let \p1 = (ending), \p2 = (pdf)
      in node at (\x1,\y2) [n3py, fill=vp1!30, minimum width = 1.2cm, minimum height=0.7cm] (tr) {$\chi^{2}_\text{tr}$};
      \path let \p1 = (ending), \p2 = (preproc)
      in node at (\x1,\y2) [n3py, fill=vp1!30, minimum width = 1.2cm, minimum height=0.7cm] (vl) {$\chi^{2}_\text{val}$};
      
      \draw[myarrow] (xinput) -- (pdf);
      \draw[myarrow] (xinput) -- (preproc);
      \draw[myarrow] (pdf) -- (fitbasis);
      \draw[myarrow] (preproc) -- (fitbasis);
      \draw[myarrow] (fitbasis) -- (normalizer);
      
      \draw[myarrow] (pdf1) -- (conv1);
      \draw[myarrow] (pdfn) -- (convn);
      \draw[myarrow] (conv1) -- ($(f1.west)-(0.2,0.0)$) ;
      \draw[myarrow] (convn) -- ($(fn.west)-(0.2,0.0)$) ;
      \draw[myarrow] ($(f1.east)+(0.2,0.0)$) -- (o1);
      \draw[myarrow] ($(fn.east)+(0.2,0.0)$) -- (on);
      
      \draw[myarrow] (trvl) -- (tr);
      \draw[myarrow] (trvl) -- (vl);
      
      \draw[decorate, decoration={brace}, thick] (pdfn.south west) -- (pdf1.north west);
      \draw[decorate, decoration={brace},thick] (o1.north east) -- (on.south east);
    \end{tikzpicture}
  }
  \caption{Diagrammatic representation of the calculation of the $\chi^{2}$
    in the NNPDF fitting framework as a function of the values of $\{x_n^{(k)}\}$
    for the different datasets.
    Each block indicates an independent component.}
  \label{fig:n3fit}
\end{figure}

Figure~\ref{fig:n3fit} illustrates the structure of the part of
NNPDF4.0 fitting code 
that evaluates the physical observables in terms
of the input PDFs and then computes the associated figure of merit to be used 
for the fitting. This is at the core of the minimization procedure, 
indicated by a blue box in Fig.~\ref{fig:fullfitfram}.
Starting from a matrix of momentum fraction $x$ values, $\{x_n^{(k)}\}$, the code
first evaluates the neural network and the preprocessing factors to construct
unnormalized PDFs which are
then normalized according to Eqs.(\ref{eq:mom_sum_rule},\ref{eq:valence_sum_rules})
in order
to produce the PDFs at the input scale,
\be
f_{jn}^{(k)} \equiv f_{j}\lp x_{n}^{(k)},Q_0\rp \,,
\ee
where $j$, $n$, and $k$ label the PDF flavor, the experimental dataset, and the node in the
corresponding $x$-grid respectively.
These PDFs 
are those listed  in Eqs.~\eqref{eq:flav_basis} and~(\ref{eq:evolution_basis}) in the
evolution and flavor bases respectively,
and are related to the neural network output by Eq.~\eqref{eq:evolution_basis_param}.

The input scale PDFs are convoluted with
partonic scattering cross-sections (including perturbative
QCD evolution); these are encoded in precomputed grids called
FK-tables (see Refs.~\cite{Ball:2010de, Bertone:2016lga})
resulting in the corresponding physical observables $\{\mathcal{O}_n\}$.
Observables  are split into a training and a validation set  and 
cost functions $\chi^2_{\rm tr}$ and $\chi^2_{\rm val}$ are computed for
each set. The  $\chi^2$ is defined as in previous NNPDF
determinations, and in particular it uses the  $t_0$
method~\cite{Ball:2009qv} for the computation of multiplicative uncertainties.

Note that  each block in Fig.~\ref{fig:n3fit} is fully
independent, so that its settings can be modified or the whole block
can be replaced as required. This characterizes the modular structure of the code.
For instance, the block
{``Neural Net''} implements by default the neural network which after
hyperoptimization has the architecture
displayed in Fig.~\ref{fig:NNarch}, but it could be replaced by any other
parametrization, even by a quantum
circuit~\cite{Perez-Salinas:2020nem} based on the {\tt QIBO}
library~\cite{efthymiou:2020qibo}.
Similarly, the $\chi^2$ with $t_0$ uncertainties could be replaced by
any other cost function.

\subsubsection{Optimization strategy}

Previous NNPDF determinations used stochastic algorithms for the
training of neural networks, and in particular in NNPDF3.1 nodal
genetic algorithms were used.
Stochastic minimization  algorithms are less prone to end up trapped in local minima, but
are generally less efficient than deterministic minimization techniques, such as
backpropagation combined with
stochastic gradient descent (SGD).
In the approach adopted here~\cite{Carrazza:2019mzf}, the optimizer is just another
modular component of the code, to be chosen through a
hyperoptimization as we discuss shortly. The algorithms that we
consider are  SGD algorithms
implemented in the {\tt Tensorflow}~\cite{tensorflow2015:whitepaper}
package. Restricting to gradient descent algorithms ensures greater
efficiency, while the use of hyperoptimization guarantees against the
risk of missing the true minimum or overfitting.
The {\tt TensorFlow} library provides automated differentiation capabilities,
which enables the use of arbitrarily complex network
 architectures without having to provide analytical expressions
for their gradients.
However, the whole convolution between input PDFs and FK-tables,
indicated in Fig.~\ref{fig:n3fit} between brackets, needs to be
provided to the optimization library in order to use gradient based
algorithms.
The specific SGD optimizer and its settings are determined via the
hyperoptimization procedure described in Sect.~\ref{sec:hyperparam}.
In comparison to the genetic algorithms used in previous
NNPDF releases,
the hyperoptimized SGD-based optimizers improve both replica stability
and computational efficiency, as we demonstrate in Sect.~\ref{sec:benchmark}
below.

\subsubsection{Stopping criterion and post-fit selection}

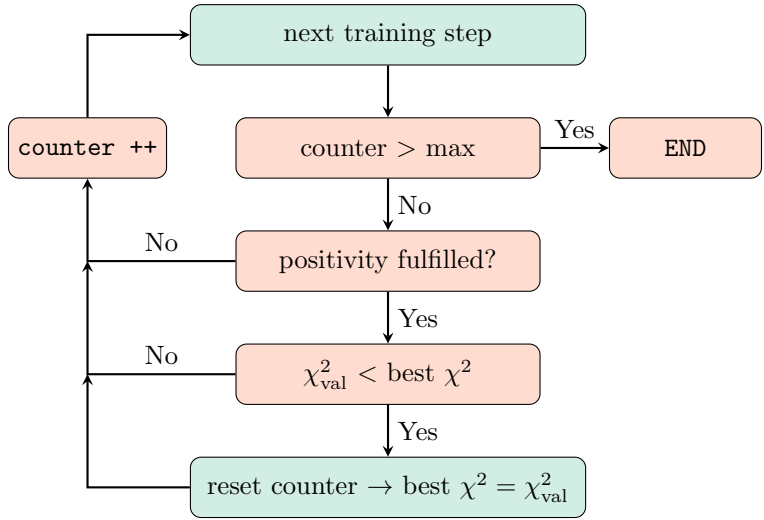
\begin{figure}[!t]
  \centering
  \begin{tikzpicture}[node distance = 1.5cm]\small

    \definecolor{vp1}{RGB}{102,194,165}
    \definecolor{vp2}{RGB}{252,141,98}
    \node (init) [roundtext, fill=vp1!30, minimum width=5.2cm] {next training step};
    \node (ccheck) [roundtext, fill=vp2!30, minimum width=4cm, minimum height=0.8cm, below of = init] {counter
      $>$ max};
    \node (pcheck) [roundtext, fill=vp2!30, minimum width=4cm, minimum height=0.8cm, below of = ccheck]
	    {positivity fulfilled?};
	\node (xcheck) [roundtext, fill=vp2!30, minimum width=4cm, minimum height=0.8cm, below of = pcheck]
	    {$\chi^2_\text{val}$ $<$ best $\chi^2$};
    \node (reset) [roundtext, fill=vp1!30, minimum width=5.2cm, below of = xcheck] {reset counter $\rightarrow$  best
		$\chi^2 = \chi^2_\text{val}$};Q

	\node (cplus) [roundtext, fill=vp2!30, left = 0.9cm of ccheck]
        {\tt counter ++};
	\node (end) [roundtext, fill=vp2!30, right = 0.9cm of ccheck]
        {\tt END};

	\coordinate[above of = cplus] (li);
	\coordinate[below of = cplus] (lp);
	\coordinate[below of = lp] (lx);
	\coordinate[below of = lx] (lr);

	\draw[myarrow] (init) -- (ccheck);
	\draw[myarrow] (ccheck) -- node[right] {No} (pcheck);
	\draw[myarrow] (pcheck) -- node[right] {Yes}(xcheck);
	\draw[myarrow] (xcheck) -- node[right] {Yes}(reset);

	\draw[myarrow] (ccheck) --  node[above] {Yes} (end);

	\draw[myarrow] (reset) -- (lr) -- (lx);
	\draw[myarrow] (xcheck) -- node[above] {No} (lx) -- (lp);
	\draw[myarrow] (pcheck) --  node[above] {No} (lp) -- (cplus);
	\draw[myarrow] (cplus) -- (li) -- (init);
  \end{tikzpicture}
  \caption{Flowchart describing the patience algorithm used in NNPDF4.0 to determine
    the optimal length of the fit based on the look-back cross-validation stopping method.}
  \label{fig:stopping}
\end{figure}

As in previous NNPDF releases, a cross-validation method is used in
order to avoid overfitting, which could lead the
 neural networks to learn noise (such as  statistical fluctuations) in the
 data, rather than the underlying law. This is done through the patience algorithm shown diagrammatically in
Fig.~\ref{fig:stopping}.
This algorithm is based on the look-back cross-validation stopping method~\cite{Ball:2014uwa}, whereby the optimal
length of the fit is determined by the absolute minimum of $\chi^2_{\rm val}$ evaluated
over a sufficiently large number of iterations of the minimizer.
Specifically, the stopping algorithm keeps track of the training step with the lowest $\chi^2_{\rm val}$,
and as soon as this value does not improve for a given number of steps (set equal to
a percentage of the maximum number of training epochs), the fit is finalized.

There are three main differences between the stopping criterion used in
NNPDF4.0 and that of its predecessor used for NNPDF3.1.
First, the patience parameter is hyperoptimized, while 
previously
it was set to be infinity, i.e., the values of $\chi^2_{\rm val}$ were monitored
until the maximum number of iterations was reached.
Second, the percentage of data that enters the training set has been increased to 75\%
for all datasets. This is motivated by the observation that the
current dataset is so wide that even with just 25\% validation
overlearning does not occur in practice. In fact, even with the previous
NNPDF3.0 dataset  it was observed in the framework of
closure testing in Ref.~\cite{Ball:2014uwa}  that larger training fractions
lead to essentially equivalent results. The faithfulness of
results found with this training fraction will be confirmed by  closure
test studies in  Sect.~\ref{sec:closure} below.
Third, the stopping algorithm now also tracks the positivity requirement
so that a fit cannot stop if the positivity condition is not satisfied.
Instead in NNPDF3.1 replicas which were not fulfilling positivity could be generated
and had to be discarded a posteriori. This is now done by verifying that the penalty term of Eq.~\eqref{eq:chi2pos_k}
      remains below the  threshold value $10^{-6}$ (numerically zero).

Once the optimal stopping point for a given fit has been identified,
the same post-fit quality checks that were imposed in
NNPDF3.1 are still enforced.
Specifically, we remove replicas with too large $\chi^2$ values or with too large
arc-lengths: in both cases, defined as replicas outside the $4\sigma$ interval of their distribution.
The post-fit selection algorithm also removes replicas
 that do not satisfy either the positivity or the integrability
 conditions. Imposing positivity and integrability constraints through post-fit
 selection has the consequence of making the fit results independent
 of the way the constraints are imposed: for instance, a looser
 constraint will simply have the effect of increasing the number of
 replicas that are discarded. 

It is interesting to note that while previously on average around 30\% of the
fitted replicas were discarded upon applying these criteria, in NNPDF4.0 this fraction has
been reduced to around 1\%.
This improvement is largely the result of the improved handling of these
constraints during the fit as well as of the higher stability of the new
SGD-based optimization strategy, which results in smoother PDFs with fewer
outliers.

%% file: tables/tab-31vs40.tex
\begin{tabularx}{\textwidth}{XX}
  \toprule
  NNPDF3.1
  & NNPDF4.0\\
  \midrule
 Genetic Algorithm optimizer
 & Gradient Descent optimization\\

 one network per PDF & one network for all PDFs\\

 sum rules imposed outside optimization & sum rules imposed during
 optimization\\

 C++ monolithic codebase & Python object-ordiented codebase \\

 fit parameters manually chosen
 (manual optimization) & fit parameters automatically chosen
 
 (hyperoptimization)\\
 in-house ML framework & complete freddom in ML library choice
 (e.g. tensorflow)\\

 private code & fully public open-source code \\
 
  \bottomrule
\end{tabularx}

%% file: subsec-hyperopt.tex
\subsection{Hyperparameter optimization}
\label{sec:hyperparam}

Hyperoptimization is at the heart of the construction of the NNPDF4.0
methodology. In brief, hyperoptimization selects the methodology, just like
gradient descent selects the values of weights and thresholds of the
neural net. The  $k$-folding method, to be discussed below,
ensures that a proper fitting (i.e. not
over- or under-fitting methodology) is arrived at, just like
cross-validation achieves the same goal for neural network training.

Indeed, the optimization procedure (neural network training) described in \secref{sec:methimplementationdetails}
requires as input a number of methodological choices, such as the neural network
architecture, the training rate, and the specific SGD variant to be
used.
We can view these choices as the set of hyperparameters that defines
a specific fitting strategy.
While in many ML studies (including previous NNPDF determinations)
these hyperparameters are determined by trial and error,
here we implement an automated algorithmic procedure to  scan the space of hyperparameters
and determine the optimal configuration according to a figure of merit.

In this work, the implementation of the hyperparameter scan is based on the
\texttt{hyperopt} library~\cite{Bergstra:2013}, which uses a Bayesian optimization
algorithm~\cite{Bergstra:2011:AHO:2986459.2986743} to identify the best configuration.

\begin{figure}[!t]
  \centering
  \includegraphics[width=0.49\textwidth]{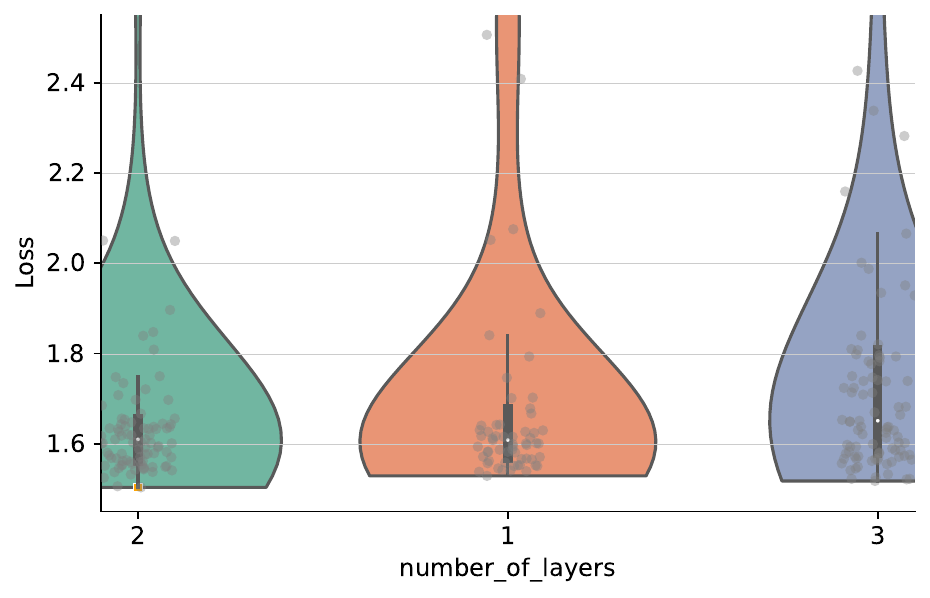}
  \includegraphics[width=0.49\textwidth]{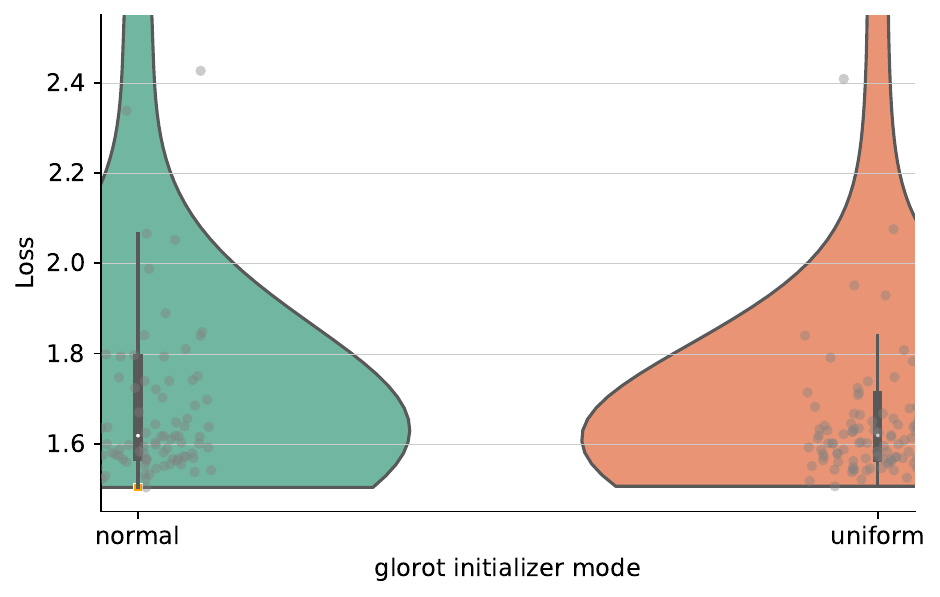}\\
  \includegraphics[width=0.49\textwidth]{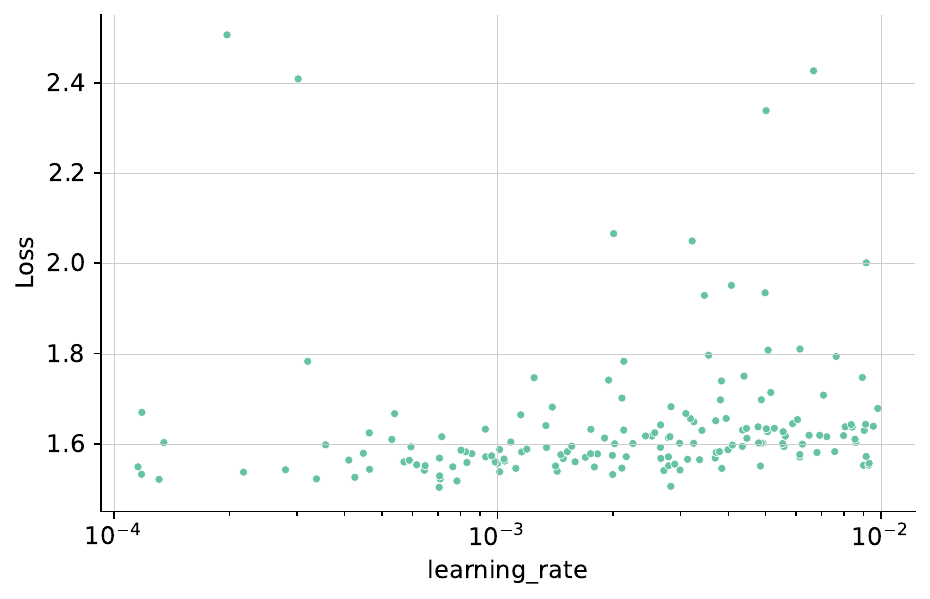}
  \includegraphics[width=0.49\textwidth]{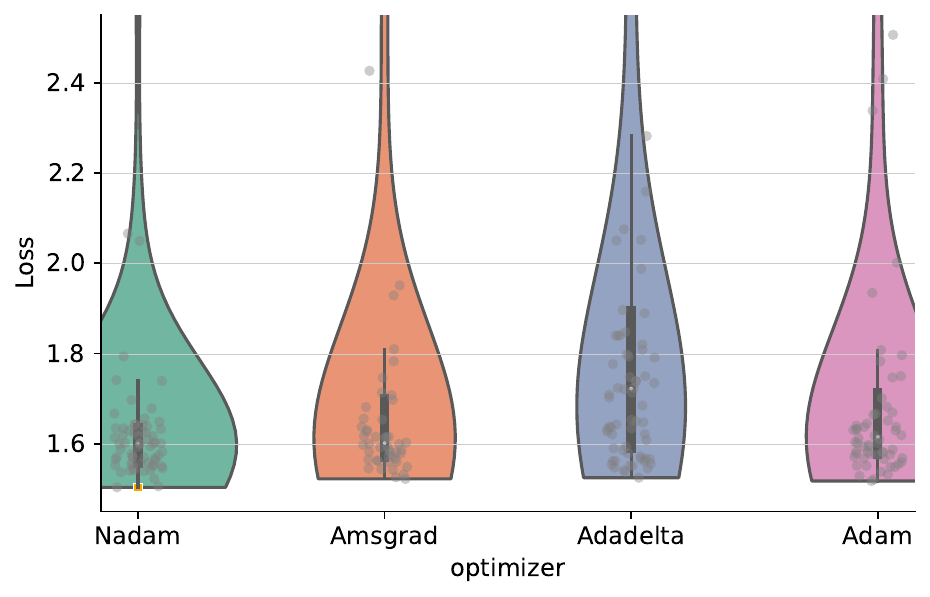}\\
  \caption{Graphical representation of the hyperoptimization loss function $L$
    corresponding to a subset of the hyperparameters in a scan based on 1500 configurations.}
  \label{fig:hyperplots}
\end{figure}

In order to visualize a typical output of a hyperparameter scan, we show in
Fig.~\ref{fig:hyperplots} the result of a scan based on 1500
independent configurations. 
We display the hyperoptimization loss function $L$ (figure of merit), to be defined below,
for a representative subset of hyperparameters: the depth of the network,
the algorithm for the initialization of the network weights, the
learning rate and the SGD optimizer variant.
The smaller the value of the loss function $L$, the better this specific point
is in the hyperparameter space.
The full list  of  hyperparameters is given in \tableref{tab:setup}.
Note that here we only display the outcome of hyperparameter configurations
that satisfy the post-fit selection cuts.
The shape of the reconstructed probability
distributions provides an indication of the stability of the
results, with a wider distribution corresponding to a higher stability with respect
to this specific hyperparameter.

In the specific case of the  number of hidden layers of the network,
one observes that the
hyperoptimization algorithm identifies that it cannot further improve the figure of merit with one single
layer, and accordingly it tests more configurations with two and three layers.
The hyperparameter configurations corresponding to two and three layers appear
to be equivalent
in terms of the loss $L$, with a slightly better stability towards lower values in the two-layer case.
No clear preference for a specific SGD variant is observed.

\subsubsection{Figure of merit and stability}
The complex interplay between hyperparameters indicates that a judicious choice
of the figure of merit $L$
is crucial for the success of the hyperoptimization procedure.
This figure of merit must relate to the quality of the fit: a possible
choice would be setting the hyperoptimization loss to the validation $\chi^2$, that is,
$L=\chi^{2}_\text{val}$.
However, this quantity is  already used in the stopping algorithm (Fig.~\ref{fig:stopping})
and hence using it may  lead to hyperparameter configurations prone to
over fitting~\cite{Carrazza:2019mzf} (``Goodhart's law'', see
Ref.~\cite{Hawkins}) .
Rather, we define the loss $L$ through a $k$-fold cross validation method~\cite{Schaffer93selectinga}.

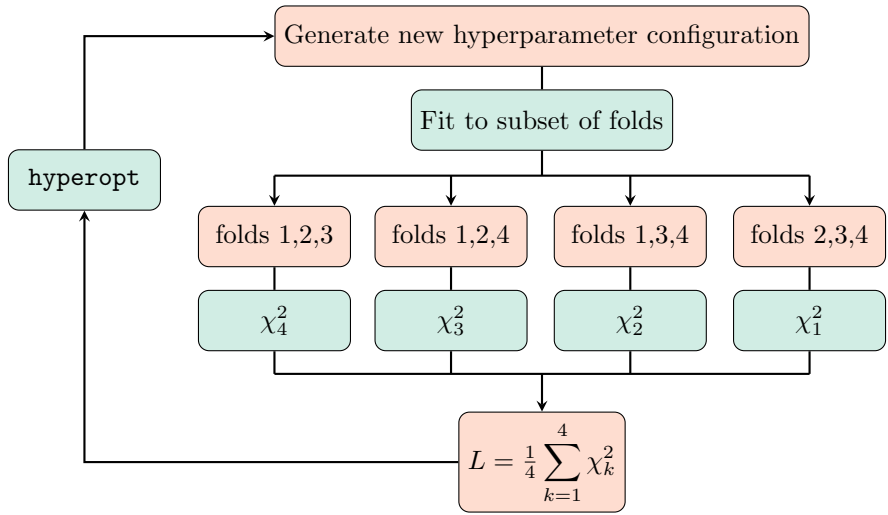
\begin{figure}[!t]
  \centering
  \begin{tikzpicture}[node distance = 1.0cm]\small
    \definecolor{vp1}{RGB}{102,194,165}
    \definecolor{vp2}{RGB}{252,141,98}
    \node[roundtext, fill=vp1!30] (hyperopt) {\texttt{hyperopt}};
    \coordinate [above = 1.5cm of hyperopt] (abovehyperopt) {};
    
    \node[roundtext, fill=vp2!30, right = 2.5cm of abovehyperopt] (xplain) {Generate new hyperparameter configuration};
    \draw[myarrow] (hyperopt) -- (abovehyperopt) -- (xplain);
    
    \coordinate [below = 1.85cm of xplain.west] (fold4v) {};
    \coordinate [below = 1.85cm of xplain.east] (fold1v) {};
    \coordinate (arrowcenter) at ($(fold4v)!0.5!(fold1v)$);
    \coordinate (fold3v) at ($(fold4v)!0.66!(arrowcenter)$);
    \coordinate (fold2v) at ($(arrowcenter)!0.33!(fold1v)$);

    \node [roundtext, fill=vp1!30, above = 0.33cm of arrowcenter] (fitto) {Fit to subset of folds};
    \draw[thick] (xplain) -- (fitto);

    \draw[thick] (fold4v) -- (fold1v);
    \draw[thick] (fitto) -- (arrowcenter);

    \node[roundtext, fill=vp2!30, below = 0.4cm of fold4v] (fold4) {folds 1,2,3};
    \node[roundtext, fill=vp2!30, below = 0.4cm of fold1v] (fold1) {folds 2,3,4};
    \node[roundtext, fill=vp2!30, below = 0.4cm of fold3v] (fold3) {folds 1,2,4};
    \node[roundtext, fill=vp2!30, below = 0.4cm of fold2v] (fold2) {folds 1,3,4};
    \draw[myarrow] (fold1v) -- (fold1);
    \draw[myarrow] (fold4v) -- (fold4);
    \draw[myarrow] (fold2v) -- (fold2);
    \draw[myarrow] (fold3v) -- (fold3);

    \node[roundtext, fill=vp1!30, below = 0.30cm of fold4] (chi24) {$\chi^{2}_{4}$};
    \node[roundtext, fill=vp1!30, below = 0.30cm of fold3] (chi23) {$\chi^{2}_{3}$};
    \node[roundtext, fill=vp1!30, below = 0.30cm of fold2] (chi22) {$\chi^{2}_{2}$};
    \node[roundtext, fill=vp1!30, below = 0.30cm of fold1] (chi21) {$\chi^{2}_{1}$};

    \draw[thick] (fold1) -- (chi21);
    \draw[thick] (fold2) -- (chi22);
    \draw[thick] (fold3) -- (chi23);
    \draw[thick] (fold4) -- (chi24);

    \coordinate [below = 0.3cm of chi24] (below4) {};
    \coordinate [below = 0.3cm of chi21] (below1) {};
    \coordinate [below = 0.3cm of chi22] (below2) {};
    \coordinate [below = 0.3cm of chi23] (below3) {};

    \draw[thick] (below1) -- (below4);
    \draw[thick] (chi24) -- (below4);
    \draw[thick] (chi23) -- (below3);
    \draw[thick] (chi22) -- (below2);
    \draw[thick] (chi21) -- (below1);

    \coordinate (belowcenter) at ($(below4)!0.5!(below1)$);
    \node[operations, fill=vp2!30, below = 0.5cm of belowcenter] (loss) {$L = \frac{1}{4}\displaystyle\sum^{4}_{k=1}\chi^{2}_{k}$};
    \draw[myarrow] (belowcenter) -- (loss);
    \path let \p1 = (hyperopt), \p2 = (loss)
      in coordinate (lleft) at (\x1,\y2);

    \draw[myarrow] (loss) -- (lleft) -- (hyperopt);

  \end{tikzpicture}
  \caption{Diagrammatic representation of the $k$-fold algorithm
    used for the hyperparameter optimization.
   }
  \label{fig:kfolds}
\end{figure}
A diagrammatic representation of the $k$-fold algorithm
used for the hyperparameter optimization is displayed
in Fig.~\ref{fig:kfolds}.
The \texttt{hyperopt} library generates a large number of
hyperparameter configurations, and each of these is then
used to produce fits to  subsets  of the
experimental data.
Specifically, for each point in the hyperparameter space we run $n_\text{fold}$ fits
to the central experimental data, where $n_\text{fold}$ is
the number of sets (folds) in which
the data are being divided.  We run  a single fit to central
data, rather than the
standard set of around 100 replicas, because
we prefer to scan over a very large number of
hyperparameters, and fitting many replicas in each case would be
computationally too intensive. 
In each of these $n_\text{fold}$ fits, the $k$-th fold
is left out; the remaining  folds are combined in a dataset which is
then separated into training and validation in the usual way, such
that the patience stopping of 
Fig.~\ref{fig:stopping} can be tested.

The loss figure of merit $L$ is then defined as the average of the $\chi^2$ for
the $k$-th, fold evaluated with the  PDFs obtained in the $k$-th fit, in which
this specific fold was left out, dubbed $\chi^2_k$ as illustrated in
Fig.~\ref{fig:kfolds}; that is
\begin{equation}
  L =  \frac{1}{n_\text{fold}}
  \displaystyle\sum^{n_\text{fold}}_{k=1} \chi_{k}^2 \, .
  \label{eq:hyperoptloss_v2}
\end{equation}
We use the $n_{\rm fold}=4$ folds defined in
\tableref{table:kfolds}. These  are chosen in such a way
that each fold is representative of the global dataset,
both in terms of process type  and kinematic coverage.
The optimal hyperparameter set ${\boldsymbol{ \hat{\theta}} }$
is then
selected to be those that produce the lowest average loss computed using
Eq.~(\ref{eq:hyperoptloss_v2}),
\begin{equation}
\boldsymbol{\hat{\theta}} = \underset{\boldsymbol{\theta} \in {\boldsymbol{\Theta}}}{\text{arg min}}\left( \frac{1}{n_\text{fold}}
	\displaystyle\sum^{n_\text{fold}}_{k=1}  \chi_{k}^2({\boldsymbol{\theta}})  \right).
	\label{eq:hyperoptloss}
\end{equation}

\begin{figure}[t]
  \centering
  \includegraphics[width=0.48\linewidth]{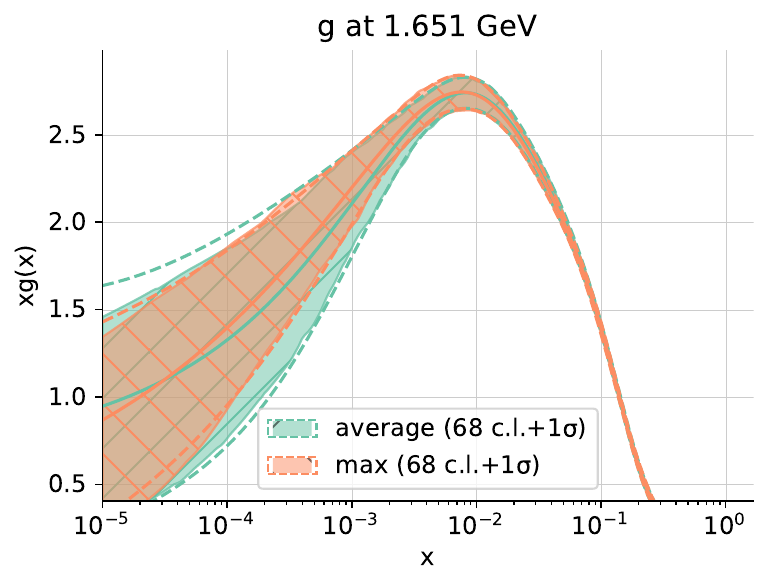}
  \includegraphics[width=0.48\linewidth]{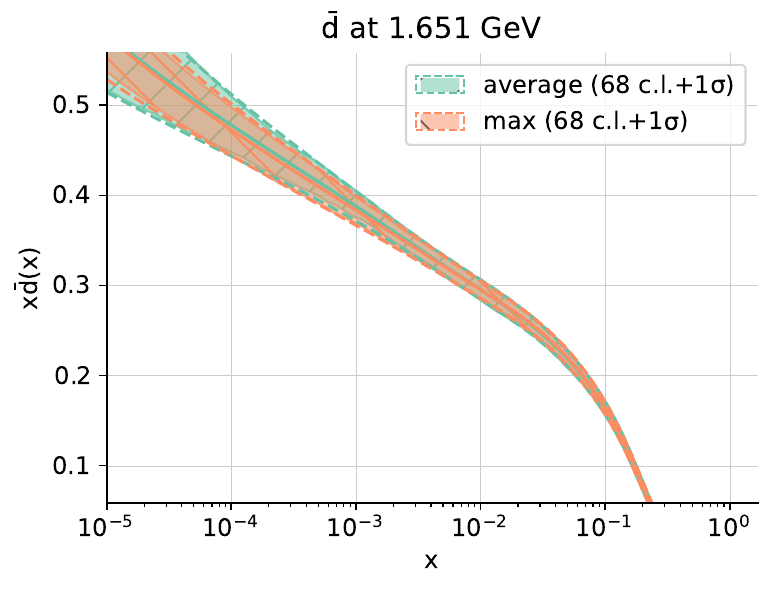}
  \caption{Comparison between the gluon (left) and
    antidown (right) PDFs  at $Q=1.65$~GeV found by using methodologies
    in which  hyperparameters are selected based on the ``average'' loss
  function Eq.~\eqref{eq:hyperoptloss_v2} (green) or the  ``max'' loss
  function Eq.~\eqref{eq:Lmax=} (orange).}
  \label{fig:compare_hyperopt_setups}
\end{figure}

We note that other choices of the loss function would be possible, such as
\be
\label{eq:Lmax=}
L =  {\rm max}\lp  \chi_{1}^2,  \chi_{2}^2, \chi_{3}^2,\ldots, \chi_{n_{\rm fold}}^2 \rp,
\ee
namely, the maximum value of $\chi_{k}^2$ evaluated over the $n_{\rm
  fold}$ folds. We checked that results obtained with either choice
are completely equivalent. In  Fig.~\ref{fig:compare_hyperopt_setups}
we compare PDFs obtained by methodologies found by hyperoptimizing
either with the  ``average'' loss function of
Eq.~\eqref{eq:hyperoptloss_v2}, or the ``max'' loss function of
Eq.~\eqref{eq:Lmax=}. The final hyperparameter values found in either
case are provided in \tableref{tab:setup}.
It is clear that these final setups are quite different, yet  the
PDFs found with either methodology are indistinguishable.
The fact that different choices for the hyperopt loss function $L$
result in rather different hyperparameter configurations that still produce
indistinguishable PDFs demonstrates the stability of our methodology with
respect to variations of the hyperoptimization procedure.

\input{tables/tab-hyperopt_folds} 

\subsubsection{Hyperparameter correlation}
An important motivation for  the automated hyperparameter optimization procedure is the fact that
the best value for a single hyperparameter cannot be determined independently of all the others, since
there is a high degree of correlation between them.
For instance, each variant of the SGD optimizer will have a
different optimal value
of the learning rate.
We illustrate this interdependence with a specific hyperparameter,
the \texttt{clipnorm} parameter of {\tt TensorFlow} optimizers,
for which a wrong choice can lead
to significant overfitting even when all other hyperparameters are optimized.
This parameter specifies the value at which to clip the norm of the gradient during a gradient descent step.
That is, if the norm of the gradient at a given epoch is larger than the value
of the \texttt{clipnorm} parameter, it will be rescaled such that the norm of
the gradient used to update the neural network parameters has the
\texttt{clipnorm} value.

\begin{figure}[!t]
  \centering
  \includegraphics[width=0.49\linewidth]{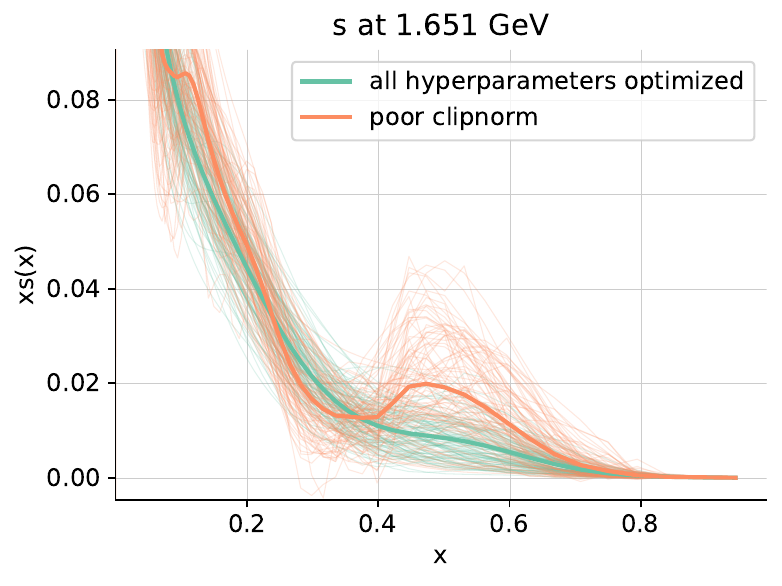}\\
  \caption{Comparison between the results for the strange
    PDF and large $x$ in two fits, one with all hyperparameters optimized and
    another where the \texttt{clipnorm} one is not hyperoptimized.}
  \label{fig:clipstrange}
\end{figure}

The choice of \texttt{clipnorm} will affect the results of the optimization algorithm:
if it is too small it can prevent convergence, while
if it is too large the training will be unstable often leading to overfitting.
In Fig.~\ref{fig:clipstrange} we compare the strange PDF $xs(x,Q)$ at $Q=1.7$~GeV
in the large-$x$ region for two variants of the NNPDF4.0 fit.
In the first one, all the hyperparameters listed in \tableref{tab:setup} enter the hyperopt
procedure, while in the second
\texttt{clipnorm} is excluded and fixed by hand to an arbitrary value.
While the two resulting hyperparameter configurations lead to
similar values of the optimization figure of merit, the PDFs
obtained in the latter case
display undesirable overfitting behavior.
This comparison illustrates the importance of including all relevant
hyperparameters in the automated optimization.

\subsubsection{Baseline hyperparameters for NNPDF4.0}

We have performed a $k$-folding hyperoptimization, as described
above, and
we have determined
the best values of the  hyperparameters that will be used for the
NNPDF4.0 determination. These 
are listed in \tableref{tab:setup}.
The hyperparameters include the network architecture,
the type of activation function,
the Glorot-type~\cite{glorot:2010} initializer, the optimizer, the values of the learning rate and of {\tt clipnorm},
the maximum number of iterations and the stopping patience, and the initial values of the Lagrange
multipliers for the PDF positivity and integrability constraints. The ranges of the
hyperparameters that are sampled by the hyperoptimization algorithm are chosen empirically:
we start out conservatively with very wide ranges, and
once we are confident that the optimal value
of a given hyperparameter falls within a sub-domain of this (conservative) range, we adjust the
sampled domain accordingly to limit the runtime and computational resources of the hyperparameter
scan.

\begin{table}[!t]
  \scriptsize
  \centering
  \renewcommand{\arraystretch}{1.4}
  \input{tables/tab-setup.tex}
  \vspace{0.2cm}
  \caption{The baseline hyperparameter configuration (left) selected using the
    $k$-folds hyperoptimization procedure with hyperoptimization loss
    Eq.~\eqref{eq:hyperoptloss} and used to perform the NNPDF4.0 fits
    in the evolution basis.
    We also show
    a configuration selected using the alternative hyperoptimization
    loss Eq.~\eqref{eq:Lmax=} (center) and the hyperparameter
    configuration employed to perform fits in the flavor basis,
    Eq.~\eqref{eq:flav_basis} (right).}
  \label{tab:setup}
\end{table}

In \tableref{tab:setup} we show both the optimal hyperparameters for
our default methodology, based on the evolution basis and the hyperoptimization
loss defined in Eq.~\eqref{eq:hyperoptloss}, as well as the hyperparameter values obtained 
with the different choice of loss function Eq.~\eqref{eq:Lmax=},
or with the same loss function but in the flavor basis. As mentioned
both different  choices of loss function (see
Fig.~\ref{fig:compare_hyperopt_setups}) or a different choice of basis
(see Sect.~\ref{subsec:flavbasis} below) lead to equivalent results, but the
corresponding hyperparameter values can be quite different.
For instance, the optimal architecture for fits based
on the alternative loss function Eq.~\eqref{eq:Lmax=} has more than twice
the number of neurons in the hidden layers compared to the baseline settings.

\begin{figure}[t!]
  \begin{center}
\includegraphics[width=0.9\linewidth]{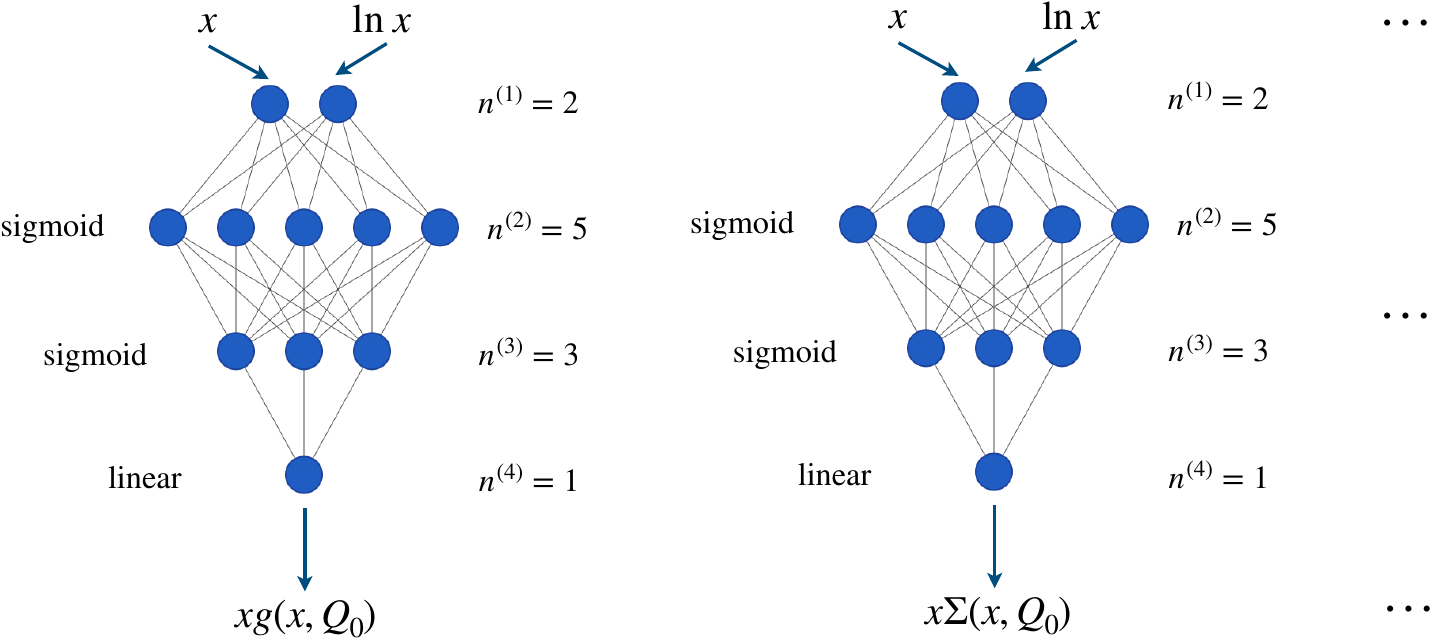}
\caption{\small The neural network architecture adopted in all
  previous NNPDF determinations up to NNPDF3.1.
  Each independent PDF  combination is parametrized
  by a separate neural network, all sharing a common architecture.
  \label{fig:NNarch_31} }
\end{center}
\end{figure}

\begin{figure}[t!]
  \begin{center}
\includegraphics[width=0.9\linewidth]{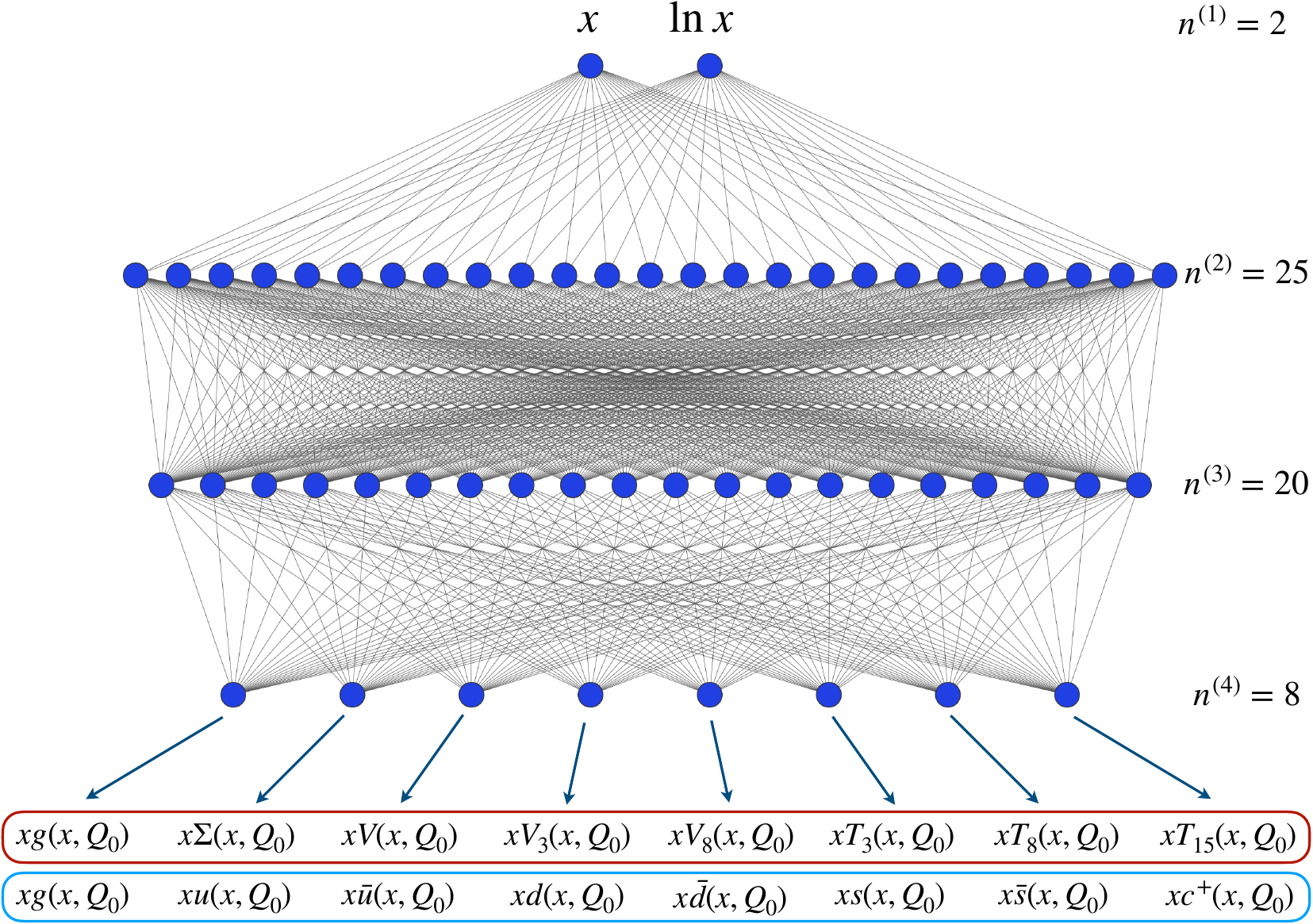}
\caption{\small The neural network architecture adopted
  for NNPDF4.0. 
  A single network is used, whose  eight output values are
  the PDFs in the evolution 
  (red) or the flavor basis (blue box). The architecture displayed
  corresponds to the optimal choice in the evolution basis; the
  optimal architecture in the flavor basis is different as indicated
  by Table~\ref{tab:setup}). 
  \label{fig:NNarch} }
\end{center}
\end{figure}

We now specifically discuss the hyperoptimization and its results for
our default choice.
Concerning the network architecture, until NNPDF3.1, each PDF was
parametrized with an individual neural network. While the number of
independently parametrized PDFs was gradually increased, this
 remained unchanged since NNPDF1.0~\cite{Ball:2008by}.
Now the
hyperoptimization scan is run with a  single
network which outputs the value of all PDFs. So while in all NNPDF fits 
up to and including NNPDF3.1 ${\rm NN}_k(x; {\boldsymbol\theta})$ in
Eq.~(\ref{eq:evolution_basis_param}) denotes the $k$-th neural
network, in NNPDF4.0 it indicates the activation state of the $k$-th neuron
in the last layer of the neural net.
The architecture used in all previous NNPDF releases, namely
 2-5-3-1 with  sigmoid activation functions and a last linear layer is depicted in
 Fig.~\ref{fig:NNarch_31}. The architecture 
 selected by the hyperoptimization 
is 2-25-20-8 with
hyperbolic activation functions
except for the final linear layer, and it is shown in Fig.~\ref{fig:NNarch}.

The NNPDF4.0 architecture has 763 free parameters, to be  compared
to a total of 296 parameters for the NNPDF3.1 neural nets.
We emphasize however that a larger network does not necessarily imply better performance,
and that for a given dataset there exists a lower bound to the
number of required free network parameters but probably not an upper one.
Given comparable performance, smaller networks are preferred in order to reduce the computational costs.

The differences between the optimizer variants are quite subtle.
While all optimizers exhibit a reasonable performance,
it is also found that after  hyperoptimization
\texttt{Nadam} results in lower absolute losses $L$ than the other
optimizers, while also appearing to be more stable.
This further illustrates the benefits of hyperoptimization.  Indeed,
separately, the stability and general performance of all optimizers is quite
similar, as can be seen in Fig.~\ref{fig:hyperplots}. This is something one
might have also found by trial and error. However, a configuration
\texttt{Nadam} that outperforms the other optimizers can be found thanks to the
simultaneous sampling of different hyperparameters. This is something that
cannot be concluded based on visual inspection of Fig.~\ref{fig:hyperplots} and
that would have been very difficult to establish by trial and error. It is
supported by the fact that the top of the ranking of setups with the smallest
losses is dominated by setups that use the \texttt{Nadam} optimizer.

\subsubsection{Hyperoptimization stability}
The main goal of the hyperoptimization procedure is to identify the best
optimization settings for the current problem of determining the PDFs. This
raises the question of deciding in which cases a new hyperoptimization would be
required.  Our current understanding encompasses changes to the experimental
data, the theoretical description, and methodological choices (such as the
choice of PDF basis).

We have checked that the procedure is quite stable
upon reasonably small changes of the dataset. For instance, the
appraisal and selection of the final dataset, see
Sect.~\ref{sec:dataselection} below, did not require any new
hyperoptimization. In fact,  the datasets included in \tableref{table:kfolds}
do not correspond exactly to the datasets included in the final
dataset, since  the final appraisal of the data to be included was 
performed after the methodology was set.
Furthermore, when removing datasets the given methodology remains
viable, though in principle there might be a computationally more
efficient one giving the same results for the small datasets. This
will be seen explicitly in the context of  ``future tests'' in
\secref{sec:futuretest} below. Of course in principle the only way of
being absolutely certain whether a new hyperoptimization is  needed or
not is to
actually perform it.

On the other hand, a substantial change in methodology or 
dataset generally needs a new hyperoptimization. This is
illustrated by the fact (see Tab.~\ref{tab:setup}) that the optimal
settings for fitting in the flavor basis differ substantially from
those of the evolution basis.
Likewise, the addition of a large number of new datasets affecting
kinematic regions or PDF combinations for which currently there is
little or no information might have an impact on the
fit sufficient to warrant a new run of the hyperoptimization procedure.

The open source NNPDF4.0 fitting framework released with this paper
includes all necessary tools to carry out an automatic scan
of hyperparameters, which means it can be readily used in situations which are
very wildly different from the specific scenario considered in this work,
be it in terms of the experimental data available
or the theoretical framework being considered.

%% file: tables/tab-hyperopt_folds.tex
\newenvironment{foldstabular}[1]{
    \begin{tabular}{ C{5.7cm}  C{5.45cm}  C{5.45cm} } \toprule
        \multicolumn{3}{ c }{\bf Fold #1} \\ \midrule
    } 
    { \bottomrule
    \end{tabular}
}
\begin{table}[!t]
    \centering
    \scriptsize
     \renewcommand{\arraystretch}{1.40}
    \begin{foldstabular}{1}
CHORUS $\sigma_{CC}^{\nu}$                              & HERA I+II $\sigma_{\rm NC}^{p}\ e^+$ (920~GeV) & BCDMS $F_2^p$ \\ 
LHCb $Z\to ee$ 7 TeV                                    & ATLAS $W,Z$ 7 TeV ($\mathcal{L}=35$~pb$^{-1}$) & CMS $Z$ $p_T$ 8 TeV\\ 
E605 $\sigma^p$                                         & CMS DY 2D 7 TeV                                & CMS 3D dijets 8 TeV\\ 
ATLAS single~$t$ 7 TeV ($1/\sigma d\sigma/dy_{\bar t}$) & ATLAS single $t$ $R_{t}$ 7 TeV                 & CMS $t\bar{t}~\ell$+jets 8 TeV ($1/\sigma d\sigma/dy_{t\bar t}$)\\ 
CMS single $t$ $R_{t}$ 8 TeV                            &                                                & \\
    \end{foldstabular}
    \vfill\vspace{0.2cm}
    \begin{foldstabular}{2}
HERA I+II $\sigma_{\rm CC}^{p}\ e^-$                    & HERA I+II $\sigma_{\rm NC}^{p}\ e^+$ (460~GeV) & HERA I+II $\sigma_{\rm NC}^{b}$\\ 
NMC $\sigma^{{\rm NC},p}$                               & NuTeV $\sigma_{CC}^{\bar{\nu}}$                & LHCb $Z\to ee$ 8 TeV\\
CMS $W$ electron asymmetry 7 TeV                        & ATLAS $Z$ $p_T$ 8 TeV ($p_T,m_{\ell\ell}$)     & D0 W muon asymmetry\\ 
E866 $\sigma^p$ (NuSea)                                 & ATLAS isolated $\gamma$ prod. 13 TeV           & ATLAS dijets 7 TeV, R=0.6\\ 
ATLAS single~$t$ 8 TeV ($1/\sigma d\sigma/dy_{\bar t}$) & CMS $\sigma_{tt}^{\rm tot}$ 7,8 TeV            & CMS single $t$ $\sigma_{t}+\sigma_{\bar{t}}$ 7 TeV\\
    \end{foldstabular}
    \vfill\vspace{0.2cm}
    \begin{foldstabular}{3}
HERA I+II $\sigma_{\rm CC}^{p}\ e^+$                    & HERA I+II $\sigma_{\rm NC}^{p}\ e^+$ (575~GeV)   & NMC $F_2^d/F_2^p$ \\ 
NuTeV $\sigma_{CC}^{\nu}$                               & LHCb $W,Z \to \mu$ 7 TeV                         & LHCb $Z\to ee$ 13 TeV \\ 
ATLAS $W,Z$ 7 TeV ($\mathcal{L}=4.6$~fb$^{-1}$) central & ATLAS $W^+$+jet 8 TeV                            & ATLAS high-mass DY 7 TeV \\ 
CMS $W$ muon asymmetry 7 TeV                            & E866 $\sigma^d/2\sigma^p$ (NuSea)                & CDF $Z$ differential\\ 
ATLAS $\sigma_{tt}^{\rm tot}$ 7,8 TeV                   & ATLAS single~$t$ 8 TeV ($1/\sigma d\sigma/dy_t$) & CMS $\sigma_{tt}^{\rm tot}$ 5 TeV\\
CMS $t\bar{t}$ 2D $2\ell$ 8 TeV ($1/\sigma d\sigma/dy_tdm_{t\bar t}$)                                &                                                  & \\
    \end{foldstabular}
    \vfill\vspace{0.2cm}
    \begin{foldstabular}{4}
CHORUS $\sigma_{CC}^{\bar{\nu}}$                        & HERA I+II $\sigma_{\rm NC}^{p}\ e^+$ (820~GeV) & LHCb $W,Z \to \mu$ 8 TeV\\ 
ATLAS $W,Z$ 7 TeV ($\mathcal{L}=4.6$~fb$^{-1}$) forward & LHCb $Z\to \mu\mu$ 13 TeV                      & ATLAS $W^-$+jet 8 TeV\\ 
ATLAS low-mass DY 7 TeV                                 & ATLAS $Z$ $p_T$ 8 TeV ($p_T,y_Z$)              & CMS $W$ rapidity 8 TeV\\ 
D0 $Z$ differential                                     & CMS dijets 7 TeV                               & ATLAS single~$t$ 8 TeV ($1/\sigma d\sigma/dy_t$)\\
ATLAS single $t$ $R_{t}$ 13 TeV                         & CMS single $t$ $R_{t}$ 13 TeV                  & \\
    \end{foldstabular}
    \vspace{0.3cm}
    \caption{\small The four folds in which the NNPDF4.0 dataset is divided for the $k$-folds
    hyperoptimisation procedure represented in Fig.~\ref{fig:hyperplots}.}
    \label{table:kfolds}	
\end{table}

%% file: tables/tab-setup.tex
\begin{tabularx}{\textwidth}{XXXX}
  \toprule
  Parameter
  & NNPDF4.0
  & $L$ as in Eq.~\eqref{eq:Lmax=}
  & Flavor basis Eq.~\eqref{eq:flav_basis} \\
  \midrule
  Architecture
  & 2-25-20-8
  & 2-70-50-8
  & 2-7-26-27-8                             \\
  Activation function
  & hyperbolic tangent
  & hyperbolic tangent
  & sigmoid                               \\
  Initializer
  & \texttt{glorot\_normal}
  & \texttt{glorot\_uniform}
  & \texttt{glorot\_normal}               \\
  Optimizer
  & \texttt{Nadam}
  & \texttt{Adadelta}
  & \texttt{Nadam}                        \\
  Clipnorm
  & 6.0$\times 10^{-6}$
  & 5.2$\times 10^{-2}$
  & 2.3$\times 10^{-5}$                   \\
  Learning rate
  & 2.6$\times 10^{-3}$
  & 2.5$\times 10^{-1}$
  & 2.6$\times 10^{-3}$                   \\
  Maximum \# epochs
  & 17$\times 10^{3}$
  & 45$\times 10^{3}$
  & 45$\times 10^{3}$                     \\
  Stopping patience
  & 10\% of max epochs
  & 12\% of max epochs
  & 16\% of max epochs                    \\
  Initial positivity $\Lambda^{(\rm pos)}$
  & 185
  & 106
  & 2                                     \\
  Initial integrability $\Lambda^{\rm (int)}$
  & 10
  & 10
  & 10                                    \\
  \bottomrule
\end{tabularx}

%% file: tables/tab-performance.tex
\begin{tabularx}{\textwidth}{XC{4cm}C{4cm}C{4cm}}
  \toprule
  & NNPDF3.1
  & NNPDF4.0 (CPU)
  & NNPDF4.0 (GPU) \\
  \midrule
  Fit timing per replica
  & 15.2~h
  & 38~min
  & 6.6~min \\
  \midrule
  Speed up factor
  & 1
  & 24
  & 140 \\
  \midrule
  RAM use
  & 1.5 GB
  & 6.1 GB
  & N/A  \\
  \bottomrule
\end{tabularx}

%% file: sec-dataselection.tex
\section{Determination of the baseline dataset}
\label{sec:dataselection}

We discuss the selection criteria that we adopt to construct
the NNPDF4.0 baseline dataset from the datasets described in
Sect.~\ref{sec:datatheory}. This baseline dataset will be used in all of the
fits presented in the sequel. In previous PDF determinations, ad-hoc
dataset selection criteria have often been applied. Here we strive to
use objective criteria, not only for imposing kinematic cuts (which
is standard), but also in order to
select an optimal dataset for PDF determination out of the global
dataset.
We explain, in turn, our choice of kinematic cuts,
our procedure to determine whether a measurement is to be included in 
the baseline dataset or not, and our selection of jet datasets, which deserve 
a separate treatment due to the need to choose the optimal observable.

\subsection{Kinematic cuts}
\label{subsec:kincuts}
As in previous NNPDF analyses, kinematic cuts are imposed to ensure that
we include only the data for which reliable predictions
can be computed with fixed-order, pure QCD theory. In NNPDF3.1,
see specifically Sect.~2 in~\cite{Ball:2017nwa},
all the data points for which NNLO QCD corrections exceeded the corresponding
experimental uncertainties were removed from the NLO fit.
Likewise, all the
data points for which electroweak (EW) corrections exceeded experimental
uncertainties were removed from the NLO and NNLO fits. Additional cuts were
also imposed on individual datasets on the basis of specific considerations.
In the NNPDF4.0 analysis, kinematic cuts are determined on the ground of
similar guiding principles, which we systematize as follows.

For the NLO fit, we discard datapoints that are subject to excessively
large corrections: specifically, we compute, for each data point,
the ratio between the absolute difference of the NNLO and NLO predictions to the
experimental uncertainty. If this quantity is smaller than a given threshold
value, the data point is retained in the NLO fit, otherwise it is discarded. We
examined two alternative values of the threshold, $1$ and $2$ respectively.
We concluded that a value of $1$ is unnecessarily aggressive, as it leads to
discarding an excessive number of data points from the NLO fit, while a
value of $2$ ensures that a reasonable number of data points are retained in
the fit with reasonable theoretical accuracy. We therefore use $2$ as our
default threshold value. On the other hand, we do not include in the NNLO fits
the data points for which NNLO theory is not available.
This is the case for the $W$+$c$ production measurements listed in
Table~\ref{tab:collider_dataset_2}. In this case, the full NNLO corrections to
the dominant CKM-diagonal contribution have been recently computed in
Ref.~\cite{Czakon:2020coa}.
However the computation of Ref.~\cite{Czakon:2020coa} uses the flavor
$k_{\perp}$ algorithm, which
is not used in the experimental measurement, thus the NNLO corrections
cannot be implemented yet in a PDF fit.

The results of Ref.~\cite{Carrazza:2020gss} allow for a more refined analysis of
cuts motivated by electroweak effects than what was possible in NNPDF3.1.
We can now evaluate
EW and mixed QCD+EW corrections in a systematic and consistent way for 
all hadronic processes included in a PDF fit, by 
taking advantage of the recent automation of these
computations in {\tt mg5\_aMC}~\cite{Frederix:2018nkq}, and using of
fast-interpolation grids with matching accuracy in the electroweak and strong
couplings produced using {\tt PineAPPL}~\cite{Carrazza:2020gss}. 
We use the NNPDF3.1QED set~\cite{Bertone:2017bme}
for the photon PDF~\cite{Carrazza:2020gss}. 
We then exclude from the  NLO and NNLO fits all data points for which the
difference between the pure NLO QCD calculation and the full NLO QCD+EW
computation (which includes the mixed corrections) exceeds the size of the
experimental uncertainty.
This strategy will also be used to investigate phenomenological implications of
the NNPDF4.0 PDF sets in Sect.~\ref{sec:pheno}.

\begin{table}[!t]
  \scriptsize
  \centering
  \renewcommand{\arraystretch}{1.4}
  \input{tables/tab-kincuts}
  \vspace{0.3cm}
  \caption{The set of kinematic cuts applied to the datasets considered in the
    NNPDF4.0 PDF determination for the NLO and NNLO fits. The kinematic cuts
    used
    in the LO fit are the same as in the NLO fit. Only the data points that
    satisfy the
    constraints listed in the table are retained. The cut on the
    HERA I+II $\sigma_{\rm NC}^{c}$ dataset at NNLO is applied, in addition to the
    other cuts for  DIS measurements, only when the charm PDF is
    independently parametrized.}
  \label{tab:kincuts}
\end{table}

Additional kinematic cuts are implemented for specific datasets, as summarized
in Table~\ref{tab:kincuts}. For datasets already included in NNPDF3.1, these
are the same as in that analysis, see Sect.~2 in~\cite{Ball:2017nwa}. For
new datasets, these follow from similar considerations. We summarize  here the
motivations. For DIS measurements
the cuts remove the low-energy ($Q^2$) region, where perturbative QCD becomes
unreliable, and the large invariant mass ($W^2$) region, where higher-twist
corrections may be non-negligible. We impose a stricter $Q^2$ cut on the
HERA I+II $\sigma_{\rm NC}^{c}$ dataset in the NNLO fit if the charm PDF is fitted
in order to minimize the possible impact of missing NNLO terms related to
initial-state charm (see Sect.~2.2 in~\cite{Ball:2017nwa}). For fixed-target DY
measurements (specifically for E866 and
E605 $\sigma^p$) the cuts remove the data points that are too
close to the production threshold, as discussed in
Ref.~\cite{Ball:2017nwa}, based on the study of Ref.~\cite{Bonvini:2015ira}. To this purpose, we define
$\tau=m_{\ell\ell}^2/s$ and $y_{\rm max}=-\frac{1}{2}\ln\tau$, where $m_{\ell\ell}$ is
the invariant mass of the dilepton pair and $\sqrt{s}$ is the center-of-mass
energy of the collision.
For collider inclusive gauge boson production, we
impose a cut on the D0 $W$ electron and muon asymmetry at NNLO because of the
difficulty in obtaining a sufficiently precise theoretical prediction when the
measured asymmetry becomes too close to zero; we exclude the lowest
lepton rapidity
bins of all of the LHCb measurements from the NNLO fit because, due to
rapidity cut on the leptons ($y_\ell>2$) in the last bin the phase
space for both leptons to pass the cut is very small, thus leading to numerical
instabilities in the computation of the NNLO $K$-factor; and we remove the large
invariant mass bins from the ATLAS low-mass DY 2D 8 TeV measurement in
order to avoid
overlap with the corresponding high-mass measurement. For $Z$ $p_T$ production
we follow Ref.~\cite{Boughezal:2017nla} and
remove the largest rapidity bins from the CMS $Z$ $p_T$ 8 TeV
measurement 
because of an apparent incompatibility with the corresponding ATLAS
measurement, while fully retaining the latter.

All the remaining cuts displayed in
Table~\ref{tab:kincuts} are imposed to remove data points for which 
$p_T$ resummation
effects (typically in the low transverse momentum tail of the various
distributions) or electroweak corrections (typically in the large transverse
momentum or invariant mass tails of the various distributions) may become large.
Finally, on top of the cuts listed in Table~\ref{tab:kincuts} we also
apply at NLO a ``similarity cut'': namely, if a datapoint is excluded
at NNLO by one of the cuts in Table~\ref{tab:kincuts}, then
it is also excluded at NLO because the NLO to NNLO difference is
unreliable so  this point is potentially subject to large NNLO
corrections. 

Kinematic cuts in the LO fit are taken to be the same as in the NLO fit.

\subsection{Baseline dataset}
\label{subsec:selection}
The datasets described in Sect.~\ref{sec:datatheory} and the 
kinematic cuts described in Sect.~\ref{subsec:kincuts} above define an extended
dataset out of which we determine a maximally
consistent baseline dataset.
This baseline
dataset is determined through a new weighted-fit procedure that we
introduce here. In this procedure, first we flag datasets that are
problematic either in terms of fit quality, or because of the
stability properties of their covariance matrix. This is done by comparing for
each measurement
respectively the value of the $\chi^2$ or the value of a 
stability indicator to a suitable threshold value. Measurements for
which thresholds are exceeded are then subject to a
dedicated weighted fit. The measurement is then retained or discarded
based on the results of this weighted fit.

Below we will first  discuss the issue of stability of covariance
matrices and  describe the stability indicator that we will use. We
will then perform an appraisal of the full dataset  of
Sect.~\ref{sec:datatheory} based on our indicators and criteria. We
will next present the weighted fit method, and finally apply it to our
dataset and perform the final dataset selection based on it.

\subsubsection{Stability of experimental covariance matrices}
Given the high precision of modern collider experiments, in particular
HERA and the LHC, many datasets are now limited by systematic, rather than
statistical, uncertainties. In these situations, the $\chi^2$ of a given
dataset often becomes extremely sensitive to small differences in the
correlation model assumed for the experimental systematic errors.
This implies that small inaccuracies in the estimate of the experimental
correlated systematic uncertainties can potentially induce spurious
disagreements between theory predictions and experimental data. 
Such spurious disagreements can complicate the interpretation of the quality of
a PDF fit. A poor $\chi^2$ may be caused solely by an instability
of the experimental covariance matrix upon its inversion, rather than by
a genuine tension with the rest of the data in the fit, or by an inaccuracy in
the theory.

In order to quantify the stability of the $\chi^2$ with respect
to potential inaccuracies affecting the experimental covariance matrices,
a new metric was derived in Ref.~\cite{COV}. This metric has the key property of
being independent of any theory predictions, and thus of the rest of the data in
the fit, as it relies exclusively on the experimental covariance matrix as
input. This property ensures it is  
independent of the actual fit quality (the value of the
$\chi^2$). The metric is derived by studying the
stability of the $\chi^2$ given ideally matching theory
predictions, that is, when these are sampled from the same multi-Gaussian
distribution as the experimental data.

Given the often limited information available on the details of
some experimental systematic errors, this metric has to rely on some 
assumptions.
The first one is that diagonal uncertainties are accurately known,
and that potential instabilities are entirely explained by an imperfect
knowledge of the correlations. The second is that
the source of inaccuracies can be traced back to a $\mathcal{O}(1)$
number of specific entries in the correlation matrix.
An example of the latter assumption would be an inaccuracy
in the estimate of the correlation between two data bins in opposite kinematic
regions.
 
Under these assumptions, one can decompose~\cite{COV} the experimental
covariance matrix $C$ as
\begin{equation} 
   C = DRD \, ,
\end{equation}
where $D$ is a diagonal matrix whose entries are
the square roots of the diagonal entries in the covariance
matrix, i.e. the  standard deviations, and $R$ is the correlation matrix.
If the smallest eigenvalue of the correlation matrix $R$ is $\lambda_0$,
then the stability of the $\chi^2$ with respect to the inaccuracies of the
experimental correlation model will be quantified by the condition number
\begin{equation}
  Z =\lambda_0^{-\frac{1}{2}} \, .
  \label{eq:Z}
\end{equation}
The value of $(\sqrt{2}Z)^{-1}$ can be related to an estimate of the required
precision at which correlations need to be determined in order to ensure that
they affect the $\chi^2$ statistic by less than one standard deviation, that is,
by less than $\sigma_{\chi^2}=\sqrt{2/N_{\rm dat}}$ when normalized by the
number of data points

For example, a value  of $Z=5$ of the metric indicates that correlations
must be estimated
with an absolute  uncertainty of less than $0.14$. This means that if the
correlation
between two bins is estimated to be 1.0 while its real value is instead 0.86,
one can expect that the $\chi^2$ may deviate significantly from
unity (by more than $\sigma_{\chi^2}$) even if the experimental data and
theory calculations are perfectly consistent.

Therefore, by evaluating the datasets in the global fit with a
relatively large value of the stability metric $Z$, one can identify
those with a potentially unstable correlation matrix.
If in addition these datasets display a poor fit quality, further investigation
is required since a high value of the $\chi^2$ does not
necessarily indicate a genuine tension in the data or a limitation
of the theory calculations, but rather it could arise from  the instability
of the experimental covariance matrix.

In the remainder of this section,  we will use the stability metric $Z$
as a diagnostic tool
to flag datasets that deserve further investigation. A regularization
procedure in order to correct a covariance matrix with large $Z$ can also
be constructed~\cite{COV}. Such a regularization procedure is not implemented
in the default NNPDF4.0 fit, rather it will be implemented in 
Sect.~\ref{sec:regcovmat} in order to assess the possible impact on the
PDFs of regularizing the covariance matrix for those datasets characterized
by large $Z$ values.

\subsubsection{Appraisal and selection criteria}
\label{subsubsection:appraisal_citeria}

We perform an appraisal of the full  dataset discussed in
Sect.~\ref{sec:datatheory} with the goal of determining its
internal consistency. Specific measurements could be inconsistent with
the rest of the dataset due to a variety of reasons of theoretical or
experimental origin, such as for
example large missing higher order QCD or electroweak corrections,
missing systematic
uncertainties, or underestimated experimental uncertainties. 
Our goal is not to attempt to have a full
understanding of the nature of the inconsistencies, but rather, to single
out and exclude from the baseline inconsistent data based on objective
criteria. These data can then be studied separately through dedicated fits.

We start by performing a NNLO fit in which the full dataset is used.
This fit adopts the theory settings discussed in
Sect.~\ref{sec:datatheory}, it implements the
kinematic cuts of Sect.~\ref{subsec:kincuts}, and it is based on the
methodology described in
Sect.~\ref{sec:methodology}. For jet observables, it is impossible to
include simultaneously dijets and single-inclusive jets because
experimental correlations between them are not available. In this
baseline fit, as well as in our default analysis,  we choose to include
dijets (and not  single-inclusive jets)
at 7~TeV and single-inclusive jet (and not dijets)
at 8~TeV. The motivation for this choice will be presented in a
separate analysis in Sect.~\ref{subsec:doublecounting}.

We then consider, for each measurement, the following indicators
and apply the following  selection criteria:

\begin{itemize}

\item The total $\chi^2$ per data point. We single out all the datasets for
  which $\chi^2> 1.5$. An excess from the expected unit value of the $\chi^2$
  could arise from dataset inconsistencies, within the dataset or between the
  dataset and the rest of the extended dataset, from inaccuracies of the
  theoretical computations, from large statistical fluctuations (especially for
  datasets with a small number of data points) or from instabilities of the
  experimental covariance matrix.
  
\item The number of standard deviations $n_\sigma$ by which the value of the
  $\chi^2$ per data point differs from the expected unit value,
  \begin{equation}\label{eq:nsigma}
    n_\sigma\equiv
    \frac{\chi^2-1}{\sigma_{\chi^2}}=\frac{\chi^2-1}{\sqrt{2/N_{\rm dat}}}.
  \end{equation}
  We single out all the datasets for which $|n_\sigma|> 2$. In these cases,
  the statistical significance of an anomalously large $\chi^2$ might not 
  be explained by a statistical fluctuation.

\item The stability metric $Z$ defined in Eq.~\eqref{eq:Z}. We single out
  the datasets with $Z> 4$. This choice is based on the regularization studies
  performed in~\cite{COV},
  which find that by minimally altering the correlation model such that they
  fulfill $Z=4$, the induced changes in the resulting covariance matrix are
  very likely within the precision to which they were determined.
  The observed differences between the regularized and unregularized
  covariance matrices
  are $5\%$ for the standard deviations and below  0.05 (in absolute units) for
  the correlation coefficients.

\end{itemize}

The first estimator flags all situations in which the
significance of the discrepancy does not depend on the number of
data points, such as for instance a missing higher order correction
that affects all data points. The latter two  instead are sensitive to
cases in which there might be issues related to systematic
uncertainties and their correlation, whose significance depends on
the number of data points.

\begin{table}[!t]
  \scriptsize
  \centering
  \renewcommand{\arraystretch}{1.4}
  \input{tables/tab-DIS_selection.tex}
  \vspace{0.3cm}
  \caption{The DIS datasets in the NNPDF4.0 fit to the extended dataset.
    For each dataset we show the number of data points, the $\chi^2$ per data
    point, the corresponding number of standard deviations $n_\sigma$ and the
    stability metric $Z$, and the value of the weight $\omega$ used in the
    definition of the weighted fit $\chi^2$ in Eq.~(\ref{eq:weighted_chi2}).
    In the last column, we also indicate whether this dataset is retained
    in the NNPDF4.0 baseline dataset.
  }
  \label{tab:dataset_selection_DIS}
\end{table}

\begin{table}[!t]
  \scriptsize
  \centering
  \renewcommand{\arraystretch}{1.4}
  \input{tables/tab-FTDY_selection.tex}
  \vspace{0.3cm}
  \caption{Same as Table~\ref{tab:dataset_selection_DIS}
  for fixed-target DY data.}
  \label{tab:dataset_selection_FTDY}
\end{table}

\begin{table}[!t]
  \scriptsize
  \centering
  \renewcommand{\arraystretch}{1.4}
  \input{tables/tab-GAUGEBOSON_selection.tex}
  \vspace{0.3cm}
  \caption{Same as Table~\ref{tab:dataset_selection_DIS}
  for collider (Tevatron, top, and LHC, bottom) inclusive gauge boson
  production data.}
  \label{tab:dataset_selection_GAUGEBOSON}
\end{table}

\begin{table}[!t]
  \scriptsize
  \centering
  \renewcommand{\arraystretch}{1.4}
  \input{tables/tab-OTHERLHCPROCESSES_selection.tex}
  \vspace{0.3cm}
  \caption{Same as Table~\ref{tab:dataset_selection_DIS}
    for other LHC processes (listed  in
    Table~\ref{tab:collider_dataset_2}).}
  \label{tab:dataset_selection_OTHERLHCPROCESSES}
\end{table}

The number of data points $N_{\rm dat}$ and the values of the three estimators
outlined above are collected, for each measurement,
in
Tables~\ref{tab:dataset_selection_DIS}-\ref{tab:dataset_selection_OTHERLHCPROCESSES}.
We flag the datasets that have both $\chi^2>1.5$ and $|n_\sigma|>2$
or $|n_\sigma|> 2$ and $Z>4$. These datasets will be investigated
through the weighted fit method presented in Sect.~\ref{sec:wfit} below.
The only exception is the ATLAS
isolated photon production measurement at 8~TeV which is discarded
given that it is superseded by the companion measurement at
13 TeV. We do not flag datasets with $\chi^2>1.5$
but with  $|n_\sigma|<2$, nor the datasets with  with $Z>4$ but  with
$|n_\sigma|<2$. In the first case the large value of the $\chi^2$ is consistent
with a statistical fluctuation. In the second case despite
its unstable covariance matrix the dataset can nevertheless be
fitted with acceptable fit quality.
Datasets characterized by large
$Z$ values will be further
investigated in Sect.~\ref{sec:regcovmat} below, where their impact
on the PDFs will be reassessed 
by means of a suitable regularization procedure that reduces their $Z$ value.

The datasets that are flagged according to these criteria  
are singled out in
Tables~\ref{tab:dataset_selection_DIS}-\ref{tab:dataset_selection_OTHERLHCPROCESSES}
by the presence of a weight in the penultimate column. These are: NMC and BCDMS
proton structure functions; combined HERA charm structure function; D0 $W$ electron asymmetry; 7 TeV ATLAS $W,Z$ central
rapidity;  8 TeV ATLAS $W$ rapidity;  7 TeV
LHCb $W$; 8 TeV LHCb electron asymmetry;  8 TeV
ATLAS lepton+jets top-pair; and 7 TeV ATLAS and CMS dijet.

These datasets are hence potentially inconsistent, and they are assessed
using the weighted fit method as discussed below. All other datasets listed in
Tables~\ref{tab:dataset_selection_DIS}-\ref{tab:dataset_selection_OTHERLHCPROCESSES}
are deemed to be consistent and thus included in the NNPDF4.0 baseline.

\subsubsection{The weighted fit method}
\label{sec:wfit}

The weighted fit method is based on the idea that in order to
determine whether a specific measurement is inconsistent with the
global dataset one should produce a PDF determination that provides the
best agreement to this dataset. One may then check whether this best
agreement does or does not lead to the deterioration of the agreement
with one or more of the other data included in the global dataset. This idea
was recently
used in Ref.~\cite{Forte:2020pyp} as a means of studying the
determination of standard model parameters, such as the
strong coupling $\alpha_s(m_Z)$, from a global PDF fit. Related
methods were previously discussed in Ref.~\cite{Collins:2001es}.

The way the idea is implemented is by performing a weighted fit, in
which the selected dataset is given a weight that is large enough for
it to carry about the same weight as the rest of the global dataset.
To this goal, the figure of merit optimized in the fit is
modified as
\be
\chi^2
=
\frac{1}{N_{\rm dat}}\sum_{i=1}^{n_{\rm exp}}N_{\rm dat}^{(i)}\chi^2_i
\qquad
\longrightarrow
\qquad
\chi^2
=
\frac{1}{N_{\rm dat}-N_{\rm dat}^{(j)}}\sum_{i\ne j}^{n_{\rm exp}}N_{\rm dat}^{(i)}\chi^2_i
+
\omega^{(j)}\chi^2_j
\,,
\label{eq:weighted_chi2}
\ee
where $N_{\rm dat}^{(i)}$ is the number of data points in the dataset $i$ and
$\chi^2_i$ is the contribution to the total $\chi^2$ from the given dataset.
The value of  $\omega^{(j)}$ is then chosen as
\begin{equation}
  \label{eq:omegaval}
  \omega^{(j)}=N_{\rm dat}/N_{\rm dat}^{(j)}.
\end{equation}
The last
column of Tables~\ref{tab:dataset_selection_DIS}-\ref{tab:dataset_selection_OTHERLHCPROCESSES}
lists the values of $\omega^{(j)}$ for the datasets that we have singled out
according to the criteria discussed above.
We have explicitly checked that the choice of the precise 
value of  $\omega^{(j)}$
does not change the general conclusions, by repeating several 
weighted
fits with two more choices of  $\omega^{(j)}$, namely,
twice or half the default value defined by Eq.~(\ref{eq:omegaval}).

The possible outcomes of a weighted fit, and the corresponding
conclusions on dataset compatibility, are the following:

\begin{itemize}

\item The value of $\chi^2_j$ does not improve significantly
  while the $\chi^2_i$ of the rest of the datasets remain essentially
  unaffected.
  In this case we conclude that the dataset $j$ exhibits internal
  inconsistencies that however do not distort the global fit.
  We keep dataset $j$ in the baseline.

\item The value of $\chi^2_j$ does not improve significantly and  the $\chi^2_i$
  of several of
  other datasets, including those belonging to the same process type of
  dataset $j$, worsen significantly. In this case we conclude that the
  internal inconsistencies of the given dataset  distort the global fit.
  We remove dataset $j$ from the baseline.

\item The value of $\chi^2_j$ improves significantly and the $\chi^2_i$ of the
  rest of the dataset is unchanged within statistical fluctuations.
  In this case we conclude that the
  dataset $j$ was not fitted properly because it carries a small weight in the
  fit. We keep dataset $j$ in the baseline.

\item The value of $\chi^2_j$ improves significantly but the $\chi^2_i$
  of several of
  other datasets, including those belonging to the same process type of
  dataset $j$, worsen significantly. In this case we conclude that the
  given dataset  is inconsistent with the global dataset.
  We remove dataset $j$
  from the baseline.
  
\end{itemize}

The appraisal, to be presented in Sect.~\ref{subsubsection:appraisal_and_selection} below,  must be done on a case-by-case basis, as there are
several factors, rather than a single figure of merit, that determine
whether or not the fit quality to other datasets worsens
significantly, such as, for instance, whether the $\chi^2$ that worsens
corresponds to data from the same process type or sensitive to the
same PDF, whether there are known issues related to missing higher
order or resummation corrections, etc.  
In all cases which are not clear-cut, we keep the dataset under consideration.

\subsubsection{Appraisal and selection}
\label{subsubsection:appraisal_and_selection}

Table~\ref{tab:weighted_fits} reports the values of the $\chi^2$ obtained in the
weighted fits for both the weighted dataset and for the rest of the datasets in
the fit, grouped by process. In the latter, the $\chi^2$
includes the contribution coming from the weighted dataset (if the weighted
dataset belongs to the process), but with
$\omega^{(i)}=1$ in Eq.~\eqref{eq:weighted_chi2}. For ease of reference, we also
reproduce (in parenthesis) the values of the $\chi^2$ in the unweighted fit originally used to assess each dataset, as given in Tables~\ref{tab:dataset_selection_DIS}-\ref{tab:dataset_selection_OTHERLHCPROCESSES}.

\begin{table}[!t]
  \scriptsize
  \centering
  \renewcommand{\arraystretch}{1.4}
  \input{tables/tab-weighted_fits.tex}
  \vspace{0.2cm}
  \caption{The $\chi^2$ obtained in the unweighted (first row) and weighted fits
    (rest of the table) to the extended dataset. In each of the weighted fits
    the dataset indicated in the first column receives the weight
    reported in Tables~\ref{tab:dataset_selection_DIS}-\ref{tab:dataset_selection_OTHERLHCPROCESSES}. For each fit, the second column reports the $\chi^2$ of
    the weighted dataset in the weighted fit. The value in the
    unweighted fit (same as in
    Tables~\ref{tab:dataset_selection_DIS}-\ref{tab:dataset_selection_OTHERLHCPROCESSES})
    is also given for reference in parenthesis. The other 
    columns display the $\chi^2$ of subsets of datasets, grouped by process,
    in the weighted fits. These values include the contribution from the
    weighted dataset but with $\omega^{(i)}=1$ in
    Eq.~\eqref{eq:weighted_chi2}.}
  \label{tab:weighted_fits}
\end{table}

Based on Table~\ref{tab:weighted_fits}, we reach to the following conclusions,
which are also summarized in the last column
of Tables~\ref{tab:dataset_selection_DIS}-\ref{tab:dataset_selection_OTHERLHCPROCESSES}.

\begin{itemize}

\item NMC $\sigma^{NC,p}$. The $\chi^2$ of this dataset improves from 1.53 to
  1.28. The $\chi^2$ of the other datasets and the total $\chi^2$ fluctuate only
  marginally. These results are consistent with those reported
  in~\cite{Pumplin:2002vw,Forte:2002fg,NNPDF:2011aa} and confirm that this dataset is 
  internally inconsistent. Because such an inconsistency does not alter the
  global fit significantly, we keep this dataset  in the baseline.

\item BCDMS $F_2^p$. The $\chi^2$ of this dataset improves from 1.42 to 1.05.
  The total $\chi^2$ worsens, however this worsening is moderate and
  it does not seem to come from
  any specific process. These
  results confirm a mild inconsistency of this dataset with the rest of the
  datasets in the fit, which however does not appear to be significant enough
  to justify its removal from the fit.
  We thus keep this dataset in the baseline.

\item HERA I+II $\sigma_{\rm NC}^c$. The $\chi^2$ of this dataset improves from
  2.03 to 1.37, but the agreement with all the other HERA data,
  driven by the inclusive structure function measurements,
  deteriorates, with
  a $\chi^2$ increase from 1.20 to 1.45. The $\chi^2$ of all of the
  other datasets fluctuate only marginally. We therefore conclude that this
  dataset is in tension with the small-$x$ HERA inclusive structure function,
  as also observed in the CT18 and MSHT20
  analyses~\cite{Hou:2019efy,Bailey:2020ooq}.
  This tension will possibly   be alleviated once small-$x$ resummation
  effects are accounted for~\cite{Ball:2017otu}, though only a
  resummed PDF
  determination could tell whether this is the case or not.
  Nevertheless the PDFs in the global fit remain unchanged if the
  dataset is removed. Furthermore, this dataset is
  required in order to stabilize the  charm PDF, especially in a
  DIS-only fit, as we will 
  discuss in Sect.~\ref{sec:dataset}.
  For these reasons we keep the
  measurement in the baseline.

\item E866 $\sigma^p$ (NuSea). The $\chi^2$ of this dataset improves
  from 1.59 to 0.90. The  $\chi^2$ of inclusive
  gauge boson production deteriorates somewhat, from 1.48 to
  1.65. A possible reason for this is the lack of large-$x$ resummation in the
  treatment of the theoretical predictions for this
  dataset~\cite{Bonvini:2015ira}.  Mild inconsistency of this
   experiment with NMC was argued in Ref.~\cite{Guzzi:2021fre}. Nevertheless,
  the fit quality of this dataset in the original unweighted fit is
  only marginally above our selection criteria, and the deterioration
  of the global $\chi^2$ is also marginal.  We keep it in the baseline.
  
\item D0 $W$ electron asymmetry. The $\chi^2$ of this dataset improves
  from 3.54 to 1.94, a value that remains sub-optimal. The $\chi^2$ of all of
  the other datasets, in particular of those belonging to the same process
  (including the D0 $W$ muon asymmetry), deteriorates very
  significantly. The dataset is surely inconsistent, though perhaps the
  inconsistency can be traced to a single data point.
  We discard the dataset from the baseline.

\item ATLAS $W,Z$ 7 TeV ($\mathcal{L}=4.6$~fb$^{-1}$) (central rapidity range).
  The $\chi^2$ of this dataset improves from 1.86 to 1.23 while
  the overall $\chi^2$ of collider gauge boson production data
  deteriorates slightly, from 1.48 to 1.60. However, this
  deterioration is very moderate, and furthermore,  as we will show in
  Sect.~\ref{sec:tests}, a small amount of regularization of experimental
  correlations significantly improve the description of the dataset
  while leaving the PDFs unchanged. There is thus no evidence that
  this dataset is inconsistent, and we keep it in the baseline.

\item LHCb $Z\to ee$ 7 TeV. The $\chi^2$ of this dataset improves from
  2.32 to 0.77. At the same time the $\chi^2$ of all collider gauge boson
  production data deteriorates slightly from 1.48 to 1.65. Given the
  moderate amount of deterioration it is unclear that this dataset is
  inconsistent and we keep it in the baseline.

\item ATLAS $W$ 8 TeV. The $\chi^2$ of this dataset improves from
  3.50 to 1.11 but the description of the other datasets, except top
  pair production, deteriorates quite significantly. As in the case of 
  the companion measurement
  at 7 TeV, given the large value of $Z$, we will investigate  in
  Sect.~\ref{sec:tests} whether
  the description of this experiment could be improved by
  regularization of its covariance matrix. However, in unregularized form it is
  inconsistent and we discard the measurement from the
  baseline.

\item LHCb $W\to e$ 8 TeV. The $\chi^2$ of this dataset improves from
  2.61 to 0.19, while the $\chi^2$ for all of the inclusive
  gauge boson production measurements (including other LHCb data) deteriorates
  significantly from 1.48 to 1.79. We discard the dataset from
  the baseline.

\item ATLAS $t\bar{t}$ $\ell$+jets 8 TeV. Here we have four different
  observables, that behave somewhat differently upon being given large
  weight. The $\chi^2$ of  any of these distributions significantly
  improves when given large weight. For the top
  transverse momentum and top pair invariant mass distributions
  this improvement is accompanied by a rather significant
  deterioration of the global fit quality, in which the agreement with
  all other datasets is spoiled by a greater or lesser extent. In the
  case of the top and top pair
  rapidity distributions the global fit quality is very similar and
  only the description of jets deteriorates moderately. This is
  consistent with the results of previous studies by
  NNPDF~\cite{Czakon:2016olj,Amoroso:2020lgh}, suggesting that the
  rapidity distributions, despite being described less well than 
  in NNPDF3.1~\cite{Ball:2017nwa}, remain largely compatible with the rest
  of the dataset. It is also consistent with previous studies concluding that 
  the simultaneous description of all of
  the ATLAS 8 TeV top distributions is problematic, possibly also because
  of ill-defined correlations within individual distributions and between
  different distributions~\cite{Bailey:2019yze,Amoroso:2020lgh}, and
  indeed other recent PDF determinations~\cite{Hou:2019efy,Bailey:2020ooq}
  include only a pair out of the four distributions (though their
  choice of pair differs from our own).  We
  thus  keep the two rapidity distributions ($y_t$ and $y_{t\bar{t}}$)
  and  discard the transverse momentum and
  invariant mass distributions from the baseline.
  
\item ATLAS and CMS dijet 7 TeV. 
  The $\chi^2$ of these datasets improves from
  2.16 to 1.84 and from 1.85 to 1.34, respectively,
  while the global fit quality is very similar and
  only the description of the top pair data deteriorates
  moderately. We accordingly keep these two datasets
  in the baseline. The reason why the
  improvement of the $\chi^2$ is moderate is likely
  related to the large value of the stability metric $Z$,
  rather than to internal inconsistencies.
  Also in this case we will
  investigate the effect of regularizing the covariance matrix  in
  Sect.~\ref{sec:tests}, where we will show that upon regularization
  the $\chi^2$ becomes close to unity but the PDFs are essentially
  unaffected. 
\end{itemize}

\begin{figure}[!t]
  \centering
  \includegraphics[width=0.49\textwidth]{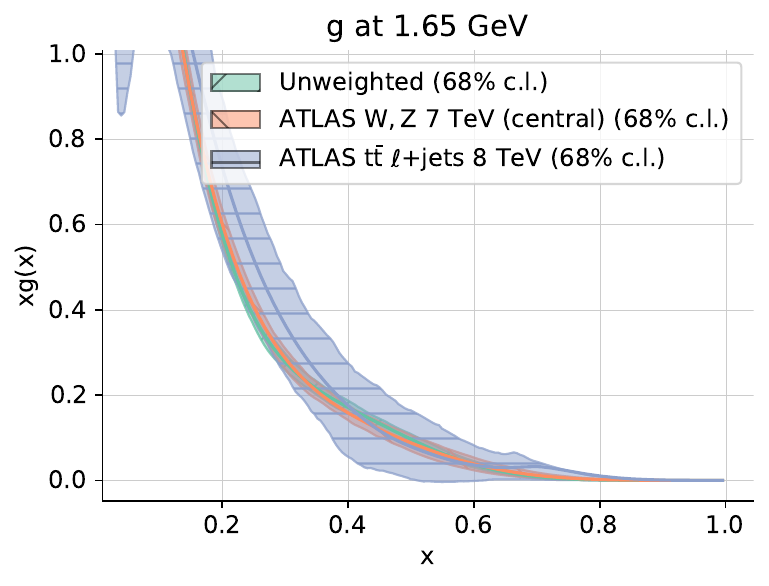}
  \includegraphics[width=0.49\textwidth]{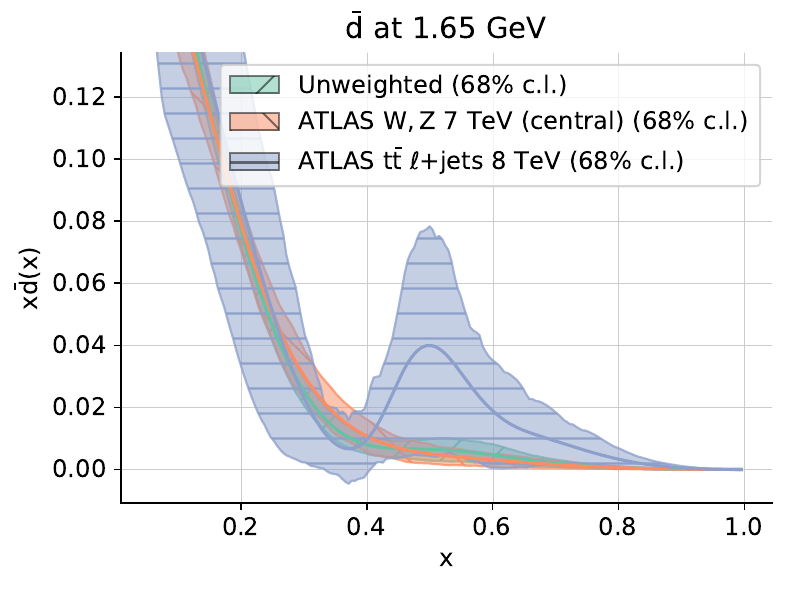}\\
  \caption{The gluon (left) and antidown  (right) PDFs at $Q=1.65$ GeV
    at large $x$,
    for the unweighted fit and the weighted fits in which the
    ATLAS $W,Z$ 7~TeV ($\mathcal{L}$=4.6~fb$^{-1}$) (central) and the 
    ATLAS $t\bar{t}$ $\ell$+jets 8 TeV datasets are assigned large weight.}
  \label{fig:PDFs_weighted_fits}
\end{figure}

Inspection of the PDFs resulting from the weighted fits can provide additional
guidance in assessing consistency.  This information is used to support,
dataset by dataset, the
conclusions summarized above. As an example we display the gluon and antidown
PDFs in Fig.~\ref{fig:PDFs_weighted_fits}. The PDFs are shown at the
input scale $Q_0=$ 1.65~GeV as a function of $x$ in linear scale for the
unweighted fit and for two weighted fits, specifically those in which the
ATLAS $W,Z$ 7~TeV ($\mathcal{L}$=4.6~fb$^{-1}$) (central) and
the ATLAS $t\bar{t}$ $\ell$+jets 8 TeV datasets are assigned large weight.
It is clear that for the ATLAS $t\bar{t}$ $\ell$+jets
8~TeV ($1/\sigma d\sigma/dp_T^t$) data, which are considered
inconsistent based on the $\chi^2$ analysis, the PDFs in the weighted fit
display a significant inflation of PDF uncertainties and
an unnatural distortion of the overall PDF shape, including
an unphysical valence-like structure of the antidown
PDF. Conversely, for  the ATLAS $W,Z$ 7~TeV
($\mathcal{L}$=4.6~fb$^{-1}$) (central) data, which are considered
consistent, the PDFs in the weighted fit have the same shape as the
default  and only moderately inflated uncertainties.
A systematic analysis  for all of the weighted fits shows that the
behavior of the best fit PDFs confirms the conclusion of the $\chi^2$ analysis.

\subsection{Choice of jet datasets}
\label{subsec:doublecounting}

As discussed in Sect.~\ref{subsubsec:jets}, in NNPDF4.0 we consider both
single-inclusive jet and dijet production datasets. However the two observables
cannot be included simultaneously in the fit because full knowledge of
experimental correlations is not available. This also means that we
cannot assess their inclusion in the dataset based on weighted fits.

We therefore select the optimal
set of jet observables by repeating the analysis carried out
in~\cite{AbdulKhalek:2020jut}. Specifically, we start from a fit based on the
baseline dataset identified above from which we remove all jet measurements.
We then compare it to a series of NNLO fits that include, one at a
time, the single-inclusive jet or dijet datasets discussed in
Sect.~\ref{subsubsec:jets}, with the theory settings discussed there.
The  decorrelation model recommended
in~\cite{Aaboud:2017dvo} is used in the case of the ATLAS 8~TeV single-inclusive
jet measurement, while systematic uncertainties are decorrelated across
rapidity bins in the case of the ATLAS 7~TeV single-inclusive jet measurement.

\begin{figure}[!t]
  \centering
  \includegraphics[width=0.49\textwidth]{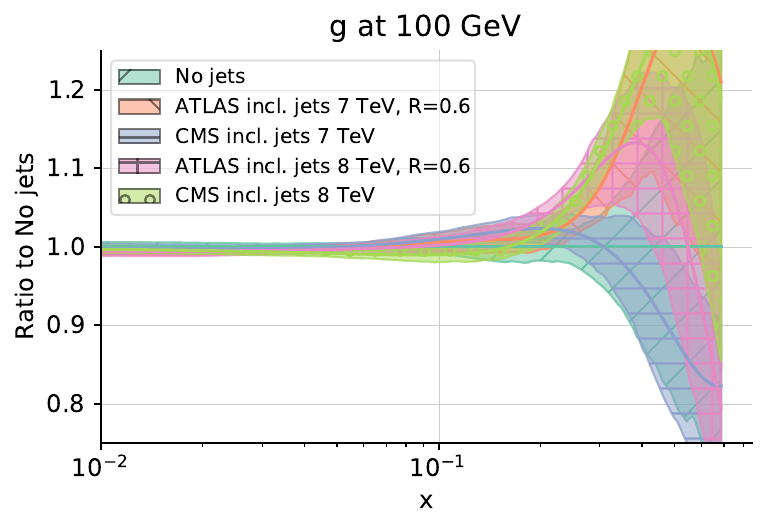}
  \includegraphics[width=0.49\textwidth]{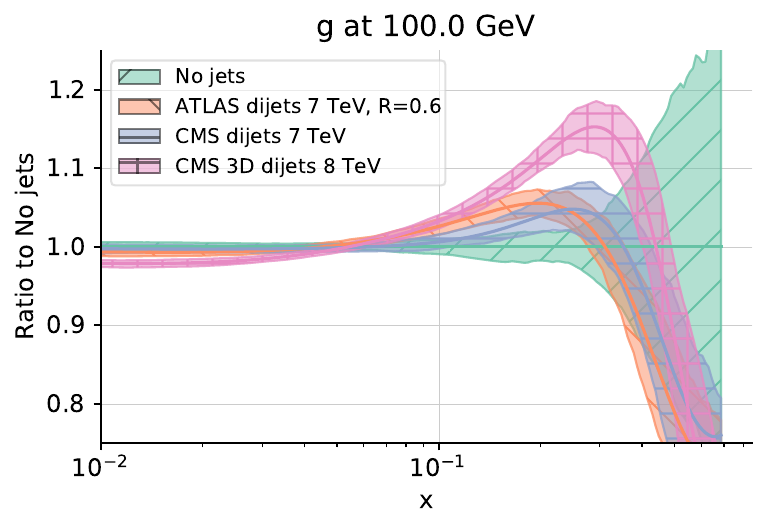}\\
  \caption{The gluon PDF, at $Q=100$~GeV, for some of the fits of
    Table~\ref{tab:chi2_jet}: the baseline variant with no jets, and
    the fits with each of the single-inclusive jet data (left) or each
    of the dijet data (right). Results are shown normalized to the central
    value of the no jets variant.}
  \label{fig:jet_fits}
\end{figure}

In Table~\ref{tab:chi2_jet} we report the values of the $\chi^2$ for all of
these fits. Values are shown for all the data grouped by process type and for
all single-inclusive jet and dijet data, for both those that are and those
that are not included in each fit. The values corresponding to the datasets
that are not included in each fit are indicated in square brackets. In
Fig.~\ref{fig:jet_fits} we compare the gluon PDF from all the fits, separately
for those that include single-inclusive jet or dijet data, at a scale
$Q=100$~GeV. The gluon PDF is normalized to the fit that does not include any
jet data. We have explicitly checked that all other PDFs are
unaffected by the inclusion of jet data.

\begin{table}[!t]
  \scriptsize
  \centering
  \renewcommand{\arraystretch}{1.4}
  \input{tables/tab-chi2_jet}
  \vspace{0.2cm}
  \caption{The $\chi^2$ for an NNPDF4.0  variant  in which all jet
    data are excluded, and a series of fits that add to this variant each of
    the jet measurements of Sect.~\ref{subsubsec:jets} one at a time. Results
    are shown for all datasets, aggregated by process type. For jet data,
    results are shown both for the sets included in each fit and also for those
    not included, which are denoted  by being enclosed in square brackets.
    Combined results for all of the jet production data (including data that
    are and that are not fitted) are also shown. The number of data points in
    each dataset is also reported.}
  \label{tab:chi2_jet}
\end{table}

Inspection of Table~\ref{tab:chi2_jet} and of Fig.~\ref{fig:jet_fits}
leads to the following conclusions.

\begin{itemize}
  
\item All of the 7~TeV data have a rather moderate impact
  and the global fit quality is essentially unchanged in comparison
  to the baseline. There is a moderate pull on the large-$x$ gluon,
  consistent between ATLAS and CMS and
  between single-inclusive jets and dijets, and also consistent with
  the baseline within uncertainties.

\item The 8~TeV single-inclusive jet data have a moderate pull on
  the large-$x$ gluon, consistent between ATLAS and CMS, and
  consistent within uncertainties with the baseline.  This pull
  is in qualitative agreement with but slightly stronger than that of the
  7~TeV jet data. The fit quality to all the other data in the global fit is
  essentially unchanged.

\item The only available 8~TeV dijet measurement, from CMS, has
  a strong pull on the gluon, leading to a result which deviates
  by about two sigma from the baseline, though the pull is
  perhaps similar in shape to that of the single-inclusive 8~TeV
  jet data. The global fit quality deteriorates, but the
  deterioration is not due  to hadron collider data that are sensitive to
  the gluon, like top and $Z$ $p_T$, whose description actually
  improves, but rather to DIS and DY data.
\end{itemize}

In general, the 8 TeV ATLAS and CMS single-inclusive jet measurements and
the 7~TeV ATLAS and CMS dijet measurements have a very similar effect on the
gluon PDF for $x\lesssim 0.2$; dijet datasets seem to suppress the gluon
PDF at slightly more moderate value of $x$ than their single-inclusive jet
counterparts. This does not seem to affect the description of the rest of the
datasets included in the fits.

However, whereas all jet data are broadly consistent with each
other, the CMS 8~TeV dijet data are somewhat problematic, as they
lead to a gluon that is in disagreement with the baseline in the region around
$x\sim0.3$ and to a visible deterioration in global fit quality.
This measurement is peculiar in that it is the only one which is associated to
a triple-differential distribution, it leads to the largest reduction of PDF
uncertainty, and it is possibly the one that carries most of the experimental
information among all of jet measurements. The fact that no corresponding ATLAS
measurement is available, and that the global $\chi^2$
deteriorates noticeably in comparison to all of the other fits, leads us to
conclude that it is more conservative to
include the companion single-inclusive jet data in the baseline. For
8~TeV data we thus include in the baseline the single-inclusive jet
measurements.

Given the fact that dijet data are  preferred on theoretical
grounds~\cite{Currie:2018xkj,Cacciari:2019qjx,AbdulKhalek:2020jut} we
include the 7~TeV dijet 
measurements in the baseline. We will investigate the effect of replacing the
7~TeV ATLAS and CMS dijet measurements with their single-inclusive jet
counterparts in Sect.~\ref{subsubsection:singleinclusivejets_7TeV}.

%% file: tables/tab-kincuts.tex
\begin{tabularx}{\textwidth}{Xcc}
\toprule
Dataset
& NLO
& NNLO \\
\midrule
DIS measurements
& $W^2\ge 12.5$~GeV$^2$; $Q^2\ge 3.5$~GeV$^2$
& $W^2\ge 12.5$~GeV$^2$; $Q^2\ge 3.5$~GeV$^2$ \\
HERA I+II $\sigma_{\rm NC}^{c}$ (in addition to the above)
& ---
& $Q^2\ge 8$~GeV$^2$ (fitted charm)\\
E866/E605 $\sigma^p$
& $\tau\le 0.08$; $|y/y_{\rm max}|\le 0.0663$
& $\tau\le 0.08$; $|y/y_{\rm max}|\le 0.0663$ \\
D0 $W$ electron/muon asymmetry
& ---
& $|A_\ell|\ge 0.03$ \\
ATLAS low-mass DY 7 TeV
& $m_{\ell\ell}>22$~GeV
& --- \\
ATLAS high-mass DY 7 TeV
& $m_{\ell\ell}<210$~GeV
& $m_{\ell\ell}<210$~GeV \\
CMS DY 2D 7 TeV
& $30 \le m_{\ell\ell}\le 200$~GeV; $|y_{\ell\ell}|\le 2.2$
& $m_{\ell\ell}\le 200$~GeV; $|y_{\ell\ell}|\le 2.2$ \\
LHCb $W,Z \to \mu$ 7 TeV
& ---
& $|\eta_\mu| / |y_{\mu\bar{\mu}}| \ge 2.25$ \\
ATLAS low-mass DY 2D 8 TeV
& $m_{\ell\ell}\le 116$~GeV
& $m_{\ell\ell}\le 116$~GeV \\
LHCb $W,Z\to \mu$ 8 TeV
& ---
& $|\eta_\mu| / |y_{\mu\bar{\mu}}| \ge 2.25$ \\
LHCb $Z\to ee$/$Z\to \mu\mu$ 13 TeV
& ---
& $|y_{\ell\ell}|\ge 2.25$ \\
ATLAS $W^\pm$+jet 8 TeV
& $p_T^W \ge 25$~GeV
& $p_T^W \ge 25$~GeV \\
ATLAS $Z$ $p_T$ 8 TeV ($p_T,m_{\ell\ell}$)
& $p_T^Z\ge 30$~GeV                      
& $p_T^Z\ge 30$~GeV \\                   
ATLAS $Z$ $p_T$ 8 TeV ($p_T,y_Z$)
& $30 \le p_T^Z\le 150$~GeV               
& $30 \le p_T^Z\le 150$~GeV\\               
CMS $Z$ $p_T$ 8 TeV
& $30 \le p_T^Z\le 170$~GeV; $|y_Z|\le 1.6$    
& $30 \le p_T^Z\le 170$~GeV; $|y_Z|\le 1.6$ \\ 
CMS incl.\ jets 8 TeV
& $p_T^{\rm jet}\ge 74$~GeV     
& $p_T^{\rm jet}\ge 74$~GeV \\  
\bottomrule
\end{tabularx}

%% file: tables/tab-DIS_selection.tex
\begin{tabularx}{\textwidth}{Xcccccl}
  \toprule
  Dataset
  & $N_{\rm dat}$
  & $\chi^2$
  & $n_\sigma$
  & $Z$
  & $\omega$
  & decision\\
  \midrule
  NMC $F_2^d/F_2^p$
  & 121
  & 0.87
  & $-1.02$
  & 1.098
  & ---
  & \\
  NMC $\sigma^{{\rm NC},p}$
  & 204
  & 1.53
  & $+5.33$
  & 2.743
  & 23
  & retain \\
  SLAC $F_2^p$
  & 33
  & 0.96
  & $-0.16$
  & 1.731
  & ---
  & \\
  SLAC $F_2^d$
  & 34
  & 0.62
  & $-1.55$
  & 1.566
  & ---
  & \\
  BCDMS $F_2^p$
  & 333
  & 1.42
  & $+5.42$
  & 4.456
  & 14
  & retain \\
  BCDMS $F_2^d$
  & 248
  & 1.01
  & $+0.13$
  & 3.468
  & ---
  & \\
  CHORUS $\sigma_{CC}^{\nu}$
  & 416
  & 0.95
  & $-0.66$
  & 4.132
  & ---
  & \\
  CHORUS $\sigma_{CC}^{\bar{\nu}}$
  & 416
  & 0.87
  & $-1.85$
  & 2.073
  & ---
  & \\
  NuTeV $\sigma_{CC}^{\nu}$ (dimuon)
  & 39
  & 0.35
  & $-2.88$
  & 1.092
  & ---
  & \\
  NuTeV $\sigma_{CC}^{\bar{\nu}}$ (dimuon)
  & 37
  & 0.58
  & $-1.80$
  & 1.043
  & ---
  & \\
  \midrule
  HERA I+II
  $\sigma_{\rm NC}^{p}\ e^-$
  & 159
  & 1.40
  & $+3.54$
  & 1.647
  & ---
  & \\
  HERA I+II
  $\sigma_{\rm NC}^{p}\ e^+$ (460~GeV)
  & 204
  & 1.08
  & $+0.78$
  & 1.739
  & ---
  & \\
  HERA I+II
  $\sigma_{\rm NC}^{p}\ e^+$ (575~GeV)
  & 254
  & 0.90
  & $-1.11$
  & 1.639
  & ---
  & \\  
  HERA I+II
  $\sigma_{\rm NC}^{p}\ e^+$ (820~GeV)
  & 70
  & 1.11
  & $+0.62$
  & 2.004
  & ---
  & \\
  HERA I+II
  $\sigma_{\rm NC}^{p}\ e^+$ (920~GeV)
  & 377
  & 1.30
  & $+4.06$
  & 1.845
  & ---
  & \\
  HERA I+II
  $\sigma_{\rm CC}^{p}\ e^+$
  & 42
  & 1.27
  & $+1.23$
  & 1.165
  & ---
  & \\
  HERA I+II
  $\sigma_{\rm CC}^{p}\ e^-$
  & 39
  & 1.23
  & $+1.01$
  & 1.101
  & ---
  & \\
  HERA I+II $\sigma_{\rm NC}^{c}$
  & 37
  & 2.03
  & $+4.45$
  & 1.629
  & 127
  & retain \\
  HERA I+II $\sigma_{\rm NC}^{b}$
  & 26
  & 1.43
  & $+1.56$
  & 1.299
  & ---
  & \\
  \bottomrule
\end{tabularx}

%% file: tables/tab-FTDY_selection.tex
\begin{tabularx}{\textwidth}{Xcccccl}
  \toprule
  Dataset
  & $N_{\rm dat}$
  & $\chi^2$
  & $n_\sigma$
  & $Z$
  & $\omega$
  & decision \\
  \midrule
  E866 $\sigma^p$ (NuSea)
  & 89
  & 1.60
  & $+3.95$
  & 1.789
  & 53
  & retain
  \\
  E866 $\sigma^d/2\sigma^p$ (NuSea)
  & 18
  & 0.53
  & $-1.30$
  & 1.428
  & ---
  &  \\
  E605 $\sigma^p$
  & 85
  & 0.45
  & $-3.56$
  & 1.627
  & ---
  & \\
  E906 $\sigma^d/2\sigma^p$ (SeaQuest)
  & 6
  & 0.87
  & $-0.22$
  & 1.963
  & ---
  & \\
  \bottomrule
\end{tabularx}

%% file: tables/tab-GAUGEBOSON_selection.tex
\begin{tabularx}{\textwidth}{Xcccccl}
  \toprule
  Dataset
  & $N_{\rm dat}$
  & $\chi^2$
  & $n_\sigma$
  & $Z$
  & $\omega$
  & decision\\
  \midrule
  CDF $Z$ differential
  & 29
  & 1.23
  & $+0.87$
  & 5.966
  & ---
  & \\
  D0 $Z$ differential
  & 28
  & 0.65
  & $-1.31$
  & 1.000
  & ---
  & \\
  D0 $W$ electron asymmetry
  & 11
  & 3.54
  & $+5.97$
  & 1.939
  & 429
  & remove \\ 
  D0 $W$ muon asymmetry
  & 9
  & 1.64
  & $+1.35$
  & 1.402
  & ---
  & \\ 
  \midrule
  ATLAS low-mass DY 7 TeV
  & 6
  & 0.89
  & $-0.18$
  & 3.696
  & ---
  & \\
  ATLAS high-mass DY 7 TeV
  & 5
  & 1.67
  & $+1.06$
  & 3.110
  & ---
  & \\
  ATLAS $W,Z$ 7 TeV ($\mathcal{L}=35$~pb$^{-1}$)
  & 30
  & 0.94
  & $-0.24$
  & 3.451
  & ---
  & \\
  ATLAS $W,Z$ 7 TeV ($\mathcal{L}=4.6$~fb$^{-1}$) central
  & 46
  & 1.86
  & $+4.13$
  & 9.013
  & 102
  & retain \\
  ATLAS $W,Z$ 7 TeV ($\mathcal{L}=4.6$~fb$^{-1}$) forward
  & 15
  & 1.04
  & $+0.10$
  & 2.838
  & ---
  & \\
  CMS $W$ electron asymmetry 7 TeV
  & 11
  & 0.97
  & $-0.07$
  & 1.061
  & ---
  & \\
  CMS $W$ muon asymmetry 7 TeV
  & 11
  & 1.69
  & $+1.61$
  & 1.246
  & ---
  & \\
  CMS DY 2D 7 TeV
  & 110
  & 1.34
  & $+2.50$
  & 8.785
  & ---
  & \\
  LHCb $Z\to ee$ 7 TeV 
  & 17
  & 1.25
  & $+0.72$
  & 1.436
  & ---
  & \\
  LHCb $W,Z \to \mu$ 7 TeV
  & 29
  & 2.32
  & $+5.04$
  & 2.890
  & 162
  & retain \\
  ATLAS $W$ 8 TeV
  & 22
  & 3.50
  & $+8.29$
  & 11.28
  & 214
  & remove \\
  ATLAS low-mass DY 2D 8 TeV
  & 60
  & 1.26
  & $+1.42$
  & 1.120
  & ---
  & \\
  ATLAS high-mass DY 2D 8 TeV
  & 48
  & 1.11
  & $+0.53$
  & 2.568
  & ---
  & \\
  CMS $W$ rapidity 8 TeV
  & 22
  & 1.20
  & $+0.65$
  & 13.51
  & ---
  & \\
  LHCb $Z\to ee$ 8 TeV
  & 17
  & 1.25
  & $+0.72$
  & 1.436
  & ---
  & \\
  LHCb $W,Z\to \mu$ 8 TeV
  & 30
  & 1.39
  & $+1.51$
  & 2.542
  & ---
  & \\
  LHCb $W \to e$ 8 TeV
  & 8
  & 2.61
  & $+3.22$
  & 1.005
  & 590
  & remove \\
  ATLAS $\sigma_{W,Z}^{\rm tot}$ 13 TeV
  & 3
  & 0.97
  & $-0.03$
  & 4.961
  & ---
  & \\
  LHCb $Z\to ee$ 13 TeV
  & 16
  & 0.94
  & $-0.16$
  & 2.354
  & ---
  & \\
  LHCb $Z\to \mu\mu$ 13 TeV
  & 15
  & 1.66
  & $+1.80$
  & 1.608
  & ---
  & \\
  \bottomrule
\end{tabularx}

%% file: tables/tab-OTHERLHCPROCESSES_selection.tex
\begin{tabularx}{\textwidth}{Xcccccl}
  \toprule
  Dataset
  & $N_{\rm dat}$
  & $\chi^2$
  & $n_\sigma$
  & $Z$
  & $\omega$
  & decision \\
  \midrule
  ATLAS $W^+$+jet 8 TeV
  & 15
  & 0.76
  & $-0.65$
  & 4.020
  & ---
  & \\
  ATLAS $W^-$+jet 8 TeV
  & 15
  & 1.50
  & $+1.36$
  & 5.679
  & ---
  & \\
  ATLAS $Z$ $p_T$ 8 TeV ($p_T,m_{\ell\ell}$)
  & 44
  & 0.91
  & $-0.42$
  & 3.325
  & ---
  & \\
  ATLAS $Z$ $p_T$ 8 TeV ($p_T,y_Z$)
  & 48
  & 0.89
  & $-0.52$
  & 8.815
  & ---
  & \\
  CMS $Z$ $p_T$ 8 TeV
  & 28
  & 1.38
  & $+1.41$
  & 9.521
  & ---
  & \\
  \midrule
  CMS $\sigma_{tt}^{\rm tot}$ 5 TeV
  & 1
  & 0.42
  & $-0.41$
  & 1.000
  & ---
  & \\
  ATLAS $\sigma_{tt}^{\rm tot}$ 7 TeV
  & 1
  & 3.66
  & $+1.88$
  & 1.000
  & ---
  & \\
  CMS $\sigma_{tt}^{\rm tot}$ 7 TeV
  & 1
  & 0.58
  & $-0.30$
  & 1.000
  & ---
  & \\
  ATLAS $\sigma_{tt}^{\rm tot}$ 8 TeV
  & 1
  & 0.03
  & $-0.71$
  & 1.000
  & ---
  & \\
  CMS $\sigma_{tt}^{\rm tot}$ 8 TeV
  & 1
  & 0.07
  & $-0.66$
  & 1.000
  & ---
  & \\
  ATLAS $\sigma_{tt}^{\rm tot}$ 13 TeV
  ($\mathcal{L}$=139~fb$^{-1}$)
  & 1
  & 0.33
  & $-0.47$
  & 1.000
  & ---
  & \\
  CMS $\sigma_{tt}^{\rm tot}$ 13 TeV
  & 1
  & 0.13
  & $-0.61$
  & 1.000
  & ---
  & \\
  ATLAS $t\bar{t}~\ell$+jets 8 TeV ($1/\sigma d\sigma/dp_T^t$)
  & 7
  & 4.11
  & $+5.82$
  & 5.165
  & 674
  & remove \\
  ATLAS $t\bar{t}~\ell$+jets 8 TeV ($1/\sigma d\sigma/dy_t$)
  & 4
  & 3.61
  & $+3.69$
  & 1.653
  & 1180
  & retain \\
  ATLAS $t\bar{t}~\ell$+jets 8 TeV ($1/\sigma d\sigma/dy_{t\bar t}$)
  & 4
  & 3.81
  & $+3.97$
  & 2.185
  & 1180
  & retain \\
  ATLAS $t\bar{t}~\ell$+jets 8 TeV ($1/\sigma d\sigma/dm_{t\bar t}$)
  & 6
  & 1.86
  & $+1.50$
  & 8.070
  & 786
  & remove \\
  ATLAS $t\bar{t}~2\ell$ 8 TeV ($1/\sigma d\sigma/dy_{t\bar t}$)
  & 5
  & 1.53
  & $+0.84$
  & 1.907
  & ---
  & \\
  CMS $t\bar{t}~\ell$+jets 8 TeV ($1/\sigma d\sigma/dy_{t\bar t}$)
  & 9
  & 1.35
  & $+0.74$
  & 1.628
  & ---
  & \\
  CMS $t\bar{t}$ 2D $2\ell$ 8 TeV ($1/\sigma d\sigma/dy_tdm_{t\bar t}$)
  & 16
  & 0.93
  & $-0.19$
  & 2.908
  & ---
  & \\
  CMS $t\bar{t}~\ell$+jet 13 TeV ($d\sigma/dy_t$)
  & 10
  & 0.53
  & $-1.05$
  & 5.163
  & ---
  & \\
  CMS $t\bar{t}~2\ell$ 13 TeV ($d\sigma/dy_t$)
  & 11
  & 0.69
  & $-0.72$
  & 7.486
  & ---
  & \\
  \midrule
  ATLAS incl. jets 8~TeV, R=0.6
  & 171
  & 0.71
  & $-2.68$
  & 5.476
  & ---
  & \\
  CMS incl. jets 8 TeV
  & 185
  & 1.20
  & $+1.96$
  & 6.273
  & ---
  & \\
  ATLAS dijets 7 TeV, R=0.6
  & 90
  & 2.16
  & $+7.76$
  & 9.936
  & 52
  & retain \\
  CMS dijets 7 TeV
  & 54
  & 1.85
  & $+4.42$
  & 4.695
  & 87
  & retain \\
  \midrule
  ATLAS isolated $\gamma$ prod. 8 TeV
  & 49
  & 2.03
  & $+5.09$
  & 7.277
  & ---
  & remove \\
  ATLAS isolated $\gamma$ prod. 13 TeV
  & 53
  & 0.75
  & $-1.31$
  & 1.304
  & ---
  & \\
  \midrule
  ATLAS single~$t$ $R_{t}$ 7 TeV
  & 1
  & 0.40
  & $-0.42$
  & 1.000
  & ---
  & \\
  CMS single~$t$ $\sigma_{t}+\sigma_{\bar{t}}$ 7 TeV
  & 1
  & 0.71
  & $-0.20$
  & 1.000
  & ---
  & \\
  CMS single~$t$ $R_{t}$ 8 TeV
  & 1
  & 0.13
  & $-0.61$
  & 1.000
  & ---
  & \\
  ATLAS single~$t$ $R_{t}$ 13 TeV
  & 1
  & 0.04
  & $-0.68$
  & 1.000
  & ---
  & \\
  CMS single~$t$ $R_{t}$ 13 TeV
  & 1
  & 0.31
  & $-0.49$
  & 1.000
  & ---
  & \\
  ATLAS single~$t$ 7 TeV ($1/\sigma d\sigma/dy_t$)
  & 3
  & 0.95
  & $-0.06$
  & 1.281
  & ---
  & \\
  ATLAS single~$t$ 7 TeV ($1/\sigma d\sigma/dy_{\bar t}$)
  & 3
  & 0.06
  & $-1.15$
  & 1.385
  & ---
  & \\
  ATLAS single~$t$ 8 TeV ($1/\sigma d\sigma/dy_t$)
  & 3
  & 0.25
  & $-0.92$
  & 1.197
  & ---
  & \\
  ATLAS single~$t$ 8 TeV ($1/\sigma d\sigma/dy_{\bar t}$)
  & 3
  & 0.19
  & $-0.99$
  & 1.230
  & ---
  & \\
  \bottomrule
\end{tabularx}

%% file: tables/tab-weighted_fits.tex
\begin{tabularx}{\textwidth}{Xccccccccccc}
  & \rotatebox{90}{Weighted dataset}
  & \rotatebox{90}{DIS (fixed-target)}
  & \rotatebox{90}{DIS (collider)}
  & \rotatebox{90}{DY (fixed-target)}
  & \rotatebox{90}{$W,Z$ prod. (inclusive)}
  & \rotatebox{90}{$W,Z$ prod. ($p_T$ and jets)}
  & \rotatebox{90}{Top pair prod.}
  & \rotatebox{90}{Jet prod.}
  & \rotatebox{90}{Isolated $\gamma$ prod.}
  & \rotatebox{90}{Single $t$ prod.}
  & \rotatebox{90}{Total}\\
  \toprule
  Unweighted fit
  & ---
  & 1.06
  & 1.20
  & 0.99
  & 1.48
  & 0.98
  & 1.51
  & 1.28
  & 1.36
  & 0.35
  & 1.20\\
  \midrule
  NMC $\sigma^{NC,p}$ 
  & 1.28 (1.53)
  & 1.07
  & 1.22
  & 0.94
  & 1.51
  & 1.00
  & 1.74
  & 1.25
  & 1.33
  & 0.36
  & 1.21\\
  BCDMS $F_2^p$ 
  & 1.05 (1.42)
  & 1.11
  & 1.29
  & 1.01
  & 1.68
  & 0.98
  & 1.50
  & 1.32
  & 1.49
  & 0.37
  & 1.28\\
  HERA I+II $\sigma_{\rm NC}^c$ 
  & 1.37 (2.03)
  & 1.07
  & 1.45
  & 0.99
  & 1.62
  & 1.01
  & 1.68
  & 1.25
  & 1.29
  & 0.38
  & 1.30\\
  E866 $\sigma^p$ (NuSea) 
  & 0.90 (1.60)
  & 1.11
  & 1.23
  & 0.72
  & 1.65
  & 0.96
  & 1.51
  & 1.30
  & 1.31
  & 0.39
  & 1.24\\
  D0 $W$ electron asymmetry 
  & 1.94 (3.54) 
  & 1.12
  & 1.23
  & 1.18
  & 1.70
  & 0.97
  & 1.52
  & 1.37
  & 1.37
  & 0.38
  & 1.28\\
  ATLAS $W,Z$ 7 TeV ($\mathcal{L}=4.6$~fb$^{-1}$) central 
  & 1.23 (1.86)
  & 1.13
  & 1.31
  & 0.96
  & 1.60
  & 1.00
  & 1.56
  & 1.33
  & 1.33
  & 0.37
  & 1.27\\
  LHCb $W,Z\to\mu$ 7 TeV 
  & 0.77 (2.32)
  & 1.13
  & 1.23
  & 1.02
  & 1.65
  & 1.04
  & 1.39
  & 1.40
  & 1.28
  & 0.42
  & 1.27\\
  ATLAS $W$ 8 TeV 
  & 1.11 (3.50)
  & 1.16
  & 1.28
  & 1.07
  & 1.59
  & 1.03
  & 1.43
  & 1.37
  & 1.33
  & 0.36
  & 1.30\\
  LHCb $W\to e$ 8 TeV 
  & 0.19 (2.61)
  & 1.11
  & 1.22
  & 1.17
  & 1.79
  & 1.00
  & 1.42
  & 1.33
  & 1.36
  & 0.34
  & 1.29\\
  ATLAS $t\bar{t}~\ell$+jets 8 TeV ($1/\sigma d\sigma/dp_T^t$) 
  & 1.21 (4.11)
  & 1.52
  & 1.40
  & 1.05
  & 2.12
  & 1.38
  & 1.82
  & 2.60
  & 2.14
  & 0.69
  & 1.73\\
  ATLAS $t\bar{t}~\ell$+jets 8 TeV ($1/\sigma d\sigma/dy_t$) 
  & 0.68 (3.61)
  & 1.09
  & 1.22
  & 1.07
  & 1.52
  & 0.97
  & 2.10
  & 1.51
  & 1.32
  & 0.37
  & 1.26\\
  ATLAS $t\bar{t}~\ell$+jets 8 TeV ($1/\sigma d\sigma/dy_{t\bar t}$) 
  & 1.67 (3.81)
  & 1.11
  & 1.24
  & 0.89
  & 1.58
  & 0.95
  & 1.63
  & 1.63
  & 1.31
  & 0.39
  & 1.28\\
  ATLAS $t\bar{t}~\ell$+jets 8 TeV ($1/\sigma d\sigma/dm_{t\bar t}$) 
  & 1.32 (1.86)
  & 1.23
  & 1.30
  & 1.15
  & 1.74
  & 1.00
  & 4.06
  & 1.73
  & 1.76
  & 0.55
  & 1.47\\
  ATLAS dijets 7 TeV, R=0.6 
  & 1.84 (2.16)
  & 1.08
  & 1.21
  & 0.99
  & 1.61
  & 0.98
  & 2.50
  & 1.39
  & 1.38
  & 0.40
  & 1.25\\
  CMS dijets 7 TeV 
  & 1.34 (1.85)
  & 1.08
  & 1.20
  & 0.93
  & 1.63
  & 0.99
  & 1.77
  & 1.28
  & 1.33
  & 0.35
  & 1.22\\
\bottomrule
\end{tabularx}

%% file: tables/tab-chi2_jet.tex
\begin{tabularx}{\textwidth}{Xccccccccc}
  Dataset
  & $N_{\rm dat}$
  & \rotatebox{90}{no jets}
  & \rotatebox{90}{ATLAS incl. jets 7 TeV, $R=0.6$}
  & \rotatebox{90}{CMS incl. jets 7 TeV}
  & \rotatebox{90}{ATLAS incl. jets 8 TeV, $R=0.6$}
  & \rotatebox{90}{CMS incl. jets 8 TeV}
  & \rotatebox{90}{ATLAS dijets 7 TeV, $R=0.6$}
  & \rotatebox{90}{CMS dijets 7 TeV}
  & \rotatebox{90}{CMS 3D dijets 8 TeV}\\
  \toprule
  DIS (fixed-target)
  & 1881 & 1.06 & 1.06 & 1.06 & 1.07 & 1.06 & 1.06 & 1.06 & 1.09 \\
  DIS (collider)
  & 1208 & 1.21 & 1.21 & 1.21 & 1.20 & 1.21 & 1.21 & 1.21 & 1.21 \\
  DY (fixed-target)
  &  195 & 0.98 & 0.99 & 0.98 & 0.99 & 1.00 & 0.98 & 0.99 & 1.01 \\
  $W,Z$ prod. (inclusive)
  &  548 & 1.34 & 1.32 & 1.33 & 1.36 & 1.32 & 1.33 & 1.34 & 1.36 \\
  $W,Z$ prod. ($p_T$ and jets)
  &  150 & 0.98 & 0.97 & 0.97 & 0.98 & 1.00 & 0.96 & 0.98 & 0.97 \\
  Top pair prod.
  &   66 & 1.24 & 1.06 & 1.16 & 1.07 & 1.14 & 1.11 & 1.12 & 1.12 \\
  Single-inclusive jets (all)
  &  629 & [1.24] & [1.14] & [1.27] & [1.02] & [1.09] & [1.31] & [1.21] & [1.14]\\
  \ \ ATLAS incl. jets 7 TeV, $R=0.6$
  &  140 & [1.32] & 1.26 & [1.33] & [1.26] & [1.25] & [1.33] & [1.32] & [1.28]\\
  \ \ CMS incl. jets 7 TeV
  &  133 & [0.93] & [0.99] & 0.86 & [0.94] & [1.04] & [0.83] & [0.83] & [0.90]\\
  \ \ ATLAS incl. jets 8 TeV, $R=0.6$
  &  171 & [1.26] & [1.07] & [1.22] & 0.73 & [1.01] & [1.10] & [1.09] & [0.65]\\
  \ \ CMS incl. jets 8 TeV
  &  185 & [1.38] & [1.22] & [1.56] & [1.17] & 1.07 & [1.82] & [1.51] & [1.65]\\
  Dijets (all)
  &  266 & [3.32] & [3.10] & [2.91] & [2.37] & [3.15] & [2.51] & [2.58] & [1.68]\\ 
  \ \ ATLAS dijets 7 TeV, $R=0.6$
  &   90 & [2.55] & [2.57] & [2.31] & [2.24] & [2.69] & 2.08 & [2.22] & [1.91]\\
  \ \ CMS dijets 7 TeV
  &   54 & [2.60] & [2.62] & [2.33] & [1.99] & [2.58] & [2.20] & 2.06 & [1.72]\\
  \ \ CMS 3D dijets 8 TeV
  &  122 & [4.21] & [3.69] & [3.61] & [2.63] & [3.75] & [2.97] & [2.09] & 1.50\\
  Isolated $\gamma$ prod.
  &   53 & 0.72 & 0.73 & 0.70 & 0.73 & 0.79 & 0.67 & 0.71 & 0.72 \\
  Single $t$ prod.
  &   17 & 0.35 & 0.36 & 0.36 & 0.36 & 0.36 & 0.36 & 0.35 & 0.38 \\
  \midrule
  Total
  &      & 1.14 & 1.14 & 1.13 & 1.13 & 1.14 & 1.16 & 1.15 & 1.18 \\
  \bottomrule
\end{tabularx}

%% file: sec-results.tex
\section{The NNPDF4.0 parton set}
\label{sec:results}

We now present the main result of this work: the NNPDF4.0 parton
set. We first discuss fit quality, then present the PDFs, and finally show a
comparison of the quality of the fit to a selection of fitted data for a 
variety of different fits.
The NNPDF4.0 PDFs presented here are determined from the baseline dataset of
Sect.~\ref{sec:dataselection} with the methodology of
Sect.~\ref{sec:methodology}. We use $\alpha_s(m_Z)=0.118$ at all perturbative
orders. All PDF sets are Monte Carlo ensembles of 100 replicas, except in the
case of the NNLO NNPDF4.0 baseline, which is a set of 1000 replicas.
Additional comparisons, beyond those reported in this section, can be obtained 
by the reader using the open source
NNPDF software framework described in~\cite{NNPDF:2021uiq},
and summarized in Appendix~\ref{sec:nnpdffitter}.
For all PDF determinations presented below a last iteration has been
performed, in which both the range of the preprocessing exponents (see
Sect.~\ref{sec:flavev}) and  the $t_0$ covariance matrix (recall
Sect.~\ref{sec:methimplementationdetails}) have been recomputed, and it has been checked explicitly
that the results for PDFs are unchanged: this ensures that iterative
procedures have achieved convergence.

\subsection{Fit quality}
\label{subsec:results_fitquality}

Table~\ref{tab:PROCESSTYPE_dataset_chi2} presents an overview of the fit
quality for the LO, NLO and NNLO NNPDF4.0 baseline fits. As in previous NNPDF
releases, $\chi^2$ values are obtained using the published experimental
covariance matrix; this is thus not the figure of merit that is minimized in
the fit, which is the $\chi^2$ computed using the $t_0$ covariance matrix (see
Ref.~\cite{Ball:2014uwa}, specifically Table~9, for a discussion of this issue).
The $\chi^2$ values that were reported for NNLO PDFs
in the NNPDF3.1 analysis of Ref.~\cite{Ball:2017nwa} are also given
for comparison. 

\begin{figure}[!t]
 \centering
 \includegraphics[width=0.7\linewidth]{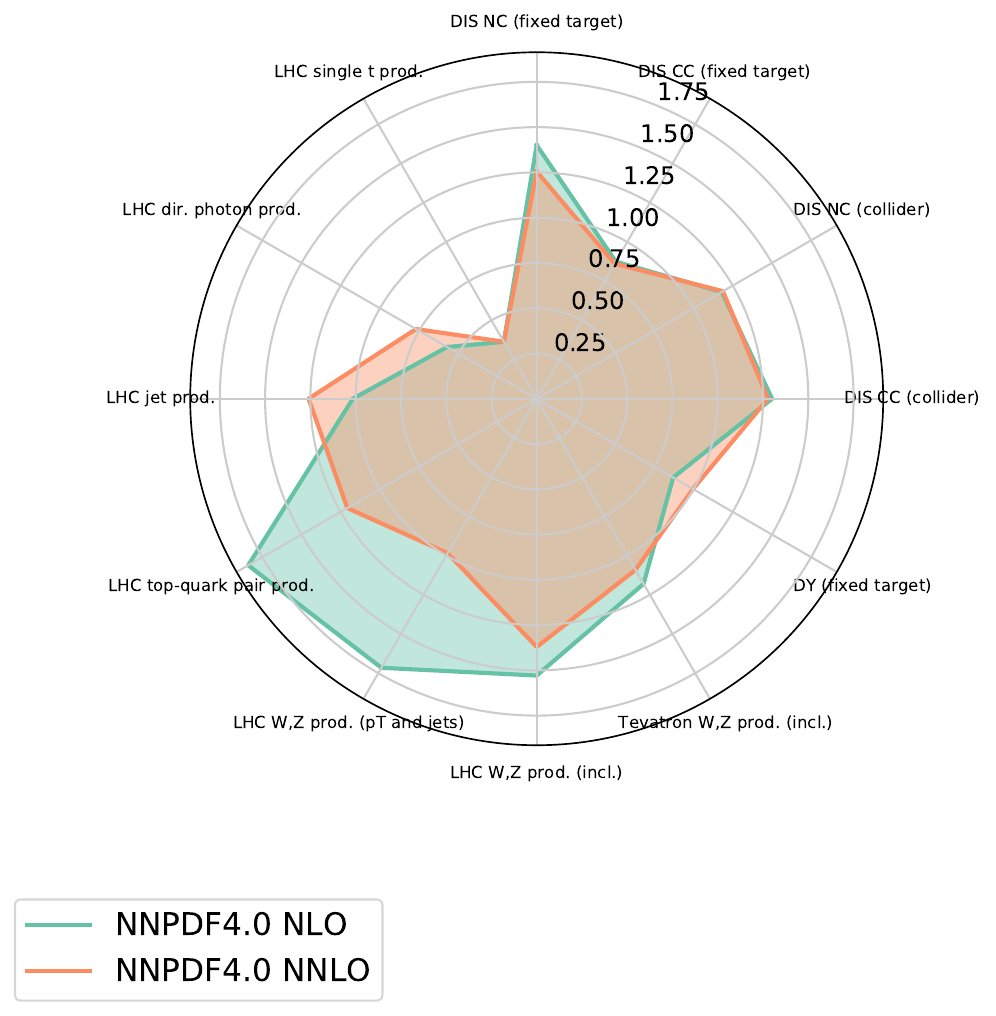}
 \caption{Graphical representation of the results of
   Table~\ref{tab:PROCESSTYPE_dataset_chi2}, comparing the $\chi^2$
 of the NNPDF4.0 NLO and NNLO baseline fits.}    
 \label{fig:plot_fits_chi2_spider}
\end{figure}

Datasets are grouped by process
type: fixed-target DIS, NC and CC; collider DIS, NC and CC; fixed-target DY;
inclusive gauge boson production, separately for the Tevatron and the
LHC; LHC gauge boson production with additional jets (including $Z$ $p_T$
and $W$+jets); LHC
single-inclusive jet and dijet production (for NNPDF3.1 this also includes
Tevatron single-inclusive jet production); LHC top pair
production; LHC direct photon production; and LHC single top production.
The number of data points included
in each fit is indicated in parentheses, and $\chi^2$ values are
provided only for fitted data. A detailed assessment of the
compatibility of the NNPDF3.1 PDFs with the full NNPDF4.0 dataset will be
presented in Sect.~\ref{sec:futuretest} below.
A graphical representation of the NLO and
NNLO values of Table~\ref{tab:PROCESSTYPE_dataset_chi2} is provided in
Fig.~\ref{fig:plot_fits_chi2_spider}.

\begin{table}[!t]
  \scriptsize
  \centering
  \renewcommand{\arraystretch}{1.4}
  \input{tables/tab-PROCESSTYPE_chi2}
    \vspace{0.2cm}
  \caption{Overview of $\chi^2$ value by process type
    for the LO, NLO, and NNLO NNPDF4.0
    baseline fits; NNLO NNPDF3.1 is also shown for comparison.}
  \label{tab:PROCESSTYPE_dataset_chi2}
\end{table}

\begin{table}[!t]
  \scriptsize
  \centering
  \renewcommand{\arraystretch}{1.4}
  \input{tables/tab-DIS_chi2}
    \vspace{0.2cm}
  \caption{Values of the $\chi^2$ for each individual experiment
    included in the NNPDF4.0 PDF determination at  LO, NLO, and NNLO;
     NNPDF3.1 NNLO is also shown for comparison. A dash denotes that
     the dataset was not included in the specific determination.}
  \label{tab:DIS_dataset_chi2}
\end{table}

\begin{table}[!t]
  \scriptsize
  \centering
  \renewcommand{\arraystretch}{1.4}
  \input{tables/tab-FTDY_chi2}
    \vspace{0.2cm}
  \caption{Same as Table~\ref{tab:DIS_dataset_chi2} for fixed-target DY
    datasets.}
  \label{tab:FTDY_dataset_chi2}
\end{table}

\begin{table}[!t]
  \scriptsize
  \centering
  \renewcommand{\arraystretch}{1.4}
  \input{tables/tab-GAUGEBOSON_chi2}
    \vspace{0.2cm}
  \caption{Same as Table~\ref{tab:DIS_dataset_chi2} for inclusive gauge
    boson production datasets.}
  \label{tab:GAUGEBOSON_dataset_chi2}
\end{table}

\begin{table}[!t]
  \scriptsize
  \centering
  \renewcommand{\arraystretch}{1.4}
  \input{tables/tab-OTHERLHCPROCESSES_chi2}
    \vspace{0.2cm}
   \caption{Same as Table~\ref{tab:DIS_dataset_chi2} for all other LHC
    datasets.}
  \label{tab:OTHERLHCPROCESSES_dataset_chi2}
\end{table}

First, one can observe how fit quality markedly
improves with perturbative order: the  $\chi^2$ decreases
from 3.35 at LO to 1.24 at NLO and 1.16 at NNLO.
The significant improvement
in fit quality from  NLO to NNLO  was already
reported in  NNPDF3.1 (see specifically Sect.~3.2
in~\cite{Ball:2017nwa}) and it is chiefly due to the large  number of
high-precision LHC data, for which the $\chi^2$ improves most:
specifically  gauge boson and top pair production.
Fit quality is generally good: specifically, both the value of
$\chi^2$ and the value of $n_\sigma$ Eq.~(\ref{eq:nsigma})
corresponding to the global fit
are similar to  those of other recent global PDF
determinations CT18~\cite{Hou:2019efy} and
MSHT20~\cite{Bailey:2020ooq}, despite the fact that this PDF
determination includes a larger number of datapoints and of different
processes. Of course, comparison of $\chi^2$ values between different
PDF sets should be taken with care, given differences in dataset and
theory settings: the recent PDF4LHC study~\cite{Cridge:2021qjj,Ball:2022hsh} has shown that fit quality
in NNPDF3.1 is similar to that of   CT18 and MSHT20.
The largest $\chi^2$ value ($\chi^2=1.37$)
is found for  LHC
inclusive gauge boson production, which has by far the highest
precision.
The opposite extreme is single top datasets, which have relatively low
precision and a very low $\chi^2$ value.

The quality of the NNLO NNPDF4.0 fit is comparable to that of its
NNPDF3.1 counterpart. This is especially remarkable in view of the
substantial extension of the dataset from NNPDF3.1 to NNPDF4.0. A comparative
analysis of the impact of different data and an assessment of the role played
by the methodology will be respectively presented in
Sect.~\ref{sec:dataset} and Sect.~\ref{sec:tests} below. Specifically,
we will see that a
NNLO fit to the NNPDF3.1-like dataset (see
Sect.~\ref{subsubsection:NNPDF31-like_dataset} below) leads to
$\chi^2=1.145$
if NNPDF4.0 methodology is used, while the significantly worse value
$\chi^2=1.186$ is found using   NNPDF3.1 methodology.

In Tables~\ref{tab:DIS_dataset_chi2}-\ref{tab:OTHERLHCPROCESSES_dataset_chi2}
we provide the details of the $\chi^2$ value for each dataset included in
each PDF determination. We make the following observations.

\begin{itemize}

\item The impact of NNLO QCD corrections is apparent for several of
  the LHC datasets, in particular for  $Z$ $p_T$ and top pair
  production, whose  $\chi^2$ improves significantly
  when moving from NLO to NNLO.
  
\item Fit quality at NNLO is good and uniform across different
  datasets, with variations compatible with statistical fluctuations.
  
\item A good description of the inclusive gauge boson
  production data is achieved, irrespective of the kinematic region
  probed by specific datasets, despite their extremely high precision.
  
\item Measurements with poor fit quality are those already singled out
  in Sect.~\ref{sec:dataselection} that have been retained for the
  reasons explained there: specifically the combined HERA charm cross
  section, the D0 muon asymmetry, the LHC $W,Z\to\mu$ 7 TeV rapidity
  distributions and the ATLAS top pair 8~TeV rapidity distributions in the
  lepton+jet final state and 7~TeV total cross-section. For some of these, fit
  quality is somewhat
  worse in NNPDF4.0 than NNPDF3.1, due to the larger number of
  competing datasets included in the NNPDF4.0 determination. We have
  checked explicitly that if we exclude in turn experiments with the
  worse fit quality, and we combine the ensuing replicas 
  into a single set, we obtain results that are compatible within
  statistical fluctuations with those of the default global fit.
\end{itemize}

\subsection{Parton distributions}
\label{subsec:PDFs}

We now examine the baseline NNPDF4.0 parton distributions. We first
show the full set of PDFs, compared to their NNPDF3.1 predecessors. We
then discuss sources of theoretical uncertainties: the dependence on
the perturbative order and on the value of the strong coupling. 
We finally compare the NNLO NNPDF4.0 baseline PDFs to 
CT18~\cite{Hou:2019efy} and MSHT20~\cite{Bailey:2020ooq}.
A further comparison with these PDF sets in terms of
phenomenology, i.e. specifically for  parton luminosities and theoretical
predictions for LHC observables, will be presented in
Sect.~\ref{sec:pheno}.

\subsubsection{Comparison to NNPDF3.1}
\label{subsubsec:NNPDF40_vs_NNPDF31_PDFs}

\begin{figure}[!t]
  \centering
  \includegraphics[width=0.45\textwidth]{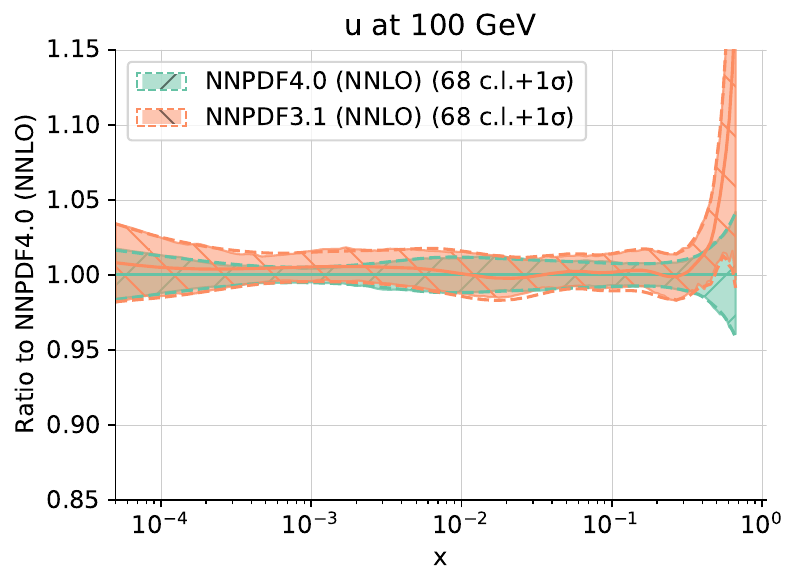}
  \includegraphics[width=0.45\textwidth]{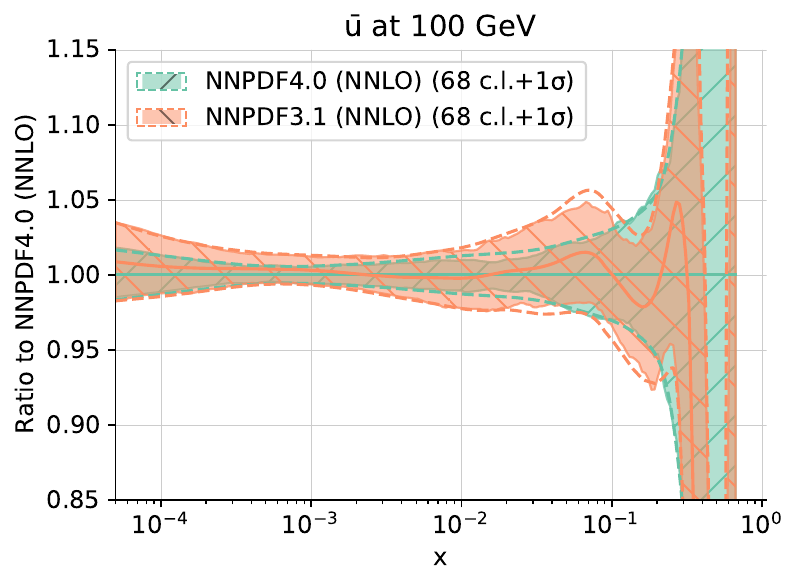}\\
  \includegraphics[width=0.45\textwidth]{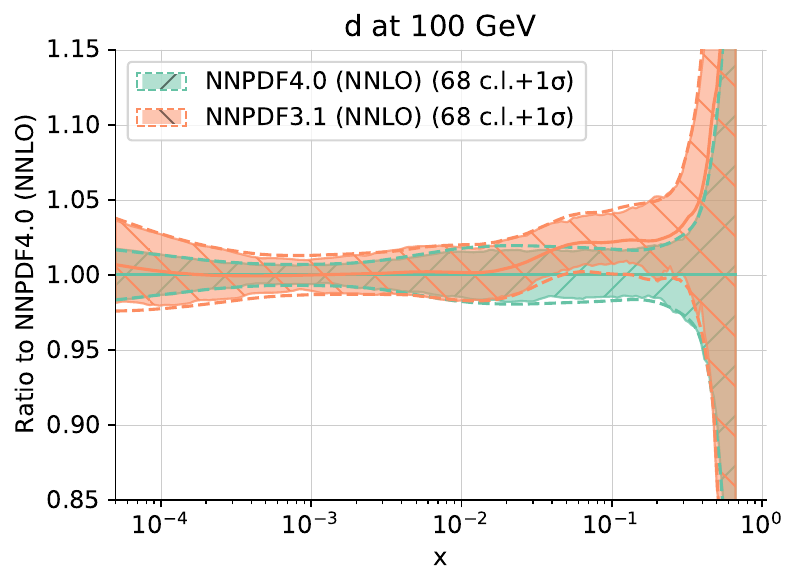}
  \includegraphics[width=0.45\textwidth]{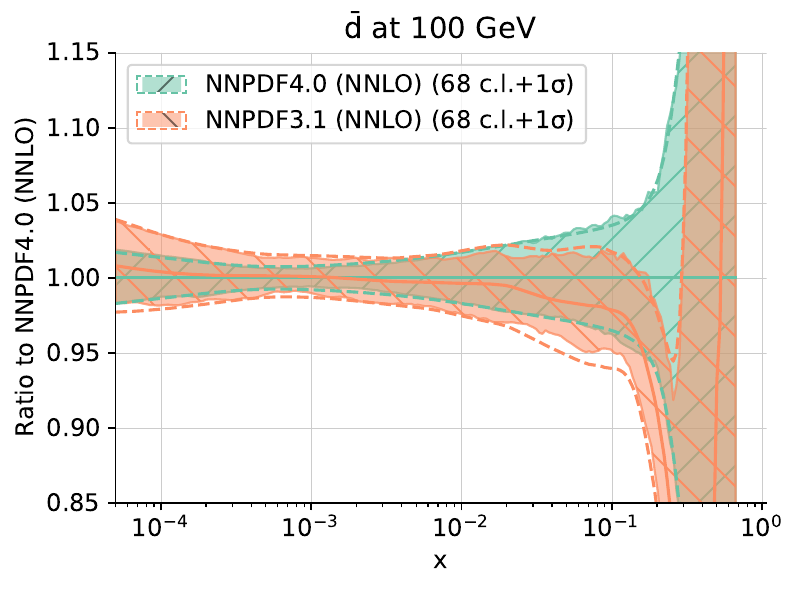}\\
  \includegraphics[width=0.45\textwidth]{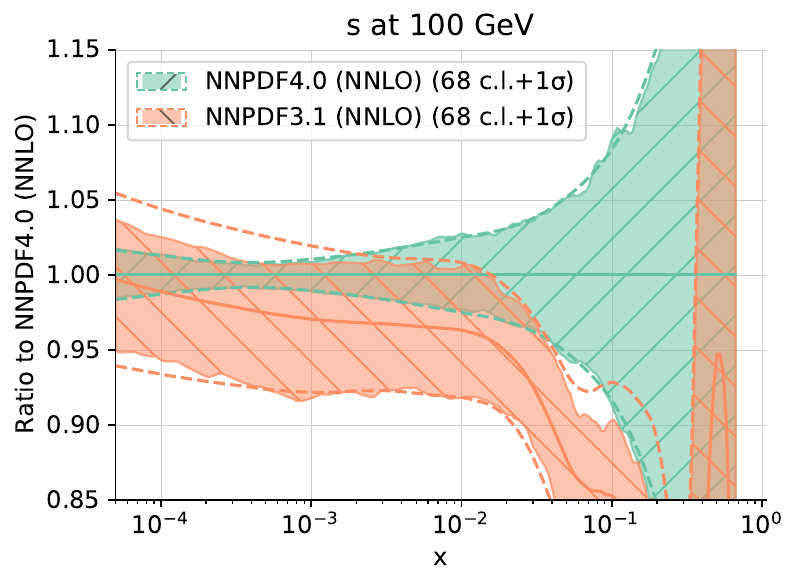}
  \includegraphics[width=0.45\textwidth]{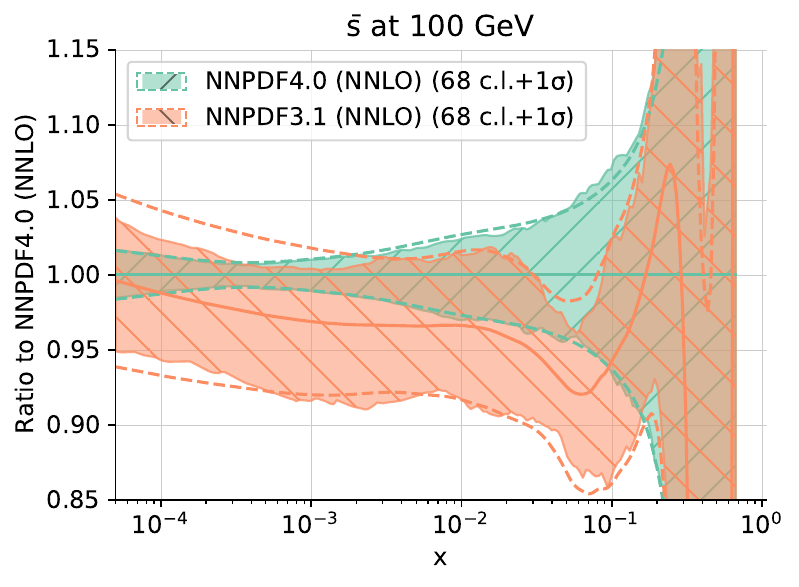}\\
  \includegraphics[width=0.45\textwidth]{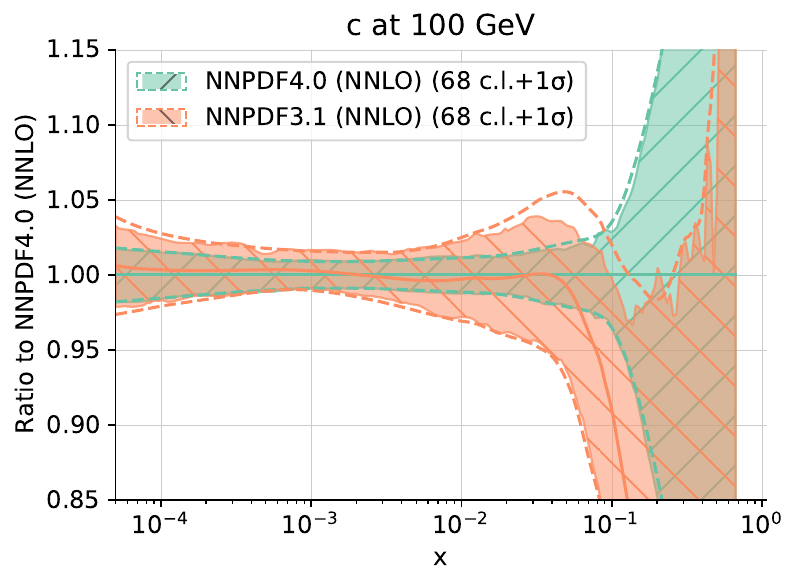}
  \includegraphics[width=0.45\textwidth]{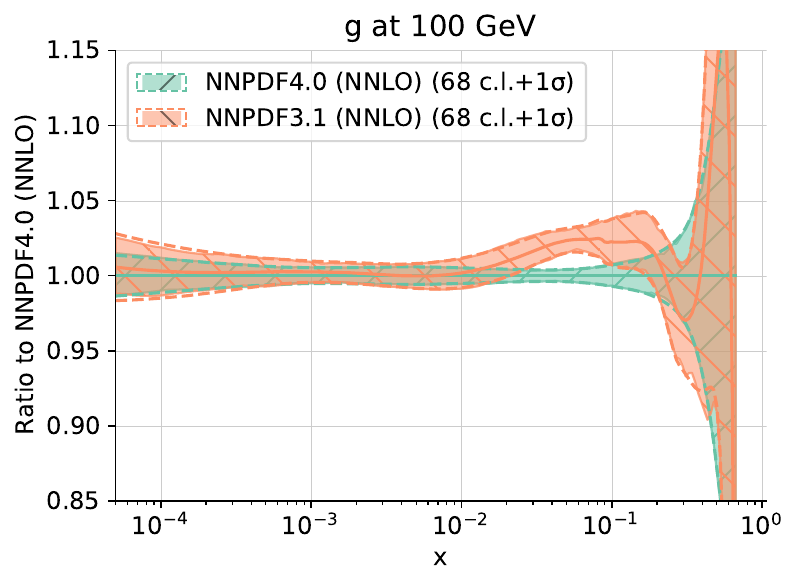}
  \caption{The full set of NNLO  NNPDF4.0 PDFs: the up, antiup, down, antidown,
    strange, antistrange, charm and gluon PDFs at $Q=100$~GeV, compared to
    NNPDF3.1. Results are normalized to the central NNPDF4.0 value. Solid and
    dashed bands correspond to 68\% c.~l. and one-sigma uncertainties, respectively.
  }
  \label{fig:40vs31_PDFs}
\end{figure}

\begin{figure}[!t]
  \centering
  \includegraphics[width=0.45\textwidth]{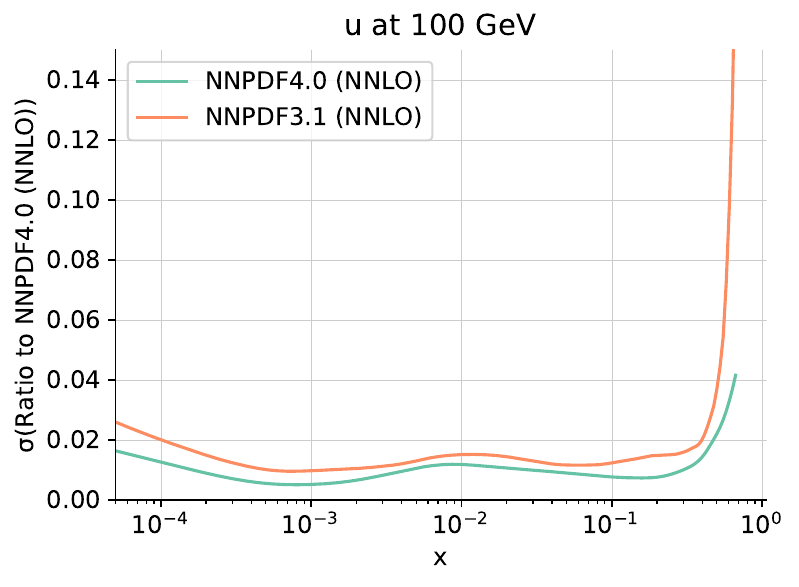}
  \includegraphics[width=0.45\textwidth]{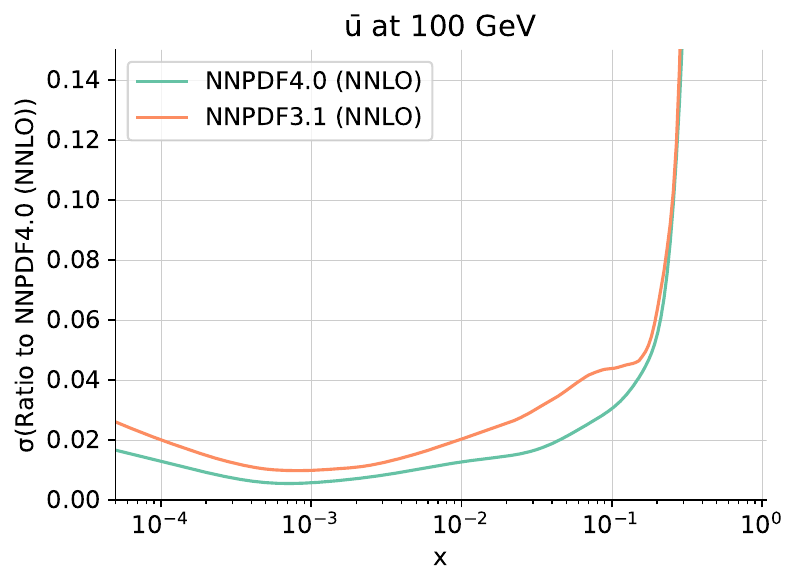}\\
  \includegraphics[width=0.45\textwidth]{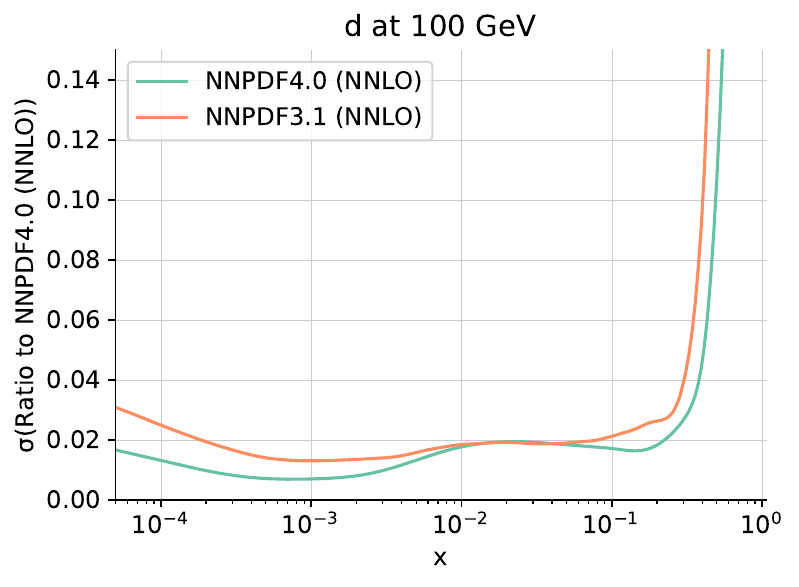}
  \includegraphics[width=0.45\textwidth]{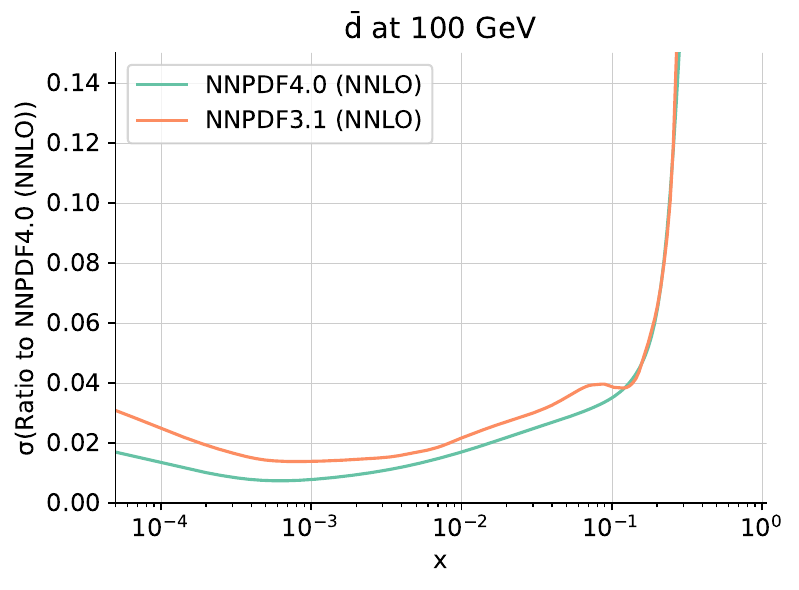}\\
  \includegraphics[width=0.45\textwidth]{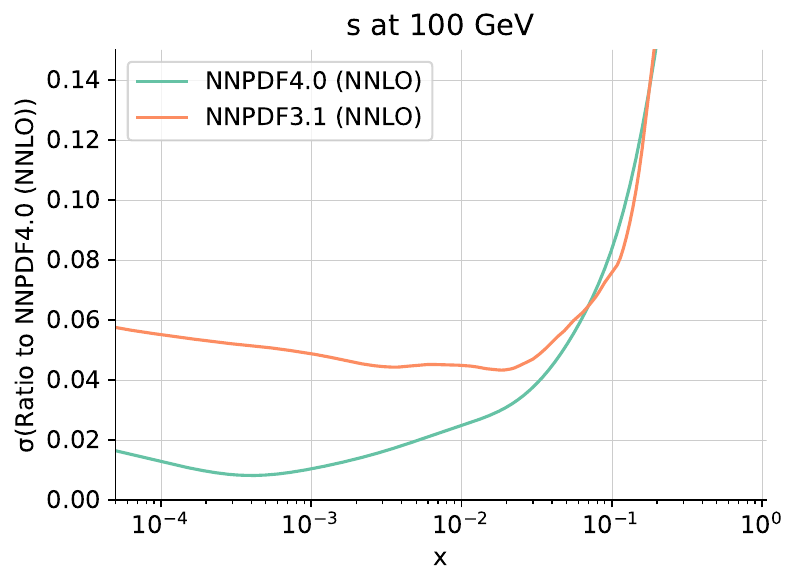}
  \includegraphics[width=0.45\textwidth]{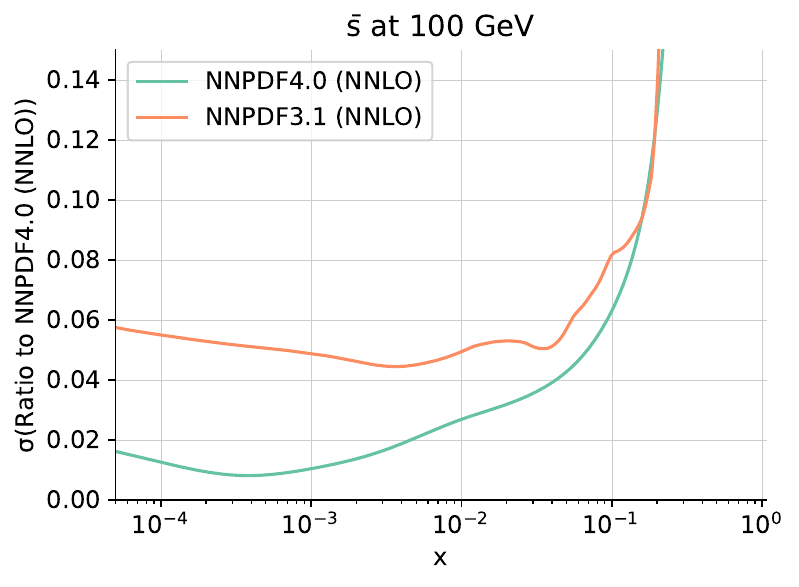}\\
  \includegraphics[width=0.45\textwidth]{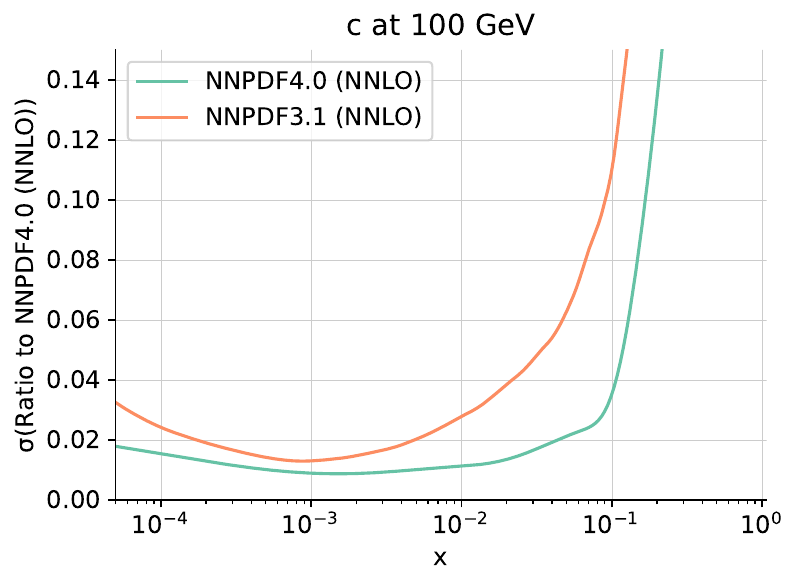}
  \includegraphics[width=0.45\textwidth]{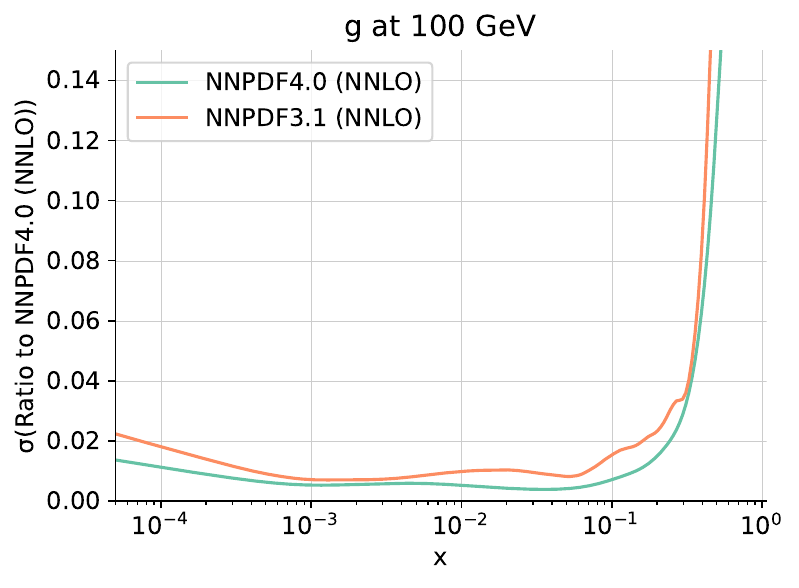}
  \caption{Same as Fig.~\ref{fig:40vs31_PDFs} but for one-sigma relative
    uncertainties.}
  \label{fig:40vs31_PDFs_uncs}
\end{figure}

The full set of NNLO NNPDF4.0 and NNPDF3.1 PDFs are shown in Fig.~\ref{fig:40vs31_PDFs},
and the associated relative one-sigma uncertainties are displayed in Fig.~\ref{fig:40vs31_PDFs_uncs}.
Specifically, we show the up, antiup, down, antidown, strange,
antistrange, charm and gluon PDFs as a function of $x$ at $Q=100$~GeV.
Results are normalized to the NNPDF4.0 central value.

There is remarkable consistency between the new NNPDF4.0 PDF set and
the previous  NNPDF3.1 analysis.
The only noticeable differences appear
in the strange and antistrange PDFs and in the gluon.
As we shall show in Sect.~\ref{subsection:updated_data},
in the former case this is mainly due to the inclusion of NNLO
corrections in the treatment of the NuTeV
data (see Sect.~\ref{subsec:dataset_overview}):
indeed, this same effect was already observed in a recent dedicated
study of strangeness~\cite{Faura:2020oom}.
In the  latter case,  the difference, i.e. the suppression of the gluon around
$x\sim 0.1$, is mainly due to the extra physical constraints provided
by  additional single-inclusive jet, dijet and top pair measurements
included in NNPDF4.0, see also the discussion of Sect.~\ref{sec:dataset}.

The precision of the
PDFs in the NNPDF4.0 set increases significantly in comparison to NNPDF3.1.
Depending on the kinematic region and on the parton, the reduction of the
PDF relative uncertainty ranges from 30\% to more than 50\%. The relative
uncertainty of almost all of the NNPDF4.0 PDFs is of the order of 1-2\% in the
region probed by experimental data. In Sects.~\ref{sec:dataset} and ~\ref{sec:tests} we will
disentangle how
much of this reduction is due to the improved fitting methodology and how much
to the extended dataset.

\subsubsection{Dependence on the perturbative order and on the strong coupling}
\label{subsubsec:NNPDF40_pert_orders}

\begin{figure}[!t]
  \centering
  \includegraphics[width=0.49\textwidth]{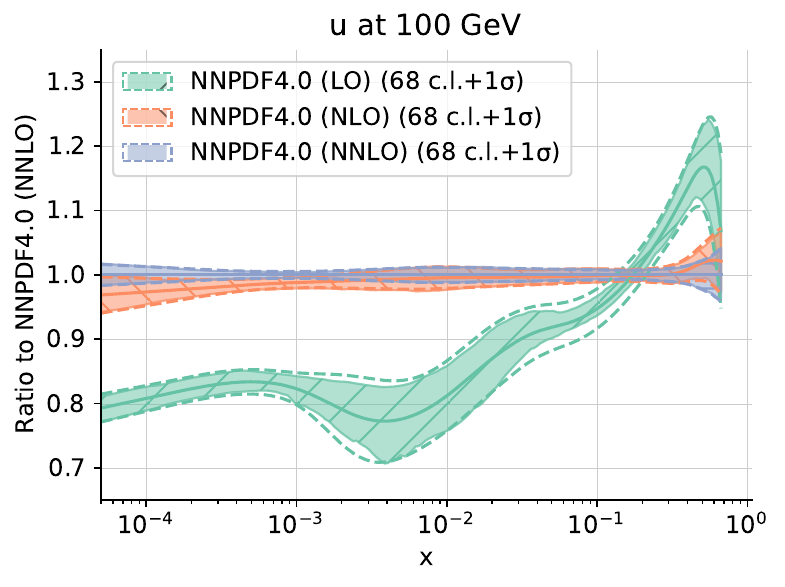}
  \includegraphics[width=0.49\textwidth]{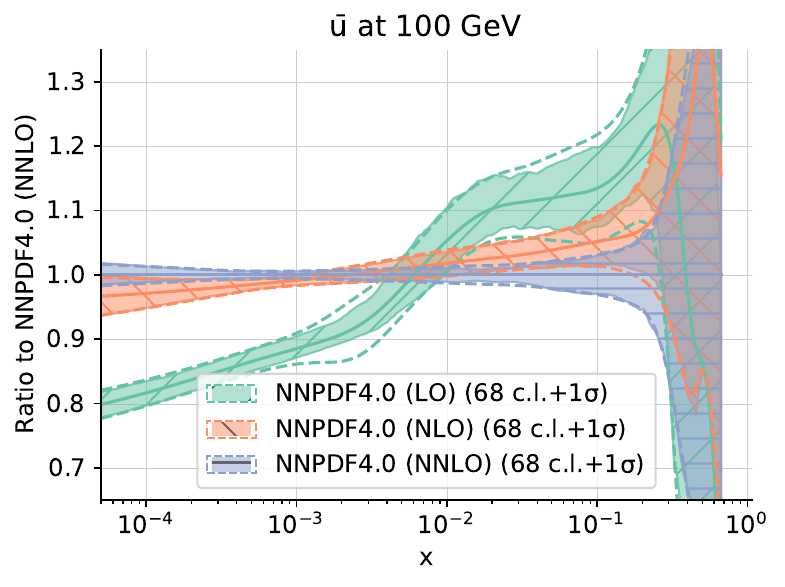}\\
  \includegraphics[width=0.49\textwidth]{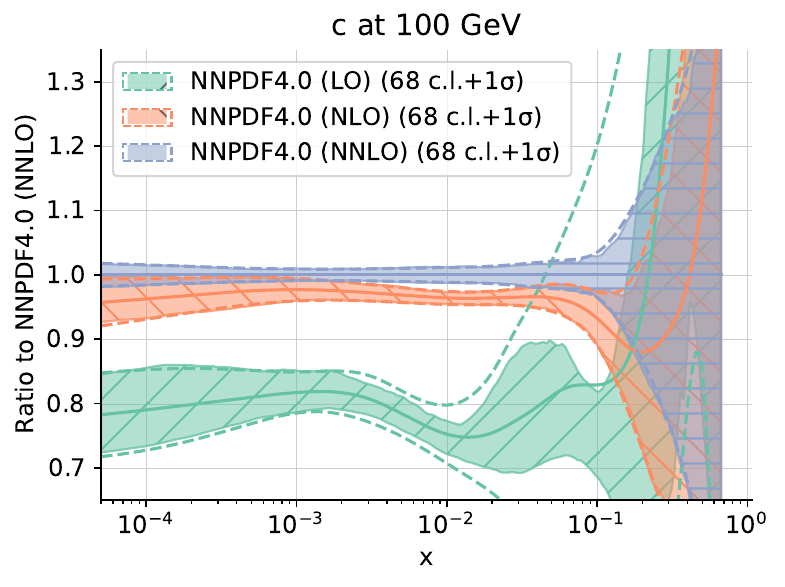}
  \includegraphics[width=0.49\textwidth]{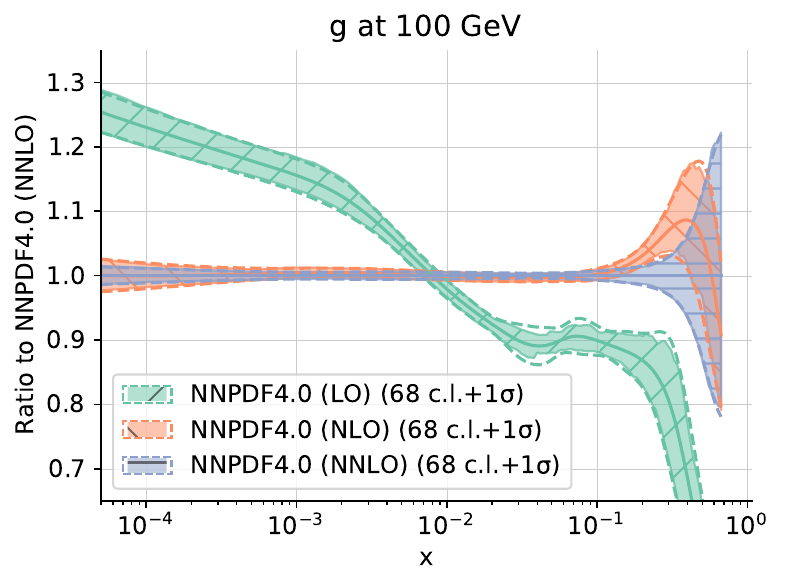}
  \caption{Comparison between the LO, NLO and NNLO NNPDF4.0 PDFs. The  up,
    antiup, charm and gluon are shown at $Q=100$~GeV.
    All results are normalized to the central value of the NNLO set. Solid and
    dashed bands correspond respectively to 68\% c.~l. and one-sigma uncertainties.}
  \label{fig:40prtord_PDFs}
\end{figure}

In Fig.~\ref{fig:40prtord_PDFs} the up, antiup, charm and gluon NNPDF4.0 PDFs are compared
for the three perturbative orders,  LO, NLO and
NNLO, as a function of $x$ at $Q=100$~GeV. Results
are normalized to the central value of the NNLO set.
As expected, a large shift is observed from LO to NLO due to the large NLO 
corrections, as is also clear from the poor quality of
the LO fit seen in Tabs.~\ref{tab:PROCESSTYPE_dataset_chi2}-\ref{tab:OTHERLHCPROCESSES_dataset_chi2}. This is consistent with previous NNPDF studies.

However, the difference between NLO and NNLO PDFs is also noticeable. While
the NLO and NNLO PDFs are very compatible within uncertainties for the up quark,
in the case of the charm quark PDF at
intermediate values of $x$ and in the case of the gluon PDF at large values of
$x$ the shift in central value is comparable or even somewhat larger
than the uncertainty band. This means that at NLO the missing higher
order uncertainty is no longer negligible in comparison to the PDF
uncertainty, unlike in previous PDF determinations, including NNPDF3.1
(see Fig.~3.12 in~\cite{Ball:2017nwa}), where NLO and NNLO
PDFs generally agreed within their larger errors.
Interestingly, the shift in central value in the NLO PDFs observed in
Refs.~\cite{AbdulKhalek:2019bux,AbdulKhalek:2019ihb} when missing higher order
corrections are added during the fit seems to be
of the same size and sign as the shift between NLO and NNLO results
seen in Fig.~\ref{fig:40prtord_PDFs}.
This suggests that the inclusion of the missing higher order
uncertainty along the lines of
Refs.~\cite{AbdulKhalek:2019bux,AbdulKhalek:2019ihb} would be highly
desirable also at NNLO.

\begin{figure}[!t]
  \centering
  \includegraphics[width=0.45\linewidth]{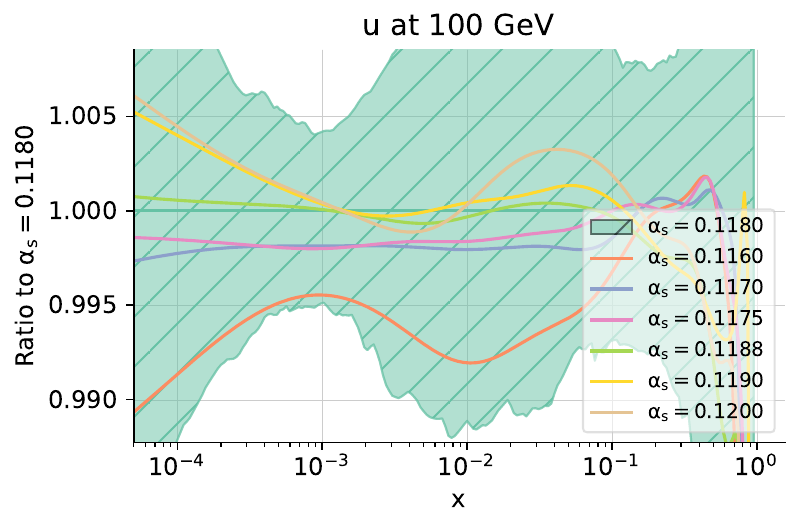}
  \includegraphics[width=0.45\linewidth]{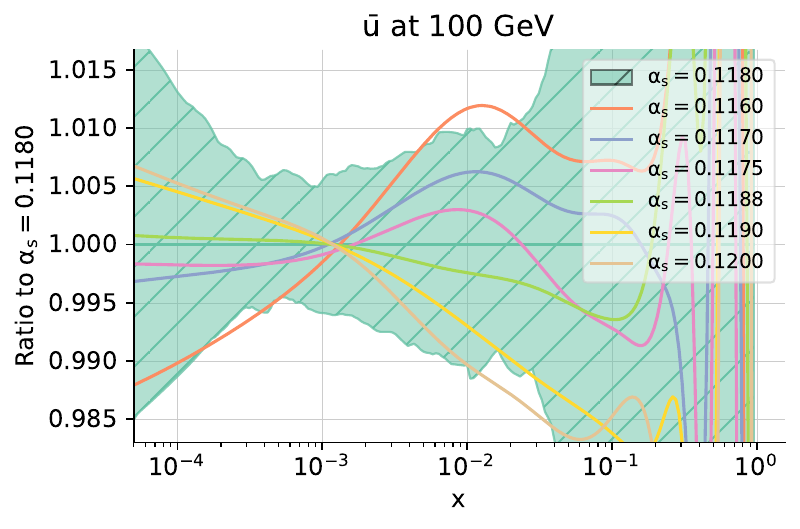}
  \includegraphics[width=0.45\linewidth]{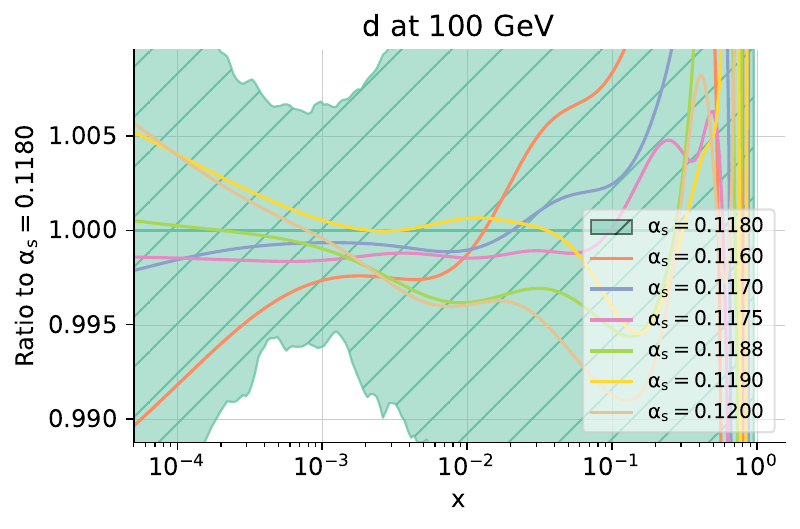}
  \includegraphics[width=0.45\linewidth]{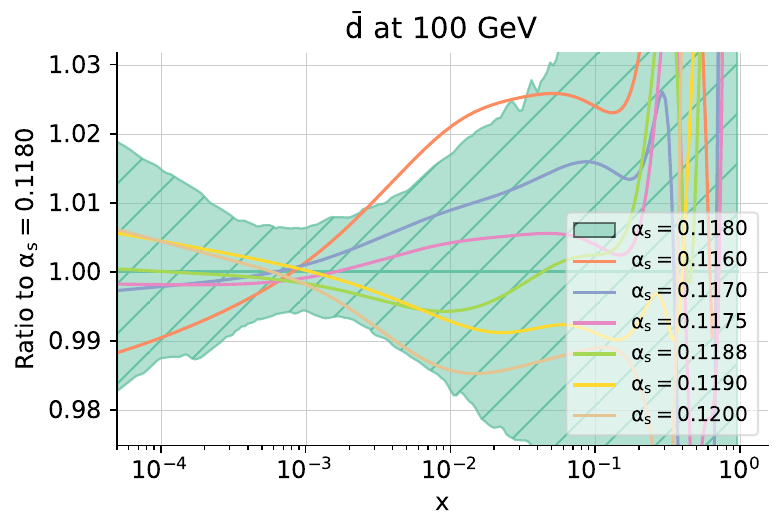}
  \includegraphics[width=0.45\linewidth]{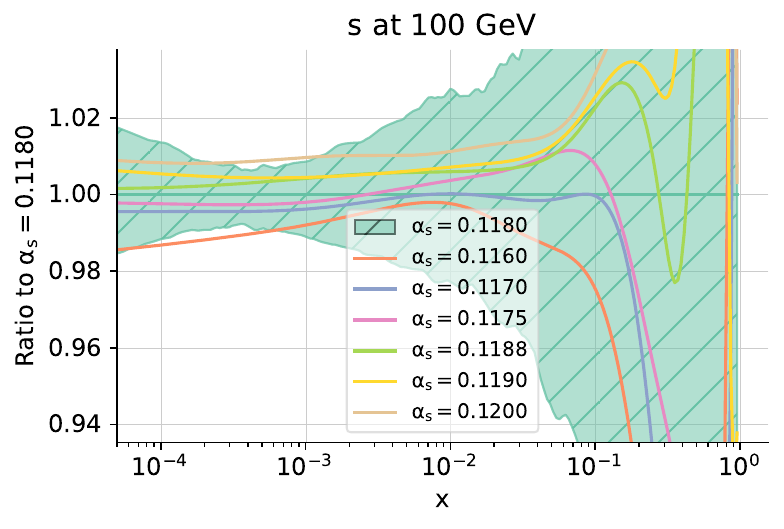}
  \includegraphics[width=0.45\linewidth]{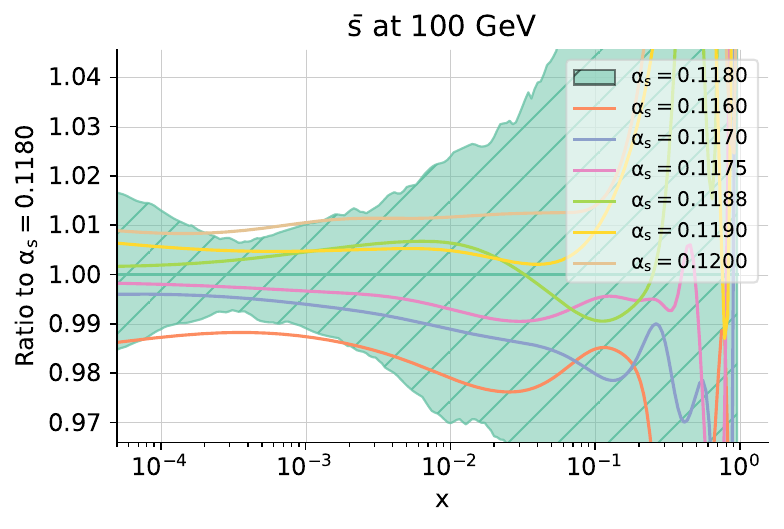}
  \includegraphics[width=0.45\linewidth]{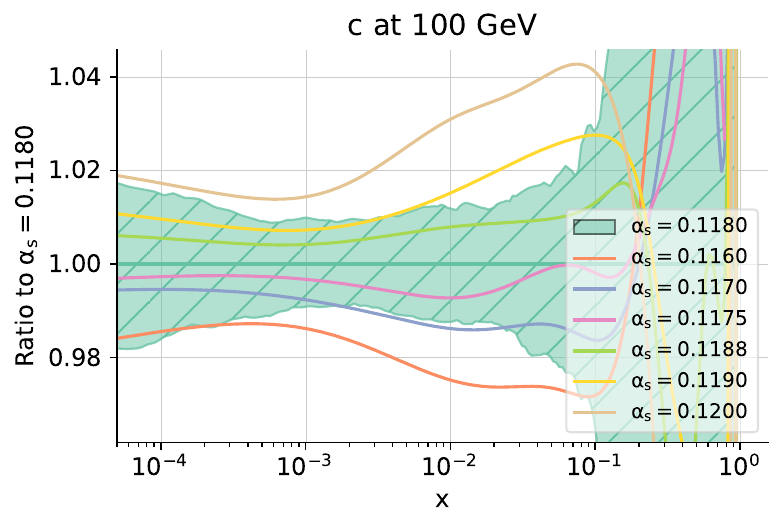}
  \includegraphics[width=0.45\linewidth]{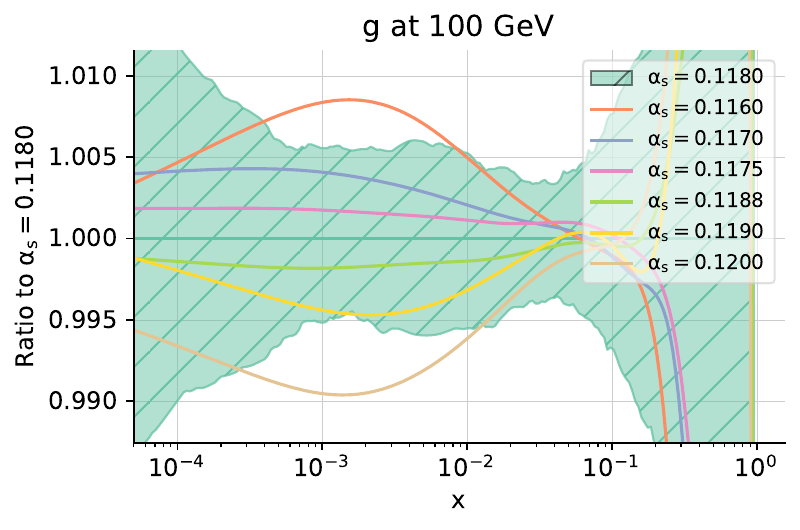}
  \caption{Same as Fig.~\ref{fig:40vs31_PDFs}, 
    now comparing PDFs obtained using different values of 
    $\alpha_s(m_Z)=0.116,\, 0.117,\,0.1175,\,0.118,\,
    0.1185,\,0.119,\,0.120$, normalized to the $\alpha_s(m_Z)=0.118$
    baseline, with only the central value shown for other sets.}
  \label{fig:pdfplot-rat-alphas_nnpdf40_q100gev_largex}
\end{figure}

An important source of theory uncertainty that is routinely included is that
related to the variation of $\alpha_s$.
The default value of the strong coupling adopted for NNPDF4.0 at all
perturbative orders is
$\alpha_s(m_Z)=0.118$, in agreement with the latest PDG value of
$\alpha_s(m_Z)=0.1179 \pm 0.0010$~\cite{Zyla:2020zbs}. In order to
properly include correlated PDF+$\alpha_s$ uncertainties~\cite{Demartin:2010er}
in the computation of LHC observables, we
also provide sets corresponding to different values of $\alpha_s$.
Specifically, we provide  PDFs obtained with
$\alpha_s(m_Z)=0.116,\, 0.117,\,0.1175,\,
0.1185,\,0.119,\,0.120$. They are shown in 
Fig.~\ref{fig:pdfplot-rat-alphas_nnpdf40_q100gev_largex}, along with
the baseline,  normalized to the  central value of the latter.
Only the change in central value is shown: 
relative PDF uncertainties are essentially unchanged when $\alpha_s$ is varied.
Note that the change in central value as $\alpha_s$ is varied by one-sigma is 
smaller or much smaller than the PDF
uncertainty. Of course, the gluon displays the strongest dependence on
$\alpha_s$, and it decreases at small $x$ and increases at large $x$
as the value of $\alpha_s$ is increased.

In Table~\ref{table:alphaschi} we show the value of the $\chi^2$ per
data point obtained in the NNLO fit corresponding to each value of
$\alpha_s$. Whereas a full determination of $\alpha_s$ should be
done~\cite{Forte:2020pyp} 
by using the correlated replica method of Ref.~\cite{Ball:2018iqk},
and also including theory uncertainties,
these values suggest that the best-fit value of $\alpha_s$ 
within the NNPDF4.0 framework is consistent with the NNPDF3.1-based
determination of
Ref.~\cite{Forte:2020pyp} and with the current PDG value.

\begin{table}[!t]
  \scriptsize
  \centering
  \renewcommand{\arraystretch}{1.4}
  \input{tables/tab-alphas.tex}
  \caption{Values of the total $\chi^2$ per data point for the NNLO
    global fit with different values of $\alpha_s(m_Z)$.}
  \label{table:alphaschi}
\end{table}

As already discussed in Ref.~\cite{Ball:2017nwa}, the remaining
parametric uncertainties, related to the values of the quark masses,
are expected to be very small, since the dependence on the charm mass
is almost entirely removed by parametrizing the charm PDF, and the
dependence on the bottom quark mass is very small 
(except on the $b$-PDF itself and processes specifically sensitive to it).

\subsubsection{Comparison to  other PDF sets}
\label{subsubsec:NNPDF40_vs_others_PDFs}

\begin{figure}[!t]
  \centering
  \includegraphics[width=0.45\textwidth]{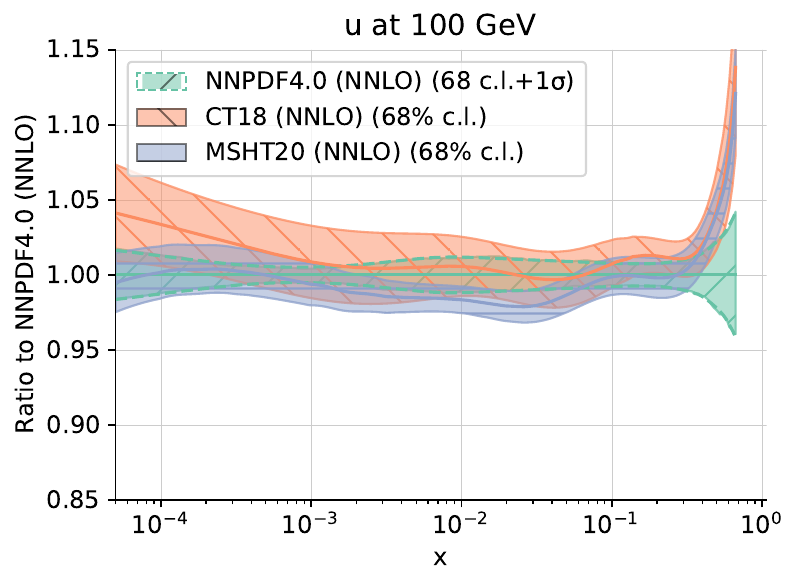}
  \includegraphics[width=0.45\textwidth]{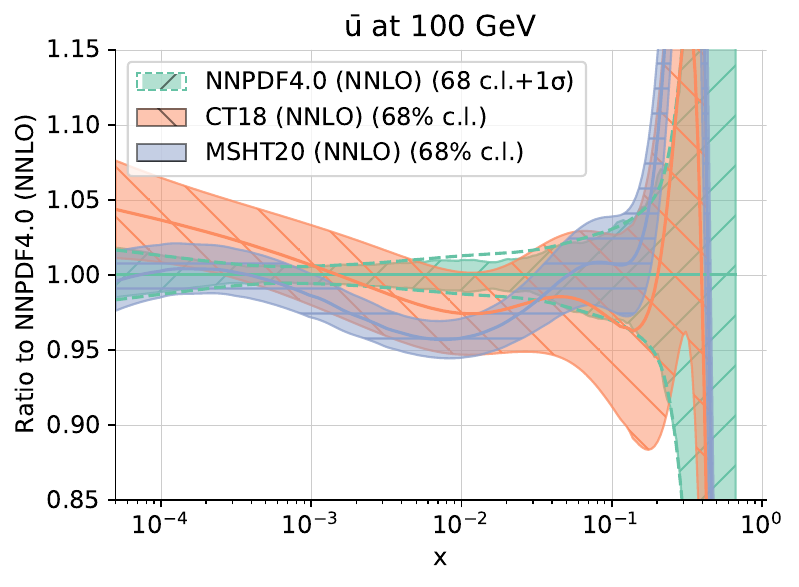}\\
  \includegraphics[width=0.45\textwidth]{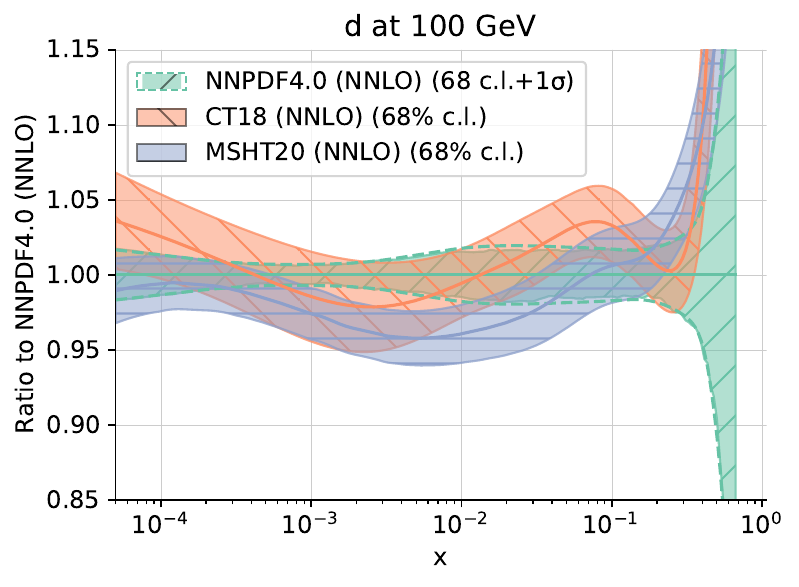}
  \includegraphics[width=0.45\textwidth]{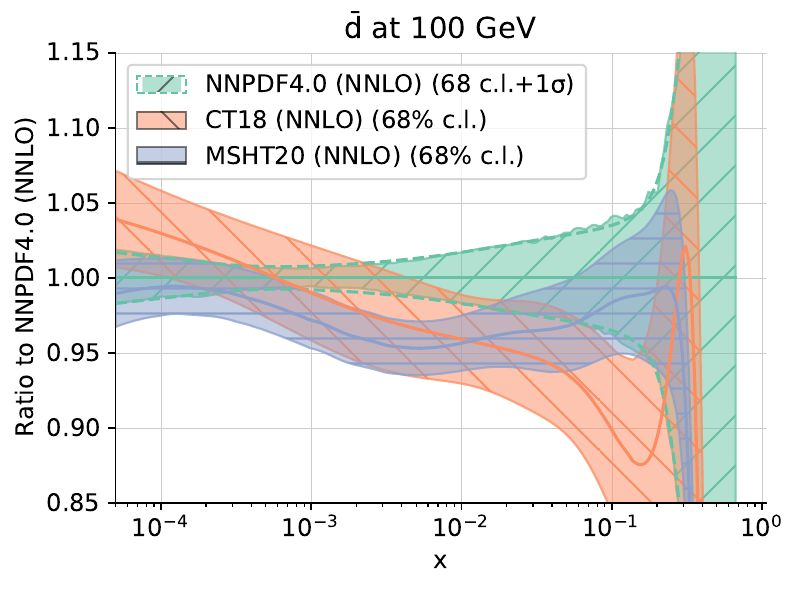}\\
  \includegraphics[width=0.45\textwidth]{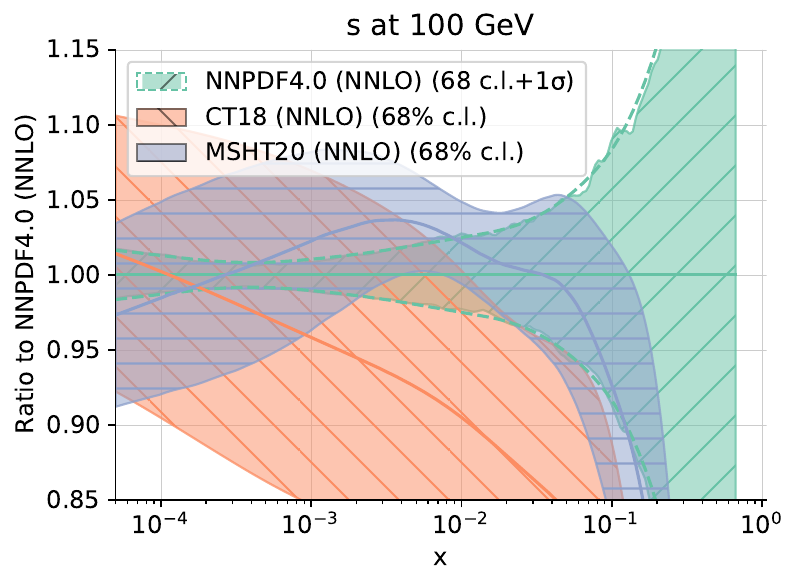}
  \includegraphics[width=0.45\textwidth]{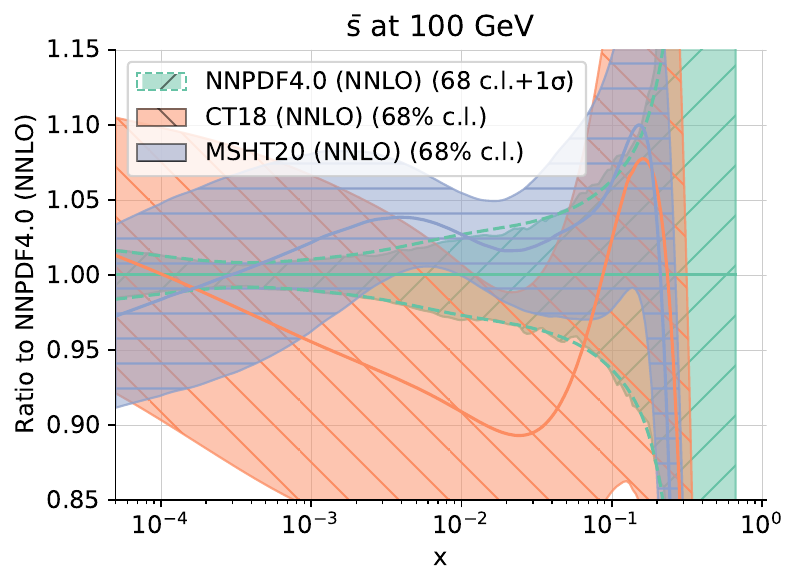}\\
  \includegraphics[width=0.45\textwidth]{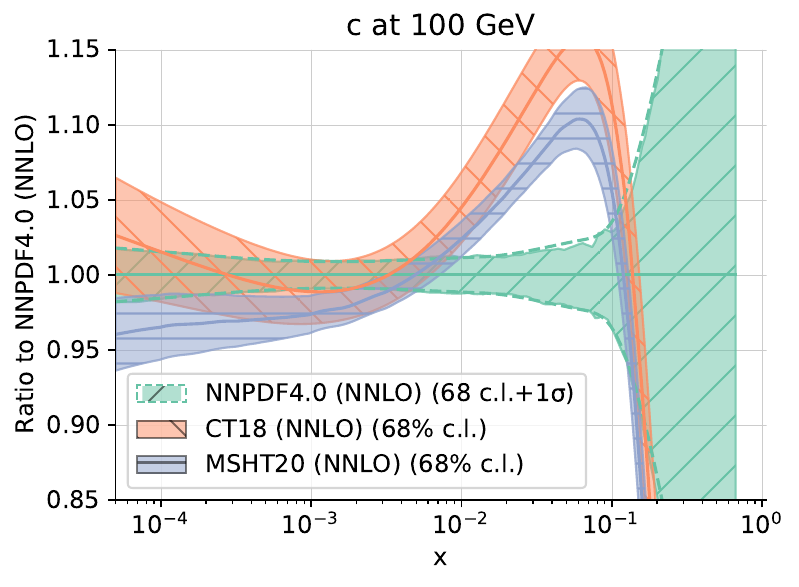}
  \includegraphics[width=0.45\textwidth]{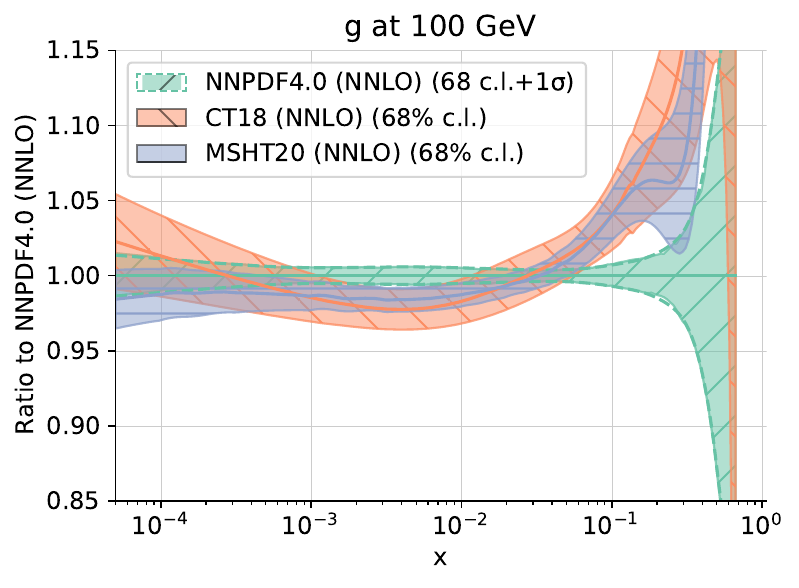}
  \caption{Comparison between the NNPDF4.0, CT18 and the MSHT20 NNLO
    PDF sets. The up,
    antiup, down, antidown, strange, antistrange, charm and gluon PDFs
    are shown at $Q=100$~GeV, normalized to
    the central NNPDF4.0 value. 
    For NNPDF4.0, solid and dashed bands correspond respectively to 68\% c.~l. and
    one-sigma uncertainties.}
  \label{fig:40vsothers_PDFs}
\end{figure}

\begin{figure}[!t]
  \centering
  \includegraphics[width=0.45\textwidth]{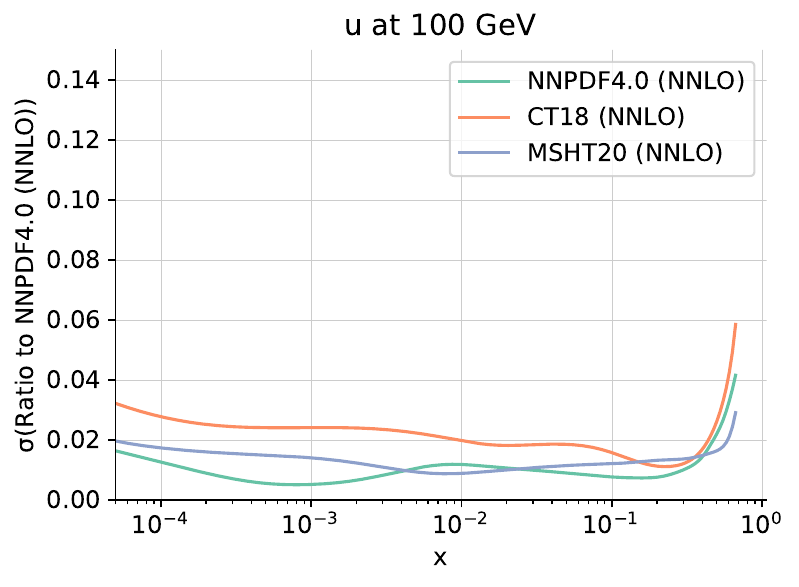}
  \includegraphics[width=0.45\textwidth]{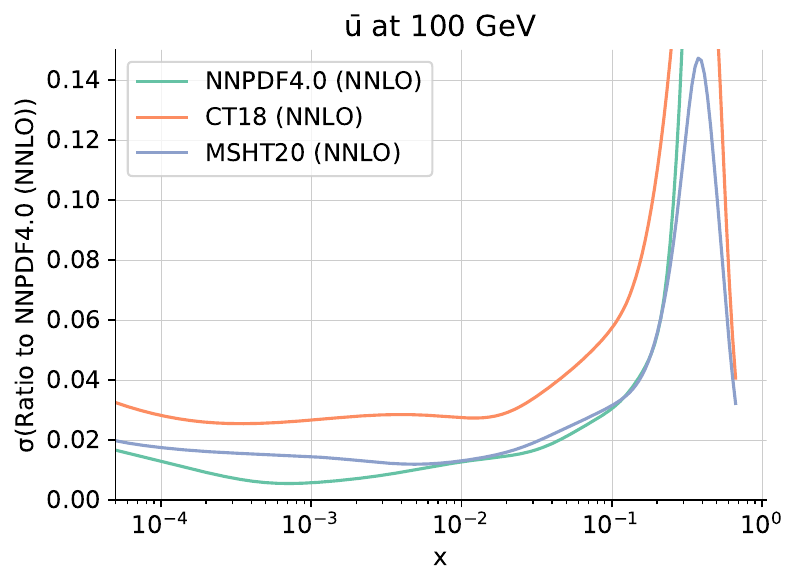}
  \includegraphics[width=0.45\textwidth]{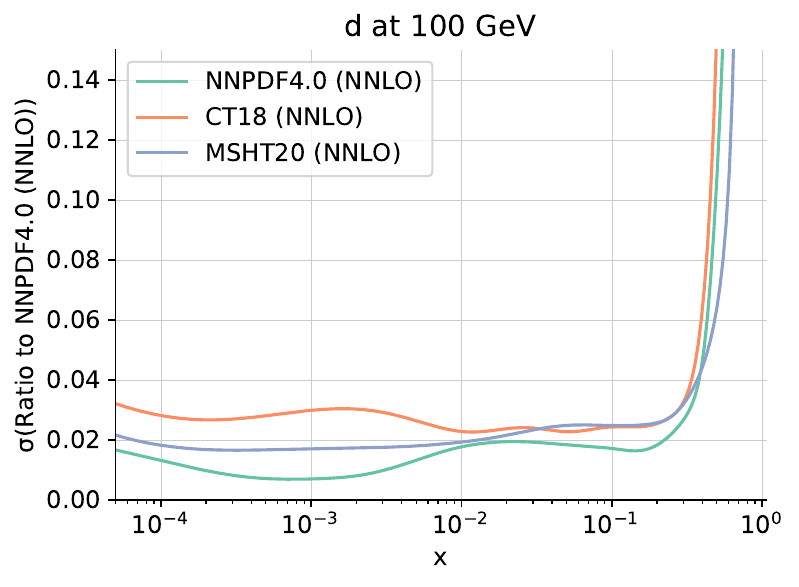}
  \includegraphics[width=0.45\textwidth]{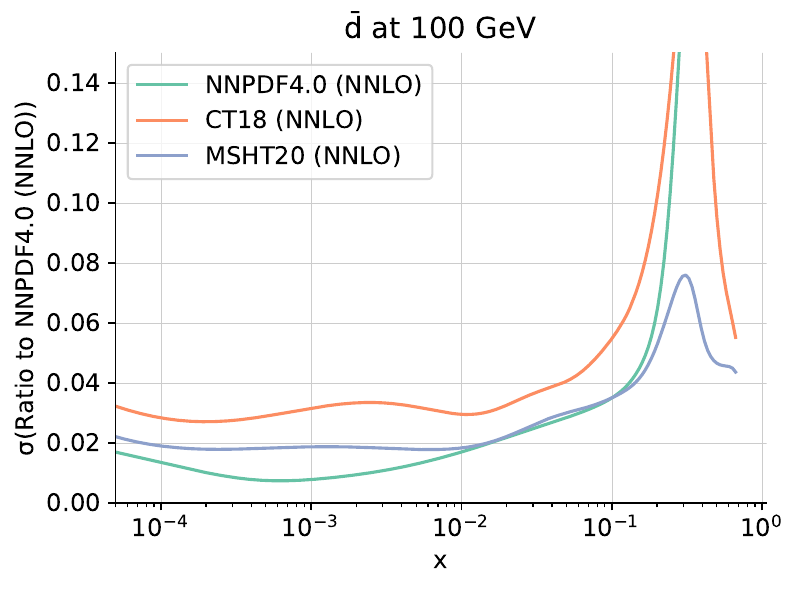}
  \includegraphics[width=0.45\textwidth]{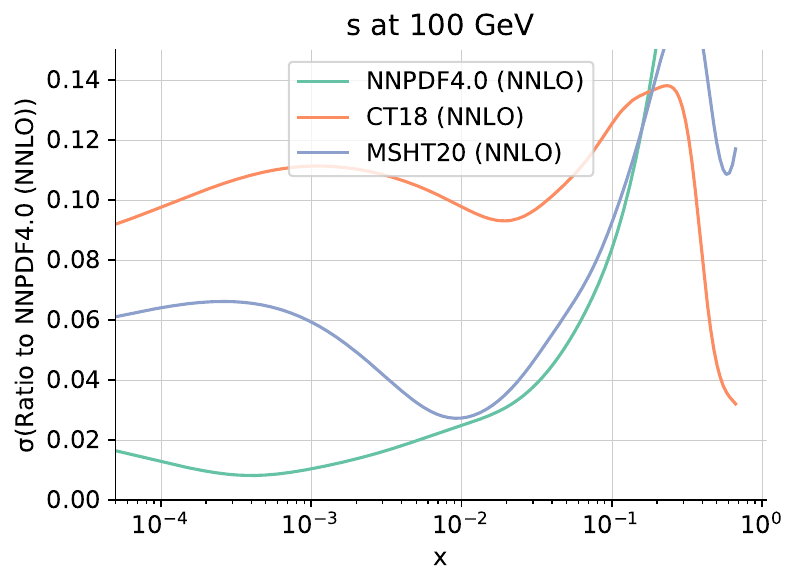}
  \includegraphics[width=0.45\textwidth]{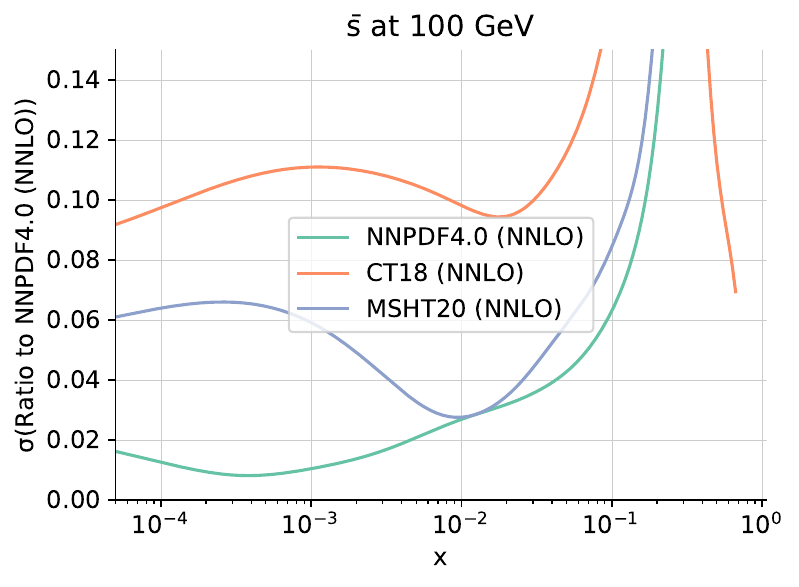}
  \includegraphics[width=0.45\textwidth]{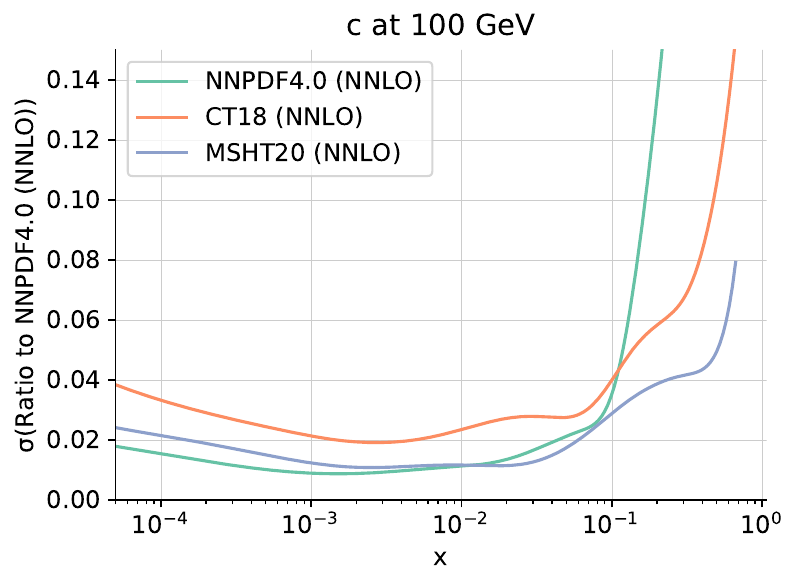}
  \includegraphics[width=0.45\textwidth]{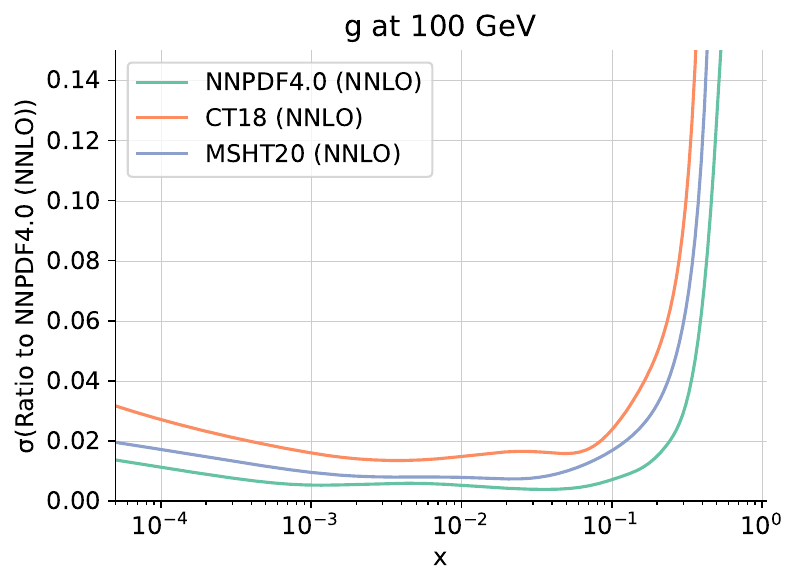}
  \caption{Same as Fig.~\ref{fig:40vsothers_PDFs} but for one-sigma relative
    uncertainties.}
  \label{fig:40vsothers_PDFs_uncs}
\end{figure}

The NNPDF4.0 NNLO PDFs are compared to other recent global sets,
namely CT18~\cite{Hou:2019efy} and
MSHT20~\cite{Bailey:2020ooq}, in Fig.~\ref{fig:40vsothers_PDFs}.
Note that there are substantial differences in the underlying dataset:  
the CT18 dataset is  very close to that of NNPDF3.1 while the MSHT20
dataset is somewhere in between NNPDF3.1 and NNPDF4.0 (see
Appendix.~\ref{app:datacomp} for a detailed comparison). All results
are shown at  $Q=100$~GeV, normalized to the central
NNPDF4.0 value. Relative uncertainties are compared  in
Fig.~\ref{fig:40vsothers_PDFs_uncs}. Note that while for NNPDF4.0
there are eight independently parametrized PDFs, for CT18 the strange
and antistrange are not independently parametrized, and for both CT18
and MSHT20 charm is not independently parametrized.

The three parton sets are overall in fair agreement within their respective
uncertainties, though some differences in shape are
observed. Interestingly, these
follow the pattern already observed in~\cite{Ball:2017nwa} when comparing
NNPDF3.1~\cite{Ball:2017nwa} to
CT14~\cite{Dulat:2015mca} and MMHT2014~\cite{Harland-Lang:2014zoa} (see in
particular Fig.~12 in Ref.~\cite{Ball:2017nwa}) .
The up and down PDFs are in  good agreement, in particular the NNPDF4.0 result
is always within the envelope of the CT18 and MSHT20 uncertainties. More
marked differences are observed for the antiup and antidown PDFs: note,
however, that the CT18 and MSHT20 PDF sets do not include
the E906/SeaQuest and the LHCb 13 TeV measurements, which provide additional
constraints on sea quark flavor separation at mid- and large-$x$
values, as discussed in
Sect.~\ref{sec:dataset} (see Ref.~\cite{Guzzi:2021fre} for a
discussion of the SeaQuest data in the CT18 framework). The NNPDF4.0 strange and antistrange PDFs agree
very well with MSHT20: in both these PDF sets, strangeness is enhanced in
comparison to CT18. As suggested
in~\cite{Bailey:2020ooq,Faura:2020oom}, this is likely due to the fact
that the ATLAS $W,Z$ 7~TeV data are not
included in the default CT18 fit (though they are included in the CT18A
variant set), and that NNLO massive corrections to the neutrino DIS
dimuon cross-sections are also not accounted for.

The NNPDF4.0 charm PDF is suppressed at intermediate values of $x$ in comparison
to CT18 and MSHT20, as a consequence of the fact that charm in CT18 and
MSHT20 is determined by perturbative matching conditions and is not
independently parametrized.
The  gluon  is in fair agreement in the region of $x\lesssim 0.03$
which is relevant for Higgs production though the NNPDF result is at
the upper edge of the MSHT20 and CT18 uncertainty; this was already the
case when comparing NNPDF3.1 to CT14 and MMHT2014.
At larger values of $x$, the NNPDF4.0 gluon 
is suppressed in comparison to CT18 and MSHT20. This behavior
is likely due to the  LHC top pair and jet data
that are included in NNPDF4.0 but not in the other sets.

Concerning the associated PDF  uncertainties, 
NNPDF is generally more precise, while CT18 has
generally larger uncertainties. This is consistent with the
observation that CT18 is based on a somewhat smaller dataset than
NNPDF4.0, with MSHT20 being in between, see
Appendix~\ref{app:datacomp} for more details.

\subsection{Comparison to experimental data}
\label{subsec:comparisondata}

\begin{figure}[!t]
 \centering
 \includegraphics[width=0.49\linewidth]{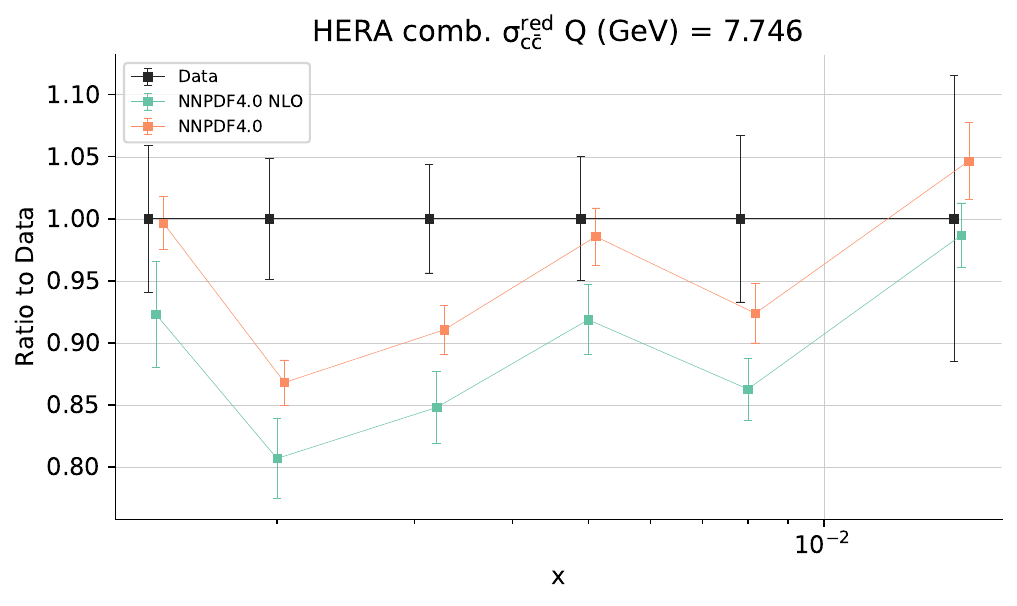}
 \includegraphics[width=0.49\linewidth]{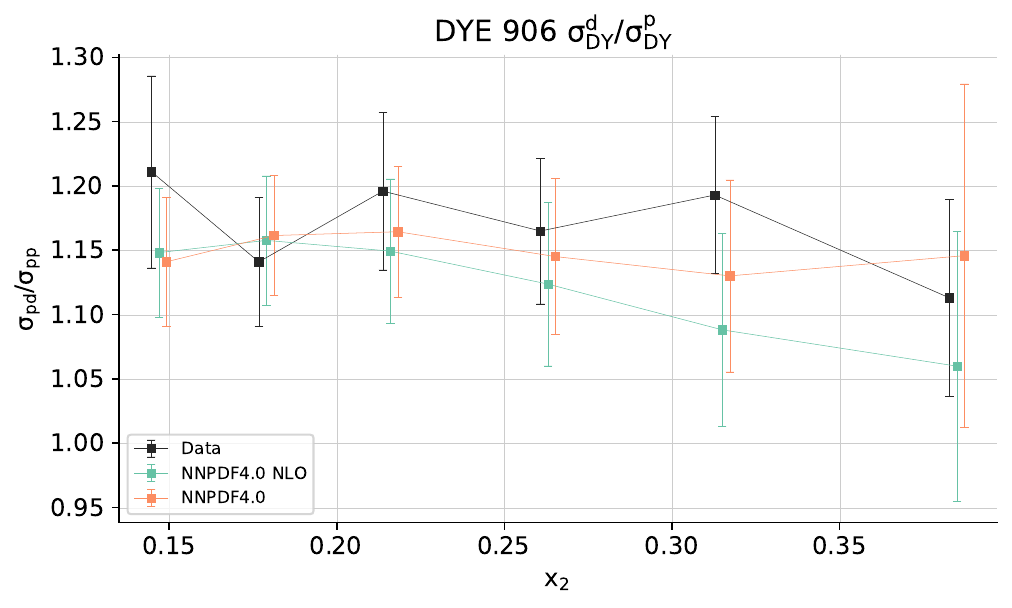}\\
 \includegraphics[width=0.49\linewidth]{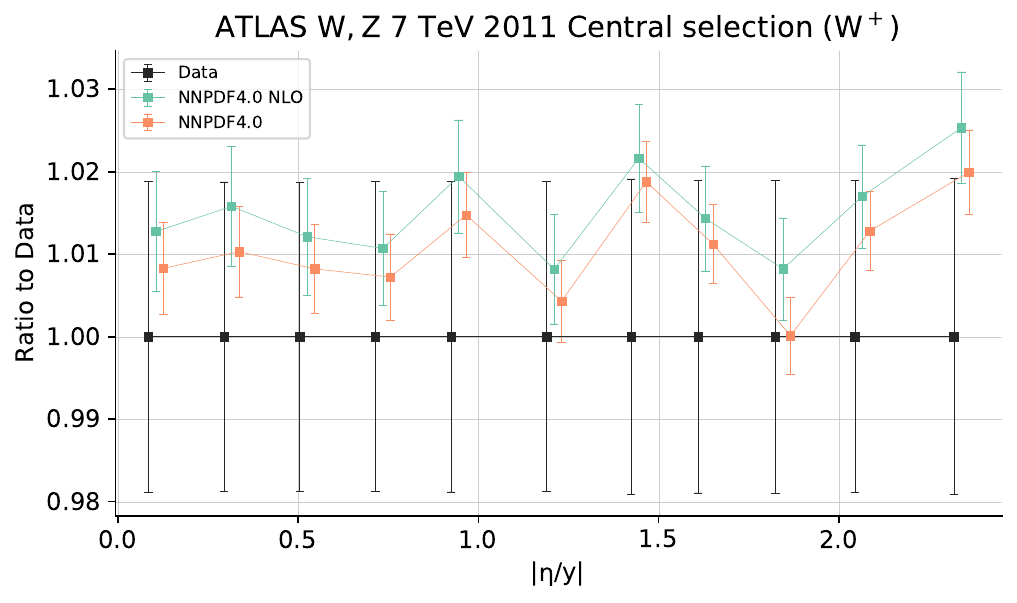}
 \includegraphics[width=0.49\linewidth]{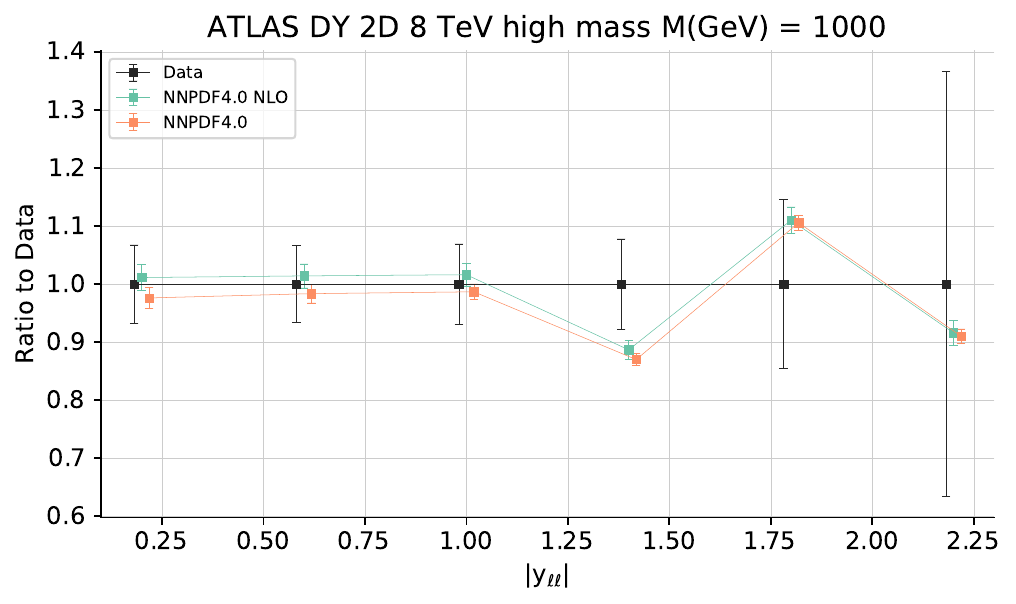}\\
 \includegraphics[width=0.49\linewidth]{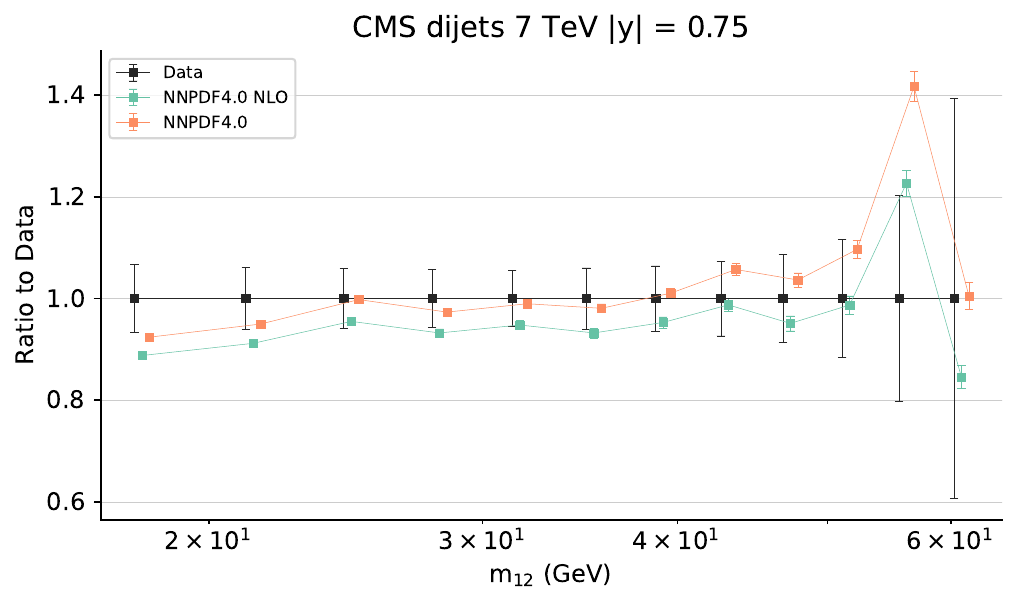}
 \includegraphics[width=0.49\linewidth]{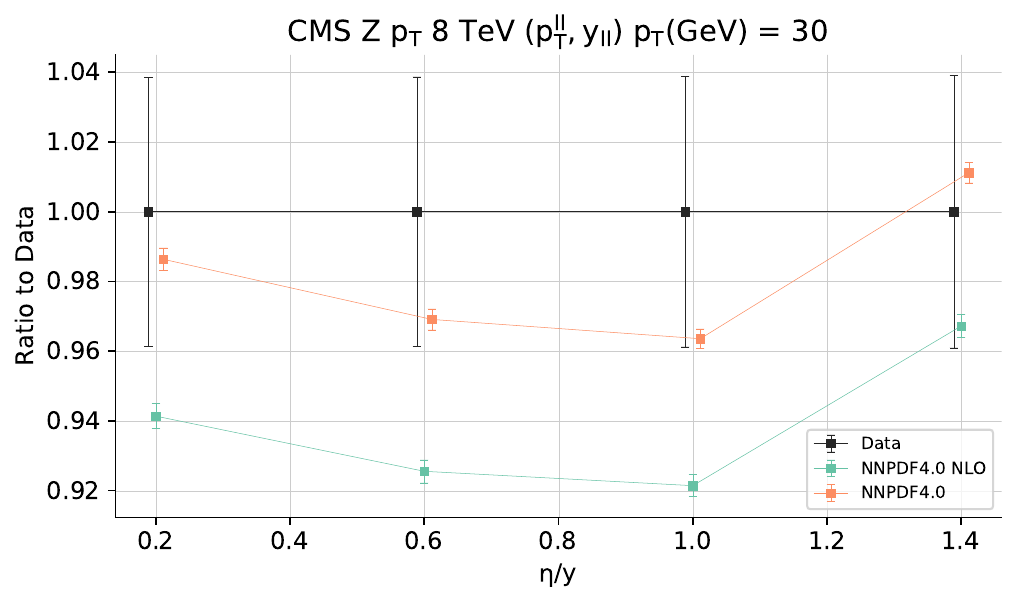}\\
 \includegraphics[width=0.49\linewidth]{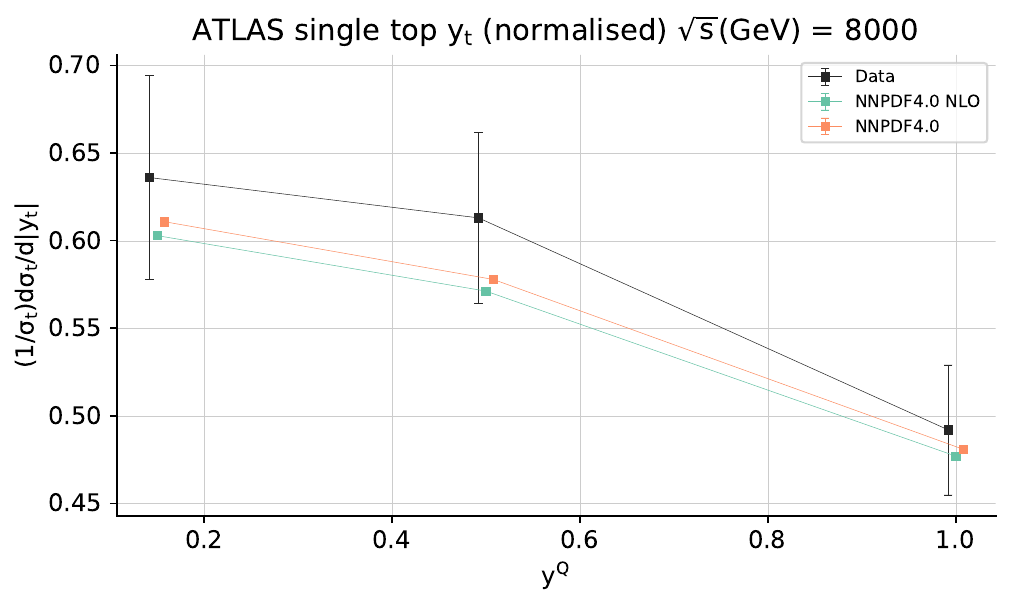}
 \includegraphics[width=0.49\linewidth]{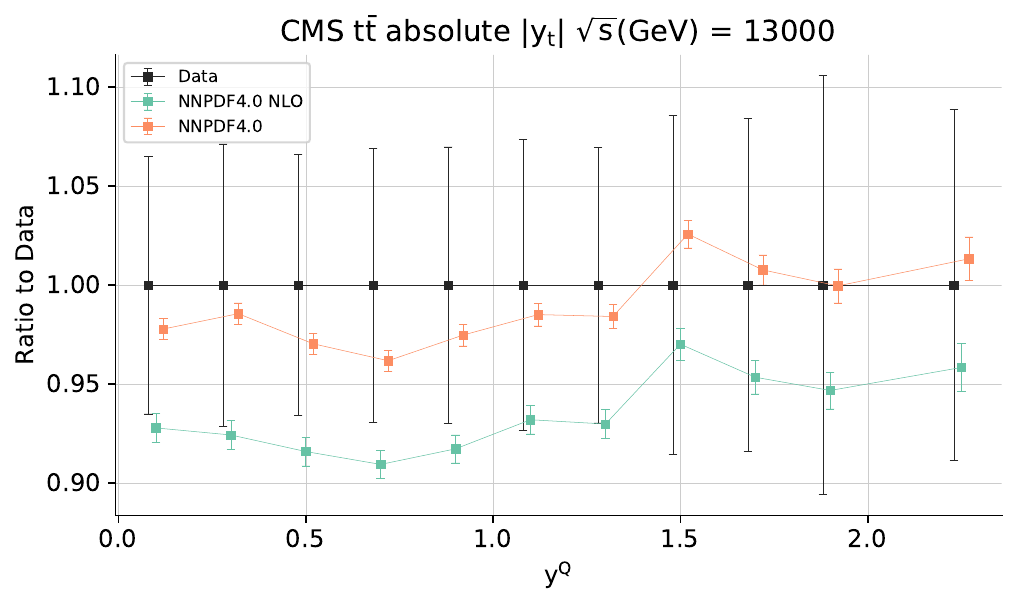}\\
 \caption{Comparison between data points and NLO and NNLO best-fit
   results for a selection of fitted data points (see text).
   Results are shown as ratios to the central experimental value, with 
   one-sigma experimental and PDF uncertainties. The
   experimental uncertainty is the sum in
   quadrature of all statistical and systematic uncertainties.}
 \label{fig:data_vs_theory_nnpdf40}
\end{figure}

In Fig.~\ref{fig:data_vs_theory_nnpdf40}
we present for illustrative purposes
a comparison between a selection of data
included in the NNPDF4.0 baseline fits and the corresponding NLO and
NNLO best-fit results, with the main goal of providing a visual
assessment of the fit quality and of the relative size of the data and
PDF uncertainties. The data shown are selected as representative of the global
dataset; specifically we show results for the 
following data: the lowest $Q$ bin of the combined HERA charm
cross-section~\cite{H1:2018flt}; the SeaQuest (DYE906) differential cross
section~\cite{Dove:2021ejl}; the central rapidity bin of the ATLAS 7~TeV $W^+$
rapidity distribution~\cite{Aaboud:2016btc}; the highest dilepton invariant
mass bin for  ATLAS 8~TeV high-mass DY~\cite{Aad:2016zzw};
the $0.5 \le |y| \le 1.0$ dijet rapidity bin for the CMS 7 TeV
dijets~\cite{Chatrchyan:2012bja}; the lowest $p_T^Z$ bin of the CMS 8~TeV
$Z$ $p_T$ distribution~\cite{Khachatryan:2015oaa}; the ATLAS 8~TeV normalized
single top rapidity distribution~\cite{Aaboud:2017pdi};
and the top rapidity distribution for CMS 13~TeV top pairs
in the lepton+jets final state~\cite{Sirunyan:2018wem}.
All results are normalized to the central experimental value. Data
error bars correspond to the sum in quadrature of all uncertainties.
Correlated systematic uncertainties are large or even dominant in several cases,
therefore the plots displayed in Fig.~\ref{fig:data_vs_theory_nnpdf40} should
be viewed as  a qualitative indication, while a quantitative
assessment is provided by  the $\chi^2$ values of
Tables~\ref{tab:DIS_dataset_chi2}-\ref{tab:OTHERLHCPROCESSES_dataset_chi2}. A 
full set of comparisons of the NNLO PDF
to all the data included in the fit are linked to the NNPDF website
\url{https://nnpdf.mi.infn.it/nnpdf4-0/} and can be found 
in~\cite{NNPDF:datatheory}. 

It is clear that  NNLO corrections are significant in many cases as
already noticed: specifically  for  combined HERA charm,
SeaQuest, the CMS 7 TeV dijets,
CMS 8~TeV $Z$ $p_T$ and the CMS 13~TeV top pairs.
In all these cases, the quality of the best fit visibly
improves  at NNLO.
PDF uncertainties are generally smaller than data
uncertainties. This is in part due to the fact that experimental
uncertainties are correlated while the diagonal uncertainty is shown
in the plots, but also to the fact that PDFs are simultaneously
constrained by several datasets.
Indeed, PDF uncertainties become
comparable to data uncertainties when the data shown are the only
ones to constrain the relevant PDFs: an example is the SeaQuest
data at very large  $x_2$ (momentum fraction of  the struck parton),
which is essentially the only dataset that constrains the
$\bar{d}/\bar{u}$ ratio in this region.

%% file: tables/tab-PROCESSTYPE_chi2.tex
\begin{tabularx}{\textwidth}{Xcrcrcrcr}
  \toprule
  \multirow{2}{*}{Dataset}
  & \multicolumn{6}{c}{NNPDF4.0}
  & \multicolumn{2}{c}{NNPDF3.1}
  \\
  & \multicolumn{2}{c}{LO}
  & \multicolumn{2}{c}{NLO}
  & \multicolumn{2}{c}{NNLO}
  & \multicolumn{2}{c}{NNLO}
  \\
  \midrule
  DIS NC (fixed-target)
  & 3.63 & (973)
  & 1.40 & (973)
  & 1.26 & (973)
  & 1.21 & (973)
  \\
  DIS CC (fixed-target)
  & 1.87 & (907)
  & 0.87 & (907)
  & 0.86 & (908)
  & 1.08 & (908)
  \\
  DIS NC (collider)
  & 2.26 & (930)
  & 1.18 & (980)
  & 1.19 & (1127)
  & 1.15 & (1130)
  \\
  DIS CC (collider)
  & 2.48 & (81)
  & 1.30 & (81)
  & 1.28 & (81)
  & 1.18 & (81)
  \\
  Drell-Yan (fixed-target)
  & 0.74 & (195)
  & 0.87 & (195)
  & 1.00 & (195)
  & 1.25 & (189)
  \\
  Tevatron $W,Z$ production (inclusive) 
  & 2.11 & (64)
  & 1.18 & (64)
  & 1.09 & (65)
  & 1.29 & (74)
  \\ 
  LHC $W,Z$ production (inclusive) 
  & 7.44 & (437)
  & 1.53 & (437)
  & 1.37 & (483)
  & 1.37 & (314)
  \\
  LHC $W,Z$ production ($p_T$ and jets)
  & 4.17 & (37)
  & 1.72 & (153)
  & 0.98 & (150)
  & 1.00 & (120)
  \\
  LHC top-quark pair production
  & 13.4 & (66)
  & 1.84 & (66)
  & 1.21 & (66)
  & 1.08 & (19)
  \\
  LHC jet production
  & 1.95 & (500)
  & 1.02 & (500)
  & 1.26 & (500)
  & 0.94 & (470)
  \\
  LHC isolated $\gamma$ production
  & 9.95 & (53)
  & 0.57 & (53)
  & 0.77 & (53)
  & \multicolumn{2}{c}{---}
  \\
  LHC single $t$ production
  & 0.82 & (17)
  & 0.36 & (17)
  & 0.36 & (17)
  & \multicolumn{2}{c}{---}
  \\
  \midrule
  Total
  & 3.35 & (4260)
  & 1.24 & (4426)
  & 1.16 & (4618)
  & 1.15 & (4285)
  \\
\bottomrule
\end{tabularx}

%% file: tables/tab-DIS_chi2.tex
\begin{tabularx}{\textwidth}{Xcrcrcrcr}
  \toprule
  \multirow{2}{*}{Dataset}
  & \multicolumn{6}{c}{NNPDF4.0}
  & \multicolumn{2}{c}{NNPDF3.1}
  \\
  & \multicolumn{2}{c}{LO}
  & \multicolumn{2}{c}{NLO}
  & \multicolumn{2}{c}{NNLO}
  & \multicolumn{2}{c}{NNLO}
  \\
  \midrule
  NMC $F_2^d/F_2^p$
  & 0.93 & (121)
  & 0.87 & (121)
  & 0.87 & (121)
  & 0.94 & (121)
  \\
  NMC $\sigma^{{\rm NC},p}$
  & 6.60 & (204)
  & 1.82 & (204)
  & 1.56 & (204)
  & 1.51 & (204)
  \\
  SLAC $F_2^p$
  & 4.87 & (33)
  & 1.67 & (33)
  & 0.95 & (33)
  & 0.81 & (33)
  \\
  SLAC $F_2^d$
  & 2.00 & (34)
  & 1.01 & (34)
  & 0.63 & (34)
  & 0.71 & (34)
  \\
  BCDMS $F_2^p$
  & 3.88 & (333)
  & 1.52 & (333)
  & 1.41 & (333)
  & 1.28 & (333)
  \\
  BCDMS $F_2^d$
  & 1.76 & (248)
  & 1.02 & (248)
  & 1.01 & (248)
  & 1.11 & (248)
  \\
  CHORUS $\sigma_{CC}^{\nu}$
  & 1.26 & (416)
  & 0.96 & (416)
  & 0.96 & (416)
  & 1.12 & (416)
  \\
  CHORUS $\sigma_{CC}^{\bar{\nu}}$
  & 1.60 & (416)
  & 0.90 & (416)
  & 0.87 & (416)
  & 1.06 & (416)
  \\
  NuTeV $\sigma_{CC}^{\nu}$ (dimuon)
  & 5.54 & (39)
  & 0.24 & (39)
  & 0.35 & (39)
  & 0.60 & (39)
  \\
  NuTeV $\sigma_{CC}^{\bar{\nu}}$ (dimuon)
  & 12.0 & (36)
  & 0.43 & (36)
  & 0.56 & (37)
  & 1.06 & (37)
  \\
  \midrule
  HERA I+II
  $\sigma_{\rm NC,CC}^{p}$
  & 2.29 & (1011)
  & 1.18 & (1011)
  & 1.17 & (1145)
  & 1.16 & (1145)
  \\
  HERA I+II $\sigma_{\rm NC}^{c}$
  & \multicolumn{2}{c}{---}
  & 2.04 & (24)
  & 2.03 & (37)
  & 1.45 & (37)
  \\
  HERA I+II $\sigma_{\rm NC}^{b}$
  & \multicolumn{2}{c}{---}
  & 1.38 & (26)
  & 1.43 & (26)
  & 1.11 & (29)
  \\
\bottomrule
\end{tabularx}

%% file: tables/tab-FTDY_chi2.tex
\begin{tabularx}{\textwidth}{Xcrcrcrcr}
  \toprule
  \multirow{2}{*}{Dataset}
  & \multicolumn{6}{c}{NNPDF4.0}
  & \multicolumn{2}{c}{NNPDF3.1}
  \\
  & \multicolumn{2}{c}{LO}
  & \multicolumn{2}{c}{NLO}
  & \multicolumn{2}{c}{NNLO}
  & \multicolumn{2}{c}{NNLO}
  \\
  \midrule
  E866 $\sigma^d/2\sigma^p$ (NuSea)
  & 0.68 & (15)
  & 0.60 & (15)
  & 0.49 & (15)
  & 0.41 & (15)
  \\
  E866 $\sigma^p$ (NuSea)
  & 1.14 & (89)
  & 1.30 & (89)
  & 1.60 & (89)
  & 1.43 & (89)
  \\
  E605 $\sigma^p$
  & 0.35 & (85)
  & 0.43 & (85)
  & 0.46 & (85)
  & 1.21 & (85)
  \\
  E906 $\sigma^d/2\sigma^p$ (SeaQuest)
  & 0.55 & (6)
  & 1.28 & (6)
  & 0.93 & (6)
  & \multicolumn{2}{c}{---} \\
\bottomrule
\end{tabularx}

%% file: tables/tab-GAUGEBOSON_chi2.tex
\begin{tabularx}{\textwidth}{Xcrcrcrcr}
  \toprule
  \multirow{2}{*}{Dataset}
  & \multicolumn{6}{c}{NNPDF4.0}
  & \multicolumn{2}{c}{NNPDF3.1}
  \\
  & \multicolumn{2}{c}{LO}
  & \multicolumn{2}{c}{NLO}
  & \multicolumn{2}{c}{NNLO}
  & \multicolumn{2}{c}{NNLO}
  \\  
  \midrule
  CDF $Z$ differential
  & 2.52 & (28)
  & 1.27 & (28)
  & 1.28 & (28)
  & 1.48 & (29)
  \\
  D0 $Z$ differential
  & 1.35 & (28) 
  & 0.69 & (28)
  & 0.64 & (28)
  & 0.60 & (28) 
  \\
  D0 $W$ electron asymmetry
  & \multicolumn{2}{c}{---}
  & \multicolumn{2}{c}{---}
  & \multicolumn{2}{c}{---}
  & 2.71 & (8)
  \\ 
  D0 $W$ muon asymmetry
  & 3.31 & (8)
  & 2.58 & (8)
  & 1.91 & (9)
  & 1.56 & (9)
  \\  
  \midrule
  ATLAS low-mass DY 7 TeV
  & 4.14 & (4)
  & 0.70 & (4)
  & 0.88 & (6)
  & 0.90 & (6)
  \\
  ATLAS high-mass DY 7 TeV
  & 4.29 & (5)
  & 1.82 & (5)
  & 1.68 & (5)
  & 1.54 & (5)
  \\
  ATLAS $W,Z$ 7 TeV ($\mathcal{L}=35$~pb$^{-1}$)
  & 3.92 & (30)
  & 1.09 & (30)
  & 0.98 & (30)
  & 0.96 & (30)
  \\
  ATLAS $W,Z$ 7 TeV ($\mathcal{L}=4.6$~fb$^{-1}$)
  & 17.2 & (53)
  & 2.91 & (53)
  & 1.67 & (61)
  & 2.14 & (34)
  \\
  CMS $W$ electron asymmetry 7 TeV
  & 2.50 & (11)
  & 0.92 & (11)
  & 0.84 & (11)
  & 0.78 & (11)
  \\
  CMS $W$ muon asymmetry 7 TeV
  & 2.10 & (11)
  & 1.98 & (11)
  & 1.70 & (11)
  & 1.75 & (11)
  \\
  CMS DY 2D 7 TeV
  & 4.08 & (88)
  & 1.31 & (88)
  & 1.36 & (110)
  & 1.27 & (110)
  \\
  LHCb $Z\to ee$ 7 TeV 
  & 3.38 & (9)
  & 1.47 & (9)
  & 1.65 & (9)
  & 1.48 & (9)
  \\
  LHCb $W,Z \to \mu$ 7 TeV
  & 5.38 & (29)
  & 1.59 & (29)
  & 1.97 & (29)
  & 1.76 & (29)
  \\ 
  ATLAS low-mass DY 2D 8 TeV
  & 19.4 & (47)
  & 1.37 & (47)
  & 1.22 & (60)
  & \multicolumn{2}{c}{---}
  \\
  ATLAS high-mass DY 2D 8 TeV
  & 5.25 & (48)
  & 1.53 & (48)
  & 1.11 & (48)
  & \multicolumn{2}{c}{---}
  \\
  CMS $W$ rapidity 8 TeV
  & 3.86 & (22)
  & 0.93 & (22)
  & 1.38 & (22)
  & 1.00 & (22)
  \\
  LHCb $Z\to ee$ 8 TeV
  & 6.97 & (17)
  & 1.64 & (17)
  & 1.33 & (17)
  & 1.14 & (17)
  \\
  LHCb $W,Z\to \mu$ 8 TeV
  & 5.37 & (29)
  & 1.06 & (29)
  & 1.42 & (30)
  & 1.37 & (30)
  \\
  ATLAS $\sigma_{W,Z}^{\rm tot}$ 13 TeV
  & 2.69 & (3)
  & 0.26 & (3)
  & 0.80 & (3)
  & \multicolumn{2}{c}{---}
  \\
  LHCb $Z\to ee$ 13 TeV
  & 2.36 & (15)
  & 1.70 & (15)
  & 1.72 & (15)
  & \multicolumn{2}{c}{---}
  \\
  LHCb $Z\to \mu\mu$ 13 TeV
  & 2.85 & (16)
  & 1.17 & (16)
  & 0.99 & (16)
  & \multicolumn{2}{c}{---}
  \\
  \bottomrule
\end{tabularx}

%% file: tables/tab-OTHERLHCPROCESSES_chi2.tex
\begin{tabularx}{\textwidth}{Xcrcrcrcr}
  \toprule
  \multirow{2}{*}{Dataset}
  & \multicolumn{6}{c}{NNPDF4.0}
  & \multicolumn{2}{c}{NNPDF3.1}
  \\
  & \multicolumn{2}{c}{LO}
  & \multicolumn{2}{c}{NLO}
  & \multicolumn{2}{c}{NNLO}
  & \multicolumn{2}{c}{NNLO}
  \\
  \midrule
  ATLAS $W^\pm+c$ 7 TeV
  & 3.24 & (22)
  & 0.61 & (22)
  & \multicolumn{2}{c}{---}
  & \multicolumn{2}{c}{---}
  \\
  CMS $W^\pm+c$ 7 TeV
  & 7.52 & (10)
  & 1.42 & (10)
  & \multicolumn{2}{c}{---}
  & \multicolumn{2}{c}{---}
  \\
  CMS $W^\pm+c$ 13 TeV
  & 1.60 & (5)
  & 0.74 & (5)
  & \multicolumn{2}{c}{---}
  & \multicolumn{2}{c}{---}
  \\
  ATLAS $W^\pm$+jet 8 TeV
  & \multicolumn{2}{c}{---}
  & 1.21 & (30)
  & 0.96 & (30)
  & \multicolumn{2}{c}{---}
  \\
  \midrule
  ATLAS $Z$ $p_T$ 8 TeV ($p_T,m_{\ell\ell}$)
  & \multicolumn{2}{c}{---}
  & 1.11 & (40)
  & 0.91 & (44)
  & 0.93 & (44)
  \\
  ATLAS $Z$ $p_T$ 8 TeV ($p_T,y_Z$)
  & \multicolumn{2}{c}{---}
  & 2.78 & (18)
  & 0.90 & (48)
  & 0.94 & (48)
  \\
  CMS $Z$ $p_T$ 8 TeV
  & \multicolumn{2}{c}{---}
  & 4.14 & (28)
  & 1.41 & (28)
  & 1.32 & (28)
  \\
  \midrule
  CMS $\sigma_{tt}^{\rm tot}$ 5 TeV
  & 16.7 & (1)
  & 1.95 & (1)
  & 0.54 & (1)
  & \multicolumn{2}{c}{---}
  \\
  ATLAS $\sigma_{tt}^{\rm tot}$ 7 TeV
  & 117. & (1)
  & 10.3 & (1)
  & 4.59 & (1)
  & 2.06 & (1) 
  \\
  CMS $\sigma_{tt}^{\rm tot}$ 7 TeV
  & 127. & (1)
  & 5.19 & (1)
  & 1.06 & (1)
  & 0.04 & (1)
  \\
  ATLAS $\sigma_{tt}^{\rm tot}$ 8 TeV
  & 81.9 & (1)
  & 1.70 & (1)
  & 0.02 & (1)
  & 0.27 & (1) 
  \\
  CMS $\sigma_{tt}^{\rm tot}$ 8 TeV
  & 119. & (1)
  & 3.43 & (1)
  & 0.26 & (1)
  & 0.08 & (1)
  \\
  ATLAS $\sigma_{tt}^{\rm tot}$ 13 TeV
  ($\mathcal{L}$=139~fb$^{-1}$)
  & 82.7 & (1)
  & 3.75 & (1)
  & 0.51 & (1)
  & 0.01 & (1)
  \\
  CMS $\sigma_{tt}^{\rm tot}$ 13 TeV
  & 52.4 & (1)
  & 0.72 & (1)
  & 0.06 & (1)
  & 0.45 & (1)
  \\
  ATLAS $t\bar{t}~\ell$+jets 8 TeV ($1/\sigma d\sigma/dy_t$)
  & 7.43 & (4)
  & 3.68 & (4)
  & 3.22 & (4)
  & 1.45 & (4)
  \\
  ATLAS $t\bar{t}~\ell$+jets 8 TeV ($1/\sigma d\sigma/dy_{t\bar t}$)
  & 17.8 & (4)
  & 6.88 & (4)
  & 3.77 & (4)
  & \multicolumn{2}{c}{---}
  \\
  ATLAS $t\bar{t}~2\ell$ 8 TeV ($1/\sigma d\sigma/dy_{t\bar t}$)
  & 2.51 & (5)
  & 1.85 & (5) 
  & 1.61 & (5)
  & \multicolumn{2}{c}{---}
  \\
  CMS $t\bar{t}~\ell$+jets 8 TeV ($1/\sigma d\sigma/dy_{t\bar t}$)
  & 4.40 & (9)
  & 1.89 & (9)
  & 1.23 & (9)
  & 0.94 & (9)
  \\
  CMS $t\bar{t}$ 2D $2\ell$ 8 TeV ($1/\sigma d\sigma/dy_tdm_{t\bar t}$)
  & 2.11 & (16)
  & 1.06 & (16)
  & 1.03 & (16)
  & \multicolumn{2}{c}{---}
  \\
  CMS $t\bar{t}~\ell$+jet 13 TeV ($d\sigma/dy_t$)
  & 5.50 & (11)
  & 0.71 & (11)
  & 0.63 & (11)
  & \multicolumn{2}{c}{---}
  \\
  CMS $t\bar{t}~2\ell$ 13 TeV ($d\sigma/dy_t$)
  & 5.47 & (10)
  & 0.79 & (10)
  & 0.52 & (10)
  & \multicolumn{2}{c}{---}
  \\
  \midrule
  CDF incl. jets 
  & \multicolumn{2}{c}{---}
  & \multicolumn{2}{c}{---}
  & \multicolumn{2}{c}{---}
  & 0.87 & (76)
  \\
  ATLAS incl. jets 2.76~TeV, $R=0.6$
  & \multicolumn{2}{c}{---}
  & \multicolumn{2}{c}{---}
  & \multicolumn{2}{c}{---}
  & 1.03 & (59)
  \\
  CMS incl. jets 2.76~TeV
  & \multicolumn{2}{c}{---}
  & \multicolumn{2}{c}{---}
  & \multicolumn{2}{c}{---}
  & 1.02 & (81)
  \\
  ATLAS incl. jets 7~TeV, $R=0.6$ (2010)
  & \multicolumn{2}{c}{---}
  & \multicolumn{2}{c}{---}
  & \multicolumn{2}{c}{---}
  & 0.95 & (90)
  \\
  ATLAS incl. jets 7~TeV, $R=0.6$ (2011)
  & \multicolumn{2}{c}{---}
  & \multicolumn{2}{c}{---}
  & \multicolumn{2}{c}{---}
  & 1.07 & (31)
  \\
  CMS incl. jets 7 TeV
  & \multicolumn{2}{c}{---}
  & \multicolumn{2}{c}{---}
  & \multicolumn{2}{c}{---}
  & 0.84 & (133)
  \\
  ATLAS incl. jets 8~TeV, $R=0.6$
  & 1.47 & (171)
  & 0.66 & (171)
  & 0.69 & (171)
  & \multicolumn{2}{c}{---}
  \\
  CMS incl. jets 8 TeV
  & 1.36 & (185)
  & 0.96 & (185)
  & 1.19 & (185)
  & \multicolumn{2}{c}{---}
  \\
  ATLAS dijets 7 TeV, $R=0.6$
  & 3.21 & (90)
  & 1.47 & (90) 
  & 2.15 & (90)
  & \multicolumn{2}{c}{---}
  \\
  CMS dijets 7 TeV
  & 3.43 & (54)
  & 1.57 & (54)
  & 1.81 & (54)
  & \multicolumn{2}{c}{---}
  \\
  \midrule
  ATLAS isolated $\gamma$ prod. 13 TeV
  & 9.95 & (53)
  & 0.57 & (53)
  & 0.77 & (53)
  & \multicolumn{2}{c}{---}
  \\
  \midrule
  ATLAS single~$t$ $R_{t}$ 7 TeV
  & 0.02 & (1)
  & 0.04 & (1)
  & 0.50 & (1)
  & \multicolumn{2}{c}{---}
  \\
  CMS single~$t$ $\sigma_{t}+\sigma_{\bar{t}}$ 7 TeV
  & 6.63 & (1)
  & 0.88 & (1)
  & 0.73 & (1)
  & \multicolumn{2}{c}{---}
  \\
  CMS single~$t$ $R_{t}$ 8 TeV
  & 0.01 & (1)
  & 0.14 & (1)
  & 0.17 & (1)
  & \multicolumn{2}{c}{---}
  \\
  ATLAS single~$t$ $R_{t}$ 13 TeV
  & 0.22 & (1)
  & 0.05 & (1)
  & 0.06 & (1)
  & \multicolumn{2}{c}{---}
  \\
  CMS single~$t$ $R_{t}$ 13 TeV
  & 0.61 & (1)
  & 0.33 & (1)
  & 0.36 & (1)
  & \multicolumn{2}{c}{---}
  \\
  ATLAS single~$t$ 7 TeV ($1/\sigma d\sigma/dy_t$)
  & 0.52 & (3)
  & 0.83 & (3)
  & 0.96 & (3)
  & \multicolumn{2}{c}{---}
  \\
  ATLAS single~$t$ 7 TeV ($1/\sigma d\sigma/dy_{\bar t}$)
  & 0.22 & (3)
  & 0.06 & (3)
  & 0.06 & (3)
  & \multicolumn{2}{c}{---}
  \\
  ATLAS single~$t$ 8 TeV ($1/\sigma d\sigma/dy_t$)
  & 1.13 & (3)
  & 0.38 & (3)
  & 0.25 & (3)
  & \multicolumn{2}{c}{---}
  \\
  ATLAS single~$t$ 8 TeV ($1/\sigma d\sigma/dy_{\bar t}$)
  & 0.26 & (3)
  & 0.19 & (3)
  & 0.19 & (3)
  & \multicolumn{2}{c}{---}
  \\
  \bottomrule
\end{tabularx}

%% file: tables/tab-alphas.tex
\begin{tabularx}{\textwidth}{XC{1.8cm}C{1.8cm}C{1.8cm}C{1.8cm}C{1.8cm}C{1.8cm}C{1.8cm}}
\toprule
$\alpha_s(m_Z)$
& 0.1160
& 0.1170
& 0.1175
& 0.1180
& 0.1185
& 0.1900
& 0.1200
\\
\midrule
$\chi^2$
& 1.183 
& 1.169
& 1.165 
& 1.162 
& 1.161 
& 1.162 
& 1.168
\\
\bottomrule
\end{tabularx}

%% file: sec-closuretests.tex
\section{Validation of the methodology}
\label{sec:closure}

We perform here a detailed validation of the NNPDF4.0 fitting methodology, 
with the main
goal of verifying that the resulting PDF uncertainties have been 
faithfully estimated.
A validation technique
through closure tests was introduced by us in
Ref.~\cite{Ball:2014uwa}, in order to validate the NNPDF3.x
methodology. This technique checks for the faithfulness of
PDF uncertainties in the region in which PDFs are constrained by the
data. We will apply it systematically to NNPDF4.0 in
Sect.~\ref{subsec:closuresettings}: thanks to the greater
computational efficiency of the NNPDF4.0 methodology 
(see Sect.~\ref{sec:benchmark}) we can now perform much more extensive 
and systematic tests than was previously possible. Furthermore, we 
can now also test for faithfulness of
uncertainties in the extrapolation region, i.e. where PDFs are not
directly constrained by data, by means of  future tests, introduced
recently in Ref.~\cite{Cruz-Martinez:2021rgy}. Future tests of the
NNPDF4.0 methodology will be presented in Sect.~\ref{sec:futuretest}.
This extensive validation, both in the data and the extrapolation
regions, is especially desirable given the 
small, percent-level PDF uncertainties that NNPDF4.0 achieves.

\input{subsec-closuretests.tex}


\input{subsec-futuretests.tex}

%% file: subsec-closuretests.tex
\subsection{Closure testing NNPDF4.0}
\label{sec:closuresettings}

The closure testing methodology was introduced for global PDF fits in
Ref.~\cite{Ball:2014uwa}, following a suggestion in
Ref.~\cite{demortier} and previous studies in
Ref.~\cite{Watt:2012tq}. Here we follow the original approach of Ref.~\cite{Ball:2014uwa} and
supplement it with a wider variety of estimators and more systematic
studies. First, we review the closure testing methodology and
 describe the settings adopted for the closure
tests of NNPDF4.0.
Then we introduce the statistical estimators used to validate
the outcome of these tests, including the definition
of some new estimators.
Finally, we present a detailed closure test analysis of the NNPDF4.0 methodology,
based on the statistical estimators introduced previously. A
discussion of the limitations of the closure tesing methdology is also
given in conclusion.
 A more detailed theoretical discussion of the
statistical underpinnings of the closure testing methodology that we
adopt can be found in Ref.~\cite{DelDebbio:2021whr}.

\subsubsection{The closure test setup}
\label{subsec:closuresettings}

The basic idea of closure testing is to perform a PDF determination
based on artificial data that have been generated with perfect
statistical properties from a known underlying law. Comparing results
to the known truth then allows one to check for statistical
consistency.

Specifically, 
assume that we have $\ndata$ experimental measurements,
normally distributed
around the true values $\boldsymbol{\law}$ with covariance matrix $\covmat$.
The central values of the experimental data $\boldsymbol{z}$ will then be given
in terms of their true values as
\begin{equation}
  \label{eq:levelonedatagen}
    \levone_{\datind} = \law_\datind + \shift_\datind \, , \quad \datind =1\,\ldots,\ndata \, ,
\end{equation}
where the vector of shifts $\boldsymbol{\shift}$ is drawn from
a multi-Gaussian distribution with covariance  $\covmat$,
$\mathcal{N}(\boldsymbol{0}, \covmat)$.
Within the Monte Carlo replica method for error
propagation adopted in this work,
the pseudodata which are used as actual input for the PDF fit,
$\boldsymbol{\levtwo}^{(\repind)}$, are generated by adding a further layer
of fluctuations,
\begin{equation}
  \label{eq:leveltwodatagen}
    \levtwo^{(\repind)}_{\datind} = \law_\datind + \shift_\datind + \noise^{(\repind)}_{\datind}\, , \quad \datind =1\,\ldots,\ndata \, , \quad \repind =1\,\ldots,\nreps \, ,
\end{equation}
where the index $\repind$ indicates that each Monte Carlo replica is generated
by drawing an independent noise vector $\boldsymbol{\noise}$ from the same
multi-Gaussian distribution $\mathcal{N}(\boldsymbol{0}, \covmat)$.
In the NNPDF approach, for each Monte Carlo replica $k$
defined in Eq.~(\ref{eq:leveltwodatagen}) a neural network
such as that displayed in
Fig.~\ref{fig:NNarch} is trained from the minimization of a figure of merit,
see also the discussion in Sect.~\ref{sec:methodology}.
This means that the neural network parameters are chosen by optimizing
\begin{equation}
    \label{eq:cost_function_rep}
    E^{(k)} = \frac{1}{\ndata} 
    \sum_{ij} (\model^{(\repind)}_i - \levtwo^{(\repind)}_i) 
\invcov{ij} (\model^{(\repind)}_j - \levtwo^{(\repind)}_j)\,,
\end{equation}
where we denote by $\boldsymbol{\model}^{(\repind)}$ the predictions
for the experimental data obtained from the neural network model fitted to
the $k$-th replica.

In a fit to actual experimental data we have access to the measured
central values
$\boldsymbol{\levone}$ and to the covariance matrix $\covmat$ as
estimated by the experimentalists. In a closure test we instead use 
a given set of PDFs and associated theoretical calculation
as input for the central values.
Hence, the starting point of the closure test is
a known proxy of the true underlying observable values,
$\boldsymbol{\law}$.
Subsequently, a
proxy for the experimental central values is generated
following Eq.~\eqref{eq:levelonedatagen}.
A  closure test thus amounts to applying to closure test data the NNPDF
methodology as it would be used in a fit to actual experimental data.

\subsubsection{Statistical estimators}
\label{sec:closure-test-estimators}

A successful closure test must be such that the resulting PDF fit yields a
faithful statistical description of the known underlying law.
In order to assess quantitatively the degree of success
of the NNPDF4.0 closure tests presented here, we have extended and
systematized
the set
of estimators introduced in previous studies~\cite{Ball:2014uwa}.
Here we provide a summary of the  estimators and their justification;
for more detailed derivations and arguments showing  of how
they fit into a Bayesian approach to inverse
problems we refer the reader to~\cite{DelDebbio:2021whr}.

\paragraph{Bias, variance, and noise in closure tests.}
We define an error function as the expectation value across PDF replicas,
denoted as $\erep{\cdot}$, of the $\chi^2$ evaluated between the data predictions obtained
from the $k$-th PDF replica, $\boldsymbol{\model}$,
and the corresponding experimental central values, $\boldsymbol{\levone}$,
\begin{equation}
    \label{eq:chi2kerep}
    \erep{\repchis} \equiv \frac{1}{\ndata} 
    \erep{ \sum_{ij} (\model^{(\repind)}_i - \levone_i) \invcov{ij} 
(\model^{(\repind)}_j - \levone_j)}\, .
\end{equation}
It is easy to check~\cite{DelDebbio:2021whr} that  this expression can be
decomposed as 
\begin{equation}
    \label{eq:chi2decomp}
    \begin{split}
        \erep{\repchis} &= {\rm noise} + {\rm bias} + {\rm variance} - {\rm cross\,term} \\
        &=  {\rm noise} + {\rm variance} + \Delta_{\chi^2}, \\
        &= \chi^2 + {\rm variance} \, ,
    \end{split}
\end{equation}
where each of the quantities on the right-hand side is defined as follows.

First of all, the {\it noise} is defined as
\begin{equation}
    \label{eq:NoiseDef}
    \mathrm{noise} = \frac{1}{\ndata} \sum_{ij}
    \left( f_i - z_i \right)
    \invcov{ij}
    \left( f_j - z_j \right)
\end{equation}
and represents the fluctuations of the experimental data $\boldsymbol{\levone}$ around the true value $\boldsymbol{\law}$.
Eq.~(\ref{eq:NoiseDef}) is clearly independent of the model adopted, being
an intrinsic property of the experimental measurements. Note that by construction the noise will tend to 
one in the limit of large $\ndata$.

The {\em
 bias} is defined as the difference between the central value of the model replica
predictions, $\erep{g}$, and the true observable values $\boldsymbol{\law}$,
in units of the experimental covariance matrix, \ie
\begin{equation}
    \label{eq:BiasDef}
    \bias = \frac{1}{\ndata} \sum_{\datind \datindj} \diffcentunder_\datind \invcov{\datind \datindj} \diffcentunder_\datindj\, .
\end{equation}
The bias measures the deviation between the result of the fit and the
underlying law. In general, it is desirable for a PDF fit to exhibit a smaller bias because that
indicates that the fit results are closer to the truth. However,
consistency of a PDF fit does not depend on the size of the bias, but
rather, on whether the size of the bias is correctly reproduced by the
PDF uncertainty, as we discuss below. 

Finally, the {\em variance} term describes the fluctuations of the model replica predictions
around their mean value again in units of the experimental covariance matrix,
\begin{equation}
    \label{eq:VarDef}
    \var = \frac{1}{\ndata} \erep{ \sum_{\datind \datindj} \diffcentrep_{\datind} \invcov{\datind \datindj} \diffcentrep_{\datindj}}\, ,
\end{equation}
which can be interpreted as the projection of the PDF uncertainty to the space
of experimental data.
We note that this variance as defined in Eq.~(\ref{eq:VarDef}) actually corresponds to the square of the estimator
$\phi$ introduced in~\cite{Ball:2014uwa}. For a discussion of the
cross term in Eq.~(\ref{eq:chi2decomp}) we refer to~\cite{DelDebbio:2021whr}.

Since the variance can
be determined purely from the model predictions and the experimental covariance matrix,
it can also be calculated for fits to real experimental data.
This is in contrast
to the noise Eq.~(\ref{eq:NoiseDef}) and bias Eq.~(\ref{eq:BiasDef}), which depend
on the true law $\boldsymbol{\law}$ and hence can only be evaluated in closure tests.
It is also important to note here that both variance and bias can be
computed  without using any knowledge of statistical fluctuations that
enter closure tests. 

One can observe that the second line of the decomposition of the error function
in Eq.~(\ref{eq:chi2decomp}) is expressed as the
sum of the noise, the variance, and of $\Delta_{\chi^2}$.
This last quantity was
introduced in~\cite{Ball:2014uwa} and is defined as the difference between the
 $\chi^2$ evaluated from comparing  the expectation value of the model predictions  $\erep{g}$ and the level
one data $\boldsymbol{\levone}$, that is $\chi^2\lc \erep{\boldsymbol{\model}},\boldsymbol{\levone}\rc$,
and the $\chi^2$ evaluated between the underlying observable values $\boldsymbol{\law}$
and the same level one data, that is $\chi^2 \lc \boldsymbol{\law},\boldsymbol{\levone} \rc$.
We note that the latter coincides with the noise in Eq.~(\ref{eq:NoiseDef}).
Here we slightly redefine $\Delta_{\chi^2}$ as compared to~\cite{Ball:2014uwa}
by normalizing by the number of data points, such that
\begin{equation}
  \label{eq:deltachi2_def}
    \begin{split}
        \Delta_{\chi^2} &\equiv 
        \chi^2\lc \erep{\boldsymbol{\model}},\boldsymbol{\levone}\rc - 
        \chi^2 \lc \boldsymbol{\law},\boldsymbol{\levone} \rc  \\
        &= \chi^2 - {\rm noise}\, .
    \end{split}
\end{equation}
With this definition, constant values of $\Delta_{\chi^2}$ define elliptical contours in data space
centered on the  pseudodata Eq.~(\ref{eq:levelonedatagen}).

The value of $\Delta_{\chi^2}$ can be interpreted as a qualitative measure
of over- or under-fitting, when it is evaluated on data included in the fit.
In particular,  $\Delta_{\chi^2} = 0$  defines a
contour which is centered on the fitted level one data and passes through the
underlying observables. If $\Delta_{\chi^2} < 0$ then the expectation value of the model
predictions fit the level one data better than the underlying observables:
this then suggests an overfitting of the shift $\boldsymbol{\shift}$.   
Similarly, $\Delta_{\chi^2} > 0$ indicates underfitting of the level one data.
As discussed in Ref.~\cite{DelDebbio:2021whr} however, the replica distribution can be perfectly sampled from
the posterior distribution in model space and $\Delta_{\chi^2}$ can still be
negative. The overall shift of the PDF predictions is thus not an issue
as long as the uncertainties account for it.
The bottom line is that finding values of
$\Delta_{\chi^2} \leq 0$ in the closure test remains acceptable
provided their magnitude is sufficiently small,
which would indicate some
combination of a smaller correlation with the level one data and a smaller
bias.
Assuming that in such a case one finds  that the PDF uncertainties are faithful, this result
can be interpreted  as  passing the closure test.

In summary, the closure tests provide us with indicators that allow
us to assess whether PDF uncertainties are faithful, 
and furthermore how close the fit is to the truth,
i.e. whether the final result is optimal fit, or an over- or
under-fit. This provides a criterion for comparing methodologies: given
two methodologies that both produce a faithful result, an over- or under-fitted 
methodology is disfavored in comparison to one that leads to a proper
fit. We now turn to our main indicator for faithfulness, the
bias-to-variance ratio.

\paragraph{The bias-to-variance ratio for  closure tests.}
In the context of a closure test fit, the experimental central values (or level one
data) defined in Eq.~(\ref{eq:levelonedatagen}) are viewed as stochastic
variables.
When one performs fits to experimental data, $\boldsymbol{\levone}$ is fixed at the
published central
value  which will be to some extent shifted from
the true observable value due to the experimental uncertainties.
However, in
closure fits we are free to generate several instances of the shift $\boldsymbol{\shift}$, and
use this feature to design our estimators --- these would correspond
to ``runs of the universe'' in the real world. 

Considering the data which are included in the fit, the bias Eq.~(\ref{eq:BiasDef}) is potentially
driven by two methodology related features which we are aiming to
validate with the closure test.
The first mechanism is broadly described as
under-fitting, and covers inflexibility of the model or inability for the
optimization algorithm to sufficiently minimize the cost
function.
The second mechanism would be over-fitting of the level one shift,
which means that  the central value of the observables is systematically shifted
towards the level one data by an amount that is not properly accounted
for by the PDF uncertainties, which are thus underestimated.
Note that in order for the testing of these effects to be nontrivial it
is necessary to select the underlying truth as sufficiently
flexible and in a model-independent way.

Due to its dependence on the shift vector,
$\boldsymbol{\shift}$, $\Delta_{\chi^2}$ is a stochastic variable. In order to 
characterize the regime our model is in, we need to understand its probability distribution, rather than 
computing a single instance of it. For this purpose, we run multiple closure fits, 
each time with different shifts; we then reconstruct the distribution, and determine the expectation
value of $\Delta_{\chi^2}$ across fits.
It is worth noting that, compared to
previous NNPDF studies, a study using multiple full replica closure fits has
only been made possible by the computational speed up from deployment of
state-of-the-art machine learning algorithms detailed in
Sec.~\ref{sec:methodology}.
Results for the distribution of the $\Delta_{\chi^2}$ estimator over
fits are presented in Sect.~\ref{sec:ClosureTestResults}.

The main  question to be addressed by the closure test
is whether the uncertainty of the PDFs, represented by
an ensemble of PDF replicas, is a faithful propagation of the data uncertainty
into the space of PDFs. In the context of running multiple closure fits 
this question can be answered either by looking at the PDFs directly
(as was done in Ref.~\cite{Ball:2014uwa}), or by looking at
predictions for physical observables obtained using these PDFs.
The latter choice offers the
distinct advantage that the space of physical observables always
has a finite dimension, equal to the number of
data points for which predictions are computed.
In order for the test to be nontrivial, we choose to
evaluate the estimators on data which were not included in the fit, so
that we are assessing whether uncertainties are faithful on {\em new} observables.

From a Bayesian perspective, the PDF replicas obtained from a fit to a 
given set of data
can be treated as a sample from the prior model distribution for data which was
not used in that fit, similarly to the concept of Bayesian
reweighting~\cite{Ball:2010gb,Ball:2011gg}.
For the present study, we will perform fits
on a subset of the full NNPDF4.0 dataset and then calculate the estimators
discussed below on some test data which were not included in each fit.

In order to evaluate the faithfulness of the PDF uncertainties, one can
first take the expectation of the bias across fits with different
shifts in Eq.~(\ref{eq:levelonedatagen}), namely
\begin{equation}
    \label{eq:biastrace}
    \begin{split}
        \eshift{\bias} &= \frac{1}{\ndata} \eshift{ \sum_{\datind \datindj} 
        \diffcentunder_\datind \invcov{\datind \datindj} \diffcentunder_\datindj} \\
        &= \frac{1}{\ndata} \tr \left(\biascov \invcov{}\right),
    \end{split}
\end{equation}
where the subindex $\eshift{.}$ indicates that we are averaging over fits with different level-one shifts $\boldsymbol{\shift}$.
In Eq.~(\ref{eq:biastrace}) we introduced $\biascov$, the covariance matrix of the
difference between the central value of the predictions and the true observable
values estimated from the sample of fits,
\be
\label{eq:biascov}
\biascov \equiv \eshift{  \diffcentunder \diffcentunder^T } \, .
\ee
The expectation of the bias across
fits is then the expected distance between the central predictions and the true values
in units of the covariance matrix averaged across all data.
If the
fluctuations over fits reproduce the experimental covariance $C$ exactly, then the estimator
defined in Eq.~(\ref{eq:biastrace})
should be equal to one.

Similarly, we can take the expectation value of the
variance across fits with different  shifts Eq.~(\ref{eq:levelonedatagen}),
\begin{equation}\label{eq:vartrace}
    \begin{split}
        \eshift{\var} &= \frac{1}{\ndata} \eshift{ \erep{ \sum_{\datind \datindj} \diffcentrep_{\datind} 
        \invcov{\datind \datindj} \diffcentrep_{\datindj}}} \\
        &= \frac{1}{\ndata} \mathbf{E}_{\shift}\left[ \tr \left(\varcov 
        \invcov{}\right)\right],
    \end{split}
\end{equation}
which, in analogy to Eqs.~(\ref{eq:biastrace}) and~(\ref{eq:biascov}), has introduced $\varcov$ which is
the covariance of the fitted model predictions about their central value,
\be
\label{eq:varcov}
\varcov \equiv \erep{  \diffcentrep \diffcentrep^T } \, .
\ee
Since it is independent
of the shift $\boldsymbol{\shift}$,
 $\varcov$ is expected to be constant across fits. However, in
practice we prefer to take the expectation value across fits, since there
are sure to be fluctuations in the variance due to the finite number of replicas
in each fit.

We can then interpret the expectation of the variance across fits, Eq.~(\ref{eq:vartrace}), to be the
uncertainty of the predictions propagated from PDFs when averaged across all data in units
of the experimental covariance matrix.
If the
uncertainty associated to the PDF replicas is faithful, the bias-to-variance ratio (averaged over fits) is
\begin{equation}
    \label{eq:BiasVarRatio}
    \frac{\eshift{\bias}}{\eshift{\var}} = 1\, ,
\end{equation}
\ie\ the average distance between the central prediction from the replicas and the true
value is of the same order as the variance across replicas. We note that both
bias and variance are squared quantities and so in practice we shall instead
consider the square root of the ratio,
\begin{equation}
  \label{eq:BiasVarRatio_sqr}
    \biasvarratio\equiv \sqrt{\frac{\eshift{\bias}}{\eshift{\var}}}.
\end{equation}

The bias-to-variance ratio Eq.~(\ref{eq:BiasVarRatio_sqr}) is somewhat
coarse: it checks that the mean-square
difference between central predictions and underlying law is the same as the
mean-square difference between replica predictions and their central values.
The value of $\biasvarratio$ is  a measure of
how much the uncertainty
has been over- or under-estimated, {\it e.g.}, the uncertainty for a given fit is, on
average, over- or under-estimated by a factor of $1/\mathcal{R}_{bv}$.

This measure can be be made more fine-grained in two different ways.  
First,  one can evaluate Eq.~(\ref{eq:BiasVarRatio_sqr}) separately for specific subsets
or groups of processes, in addition to the total dataset: this then
effectively tests faithfulness for different PDFs or different
kinematic regions, namely, those to which the specific chosen processes
are most sensitive. Second, one can view the bias and variance as
measures of one-sigma deviations, and extend them to generic quantile
statistics measures, as we now discuss.

\paragraph{Quantile statistics in PDF and data space.}
In order to demonstrate that the PDF uncertainties were faithfully estimated,
in the NNPDF3.0 closure test studies estimators $\xi_{1\sigma}$,
$\xi_{2\sigma}$, etc. were defined, which provide the fraction of fits for which the input
PDF falls within one-sigma, two-sigma, etc. intervals of the central PDF, averaged over
PDF flavors and values of $x$, where the standard deviation is estimated
as usual from  the ensemble of PDF
replicas.
Specifically, the definition of these estimators was the following:
\begin{equation}
    \label{eq:CT_xi1_PDFs}
    \xi_{n\sigma}^{\rm (pdf)} = \frac{1}{\nflav}\frac{1}{\nx}\frac{1}{\nfits}
        \sum_{i=1}^{\nflav}\sum_{j=1}^{\nx}\sum_{l=1}^{\nfits}
        I_{\lc -n\sigma^{i(l)}(x_j), n\sigma^{i(l)}(x_j)\rc}
        \lp \erep{q^{i(l)}(x_j)} - q_{\rm in}^i(x_j)  \rp \, ,
\end{equation}
where $I_A(x)$ denotes the indicator function of the interval $A$: it is only non-zero,
and equal to one, if its argument lies in the interval $A$, while it vanishes for all other
values of its argument.
Here $q_{\rm in}^i$ indicates the true value of the $i$-th flavor PDF used to generate
the pseudodata and $q^{i(l)}$ the corresponding fitted PDF
from the $l$-th fit, and where both PDFs are evaluated at the input parametrization
scale $Q_0$.
The average is carried out over the $n_{\rm flav}$ non-zero flavors at $Q_0$
over a grid $\{ x_j\}$ with $n_x$ nodes.
 Finally, $\sigma^{i(l)}(x_j)$ is the standard deviation of the
replicas of the $l$-th fit for flavor $i$ estimated at $x_j$ from the
fitted replica distribution.

The estimators defined in Eq.~(\ref{eq:CT_xi1_PDFs}) can be evaluated
in the
closure test fits which reproduce the methodology of an actual
fit, and is thus
where the replica distribution should give faithful uncertainties.
For a successful closure test one should find that $\xi_{1\sigma}\simeq0.68$
if the PDF uncertainties are correctly estimated.
An important caveat here is that one relies on the
assumption that both the PDF replicas and expectation values of the PDFs across
fits both are distributed normally.
This assumption holds by construction for the closure test data
Eqs.~(\ref{eq:levelonedatagen},\ref{eq:leveltwodatagen}), so for PDFs it
likely only holds in
the region where the PDFs are constrained by the normally distributed
data. The measure Eq.~(\ref{eq:CT_xi1_PDFs}) is thus only significant
if computed for well constrained PDFs $q^{i(l)}(x_j)$: it can then be
defined by choosing a suitable sampling of PDFs in the relevant region.

One can also define an analogous estimator, now in the space of experimental
data as opposed to the PDF-space definition of Eq.~(\ref{eq:CT_xi1_PDFs}), as follows
\begin{equation}
    \label{eq:XiDataDef}
    \xi^{(\rm data)}_{n\sigma} =
        \frac{1}{\ndata} \frac{1}{\nfits}
        \sum_{i}^{\ndata}  \sum_{l}^{\nfits}
        I_{[-n\sigma_i^{(l)}, n\sigma_i^{(l)}]}
        \left( \erep{\model_i}^{(l)} - \law_i \right),
\end{equation}
where $\sigma^{(l)}_i$ is the standard deviation (PDF uncertainty) of the theory
predictions for the $i$-th observable estimated from the $\nreps$ replicas
of the $l$-th fit.
Here, if the test is performed by computing the estimator for data not
used for PDF fitting, in order to make sure that the Gaussianity
assumption holds one must choose testing data which are sensitive to PDF
combinations and kinematic regions that are well constrained by the
fitting data.

This $\xi^{(\rm data)}_{n\sigma}$ estimator provides the desired generalization
to quantile statistics of the bias-to-variance ratio $\mathcal{R}_{bv}$.
To see this, note first that 
we can calculate $\xi^{(\rm data)}_{n\sigma}$
in different bases and that, unlike $\chi^2$ or other quantities with
bilinear forms,
$\xi^{(\rm data)}_{n\sigma}$ is not basis independent.
Then, in order  to compare $\xi^{(\rm data)}_{n\sigma}$ to
$\biasvarratio$, compute  $\xi^{(\rm data)}_{1\sigma}$
in the basis which diagonalizes the experimental covariance
matrix. The sum across data points then becomes the sum across eigenvectors
of the experimental covariance matrix.

In this basis, one can then  evaluate~\cite{DelDebbio:2021whr}  Eq.~(\ref{eq:XiDataDef}) by
means of  the  approximation 
\begin{equation}
    \label{eq:expectedxi}
    \xi_{n\sigma}^{(\rm data)} \approx
    \erf \left( \frac{n \biasvarratio}{\sqrt{2}}\right),
\end{equation}
which is the standard result of integrating a Gaussian over some finite
symmetric interval, assuming that the ratio of uncertainties is approximately
constant across all eigenvectors of the experimental covariance matrix.
Clearly, if the distribution of central predictions about the
underlying law matches the distribution of the replica predictions around the
central predictions ($\biasvarratio \simeq 1$), then the expected value of $\xi_{1\sigma}^{(\rm data)}$
is 0.68.
This shows that the bias-to-variance ratio tests for accuracy of
quantile statistics, just like the estimator Eq.~(\ref{eq:XiDataDef}),
and its counterpart in PDF space Eq.~(\ref{eq:CT_xi1_PDFs}).

Once again, note that the computation of the estimators
Eqs.(\ref{eq:CT_xi1_PDFs},\ref{eq:XiDataDef})  requires running
multiple replica closure fits based on different underlying 
data Eq.~(\ref{eq:levelonedatagen}). This,  as mentioned, is only possible now, thanks to the much
greater computational efficiency of the current methodology. Indeed, in Ref.~\cite{Ball:2014uwa}
the estimator Eq.~(\ref{eq:CT_xi1_PDFs}) was only evaluated
approximately, based on a single closure test run and a suitable
approximation. We have in fact now verified a posteriori that the
approximation of Ref.~\cite{Ball:2014uwa} is reasonably accurate, but
only now it is possible to compute the estimator exactly.

\subsubsection{Closure test settings}

\begin{figure}[!t]
    \begin{center}
      \includegraphics[width=0.49\textwidth]{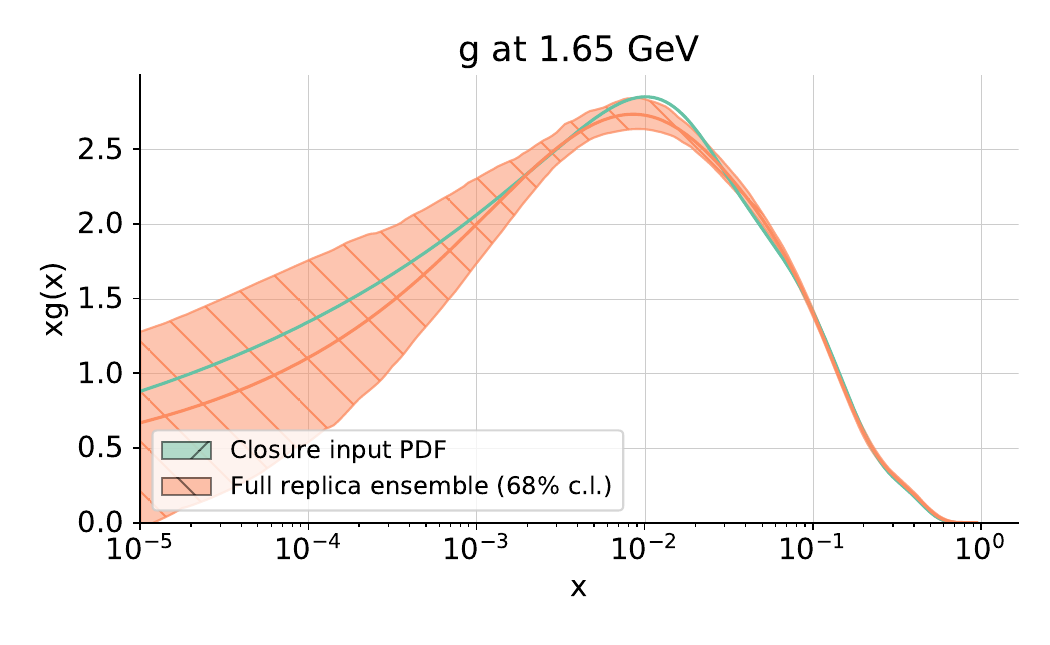}
      ~
      \includegraphics[width=0.49\textwidth]{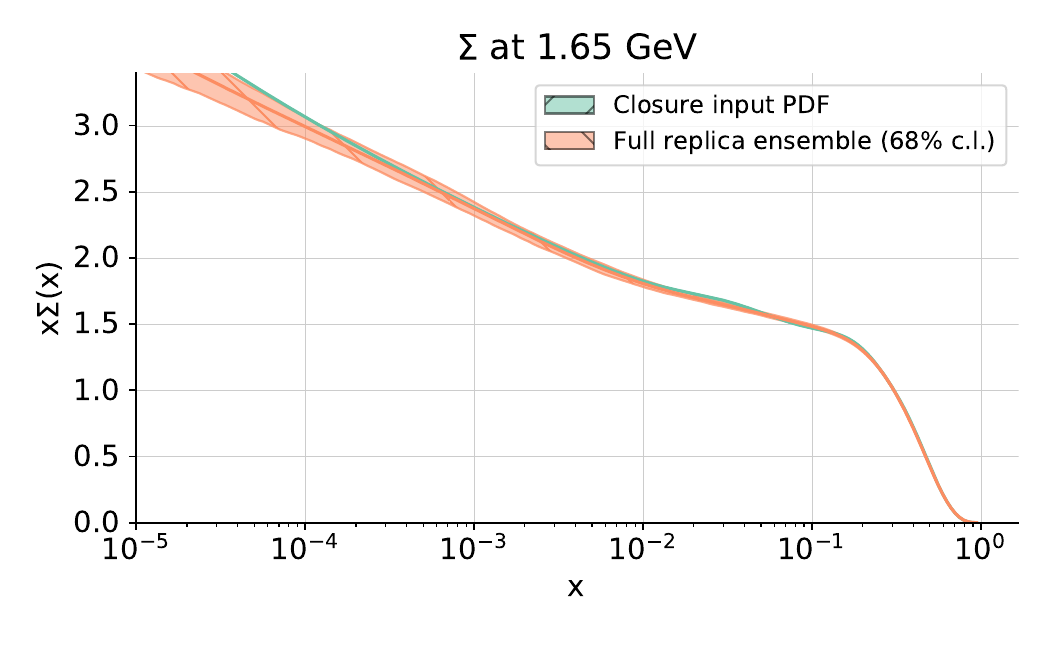}
      \caption{\small
The replica (solid green line) chosen as the true
        underlying PDF  $\boldsymbol{\law}$ for the closure test: the
        gluon (left) and quark singlet (right) are displayed. The
        NNPDF4.0 central value and  68\%
        confidence interval (same as in
        Fig.~\ref{fig:40vs31_PDFs}) are also shown for reference. 
         }
       \label{fig:ClosureInputExample}
    \end{center}
\end{figure}

We have performed a closure test by assuming
as input PDF set used to produce the true observable values $\boldsymbol{\law}$
a specific replica randomly selected out of the $N_{\rm rep}$ replicas
of the NNPDF4.0 NNLO global determination. The reason for this choice
is that on the one hand, it automatically satisfies known theoretical
constraints, such as the sum rules of Sect.~\ref{subsec:sumrules}. On
the other hand, thanks to it being randomly selected out of a
replica sample, it satisfies the criteria of flexibility and
model-independence of Sect.~\ref{sec:closure-test-estimators}. In
particular, individual replicas have generally more structure than the
final central PDF, so by
choosing a repica, rather than the central fit from either NNPDF or
any other PDF set, we are making the closure test somewhat more stringent.
The specific replica that we chose is shown in 
Fig.~\ref{fig:ClosureInputExample} (gluon and quark singlet),
together with the NNPDF3.1 central
value and uncertainty.

We have 
produced $n_{\rm fit}=25$ sets of data Eq.~(\ref{eq:levelonedatagen}), each of which has been
used to produce a fit with $N_{\rm rep}=40$ replicas.
Results are then bootstrapped~\cite{Efron:1979bxm,Efron:1986hys} in
order to improve stability. We have checked that increasing the number
of replicas or the number of fits results are unchanged within the
bootstrap uncertainty.
The fits are produced using the NNPDF3.1-like dataset discussed in Sect.~\ref{subsec:dataset_overview}.

Data space  estimators,
such as the bias-to-variance ratio $\biasvarratio$, are produced by 
selecting out of the full 
datasets that enter the NNPDF4.0 determination all data that were not
already used for fitting.
An advantage of this choice 
is that  the kinematic coverage of the fitting dataset and the testing
dataset are then reasonably
similar, thus ensuring Gaussianity, as discussed above,

In PDF space, we perform tests for PDFs in the evolution basis at the
PDF parametrization scale and over
 a grid of  $x$ points, chosen for the gluon and singlet
as
logarithmically spaced for $10^{-3} < x < 0.1$ and linearly spaced for
$0.1 < x <0.5$, and for nonsinglet quark
distributions $V$, $V_3$, $T_3$, and $T_8$ as  purely linearly spaced for
$0.1 < x <0.5$. We do not consider the $V_8$ and $T_{15}$ nonsinglet
combinations that are too noisy at the initial scale.
Furthermore,  we  evaluate $\xi_{1\sigma}$ in Eq.~(\ref{eq:CT_xi1_PDFs})
with $\nx=4$ to reduce the correlations between points, and we also rotate
into the basis which diagonalizes the covariance estimated on the PDF replicas
as an extra precaution.

\subsubsection{Validation of the NNPDF4.0 methodology}
\label{sec:ClosureTestResults}

We now turn to the validation of the NNPDF4.0 methodology.
First of all, we evaluate the expectation value of $\Delta_{\chi^2}$,
Eq.~(\ref{eq:deltachi2_def}), over the $n_{\rm fit}$ fits
that constitute the NNPDF4.0 closure tests and present in
Table~\ref{tab:delta-chi2-fitted} the results separated into groups of datasets.
As mentioned, the input dataset is NNPDF3.1-like.
One can observe how
$\eshift{\Delta_{\chi^2}} < 0$ for all datasets considered, indicating the absence of
under-fitting.
Furthermore,
its small absolute magnitude, typically at the per-mille level or at
most being a couple of percent, corresponds to a negligible amount of
overfitting, and it is thus consistent with proper learning.

\input{tables/delta_chi2_fitted_process}

We now turn to the  bias-to-variance ratio
$\biasvarratio$, Eq.~(\ref{eq:BiasVarRatio_sqr}), which is shown
in Table~\ref{table:outofsample-ratio-closure}, evaluated for
   testing datasets
   that were not used as input to the closure test fits, with results
   divided by groups of processes.  %
  The combination of the fitting set used to evaluate Table~\ref{tab:delta-chi2-fitted}
 and the testing set shown here add up to the complete NNPDF4.0 baseline dataset.
  The last column indicates the uncertainty of the 
  $\biasvarratio$,
  determined as its standard deviation 
over a bootstrap sample of both fits and replicas.

 For the total testing set, it is found that $\biasvarratio\simeq 1$ within the bootstrap error,
 demonstrating the faithfulness
 of the estimated PDF uncertainties.

\input{tables/closure-outofsample-ratio-process}

\begin{figure}[!t]
    \begin{center}
      \includegraphics[width=0.7\textwidth]{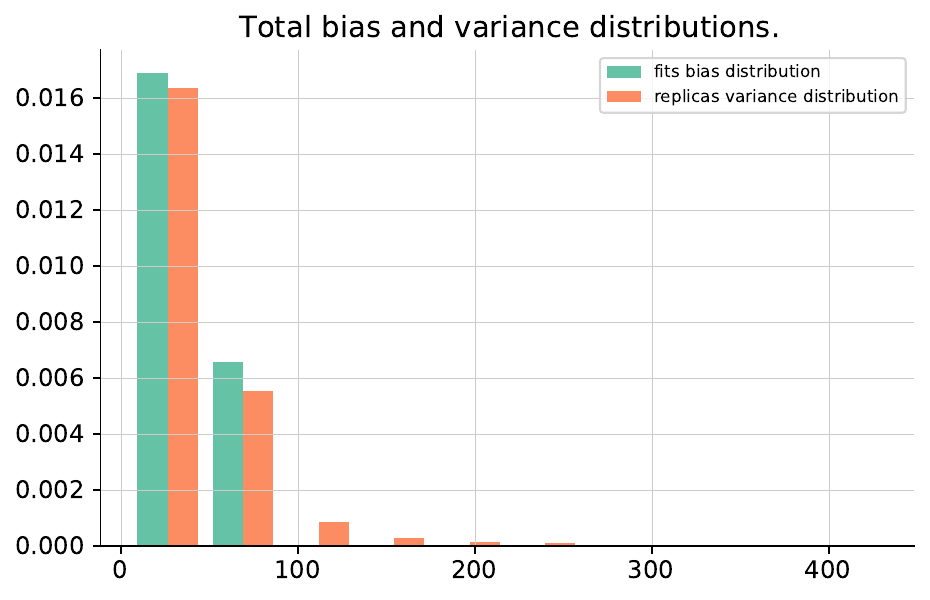}
      \caption{\small
        The normalized distribution of the total bias,
        Eq.~(\ref{eq:BiasDef}), and of total variance, Eq.~(\ref{eq:VarDef}),
        over the $n_{\rm fits}$ constituting the NNPDF4.0 closure tests.
        The square root of the mean of these two distributions
        defines  $\biasvarratio$, the bias-to-variance ratio.
      }
       \label{fig:TotalRatioDistribution}
    \end{center}
\end{figure}

In order to gain some more understanding of the results from
Table~\ref{table:outofsample-ratio-closure},
it is instructive to plot
the full distributions of both the
total bias,
Eq.~(\ref{eq:BiasDef}), and of the total variance, Eq.~(\ref{eq:VarDef}),
over the $n_{\rm fits}$ constituting the NNPDF4.0 closure tests.
From these two distributions, displayed         
in Fig.~\ref{fig:TotalRatioDistribution},
one can observe that not only are their means consistent, but
also that they exhibit a similar shape.
The only difference is that the  distribution over the
variances is somewhat broader,
with a small tail towards large values of the estimator.
Since each of the $n_{\rm fit}$ fits has 40 replicas,
one expects better statistics in the distributions
over variances as compared to that over biases,
which is why the tail of the former is better sampled.
Furthermore, we performed 
checks that the results in Table~\ref{table:outofsample-ratio-closure} are
stable upon removing selected fits and replica within the bootstrap uncertainty,
and hence we are confident that the results are not subject to  finite size effects.

The fact that the bias-to-variance ratio satisfies $\biasvarratio\simeq 1$
both for the total testing dataset and at the level of groups of processes
indicates that the PDF uncertainties in the NNPDF4.0 methodology are
being faithfully estimated.
Further confirmation of this property can be obtained by evaluating
the quantile estimators in both PDF and data space, respectively defined in
Eqs.~(\ref{eq:CT_xi1_PDFs}, \ref{eq:XiDataDef}).
First of all, Table~\ref{table:outofsample-xi-closure} displays
the one-sigma quantile estimator in the space of experimental
data, $\xi^{(\rm data)}_{1\sigma}$,  evaluated for the same
testing dataset as that used for Table~\ref{table:outofsample-ratio-closure},
together with the corresponding bootstrap error.
In addition,  we also indicate the value
of  $\erf(\biasvarratio/\sqrt{2})$ evaluated using
the corresponding bias-to-variance ratio.
As indicated by Eq.~(\ref{eq:expectedxi}), for
a successful closure test one expects that these two quantities
coincide, that is,
$\xi^{(\rm data)}_{1\sigma}\simeq \erf(\biasvarratio/\sqrt{2})$.

It is clear that  $\xi^{(\rm data)}_{1\sigma}$  and $\erf(\biasvarratio/\sqrt{2})$  agree well with each
other within the bootstrap error, which provides a non-trivial
consistency test. Furthermore,   $\xi^{(\rm data)}_{1\sigma} = 0.68$
for the total dataset as expected, with reasonable fluctuations
between different process types. 
The observed deviations between the two indicators
may be explained by
quantile statistics being more robust to outliers, or because the value
of $\erf(\biasvarratio/\sqrt{2})$ can be
dominated by a few eigenvectors
of the experimental covariance matrix.

\input{tables/closure-outofsample-xi-process}

\begin{figure}[!t]
    \begin{center}
      \includegraphics[width=0.49\textwidth]{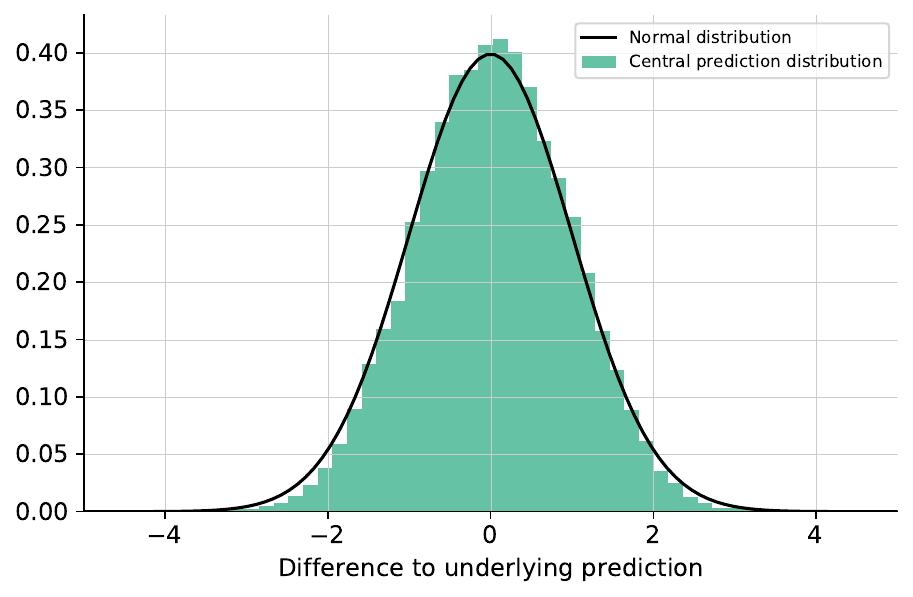}
      \includegraphics[width=0.49\textwidth]{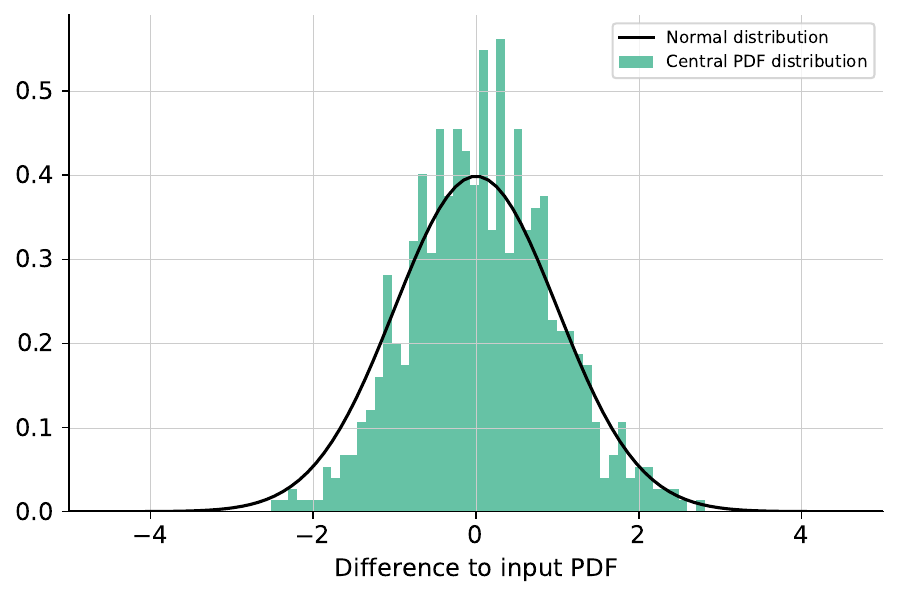}
      \caption{\small  The normalized distribution of  relative differences
        $\delta_i^{(l)}$ in data 
        data  space Eq.~(\ref{eq:diff_reps_truth} (left) or $\widetilde{\delta}_{i,j}^{(l)}$
        Eq.~(\ref{eq:diff_reps_truthPDF} in PDF space (right).
      In both cases, a univariate
      zero-mean
      Gaussian distribution is plotted for reference.
      }
       \label{fig:DataCentralDifferenceHist}
    \end{center}
\end{figure}

In order to provide a graphical representation of the information contained in
Table~\ref{table:outofsample-xi-closure}, it is instructive
to evaluate the difference between the
mean value (over replicas) of the theory predictions and the
corresponding truth observable values normalized
by the PDF uncertainties, that is
\be
\label{eq:diff_reps_truth}
\delta_i^{(l)} \equiv
\frac{\left( \erep{\model_i}^{(l)} - \law_i \right)}{\sigma_i^{(l)}}\, ,\qquad
i=1,\ldots,N_{\rm dat} \, ,\qquad l=1,\ldots,n_{\rm fit} \, .
\ee
The normalized distribution of these relative differences $\delta_i^{(l)}$
is displayed in the left panel of
Fig.~\ref{fig:DataCentralDifferenceHist} together with
a univariate zero-mean Gaussian for reference.
The fraction of the 
histogram entries which fall inside the 1-sigma confidence interval of the scaled
Gaussian is  then equal to the value of the total $\xi^{(\rm data)}_{1\sigma}$ displayed
in Table~\ref{table:outofsample-xi-closure}.

From Fig.~\ref{fig:DataCentralDifferenceHist} it is apparent that
the central values of the model predictions for
physical observables fluctuate
around the true values by an amount which is consistent with the expectations
of the associated PDF uncertainties.
Indeed, there is  excellent agreement between the
distribution of $\delta_i^{(l)}$ and that of the reference
Gaussian, consistently with the value of
$\xi^{(\rm data)}_{1\sigma}=0.68$ reported
in Table~\ref{table:outofsample-xi-closure}.

We now compute the quantile estimator in PDF space, defined in Eq.~(\ref{eq:CT_xi1_PDFs}).
This estimator, $\xi_{n\sigma}^{\rm (pdf)}$, was already introduced
as part of
the original study in~\cite{Ball:2014uwa}.
However, as mentioned it was only possible to evaluate it
 approximately, as performing multiple closure test
fits was computationally infeasible.
The values of $\xi_{1\sigma}^{\rm (pdf)}$ are presented
in Table~\ref{table:closure-xi-pdf}, along with their
bootstrap error.
In general, there is reasonable agreement within bootstrap
errors between the computed value
of $\xi_{1\sigma}^{(\rm pdf)}$ and
the expected value of 0.68.
However, in comparison  to the corresponding estimator in data space
larger fluctuations are observed, specifically
for the singlet PDF $\Sigma$, and the average value
$\xi_{1\sigma}^{(\rm pdf)}=0.71\pm 0.02$ is somewhat overestimated.
It should be noticed that the PDF-space estimator is somewhat less
stable and accurate than that in data space, due to the need to pick a
grid of points that corresponds to the measured region, and also
because of the very high correlation between PDFs at neighboring points
which may lead to an unstable covariance matrix.
The fact that the average $\xi_{1\sigma}^{(\rm pdf)}$ is slightly more
than 0.68 suggests that anyway PDF uncertainties are conservatively
estimated.

\input{tables/closure-pdf-xi}

Finally, in  Fig.~\ref{fig:DataCentralDifferenceHist}
the  histogram of relative differences is also shown using a  PDF
space definition:
\be\label{eq:diff_reps_truthPDF}
\widetilde{\delta}_{i,j}^{(l)}\equiv
\frac{\lp \erep{q^{i(l)}(x_j)} - q_{\rm in}^i(x_j)  \rp}{\sigma^{i(l)}(x_j)}\,,\quad
i=1,\ldots,n_{\rm flav} \, ,\quad
j=1,\ldots,n_{x} \, ,\quad l=1,\ldots,n_{\rm fit} \, ,
\ee
We see that, even though also in this case there is excellent
agreement with the expected univariate Gaussian behavior, results are
indeed rather noisier than in data space.

\subsubsection{Extent and limitations of closure testing.}
\label{subsubsec:extlim}

The closure tests presented in this section are entirely successful,
thereby validating the NNPDF4.0 methodology in the data region.
However, it is important to understand what the closure tests do and
do not verify.

The closure test, at least in its present incarnation, makes two assumptions.
The first is that the underlying distribution of the experimental  data is known
exactly.
Specifically, if the data are Gaussian, it is assumed that their
distribution is unbiased and that  the covariance that characterizes this multi-Gaussian
distribution  is fully known.
In realistic situations, of course, this, even in the best of
hypotheses, can only be approximately the
case, since the experimental covariance matrix is itself an observable
which is extracted from the data, and thus it is characterized by an
uncertainty on the uncertainty. Furthermore, some sources of
systematic uncertainty are based on theoretical estimates and thus
subject to theoretical uncertainties which are difficult to
estimate. Finally, in the worst case it may happen that some data or the associated
uncertainty are simply incorrect: this would correspond to a biased
distribution (wrong central value) or an incorrect uncertainty or
correlations (wrong covariance matrix).

The second assumption is that the data are obtained from the PDF
using a known underlying physical law.
In realistic situations this is surely not the
case, since theoretical predictions are computed at finite
perturbative accuracy, and thus data predictions are affected by an
uncertainty corresponding to the very least
to missing higher order perturbative corrections, and generally also to
other possible corrections such as nuclear effects, electroweak
corrections, heavy quark mass effects, limited knowledge of standard model parameters, and so
on.

Therefore, the closure test presented here checks for faithfulness of the component
of the PDF uncertainty which is induced by the data uncertainty,
assuming the latter is perfectly known.
It does not check for other sources of
uncertainty, such as theory uncertainties: this would have to be added
separately. A methodology for doing so was discussed and applied to
missing higher order perturbative uncertainties in
Refs.~\cite{AbdulKhalek:2019bux,AbdulKhalek:2019ihb}, but is not
implemented in a global NNLO PDF determination yet.
Also, it does not
account for possible ``data inconsistencies'', i.e., incorrectly
estimated experimental values and uncertainties. This motivates the
need to select a maximally consistent dataset, as we have done in
Sect.~\ref{sec:dataselection}, that guarantees that no major
inconsistencies are present in the baseline dataset. However,
remaining small inconsistencies might still lead to a certain amount
of uncertainty underestimation, whose exact assessment will require
performing closure tests with artificial inconsistent data.

%% file: tables/delta_chi2_fitted_process.tex
\begin{table}[t]
\begin{center}
  \footnotesize
   \renewcommand{\arraystretch}{1.30}
\begin{tabular}{lrr}
\toprule
{} & $\qquad$ $\ndata$ &  $\qquad$ $\eshift{\Delta_{\chi^2}}$ \\
Dataset &        &                    \\
\midrule
DIS NC     &   2100 &          -0.0059 \\
DIS CC     &    989 &          -0.0112 \\
DY         &    712 &          -0.0148 \\
Top        &     19 &          -0.0054 \\
Jets       &    273 &           0.0001 \\
Total      &   4093 &          -0.0087 \\
\bottomrule
\end{tabular}
\end{center}
\caption{\small
  The expectation value of $\Delta_{\chi^2}$,
Eq.~(\ref{eq:deltachi2_def}), evaluated over the $n_{\rm fit}$ fits
that constitute the NNPDF4.0 closure test.
Results are presented separated into different processes.
}
\label{tab:delta-chi2-fitted}
\end{table}

%% file: tables/closure-outofsample-ratio-process.tex
\begin{table}[t]
\centering
\footnotesize
   \renewcommand{\arraystretch}{1.30}
\begin{tabular}{lrr}
\toprule
{} &   $\qquad$$\biasvarratio$ & $\qquad$ bootstrap error \\
Dataset &             &                  \\
\midrule
DY         &          0.99 &             0.08 \\
Top pair      &            0.75 &             0.06 \\
Jets       &          1.14 &             0.05 \\
Dijets     &          0.99 &             0.07 \\
Direct photon           &    0.71 &             0.06 \\
Single top  &          0.87 &             0.07 \\
Total      &          1.03 &             0.05 \\
\bottomrule
\end{tabular}
\vspace{0.3cm}
\caption{
  The bias-to-variance ratio
  $\biasvarratio$,  Eq.~(\ref{eq:BiasVarRatio_sqr}),
  divided by groups of processes and evaluated for the testing datasets
  that were not used as input to the NNPDF4.0 closure test fits.
  The last column indicates the uncertainty associated to
  $\biasvarratio$,
  determined as its standard deviation 
over a bootstrap sample of both fits and replicas.
}
\label{table:outofsample-ratio-closure}	
\end{table}

%% file: tables/closure-outofsample-xi-process.tex
\begin{table}[t]
  \centering
  \footnotesize
  \renewcommand{\arraystretch}{1.30}
 \begin{tabular}{lrrrr}
\toprule
{} &   $\xi^{(\rm data)}_{1\sigma}$ &  bootstrap error &  $\erf(\biasvarratio/\sqrt{2})$ &  bootstrap error \\
experiment &      &        &                 &        \\
\midrule
DY         & 0.69 &   0.02 &            0.69 &   0.04 \\
Top        & 0.75 &   0.03 &            0.82 &   0.03 \\
Jets       & 0.63 &   0.03 &            0.62 &   0.02 \\
Dijets     & 0.70 &   0.03 &            0.69 &   0.04 \\
Direct photon     & 0.81 &   0.03 &            0.84 &   0.03 \\
Single top  & 0.69 &   0.04 &            0.75 &   0.04 \\
Total      & 0.68 &   0.02 &            0.67 &   0.03 \\
\bottomrule
\end{tabular}
\vspace{0.3cm}
\caption{
  The one-sigma quantile estimator in the space of experimental
  data, $\xi^{(\rm data)}_{1\sigma}$
  iEq.~(\ref{eq:XiDataDef}) and evaluated for the same
  testing dataset as used for Table~\ref{table:outofsample-ratio-closure},
  together with the corresponding bootstrap error.
  For each group of processes, we also display the value
  of  $\erf(\biasvarratio/\sqrt{2})$ evaluated using
  the corresponding bias-to-variance ratio.
}
\label{table:outofsample-xi-closure}
\end{table}

%% file: tables/closure-pdf-xi.tex
\begin{table}[t]
\centering
 \footnotesize
  \renewcommand{\arraystretch}{1.30}
\begin{tabular}{lrr}
\toprule
{} & $\qquad$ $\qquad$ $\xi^{\rm (pdf)}_{1\sigma}$ & $\qquad$ bootstrap error \\
flavor  &      &        \\
\midrule
$\Sigma$ & 0.82 &   0.04 \\
$g$    & 0.70 &   0.05 \\
$V$        & 0.65 &   0.05 \\
$V_3$       & 0.63 &   0.05 \\
$V_8$       & 0.72 &   0.04 \\
$T_3$      & 0.71 &   0.05 \\
$T_8$       & 0.71 &   0.05 \\
\midrule
Total    & 0.71 &   0.02 \\
\bottomrule
\end{tabular}
\vspace{0.3cm}
\caption{\small The values of the quantile estimator
  in PDF space,
  $\xi_{1\sigma}^{\rm (pdf)}$ Eq.~(\ref{eq:CT_xi1_PDFs}),
  separated into the contributions from individual flavor
  combinations together with the corresponding bootstrap
  uncertainty.
}
\label{table:closure-xi-pdf}
\end{table}

%% file: subsec-futuretests.tex
\subsection{Future testing NNPDF4.0}
\label{sec:futuretest}

The closure tests presented in Sect.~\ref{sec:closuresettings} allow
for an 
assessment of  the faithfulness of PDF uncertainties in the region
covered by available experimental data.
However, they are ill-suited for an assessment of the  behavior
of PDFs and their uncertainties in the extrapolation regions
where little or no experimental constraints are available, for a
variety of reasons, the most obvious  of which is that the multi-Gaussian
assumption is likely to fail outside the data region

Hence, closure tests  have limited applicability to study the
generalization power of the resulting PDF fit to new, unexplored kinematic regions.
A more suitable strategy to assess this generalization power are the so-called
``future tests'' proposed in~\cite{Cruz-Martinez:2021rgy}.
The main idea underlying future tests is that what we ask for in an
extrapolation region is that PDF uncertainties correctly reflect the
lack of information. Whereas in principle in the absence of 
information uncertainties are infinite, in practice PDF uncertainties not
too far from the data region are constrained by requirements of continuity
and smoothness. Whereas in the absence of direct information
we cannot expect to be able to achieve a full
and detailed
knowledge of the covariance between any two PDFs (or indeed any two
predicted data points), we do wish for PDF uncertainties to reproduce
the possible deviation of the best-fit PDFs,
and of physical predictions obtained using them,
from the true value that would be obtained  if the PDF
was known say as accurately as it is known in the data region.

The future test verifies explicitly whether this is the case:
a ``future test  PDF'' is determined
from a restricted subset of the full dataset that only covers a
limited region. This future test PDF is then  used
to predict all the rest of the dataset.
The  restricted datasets can be thought of as representative of the
limited knowledge that was available at some point in the past, 
(hence the name ``future test'') but this is of course just a manner
of speaking, as any partitioning of the dataset into restricted  and
full may be considered.
Because future tests of NNPDF3.1 were never performed, here we will
present future tests of both the NNPDF3.1 and NNPDF4.0
methodologies. This allows us to simultaneously  validate the new
methodology, and also put it in context. 

\subsubsection{Future testing methodology}

Following the discussion in~\cite{Cruz-Martinez:2021rgy} we test 
 the NNPDF3.1 and NNPDF4.0 methodologies by choosing as input
specific subsets of the complete NNPDF4.0 baseline dataset, and
determining corresponding PDF sets from them.
The predictions obtained using these PDFs are then compared to the
data not included in their fit,
in order
to assess whether the uncertainty in the prediction correctly accounts
for the correspondingly missing information.

This is done by evaluating  the $\chi^2$ to the datasets not used in the  fit
with PDF uncertainties also included along with the data uncertainties
in the $\chi^2$ definition. Indeed, we expect in general that the $\chi^2$
evaluated including data uncertainties only should be larger than one,
as soon as the deviation of the PDF from its true value is larger than
the experimental uncertainty, which is bound to happen for
sufficiently accurate data in an extrapolation region. However, if the
deviation from the true value is correctly reproduced by the PDF
uncertainty, the $\chi^2$ should then become again close to one once
the PDF uncertainty is included. Note that the test is only nontrivial
if the $\chi^2$ value before inclusion of the PDF uncertainty is
significantly larger than one: otherwise,  the data are not precise
enough to test for faithfulness of the PDF uncertainty.

Specifically the $\chi^2$  with PDF uncertainties included is computed
using the  covariance matrix
\be
\label{eq:fullcovmat}
{\rm cov}_{ij}^{\rm (tot)} = {\rm cov}_{ij}^{\rm (exp)}  + {\rm cov}_{ij}^{\rm (pdf)} \, ,
\ee
where ${\rm cov}_{ij}^{\rm (exp)}$ is the usual experimental covariance
matrix, while  the covariance matrix $ {\rm cov}_{ij}^{\rm (pdf)}$
corresponding to PDF uncertainties can be determined as
\be\label{eq:pdfcovmat}
{\rm cov}_{ij}^{\rm (pdf)} = \la \mathcal{F}_i\mathcal{F}_j  \ra_{\rm rep}
- \la \mathcal{F}_i  \ra_{\rm rep}\la \mathcal{F}_j  \ra_{\rm rep} \, ,
\ee
where $\mathcal{F}_i^{(k)}$ is  the $i$-th physical prediction found
using  the $k$-th replica of a given PDF set, and the average is
performed over replicas.
Simply combining the two covariance matrices according to
Eq.~(\ref{eq:fullcovmat}) is justified when the corresponding 
sources
of uncertainty are uncorrelated~\cite{AbdulKhalek:2019ihb}.  This is
clearly the case since the experimental uncertainty on data which are
not fitted is completely independent of the PDF uncertainty, as the
latter is
driven by the uncertainty on the fitted data.

\subsubsection{Future testing datasets}

We choose three subsets of the full NNPDF4.0 datasets,
 inspired by the chronological order in which actual
measurements became available, respectively   chosen to 
correspond approximately to a ``pre-HERA''
and ``pre-LHC'' dataset, and to the 
NNPDF3.1-like dataset that was used as fitting dataset in the closure
tests  of Sect.~\ref{sec:closure}.

They are  defined as follows:
\begin{itemize}

\item  Pre-HERA.
   Only fixed-target DIS structure function data
   and fixed-target Drell-Yan cross-sections data are included.
  
\item Pre-LHC.
  This is a superset of the pre-HERA dataset, which is extended to
  also include HERA collider inclusive and charm
  structure function data, and   Tevatron $W$ and $Z$ production data.

\item  NNPDF3.1-like.
  This is the dataset defined in Ref.~\cite{Faura:2020oom} and used as
  fitting dataset in the closure tests presented in Sect.~\ref{sec:closure}.
    
\end{itemize}

\begin{figure}[!t]
\begin{center}
   \includegraphics[width=0.49\textwidth]{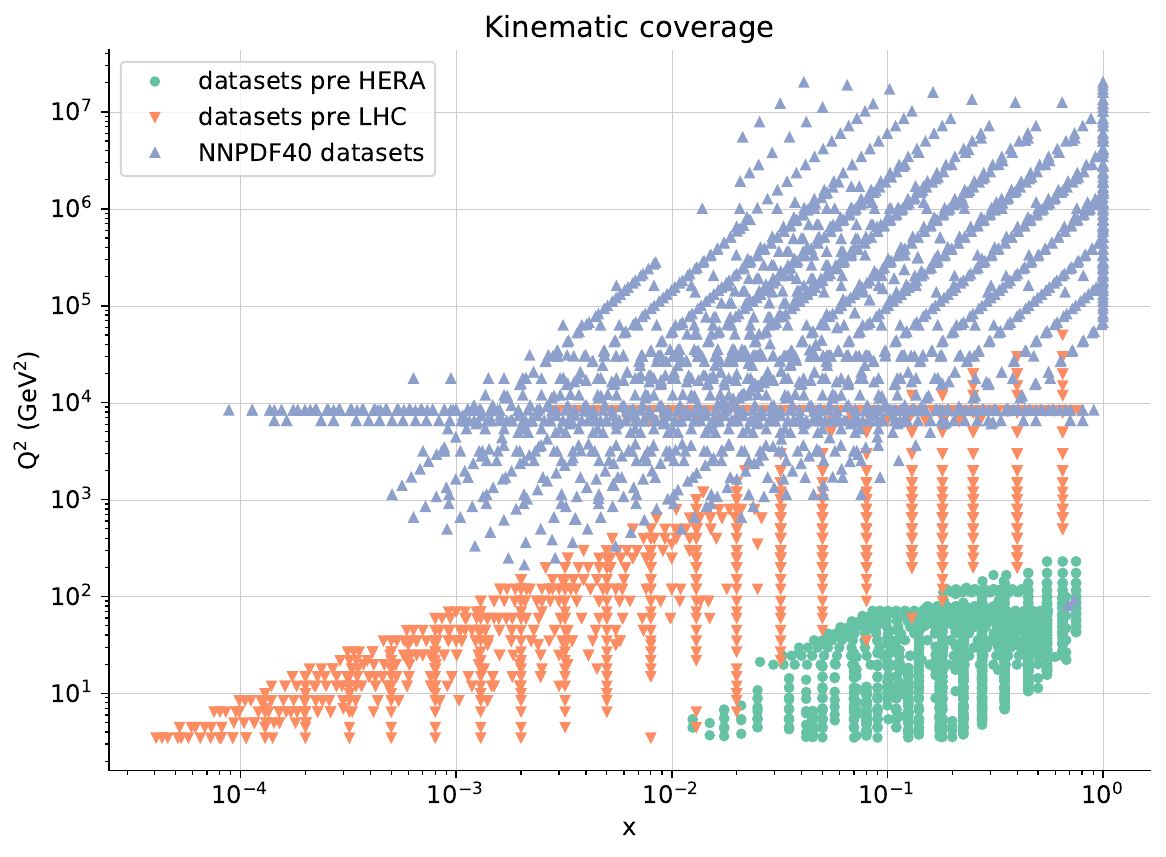}
  \includegraphics[width=0.49\textwidth]{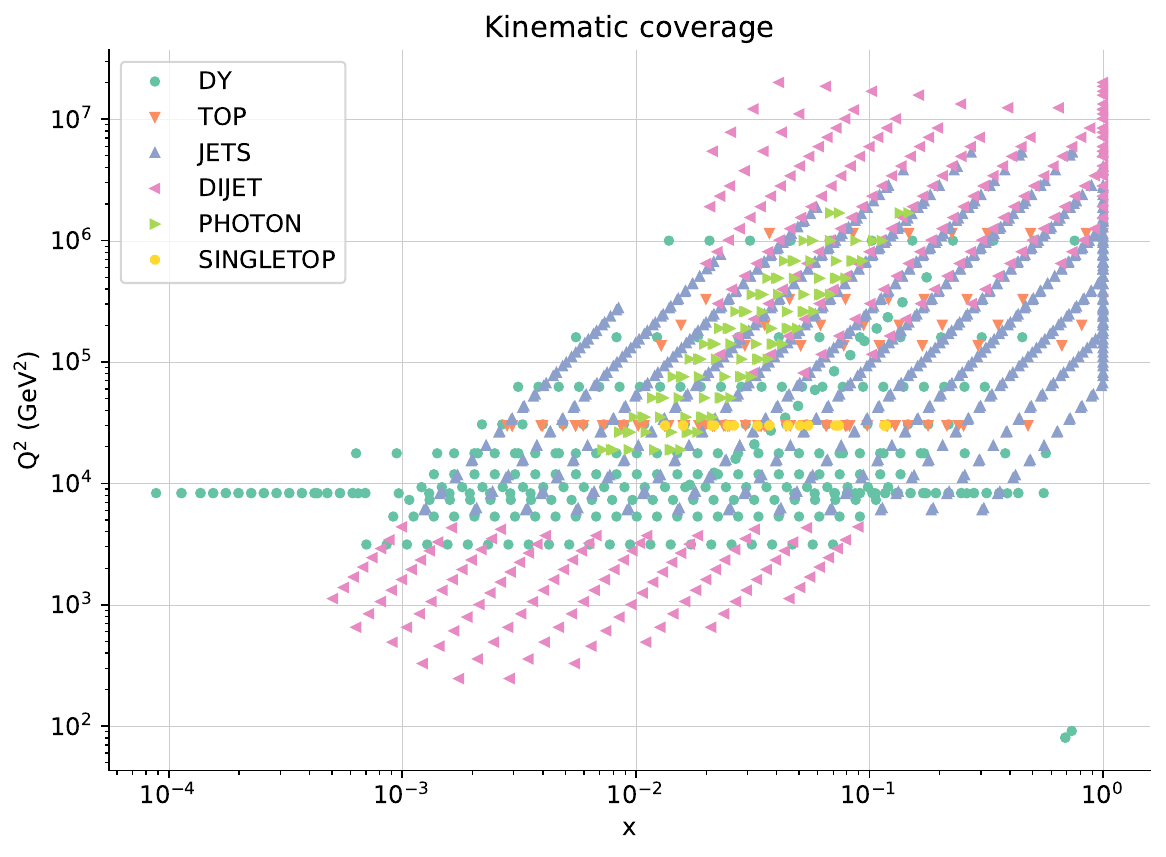}
  \caption{\small Scatter plots comparing various future test data
    subsets to the full NNPDF4.0 of Fig.~\ref{fig:kinplot}. Left:
    comparison of the 
     pre-HERA, pre-LHC and NNPDF4.0 datasets.
    Note that each dataset is a superset of the
    previous one, so all the pre-HERA data are included in the pre-LHC
    set, and all data are included in the NNPDF4.0 set.
    Right: the data points which are included in NNPDF4.0
    but not in the NNPDF3.1-like dataset, grouped by process type.
   }
   \label{fig:future_test_scatter}
\end{center}
\end{figure}

It is important to draw a distinction between the NNPDF3.1-like dataset and the other 
two subsets; while going from pre-HERA to pre-LHC to NNPDF4.0 consecutively adds data
in new kinematic regions, going from NNPDF3.1-like to NNPDF4.0 instead adds more data
points in regions for which data already exist. So we can think of the transition from 
NNPDF3.1 to NNPDF4.0 as an interpolation rather than an extrapolation. This is reflected in scatter plots in Fig.~\ref{fig:future_test_scatter}, where the difference between the first two subsets 
and NNPDF4.0 are shown on the left,
and the difference between the NNPDF3.1-like subset and NNPDF4.0 is shown on the right.

More
specifically, first (left)
we compare the pre-HERA, pre-LHC and full 
NNPDF4.0 datasets.
Note that the pre-LHC dataset contains the points marked with an orange triangle
as well as the pre-HERA points, and the  NNPDF4.0 dataset contains all
points: the three datasets are each a superset of the previous one.
It is also clear that each dataset probes an increasingly wide
kinematic region. Specifically, pre-HERA data  are restricted to
$x\gsim 0.01$ and $Q^2 \lsim 200$ GeV,
while pre-LHC data cover the complete range of $x$ but only $Q^2
\lsim 10^5$ GeV. 
Furthermore, each dataset provides an increasingly wide handle on
specific PDFs or PDF combinations: for instance, pre-HERA data provide
little handle on quark flavor decomposition, and pre-LHC data provide
essentially no handle on the large-$x$ gluon. The pre-HERA and pre-LHC
allow us to test for far-extrapolation (pre-HERA) and
near-extrapolation (pre-LHC).

Then (right) we show all the data that are included in NNPDF4.0 but
not in the NNPDF3.1-like 
dataset, classified by process type.
In this
  case the kinematic
region covered by the new datasets included in NNPDF4.0 essentially overlaps
with the NNPDF3.1-like dataset, though with a lower density. Hence, in
this case it is interpolation, rather than extrapolation, what is being tested.

\subsubsection{Future test results}

We now present results of future testing  both the NNPDF4.0 and NNPDF3.1 methodologies.
We first discuss the case of near and far extrapolation, namely, the
pre-HERA and pre-LHC datasets.
The $\chi^2$ values for all data, divided by process type, are
collected in 
Table~\ref{table:chi2-futuretest}. For either methodology we have
determined three PDF sets from the three datasets, and we show
$\chi^2$ values obtained using each of them. The process-type
classification is made in such a way that the data for any given
process type are either all included, or all excluded in the
determination of each of the three PDF sets. When a process type is
not included in the PDF determination  both the $\chi^2$ value without
PDF uncertainty (in italic) and the PDF value with PDF uncertainty (in
boldface) are shown. All other $\chi^2$ values correspond to fitted data. 
We also tabulate $\chi^2$ values for the full set of data which in
each case is not fitted, denoted as out-of-sample data.

\begin{table}[!t]
  \scriptsize
  \centering
  \renewcommand{\arraystretch}{1.4}
  \input{tables/chi2-futuretest.tex}
  \caption{Values of the  $\chi^2$ per datapoint for the total
    dataset and for specific process types obtained for NNPDF4.0 and for
    the pre-HERA and pre-LHC future test PDFs, determined using
    NNPDF3.1 or NNPDF4.0 methodology. All the data in each process type
    are either fully included or fully excluded from each PDF determination.
    Values in regular font correspond to fitted datasets, evaluated
    with the experimental covariance matrix.
    Values  in bold or italics correspond to data that are not
    fitted. The value in italic is evaluated
    with the experimental covariance matrix, while the value in bold
    also includes PDF uncertainties,
    Eqs.~(\ref{eq:fullcovmat}-\ref{eq:pdfcovmat}). Values of $\chi^2$
    for the full set of data that are not fitted  (denoted
    as total out-of-sample) is also given in each case.}
  \label{table:chi2-futuretest}	
\end{table}

First, we note that the total $\chi^2$ for out-of-sample data is very large (of order twenty) for  pre-HERA PDFs
while it is moderately large (of order three) for pre-LHC PDFs.
This shows that the test is nontrivial in both cases, and it indeed
tests for far-extrapolation for pre-HERA and near-extrapolation for
pre-LHC. A similar pattern is observed for all process types: HERA,
that probes the small $x$ gluon, top and
jets, that probe the large $x$ gluon, and Drell-Yan, that probes quark 
flavor separation.

When the PDF uncertainties are introduced, all $\chi^2$ values become
of order one, thereby showing that the future test is successful. This
is especially remarkable given that in some cases (such as HERA data
or collider Drell-Yan data for the pre-HERA PDFs) the reduction in
$\chi^2$ is by almost a factor 30. This means that the
PDF uncertainty accounts for a deviation between data and theory which
is over five times bigger than the data uncertainty.

Finally, comparing the two methodologies it is clear that both are
equally successful in satisfying the future tests. However, with
NNPDF3.1 methodology $\chi^2$ values computed without PDF uncertainty
for out-of-sample data are rather larger than with NNPDF4.0
methodology. This means that while both methodologies lead to faithful
uncertainties in the extrapolation region, NNPDF4.0 has smaller
extrapolation uncertainties, i.e., it provides a more 
efficient generalization of the fitted data.

We then turn to fits based on  the NNPDF3.1-like dataset. In this case,
each process type is represented both in the fitted and extrapolation
dataset, hence in  Table~\ref{table:futuretest-31vs40}
we only show $\chi^2$ values for the total fitted and out-of-sample
datasets. In this case, the out-of-sample $\chi^2$ is smaller than
two, and smaller than 1.5 for NNPDF4.0 methodology consistent with the
fact that the out-of-sample data are now in an interpolation region. 
Also in this case, upon inclusion of the PDF uncertainty all $\chi^2$
value become of order one, and close to the $\chi^2$ value for the
fitted dataset.

\begin{table}[!t]
 \centering
 \scriptsize
 \renewcommand{\arraystretch}{1.40}
 \begin{tabularx}{\textwidth}{XC{3cm}C{3cm}C{3cm}C{3cm}}
   \toprule
                   & \multicolumn{2}{c}{NNPDF3.1 methodology} & \multicolumn{2}{c}{NNPDF4.0 methodology}  \\ \midrule
                   & NNPDF3.1-like & Global & NNPDF3.1-like & Global \\
   \toprule
   $\chi^{2}_{\rm exp}$        &  {\it 1.74}    &  1.29    &   {\it 1.46}   & 1.20         \\
\midrule
 $\chi^{2}_{\rm exp+pdf}$  &  {\bf 1.12}    &  $-$     &   {\bf 1.17}   &  $-$    \\
  \bottomrule
 \end{tabularx} 
 \caption{Same as Table~\ref{table:chi2-futuretest}, now for the
   NNPDF3.1-like future test.
 }
\label{table:futuretest-31vs40}	
\end{table}

We conclude from this analysis that the future test is fully successful for both
methodologies, and that for the same datasets near- and far-extrapolation and
interpolation uncertainties are smaller with NNPDF4.0 methodology as compared to
 its NNPDF3.1 counterpart.

\begin{figure}[!t]
  \begin{center}
    \includegraphics[width=0.45\linewidth]{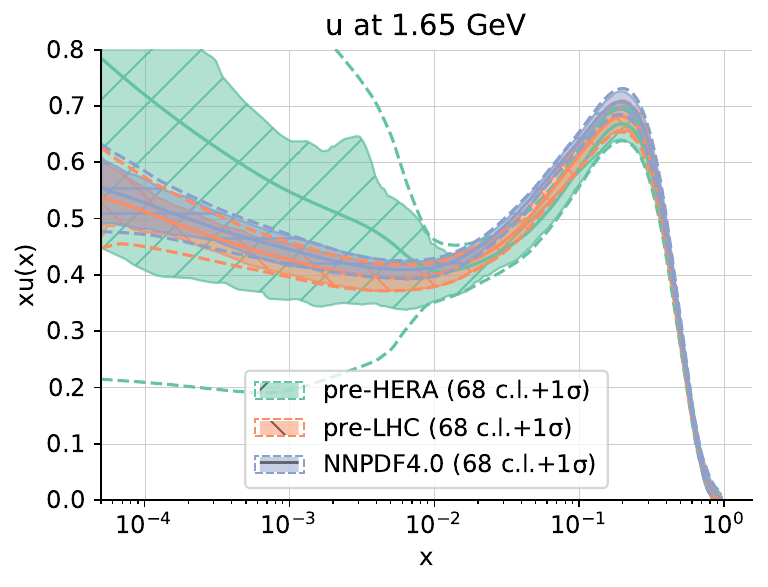}
    \includegraphics[width=0.45\linewidth]{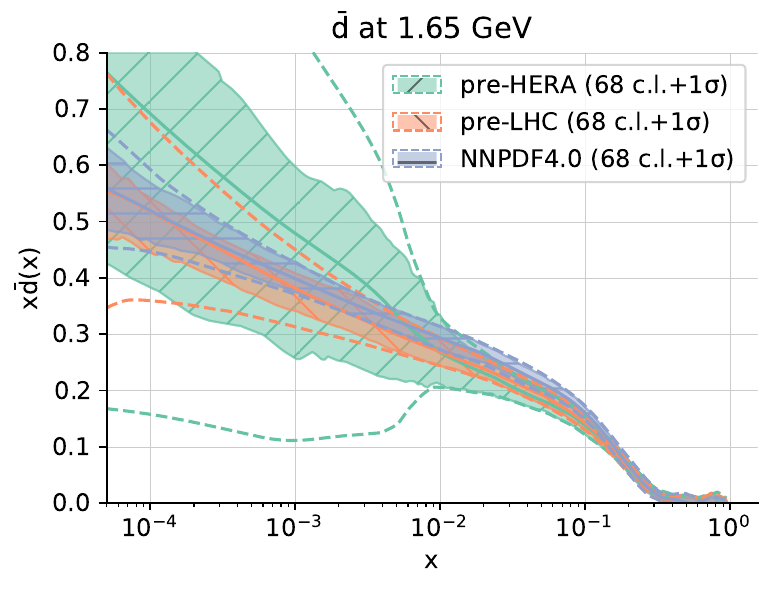}\\
    \includegraphics[width=0.45\linewidth]{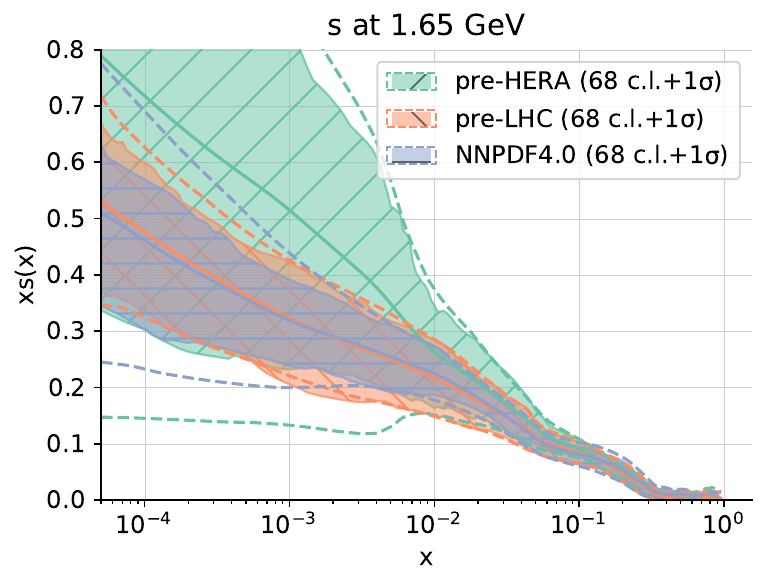}
    \includegraphics[width=0.45\linewidth]{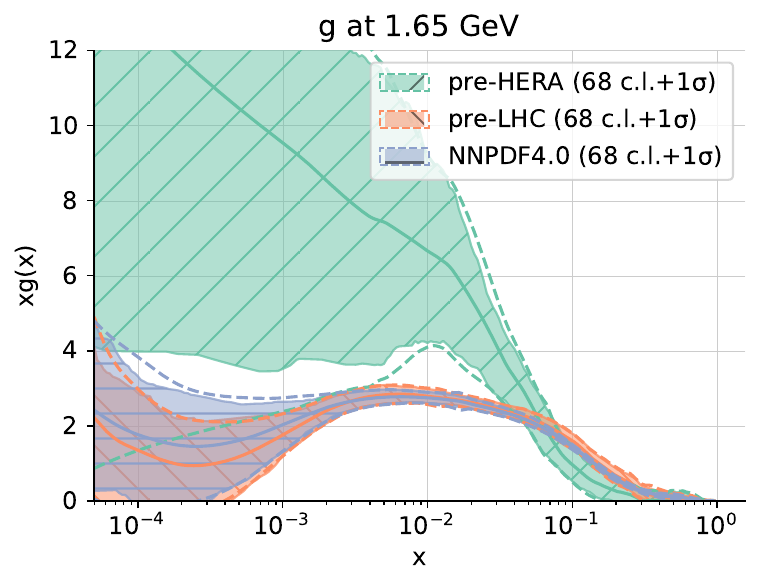}
    \caption{Some pre-HERA and pre-LHC PDFs compared to PDFs
      based on full
      NNPDF4.0 dataset, in all cases obtained using the
      NNPDF3.1 fitting methodology.
      The up (top left), antidown (top right),
      strange (bottom left) and gluon (bottom right) are shown at the
      input parametrization scale of $Q=1.65$ GeV. 
  \label{fig:pdf-future-nnfit} }

  \end{center}
\end{figure}

\begin{figure}[!t]
  \begin{center}
    \includegraphics[width=0.45\linewidth]{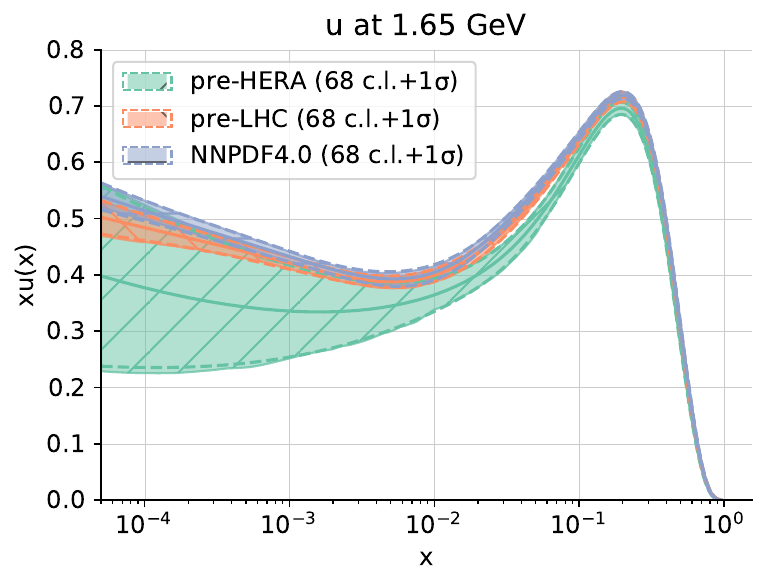}
    \includegraphics[width=0.45\linewidth]{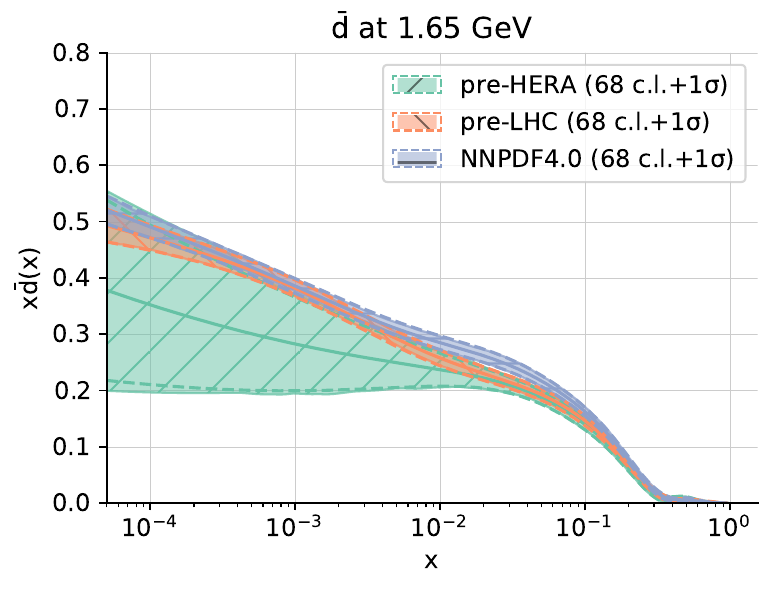}\\
    \includegraphics[width=0.45\linewidth]{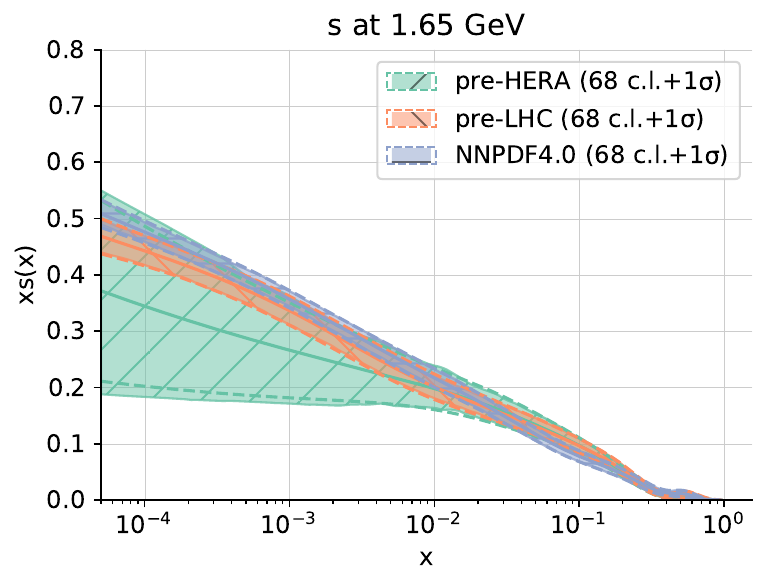}
    \includegraphics[width=0.45\linewidth]{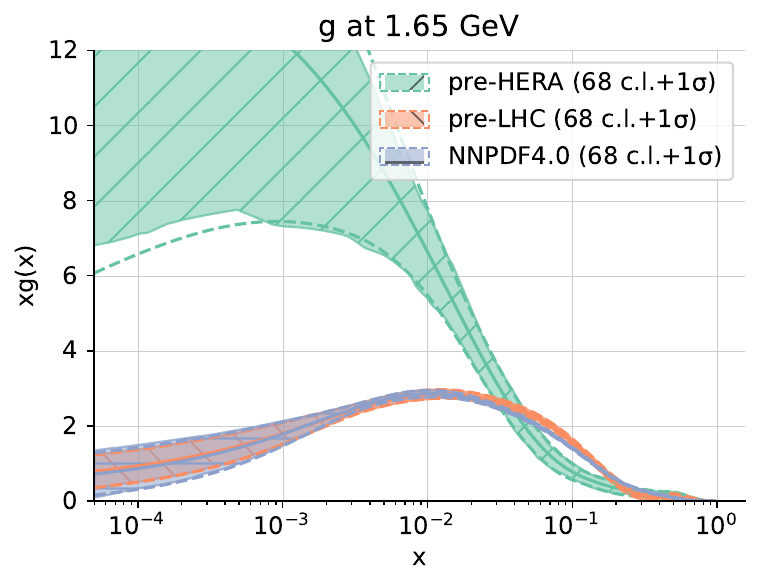}
    \caption{Same as Fig.~\ref{fig:pdf-future-nnfit}, but now
      showing PDFs determined using the NNPDF4.0 methodology. 
  \label{fig:pdf-future-n3fit} }
\end{center}
\end{figure}

By construction, the performance of
future tests should always be assessed at the level of $\chi^2$.
However,
for the sake of visualization, we also provide some comparisons, both at
the PDF level  and at the data level, between future-test PDFs and PDFs
determined from the global NNPDF4.0 baseline dataset. In
Fig.~\ref{fig:pdf-future-nnfit} and Fig.~\ref{fig:pdf-future-n3fit} we compare
future test pre-HERA and pre-LHC PDFs at the parametrization scale
to those determined using the
full dataset, using respectively the NNPDF3.1 and NNPDF4.0
fitting methodologies.
The inflation of PDF uncertainties when a particular $x$ range for a
given PDF changes from data to extrapolation between different sets is
apparent. The smaller extrapolation uncertainty found using NNPDF4.0
methodology in comparison to the NNPDF3.1 methodology is also
visible. Finally, it is clear that there is good overall compatibility of all
PDFs when comparing 
the data region of one set to the extrapolation region of a different
set, in agreement with the $\chi^2$ values of
Table~\ref{table:chi2-futuretest}. A possible exception is the gluon
from the pre-HERA future test which, while compatible with the global
result when using NNPDF3.1 methodology, disagrees with it at the two
sigma level or even more when using NNPDF4.0 methodology in the
$x\lesssim 0.002$ region. This might be due to the poor internal
consistency of the BCDMS and NMC data already noted in
Sect.~\ref{subsubsection:appraisal_and_selection}: if so, this would
indicate that the NNPDF4.0 methodology is sensitive enough to pick up
this, while the NNPDF3.1 methodology is not. 

\begin{figure}[!t]
  \begin{center}
    \includegraphics[width=0.49\linewidth]{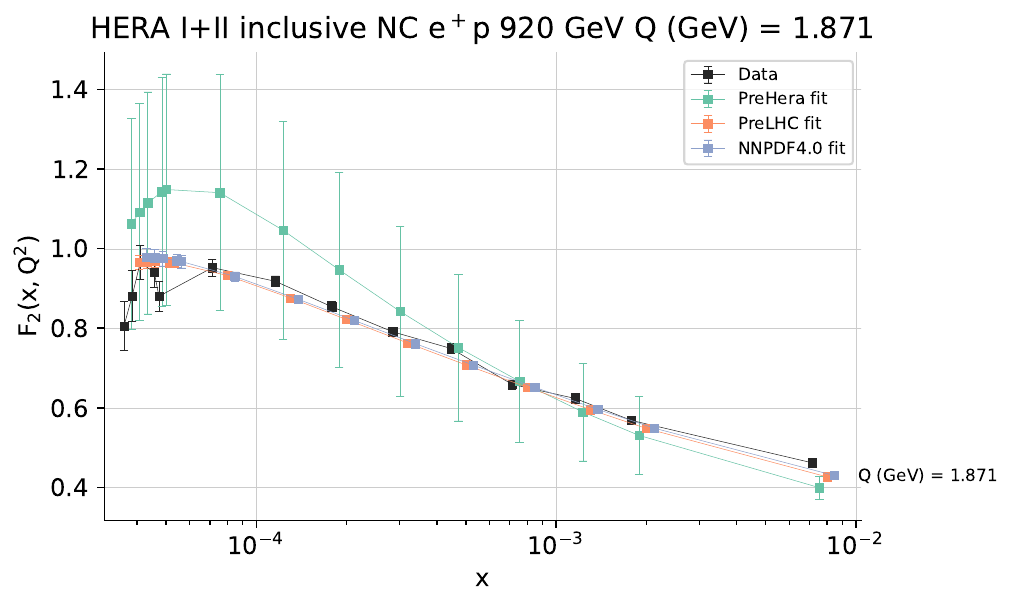}
    \includegraphics[width=0.49\linewidth]{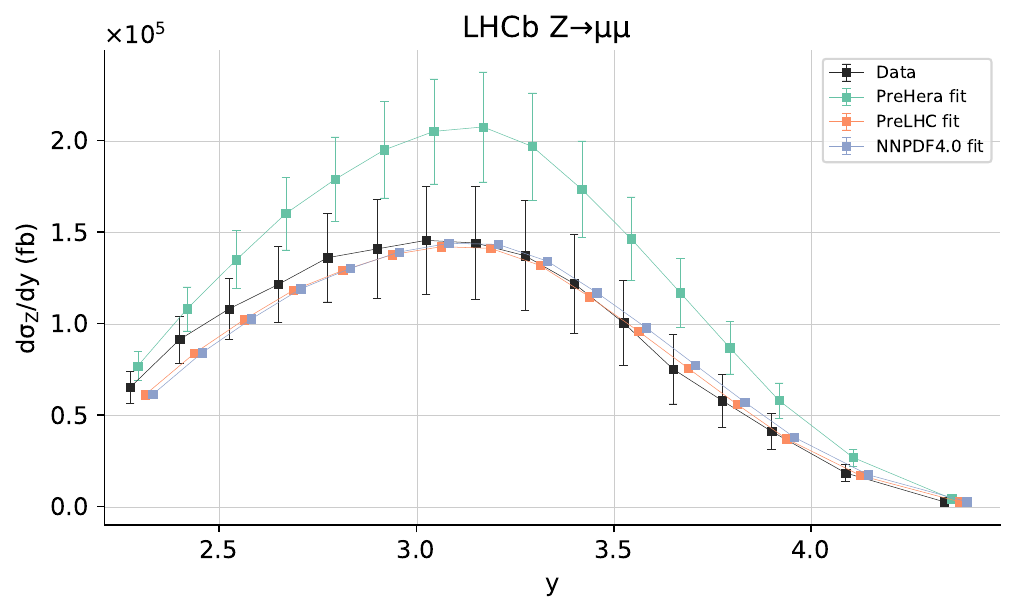}
    \includegraphics[width=0.49\linewidth]{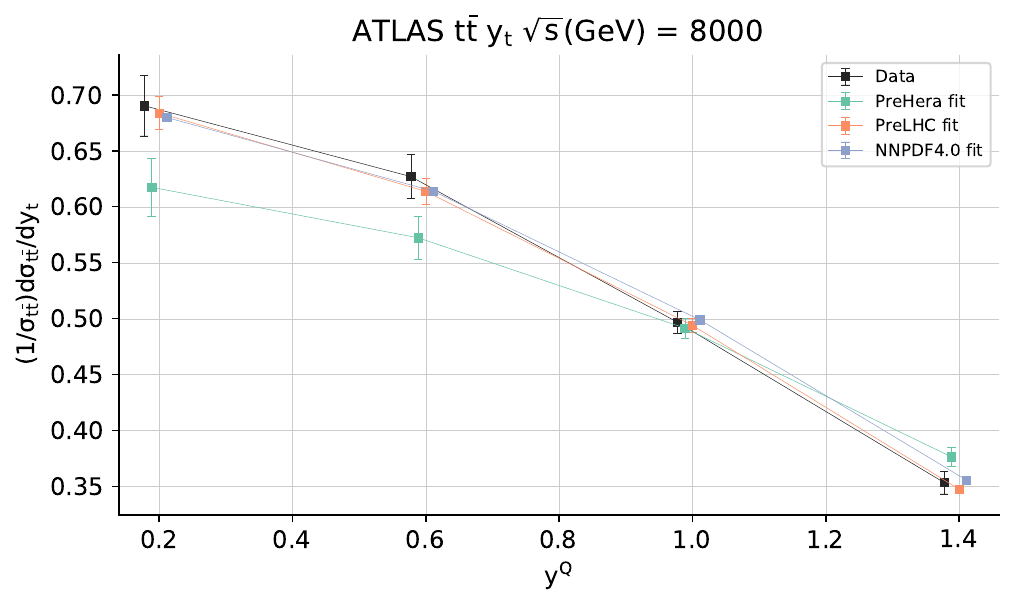}
    \includegraphics[width=0.49\linewidth]{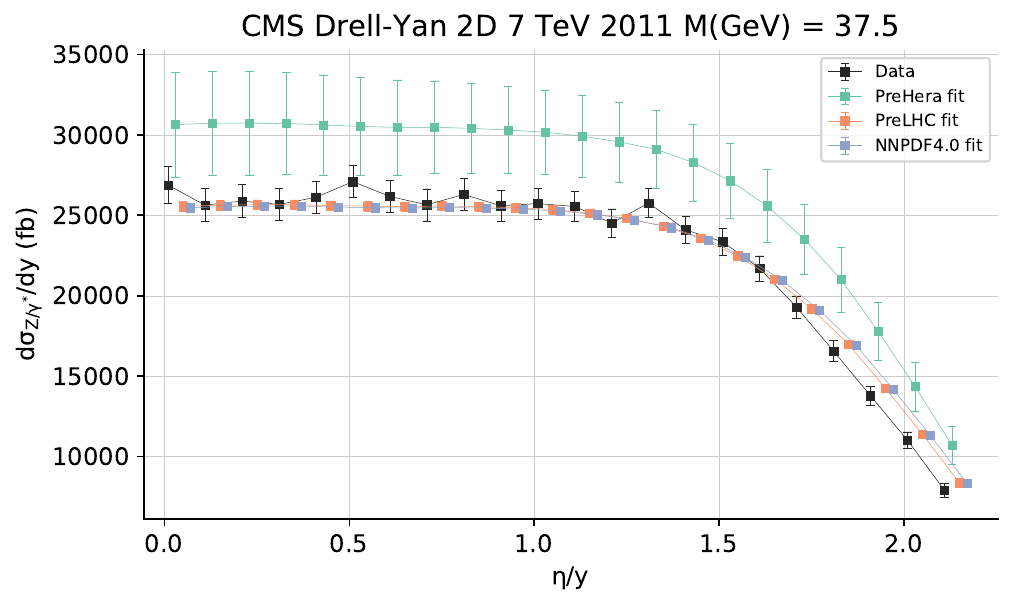}
    \caption{Comparison of the theoretical
      predictions including PDF uncertainties from the pre-HERA and
      pre-LHC PDF sets  based on
      NNPDF4.0 methodology 
      and those of the global fit to four representative measurements: t
      HERA NC structure functions,
      dimuon rapidity distributions in  $Z$ production
      at LHCb,  top rapidity distributions
      in ATLAS $t\bar{t}$ production, and the dilepton rapidity distribution for 
      CMS double-differential Drell-Yan (see text for details).
      Note that only the HERA structure function data enter  the
      pre-LHC fit (but not the pre-LHC fit), and all the remaining
      data do not enter either the pre-HERA or pre-LHC fit.
      The uncertainty in the data corresponds to the total diagonal experimental error.}
  \label{fig:futuretest-data}
\end{center}
\end{figure}

Finally, in Fig.~\ref{fig:futuretest-data} we compare predictions
obtained using the pre-HERA, pre-LHC, and
global (NNPDF4.0) PDFs to a representative selection of data included in the global fit but
not in the pre-LHC fit.
Specifically, we consider the
HERA NC structure functions at $\sqrt{s}=920$ GeV in the $Q=1.871$ GeV bin;
the dimuon rapidity distributions in forward $Z\to \mu\mu$ production
at LHCb; the top quark rapidity distributions
in the ATLAS $t\bar{t}$ lepton+jet measurement at 8 TeV;
and the dilepton rapidity distribution for $M_{\ell\ell}=25$ GeV and
the CMS double-differential Drell-Yan measurement at 7 TeV. Of
these, only the HERA structure function data are included in the
pre-LHC fit, though of course not in the pre-HERA fit, while all
other data are predictions for both future-test PDF sets.
All results displayed in these comparisons
have been obtained using NNPDF4.0 methodology. A historical curiosity
here is the observation that the rise of the $F_2$ structure function
at HERA, which came as a surprize (see
e.g. Refs.~\cite{DeRoeck:1995mt,Tung:2004ab}) is correctly reproduced
by the pre-HERA fit based on its onset in pre-HERA data. Note however
that this should not be taken as a prediction: both methodologies that
we are testing here have been developed based on later datasets, and
thus do encode to some extent some of the information contained in the later
data. 

The very large difference between fitted and extrapolation PDF
uncertainty is apparent, and so is the hierarchy  between
near-extrapolation uncertainties (pre-LHC) and far-extrapolation
uncertainties (pre-LHC), e.g. for the top pair production
data.
The good compatibility between data and predictions
including PDF uncertainties is also clear, again confirming the
success of  the future test as summarised in Table~\ref{table:chi2-futuretest}.
      
%% file: tables/chi2-futuretest.tex
\begin{tabularx}{\textwidth}{lC{0.7cm}C{1.8cm}C{1.7cm}C{1.6cm}C{1.8cm}C{1.7cm}C{1.6cm}}
  \toprule
                      &  $N_{\rm dat}$ & \multicolumn{3}{c}{NNPDF3.1 methodology} & \multicolumn{3}{c}{NNPDF4.0 methodology}  \\ \midrule
Process                       &      & pre-HERA    & pre-LHC     & Global & pre-HERA                 & pre-LHC                 & Global \\
\midrule
Fixed target NC DIS           & 973  & 1.07                     & 1.21                     & 1.27 & 1.05                     & 1.18                    & 1.23       \\
\midrule
Fixed target CC DIS           & 908  & 0.97                     & 1.06                     & 1.14 & 0.80                     & 0.85                    & 0.87       \\
\midrule
Fixed target Drell-Yan        & 89   & 0.87                     & 1.03                     & 1.28 & 0.92                     & 1.27                    & 1.59       \\
\midrule
HERA                          & 1208 & {\it 35.75} ({\bf 1.18})  & 1.21                     & 1.20 & {\it 27.20} ({\bf 1.23}) & 1.22                    & 1.20      \\
\midrule
Collider Drell-Yan (Tevatron) & 65   & {\it 16.95} ({\bf 0.98}) & 0.93                     & 1.00 & {\it 5.52} ({\bf 1.02})  & 0.99                    & 1.11       \\
\midrule
Collider Drell-Yan (LHC)      & 116  & {\it 22.96} ({\bf 1.03}) & {\it 3.39 } ({\bf 1.31}) & 1.61 & {\it18.91} ({\bf 1.31})  & {\it 2.63} ({\bf 1.58}) & 1.53         \\
\midrule
Top quark production          & 83   & {\it 9.63 } ({\bf 0.63}) & {\it 1.31 } ({\bf 0.65}) & 0.90 & {\it20.01} ({\bf 1.06})  & {\it 1.30} ({\bf 0.87}) & 1.01         \\
\midrule
Jet production                & 500  & {\it 4.92 } ({\bf 0.88}) & {\it 3.13 } ({\bf 0.99}) & 1.43 & {\it2.69} ({\bf 0.98})   & {\it 2.12} ({\bf 1.10}) & 1.26        \\
\midrule
Total out-of-sample           &      & 25.46       ({\bf 1.07}) & 2.96 ({\bf 1.00 })       & $-$   & {\it19.48} ({\bf 1.16})  & {\it 2.10} ({\bf 1.15}) & $-$ \\
\midrule
Total                         & 3942 & 13.20                    & 1.47                     & 1.24 & 10.21                    & 1.28                    & 1.15 \\
  \bottomrule
 \end{tabularx} 


%% file: sec-dataset.tex
\section{Dataset dependence of the NNPDF4.0 parton set}
\label{sec:dataset}

Having established the reliability of the NNPDF4.0 determination, we now study
in detail the impact on the PDFs of the data (in this section) and of
the methodology (in the next section). This also provides us with
further a posteriori checks of the stability and reliability of our results.

In this Section, we first assess the global impact of the change in
dataset when going from NNPDF3.1 to NNPDF4.0, and then we present variants of
the baseline NNPDF4.0 fit, in which the impact of specific datasets is
studied by removing them from the baseline. 
Finally, we assess  the impact of datasets that have not been included in the
NNPDF4.0 baseline, with the main aim of checking the stability of
our results. Whereas the analysis presented in this section gives some
indication on the pull of some data on individual PDFs, a full
assessment of the impact of various data on the PDFs would require the
use of correlation tools such as presented in
Ref.~\cite{Carrazza:2016htc}, as well as systematic studies of PDFs
based on partial datasets, such as presented in Sect.~2.3 of
Ref.~\cite{Butterworth:2014efa} in the context of the NNPDF2.3
determination.

Except otherwise stated, all the fits presented in
this section utilize the methodology discussed in Sect.~\ref{sec:methodology}
and correspond to Monte Carlo ensembles of 100 replicas.

\subsection{Impact of the updated dataset}
\label{subsection:updated_data}
As  explained in Sect.~\ref{subsec:dataset_overview}, the
NNPDF4.0 dataset 
differs from NNPDF3.1 not only because of the addition of
a significant amount of measurements not included in NNPDF3.1, but
also because of changes in the treatment of
some of the data already included in NNPDF3.1, mostly related
to updates in the data and in the corresponding theory calculations.
These changes are incorporated in a dataset called NNPDF3.1-like in
Sect.~\ref{subsec:dataset_overview} and used throughout this paper
whenever comparisons to NNPDF3.1 are required, e.g. in the code benchmarks
of Sect.~\ref{sec:benchmark} or in the future tests of
Sect.~\ref{sec:futuretest}. However, in
Sect.~\ref{subsubsec:NNPDF40_vs_NNPDF31_PDFs} (specifically
in Fig.~\ref{fig:40vs31_PDFs})  we compared NNPDF4.0 to
the published NNPDF3.1 set. We must therefore start
this discussion of dataset dependence
with an assessment of the differences between the published NNPDF3.1 fit and
its update based on this NNPDF3.1-like dataset.

\subsubsection{The NNPDF3.1-like dataset and PDFs}
\label{subsubsection:NNPDF31-like_dataset}

\begin{figure}[!t]
  \centering
  \includegraphics[width=0.45\textwidth]{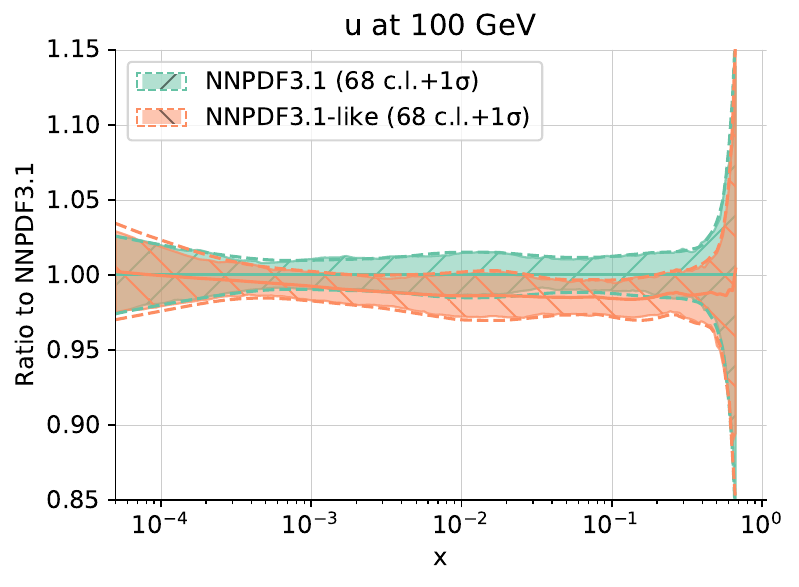}
  \includegraphics[width=0.45\textwidth]{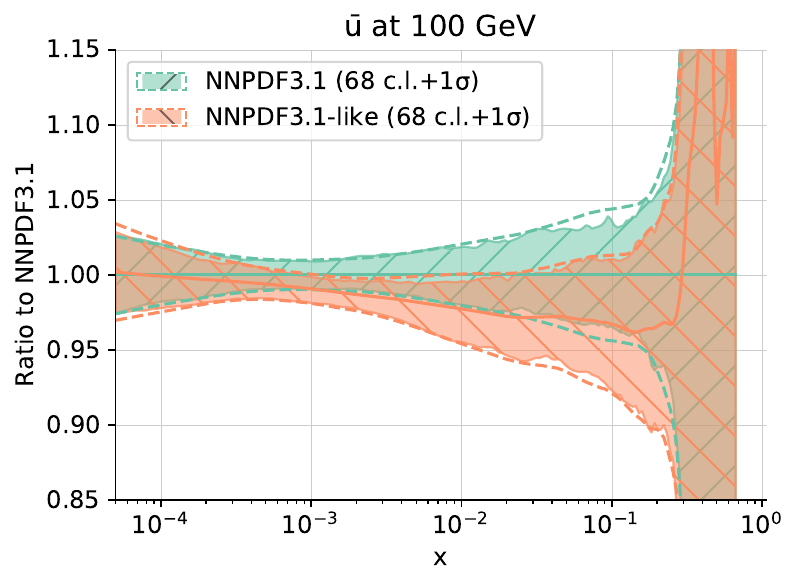}
  \includegraphics[width=0.45\textwidth]{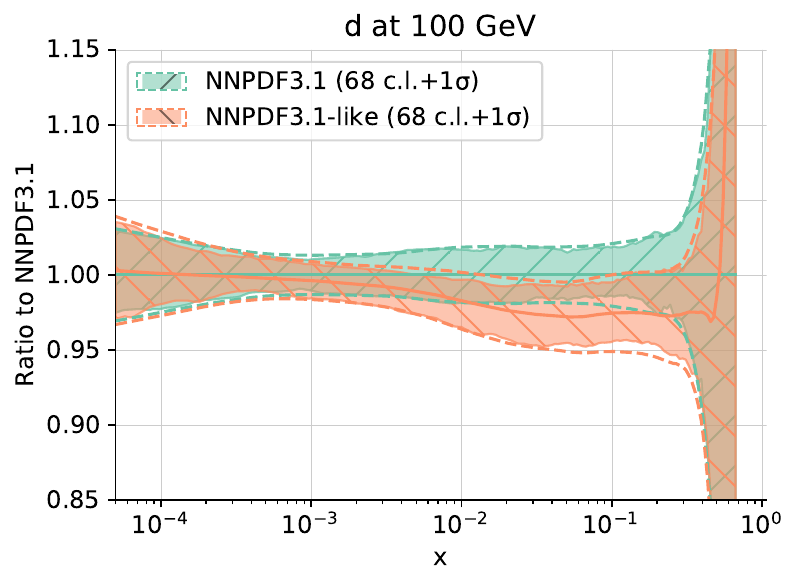}
  \includegraphics[width=0.45\textwidth]{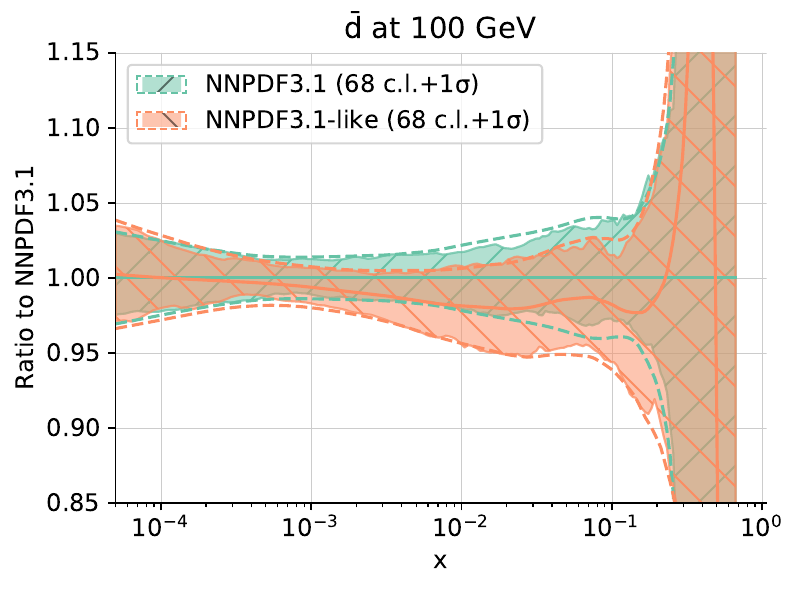}
  \includegraphics[width=0.45\textwidth]{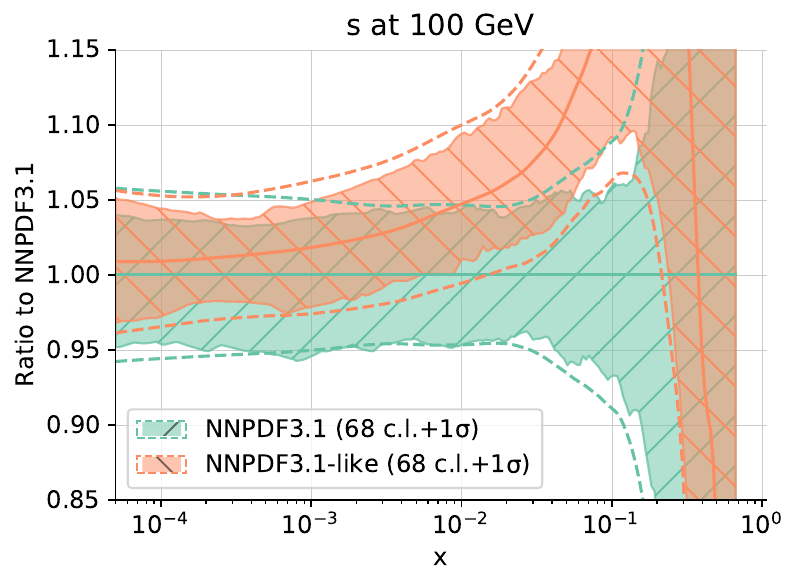}
  \includegraphics[width=0.45\textwidth]{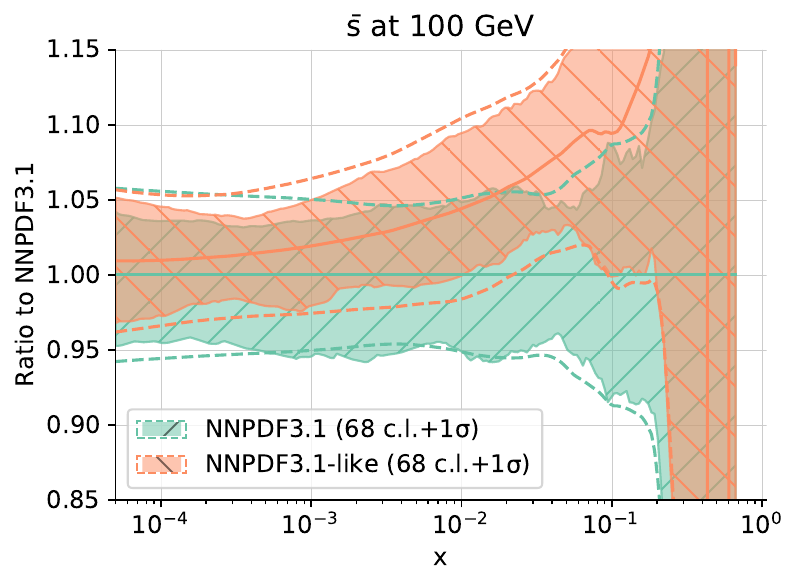}
  \includegraphics[width=0.45\textwidth]{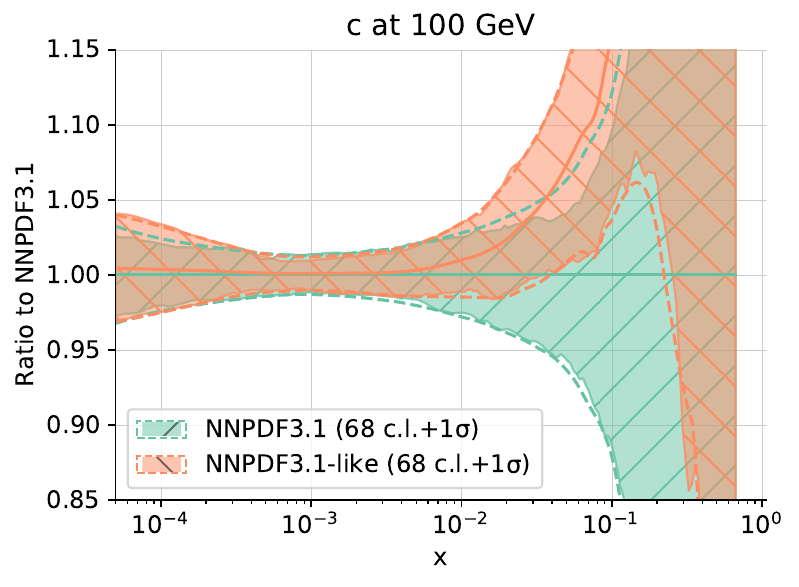}
  \includegraphics[width=0.45\textwidth]{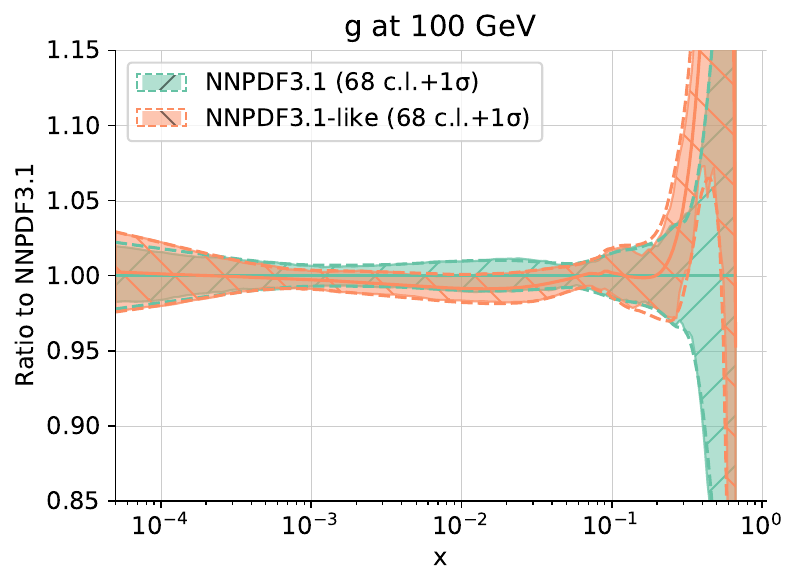}

  \caption{The up, antiup, down, antidown, strange, antistrange, charm and
    gluon PDFs from NNPDF3.1 and
    from a fit based on the same NNPDF3.1 methodology but on the
    NNPDF3.1-like dataset defined in Sect.~\ref{subsec:dataset_overview}.
    Results are displayed at $Q=100$~GeV, normalized to the NNPDF3.1
    central value. Solid and dashed bands correspond to 68\% and one-sigma
    uncertainties, respectively.}
  \label{fig:31vs31-like}
\end{figure}

The impact of the alterations made to the NNPDF3.1 dataset are studied by
comparing the original NNPDF3.1 determination~\cite{Ball:2017nwa} to a PDF
fit based on same NNPDF3.1 methodology but using the
NNPDF3.1-like dataset discussed in Sect.~\ref{subsec:dataset_overview}.
The corresponding PDFs are compared in Fig.~\ref{fig:31vs31-like}.

\begin{figure}[!t]
  \centering
  \includegraphics[width=0.45\textwidth]{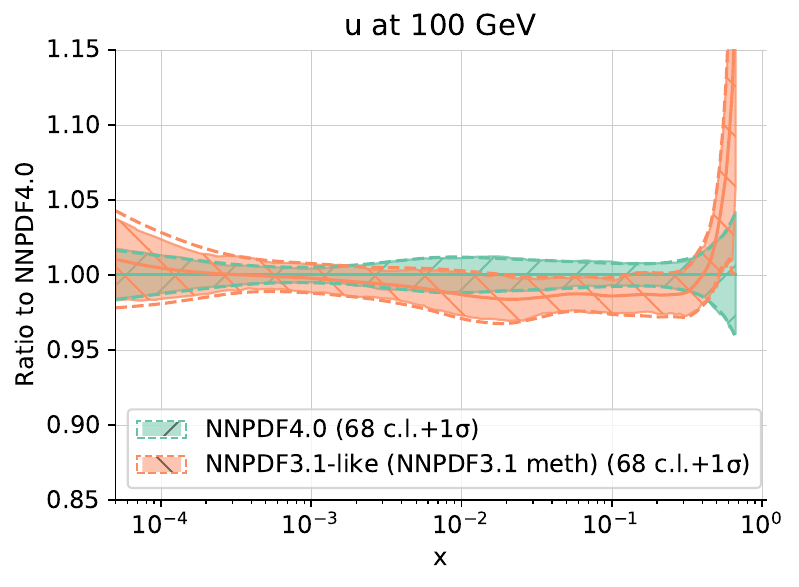}
  \includegraphics[width=0.45\textwidth]{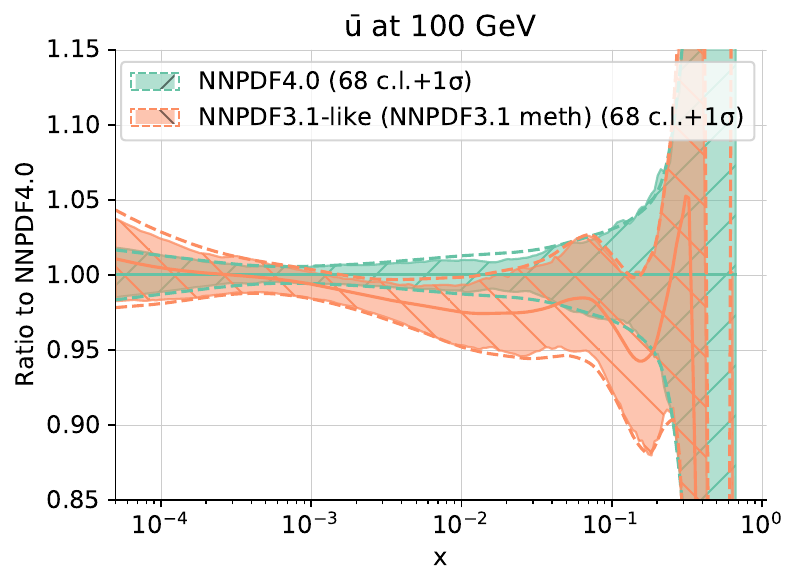}
  \includegraphics[width=0.45\textwidth]{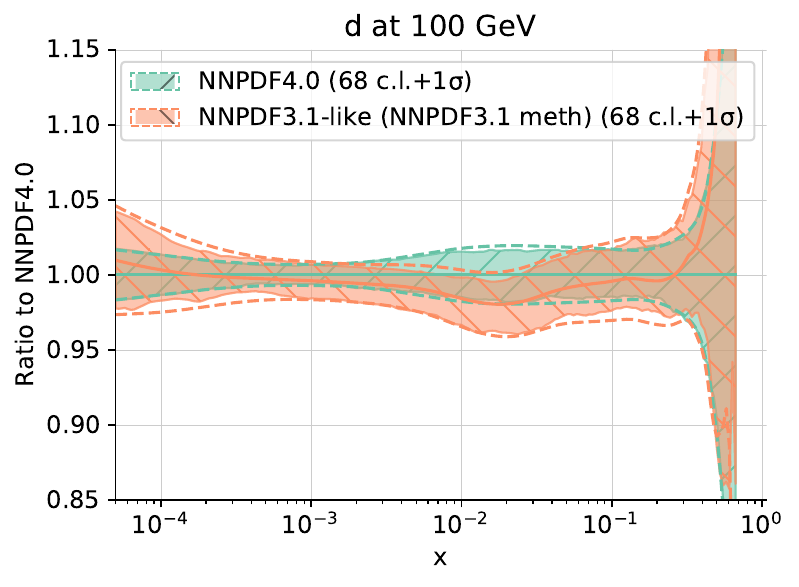}
  \includegraphics[width=0.45\textwidth]{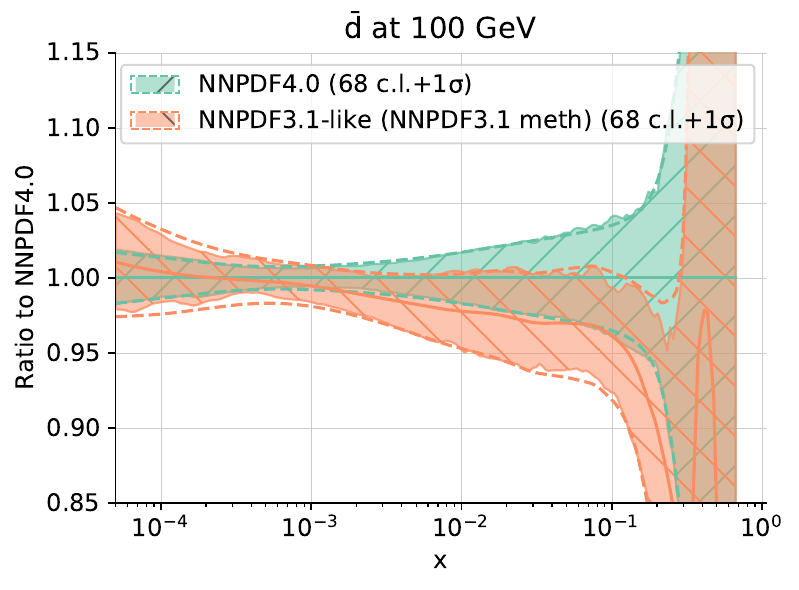}
  \includegraphics[width=0.45\textwidth]{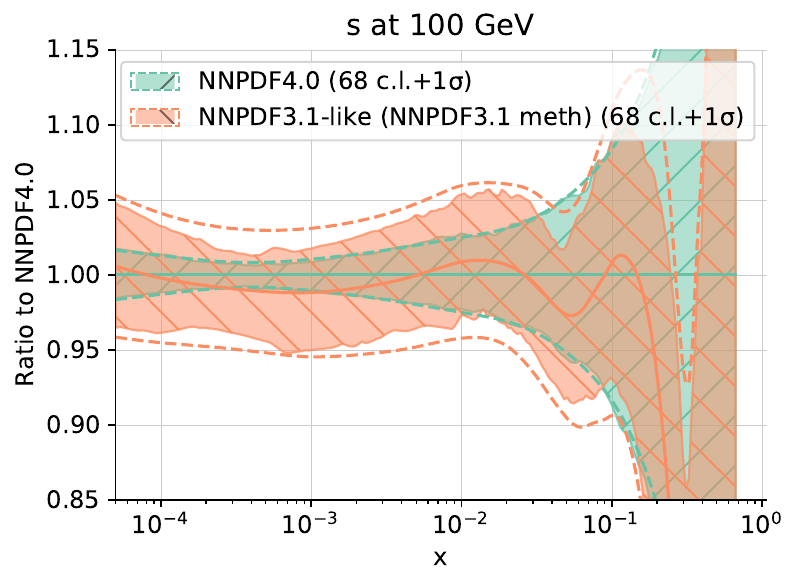}
  \includegraphics[width=0.45\textwidth]{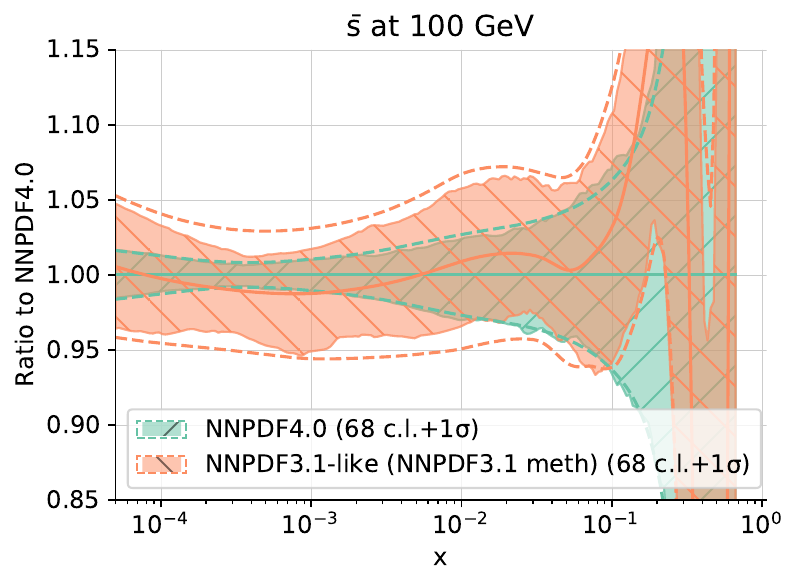}
  \includegraphics[width=0.45\textwidth]{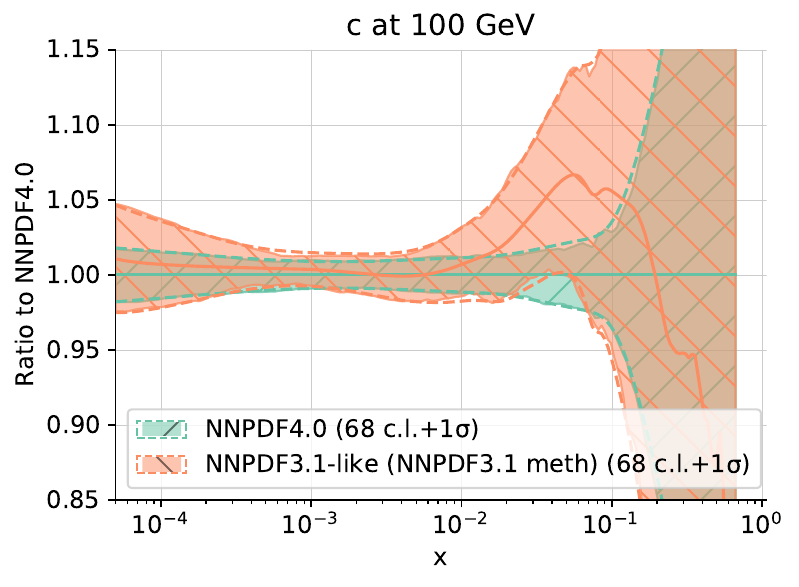}
  \includegraphics[width=0.45\textwidth]{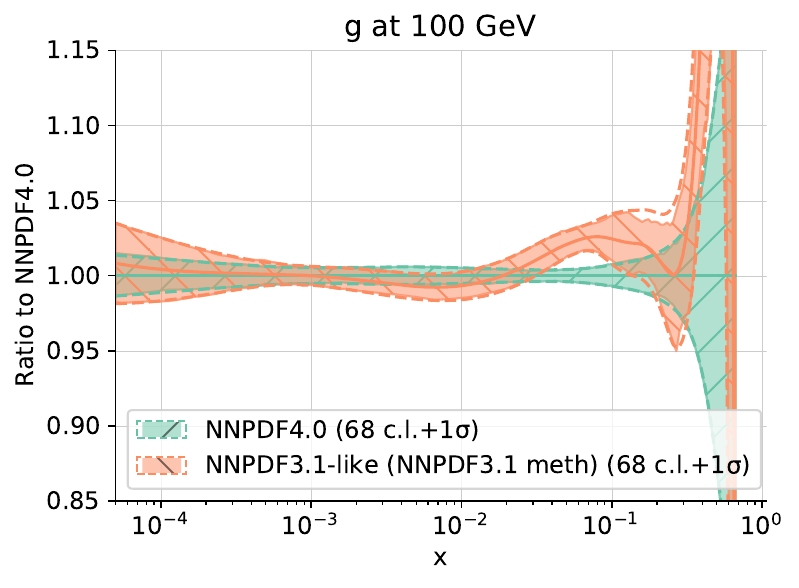}
  \caption{Same as Fig.~\ref{fig:40vs31_PDFs} but now comparing
    NNPDF4.0 to  NNPDF3.1-like instead of the published NNPDF3.1. The
    NNPDF3.1-like PDFs shown here  define the NNPDF3.1 baseline which will 
    be used in all
    subsequent plots in this section.}
  \label{fig:40vs31-like_PDFs}
\end{figure}

As expected, the two PDF sets are overall well consistent, with the PDF
central values of each set being almost always included in the PDF
uncertainties of the other across the entire range of $x$. Some differences
are nevertheless seen for individual PDFs. These are the largest for the
strange quark and antiquark PDFs. In this case, differences are mostly
explained by the improved treatment of the NuTeV data: the
NNPDF3.1-like dataset incorporates NNLO massive QCD corrections to
the dimuon cross-sections~\cite{Gao:2017kkx}, which were not available at the
time the original NNPDF3.1 set was produced,
and an update of the value for the 
branching ratio of charmed hadrons into muons (see
Sect.~\ref{subsec:dataset_overview} and the discussion of~\cite{Faura:2020oom}.)

The combined effect of these two updates is an enhancement of the
strange quark and 
antiquark PDFs in comparison to the original NNPDF3.1 analysis, already 
reported in
Ref.~\cite{Faura:2020oom}. To compensate for this effect, the down quark and
antiquark PDFs are correspondingly suppressed. In the case of the charm PDF,
a different behavior of the central value is observed for $x\gtrsim 0.01$,
possibly because of the replacement of the HERA charm cross-section
data with their final combined version, see
Sect.~\ref{sec:datatheory}. Finally, slight differences in the gluon PDF are
likely due the different treatment of single-inclusive jet data:
Tevatron and 2.76~TeV ATLAS and CMS measurements are no longer included in the
NNPDF3.1-like dataset, and NNLO $K$-factors, computed with the recommended 
choice
of scale, are incorporated for the remaining 7~TeV ATLAS and CMS measurements
(no NNLO $K$-factors were used in NNPDF3.1, as they were not yet available).
The precision of the PDFs in the two parton sets is almost
identical. We conclude that the difference in strange PDFs between the published
NNPDF3.1 and NNPDF4.0 observed in
Sect.~\ref{subsubsec:NNPDF40_vs_NNPDF31_PDFs} is due to these reasons.

We conclude that the NNPDF3.1-like dataset is compatible with NNPDF3.1
but not identical to it, with differences due to updates in
either the data, or their theoretical treatment after the original
NNPDF3.1 PDF set was produced. Henceforth, in this and the next
section, we will always compare to this updated NNPDF3.1-like PDF set
and dataset,  and by ``NNPDF3.1 baseline'' will
always refer to the PDFs obtained with NNPDF3.1 methodology and NNPDF3.1-like
dataset. For completeness, these NNPDF3.1 baseline PDFs are also shown in
Fig.~\ref{fig:40vs31-like_PDFs}, compared now
to the  NNPDF4.0 baseline. Of course, the overall pattern is
very similar to that of the comparison between NNPDF4.0 and the
published NNPDF3.1 previously shown in
Fig.~\ref{fig:40vs31_PDFs}. Fig.~\ref{fig:40vs31-like_PDFs} will
serve as a reference  when assessing the relative impact of data and
methodology in driving the differences between NNPDF3.1 and
NNPDF4.0.

\subsubsection{Impact of the new data in NNPDF4.0}
\label{subsubsection:NNPDF4.0_dataset}

\begin{figure}[!t]
  \centering
  \includegraphics[width=0.45\textwidth]{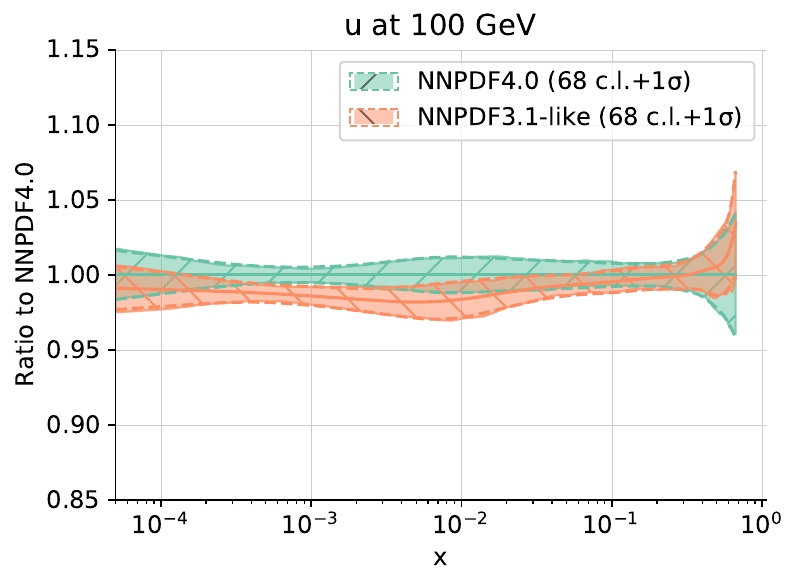}
  \includegraphics[width=0.45\textwidth]{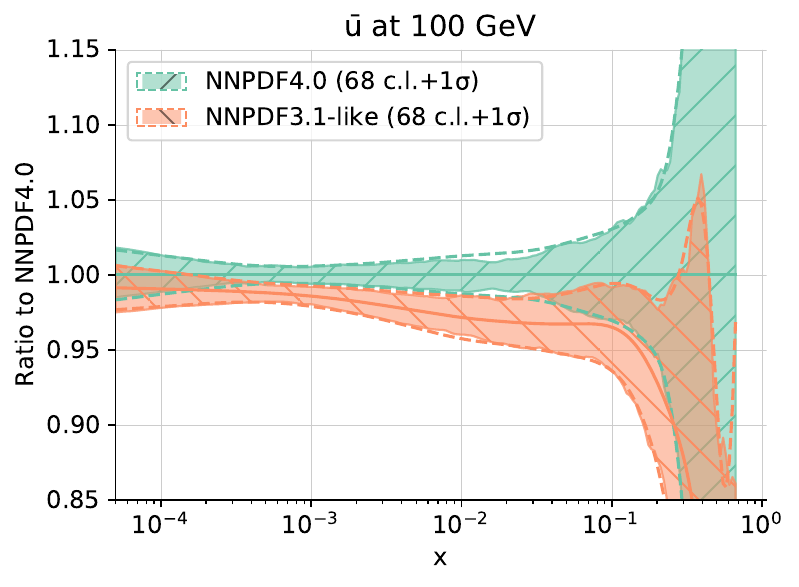}
  \includegraphics[width=0.45\textwidth]{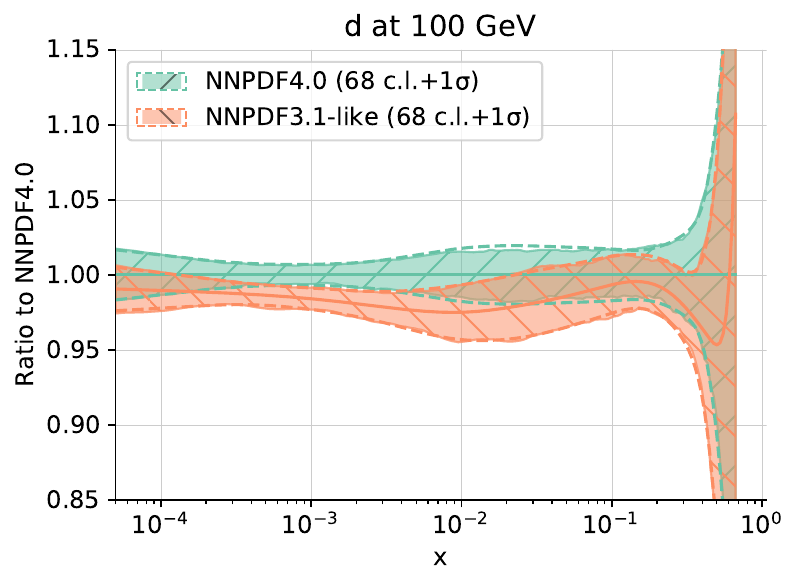}
  \includegraphics[width=0.45\textwidth]{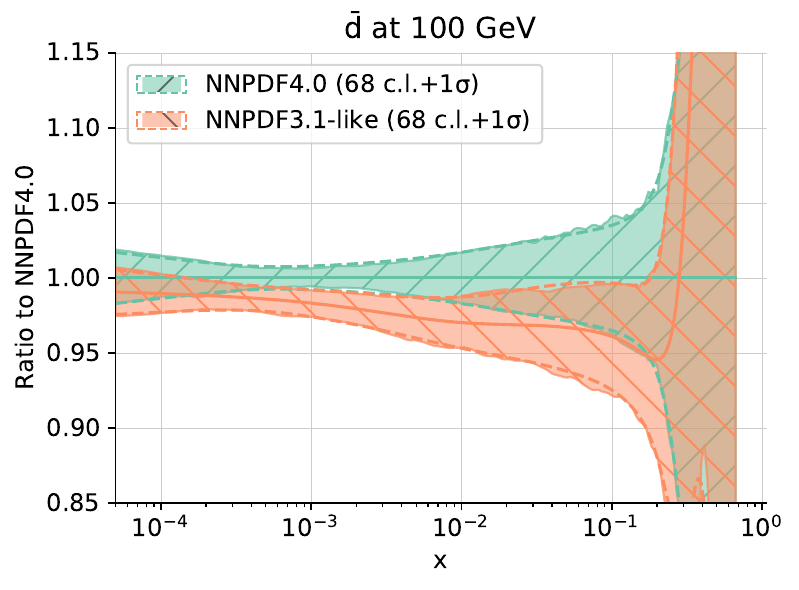}
  \includegraphics[width=0.45\textwidth]{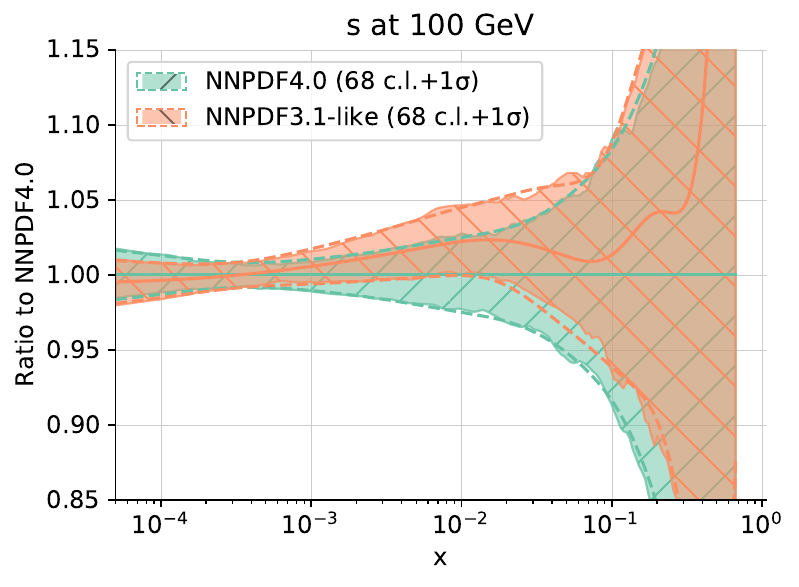}
  \includegraphics[width=0.45\textwidth]{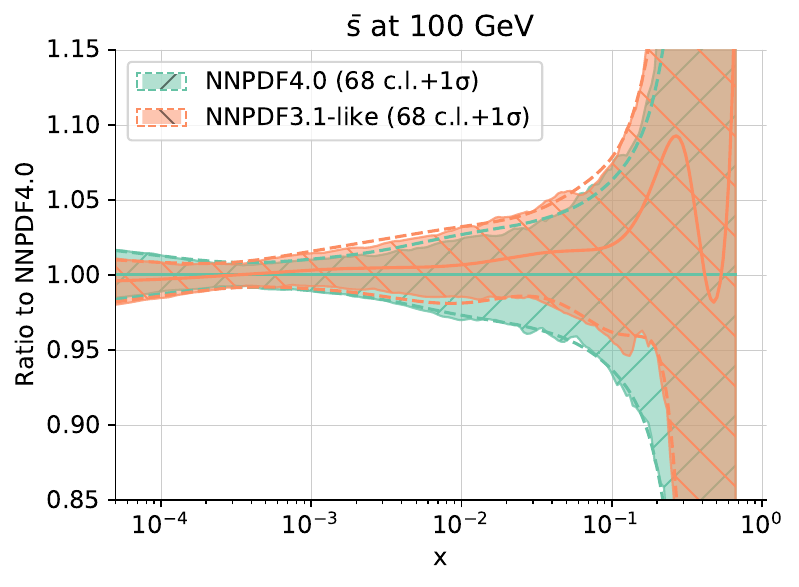}
  \includegraphics[width=0.45\textwidth]{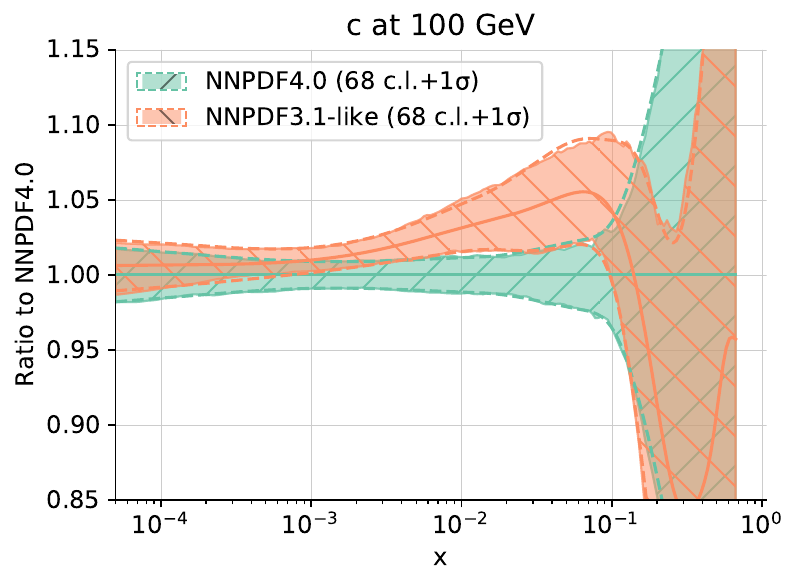}
  \includegraphics[width=0.45\textwidth]{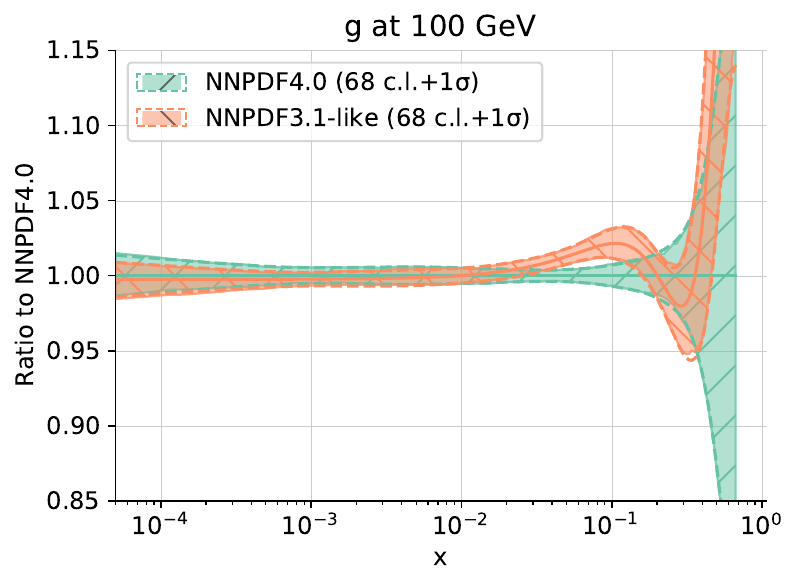}
  \caption{Same as Fig.~\ref{fig:31vs31-like} now comparing to  the
    NNPDF4.0 baseline a PDF set based on the same NNPDF4.0
    methodology but on the NNPDF3.1-like dataset defined in
    Sect.~\ref{subsec:dataset_overview}.}
  \label{fig:40vs31-like}  
\end{figure}

\begin{figure}[!t]
  \centering
  \includegraphics[width=0.45\textwidth]{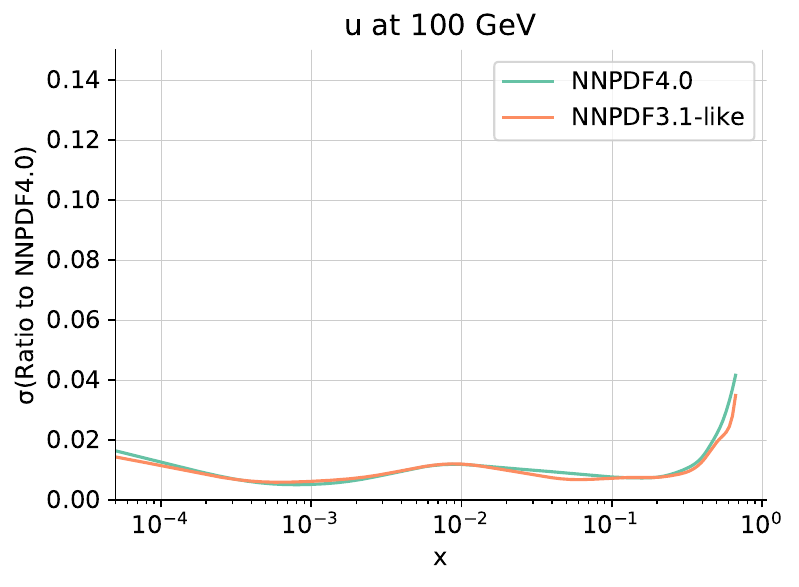}
  \includegraphics[width=0.45\textwidth]{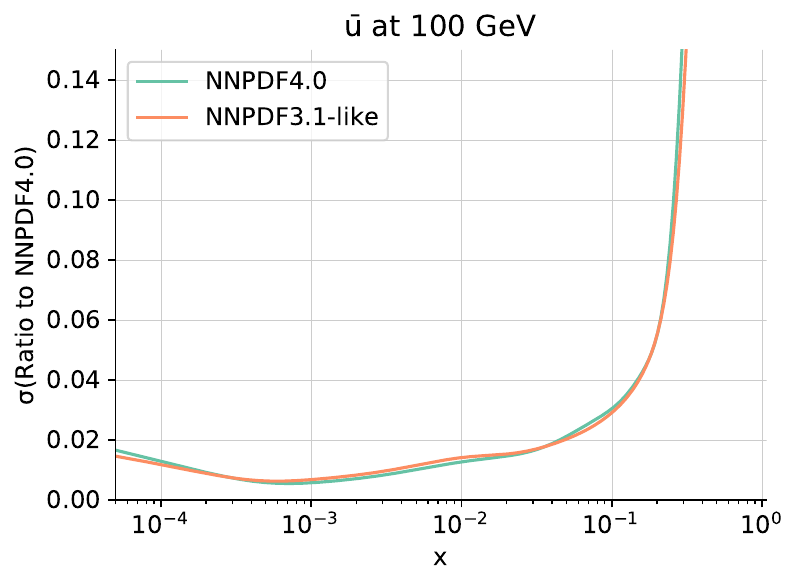}
  \includegraphics[width=0.45\textwidth]{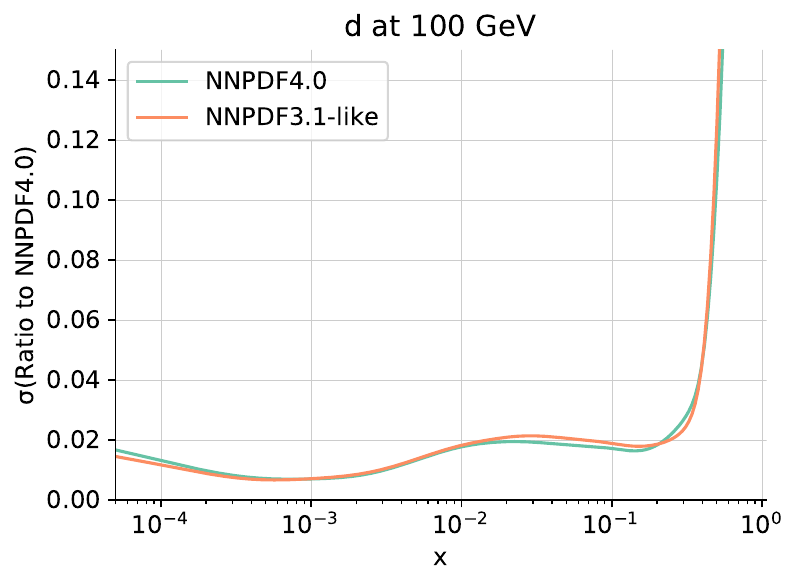}
  \includegraphics[width=0.45\textwidth]{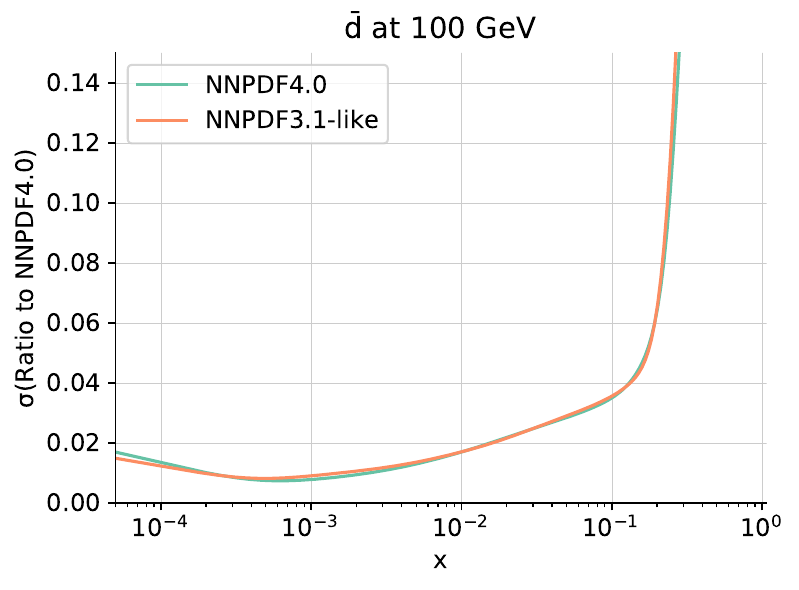}
  \includegraphics[width=0.45\textwidth]{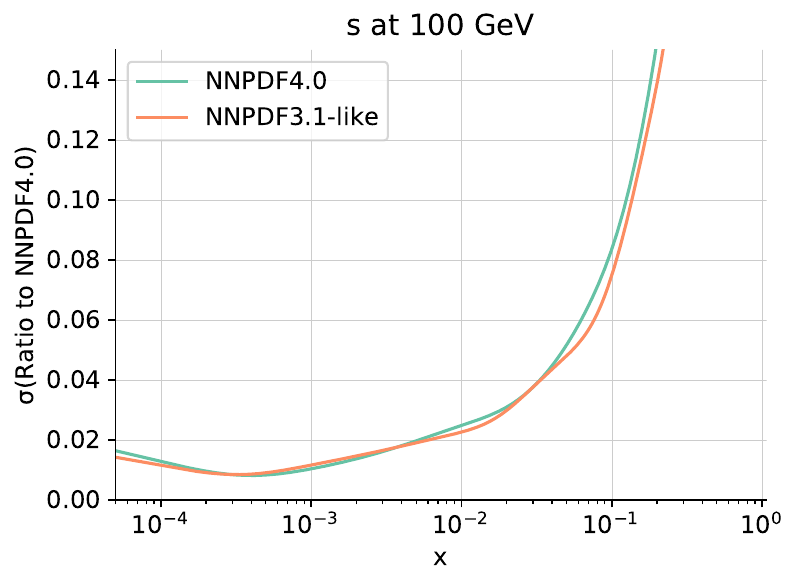}
  \includegraphics[width=0.45\textwidth]{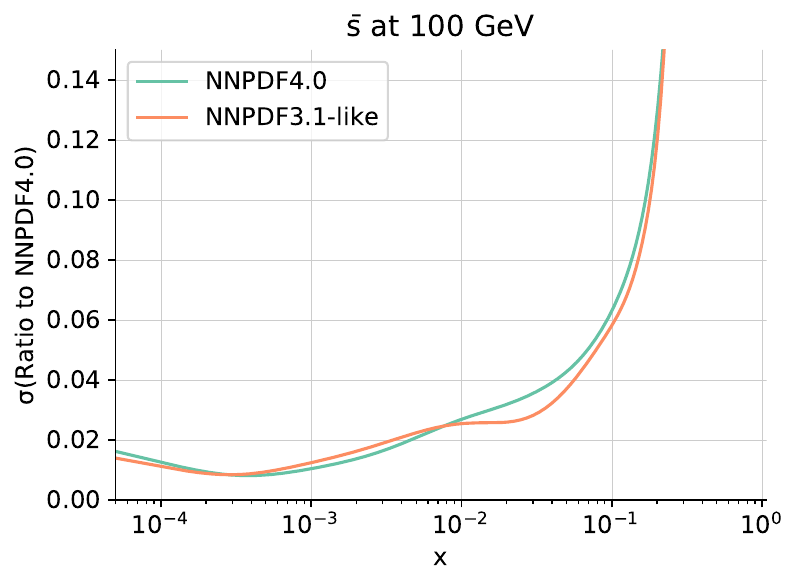}
  \includegraphics[width=0.45\textwidth]{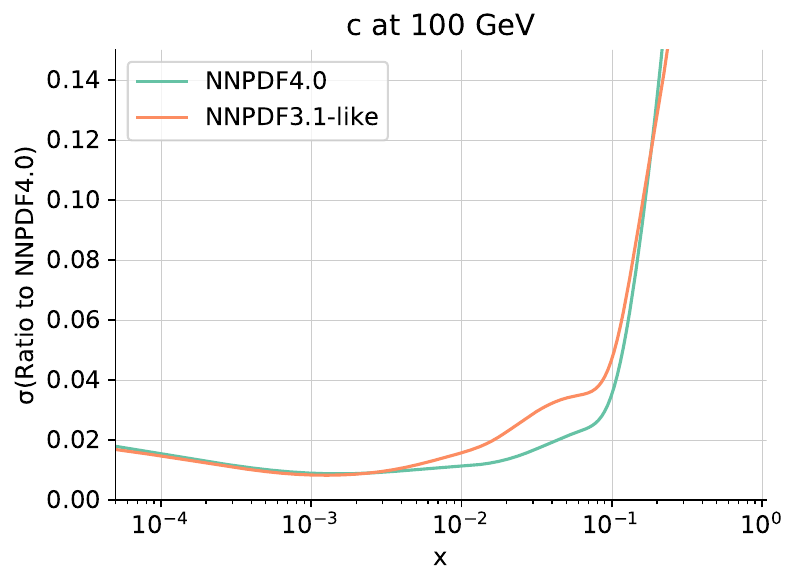}
  \includegraphics[width=0.45\textwidth]{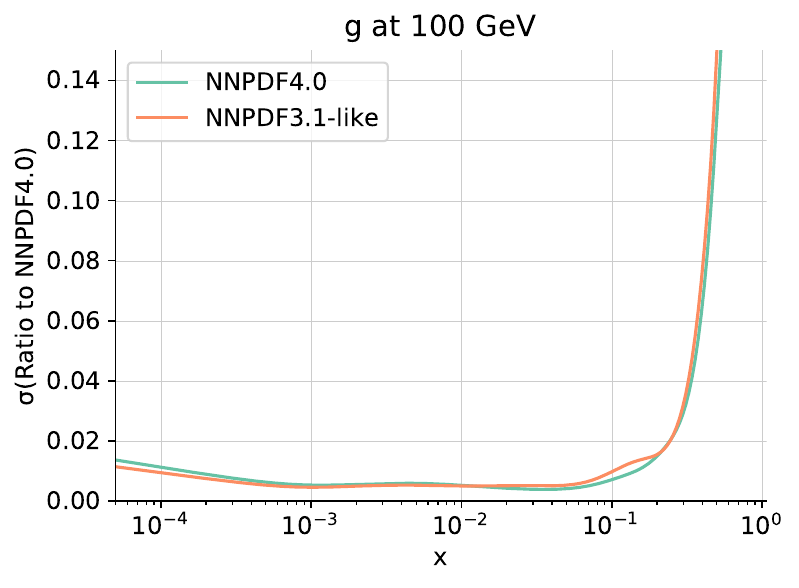}
   \caption{Same as Fig.~\ref{fig:40vs31-like} but for one-sigma relative
    uncertainties.}
  \label{fig:40vs31-like_uncs}
\end{figure}

The impact of the new measurements included in the NNPDF4.0 dataset
is studied by comparing the baseline NNPDF4.0 parton set to a PDF
determination based on the same NNPDF4.0 methodology (presented in
Sect.~\ref{sec:methodology}), but using the NNPDF3.1-like dataset defined in
Sect.~\ref{subsec:dataset_overview}. In Fig.~\ref{fig:40vs31-like} we compare
the corresponding up, antiup, down, antidown, strange, antistrange, charm and
gluon PDFs as a function of $x$ at $Q=100$~GeV.
Results are normalized to the NNPDF4.0 central value. In
Fig.~\ref{fig:40vs31-like_uncs} we compare the corresponding one-sigma
PDF uncertainties.

Interestingly, even though there is compatibility within uncertainties,
the central values of all PDFs change, often almost at the one-sigma
level, with the largest differences seen in the gluon, as noted in Sect.~\ref{subsubsec:NNPDF40_vs_NNPDF31_PDFs}. This means that the new data are bringing in  new experimental
information.
In the case of the light quark and antiquark PDFs, the new data
(mostly from LHC inclusive gauge boson production, as we will see in
Sect.~\ref{subsubsec:gaugeboson_data}) produce  an enhancement of up to 3\%
for $0.01\lesssim x\lesssim 0.1$ of the up and down PDFs, and a milder
suppression of the strange PDF.
In the case of  charm, a suppression of about 4-5\% is seen for
$0.01\lesssim x\lesssim 0.1$ and an enhancement of about 10\%
for $x\gtrsim 0.1$. In the case of the gluon, the impact 
(mostly from single-inclusive jet and dijet production, as we will see in
Sect.~\ref{subsubsec:gluon_data}) is a suppression of about 2-3\%
around $x\sim 0.1$ and a similar enhancement around $x\sim 0.3$.

While shifts of central values are typically of the size of the PDF
uncertainties, it is clear from Fig.~\ref{fig:40vs31-like_uncs} that
the uncertainties themselves are unchanged, except possibly for a reduction of
the uncertainty in charm in the region  $0.01\lesssim x\lesssim 0.1$.
On the other hand, comparing to the PDFs determined using the
NNPDF3.1-like dataset and methodology,  Fig.~\ref{fig:40vs31-like_PDFs},
one observes that the pattern in change of central values is the same.
Therefore we conclude that the differences in the shape of PDFs between
NNPDF3.1 and NNPDF4.0 are  mostly data-driven, but with little or no
impact on uncertainties. It follows that  when
comparing to the published NNPDF3.1,  the overall effect of the new data is to
improve the accuracy of the parton set while not significantly
affecting its precision.

\subsection{PDFs from  reduced datasets}
\label{subsec:reduced}

We now discuss a number of PDF sets determined by removing specific
measurements from the baseline, with the goal of assessing their
impact. We consider in turn: LHC inclusive gauge boson production; LHC single
top-quark and SeaQuest data; LHC jet, top  pair, $Z$ $p_T$, and
direct photon; all the LHC data altogether; collider data; and DIS data.

\subsubsection{The impact of LHC inclusive gauge boson production data}
\label{subsubsec:gaugeboson_data}

In the global fit, quark flavor separation is driven by charged-current
DIS structure functions and by inclusive gauge boson production in hadronic
collisions. In the latter case, the bulk of the data
comes from the  LHC. The ATLAS and CMS data are mostly in the
central rapidity region, sensitive to quarks and antiquarks in the
intermediate-$x$ region, while the LHCb data cover the forward rapidity region,
sensitive to quarks and antiquarks at large $x$ and small $x$.

\begin{figure}[!t]
  \centering
  \includegraphics[width=0.45\textwidth]{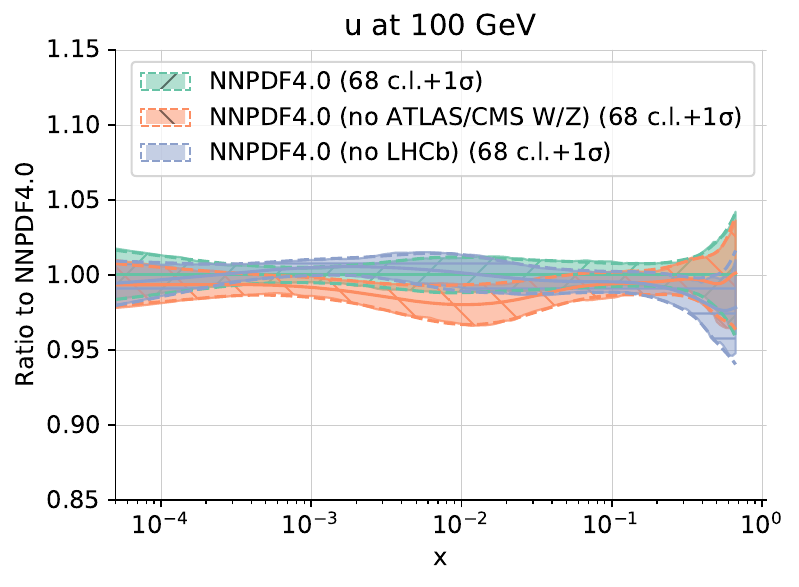}
  \includegraphics[width=0.45\textwidth]{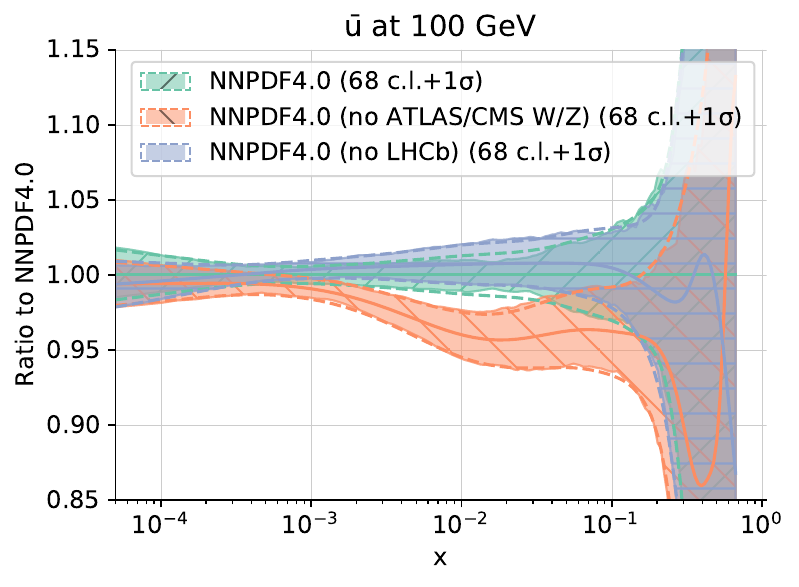}
  \includegraphics[width=0.45\textwidth]{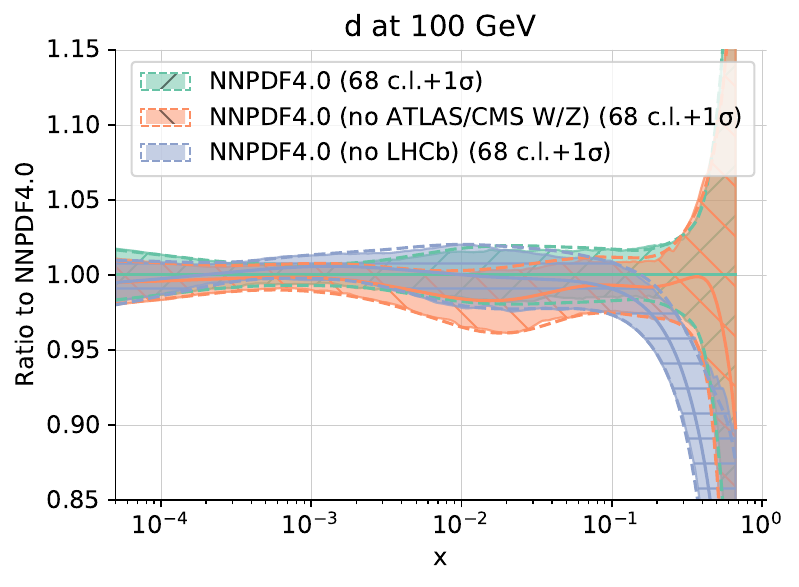}
  \includegraphics[width=0.45\textwidth]{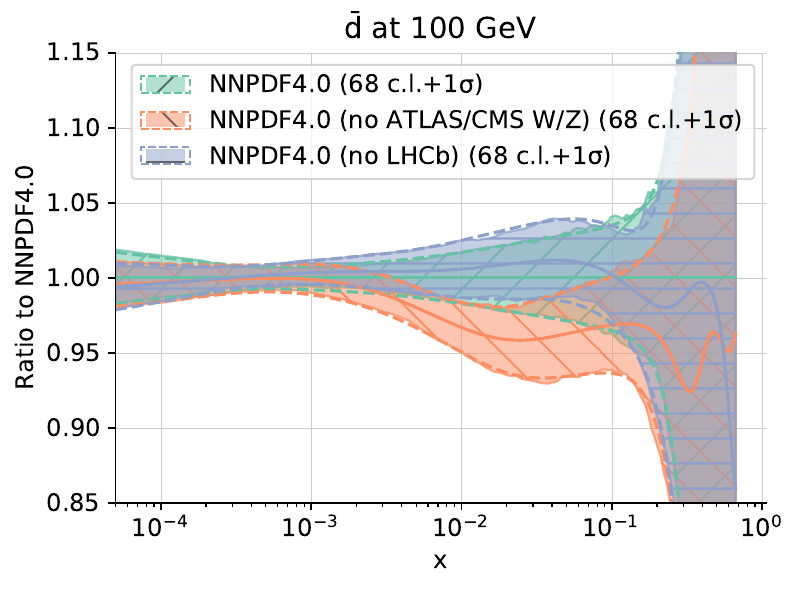}
  \includegraphics[width=0.45\textwidth]{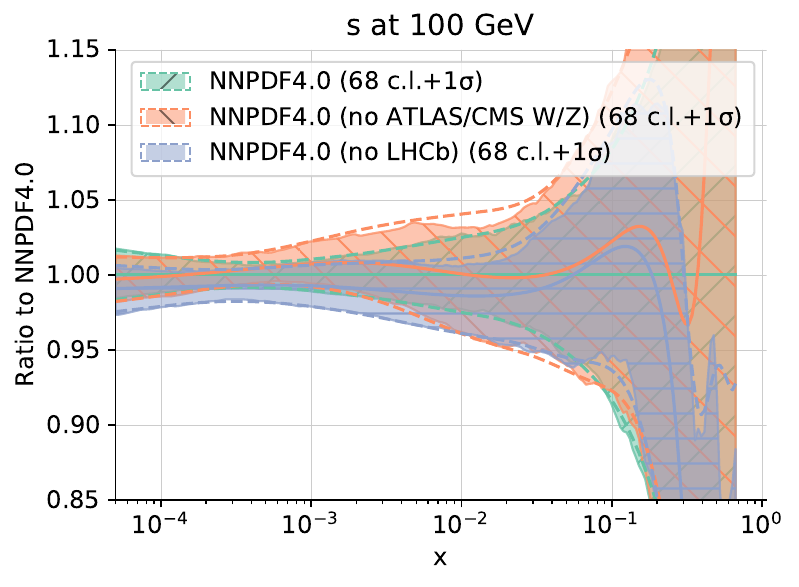}
  \includegraphics[width=0.45\textwidth]{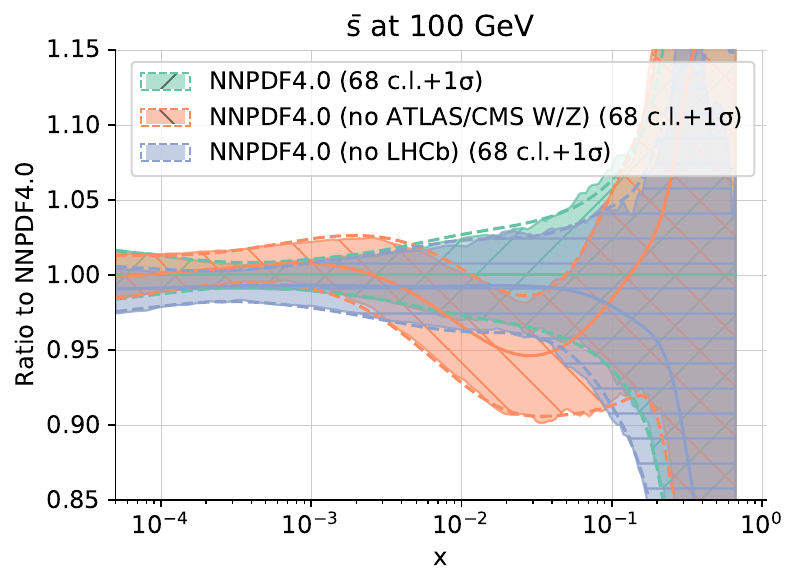}
  \includegraphics[width=0.45\textwidth]{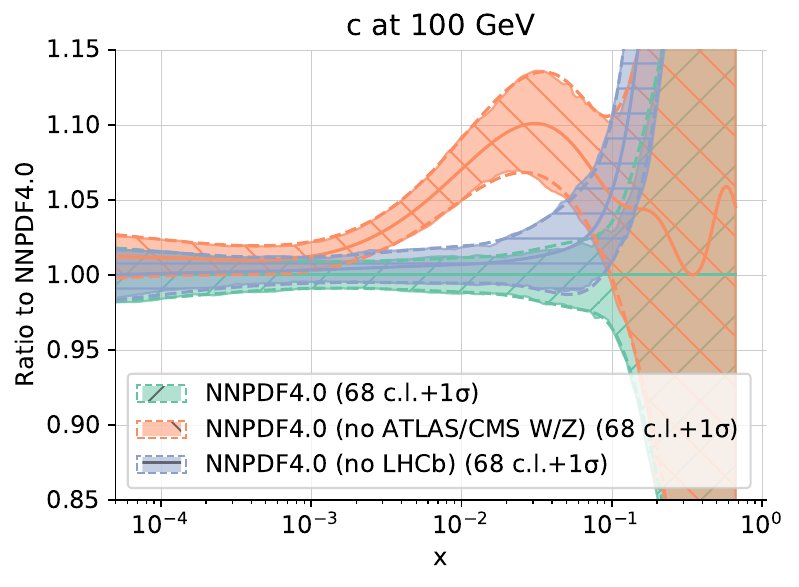}
  \includegraphics[width=0.45\textwidth]{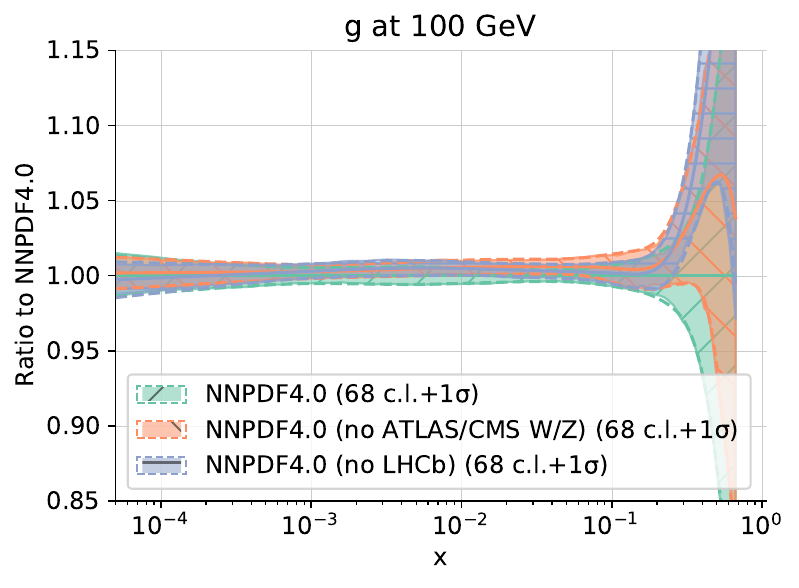}
  \caption{Same as Fig.~\ref{fig:31vs31-like} comparing the baseline
    to PDFs determined removing  either all of the ATLAS and CMS, or
    all of the LHCb inclusive gauge boson production data.}
  \label{fig:noWZ}
\end{figure}

In order to assess the impact of  this data we have produced
two PDF sets  removing from the baseline all of the inclusive gauge
boson production measurements,  either from  ATLAS and CMS, or  from 
LHCb. Fig.~\ref{fig:noWZ} compares these PDFs to the baseline.
The effect of removing the ATLAS and CMS data is a
suppression of the light quarks and antiquarks (by 2-4\%) and an enhancement of
the charm (by up to 10\%) around $0.01\lesssim x\lesssim 0.1$. The effect of
removing the LHCb data is more moderate and predominately affects the down,
charm and gluon at around $x\gtrsim 0.1$. Specifically, the former is suppressed
while the latter two  are enhanced in comparison to the baseline 
(in both cases by up to 10\%). The shift of central values is generally within
the PDF uncertainty, except for the up, antiup, antidown and
charm when excluding the ATLAS and CMS data, and for the down quark
when excluding LHCb data. 
As expected, and as mentioned in
Sect.~\ref{subsubsection:NNPDF4.0_dataset}, this data is thus responsible
for the bulk of the changes in light quark PDFs between NNPDF3.1 and NNPDF4.0.

\subsubsection{The impact of LHC single-top production data and of SeaQuest data}
\label{subsubsec:udratio}

Additional constraints on quark flavor separation at large $x$, in particular
on the $d/u$ and $\bar{d}/\bar{u}$ ratios, are in principle provided by
single top-quark production at the LHC and by fixed-target DY production
recently measured by the SeaQuest experiment. Because these measurements are
included for the first time in NNPDF4.0 (see Sects.~\ref{subsubsec:singletop}
and \ref{subsubsec:FTDY}) it is interesting to study their impact. To this
purpose, we have produced two PDF sets, respectively removing from the baseline
 either all of the single top data, or the SeaQuest
measurement.

\begin{figure}[!t]
  \centering
  \includegraphics[width=0.49\textwidth]{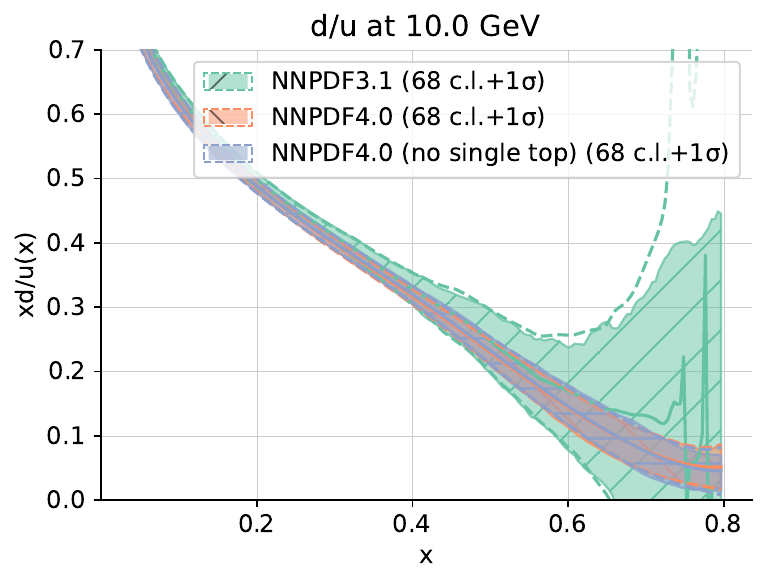}
  \includegraphics[width=0.49\textwidth]{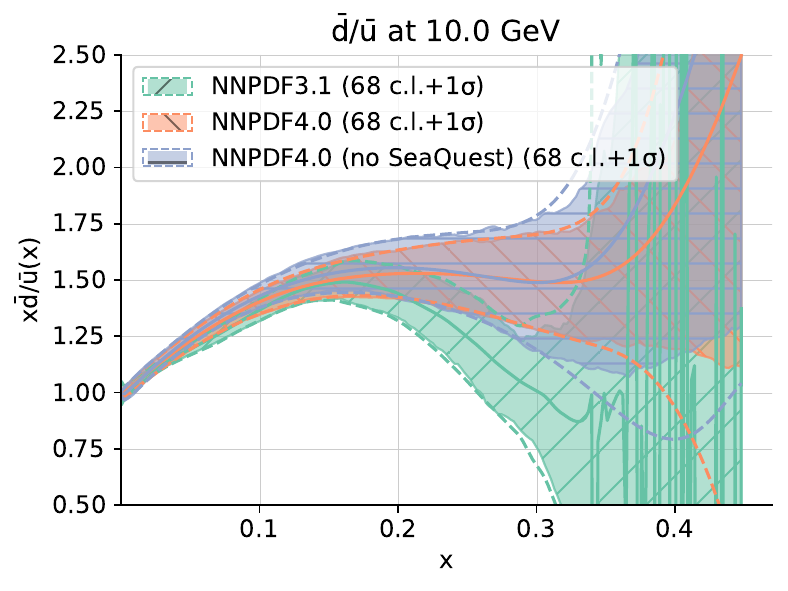}\\
  \caption{The $d/u$ (left) and $\bar{d}/\bar{u}$ (right) ratios,
    at $Q=10$~GeV, computed, respectively, from a NNPDF4.0 fit without single
    top-quark data or without SeaQuest data. In both cases we show results
    obtained with the NNPDF3.1 and NNPDF4.0 baseline fits.}
  \label{fig:udratio}
\end{figure}

In Fig.~\ref{fig:udratio} we compare the $d/u$ and $\bar{d}/\bar{u}$ ratios,
at $Q=10$~GeV; in the former case we show  results obtained from the fit
without single top data, and in the latter  we show  results obtained
omitting SeaQuest data, both compared to the NNPDF3.1 and
NNPDF4.0 baselines.

Single-top data have essentially no impact on  the $d/u$ ratio and
more generally on the whole PDF determination.  This is due to the
relatively large experimental uncertainties of the corresponding measurements,
as already noted in Sect.~\ref{subsec:results_fitquality} and in
Ref.~\cite{Nocera:2019wyk}. The significant reduction in uncertainty
on  the $d/u$ ratio between NNPDF3.1 and NNPDF4.0  is
methodology-driven, as  we will show explicitly
in Sect.~\ref{sec:tests} below. Indeed, we will show (see
Fig.~\ref{fig:40vs31_meth_uncs}) that the uncertainty on the large-$x$
up and down quark distributions is significantly reduced when switching
from NNPDF3.1 to NNPDF4.0 methodology with fixed NNPDF4.0 data, while we have
seen (compare Fig.~\ref{fig:40vs31-like_uncs}) that the same
uncertainty is essentially unchanged when reducing the dataset to the
NNPDF3.1 one with fixed NNPDF4.0 methodology. It is interesting to
note that the expectation for the $d/u$ ratio seems to converge to a finite
value between 0 and 1. This result  may be used to discriminate
non-perturbative models of nucleon structure~\cite{Ball:2016spl}.

The SeaQuest data have a moderate impact on
the $\bar{d}/\bar{u}$ ratio, and essentially no impact on other
PDFs. They lead to 
a moderate reduction in the PDF uncertainty, but they leave the
baseline central value almost
unchanged. In comparison to NNPDF3.1,
the $\bar{d}/\bar{u}$ ratio is enhanced by 50\% around $x\sim 0.3$ but remains
compatible with the larger NNPDF3.1 uncertainties. We therefore
conclude that the SeaQuest data  have very little impact on NNPDF4.0
due to their overall consistency with other data. Interestingly, the
$\bar{d}/\bar{u}$  ratio in NNPDF4.0 differs somewhat
from that in NNPDF3.1, due to the updated flavor separation driven by
the gauge boson production data discussed in
Sect.~\ref{subsubsec:gaugeboson_data}. The SeaQuest data thus provide, for the
particular case of the  $\bar{d}/\bar{u}$ ratio,
an independent confirmation of the improved knowledge on flavor
separation obtained in NNPDF4.0 thanks to LHC data.

\subsubsection{The impact of LHC jet, top-quark pair, $Z$ $p_T$ and direct photon data}
\label{subsubsec:gluon_data}

Various LHC processes in the NNPDF4.0  dataset constrain
the gluon PDF: top  pair and single-inclusive jet or dijet
production, at large values of $x$; and $Z$ $p_T$ and direct photon production
at intermediate values of $x$. In order to assess the impact of these
measurements, we have produced four fits by removing each of them in turn from
the baseline.

\begin{figure}[!t]
  \centering
  \includegraphics[width=0.49\textwidth]{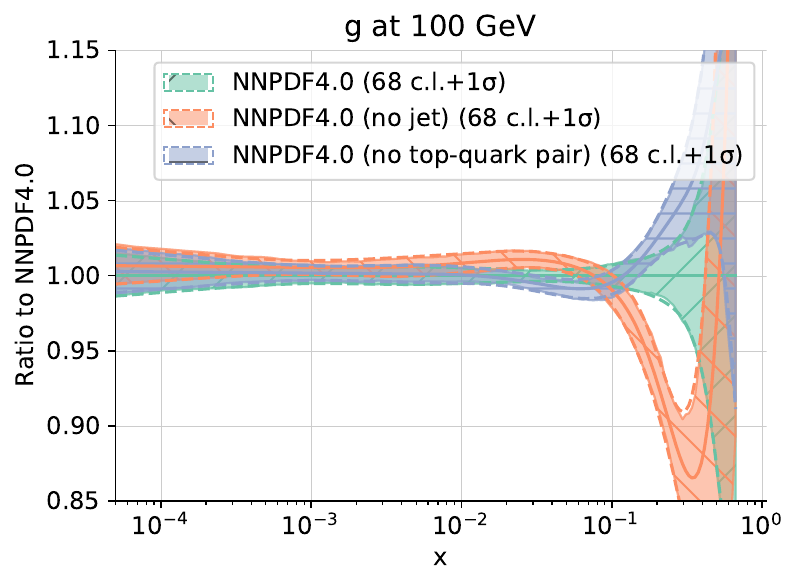}
  \includegraphics[width=0.49\textwidth]{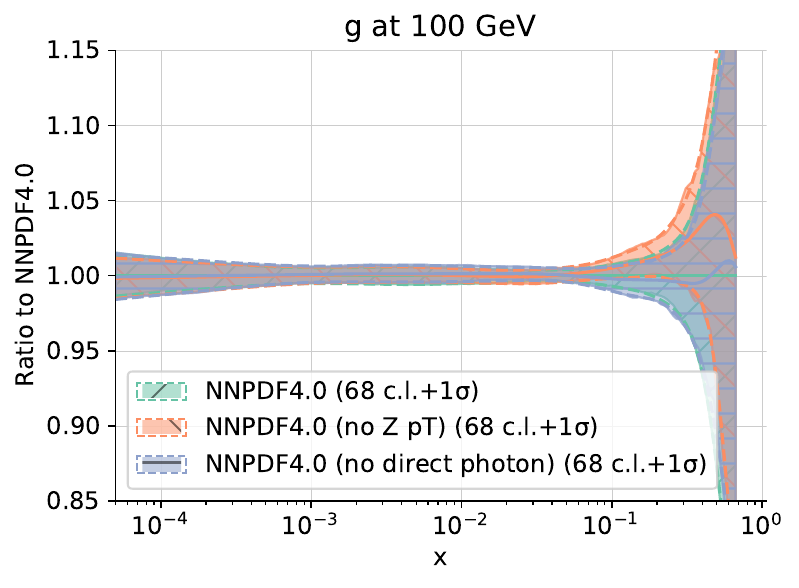}\\
  \caption{The gluon PDF obtained removing 
    single-inclusive jet and dijet data or top pair data (left), or
    $Z$ $p_T$ data or direct photon data (right).}
  \label{fig:nojetsnottbarnoZpTnophoton}
\end{figure}

In Fig.~\ref{fig:nojetsnottbarnoZpTnophoton} we compare to the
baseline the gluon from
each of these determinations. All other PDFs are essentially unaffected by
these changes in dataset, with only small changes in the quark PDFs
when removing the jet observables.
For clarity, we display separately PDFs  without 
top pair production and jet data, and PDFs without $Z$ $p_T$ and direct photon data. Only the gluon PDF is shown, normalized
to the central value of the NNPDF4.0 baseline.

The effect of the data is hierarchical. Single-inclusive jet and dijet data have
the largest impact: if they are removed, the gluon is slightly enhanced (by
2-3\%) around $0.01\lesssim x\lesssim 0.1$ and then more strongly suppressed
(by up to 15\%) for $x\gtrsim 0.1$. This suggests that the other datasets,
specifically  top pair data, tend to pull in the opposite direction,
suppressing  somewhat the gluon at large $x$. Top pair data have a moderate
impact: if they
are removed, the gluon is slightly enhanced for $x\gtrsim 0.1$, but within
the baseline uncertainty. $Z$ $p_T$ data have a yet smaller  impact:
if they are removed, the gluon is again a little enhanced for $x\gtrsim 0.1$.
The size of this shift is smaller than that observed in the case of the fit
without top-quark pair data and it remains compatible with baseline
uncertainty. Direct photon data have no effect: if they are removed,
the gluon does not change at all.

These results indicate that single-inclusive and dijet production data, which
are the most abundant and precise, drive the features of the gluon in the
global fit. Other data provide some generally consistent and complementary
information, particularly the top pair production data.

\subsubsection{The impact of LHC data}
\label{subsubsec:LHC_data}

\begin{figure}[!t]
  \centering
  \includegraphics[width=0.45\textwidth]{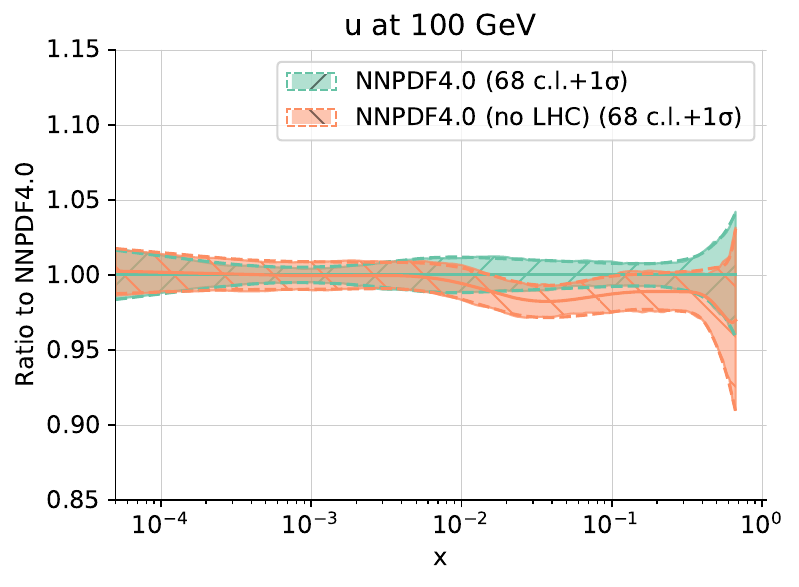}
  \includegraphics[width=0.45\textwidth]{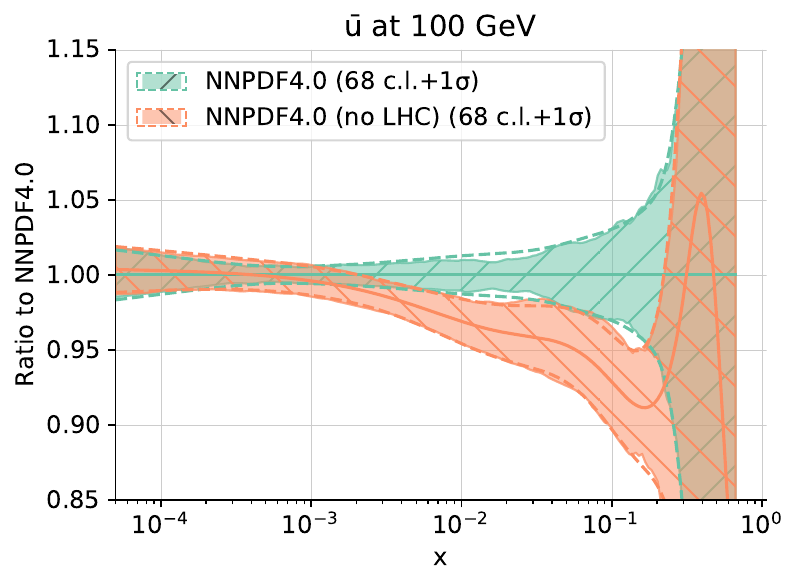}
  \includegraphics[width=0.45\textwidth]{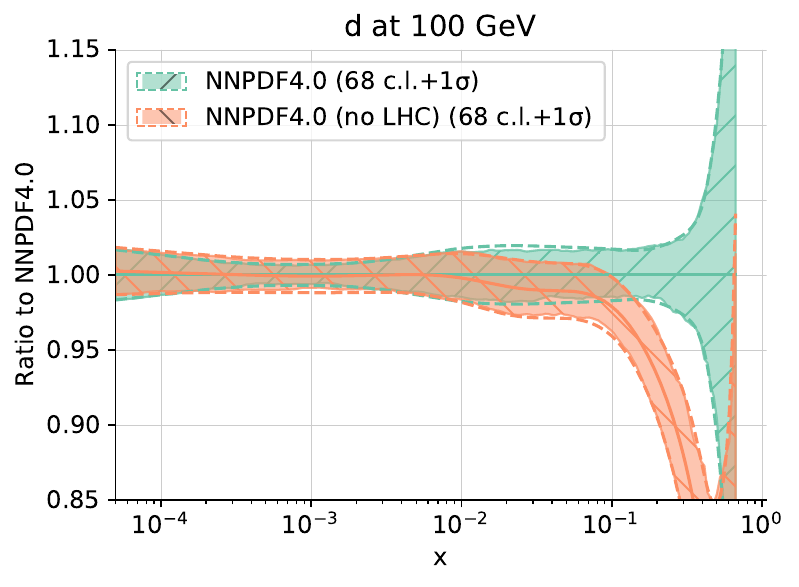}
  \includegraphics[width=0.45\textwidth]{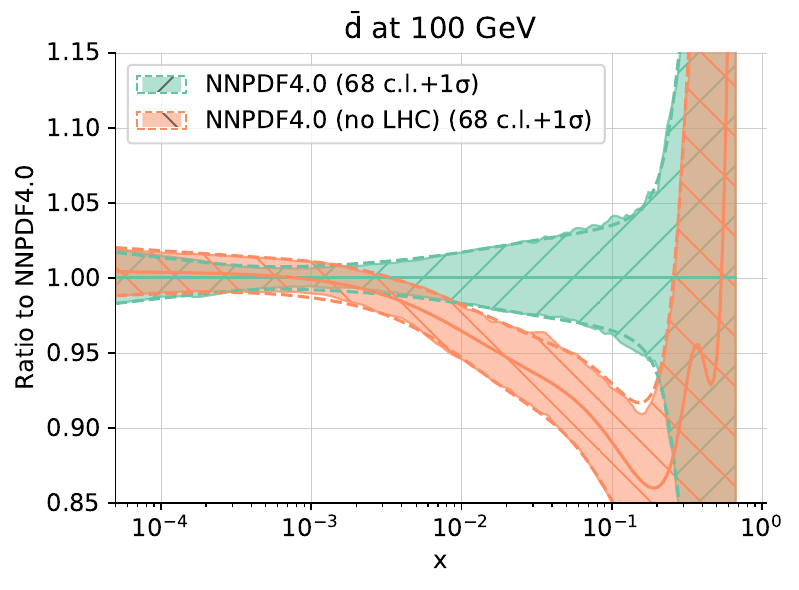}
  \includegraphics[width=0.45\textwidth]{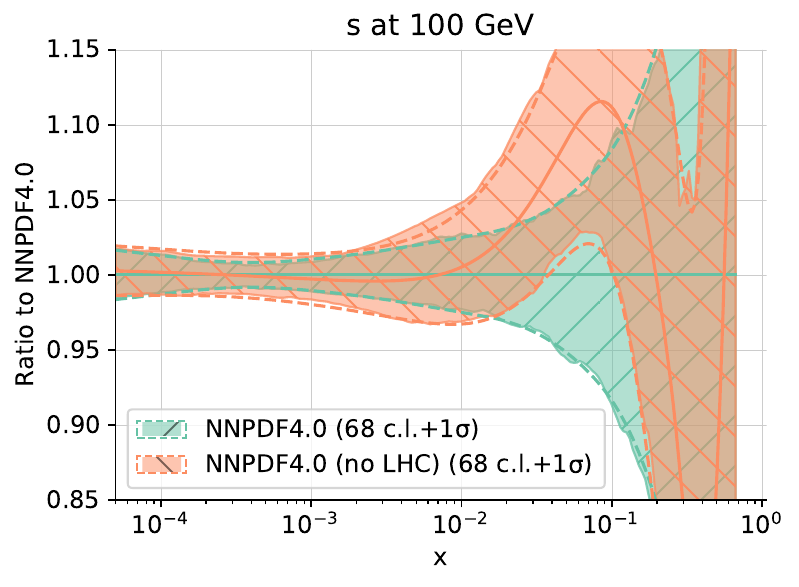}
  \includegraphics[width=0.45\textwidth]{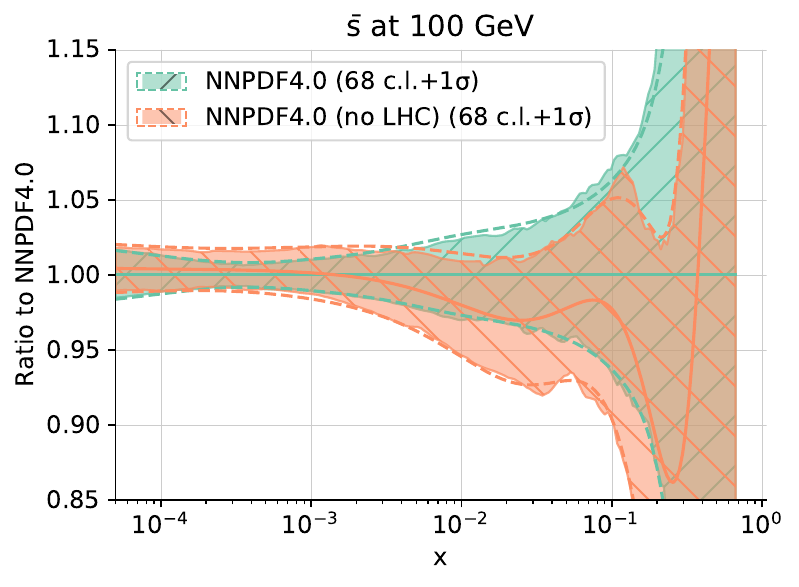}
  \includegraphics[width=0.45\textwidth]{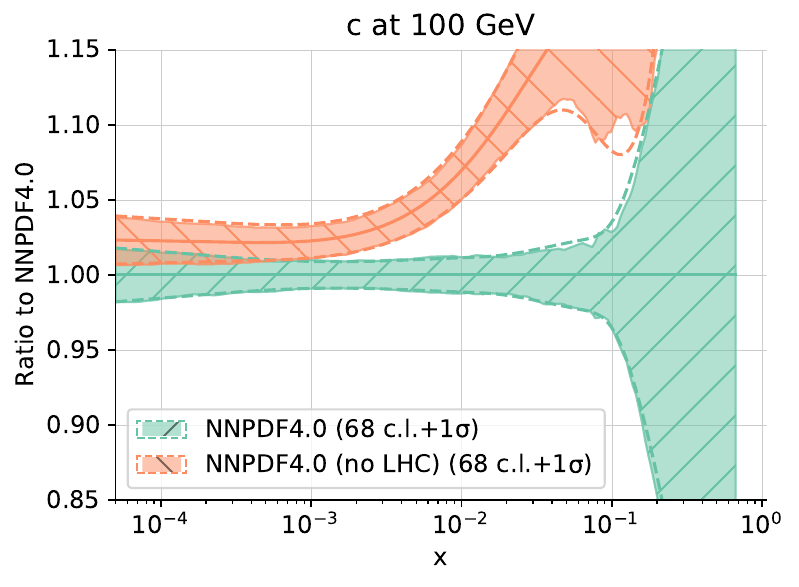}
  \includegraphics[width=0.45\textwidth]{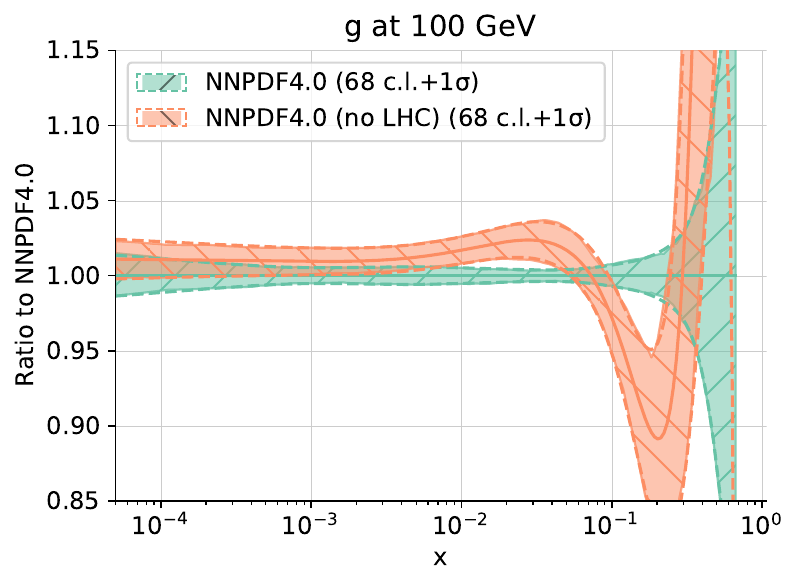}
  \caption{Same as Fig.~\ref{fig:31vs31-like} now comparing the
    baseline to PDFs determined removing from the dataset all LHC data.}
  \label{fig:noLHC}
\end{figure}

It is clear from
Sects.~\ref{subsubsec:udratio}-\ref{subsubsec:gluon_data} that the
impact of LHC data on NNPDF4.0 is non-negligible.
In order to assess their cumulative effect, we have produced a PDF set
by removing all of the LHC measurements.
Fig.~\ref{fig:noLHC} compares this PDF set to the baseline.
It is clear that the LHC data have a substantial impact, both on
central values and uncertainties:  PDF central values change  by up to
two sigma in the region
$0.01\lesssim x\lesssim 0.4$. This change is qualitatively similar to,
but rather more significant than, the change when removing LHC data
from NNPDF3.1 (see Sect.~4.10 in
Ref.~\cite{Ball:2017nwa}). The change in central value is well within
the PDF uncertainty in
the large-$x$ region, $x\gtrsim 0.4$, except for the charm PDF. We 
conclude that NNPDF4.0 PDFs are significantly more accurate than PDFs
obtained omitting  LHC data, except
at very large-$x$, where the loss of precision may be not greater
than the loss of accuracy.

It is clear  that the role of the LHC data has now
substantially changed in comparison to PDFs determined before the LHC
Run II. Indeed, for NNPDF3.0 the impact of the LHC data was still
moderate, and subdominant in comparison to that of the combined HERA
data (see in particular Sect.~5.2.2 of Ref.~\cite{Ball:2014uwa}).

\subsubsection{The impact of collider data}
\label{subsubsec:collider_data}

We have previously~\cite{Ball:2012cx,Ball:2014uwa,Ball:2017nwa} suggested that
collider-only PDFs could be more
accurate than global PDFs: retaining only collider data 
excludes low-energy datasets, which may be subject to potentially large
perturbative and non-perturbative corrections, and datasets for which
the reliability of experimental uncertainties has sometimes been questioned.
However, in the NNPDF3.1 analysis (see Sect.~4.12 in Ref.~\cite{Ball:2017nwa}) 
it was observed that in practice collider-only PDFs are not competitive due to
their very large uncertainties: the increase in uncertainty when
fitting only collider data was generally much larger than the change
in central value, thus suggesting that the loss of precision was much
greater than any possible gain in accuracy.

\begin{figure}[!t]
  \centering
  \includegraphics[width=0.45\textwidth]{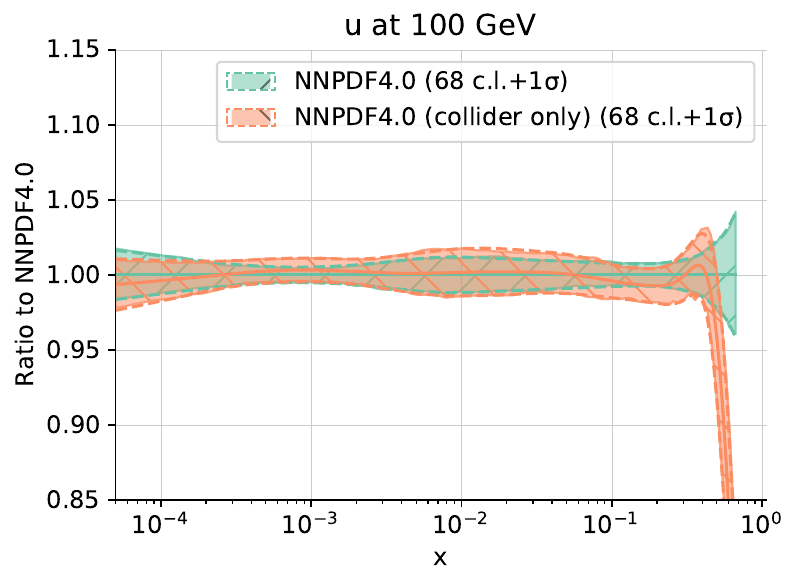}
  \includegraphics[width=0.45\textwidth]{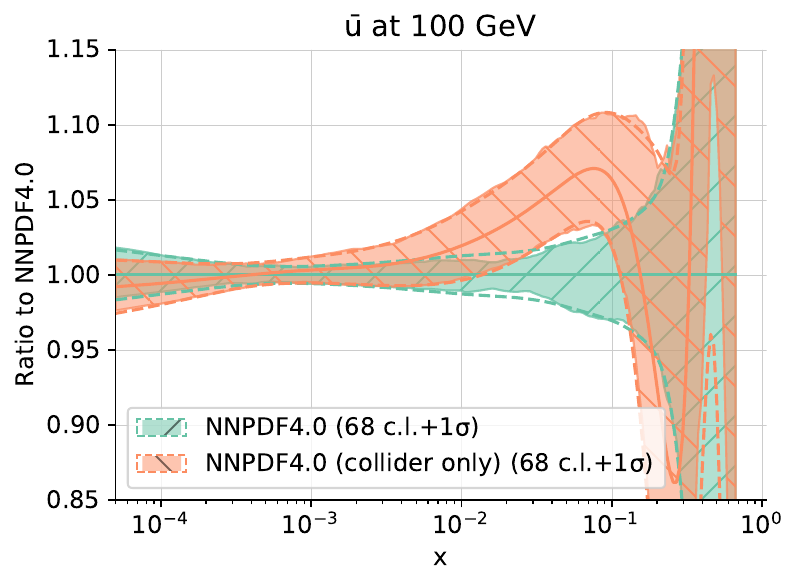}
  \includegraphics[width=0.45\textwidth]{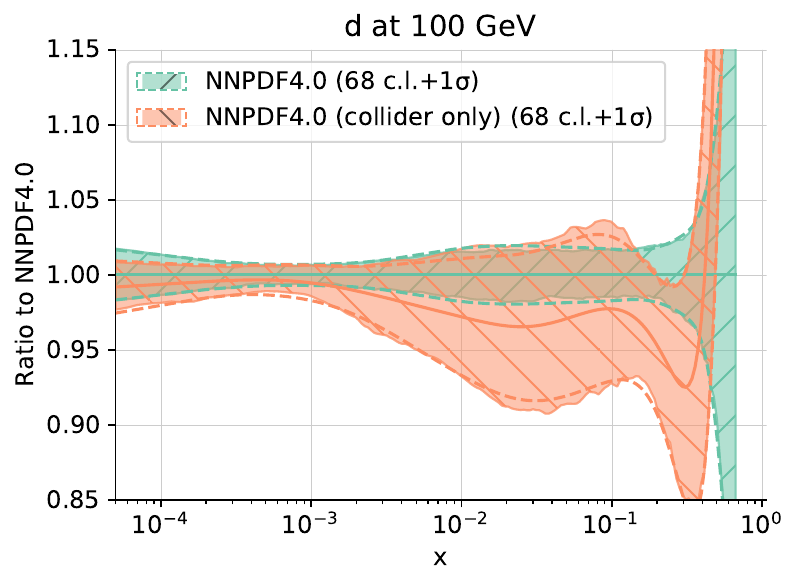}
  \includegraphics[width=0.45\textwidth]{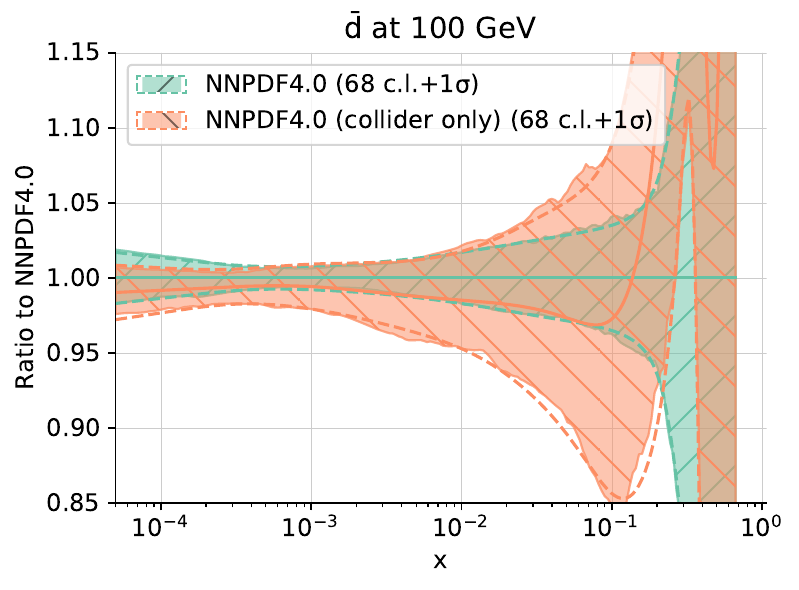}
  \includegraphics[width=0.45\textwidth]{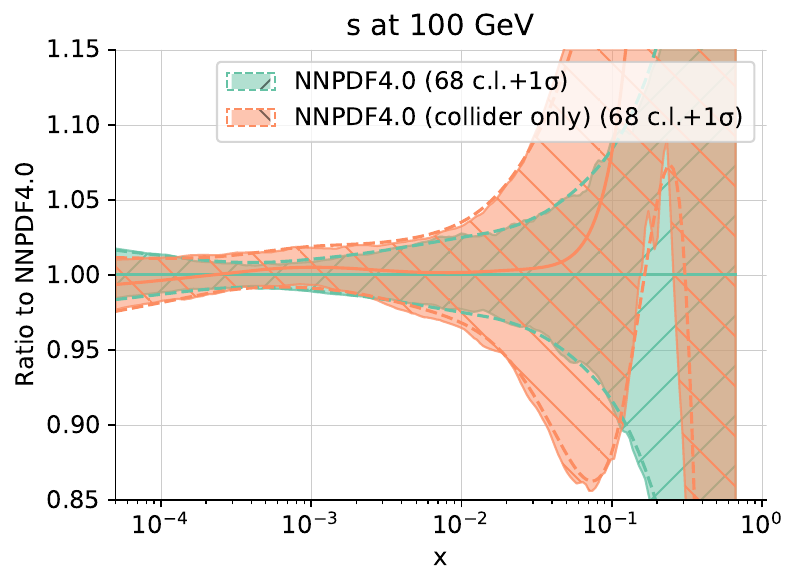}
  \includegraphics[width=0.45\textwidth]{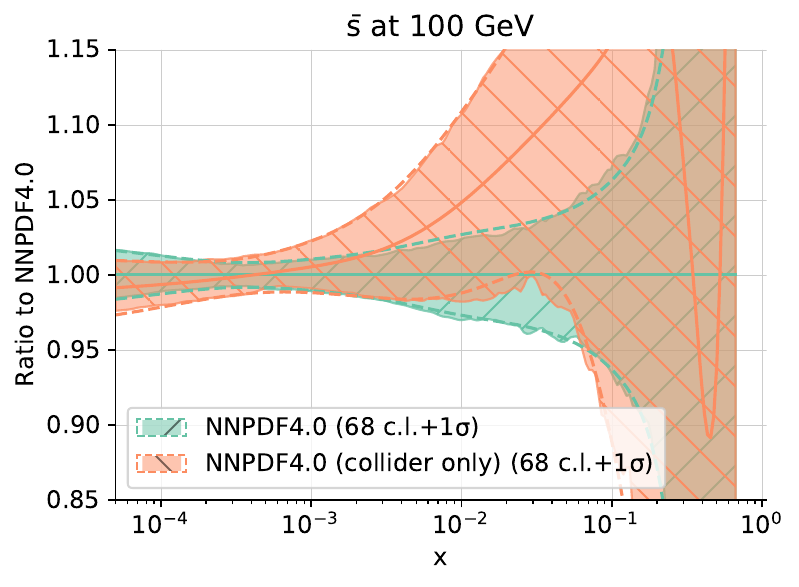}
  \includegraphics[width=0.45\textwidth]{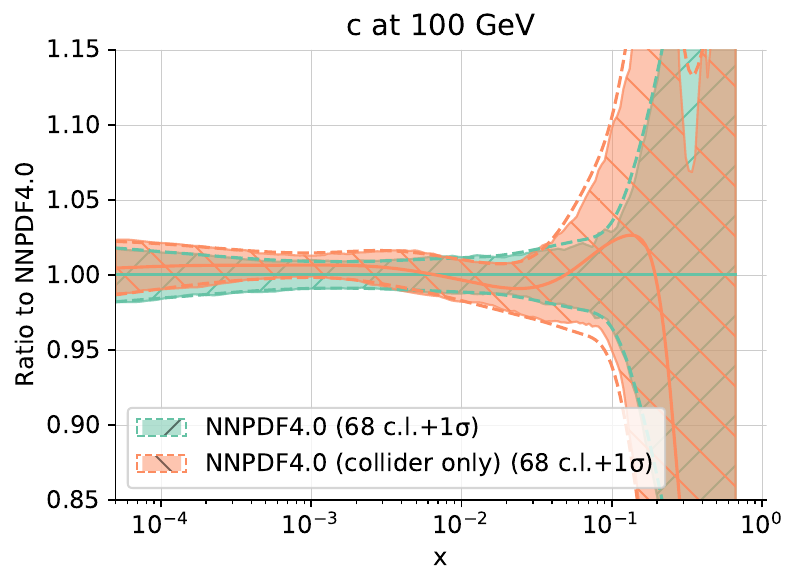}
  \includegraphics[width=0.45\textwidth]{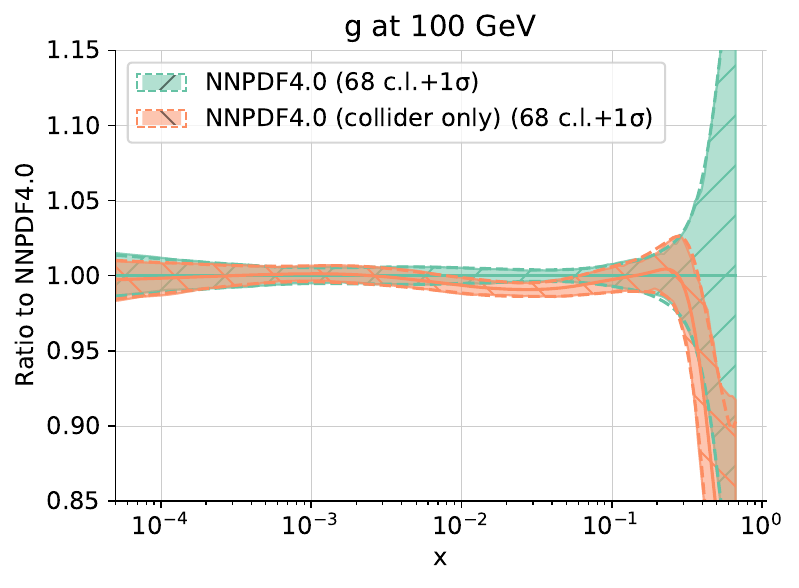}
  \caption{Same as Fig.~\ref{fig:31vs31-like} now comparing the
    baseline to PDFs determined excluding all fixed-target data from
    the dataset (collider-only PDFs).}
  \label{fig:collider_only}
\end{figure}

We revisit this state of affairs in the context of NNPDF4.0, where
the amount of LHC data has been significantly expanded. The
collider-only PDFs are compared to the baseline in Fig.~\ref{fig:collider_only}.
It is clear that now, unlike in the case of NNPDF3.1, some PDFs are
almost as precise in the collider only and global fit: this is the
case for the up, charm, and gluon. However, there is still a very
considerable loss of precision on the other PDFs at large $x$, most
likely due to the impact of neutrino data and of data with deuterium
targets on the down and strange quark and antiquark PDFs. We conclude
that even though we are approaching a situation in which collider-only
PDFs might be  competitive, we are not quite there yet.

\subsubsection{The impact of DIS data}
\label{subsubsection:DIS_data}

\begin{figure}[!t]
  \centering
  \includegraphics[width=0.45\textwidth]{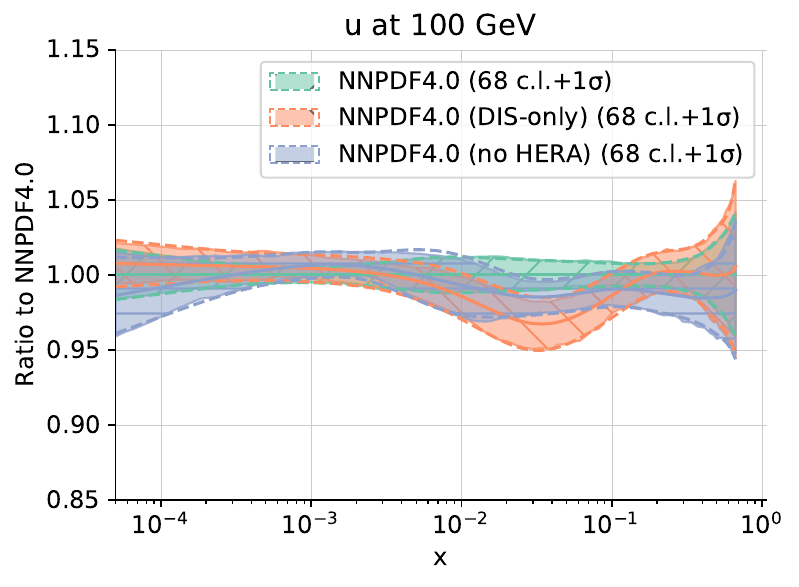}
  \includegraphics[width=0.45\textwidth]{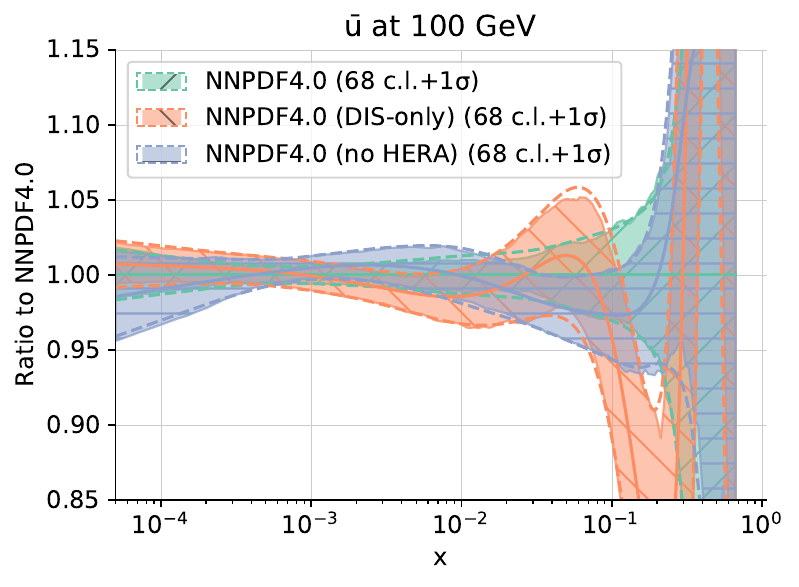}
  \includegraphics[width=0.45\textwidth]{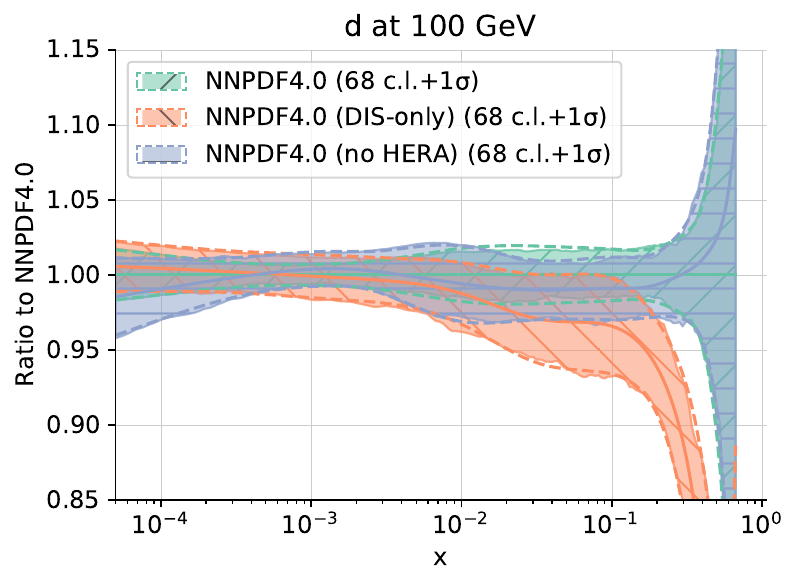}
  \includegraphics[width=0.45\textwidth]{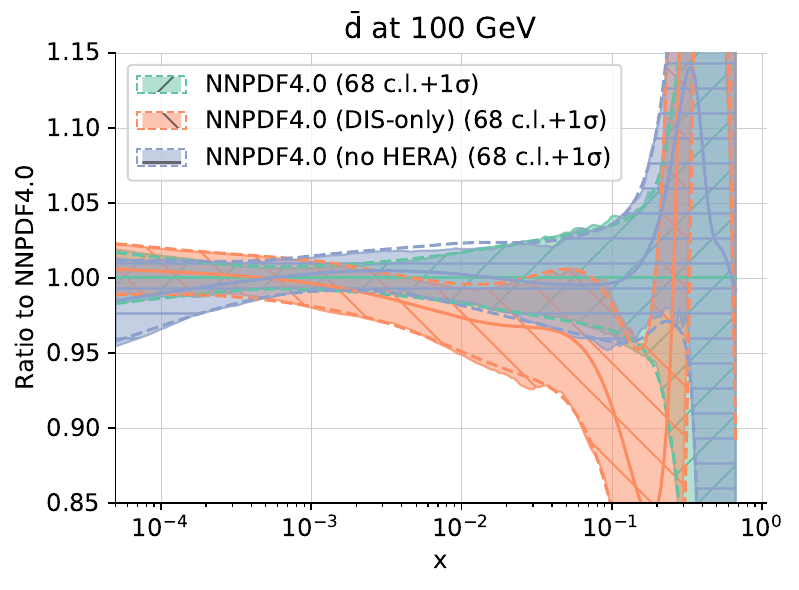}
  \includegraphics[width=0.45\textwidth]{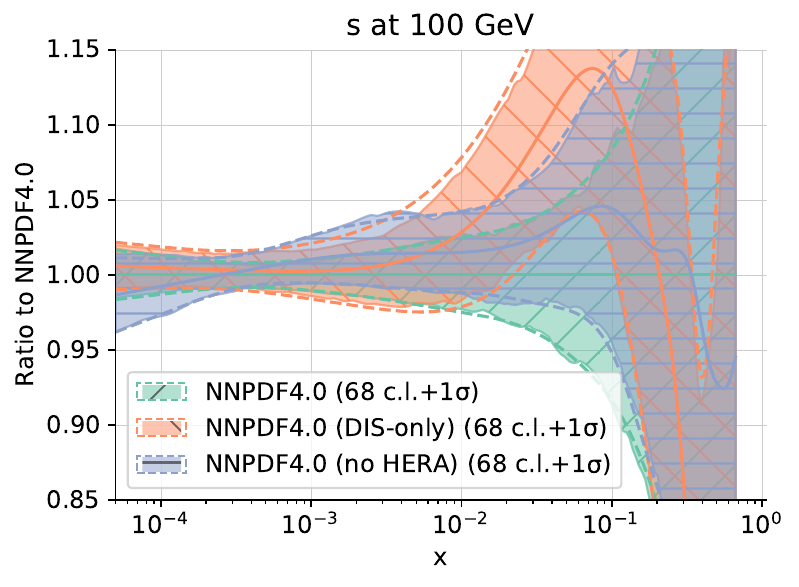}
  \includegraphics[width=0.45\textwidth]{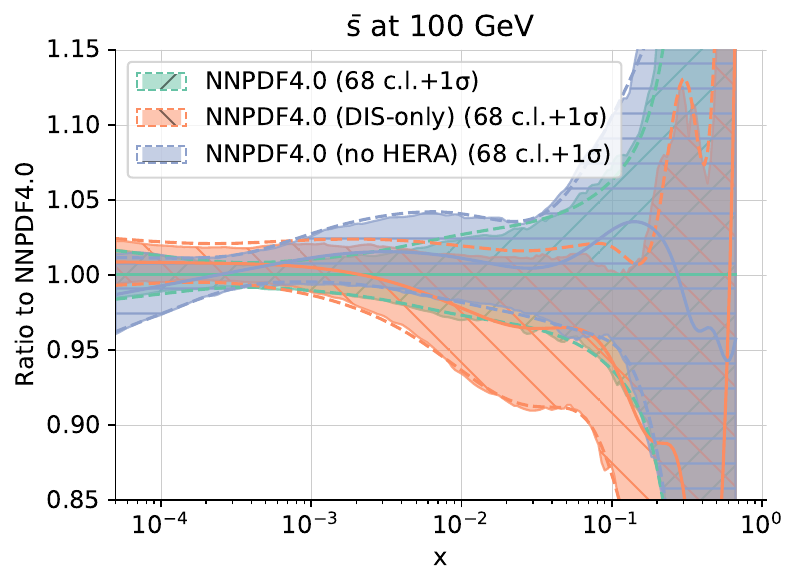}
  \includegraphics[width=0.45\textwidth]{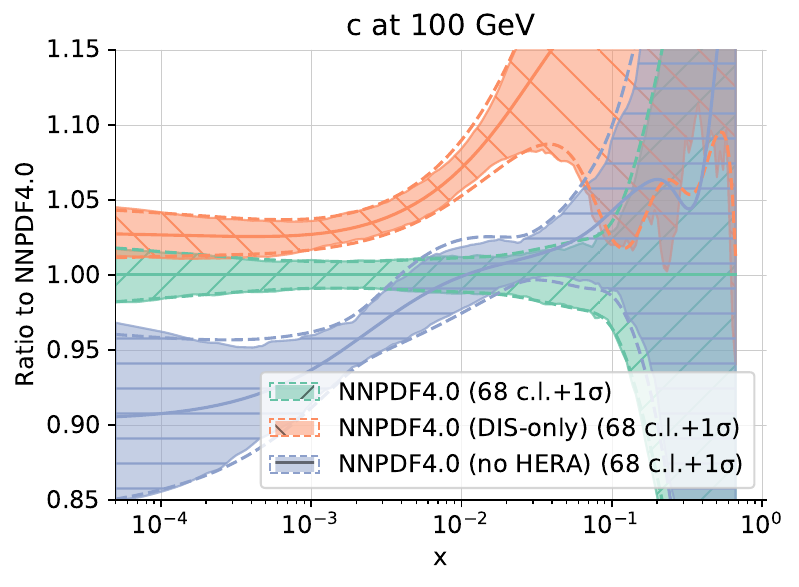}
  \includegraphics[width=0.45\textwidth]{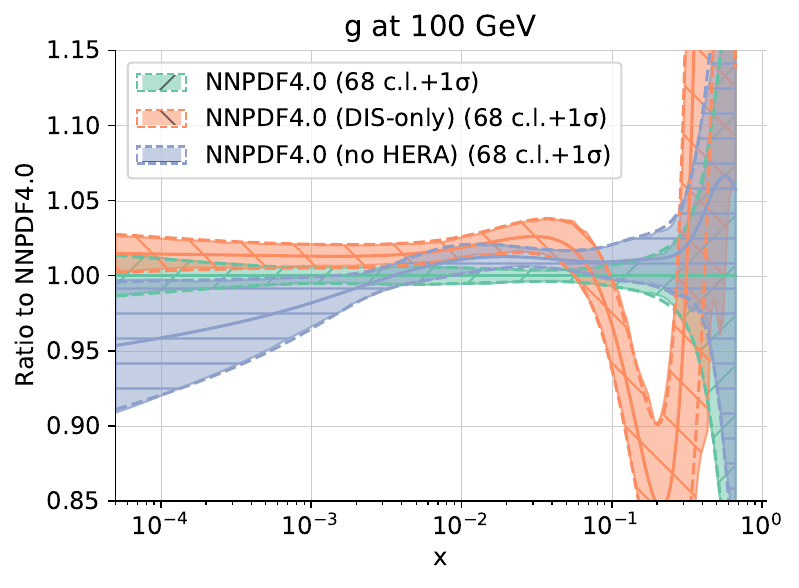}
  \caption{Same as Fig.~\ref{fig:31vs31-like} now comparing the
    baseline to PDFs determined from DIS data only, or removing all
    HERA data.}
  \label{fig:DIS}
\end{figure}

Deep-inelastic scattering measurements have provided the bulk of the
experimental information in global fits for a long time, and DIS-only
PDFs have been widely used as a possibly more accurate and only
marginally less precise alternative to global fits.
As with collider-only PDFs, the situation is now worth revisiting.
To this purpose, we have produced a PDF determination in which only DIS
data are retained; and one in which all the HERA data are removed from
the dataset. They are compared to the baseline in Fig~\ref{fig:DIS}.

Comparing the DIS-only PDFs to the baseline, large
differences are seen, for both central values and
uncertainties. It is only in the small $x$ region, where quark PDFs
are controlled by the mixing of the dominant singlet component with
the gluon, that there is good agreement between  DIS-only and global
PDFs. The only PDFs which remain essentially unchanged are 
the strange quark and antiquark.  This confirms the key role played by
neutrino DIS (dimuon) data in  constraining them.

Interestingly however, the no-HERA PDFs are in perfect agreement with
the baseline, with only a moderate increase in uncertainty,
with the exception of charm. This means that whereas the small-$x$
behavior of the gluon and singlet determined from HERA is in
agreement with that coming from the LHC data, the HERA data are no
longer required in order to determine the correct behavior of 
PDFs at small $x$.
An exception is charm, which at small $x$ is constrained by the
combined HERA $\sigma_{\rm NC}^c$ data. As mentioned in
Sect.~\ref{subsubsection:appraisal_and_selection} this is the reason
why this data is retained in the baseline, despite its poor fit
quality, which is possibly due to missing higher order corrections.

We conclude that on the one hand, unlike in previous NNPDF
determinations, for NNPDF4.0 it is no longer true that a DIS-only
fit is competitive, and on the other hand the HERA data are no longer
needed in order to fix the small $x$ behavior of PDFs (with the
exception of charm). This is consistent with our previous conclusion
in Sect.~\ref{subsubsec:LHC_data} that the NNPDF4.0 PDF determination is
largely controlled by LHC data.

\subsection{PDFs from extended datasets}

We now discuss a number of PDF sets determined by adding specific
measurements to the baseline. We consider in turn:
the ATLAS 8~TeV $W^\pm$ lepton rapidity distributions~\cite{Aad:2019rou};
the EMC charm structure function data~\cite{Aubert:1982tt}; the
7~TeV ATLAS and CMS single-inclusive jet
data~\cite{Aad:2014vwa,Chatrchyan:2014gia} (in lieu of dijets);
the NOMAD neutrino dimuon data~\cite{Samoylov:2013xoa}; and the HERA
single-inclusive and dijet
data~\cite{ZEUS:2002nms,ZEUS:2006xvn,H1:2016goa,H1:2014cbm}.
In the last two cases, the impact of the additional measurements is
studied by means of Bayesian reweighting~\cite{Ball:2010gb,Ball:2011gg},
for the reasons explained in Sect.~\ref{sec:datatheory}, starting from a
prior PDF ensemble of 1000 replicas.

\subsubsection{The ATLAS 8 TeV $W^\pm$ data}
\label{subsubsec:ATLASWpm8TeV}

\begin{figure}[!t]
  \centering
  \includegraphics[width=0.49\textwidth]{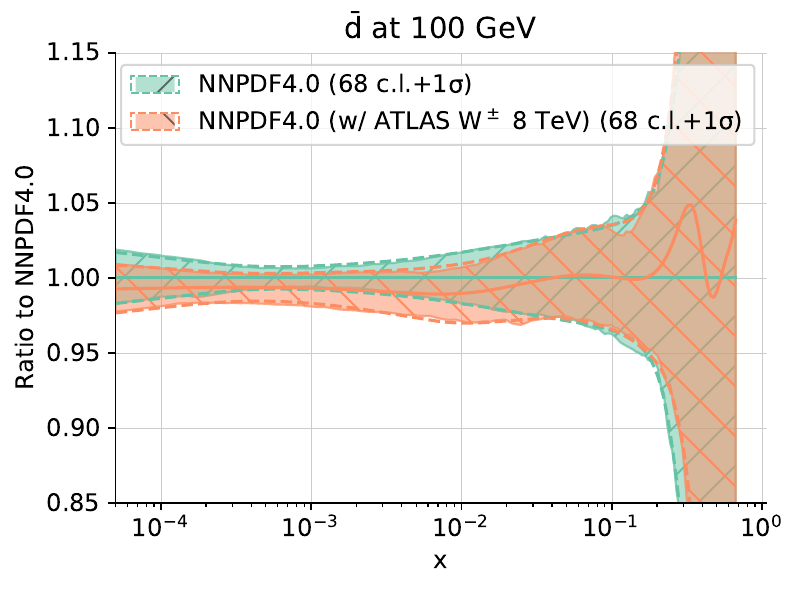}
  \includegraphics[width=0.49\textwidth]{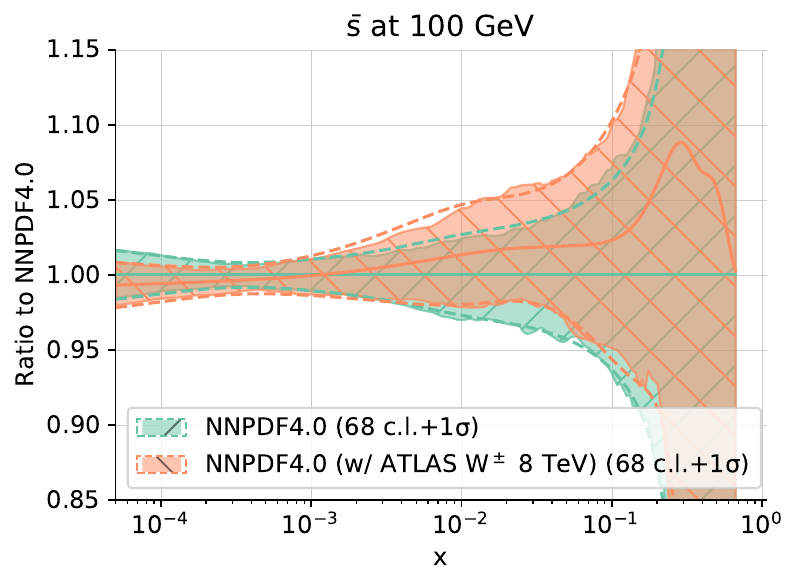}
  \caption{Comparison to the baseline of the antidown and antistrange  PDFs obtained adding to the baseline
 the ATLAS lepton
    rapidity distributions from $W^\pm$ production at 8~TeV~\cite{Aad:2019rou}.}
  \label{fig:ATLASW}
\end{figure}

As discussed in Sect.~\ref{sec:dataselection}, the ATLAS measurement of the
8~TeV lepton rapidity differential cross-section for $W^\pm$
production~\cite{Aad:2019rou} is not included in the
baseline dataset because it does not pass our selection criteria.
Nevertheless we study its impact by performing a fit in which it is
added to the NNPDF4.0 baseline dataset.  It turns out that the impact
on PDFs of these data is tiny. The down and strange antiquarks are the
most affected: their central values are respectively suppressed and enhanced 
by half a sigma in the region $0.01\lesssim x\lesssim 0.1$. The PDFs, 
normalized to the
central value of the NNPDF4.0 baseline, are displayed at $Q=100$~GeV
in Fig.~\ref{fig:ATLASW}. We conclude that this dataset is in fact
consistent with the baseline, and its pathological behavior upon being
given a large weight is likely related to its poorly behaved
covariance matrix. This will be shown to be indeed the case in
Sect.~\ref{sec:regcovmat} below. A poor fit quality to this dataset
was also found in the MSHT20 analysis~\cite{Bailey:2020ooq}.

\subsubsection{The EMC charm structure function data}
\label{subsubsec:EMC_F2c}

In previous NNPDF studies~\cite{Ball:2016neh,Ball:2017nwa}, it was
found that EMC charm structure function data~\cite{Aubert:1982tt}
significantly reduce the uncertainty on the  charm PDF at large $x$,
which in this region, upon inclusion of this data, deviates significantly 
from the result (compatible with zero) of  perturbative matching,
and exhibits a behavior similar to models of
intrinsic charm~\cite{Brodsky:2015fna}. These data
however have not been included in the baseline because the reliability
of the EMC estimate of systematic uncertainties has been questioned,
even though not for this specific measurement (see
Refs.~\cite{Ball:2016neh,Rottoli:2016lsg} for details).
We revisit this issue here by adding  the EMC data to the baseline dataset.
Furthermore, nuclear uncertainties related to the use of a Fe target are now
taken into account following
the procedure explained in Sect.~\ref{subsec:nuclear}.

\begin{figure}[!t]
  \centering
  \includegraphics[width=0.49\textwidth]{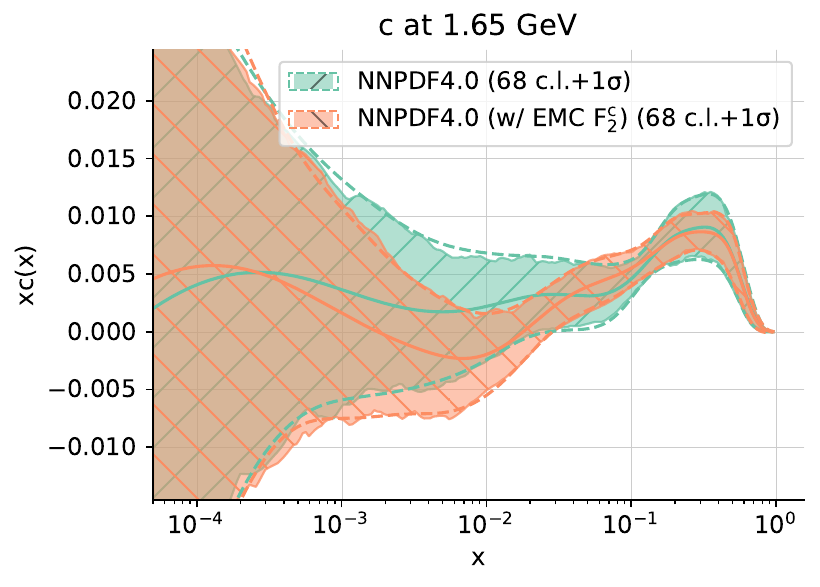}
  \includegraphics[width=0.49\textwidth]{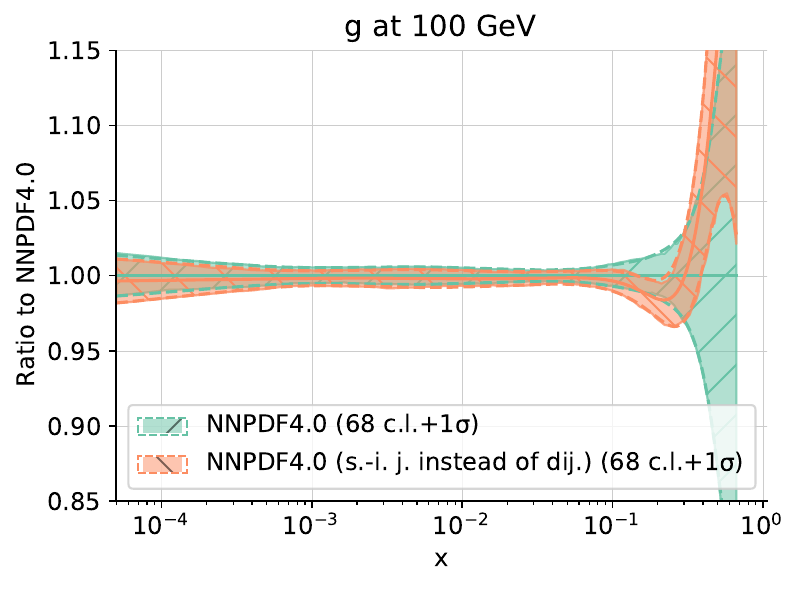}\\
  \caption{(Left) Comparison to the baseline of the charm PDF at
    $Q=1.65$~GeV from a determination in which the EMC
    charm structure function data ~\cite{Aubert:1982tt} are included.
    (Right) The gluon PDF at $Q=100$~GeV compared and normalized the
    baseline from a determination replacing   7~TeV  ATLAS and CMS
    dijet data with single-inclusive jets.}
  \label{fig:EMC_and_jets}
\end{figure}

A good fit quality is obtained overall
and specifically for the EMC measurement, with a value of the $\chi^2$ of 0.62.
The charm PDFs for this
determination is compared to the baseline in
Fig.~\ref{fig:EMC_and_jets} (left)
at $Q=1.65$~GeV, just  above
the charm threshold. Remarkably, the inclusion of this data leaves the
central charm PDF unchanged: there is perfect consistency between the
EMC data
and the global dataset. Thanks to this consistency,
a  reduction of the charm PDF uncertainty is found
around $x\sim 0.03$ and $x\sim 0.3$, by a moderate amount.
A much more significant uncertainty reduction upon the inclusion of the
EMC data was observed in Ref.~\cite{Ball:2017nwa} (see Sect.~4.9).
This means that the extension of the dataset from NNPDF3.1 to NNPDF4.0
leads to a charm PDF whose
uncertainty is greatly reduced, and whose central value is in perfect
agreement with that determined by the EMC data.

These findings suggests that the NNPDF4.0 analysis favors a non-zero intrinsic
charm component in the proton. A more quantitative assessment of this statement
requires however a determination of the
PDFs in the $n_f=3$ scheme, which is left to future studies~\cite{ICpaper}.

\subsubsection{ATLAS and CMS single-inclusive jet data}
\label{subsubsection:singleinclusivejets_7TeV}

In Sect.~\ref{subsec:doublecounting} as a part of dataset selection we
had to choose between single-inclusive jets and dijets, given that the
lack of information on their correlation prevents their simultaneous
inclusion. Whereas we concluded that 8~TeV CMS dijet data has
potential issues and thus decided in favor of the inclusion of
single-inclusive jets, for 7~TeV data we concluded that the
single-inclusive jets and dijets are consistent and we decided for
the inclusion of dijets due to the fact that the dijet observable is
favored theoretically~\cite{Currie:2018xkj,AbdulKhalek:2020jut}.

We now consider a variant of the baseline in which the  7~TeV dijet
data are replaced by 
single-inclusive jets. In the case of the ATLAS data, we decorrelate
systematic uncertainties across different rapidity bins according to
the procedure recommended in~\cite{Aaboud:2017dvo}. Results remain
unchanged if we include any of the individual rapidity bins, as we had
already observed in the context of NNPDF3.1~\cite{Nocera:2017zge}. The fit
quality is as good as the baseline, with statistically equivalent
PDFs. The gluon from this set is compared to the baseline in
Fig.~\ref{fig:EMC_and_jets} (right). We observe a mild distortion of
the large-$x$ shape: a slight suppression around $x\simeq 0.3$
followed by an enhancement at larger $x$, well within the PDF
uncertainty. We thus confirm compatibility between jets and dijets at 7~TeV.

\subsubsection{The NOMAD neutrino dimuon data}
\label{subsubsec:NOMAD}

As discussed in Sect.~\ref{subsubsection:DIS_data}, the strange quark PDF is
mostly constrained by the neutrino-DIS charm dimuon data from 
NuTeV. LHC data, namely $W$ and $Z$ boson production, possibly in
association with jets, provide additional, consistent constraints.
In Ref.~\cite{Faura:2020oom}, the NOMAD measurement~\cite{Samoylov:2013xoa}
of the dimuon to inclusive neutrino-nucleus CC DIS cross-section ratio,
$\mathcal{R}_{\mu\mu}$, was shown to further pin down the uncertainty of the
strange quark PDF.

\begin{figure}[!t]
  \centering
  \includegraphics[width=0.49\textwidth]{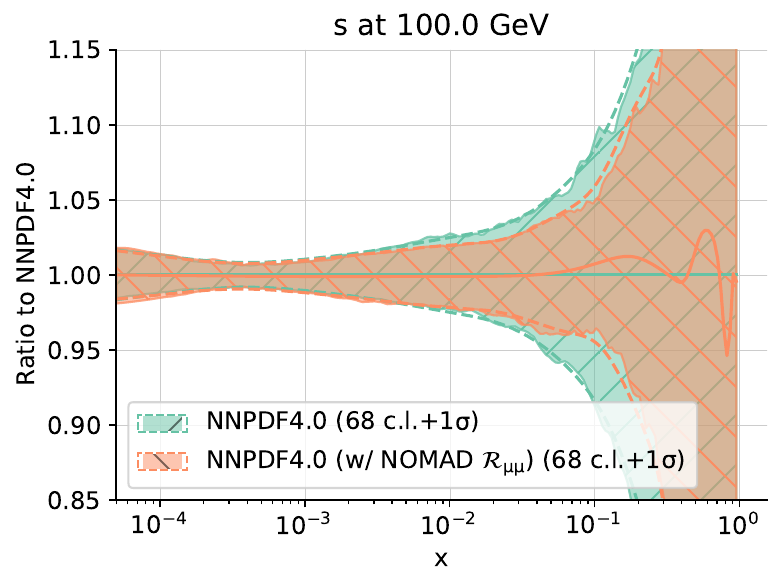}
  \includegraphics[width=0.49\textwidth]{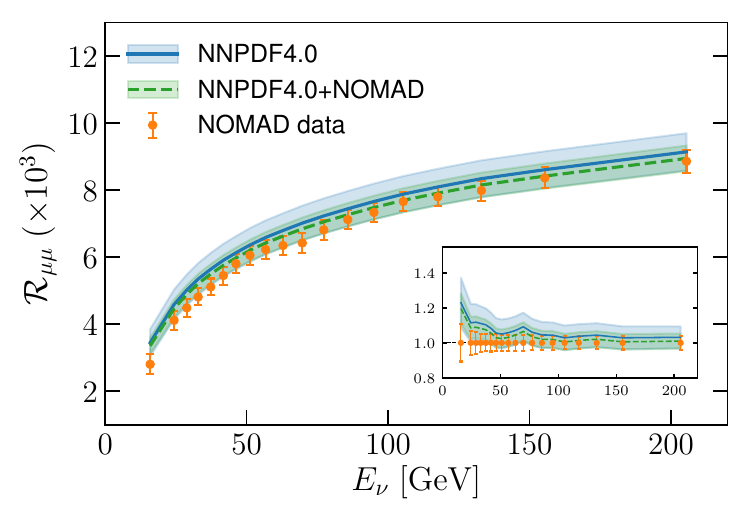}
  \caption{(Left) Comparison between the baseline and PDFs in which
    the  NOMAD neutrino DIS data are included
    by reweighting. The strange PDF is shown at $Q=100$~GeV.
    (Right) The same comparison for the measured ratio $\mathcal{R}_{\mu\mu}$
    as a function of the neutrino energy $E_\nu$.
    The inset displays quantities normalized to the central experimental value.
  \label{fig:NOMAD}}
\end{figure}

Here we  assess whether or not the same conclusion holds within the reduced 
uncertainties of the NNPDF4.0 determination. To this purpose, we repeat the
reweighting analysis of Ref.~\cite{Faura:2020oom}, but now starting from the
NNPDF4.0 baseline as a prior. No nuclear corrections are taken into account,
despite the fact that the NOMAD experiment utilized a Fe target, as nuclear
corrections cancel in the cross-section ratio measured by this experiment
(see Ref.~\cite{Faura:2020oom}).
We find that the NOMAD data are very well described by the NNPDF4.0
prior before reweighting: the $\chi^2$ per data point is equal to 0.66.
The impact of the data is therefore expected to be limited.
After reweighting, the $\chi^2$ improves to
0.61. The number of effective replicas is $N_{\rm eff}=622$, out of
$N_{\rm rep}=1000$ in the prior set. The strange quark PDF, the only one
to  to be affected, is displayed before and after reweighting in
Fig.~\ref{fig:NOMAD} (left). It is clear that the NOMAD data leave unchanged the
central value and only contribute to a moderate uncertainty reduction in the
region around $x\sim 0.1$. Similar conclusions can be drawn from the
comparison of the ratio
$\mathcal{R}_{\mu\mu}$  as a function of the neutrino energy
$E_\nu$, also shown in Fig.~\ref{fig:NOMAD} (right).

\subsubsection{The HERA DIS jet data}
\label{subsubsec:HERAjets}

Additional constraints on the gluon are provided  by deep-inelastic jet production. We study the impact of the selection of available
measurements performed by ZEUS and H1 discussed in
Sect.~\ref{subsubsection:DISjet}, by means of Bayesian reweighting,
for the reasons discussed there. All the datasets are included at once
in the reweighting; results are given in 
Table~\ref{tab:chi2DISjet}, where for each dataset we give the number
of data points and the $\chi^2$ value before and after reweighting,
along with  the
total $\chi^2$ values for the full DIS jet dataset.
Experimental correlations between single-inclusive jet and dijet production
measurements are taken into account whenever provided (specifically for
Refs.~\cite{H1:2016goa,H1:2014cbm}). However, because  DIS jet data
are included via reweighting, their correlations with the inclusive
DIS data used in the baseline fit cannot be included. This is a
partial limitation of the reweighting analysis. 

\begin{figure}[!t]
  \centering
  \includegraphics[width=0.49\textwidth]{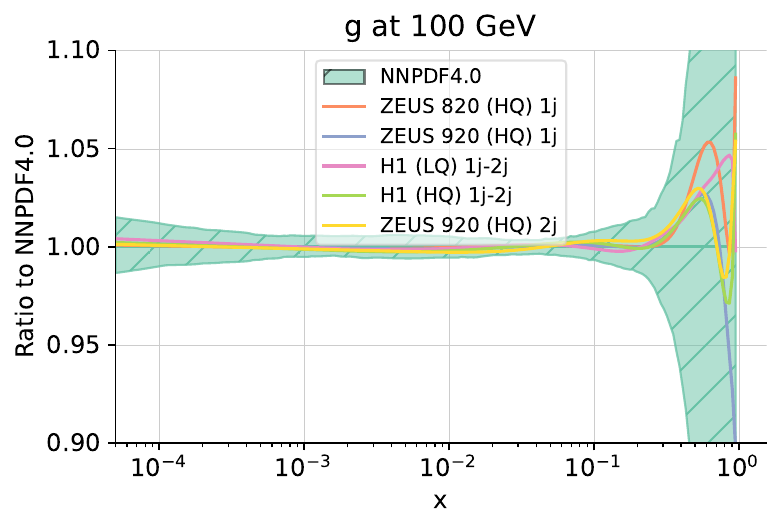}
  \includegraphics[width=0.49\textwidth]{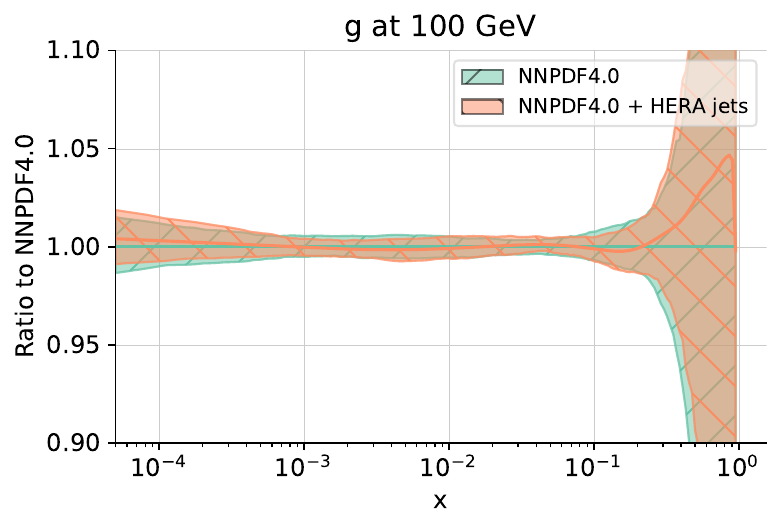}
  \caption{The gluon PDF obtained reweighting the NNPDF4.0
    baseline with DIS jet data, shown as a ratio to the former at
    $Q=100$~GeV. We show the central gluon obtained when reweighting
    with each of the DIS jet data of   Table~\ref{tab:chi2DISjet} in
    turn (left), and the central gluon and uncertainty obtained when
    reweighting with the full DIS jet dataset considered here (right).}
  \label{fig:DISjets}
\end{figure}

The number of effective replicas after reweighting is
$N_{\rm eff}=530$, out of $N_{\rm rep}=1000$ in the prior set.
In Fig.~\ref{fig:DISjets} we compare the reweighted
gluon PDF to the baseline 
NNPDF4.0 result, shown as a ratio to the latter at $Q=100$~GeV.
We show both the central gluon obtained when
reweighting  with each of the datasets listed in
Table~\ref{tab:chi2DISjet} (left) and the central value and
uncertainty obtained when reweighting with the full set of DIS jet
data (right).
Single-inclusive jet and dijet measurements from H1 (separately for
low-$Q$ and high-$Q$) are considered as a single dataset, given that
experimental correlations are completely known.
\begin{table}[!t]
  \scriptsize
  \centering
  \renewcommand{\arraystretch}{1.4}
  \input{tables/tab-chi2_disjet.tex}
  \caption{The number of data points $N_{\rm dat}$ and the 
    $\chi^2$ value before and after reweighting
    the NNPDF4.0 baseline PDF set with the full set of DIS jet  data
    (see Sect.~\ref{subsubsection:DISjet} for
    details). The total $\chi^2$ values are also shown.}
  \label{tab:chi2DISjet}
\end{table}
  
It is clear that the impact of the DIS jet data is very
moderate. Indeed, the fit quality of this data is already quite good
before their inclusion and does not change substantially: this is also
apparent from the small reduction of the effective number of replicas
upon reweighting. We conclude that this data is consistent with the
baseline and, if fully included in the baseline dataset would not affect
significantly the outcome of the PDF determination.

%% file: tables/tab-chi2_disjet.tex
\begin{tabularx}{\textwidth}{Xllllllll}
  & \rotatebox{70}{ZEUS 820 (HQ) (1j)}
  & \rotatebox{70}{ZEUS 920 (HQ) (1j)}
  & \rotatebox{70}{H1 (LQ) (1j)}
  & \rotatebox{70}{H1 (HQ) (1j)}
  & \rotatebox{70}{ZEUS 920 (HQ) (2j)}
  & \rotatebox{70}{H1 (LQ) (2j)}
  & \rotatebox{70}{H1 (HQ) (2j)}
  & \rotatebox{70}{Total DIS+jets}\\
  \toprule
  $N_{\rm dat}$
  & 30 & 30 & 48 & 24 & 22 & 48 & 24 & 226\\
  \midrule
  $\chi^2$ (NNPDF4.0, before reweighting)
  & 0.96 & 1.75 & 1.86 & 1.78 & 1.82 & 1.62 & 1.98 & 1.80 \\
  $\chi^2$ (NNPDF4.0, after reweighting)
  & 0.96 & 1.45 & 1.59 & 1.62 & 1.67 & 1.53 & 1.65 & 1.68 \\
  \bottomrule
\end{tabularx}

%% file: sec-tests.tex
\section{Methodology dependence and stability}
\label{sec:tests}

After assessing, in the previous section, the impact of the new data
on NNPDF4.0, we now turn to the corresponding assessment of the impact of
the new methodology. This has the dual aim of, on the one hand, complementing
the analysis of the previous section and providing a full
understaning of the differences between NNPDF4.0 and previous PDF
sets, specifically NNPDF3.1, and on the other hand, providing
detailed tests of the stability and robustness of our results.

We first assess the impact of the new NNPDF4.0
methodology, by comparing PDF sets based on the same underlying dataset,
but using either the new NNPDF4.0 or the previous NNPDF3.1 methodology.
We then study specifically the impact of the new positivity and integrability
constraints, respectively discussed in Sect.~\ref{sec:positivity} and
Sect.~\ref{sec:pdfint}. Next,  we then turn to the explicit demonstration of the
independence of results on the choice of parametrization basis of
Sect.~\ref{sec:flavev},  we discuss the impact of independently parametrizing
the charm PDF (which is the NNPDF default since NNPDF3.1), and we study the
impact of the new implementation of nuclear corrections presented in
Sect.~\ref{subsec:nuclear}. Finally, we study the possibility of regularizing
the covariance matrix for datasets for which it is poorly conditioned, and use
the result to reassess the impact of some of the problematic datasets considered in
Sect.~\ref{subsubsection:appraisal_and_selection}.

\input{subsec-newmethodology.tex}
\input{subsec-integrability.tex}
\input{subsec-flavourbasis.tex}

\input{subsec-charmfit.tex}
\input{subsec-nuclearimpact}

\input{subsec-regcovmat}

%% file: subsec-newmethodology.tex
\subsection{Impact of the NNPDF4.0  methodology}
\label{sec:comparison_31_methodology}

\begin{figure}[!t]
  \centering
  \includegraphics[width=0.45\textwidth]{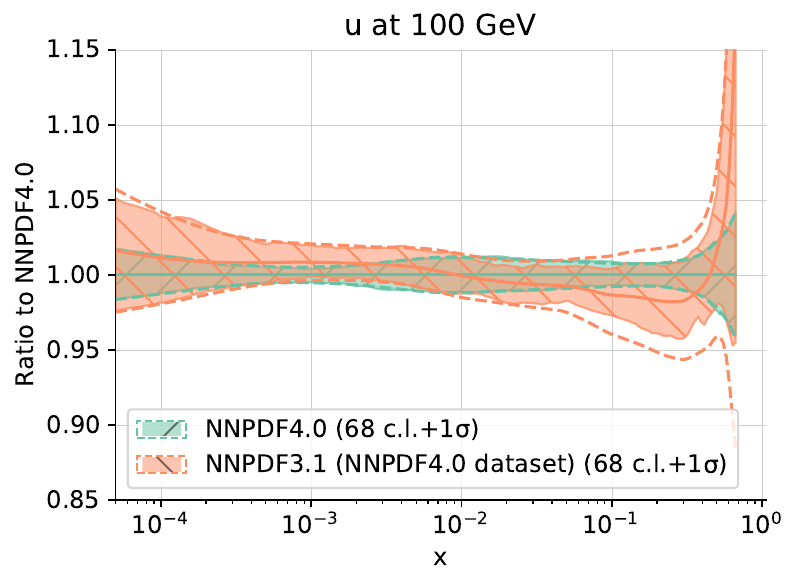}
  \includegraphics[width=0.45\textwidth]{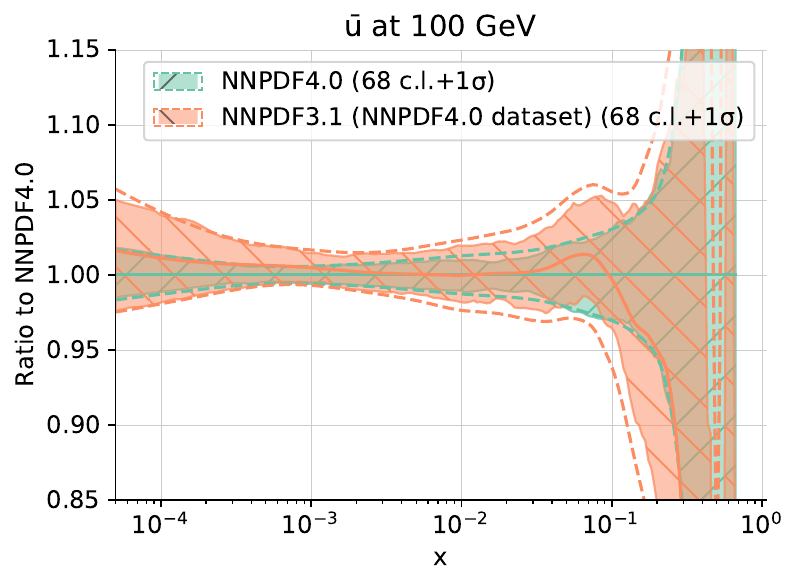}
  \includegraphics[width=0.45\textwidth]{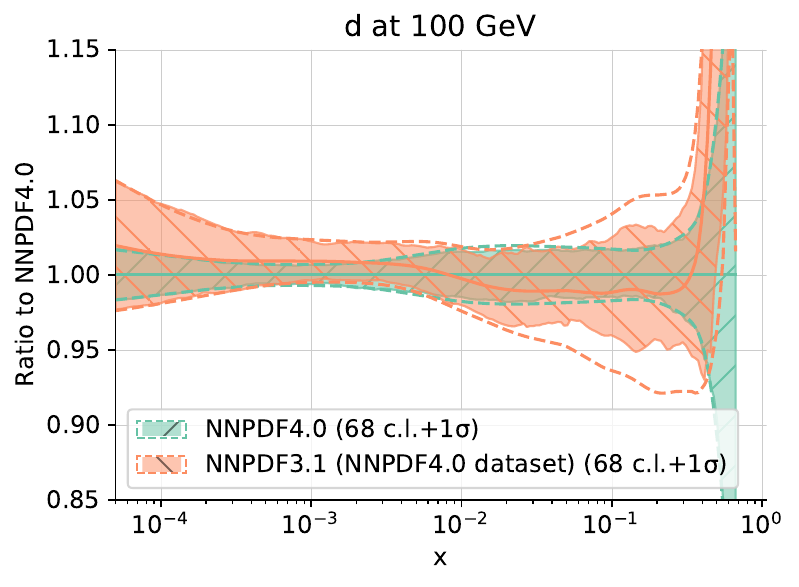}
  \includegraphics[width=0.45\textwidth]{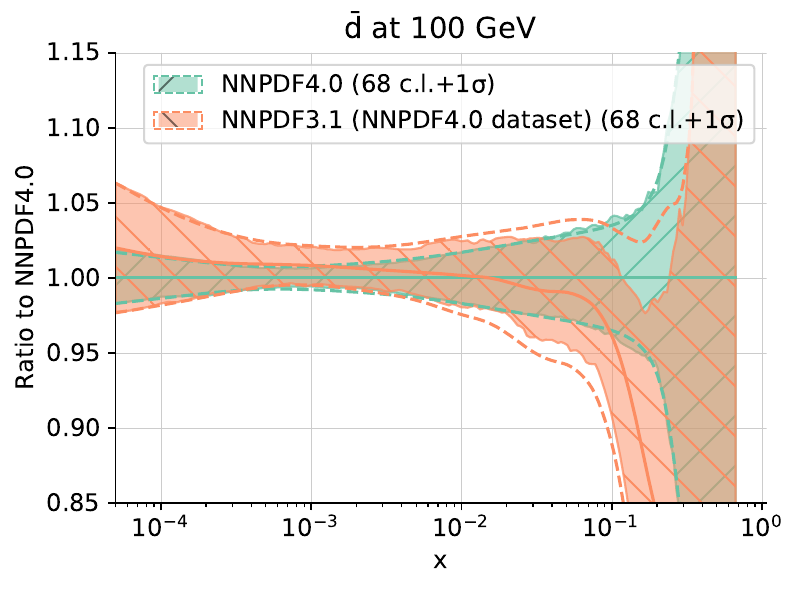}
  \includegraphics[width=0.45\textwidth]{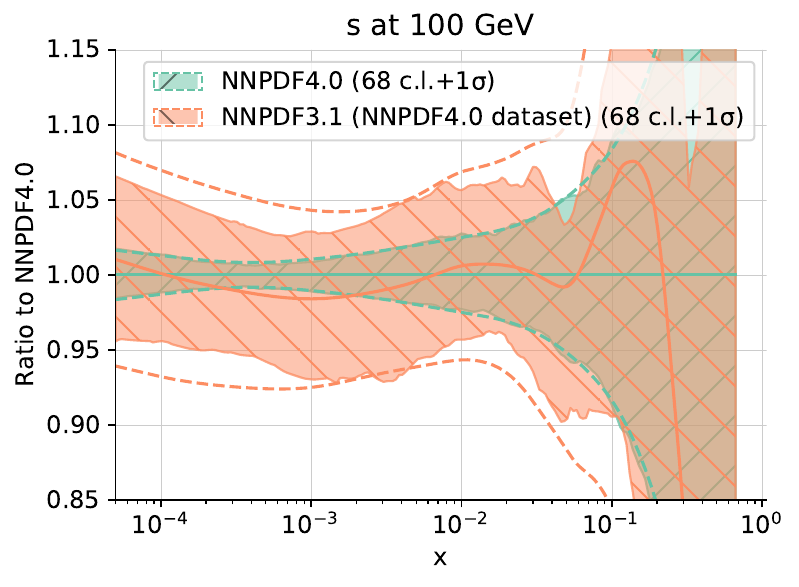}
  \includegraphics[width=0.45\textwidth]{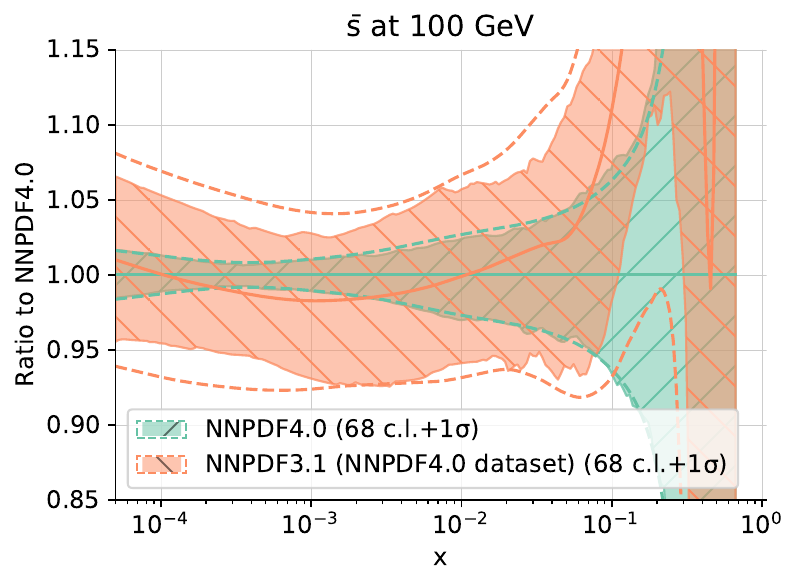}
  \includegraphics[width=0.45\textwidth]{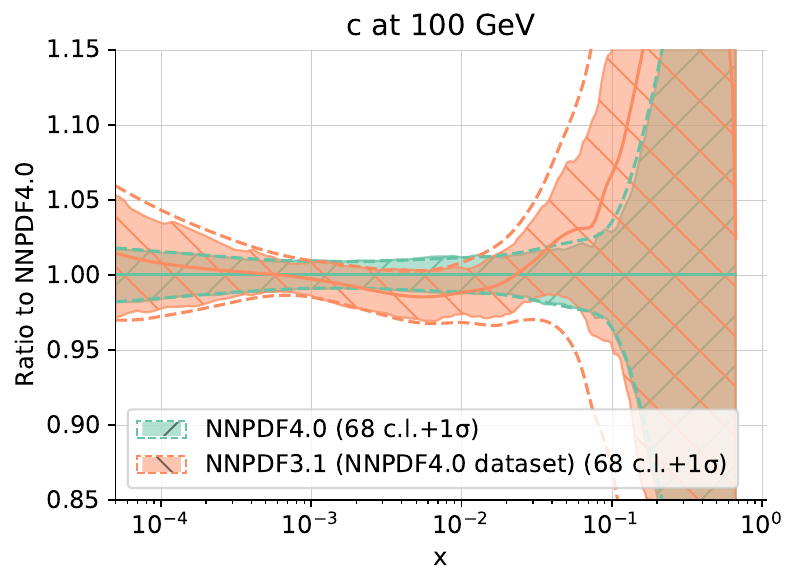}
  \includegraphics[width=0.45\textwidth]{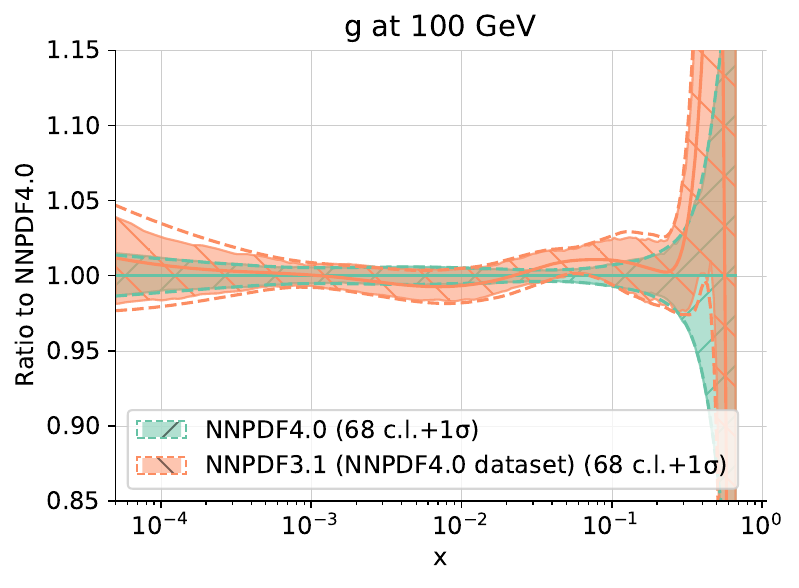}
  \caption{ Same as Fig.~\ref{fig:40vs31-like}  but now presenting the
    complementary comparison of the baseline of PDFs to a set based on the 
    same NNPDF4.0 dataset, but using the old NNPDF3.1 methodology.}
  \label{fig:40vs31_meth}
\end{figure}

\begin{figure}[!t]
  \centering
  \includegraphics[width=0.45\textwidth]{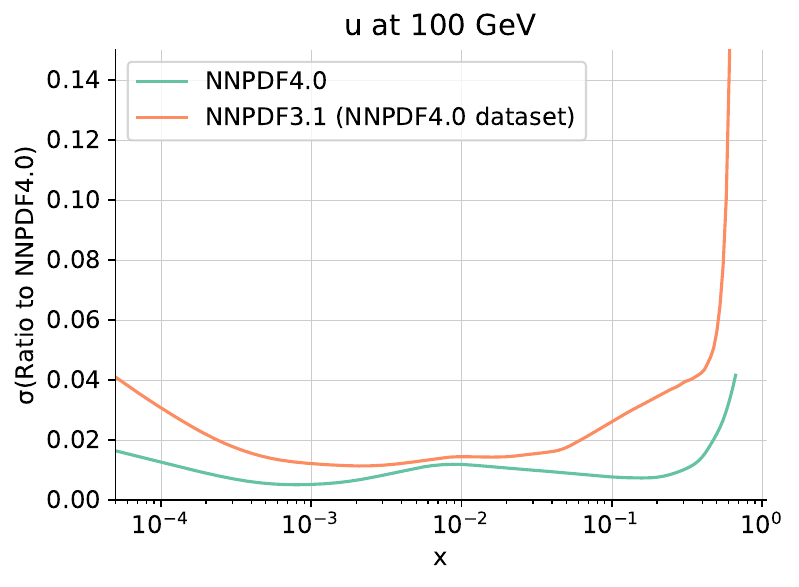}
  \includegraphics[width=0.45\textwidth]{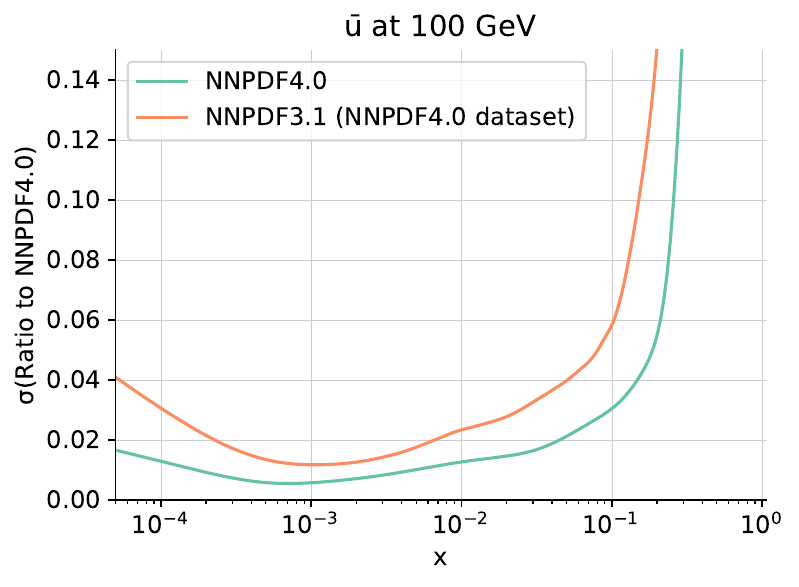}
  \includegraphics[width=0.45\textwidth]{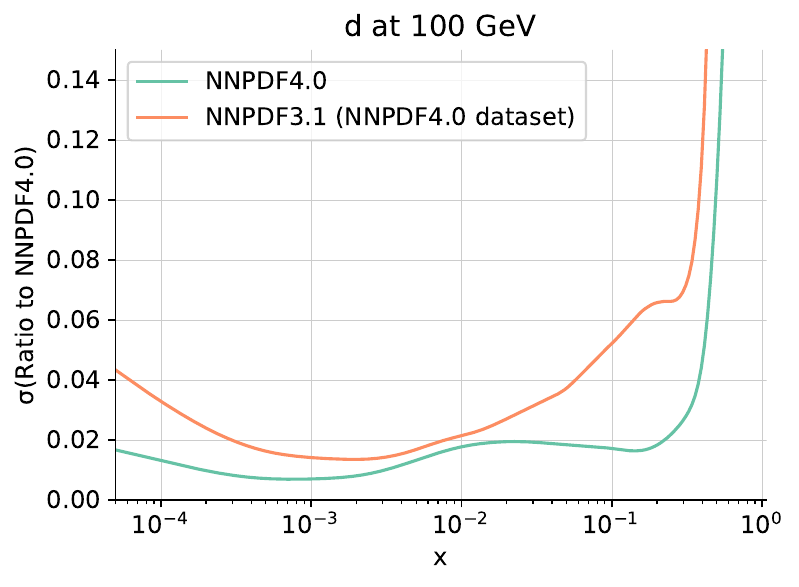}
  \includegraphics[width=0.45\textwidth]{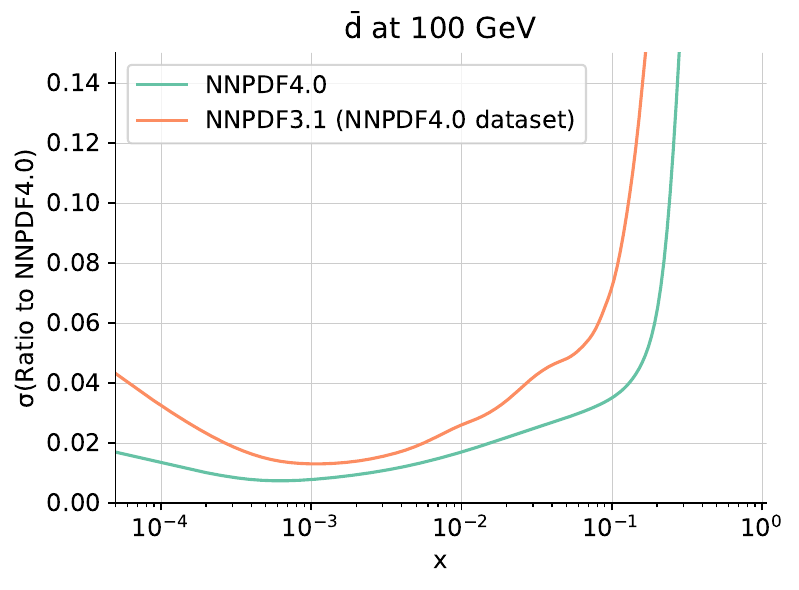}
  \includegraphics[width=0.45\textwidth]{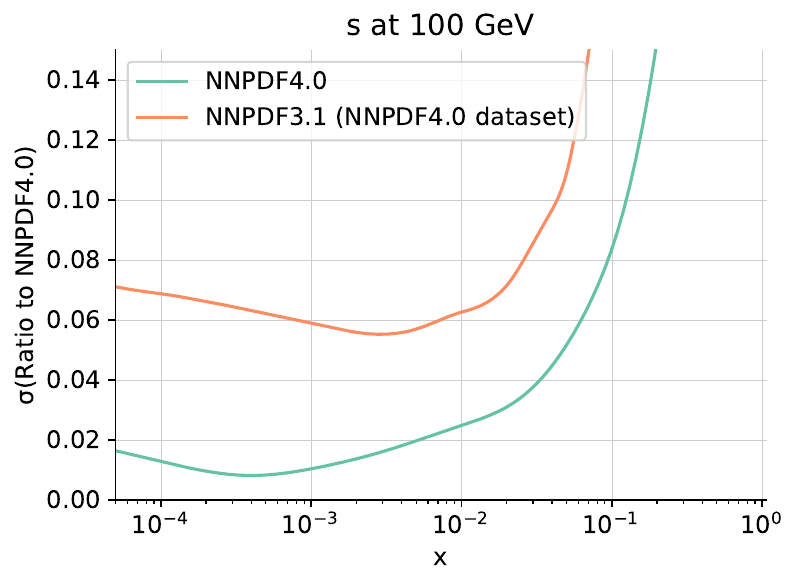}
  \includegraphics[width=0.45\textwidth]{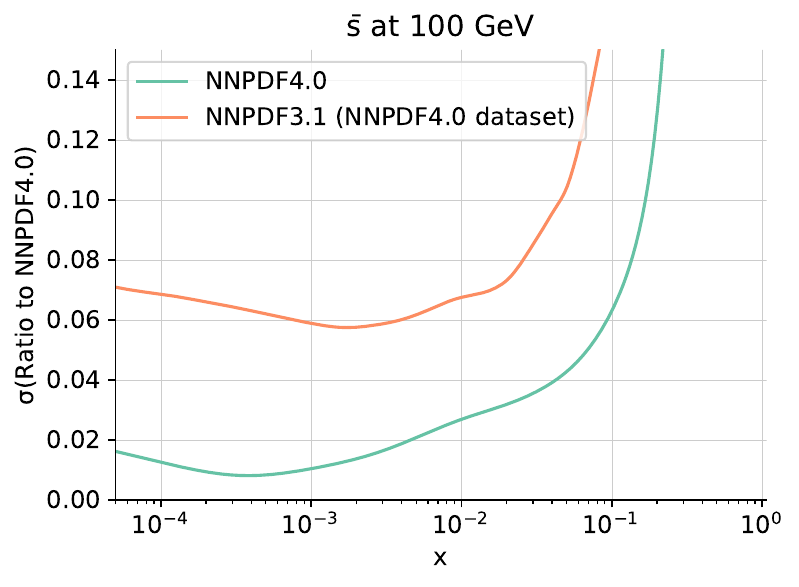}
  \includegraphics[width=0.45\textwidth]{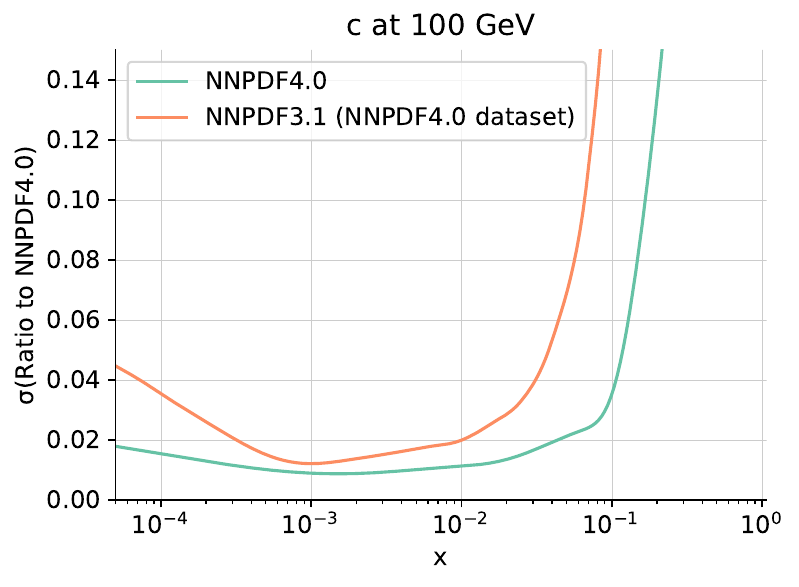}
  \includegraphics[width=0.45\textwidth]{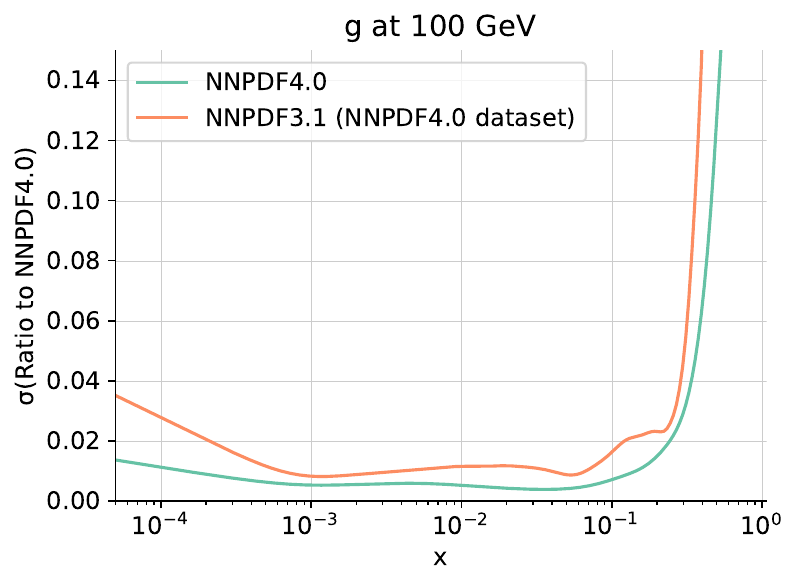}
  \caption{Same as Fig.~\ref{fig:40vs31_meth} but showing the one-sigma relative
    uncertainties. }
  \label{fig:40vs31_meth_uncs}
\end{figure}

We complement the comparison between NNPDF3.1 and NNPDF4.0 presented
in Sect.~\ref{subsubsection:NNPDF4.0_dataset}, where the impact of the
NNPDF4.0 dataset was analyzed, by now studying the impact of the
NNPDF4.0 methodology. This is done by comparing to the NNPDF4.0
baseline a PDF set determined from the NNPDF4.0 dataset, but using
NNPDF3.1 methodology.
Results are shown in Figs.~\ref{fig:40vs31_meth}-\ref{fig:40vs31_meth_uncs}.

It is clear that PDFs obtained by the two methodologies are
in perfect agreement: given a common dataset, the NNPDF4.0 and NNPDF3.1
methodologies produce consistent results. This confirms the 
conclusions of Sect.~\ref{sec:closure}, where the two methodologies
were compared specifically in the framework of closure and future tests.
However, it is clear that the NNPDF4.0 methodology leads to significantly more
precise results, as is apparent from
Fig.~\ref{fig:40vs31_meth_uncs}. 
This also agrees with the conclusions  of
Sect.~\ref{sec:closure}: the old and new methodology are both 
faithful (accurate within their stated precision), but the new methodology 
is more precise.

Putting this together with the results of
Sect.~\ref{subsubsection:NNPDF4.0_dataset} we conclude that the change
in PDF central values from NNPDF3.1 to NNPDF4.0 is due to the much
expanded dataset, especially because of LHC data, but the reduction
in uncertainty is almost entirely due to the improved methodology.

%% file: subsec-integrability.tex
\subsection{Impact of PDF positivity}
\label{subsec:positivity}

As discussed in Sect.~\ref{sec:positivity}, strict positivity
of the gluon and the light quarks and antiquarks PDFs is
enforced in NNPDF4.0, based on the results of
Ref.~\cite{Candido:2020yat}.
This implies that there is an extra set of
positivity constraints, on top of those that were already implemented
in NNPDF3.1 where positivity of several observables or
pseudo-observables (such as DIS structure functions for individual
quark flavors) was required.
In order to assess the impact of these new PDF positivity constraints,
we have produced a PDF determination in which only the previous NNPDF3.1
positivity constraints are implemented, while everything else is
identical to the NNPDF4.0 baseline in terms of both data and methodology.

\begin{figure}[!t]
  \centering
  \includegraphics[width=0.49\textwidth]{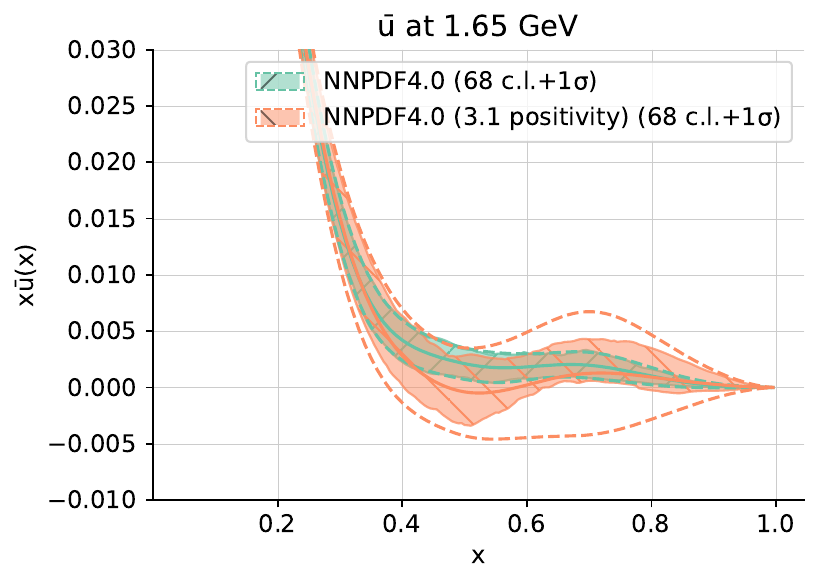}
  \includegraphics[width=0.49\textwidth]{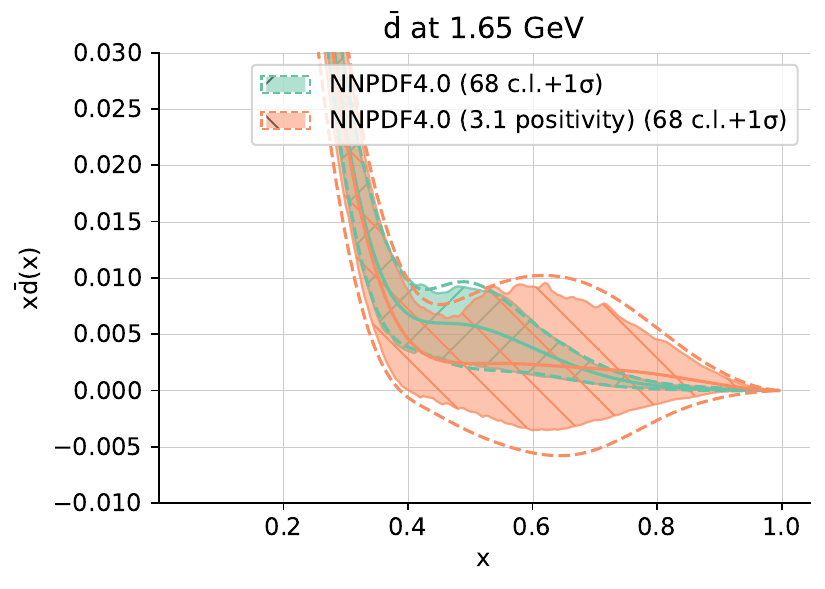}
  \includegraphics[width=0.49\textwidth]{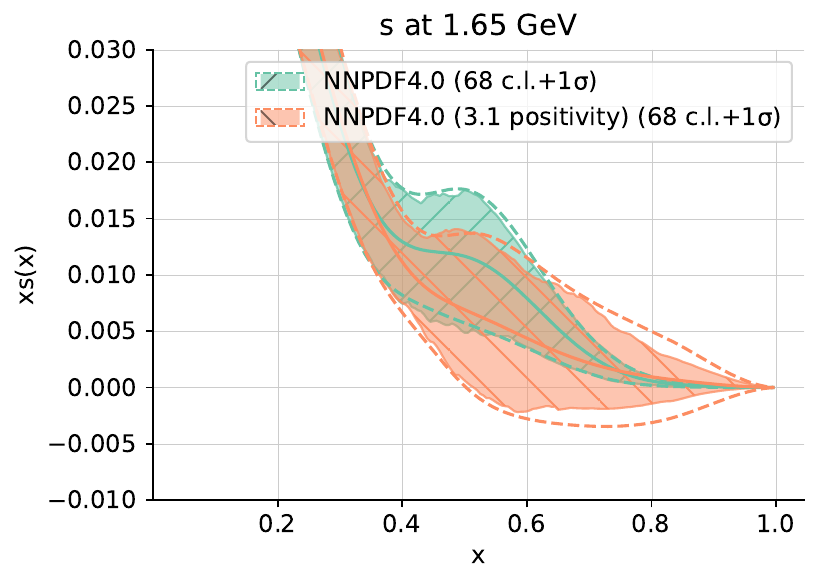}
  \includegraphics[width=0.49\textwidth]{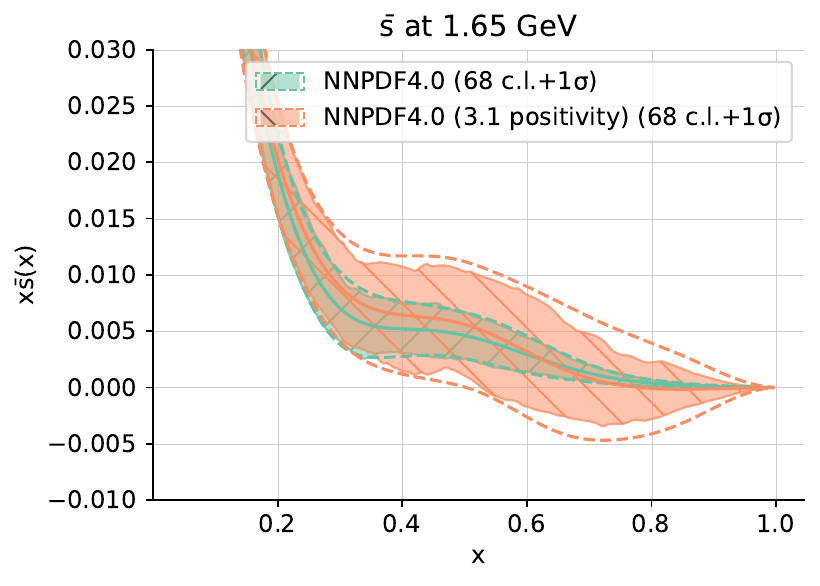}
  \caption{Comparison to the baseline NNPDF4.0 fit of the PDFs determined by
    removing the new PDF positivity constraints, and hence using only 
    the NNPDF3.1 positivity conditions. The antiup,
  antidown, strange and antistrange PDFs are shown at the
  input parametrization scale $Q=1.65$~GeV.}
  \label{fig:40vs40nopos}
\end{figure}

In Fig.~\ref{fig:40vs40nopos}  we compare to the NNPDF4.0 baseline fit some of the
ensuing PDFs: we show the antiup, antidown, strange and antistrange at
the parametrization scale $Q=1.65$~GeV. It is clear that the new PDF
positivity constraints have a substantial impact in the large-$x$ region,
$x\gtrsim0.3$,
both in terms of reducing the uncertainty and of preventing PDF replicas from going negative.
This latter property ensures positivity of cross-sections for
the production of final states even for very 
large invariant masses $m_X$.

\subsection{Impact of nonsinglet integrability}
\label{subsec:integrability}

As explained in Sect.~\ref{sec:pdfint}, in NNPDF4.0 additional
integrability constraints are added to those already implemented in
NNPDF3.1.
First, integrability of
the Gottfried and strangeness sums, i.e. integrability of $T_3$ and
$T_8$, is imposed through Lagrange multipliers. Second, the  range
of preprocessing exponents is determined self-consistently as for
NNPDF3.1, but it is no longer allowed to extend into the
non-integrable region. Finally, integrability is imposed at the
post-fit selection level. This ensures that all replicas remain
integrable, so nonsinglet sum rules are finite and with finite
uncertainty.

We assess the impact of these new integrability constraints by
comparing to the NNPDF4.0 baseline the PDFs obtained by removing both
of them, i.e. with no Lagrange multipliers for $T_3$ and $T_8$ and
unconstrained preprocessing 
range, and the PDFs determined by keeping the constraint on the preprocessing
range but removing the Lagrange multipliers for $T_3$ and $T_8$.

\begin{figure}[!t]
  \centering
  \includegraphics[width=0.49\textwidth]{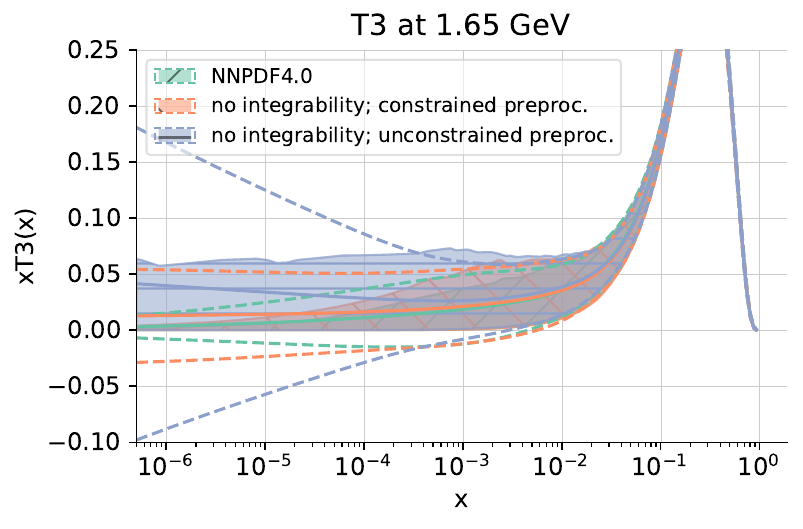}
  \includegraphics[width=0.49\textwidth]{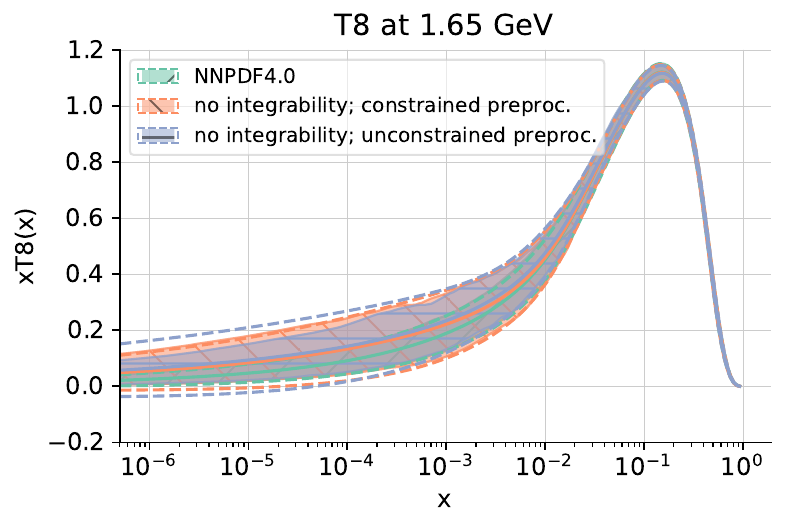}\\
  \caption{Comparison to the baseline of PDFs obtained removing either
    or both the new integrability constraints on the triplet and octet
    PDFs (see text). The triplet $T_3$ and octet $T_8$ are shown at
    $Q=1.65$~GeV.}
  \label{fig:pdfplot-abs-nnpdf40-nointeg-q1p65gev}
\end{figure}

In Fig.~\ref{fig:pdfplot-abs-nnpdf40-nointeg-q1p65gev} we compare the PDFs obtained in this way to the NNPDF4.0 baseline: we show the  $T_3$ and $T_8$
nonsinglet PDFs at the parametrization scale $Q=1.65$~GeV.
It is clear that the effect of the
new constraints is seen only in the small $x\lesssim10^{-3}$ region,
where there is limited experimental information on quark flavor
separation (see Fig.~\ref{fig:kinplot}).
The effect of the new integrability constraints is significant for $T_3$, but
moderate for $T_8$: in particular, $T_8$ remains integrable even when
both constraints are removed, while integrability of $T_3$ is enforced
when constraining the preprocessing, but would otherwise fail. The effect
of the Lagrange multiplier is mostly to reduce somewhat the small-$x$
uncertainties by removing some outliers.
It is important to note, however, that these constraints can be rather
more significant when PDFs are determined from a restricted dataset,
such as those considered in Sect.~\ref{sec:dataset}. Indeed,
inspection of   $T_8$ in the no-LHC  and  DIS-only  fits respectively
discussed in Sect.~\ref{subsubsection:DIS_data}
and~\ref{subsubsec:LHC_data} shows a rather different small-$x$
behavior and larger uncertainties, that could well extend into the
nonintegrable region in the absence of an explicit constraint.

\begin{figure}[!t]
  \centering
  \includegraphics[width=0.49\textwidth]{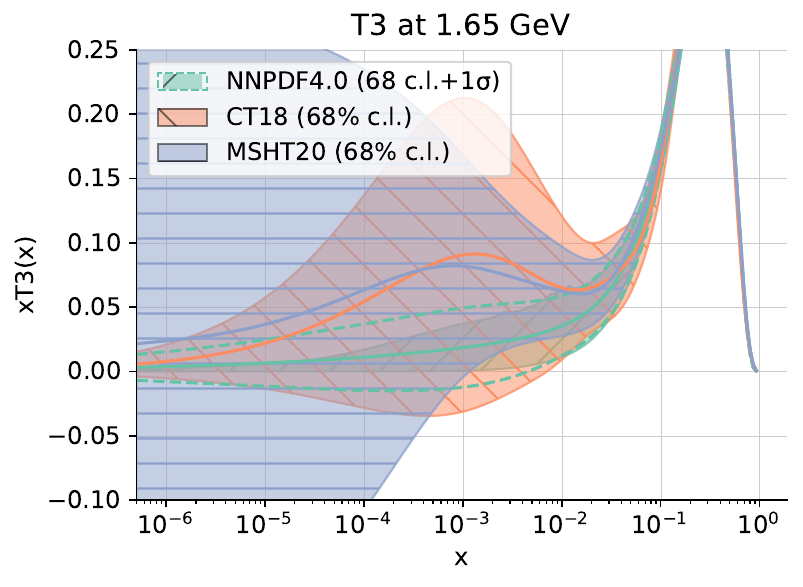}
  \includegraphics[width=0.49\textwidth]{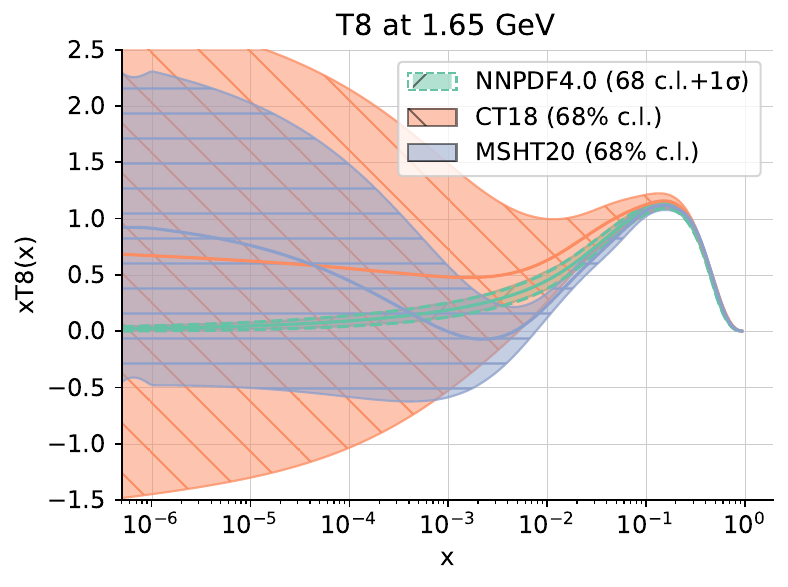}
  \caption{Same as
    Fig.~\ref{fig:pdfplot-abs-nnpdf40-nointeg-q1p65gev} now
    comparing the NNPDF4.0 baseline to  CT18 and MSHT20.}
  \label{fig:pdfplot-abs-nnpdf40-nointeg-q1p65gev-global}
\end{figure}

It is interesting to compare these results to those of the CT18 and MSHT20
determinations, shown in Fig.~\ref{fig:pdfplot-abs-nnpdf40-nointeg-q1p65gev-global}.
In the case of the triplet $T_3$, the central CT18
and MSHT20 $xT_3$ PDF combination also vanishes as $x\to 0$, but
for MSHT20 the uncertainty band extends into the nonvanishing (positive) range.
In the case of the  octet, for both CT18 and MSHT20  $xT_8$ does not
vanish as  $x\to 0$, resulting in substantially larger PDF uncertainties for light flavor 
separation in the small-$x$ region.

%% file: subsec-flavourbasis.tex
\subsection{Parametrization basis independence}
\label{subsec:flavbasis}

As discussed in  Sect.~\ref{sec:methparametrisation},
in the NNPDF4.0 determination the PDFs are
parametrized by default in the evolution basis at the input scale
$Q_0=1.65$ GeV.
This means that  the eight neurons of the final layer of the
neural network displayed 
in Fig.~\ref{fig:NNarch} correspond to the eight basis PDFs $f_k$ listed in
Eq.~(\ref{eq:evolution_basis}), up to  preprocessing and normalization
prefactors as given in
Eq.~(\ref{eq:evolution_basis_param}).
However, results should be completely independent of this basis
choice.
An alternative option, also discussed in
Sect.~\ref{sec:methparametrisation}, is to use the flavor basis, in
which the eight neurons of the final layer
now correspond instead to the eight basis PDFs $\tilde{f_k}$
of Eq.~\eqref{eq:flav_basis}.
The results of a global PDF analysis should in 
principle be the same
irrespective of whether PDFs are parametrized
in the evolution basis, Eq.~\eqref{eq:evol_basis}, or in the flavor
basis, Eq.~\eqref{eq:flav_basis}, or indeed in any other basis.

\begin{figure}[!t]
  \centering
  \includegraphics[width=0.45\textwidth]{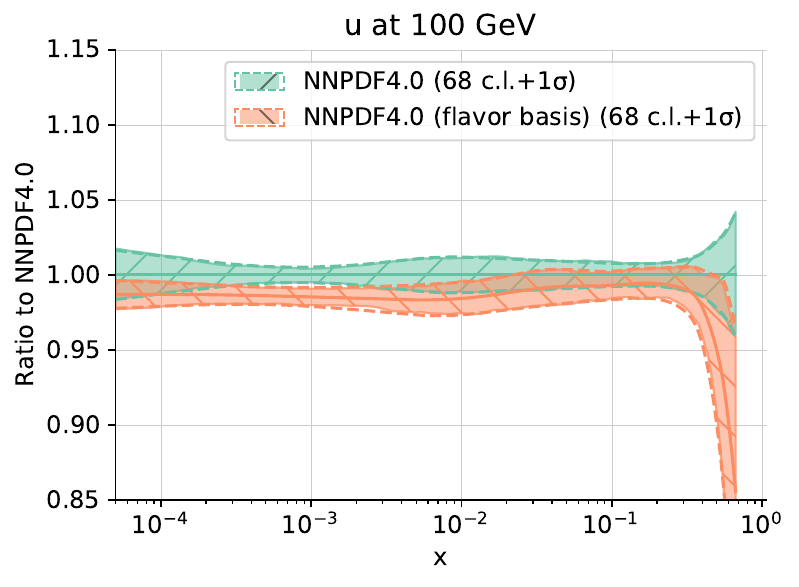}
  \includegraphics[width=0.45\textwidth]{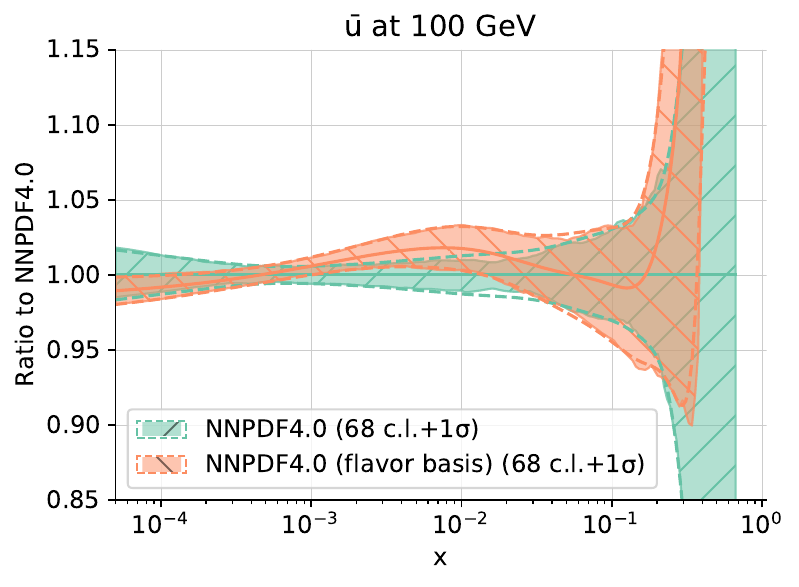}\\
  \includegraphics[width=0.45\textwidth]{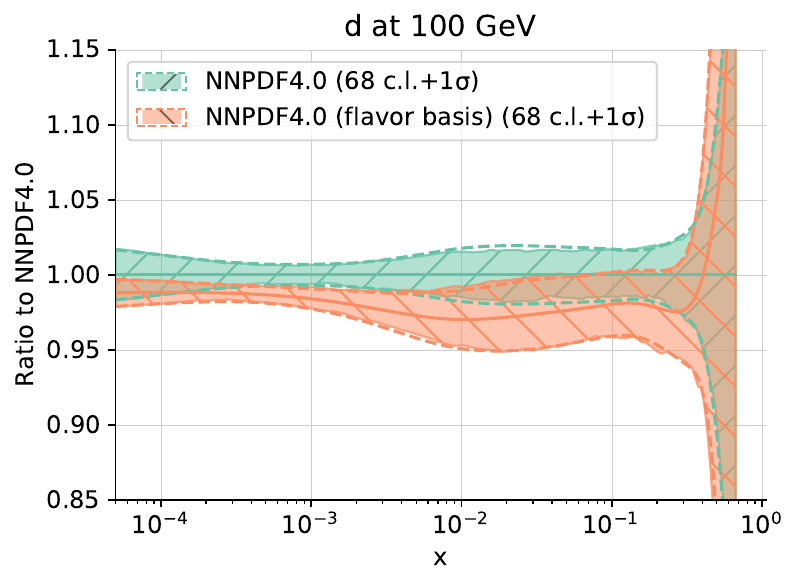}
  \includegraphics[width=0.45\textwidth]{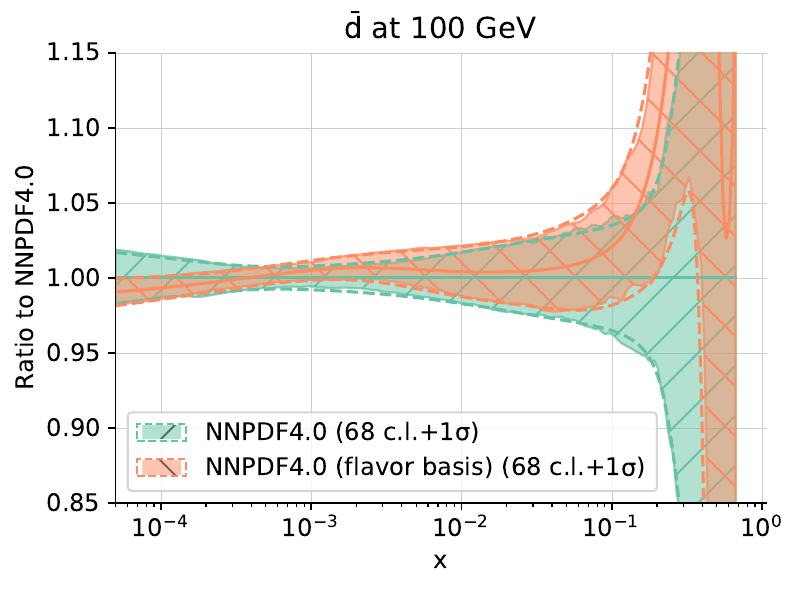}\\
  \includegraphics[width=0.45\textwidth]{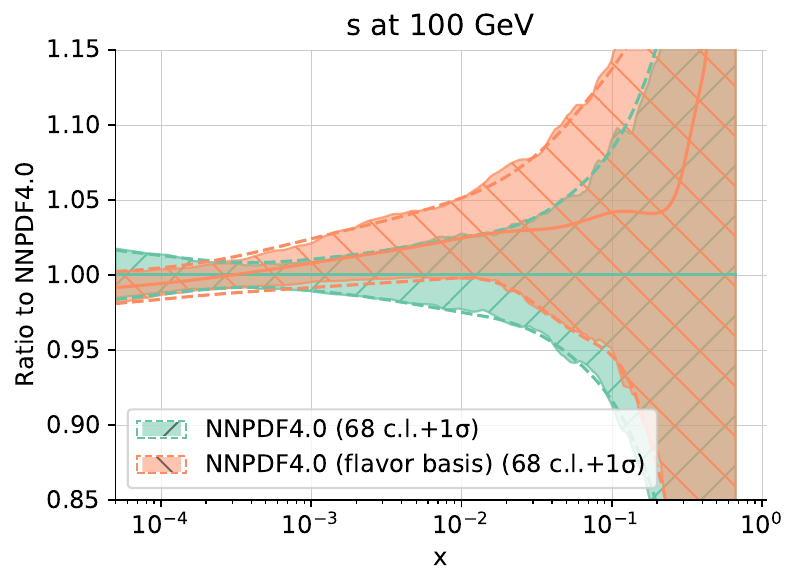}
  \includegraphics[width=0.45\textwidth]{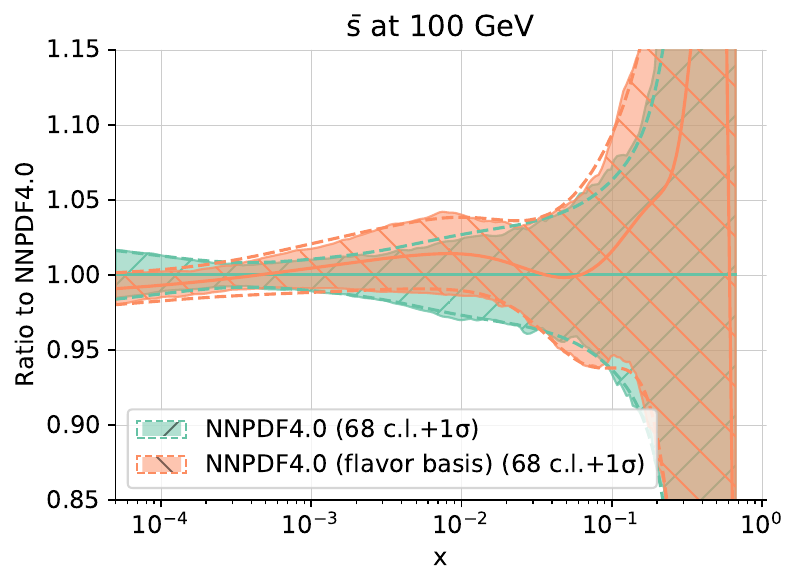}\\
  \includegraphics[width=0.45\textwidth]{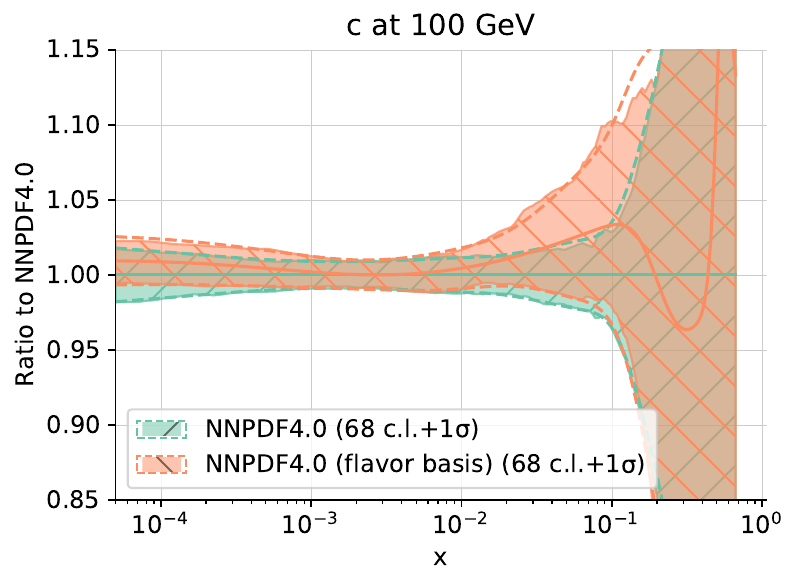}
  \includegraphics[width=0.45\textwidth]{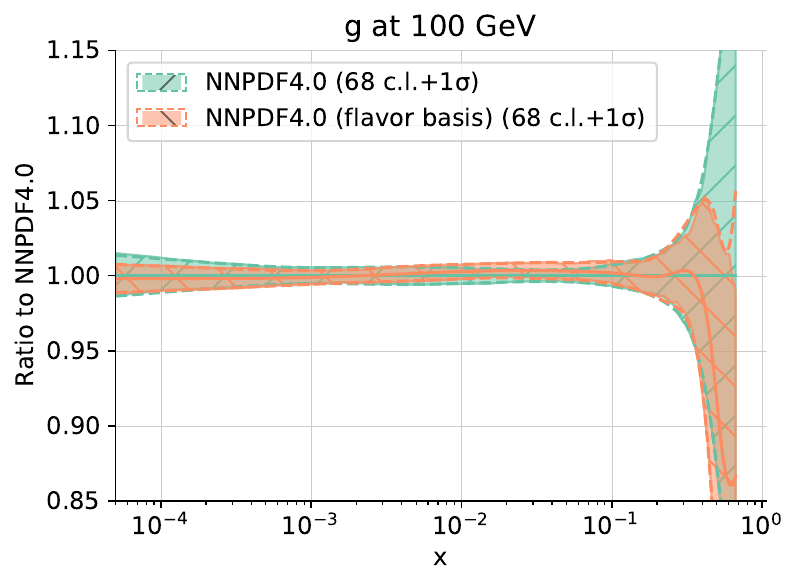}
  \caption{Same as Fig.~\ref{fig:40vs31-like}, but now comparing the
    baseline PDFs, parametrized in the evolution basis, to PDFs
    parametrized in the flavor basis and determined with the
    corresponding hyperparameter settings of Table~\ref{tab:setup}.}
  \label{fig:nnpdf40_evol_vs_flav.pdf} 
\end{figure}

To demonstrate explicitly that this is the case for NNPDF4.0,
we have carried out a PDF determination in the flavor basis.
This is a
significant modification of the fitting methodology, so the hyperoptimization procedure has
been repeated.
The final methodology settings  in this case are provided in
Table~\ref{tab:setup}, along with the baseline (evolution basis) settings.
The ensuing
PDFs are compared to the baseline in
Fig.~\ref{fig:nnpdf40_evol_vs_flav.pdf}. PDFs are not shown in the
far small-$x$ extrapolation region where, as discussed in
Sect.~\ref{sec:flavev}, the  behavior of flavor-basis PDFs is the
superposition of different powers and cannot be preprocessed as in the
evolution basis, and hence the corresponding integrability constraints cannot be enforced,
see Sects.~\ref{sec:pdfint} and~\ref{subsec:integrability}.

It is clear from Fig.~\ref{fig:nnpdf40_evol_vs_flav.pdf} that PDFs
in the two bases are in excellent agreement, with
differences fully compatible within the PDF uncertainties. 
It is important to understand that the results obtained from a flavor
basis parametrization  correspond
to an entirely new methodology: specifically, as discussed in Sect.~\ref{sec:methparametrisation} they do not contain any
small--$x$ preprocessing, and indeed this requires a considerably 
larger neural net architecture, compare the first and third column in
Tab.~\ref{tab:setup}. Hence we do not expect them to be trivially
identical to those obtained from the evolution basis parametrization,
but rather statistically compatible with them, as it is indeed the
case. We have in fact verified that if we combine replicas obtained
using the flavor basis and evolution basis parametrization in a
single replica set uncertainties are essentially unchanged, thus
confirming compatibility of the two results.

The flavor basis parametrization is more unstable due to the need of
using a larger neural network architecture, and it becomes
unreliable at small $x$ because of the difficulty of enforcing the
correct subleading Regge behavior, as discussed in
Sect.~\ref{sec:methparametrisation}. Therefore, we have not pursued
the flavor basis parametrization further for the sake of precision
phenomenology. However, the results presented here demonstrate
independence  of the choice of the
parametrization and provide  a highly nontrivial test of the robustness of
the NNPDF4.0 framework.

%% file: subsec-charmfit.tex
\subsection{Treatment of the charm PDF}

Since the NNPDF3.1 analysis, in the NNPDF baseline fits the charm PDF is parametrized alongside the light quark PDFs.
This has various advantages, specifically in absorbing into the initial PDF possible higher-order contributions
to perturbative matching conditions, thereby greatly reducing the
dependence of results on the value of the charm mass~\cite{Ball:2016neh}, and also allowing for a possible non-perturbative intrinsic charm component.

Here we assess the impact of parametrizing charm by comparing the baseline PDFs to PDFs in which charm is determined using standard NNLO perturbative
matching.
The fit quality deteriorates somewhat, with the total
$\chi^2$ per data point increasing from the value 1.16 of
Table~\ref{tab:PROCESSTYPE_dataset_chi2} to 1.18. The datasets that show a
more marked deterioration are gauge boson production and
deep-inelastic scattering, which are those most sensitive to quark flavor
decomposition. 

\begin{figure}[!t]
  \centering
  \includegraphics[width=0.45\textwidth]{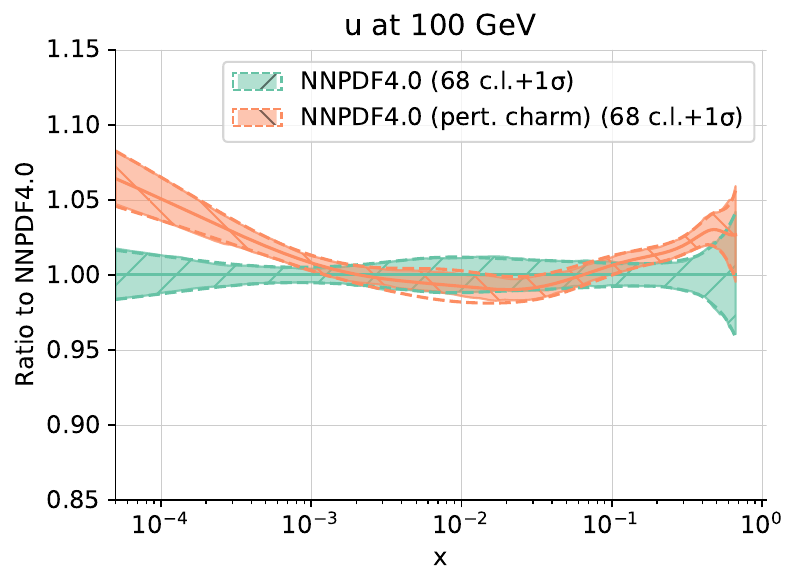}
  \includegraphics[width=0.45\textwidth]{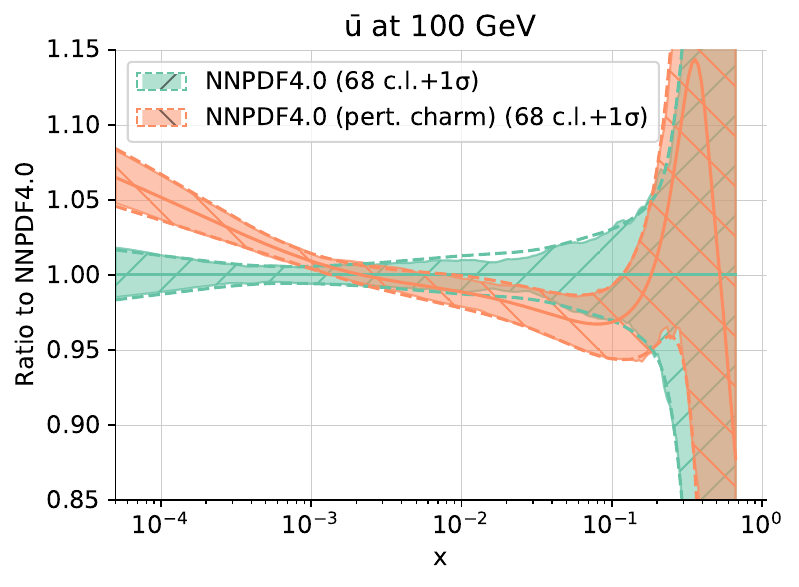}\\
  \includegraphics[width=0.45\textwidth]{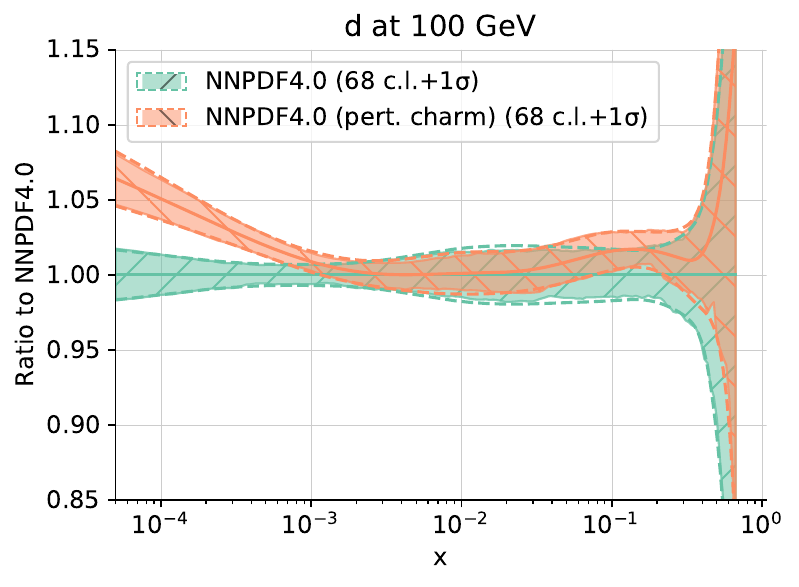}
  \includegraphics[width=0.45\textwidth]{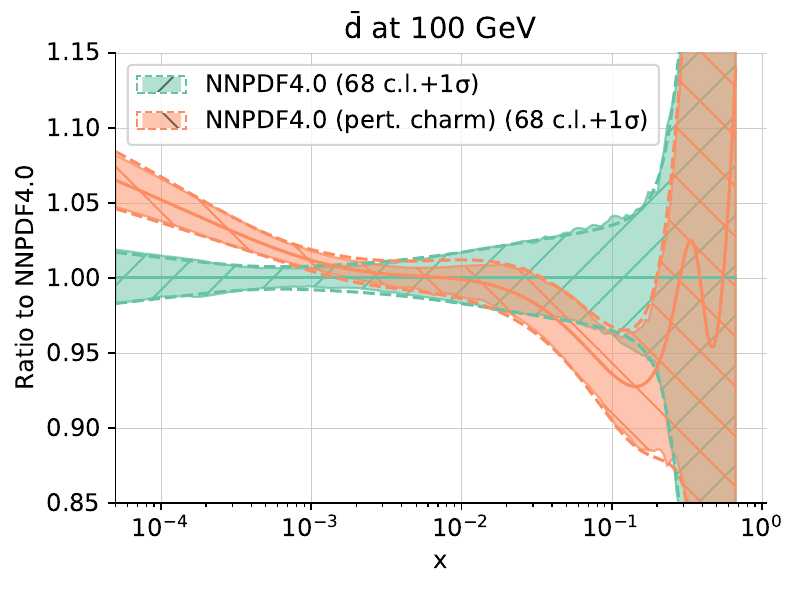}\\
  \includegraphics[width=0.45\textwidth]{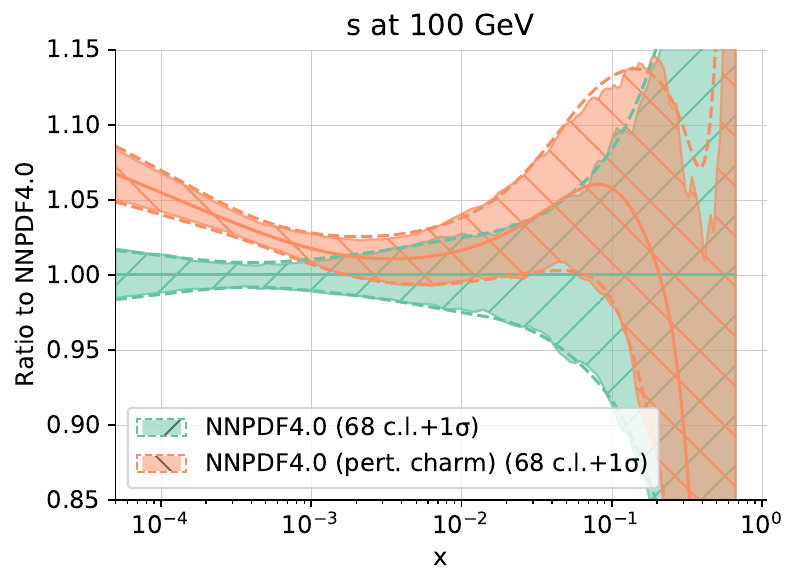}
  \includegraphics[width=0.45\textwidth]{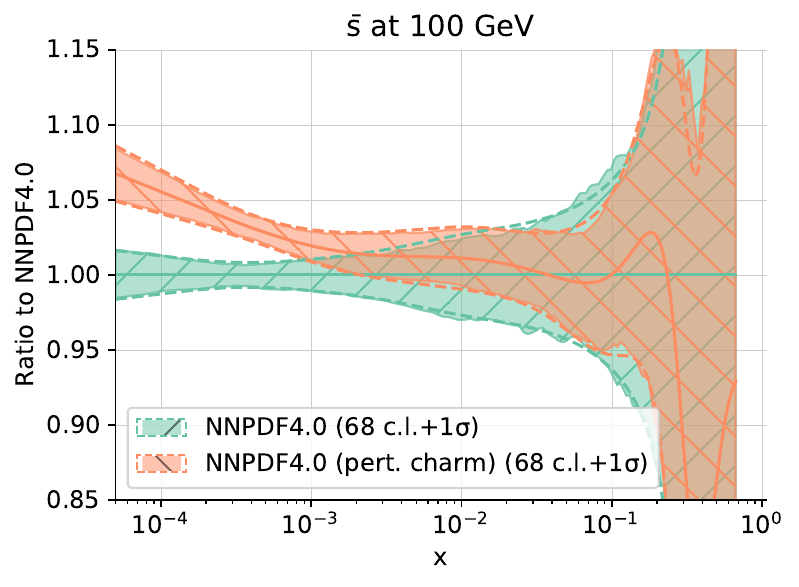}\\
  \includegraphics[width=0.45\textwidth]{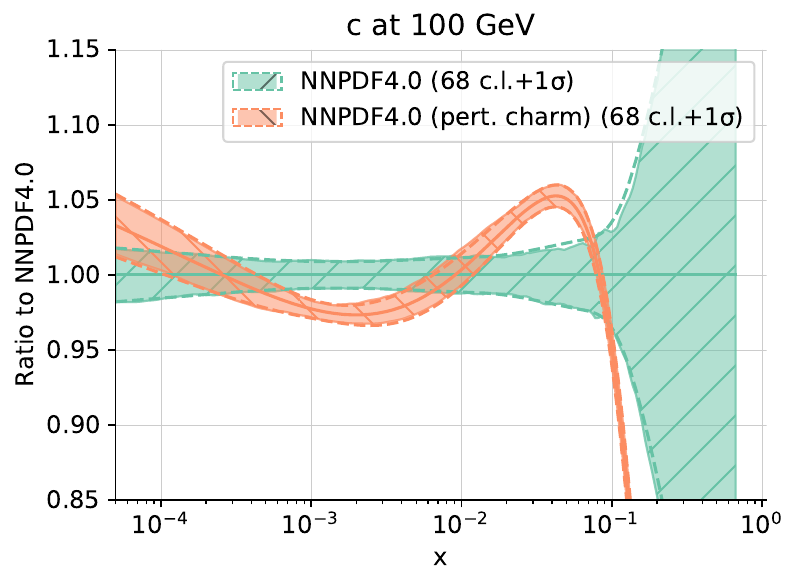}
  \includegraphics[width=0.45\textwidth]{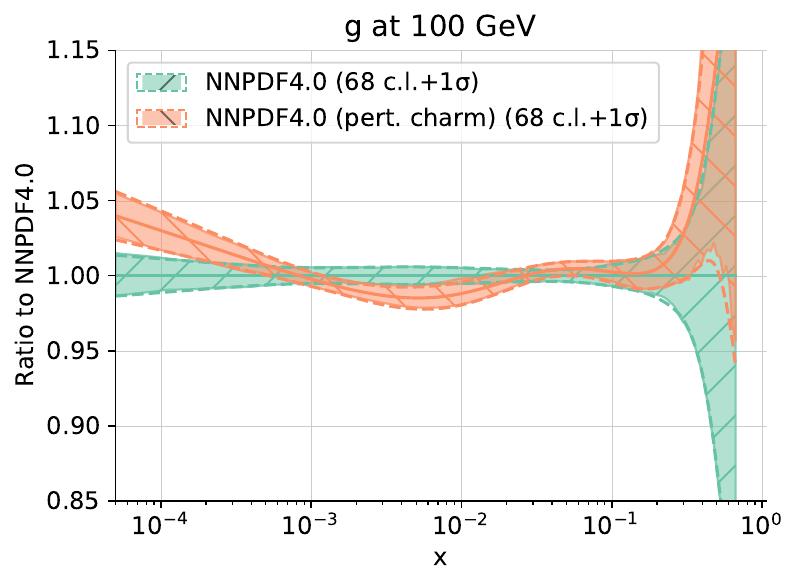}
  \caption{Same as Fig.~\ref{fig:40vs31-like}, comparing to the
    baseline PDFs in which 
  charm is not independently parametrized but
    rather determined by perturbative matching. The charm mass is 
     taken to be $m_c=1.51$ GeV in both fits.}
  \label{fig:pdfplot-rat-nnpdf40-vs-pertcharm}
\end{figure}

The PDFs obtained when charm is determined by perturbative matching
are compared to the baseline in
Fig.~\ref{fig:pdfplot-rat-nnpdf40-vs-pertcharm}. Results are
qualitatively similar to those already observed when the same
comparison was performed in  NNPDF3.1~\cite{Ball:2017nwa}.
It is particularly interesting to note the stability of the gluon PDF, which in the perturbative
charm approach is directly responsible for driving the charm
PDF.
Light quark PDFs are generally larger at small $x\lesssim0.003$
and smaller at larger $x\sim0.1$ when charm is not parametrized.
The charm PDF is of course most affected, with the PDF, when parametrized, being
rather larger at large $x\gtrsim0.1$,
smaller for $0.01 \lsim x \lsim 0.1$,
and then larger again for  $2\times 10^{-4} \lsim x \lsim 0.01$ as compared
to its perturbatively determined counterpart.
Note however that if charm is not parametrized, its
value in
the region $0.01\lsim x\lsim 0.1$ depends very strongly on the value
of the charm  mass $m_c$.

It is interesting to observe that  the uncertainties of all PDFs other
than charm  are quite similar whether or not charm is parametrized.
In fact, in several  cases, such as the gluon at small $x\lsim 10^{-3}$ and light
antiquark PDFs at intermediate $x\lsim 0.1$, the PDF uncertainties
are actually smaller when charm is parametrized.
This demonstrates the
improved overall consistency of the global PDF determination when charm is
parametrized.
Of course, the uncertainty on the charm PDF itself is significantly
larger when it is parametrized.

\begin{figure}[!t]
  \centering
  \includegraphics[scale=0.6]{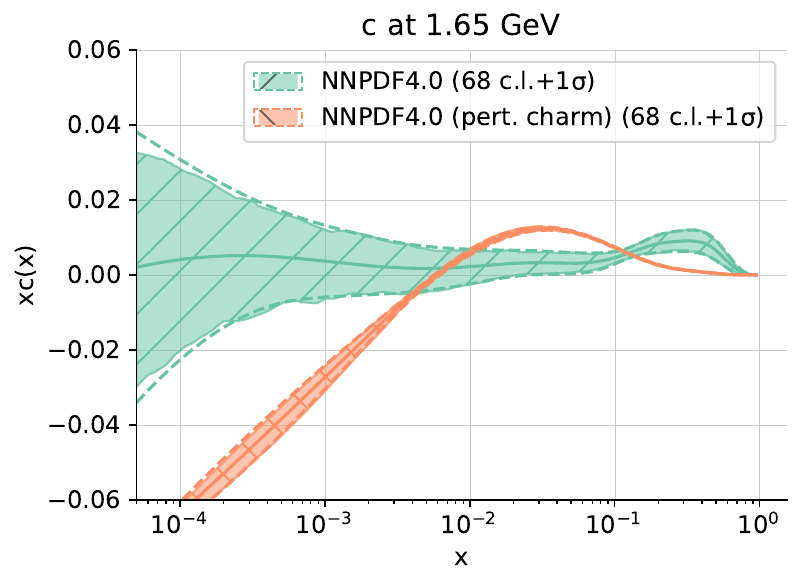}
  \includegraphics[scale=0.6]{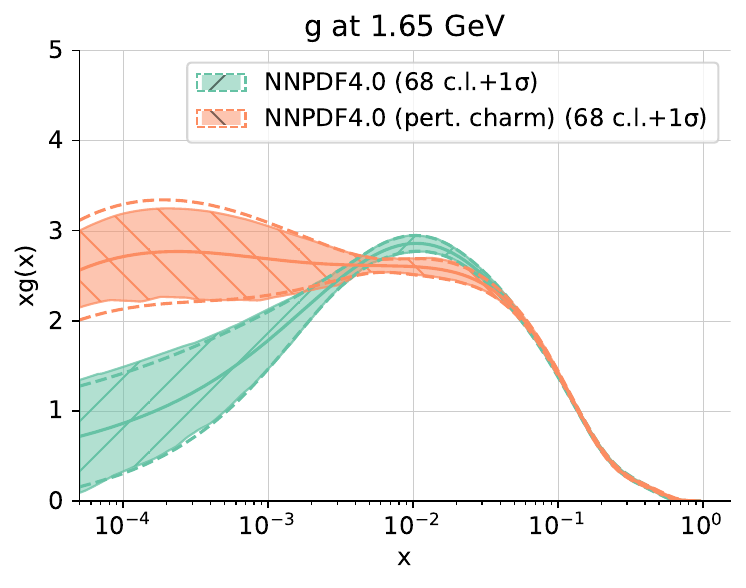}
  \caption{ 
    Same as Fig.~\ref{fig:pdfplot-rat-nnpdf40-vs-pertcharm} for the gluon
    and the charm PDFs at the parametrization scale $Q=1.65$.}
  \label{fig:pdfplot-abs-nnpdf40-vs-pertcharm_lowq}
\end{figure}

The charm PDF at the parametrization scale of
$Q_0=1.65$~GeV is directly compared in
Fig.~\ref{fig:pdfplot-abs-nnpdf40-vs-pertcharm_lowq} to its
perturbatively generated counterpart, along with the gluon PDF that
drives the latter. The stability of the gluon PDF can be directly
appreciated, in particular for $x\gsim 10^{-3}$.
The charm PDF, when independently parametrized, displays clear evidence for a valence-like component
at low scales and for $x\gsim 0.1$, with a statistical significance approaching
the $3\sigma$ level, while in the $x\lsim 0.1$ region
it is consistent with zero within uncertainties. The shape of the perturbatively generated charm is very different, and its very small uncertainty (which does not include the charm mass uncertainty or missing higher order corrections) looks unrealistic.

We conclude that parametrizing charm has a moderate but non-negligible
effect, especially on the light flavor separation, and it improves the 
overall fit quality and consistency.
The best-fit parametrized charm displays
evidence for a valence-like component at large $x$ and low scale, which
could be identified with an intrinsic charm component of the proton.
A dedicated investigation of this issue will be presented
in a follow-up publication~\cite{ICpaper}.

%% file: subsec-nuclearimpact.tex
\subsection{Impact of nuclear corrections}
\label{subsec:nuclear_impact}
\label{subsec:nuclearimpact}

As discussed in  Sect.~\ref{subsec:nuclear}, the baseline NNPDF4.0 determination
includes nuclear uncertainties as an extra contribution to the
covariance matrix, both for data taken on deuteron and heavy nuclei targets.
The impact of
these corrections is assessed here.
To this purpose, we have
produced dedicated PDF sets with different settings
for the treatment of deuteron and heavy nuclear uncertainties,
summarized in Table~\ref{tab:totchi2}. 
These correspond to including nuclear effects in either the
default way, as additional theory uncertainties (denoted as ``unc''), or in
the alternative way briefly discussed in  Sect.~\ref{subsec:nuclear}
in which they are included 
as a correction to the
experimental data, with a correspondingly reduced uncertainty, 
(denoted as ``shift'') or not at all,
for either or both deuterium or heavy nuclei. 

\begin{figure}[!t]
  \centering
  \includegraphics[width=0.3\linewidth]{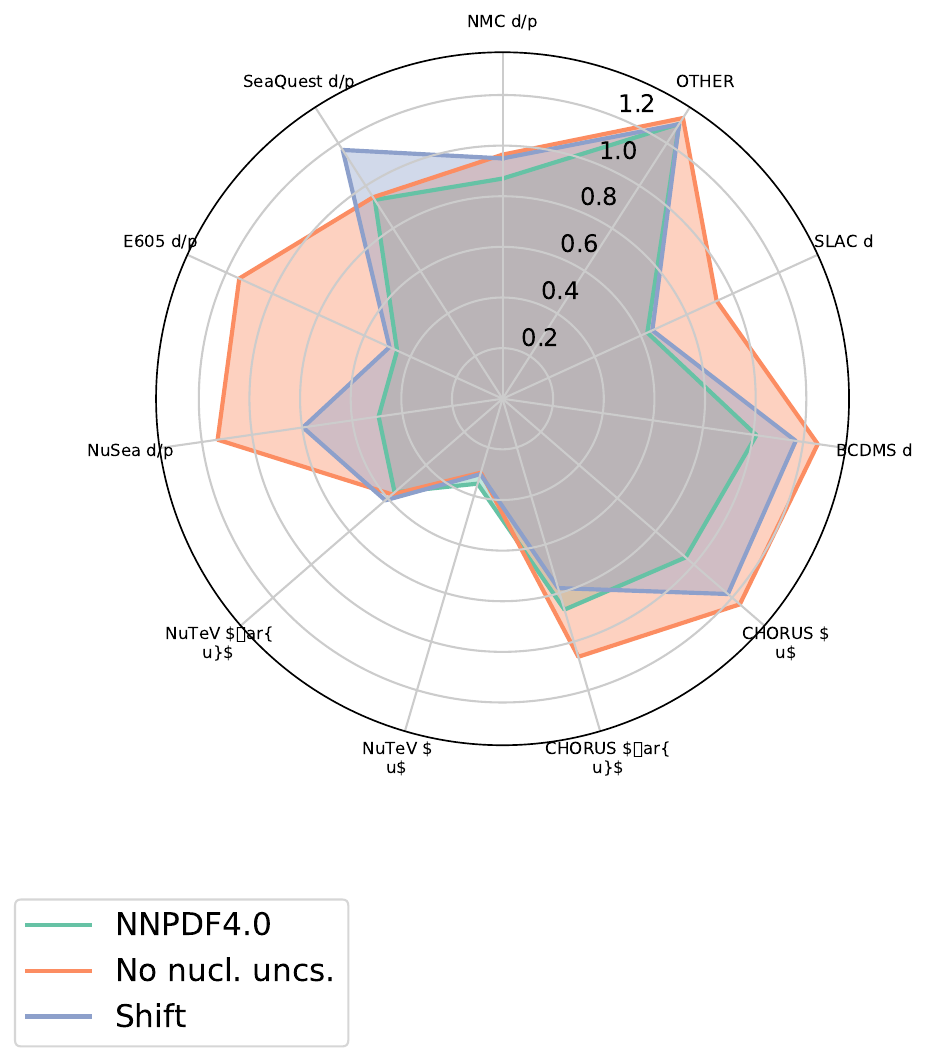}
  \includegraphics[width=0.3\linewidth]{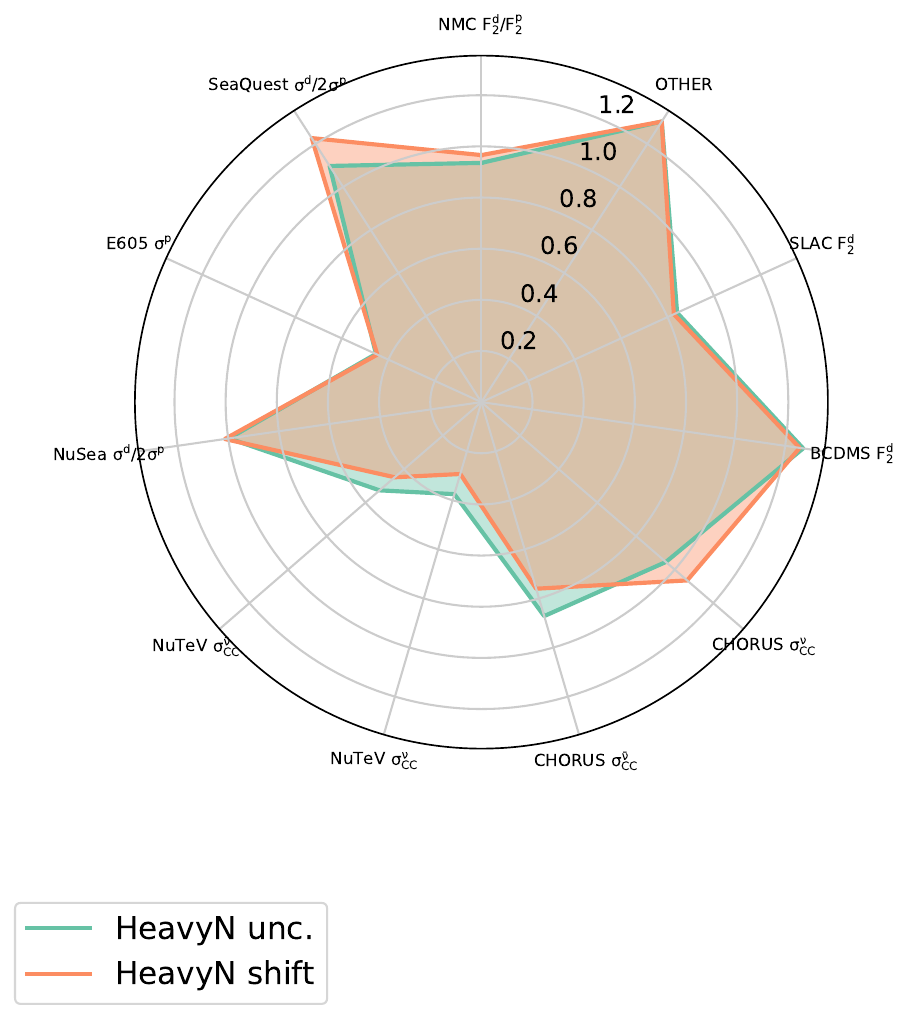}
  \includegraphics[width=0.3\linewidth]{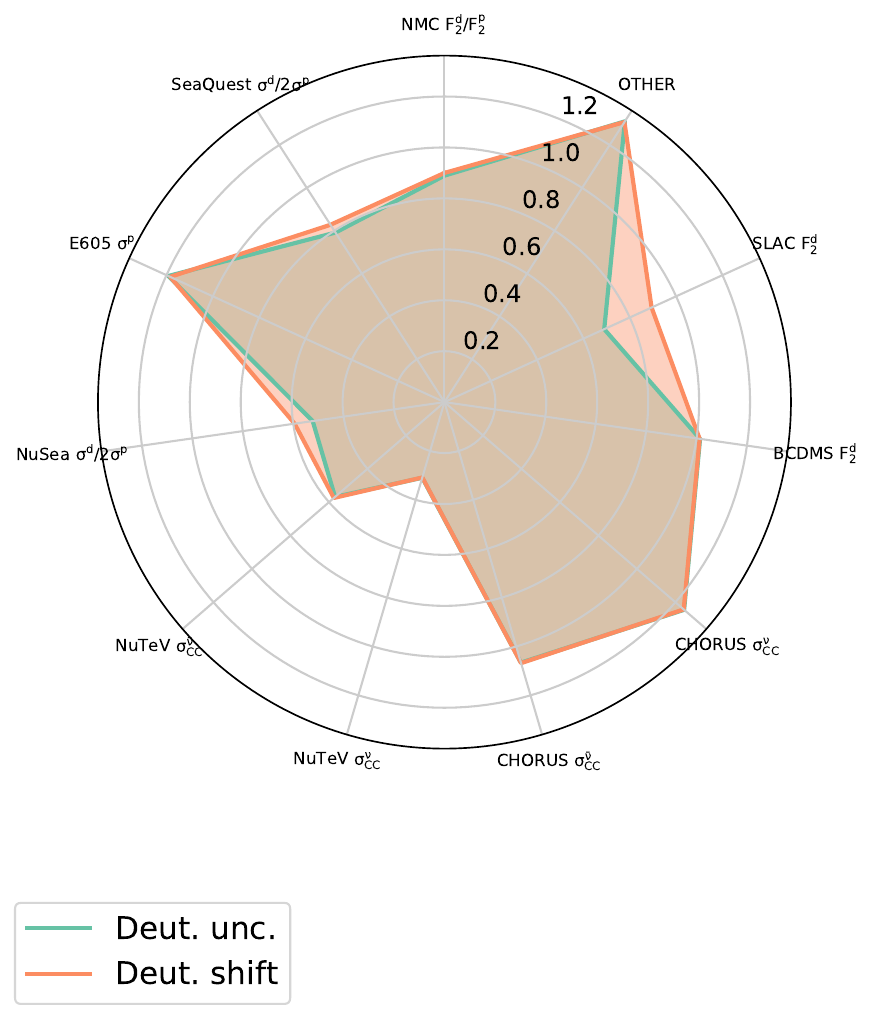}
  \caption{The values of the $\chi^2$ for individual datasets
    for the PDF fits listed in Table~\ref{tab:totchi2}.
    The datasets unaffected by nuclear corrections are grouped in the 
    ``other'' category.}
  \label{fig:chi2}
\end{figure}

The values of the $\chi^2$ per data point, for each process type
and for the complete
dataset, for each of these PDF determinations are collected in
Table~\ref{tab:totchi2}. The value of the $\phi$ estimator (as defined
in Eq.~(4.6) of Ref.~\cite{Ball:2014uwa}, also equal to the
square-root of the variance Eq.~(\ref{eq:VarDef}))  is also given. This is a
measure of the (correlated) PDF uncertainty in units of the data uncertainty.
A graphical representation of the results of Table~~\ref{tab:totchi2}
is provided in Fig.~\ref{fig:chi2}, where all datasets that are
unaffected by nuclear corrections are grouped as in the ``other'' category.

Upon including nuclear uncertainties,
the $\chi^2$ for the global fit improves rather significantly,
from 1.27 to 1.17.
This better fit quality can be traced to the improved description of the fixed-target CC DIS  and
Drell-Yan datasets, with similar outcomes for the ``unc'' and
``shift'' options.
This decrease in $\chi^2$  may look unsurprising, since
an extra source 
of  uncertainty is being added, which affects around one third
of the global dataset.
However, note that
the $\phi$ estimator is almost unchanged: this means
that PDF uncertainties remain 
almost the same.
The lowest total $\chi^2$ value is found for the
baseline fit.
Indeed, the reduction in $\chi^2$ is a little more
marked when nuclear corrections
are added as an extra uncertainty, rather than a shift. In the latter
case, the extra contribution to the uncertainty only corresponds to
the uncertainty in the shift itself. This suggests that the baseline
treatment of nuclear corrections as uncertainties  is a little more
conservative than the shift option.
The reduction in $\chi^2$ from the fit with no nuclear
corrections to the baseline is roughly the sum of the decreases
observed when either the deuteron or the
heavy nuclear datasets are corrected.

\begin{table}[!t]
  \scriptsize
  \centering
  \renewcommand{\arraystretch}{1.4}
  \input{tables/nuc-tab-total-chi2.tex}

  \vspace{0.2cm}
  \caption{The value of the $\chi^2$ per data point for the NNPDF4.0
    baseline and its variants with different treatments of nuclear
    corrections.
    Values are shown for each process type and for the complete
    dataset. The value of the $\phi$ estimator for the complete dataset is also provided
    (see text).}
  \label{tab:totchi2}
\end{table}

\begin{figure}[!t]
  \centering
  \includegraphics[width=0.49\linewidth]{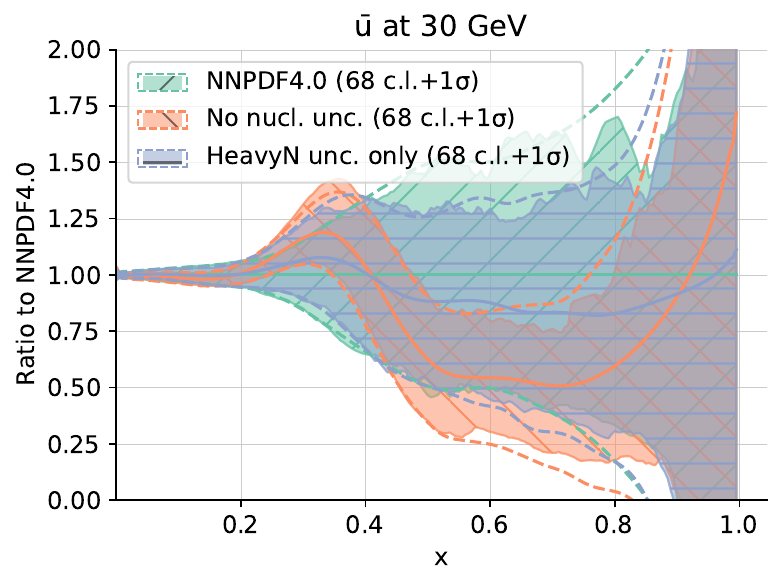}
  \includegraphics[width=0.49\linewidth]{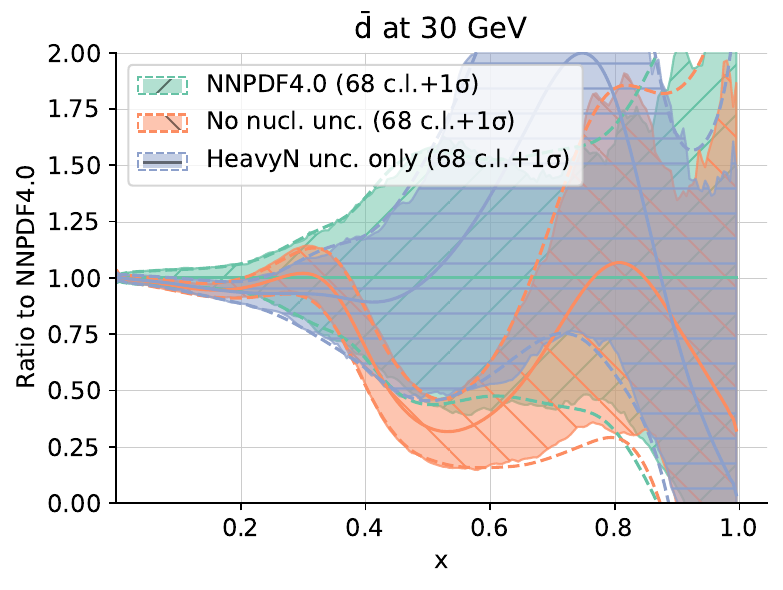}\\
  \caption{The antiup and antidown PDFs at $Q=30$~GeV from the ``No nucl. unc.''
    and ``HeavyN unc.'' PDF sets of Tab.~\ref{tab:totchi2} compared to the
    baseline.}
    \label{fig:pdfs1}
\end{figure}

The effect of nuclear corrections on PDFs is non-negligible, in
particular in the large-$x$ region.
In Fig.~\ref{fig:pdfs1} the antiup and antidown PDFs
at $Q=30$~GeV determined without nuclear corrections, or with heavy nuclear
corrections only, are compared to the baseline (with the default
treatment of nuclear corrections). Inclusion of nuclear corrections
leads to an increase in uncertainty at large $x\gsim 0.2$,
and also a different shape,
with in particular a significant enhancement around $x\simeq 0.5$.
Heavy nuclear corrections have the largest impact, especially on the
antidown PDF.
Nevertheless, all PDFs agree well within their respective uncertainty bands.
This suggests that neglecting  deuteron and heavy nuclear uncertainties
could distort the determination of the sea quark PDFs at large-$x$.

\begin{figure}[!t]
  \centering
  \includegraphics[width=0.49\linewidth]{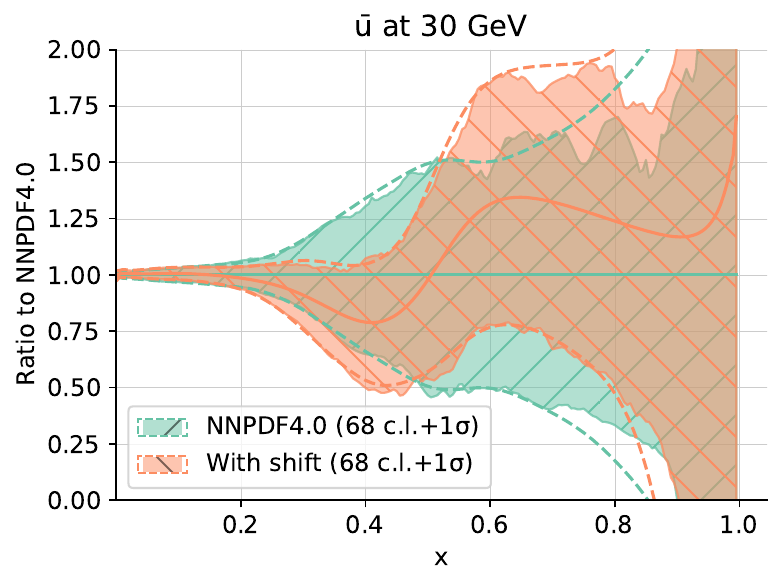}
  \includegraphics[width=0.49\linewidth]{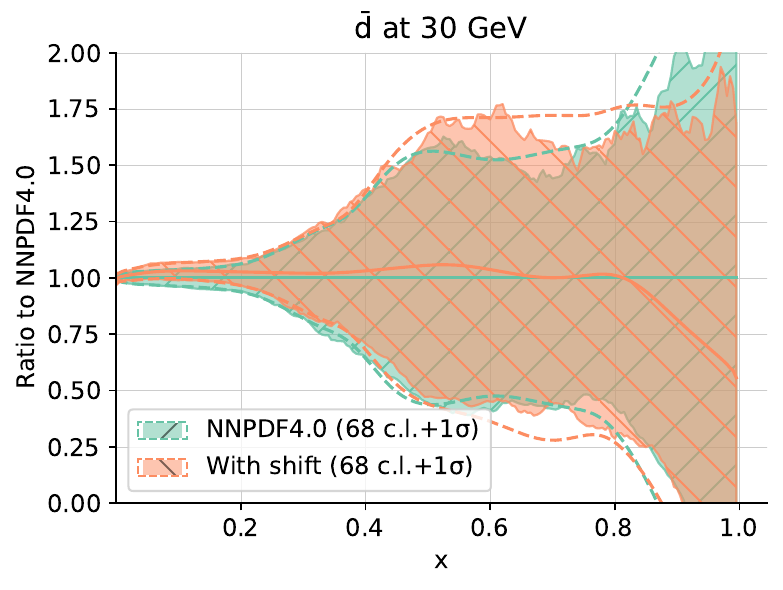}
  \includegraphics[width=0.49\linewidth]{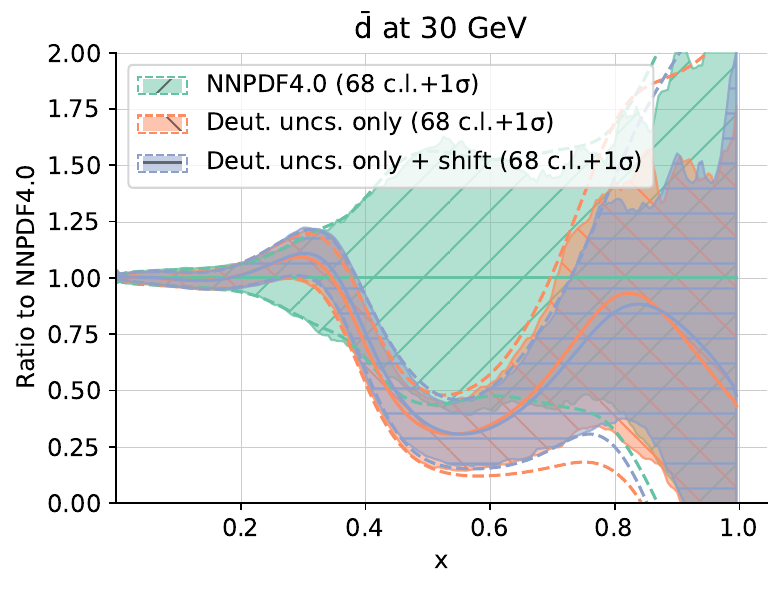}
  \includegraphics[width=0.49\linewidth]{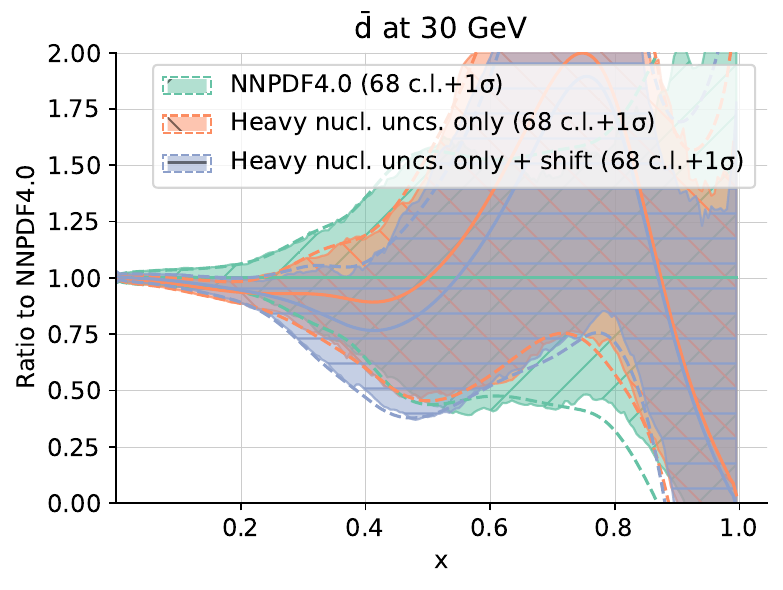}
  \caption{Top: same as Fig.~\ref{fig:pdfs1}, but now with
    PDFs from the ``Shift'' set.
    Bottom: comparison to the baseline of
    the antidown PDF at $Q=30$ from the ``Deut Unc'' and ``Deut
    shift'' sets (left) or from the ``HeavyN unc'' and the ``HeavyN
    shift'' sets (right).}
    \label{fig:pdfs2}
\end{figure}

PDFs obtained with either of the two
alternative treatments of nuclear corrections are compared in Fig.~\ref{fig:pdfs2}. First (top), we compare to the
baseline the antiup and antidown PDFs as in Fig.~\ref{fig:pdfs1} but
now with all nuclear and deuterium corrections included as shifts, and
then (bottom) we compare directly the antiup PDF when either
the deuterium or the nuclear corrections are included with either
the uncertainty or the shift method.
It is clear that the impact of the nuclear corrections on the PDF
with either method is quite similar, the only difference being that
uncertainties are somewhat smaller when the shift method is
adopted.
This is in agreement with the behavior of the $\chi^2$ values
observed previously, and confirms that the baseline  prescription is 
somewhat more conservative.

As mentioned in Sect.~\ref{subsec:nuclear}, the evaluation of
the deuterium corrections with the method of Ref.~\cite{Ball:2020xqw}
requires a self-consistent determination of the deuterium PDF, which
has been performed here starting  with the NNPDF4.0 set and then
proceeding as was done in Ref.~\cite{Ball:2020xqw} for NNPDF3.1.
A byproduct of this procedure is then, of course,
an independent determination of the
deuterium PDFs and thus of deuterium structure functions, with
corresponding correlated uncertainties, which we now discuss briefly.

\begin{figure}[!t]
  \centering
  \includegraphics[width=0.49\linewidth]{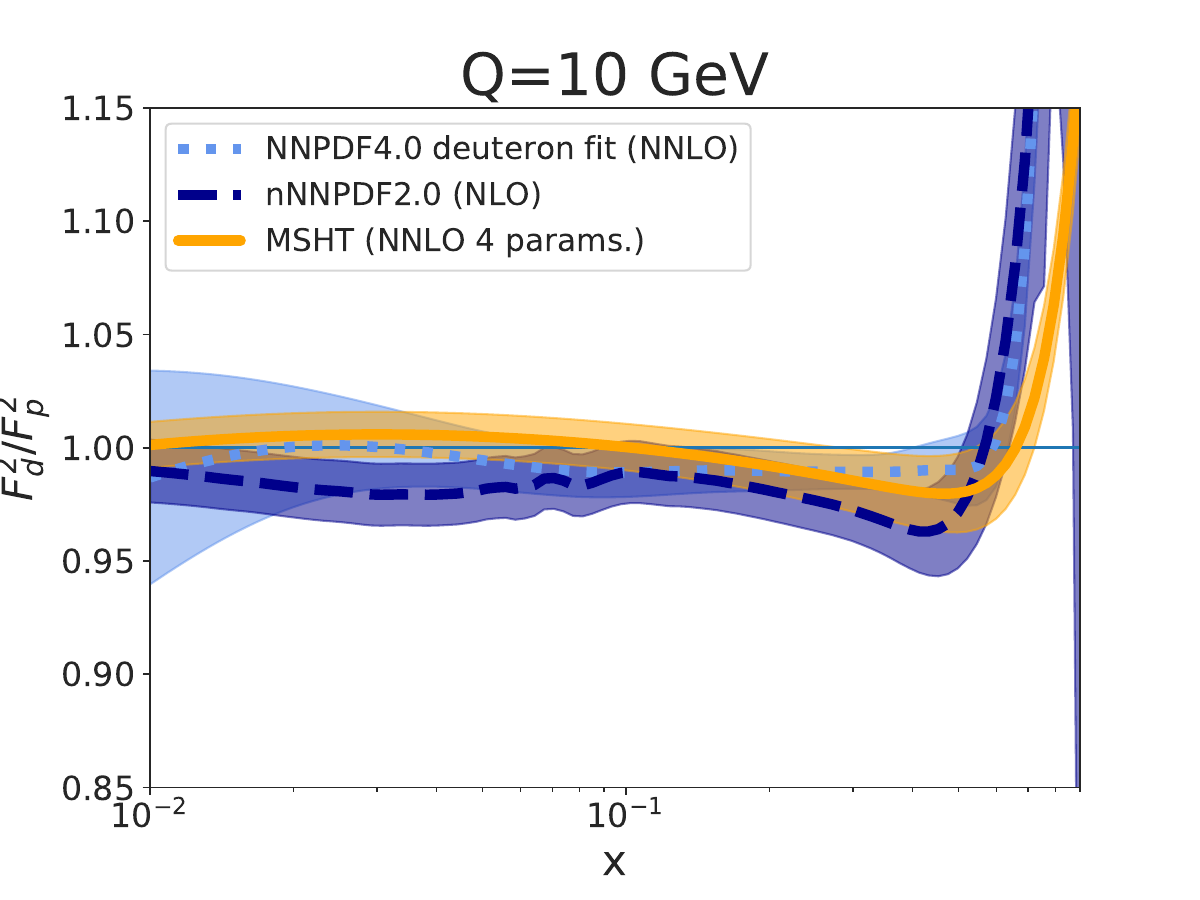}
  \caption{The ratio of deuteron to the iso-singlet
    proton structure functions,
    $F_2^d//F_2^{p,0}$, evaluated using the proton and deuteron PDFs obtained
    in the present NNPDF4.0 analysis at $Q=10$ GeV as a function of $x$.
    Results are compared to the nNNPDF2.0 nuclear PDF fit and
    the phenomenological correction factor from MSHT20.}
    \label{fig:deutpdfcomp}
\end{figure}

In Fig.~\ref{fig:deutpdfcomp} we display the  $F_2^d/F_2^{p,0}$ structure
function ratio at $Q$ = 10 GeV, where by $F_2^{p,0}$ we denote the
isospin singlet component of the proton structure function, so 
$F_2^d/F_2^{p,0}=1$ in the absence of nuclear corrections. The 
associated one-sigma PDF uncertainty band is
also shown, with correlations
between deuteron and proton PDFs taken into account.
The results from  the nNNPDF2.0
nuclear PDF fit and from a phenomenological determination in
MSHT20~\cite{Bailey:2020ooq} are also shown for comparison.

The deuteron corrections to  $F_2^d/F_2^{p,0}$ are seen in
Fig.~\ref{fig:deutpdfcomp} to be quite  small, as expected
since the  deuteron is a loosely bound nucleus.
The three estimates for  $F_2^d/F_2^{p,0}$  are consistent with each
other and agree within uncertainties.
In all three cases, one finds that the correction is only important at large-$x$,
with a dip of a couple percent for $x\simeq 0.4$ and then an enhancement at larger values of $x$.
The uncertainties for the NNPDF4.0-based determination are
slightly larger in the low-$x$ region, reflecting that this
determination is a somewhat more conservative. This determination also has
the smallest correction factor, which is in general very close to one
except for $x\gsim 0.6$.

%% file: tables/nuc-tab-total-chi2.tex
\begin{tabularx}{\textwidth}{Xccccccc}
  \toprule
  & No nucl. unc.
  & Deut. unc.
  & Deut. shift
  & HeavyN unc.
  & HeavyN shift
  & Shift
  & NNPDF4.0 \\
  \midrule
  DIS NC (fixed-target)
  & 1.27
  & 1.27
  & 1.27
  & 1.26
  & 1.27
  & 1.26
  & 1.26
  \\
  DIS CC (fixed-target)
  & 0.86
  & 0.86
  & 0.86
  & 0.86
  & 0.87
  & 0.86
  & 0.86
  \\
  DIS NC (collider)
  & 1.20
  & 1.19
  & 1.19
  & 1.19
  & 1.19
  & 1.19
  & 1.19
  \\
  DIS CC (collider)
  & 1.27
  & 1.299
  & 1.29
  & 1.26
  & 1.25
  & 1.28
  & 1.28
  \\
  \midrule
  Drell-Yan (fixed-target)
  & 0.94
  & 0.93
  & 0.92
  & 1.00
  & 1.01
  & 0.96
  & 0.98
  \\
  Tevatron $W,Z$ prod. (incl.)
  & 1.22
  & 1.25
  & 1.27
  & 1.15
  & 1.14
  & 1.12
  & 1.10
  \\
  \midrule
  LHC inclusive $W,Z$ prod. (incl.)
  & 1.51
  & 1.49
  & 1.48
  & 1.40
  & 1.40
  & 1.37
  & 1.37
  \\
  LHC $W,Z$ production ($p_T$ and jets)
  & 1.13
  & 1.14
  & 1.14
  & 1.14
  & 1.14
  & 1.14
  & 1.14
  \\
  LHC top-quark pair production
  & 1.23
  & 1.21
  & 1.20
  & 1.18
  & 1.20
  & 1.18
  & 1.20
  \\
  LHC jet production
  & 1.30
  & 1.27
  & 1.26
  & 1.30
  & 1.30
  & 1.25
  & 1.26
  \\
  LHC isolated $\gamma$ production
  & 0.74
  & 0.74
  & 0.73
  & 0.77
  & 0.76
  & 0.76
  & 0.76
  \\
  LHC single $t$ production
  & 0.36
  & 0.37
  & 0.36
  & 0.37
  & 0.36
  & 0.36
  & 0.36
  \\
  \midrule
  Total $\chi^2$  & 1.27 & 1.25 & 1.25 & 1.19 & 1.21 & 1.19 & 1.17 \\
  \midrule
  Total $\phi$  & 0.15 & 0.16  & 0.16 & 0.16   & 0.16  &  0.16 & 0.16\\
  \bottomrule
\end{tabularx}

%% file: subsec-regcovmat.tex
\subsection{Regularized covariance matrices}
\label{sec:regcovmat}

The selection procedure of Sect.~\ref{sec:dataselection} revealed that several
of the datasets considered as potential candidates for the inclusion
in the global PDF analysis exhibit a large
value of the stability metric $Z$, Eq.~(\ref{eq:Z}),
which may lead to artificially high $\chi^2$ values due
to ill-defined covariance matrices.
As discussed there, the value of $(\sqrt{2}Z)^{-1}$ can be interpreted as the
precision at which 
 correlations need to be estimated in order to ensure that they
 affect the $\chi^2$ by less than one standard deviation.
 This implies e.g. that a dataset with $Z=10$ requires
 correlations to be estimated with
 an absolute uncertainty of less than $0.07$, else the $\chi^2$ will be 
 inflated.
 The potentially problematic nature  of publicly released experimental covariance matrices is sometimes acknowledged by the experimental
 collaborations, and alleviated by their provision of alternative decorrelation 
models characterized by a different pattern of correlated systematics.

The stability analysis carried out in Sect.~\ref{sec:dataselection}
focused on the impact of large weight fits at the  PDF level, and
based on the results of these fits, it
established which datasets were  suitable for inclusion in the
baseline dataset, essentially by making sure that they would not
distort the global fit.
Here we assess the effect on the global PDF fit 
when datasets exhibiting large values of $Z$ have their covariance matrices
regularized by means of a tailored procedure.
For datasets that we did decide to include in NNPDF4.0, the purpose 
of this is to confirm that our best-fit PDFs are indeed not distorted 
by the inclusion of this data.
For datasets that were not included, the aim is to assess
what would be their impact if it was possible to safely include them.

The  decorrelation procedure that we apply here is described
in more detail in Ref.~\cite{COV}. It is based on  clipping
the eigenvectors until a target value of the stability metric, 
$Z_{\rm reg}$, is achieved.
For instance, if the target value is chosen to be  $Z_{\rm reg}=4$, then
the clipping algorithm  transforms the original experimental correlation matrix into
a different matrix with the same eigenvectors as the original one but such that the
eigenvalues that were smaller than $1/Z_{\rm reg}^2=1/16$ are replaced by 1/16.
The motivation for this decorrelation procedure is to give a decorrelated 
covariance matrix which is as close as possible
to the original one provided by the experiments. This is in contrast to 
other approaches such as adding a small diagonal contribution, or varying 
ad hoc the pattern of correlations for specific sources of systematic 
uncertainties.

\begin{table}[!t]
  \scriptsize
  \centering
  \renewcommand{\arraystretch}{1.4}
  \input{tables/tab-regcovmat}
  \caption{The values of the  $\chi^2$ in NNPDF4.0
    variants in which the covariance matrices for selected datasets
    have been  regularized following the procedure discussed in the text.
    For each dataset, we indicate the number of data points, the original values
    of the fit quality $\chi^2_{\rm orig}$ and of the stability metric
    $Z_{\rm orig}$,
    and then the values of the $\chi^2_{\rm reg}$ obtained by repeating the fit
    with the regularized covariance matrix for this dataset, for a choice
    of the target metric of $Z_{\rm reg}=4$. Datasets denoted by (*) are not
    part of the baseline and have been obtained from dedicated PDF
    fits (see text).}
  \label{tab:tab-regcovmat}
\end{table}

We have repeated the global NNPDF4.0 NNLO
determination, but now regularizing in turn the covariance matrix of
those datasets that exceeded the threshold value of the stability metric
(See
Tabs.~\ref{tab:dataset_selection_DIS}-\ref{tab:dataset_selection_OTHERLHCPROCESSES}
in Sect.~\ref{sec:dataselection}), with the threshold value  $Z_{\rm reg}=4$
now chosen as target clipping value.
 Results are shown in
Table~\ref{tab:tab-regcovmat}:  in each case we display the number of
data points, the value of
 $Z$ for the given experiment before
regularization ($Z_{\rm orig}$), and the $\chi^2$ for the experiment before and
after regularization.
Note that, based on the dataset selection procedure of Sect.~\ref{sec:dataselection},
the ATLAS $W$ 8 TeV and CMS 3D dijets 8 TeV datasets are
not part of the NNPDF4.0 baseline. In the former case,
the regularization has been applied to
a dedicated PDF determination in which the ATLAS data  have been
added to the baseline. In the latter case, the regularization has been
applied to the the PDF
determination shown in Table~\ref{tab:chi2_jet}, ``CMS 3D
dijets~8~TeV'' entry.
All other datasets listed in Table~\ref{tab:tab-regcovmat} are already part of the baseline. It is clear from Table~\ref{tab:tab-regcovmat} that after
regularization all $\chi^2$
values are of order unity, with the possible
exception of CMS~7~TeV dijets.
Note that 
the improvement in the values of the $\chi^2$ is not driven by an
increase in the diagonal 
elements of the covariance matrix, which remains smaller than 5\%,  but rather from
the regularization of the smallest eigenvectors.
It thus amounts to a minimal modification of the covariance matrix.

\begin{figure}[!t]
  \centering
  \includegraphics[width=0.49\textwidth]{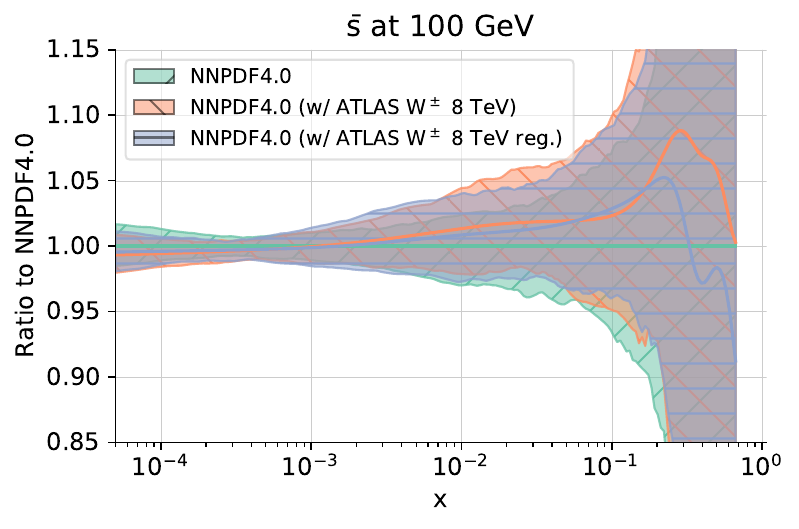}
  \includegraphics[width=0.49\textwidth]{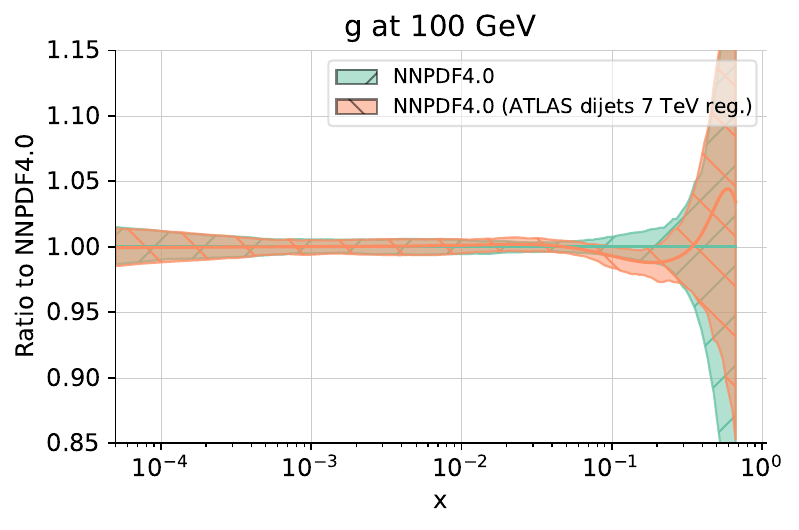}
  \caption{Comparison to the baseline of PDFs obtained by regularizing
    the covariance matrix for 
    the ATLAS $W$ 8 TeV (left) and ATLAS 7 TeV dijet dataset
    (right). In each case, the PDF which is most affected is shown:
    antistrange (left) and gluon (right).
    In the left plot both the default baseline and a baseline
    with the unregularized data are shown.}
  \label{fig:pdfplot-rat-nnpdf40-regularised_atlasw8tev}
\end{figure}

Interestingly, one also finds that the
best-fit PDFs are left almost unchanged by the regularization procedure.
Specifically, 
in  Fig.~\ref{fig:pdfplot-rat-nnpdf40-regularised_atlasw8tev} we
compare PDFs obtained regularizing
the ATLAS $W$ 8 TeV (left) and ATLAS 7~TeV dijet data (right) to the 
baseline PDFs. In the
former case, since the ATLAS $W$ 8~TeV are not part of the default
dataset, we show both the default baseline, and a modified version in
which this data has been added in unregularized form.
In each case, we show the PDF that is most affected by the 
regularization, respectively the antistrange and the gluon.
It is clear that, despite the large differences at the $\chi^2$ level,
the regularization procedure leaves the PDFs mostly unaffected.
This said, the effects of regularization are not completely negligible in all cases:
for example, for the ATLAS 7 TeV dijets at $x\simeq 0.2$ the gluon PDF
is suppressed by around one-sigma as compared to the baseline
in the regularized fits.
Nevertheless, these remain quite moderate effects, a feature which might appear somewhat counterintuitive given the large reduction in the $\chi^2$ values.

Our general conclusion is that a poor $\chi^2$ does not necessarily imply
a genuine inconsistency, since it can arise from ill-defined 
(unstable) covariance matrices. The specific conclusion for the datasets 
that have been examined here is that
we observe almost no difference between
PDFs determined with and without regularizing the corresponding covariance
matrices.
For the datasets that we retained in the baseline dataset, this
analysis confirms that the global fit
is not distorted by the poorly behaved nature of their covariance
matrices.

For the two datasets that we did not retain, the situation is
somewhat different. In the case of the  ATLAS $W$ 8 TeV data shown
in Fig.~\ref{fig:pdfplot-rat-nnpdf40-regularised_atlasw8tev}, there is
essentially no difference between PDFs determined including or not
including this dataset in regularized or unregularized form. For the
CMS~3D 8~TeV dijets, we see no difference between PDFs determined with
regularized or unregularized covariance matrix, but both differ
significantly from the baseline, as discussed in
Sect.~\ref{subsec:doublecounting}. Hence, in both cases the poor
$\chi^2$ is due to the properties of the covariance matrix, and we
confirm our decision not to include these datasets in the baseline: in
the former case on the grounds that it would make no difference, and
in the latter case for the reasons discussed in 
Sect.~\ref{subsec:doublecounting}.

%% file: tables/tab-regcovmat.tex
\begin{tabularx}{\textwidth}{XC{1.0cm}C{1.2cm}C{1.2cm}C{1.2cm}}
  \toprule
  Dataset  &  $N_{\rm dat}$   &   $Z_{\rm orig}$   &   $\chi^2_{\rm orig}$  &  $\chi^2_{\rm reg}$ 
  \\
  \midrule
  ATLAS $W,Z$ 7 TeV CC ($\mathcal{L}=4.6$~fb$^{-1}$)
  & 46
  &  9.01
  & 1.89
  & 0.93
  \\
   ATLAS $W$ 8 TeV (*) 
  & 22
  & 11.28
  & 3.50
  & 1.15
  \\
    CMS dijets 7 TeV
  & 54
  & 4.70
  & 1.81
  & 1.73
  \\
    ATLAS dijets 7 TeV
  & 90
  & 9.93
  & 2.14
  & 0.92
    \\
    CMS 3D dijets 8 TeV (*)
  & 122
  & 4.47
  & 1.50
  & 0.92
  \\
  \bottomrule
\end{tabularx}

%% file: sec-pheno.tex
\section{Phenomenology}
\label{sec:pheno}

We present a first study of the implications
of the NNPDF4.0 PDFs for hadron collider phenomenology.
Specifically,  we  compare the PDF luminosities at $\sqrt{s}= 14$~TeV from
NNPDF4.0 to other available PDF sets, and we then present
theoretical predictions obtained  using these PDF sets
for representative  LHC inclusive cross-sections and differential distributions.
Specifically, we consider inclusive gauge boson production, Higgs boson
production in different channels, and top quark pair production.
As we shall see,  PDF uncertainties found using
NNPDF4.0 are typically of the order of one percent 
for a broad range of observables and in a wide
kinematic region.

\input{subsec-pdflumis.tex}
\input{subsec-lhcpheno}

\input{subsec-lhcpheno-diff}

%% file: subsec-pdflumis.tex
\subsection{PDF luminosities}
\label{sec:pdflumis}

We evaluate here  PDF luminosities for different
parton initial state combinations.
We consider the parton luminosities  as a function of
the invariant mass of the final state $m_X$, both integrated
over rapidity and differential in rapidity, as defined in Eqs.~(1-4)
of Ref.~\cite{Mangano:2016jyj}.
 
\begin{figure}[p]
  \centering
  \includegraphics[width=.43\textwidth]{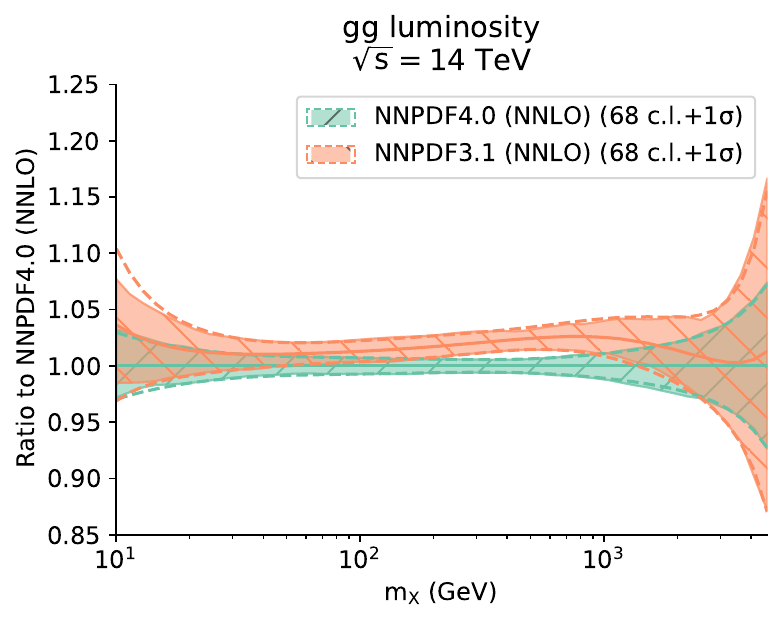}
  \includegraphics[width=.43\textwidth]{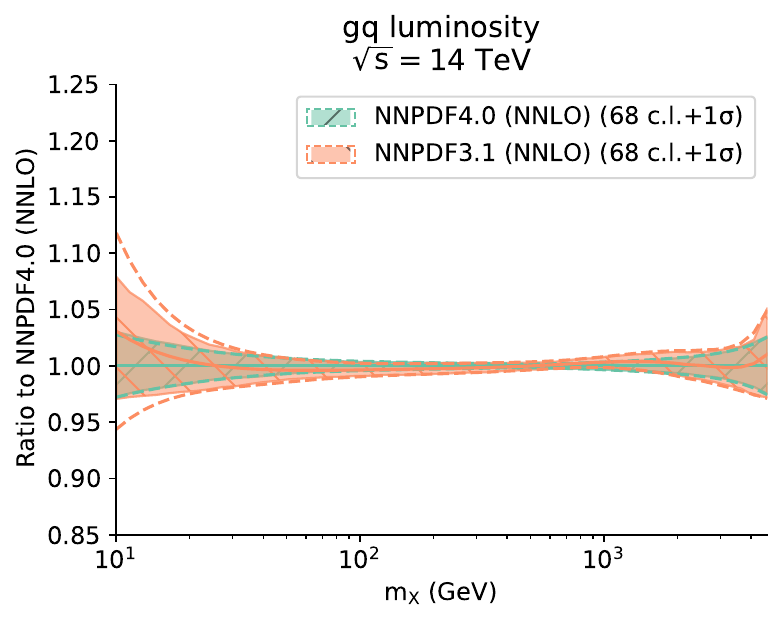}\\
  \includegraphics[width=.43\textwidth]{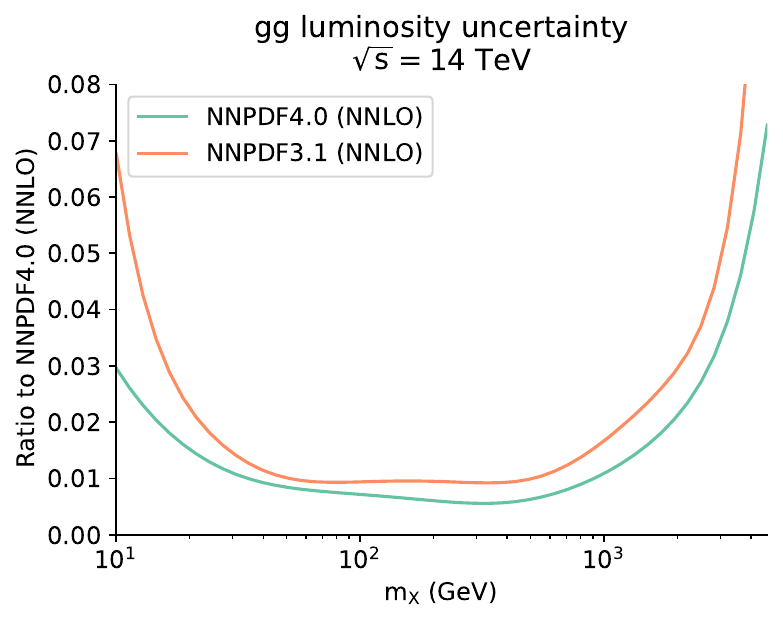}
  \includegraphics[width=.43\textwidth]{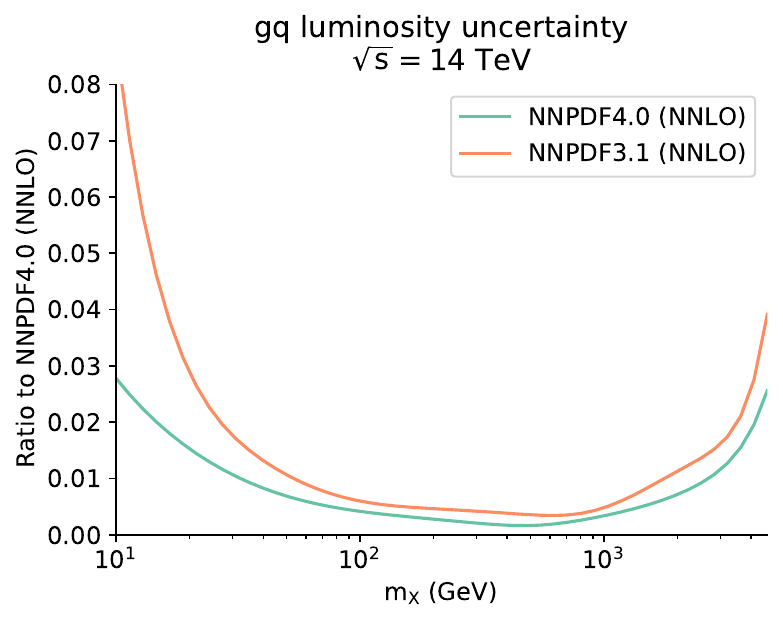}\\
  \includegraphics[width=.43\textwidth]{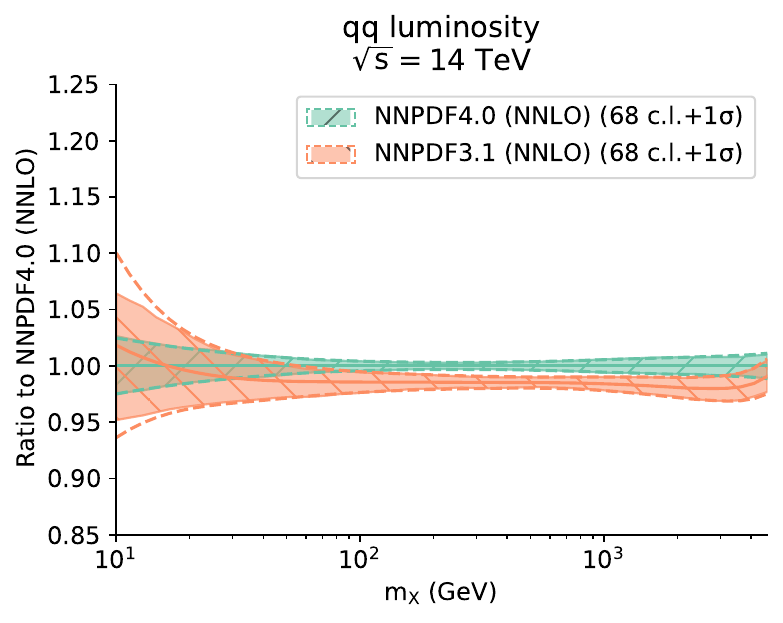}
  \includegraphics[width=.43\textwidth]{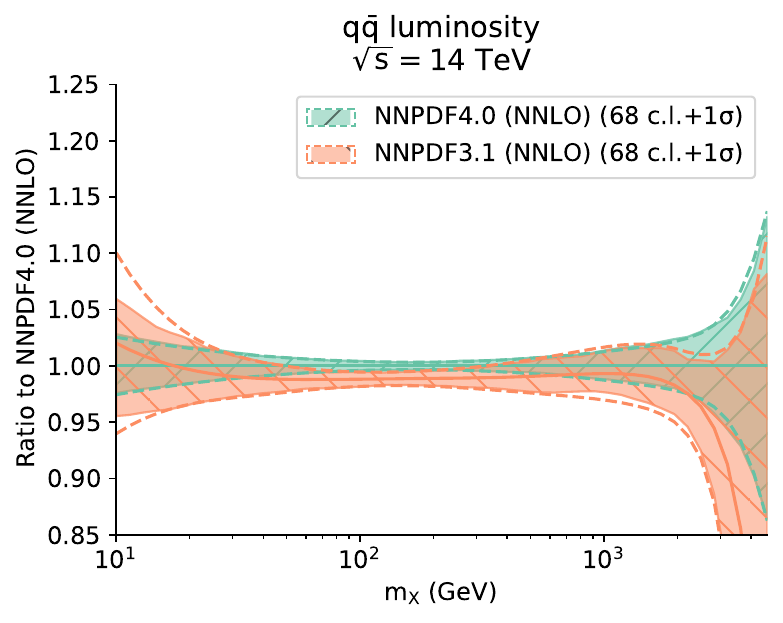}\\
  \includegraphics[width=.43\textwidth]{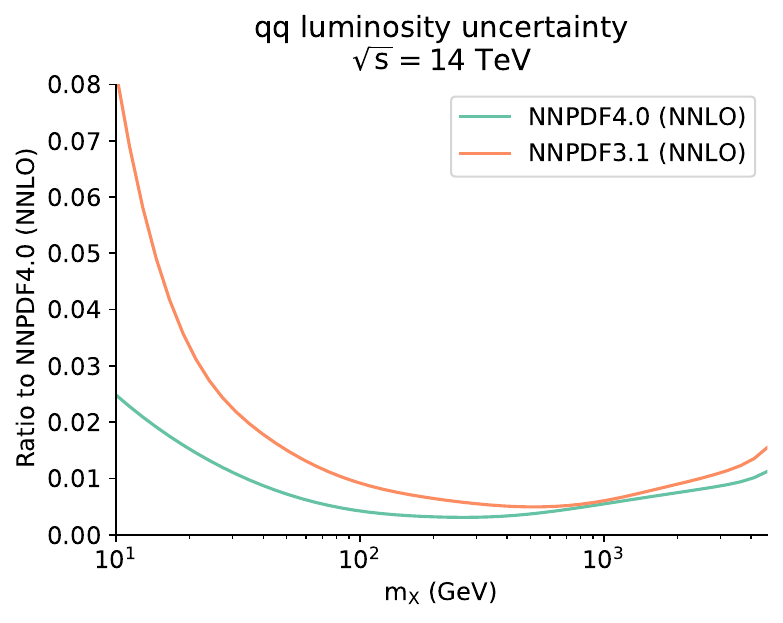}
  \includegraphics[width=.43\textwidth]{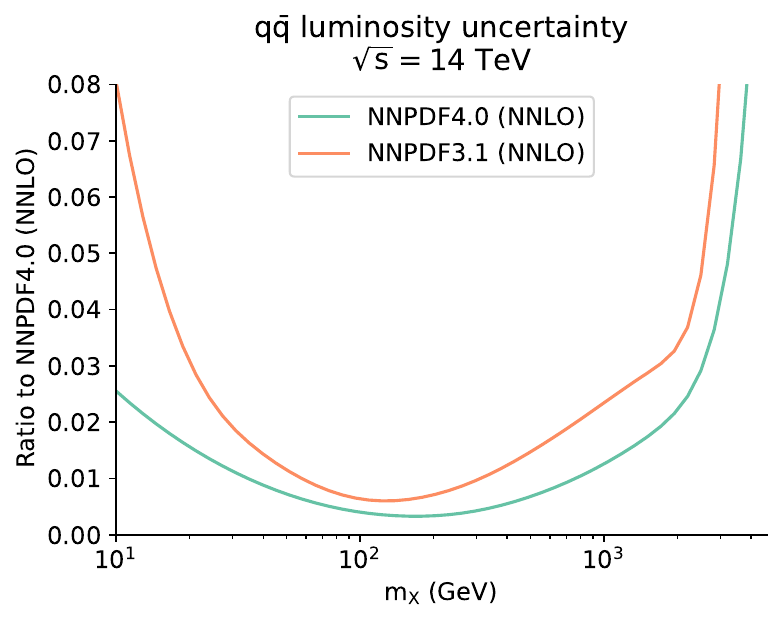}\\
  \caption{\small Comparison, as a function of the invariant mass $m_X$,  of
    the parton luminosities at $\sqrt{s}=14$~TeV
    computed using NNLO NNPDF4.0 and  NNPDF3.1 PDFs, where the luminosities have been integrated
    over the final-state rapidity $y$.
    The ratio to the
    NNPDF4.0 central value and the relative one-sigma uncertainty are shown for each
    parton combination.}
  \label{fig:nnpdf40_lumis_14tev} 
\end{figure}

\begin{figure}[p]
  \centering
  \includegraphics[scale=0.46]{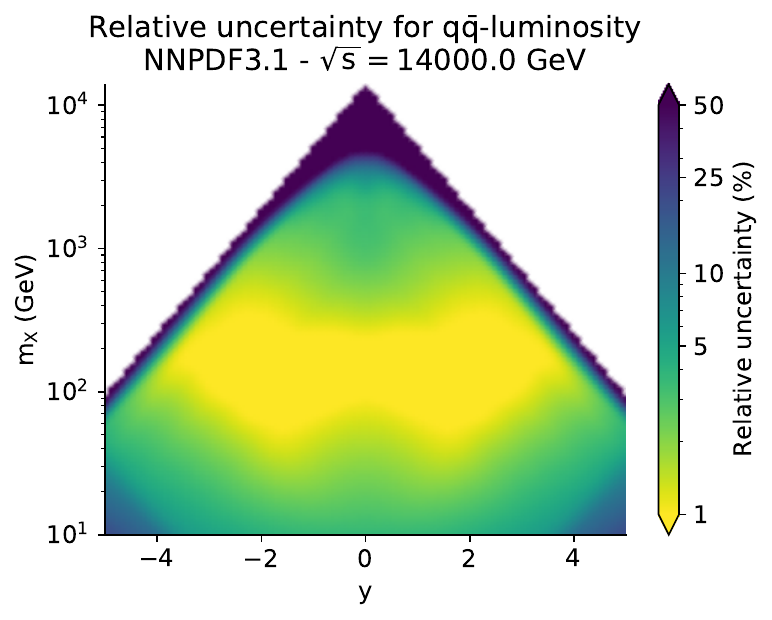}   
  \includegraphics[scale=0.46]{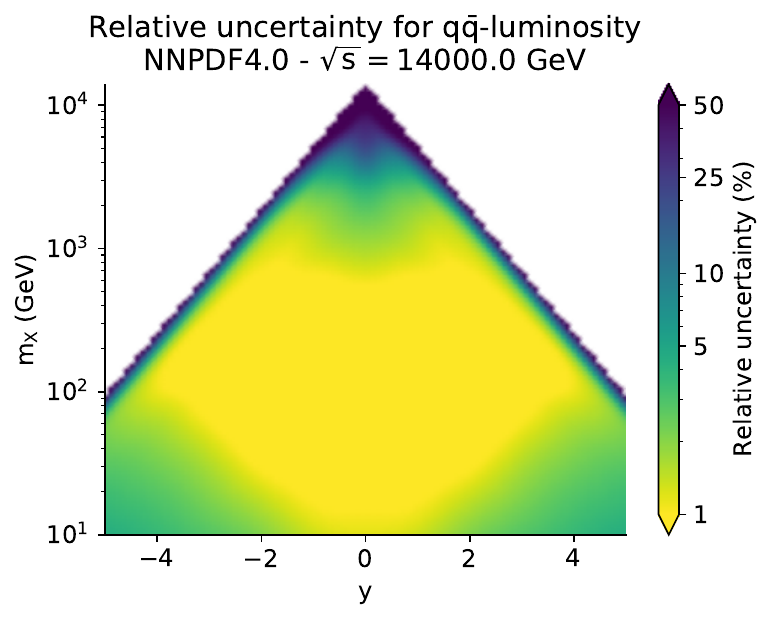}\\
  \includegraphics[scale=0.46]{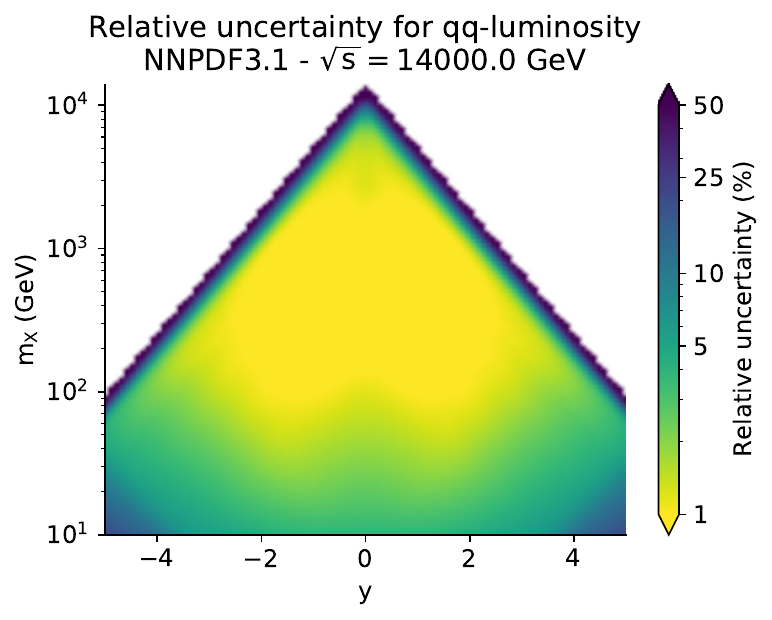}
  \includegraphics[scale=0.46]{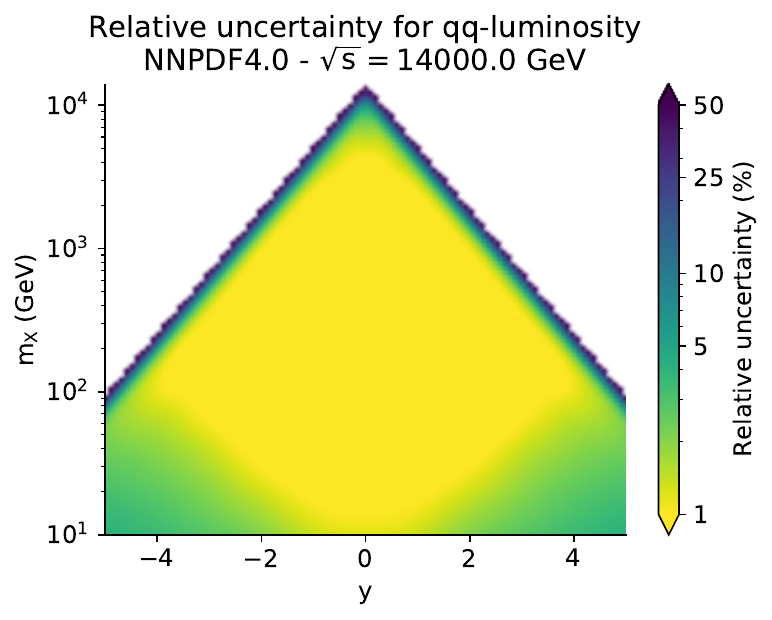}\\
  \includegraphics[scale=0.46]{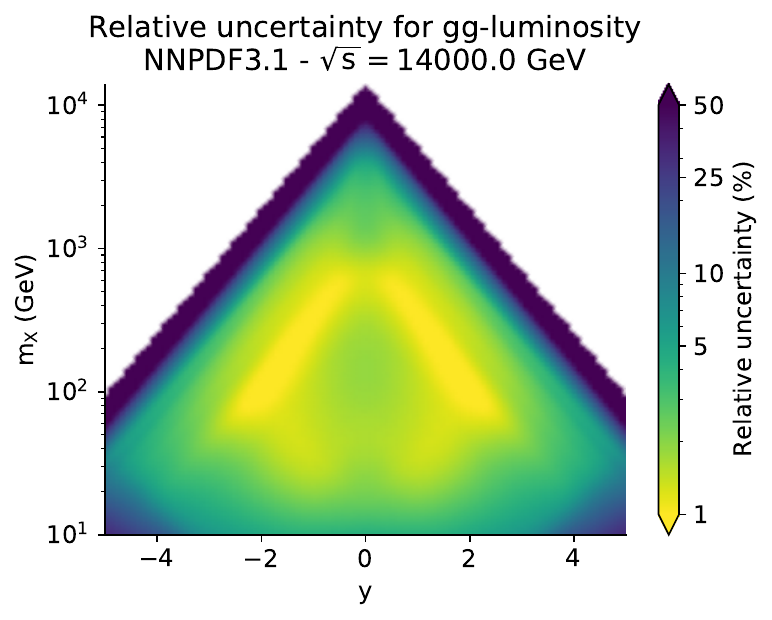}
  \includegraphics[scale=0.46]{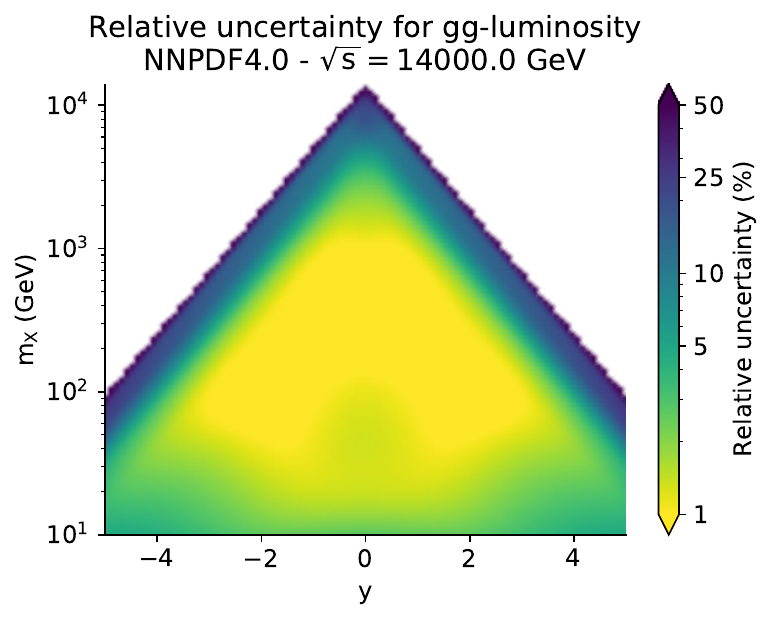}\\
  \includegraphics[scale=0.46]{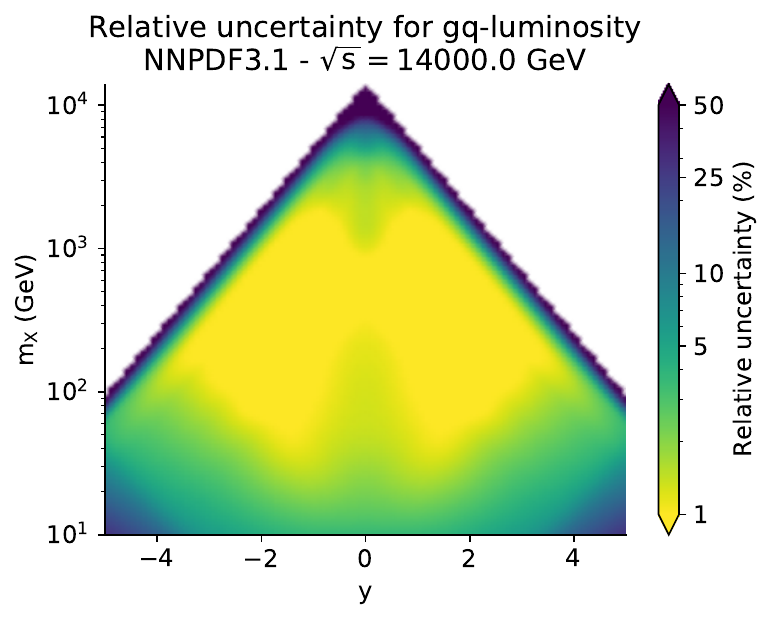}
  \includegraphics[scale=0.46]{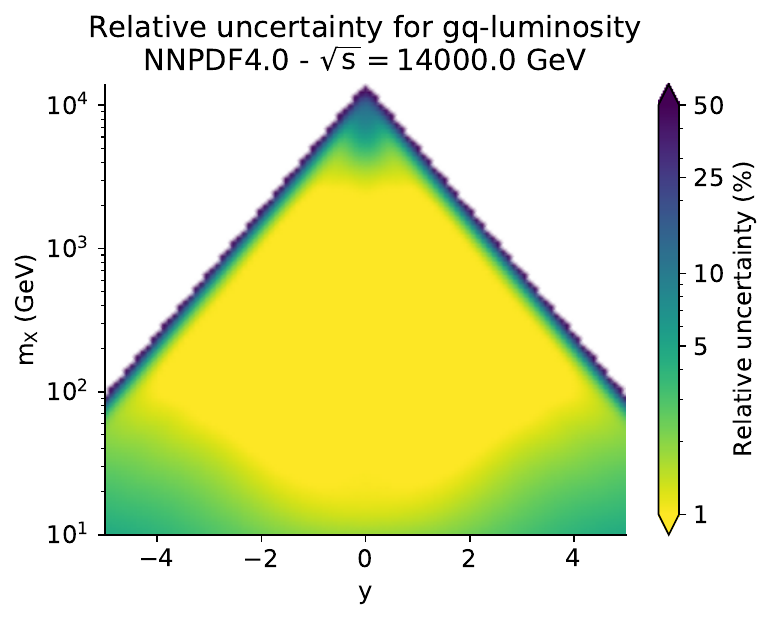}\\
  \includegraphics[scale=0.46]{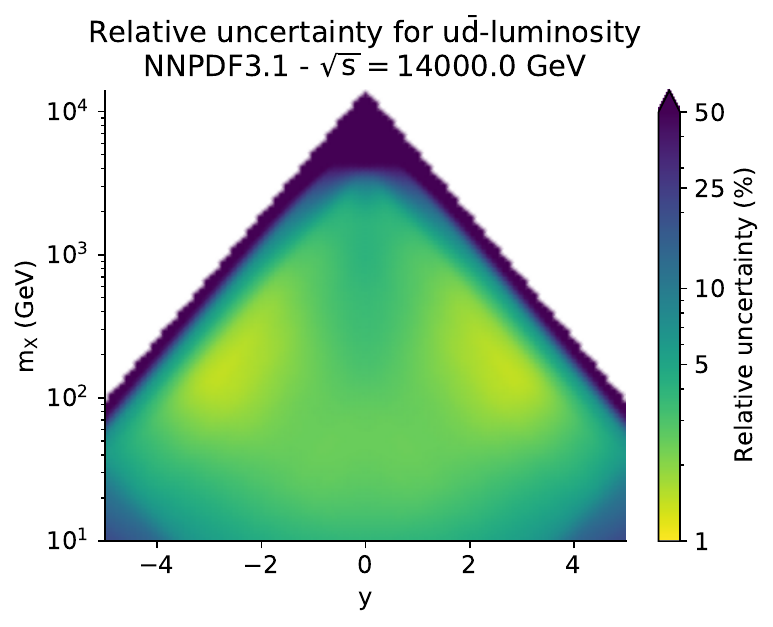}
  \includegraphics[scale=0.46]{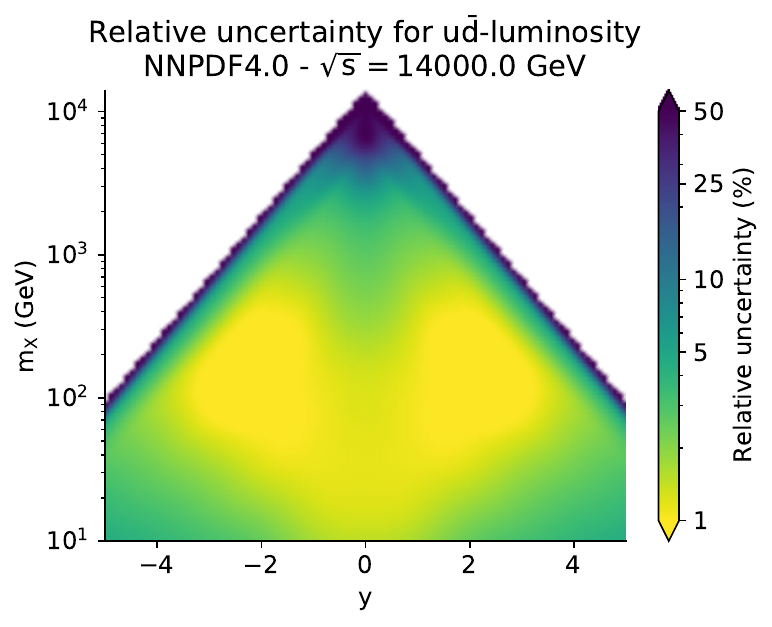}
  \caption{The relative uncertainty on the parton luminosities of
    Fig.~\ref{fig:nnpdf40_lumis_14tev}, now plotted as  a function of
    the invariant mass $m_X$ and the rapidity $y$ of the final
    state; the left plots show results for NNPDF3.1 and the right
    plots for NNPDF4.0; results for the up-antidown luminosity are
    also shown in the last row.}
  \label{fig:lumi-nnpdf40-2d}
\end{figure}

In Fig.~\ref{fig:nnpdf40_lumis_14tev} we compare
the luminosities integrated over rapidity, computed at $\sqrt{s}=14$~TeV using
NNPDF4.0 and NNPDF3.1 PDFs, as a function
of the final-state invariant mass $m_X$.
For each parton combination,
we show the ratio to the central NNPDF4.0 and the relative
one-sigma PDF uncertainty.
Then in Fig.~\ref{fig:lumi-nnpdf40-2d} percentage
uncertainties on the
parton luminosities differential in rapidity are shown
as a
two-dimensional contour plot as a function of the invariant mass $m_X$
and rapidity $y$ of the final state. In this case,  we also show
for reference the up-antidown luminosity (relevant e.g.\ for $W^+$
production). 

The first obvious observation is the significant reduction of PDF
uncertainties that was already observed in
Sect.~\ref{subsec:PDFs}. Indeed, it is clear, especially from
Fig.~\ref{fig:lumi-nnpdf40-2d}, that the uncertainty is now around 1\%
in a wide kinematic region and for several parton channels.
In terms of overall compatibility, all  luminosities 
agree at the one sigma level. 
While central values for the
quark-gluon and quark-antiquark luminosities are almost unchanged,
the quark-quark luminosity is somewhat enhanced  and the
gluon-gluon luminosity somewhat suppressed in NNPDF4.0
compared to NNPDF3.1, in the region $m_X\lesssim 3$~TeV.

\begin{figure}[p]
  \centering
  \includegraphics[width=.43\textwidth]{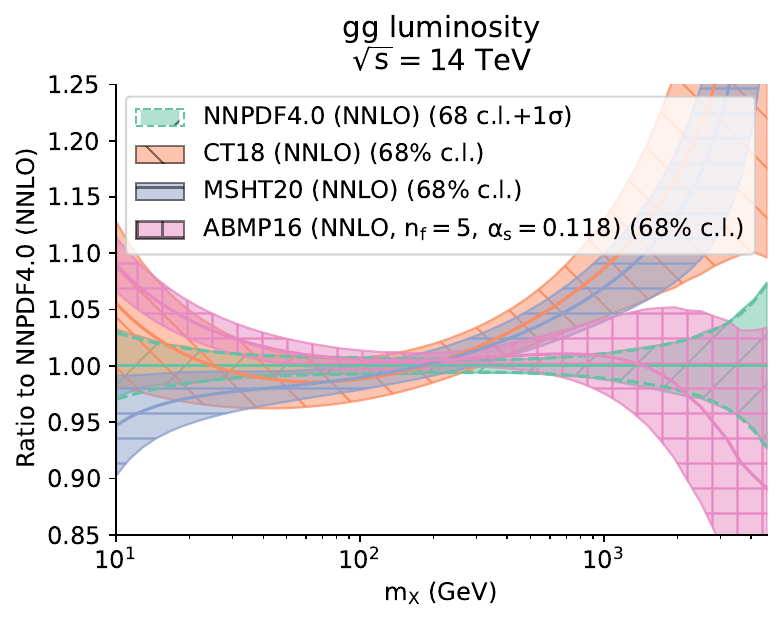}
  \includegraphics[width=.43\textwidth]{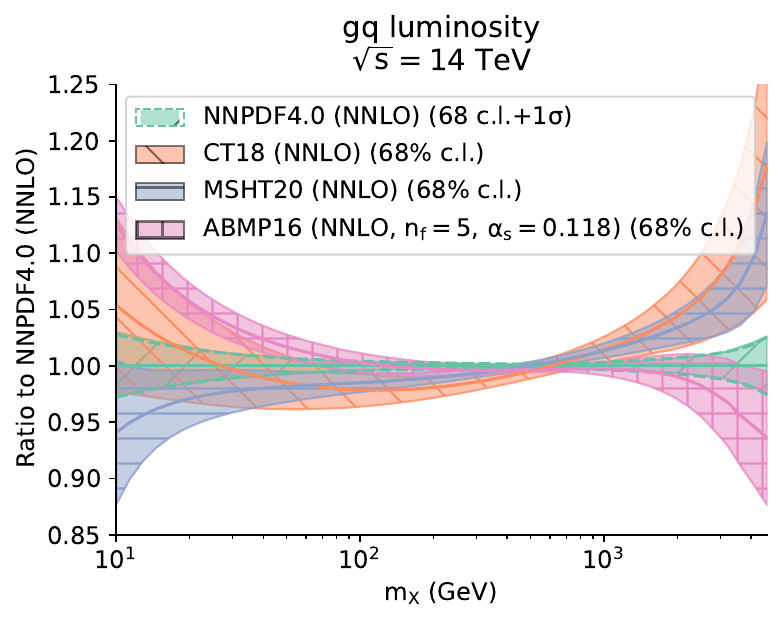}\\
  \includegraphics[width=.43\textwidth]{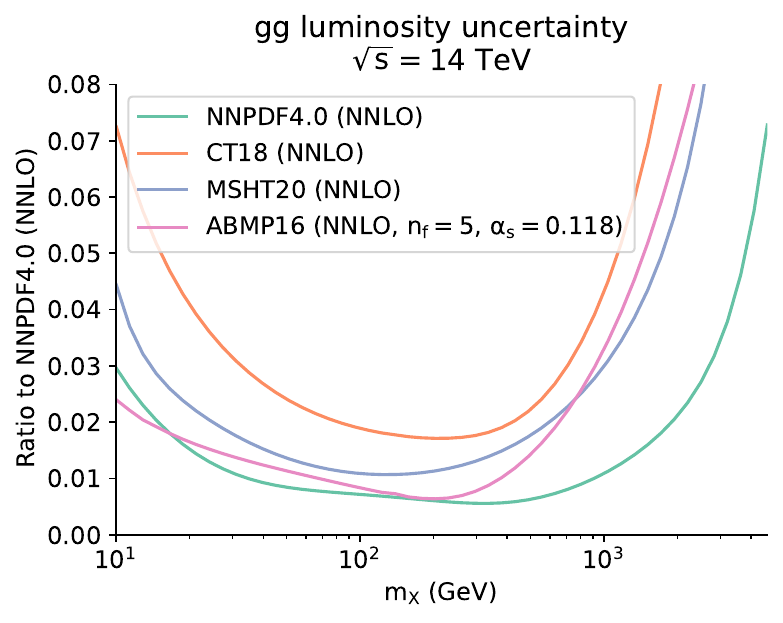}
  \includegraphics[width=.43\textwidth]{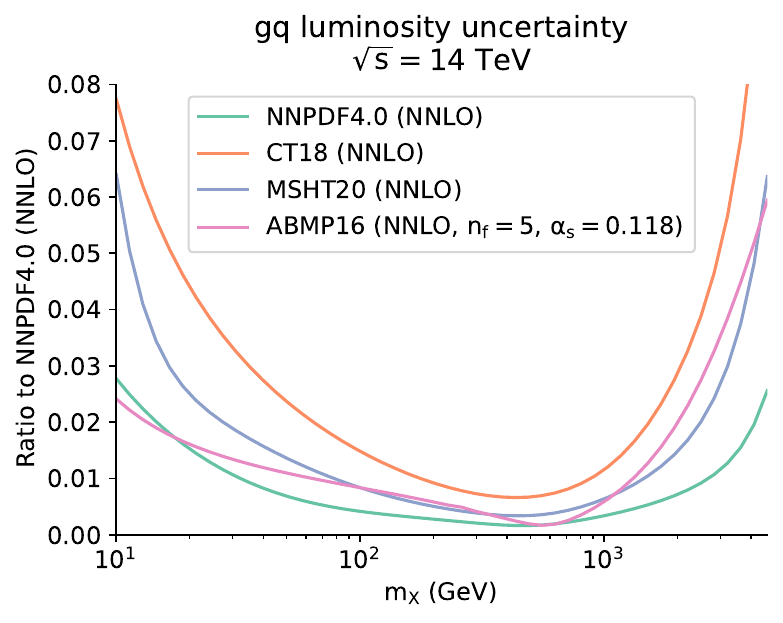}\\
  \includegraphics[width=.43\textwidth]{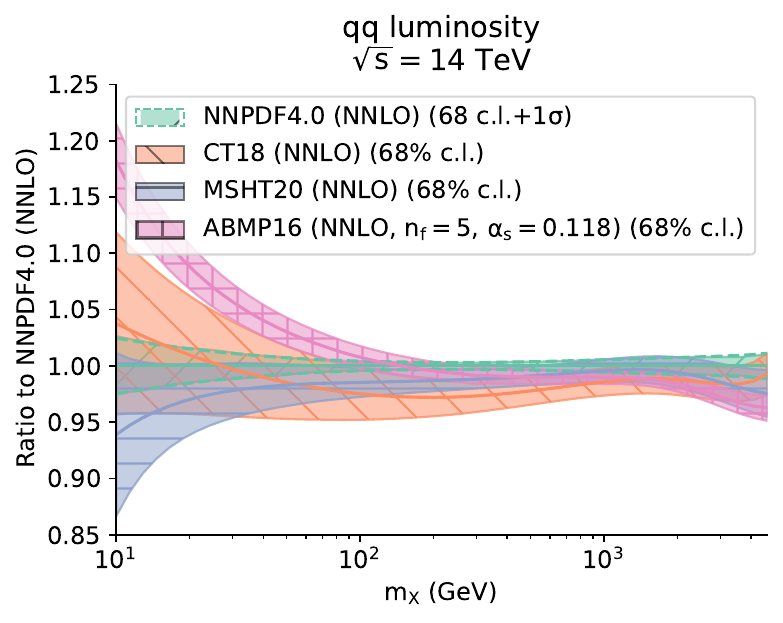}
  \includegraphics[width=.43\textwidth]{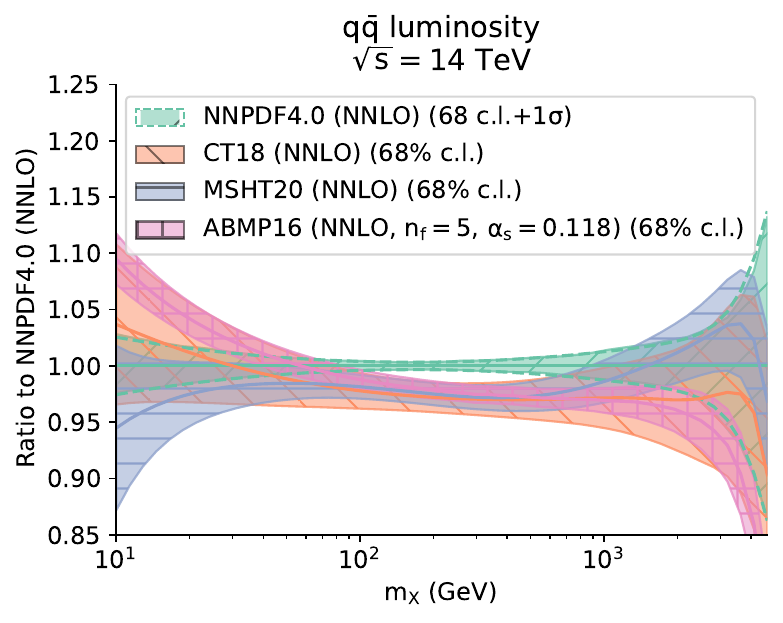}\\
  \includegraphics[width=.43\textwidth]{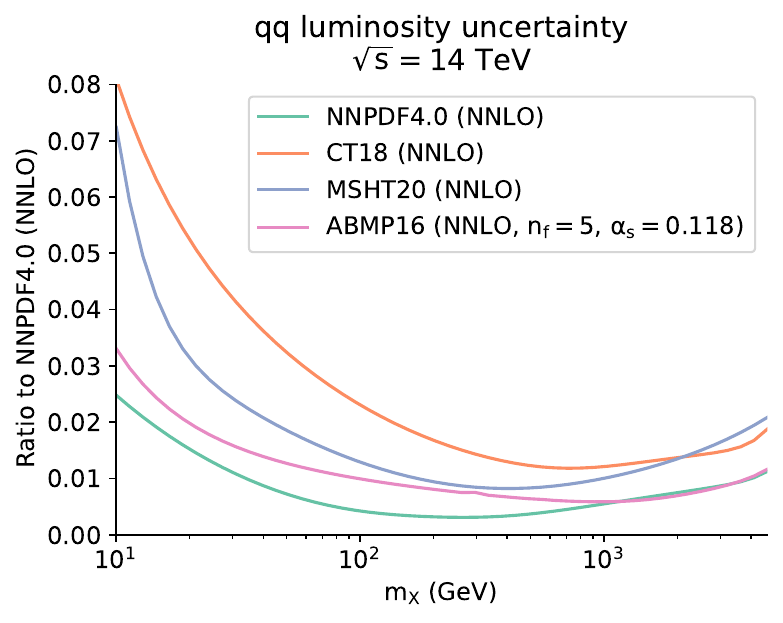}
  \includegraphics[width=.43\textwidth]{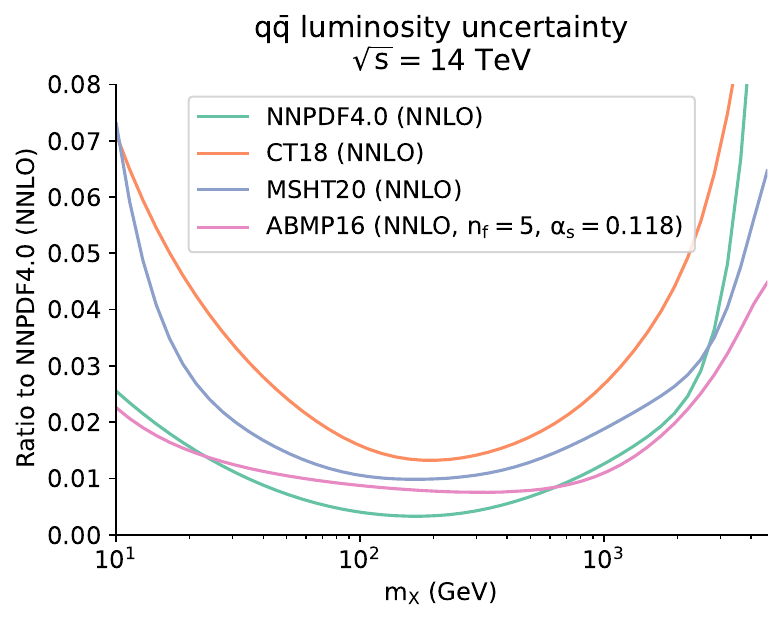}\\
  \caption{Same as Fig.~\ref{fig:nnpdf40_lumis_14tev} but now
    comparing NNPDF40, ABMP16, CT18, and MSHT20 PDFs.}
  \label{fig:nnpdf40_lumis_14tev_global}
\end{figure}

We next compare the NNPDF4.0 luminosities integrated in rapidity
to those obtained using PDFs from the
CT18~\cite{Hou:2019efy}, MSHT20~\cite{Bailey:2020ooq} and
ABMP16~\cite{Alekhin:2017kpj} sets.
When comparing uncertainties, it should
be kept in mind that while CT18, MSHT20 and
    ABMP16 all adopt a Hessian methodology with a fixed functional
    form,
    their respective treatments of
    uncertainties differ. Specifically, both CT18 and MSHT20 adopt
    a ``tolerance''~\cite{Pumplin:2002vw,Martin:2003tt} criterion and
    further study functional form dependence in
    order to span adequately the space of parametrizations, while ABMP do not. Hence, CT18 and MSHT20 uncertainties
    are directly comparable to those of NNPDF (which adopts a very general
    neural network parametrization), while ABMP uncertainties
    generally are not.
The same common value of $\alpha_s(m_Z)=0.118$
is used in all cases. Note that this significantly differs from the value
$\alpha_s(m_Z)=0.113$ adopted as default in Ref.~\cite{Alekhin:2017kpj}.
Again as already observed in
Sect.~\ref{subsec:PDFs}, it is clear that NNPDF4.0 generally has the
smallest uncertainty. An exception is ABMP16 in some regions (such as
the gluon-gluon luminosity for low invariant mass), possibly for the
reason mentioned above (as
already pointed out in Ref.~\cite{Ball:2017nwa}.

All luminosities agree within uncertainties in the region around
$m_X\sim 100$ GeV,
relevant e.g. for Higgs and gauge boson production.
Furthermore, the
quark-quark and quark-antiquark luminosity are in good agreement
within uncertainties over 
the full mass range. For the gluon sector luminosities (gluon-gluon and
gluon-quark), however, differences are seen at large mass.
Specifically, in the high-mass region, $m_X\gsim 1$ TeV, the gluon-gluon  and
quark-gluon 
luminosities for  NNPDF4.0 are rather smaller than  MSHT20 and CT18,
though they  agree with ABMP16.
These differences are possibly a consequence of the fact that 
NNPDF4.0 includes a
variety of data which are sensitive to the  gluon and are not
used by other groups,  in particular
the dijet cross-sections at 7 TeV and the $t\bar{t}$ differential distributions from
the LHC Run II.

A full understanding of the origin of the differences between PDFs
determined by different groups and their impact on LHC phenomenology
would require a  dedicated benchmark study,
such as the ones carried out for the
PDF4LHC15~\cite{Butterworth:2015oua} and
PDF4LHC21~\cite{Cridge:2021qjj,Ball:2022hsh} combinations for NNPDF3.0
and NNPDF3.1 respectively.
In the remainder of this section,
we will assess how these differences at the level
of parton luminosities translate into LHC cross-sections and distributions.

%% file: subsec-lhcpheno.tex
\subsection{Inclusive cross-sections}
\label{sec:pheno-integrated-xsecs}

We present theory predictions for 
representative LHC processes, first for
integrated cross-sections and then for the corresponding differential
distributions, based on the luminosities
discussed in Sect.~\ref{sec:pdflumis}.
In all cases, realistic acceptance requirements and final state kinematic cuts are imposed,
in order to provide theoretical predictions which are as close as possible to the
associated experimental measurements.
All cross-sections are evaluated at $\sqrt{s}=14$~TeV.

We consider the following processes: neutral and charged current Drell--Yan
production in the leptonic final state,  top pair production,
gauge boson pair production (both in the $\mathrm{W}^+\mathrm{W}^-$ and the $\mathrm{W}^\pm\mathrm{Z}$ channels),
inclusive Higgs  production
via gluon fusion or  vector boson fusion, and the associated
production of Higgs and $\mathrm{W}^{\pm}$.
Note that some of these processes are already part of the NNPDF4.0
determination, but at a different center-of-mass energy:
specifically,  neutral current (dilepton) Drell--Yan production and top pair
production data are included for center-of-mass energies of $7$, $8$ and 
$13$~TeV.

\paragraph{Calculational settings.}
Results presented in this section have been produced using  \texttt{MadGraph5\_aMC@NLO}~\cite{Alwall:2014hca,Frederix:2018nkq}
and account for complete NLO corrections both in the QCD and electroweak couplings.
These \texttt{mg5\_aMC} calculations have been interfaced
to \texttt{PineAPPL}~\cite{Carrazza:2020gss}, which produces interpolation 
grids so that the LHC predictions
can be quickly evaluated for arbitrary PDF sets without redoing the MC integration.
In the specific case of top pair production,
our calculation include only the $\mathcal{O} (\alpha_\mathrm{s}^2)$ and $\mathcal{O} (\alpha_\mathrm{s} \alpha)$ terms at LO and the $\mathcal{O} (\alpha_\mathrm{s}^3)$ and $\mathcal{O} (\alpha_\mathrm{s}^2 \alpha)$ corrections at NLO.
This is justified since the pure-EW and mixed corrections that we
neglect, namely  $\mathcal{O} (\alpha^2)$, $\mathcal{O} (\alpha_\mathrm{s} \alpha^2)$ and $\mathcal{O} (\alpha^3)$, are very small in the kinematic regions under consideration~\cite{Pagani:2016caq}.

For electroweak gauge boson production,
we account for their decays into leptons.
In order to simplify the calculation, we choose the flavors of these
final-state leptons 
to be different from each other, so as to minimize  the number of
Feynman diagrams.
This, for example, avoids the overlap of $\mathrm{Z}\mathrm{Z}$ with
$\mathrm{W}^+\mathrm{W}^-$ diboson production, both of which can decay
into the $\ell\bar{\ell}\nu_\ell\bar{\nu}_\ell$ final state, while
only the later can decay into the
$\ell\bar{\ell}'\nu_{\ell'}\bar{\nu}_\ell$ final state, which is the
one we have selected. 

For all calculations except Higgs production,
we use the model \texttt{loop\_qcd\_qed\_sm\_Gmu} with enabled
complex-mass scheme~\cite{Denner:1999gp,Denner:2005fg,Denner:2006ic} as implemented by Ref.~\cite{Frederix:2018nkq}.
For Higgs production we use the UFO model
of~\cite{Artoisenet:2013puc,Demartin:2014fia} with an effective
Higgs--gluon--gluon coupling, for which EW corrections vanish.
The following are taken as   independent input parameters:
\begin{equation}
\begin{aligned}
     m_\mathrm{W}  &= 80.352~{\rm GeV} \text{,} &
\Gamma_\mathrm{W}  &= 2.084~{\rm GeV} \text{,} &
     m_t           &= 172.5~{\rm GeV} \text{,} \\
m_\mathrm{Z}       &= 91.1535~{\rm GeV} \text{,} &
\Gamma_\mathrm{Z}  &= 2.4943~{\rm GeV} \text{,} &
\Gamma_t           &= 1.37758~{\rm GeV} \text{,} \\
m_\mathrm{H}       &= 125.0~{\rm GeV} \text{,} &
\Gamma_\mathrm{H}  &= 4.07468 \times 10^{-3}~{\rm GeV} \text{,} &
G_\mu              &= 1.166378 \times 10^{-5}~/{\rm GeV}^2 \text{,}
\end{aligned}
\end{equation}
which are directly fed into the \texttt{mg5\_aMC} calculation.
In the case of top pair production we assume 
stable top quarks in the final state, which corresponds to setting the
top-quark width to $\Gamma_\mathrm{t} = 0$. 
All calculations with final-state leptons employ a dressed lepton definition which recombines leptons with photons
if their separation is smaller than $\Delta R_{\ell\gamma} < 0.1$.
Furthermore, in all cases
each process is defined inclusively with respect to additional particles such as jets and photons.

While all the results presented in this section have been obtained
using NNLO PDF sets, we note that they are based on matrix
elements evaluated at NLO accuracy in the QCD coupling,
i.e.\ NNLO QCD corrections are not included.
This procedure is adequate in order
to discuss features and differences of PDF sets, which is our main goal
here, but of course not for precision phenomenology.\\

\noindent
We now provide in turn specific information about the calculational
settings, acceptance requirements, and final-state selection cuts
for each of the processes under consideration.

\paragraph{Drell--Yan lepton-pair production.}
For this process, mostly dominated by the exchange of an off-shell $\mathrm{Z}$,
 we require exactly two same-flavor opposite-sign leptons.
These two leptons must satisfy the central acceptance cuts of
$p_\mathrm{T}^\ell > 15~{\rm GeV}$ and $|\eta_\ell| < 2.4$,
while their invariant mass must fulfill $40~{\rm GeV} < m_{\ell\bar{\ell}} < 3000~{\rm GeV}$, similar to the CMS 13~TeV analysis~\cite{Sirunyan:2018owv}.
The factorization and renormalization scales are set dynamically to
$\mu = \langle m_{\ell \ell} \rangle$, where $\langle
m_{\ell \ell} \rangle$ represents the center of each bin in the
dilepton invariant mass distribution (see Fig.~\ref{fig:pheno-dy}). 

\paragraph{Charged vector-boson production.}
This process is dominated by the exchange of an off-shell $\mathrm{W}$ boson,
hence the acceptance cuts imposed on the final-state charged lepton (of any flavor) are
$p_\mathrm{T}^\ell > 20~{\rm GeV}$ and $|\eta_\ell| < 2.5$.
In this case we adopt  fixed factorization and renormalization scales,
set to the 
value of the $\mathrm{W}$-boson mass $\mu = m_\mathrm{W}$.

\paragraph{Diboson production.}
We consider gauge boson pair production in the $\mathrm{Z} \mathrm{W}^\pm$ and $\mathrm{W}^+\mathrm{W}^-$ channels,
with bosons subsequently decaying leptonically.
We impose cuts on the final-state
leptons of $p_\mathrm{T}^\ell > 20~ {\rm GeV}$ and $|\eta_\ell| < 2.5$.
In the $\mathrm{W}^+ \mathrm{W}^-$ channel, we require two opposite-sign charged leptons from the boson decays
with different lepton flavors.
Also for this process we set $\mu = m_\mathrm{W}$.

\paragraph{Top pair production.}
The simulation of this process is carried out at the level of stable top quarks.
We impose that the invariant mass $m_{\mathrm{t}\bar{\mathrm{t}}}$
of the top-quark pair system be within the range
$300~{\rm GeV} < m_{\mathrm{t}\bar{\mathrm{t}}} < 2500~{\rm GeV}$,
and adopt the same choice of binning as that used for  CMS in their 13~TeV
analysis~\cite{Sirunyan:2018wem} based on the lepton+jet final state.
The factorization and renormalization scales depend on the event kinematics and are set dynamically to
\begin{equation}
\mu = H_\mathrm{T}/4 = \frac{1}{4} \left[ \sqrt{m_\mathrm{t}^2 + (p_\mathrm{T}^\mathrm{t})^2} + \sqrt{m_\mathrm{t}^2 + (p_\mathrm{T}^{\bar{\mathrm{t}}})^2} \right] \text{,}
\end{equation}
where $p_\mathrm{T}^{\mathrm{t}}$ and $p_\mathrm{T}^{\bar{\mathrm{t}}}$ indicate 
the transverse momentum of the top and antitop quarks, respectively.

\paragraph{Higgs production via gluon fusion.}
For the simulation of all the Higgs production processes we consider a
stable Higgs, since its decays do not 
contain relevant information on the PDF sensitivity of the process.
We require the Higgs  to be produced in the central region, $|y_\mathrm{H}| < 2.5$,
and use fixed scales of $\mu = m_\mathrm{W}$.

\paragraph{Higgs production with associated $\mathrm{W}^\pm$ boson.}
For this Higgs production channel, in addition to the central production
requirement $|y_\mathrm{H}| < 2.5$ we impose the same cuts on the charged lepton arising
from the $\mathrm{W}^{\pm}$ decay as for charged-current Drell--Yan production, namely
 $p_\mathrm{T}^\ell > 20~{\rm GeV}$ and $|\eta_\ell| < 2.5$.
 Here we also set the scales to $\mu = m_\mathrm{W}$.

\paragraph{Higgs production in vector boson fusion.}
In this case, in addition to the centrally produced Higgs we require a final state
with (at least) two anti-$k_\mathrm{t}$ jets of radius $R = 0.4$.
These forward tagging jets must satisfy $p_\mathrm{T}^\mathrm{j} > 20~{\rm GeV}$, $|y_\mathrm{j}| < 4.5$,
with a dijet invariant mass of $m_{\mathrm{j}_1\mathrm{j}_2} > 500~{\rm GeV}$ and
a rapidity separation of $|y_{\mathrm{j}_1} - y_{\mathrm{j}_2}| > 2.5$, where $\mathrm{j}_1$ is the leading, and $\mathrm{j}_2$ the subleading jet (ordered in $p_T$).
As for the other Higgs production processes,
the scale is set to $\mu = M_\mathrm{W}$.

\begin{figure}[t]
  \begin{center}
\includegraphics[width=0.45\textwidth]{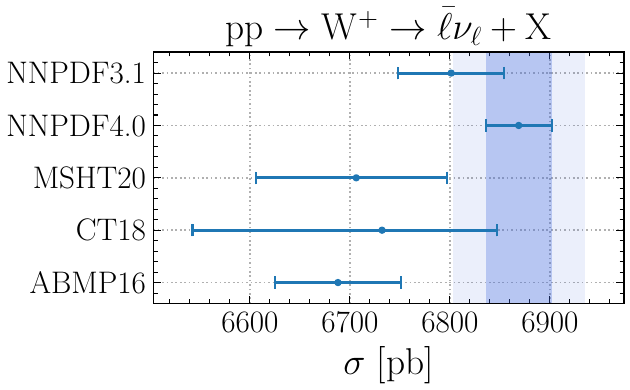}
\includegraphics[width=0.45\textwidth]{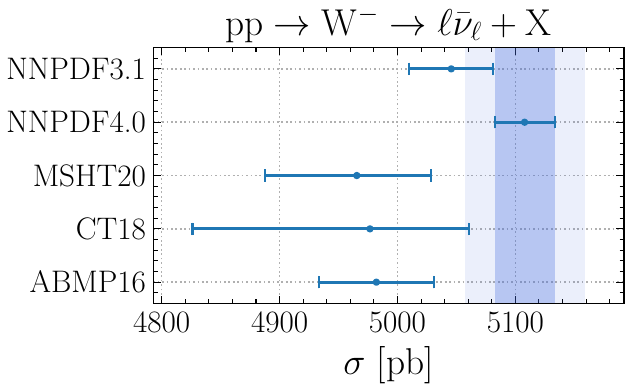}\\
\includegraphics[width=0.45\textwidth]{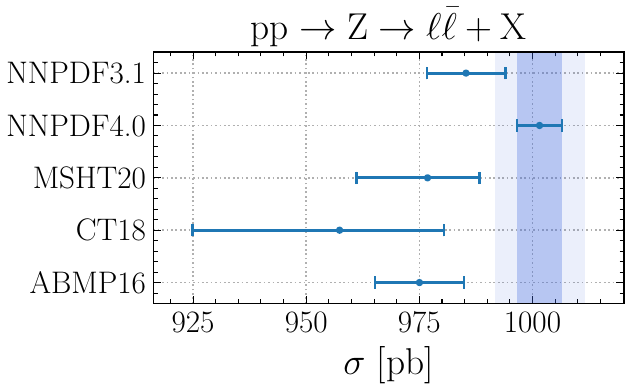}
\includegraphics[width=0.45\textwidth]{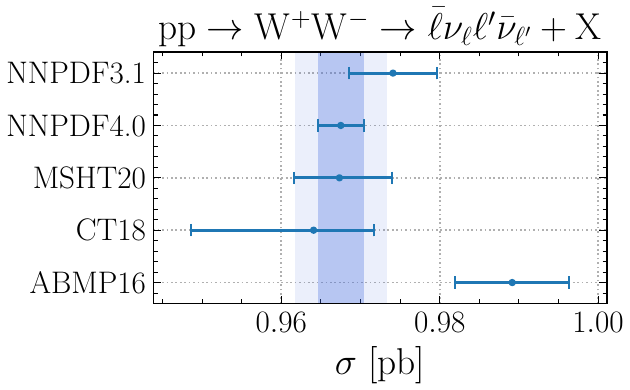}\\
\includegraphics[width=0.45\textwidth]{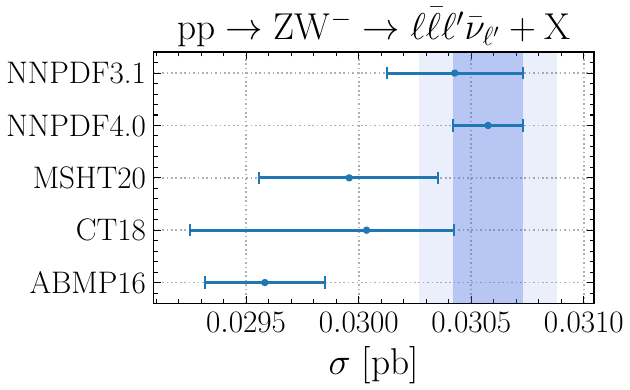}
\includegraphics[width=0.45\textwidth]{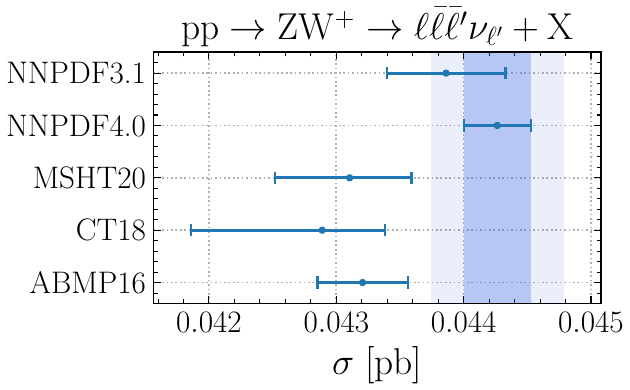}
\caption{\label{fig:LHC_integrated_xsecs_1}
  Integrated LHC cross-sections at $14$~TeV for neutral and charged-current Drell--Yan production
(top) and gauge boson pair production (bottom) obtained with a
variety of different PDF sets, all with  $\alpha_s(m_Z)=0.118$.
The edges of $1\sigma$ and $2\sigma$ PDF uncertainty bands for NNPDF4.0 are
indicated by dark and light lines respectively.
}
\end{center}
\end{figure}

\begin{figure}[t]
  \begin{center}
\includegraphics[width=0.45\textwidth]{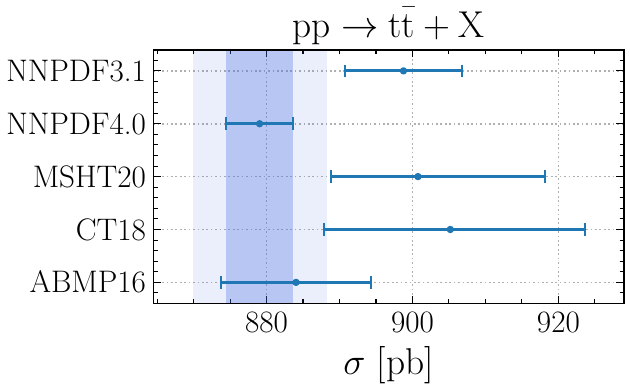}%
\includegraphics[width=0.45\textwidth]{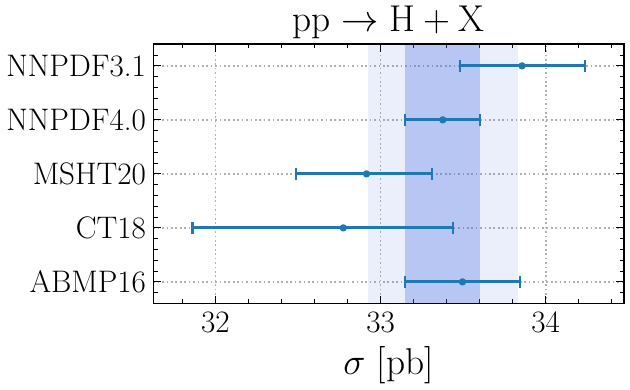}\\
\includegraphics[width=0.45\textwidth]{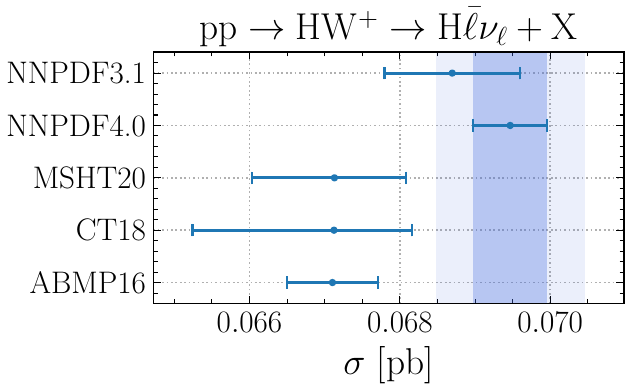}
\includegraphics[width=0.45\textwidth]{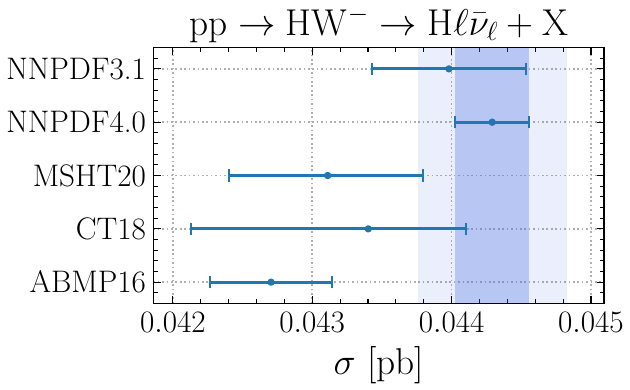}\\
\includegraphics[width=0.45\textwidth]{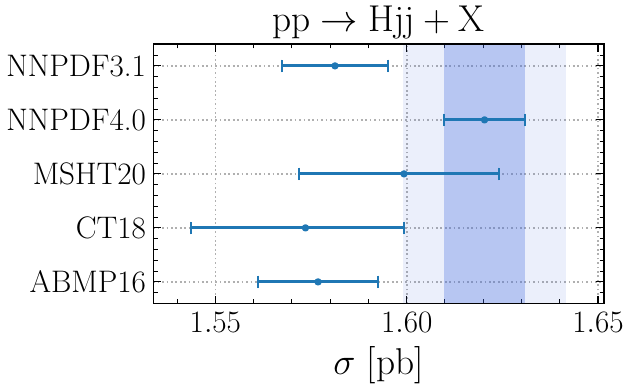}%
\caption{\label{fig:LHC_integrated_xsecs_2} Same as Fig.~\ref{fig:LHC_integrated_xsecs_1} for top
pair production and for Higgs production in different channels: gluon fusion, associated production
with a $\mathrm{W}^\pm$ boson, and vector-boson fusion.}
\end{center}
\end{figure}

\paragraph{Results.}
Using the calculational settings described above, we have computed differential distributions
(to be discussed below) and then combined the bins into integrated cross-sections.
Figs.~\ref{fig:LHC_integrated_xsecs_1} and~\ref{fig:LHC_integrated_xsecs_2}
display the integrated LHC cross-sections at $14$~TeV for the  processes under consideration:
neutral and charged-current Drell--Yan production,
gauge boson pair production,
top-quark
pair production, and Higgs production in different channels: gluon fusion, associated production
with a $\mathrm{W}^\pm$ boson, and vector-boson fusion.

We compare results obtained using the NNPDF3.1, NNPDF4.0, CT18, MSHT20,
and ABMP16 PDFs, in all cases with a common value of $\alpha_s(m_Z)=0.118$.
In order to facilitate visualization of the statistical compatibility
between results obtained NNPDF4.0 and all other PDF sets, we display
vertical bands indicating
the $1\sigma$~(dark) and $2\sigma$ (light) uncertainty ranges of the NNPDF4.0
prediction. 
For CT18 and MSHT20,  PDF uncertainties are computed with the asymmetric Hessian prescription
so positive and negative uncertainties generally differ.

For charged- and neutral-current DY production, we observe
good agreement at the $1\sigma$
level between NNPDF3.1 and NNPDF4.0, consistent with the comparisons
at the luminosity level reported
in Sect.~\ref{sec:pdflumis}.
The NNPDF4.0 cross-sections are found to be higher than those of MSHT20 and CT18, as expected
given the larger $\mathrm{q}\bar{\mathrm{q}}$ luminosity in the $m_X\simeq m_V$ region which dominates
the integrated cross-section shown in Fig.~\ref{fig:nnpdf40_lumis_14tev_global},
with central values in agreement at the $1\sigma$ or at most $2\sigma$ level.
For these three cross-sections, the smaller PDF uncertainties of NNPDF4.0 compared to MSHT20
and especially CT18 that was observed in Sect.~\ref{sec:pdflumis} is
clearly visible.

For the diboson production cross-sections, the comparison
presents different features according to the specific final state.
Indeed, diboson production is generally dominated by quark-quark
scattering, so the specific
partonic  combination depends on the final state, and the
total quark-quark luminosity $\mathcal{L}_{qq}$  only provides a crude
average measure.
While there is always excellent compatibility between NNPDF3.1 and NNPDF4.0,
the comparison to other groups differs according to the specific process.
For $\mathrm{W}^+\mathrm{W}^-$ production, NNPDF4.0 agrees well  with
CT18 and MSHT20, while the ABMP16 result is significantly larger.
For $\mathrm{W}^-\mathrm{Z}$, NNPDF4.0 is somewhat higher 
than the  other groups, though all except ABMP16 agree within uncertainties. For
$\mathrm{W}^+\mathrm{Z}$ NNPDF4.0 is higher than the other groups by
about $2\sigma$.
NNPDF4.0 uncertainties are in general markedly smaller than those of the 
other groups, just as for DY production.

For top quark
pair production, shown  in Fig.~\ref{fig:LHC_integrated_xsecs_2}, 
there is general agreement at the $1\sigma$ level, with 
NNPDF4.0 somewhat lower than CT18 and MSHT10, as expected from the
luminosity comparison in the relevant  invariant mass
$m_X \simeq 450~{\rm GeV}$ region. Here too NNPDF4.0
leads to rather smaller PDF uncertainties.

For Higgs production we consider gluon fusion, associated production
with vector bosons, and vector-boson fusion. For
gluon fusion there is excellent agreement within uncertainties 
between all the groups.
Interestingly, the NNPDF4.0 result, while still in excellent agreement with
its NNPDF3.1 predecessor, now has a central value rather closer to that of the
other groups.
For associated production with gauge bosons, $\mathrm{H}\mathrm{W}^+$ and
$\mathrm{H}\mathrm{W}^-$, the observed pattern is similar to
charged-current DY, as expected due to the closely related
underlying luminosities,
with  NNPDF3.1 and NNPDF4.0 in agreement and
higher than other groups.
For vector-boson-fusion, the NNPDF4.0 cross-section is higher than all the 
earlier determinations, and agrees best within uncertainties with MSHT20.
In this case, NNPDF3.1 agreed better with other groups.
Here too NNPDF4.0 uncertainties are the smallest.

%% file: subsec-lhcpheno-diff.tex
\subsection{Differential distributions}
\label{sec:pheno-differential-xsecs}

The integrated fiducial cross-sections discussed in the previous section
are typically dominated
by a localized region of the phase space corresponding to the
bulk of the distribution, and
hence they are only sensitive to PDFs in a
narrow range
of $x$ and $Q$.
Differential distributions, that we now discuss,
allow us to assess the compatibility between PDF sets
also in regions where experimental constraints are scarce, such
 as the large $x$ region, relevant for searches of new massive
 particles, and the  small-$x$ region, relevant for
calculations of neutrino cross-sections for high-energy 
astrophysics~\cite{Bertone:2018dse}.

For each differential distribution, we
provide the absolute cross-sections obtained using NNPDF4.0, with theory
uncertainties found by standard seven-point scale variation shown as a
band. As mentioned, all computations are performed with NLO QCD accuracy:
hence scale uncertainties would be smaller at NNLO.
We then display the percentage shift between the pure QCD and the full
QCD+EW computation, compared to the PDF and scale variation uncertainties.
We next compare the relative PDF uncertainty found using all the PDF sets
discussed in this Section. Finally we show the pull in units of the
PDF uncertainty only between the result
found using NNPDF4.0 and any of the other PDF sets, defined as
    \be
\label{eq:pulldef_xsec}
P\lp  \sigma_{2,i}, \sigma_{1,i}\rp\equiv \frac{ \sigma^{(0)}_{2,i} -\sigma^{(0)}_{1,i} }{
  \sqrt{ \lp  \delta \sigma_{2,i}\rp^2+\lp  \delta \sigma_{1,i}\rp^2 }} \, , \qquad i=1,\ldots,n_{\rm bin} \, ,
\ee
where  $\sigma^{(0)}_{1,i}$ and $\sigma^{(0)}_{2,i}$ are the central values of the
theory predictions in the $i$-th bin and $\delta \sigma_{1,i}$, $\delta \sigma_{2,i}$ are
the corresponding PDF uncertainties.

\begin{figure}[t]
    \centering
    \includegraphics[width=0.5\textwidth]{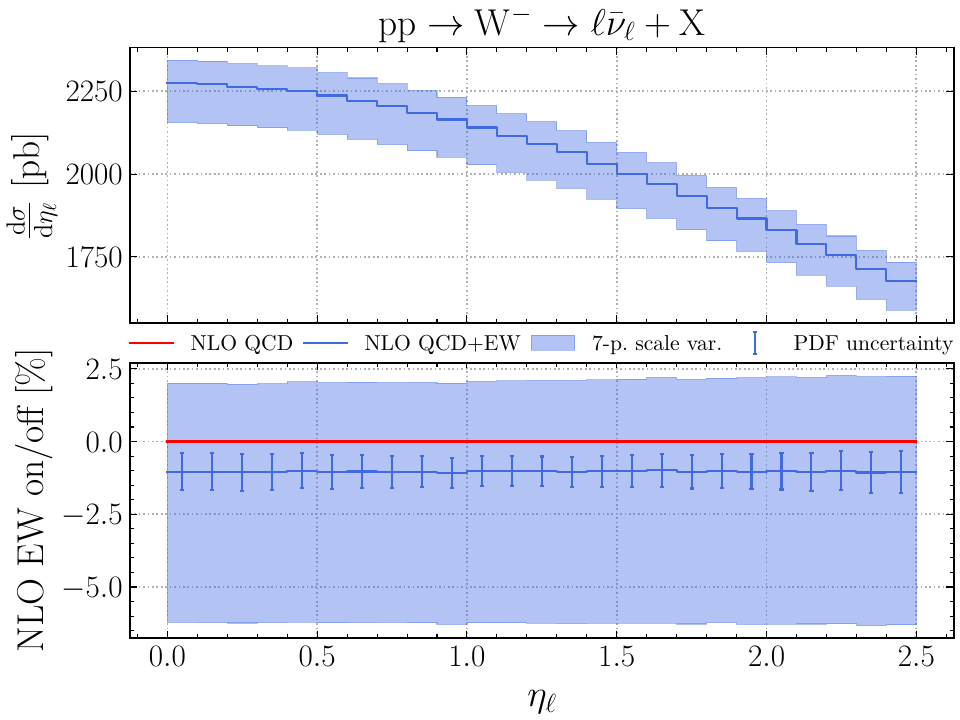}%
    \includegraphics[width=0.5\textwidth]{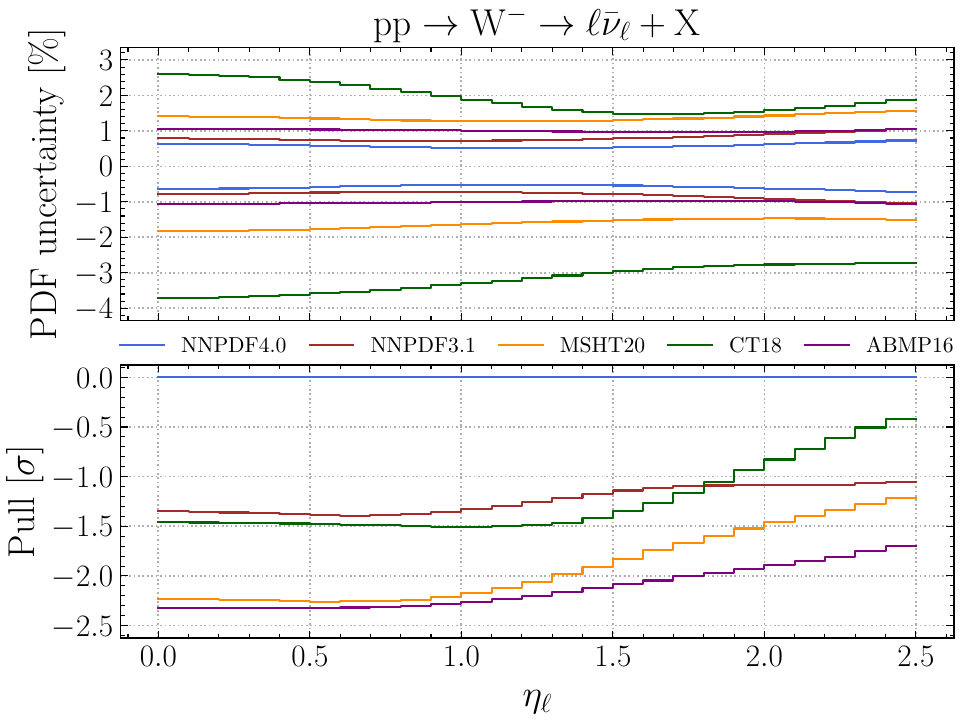}%
    \caption{The differential distribution in charged lepton rapidity, $\eta_\ell$,
    for inclusive $l\bar{\nu}_\ell$ production. Note that
    the result of a fully off-shell calculation is presented;
    the heading in the plot
    indicates the dominant $W^-$ intermediate state.
    Predictions obtained using NNPDF3.1, NNPDF4.0,  CT18, MSHT20,
    and ABMP16 are compared. We show the  NNPDF4.0 absolute cross-sections (top
    left) with the band indicating the 7-point scale variation
    uncertainties; the percentage shift in central values between pure
    QCD and QCD+EW along with the PDF and scale variation
    uncertainties (bottom left), all for NNPDF4.0;
    the relative PDF uncertainties for all PDF sets (top right); and
    the pull defined in Eq.~\eqref{eq:pulldef_xsec} between results obtained
    using NNPDF4.0 and each of the other PDF sets (bottom right).
    }
    \label{fig:pheno-wm}
\end{figure}
    
\begin{figure}[t]
    \centering
    \includegraphics[width=0.5\textwidth]{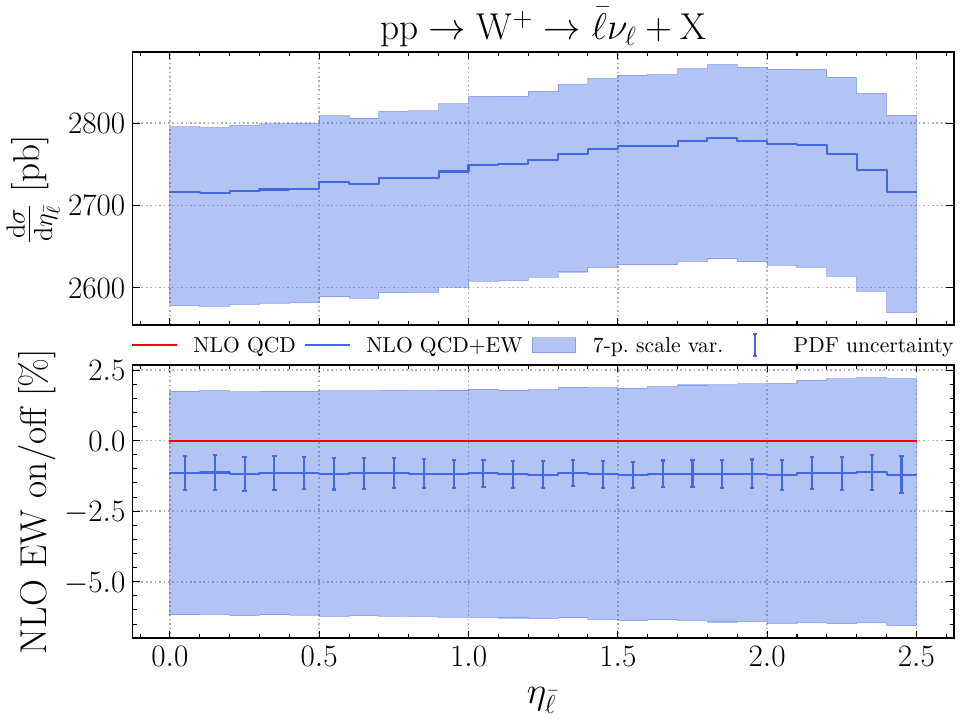}%
    \includegraphics[width=0.5\textwidth]{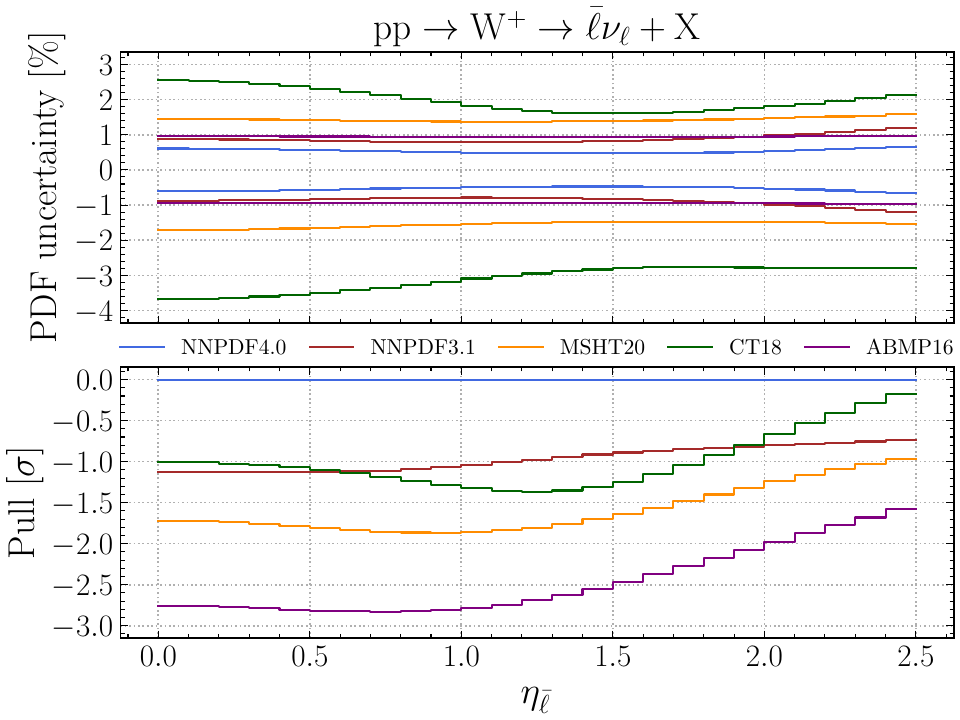}
    \caption{Same as Fig.~\ref{fig:pheno-wm} for $\mathrm{p}\mathrm{p} \to \bar{\ell} \nu_\ell + \mathrm{X}$.}
    \label{fig:pheno-wp}
\end{figure}

\begin{figure}[t]
    \centering
    \includegraphics[width=0.5\textwidth]{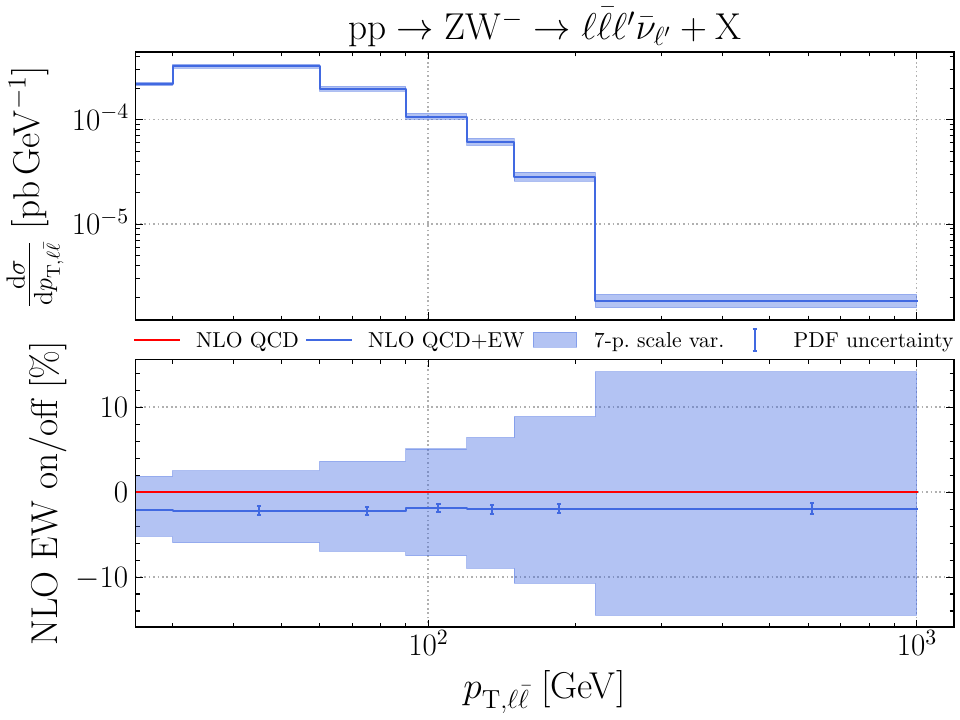}%
    \includegraphics[width=0.5\textwidth]{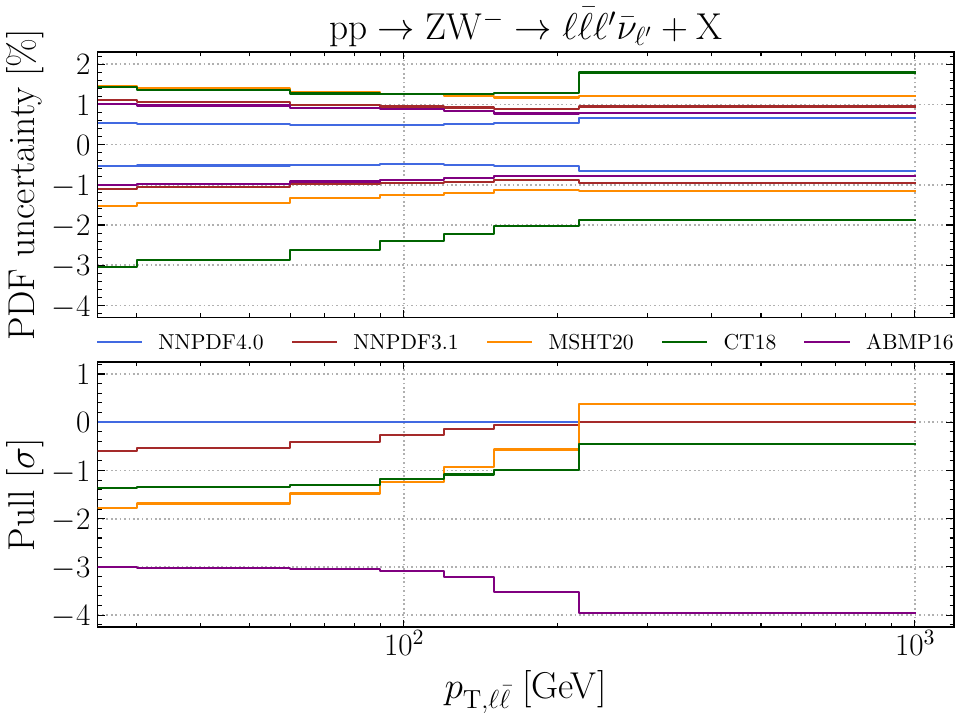}
    \caption{Same as Fig.~\ref{fig:pheno-wm} for $\mathrm{p}\mathrm{p} \to \ell \bar{\ell} \ell' \bar{\nu}_{\ell'} + \mathrm{X}$.}
    \label{fig:pheno-diboson-wmz}
\end{figure}

\begin{figure}[t]
    \centering
    \includegraphics[width=0.5\textwidth]{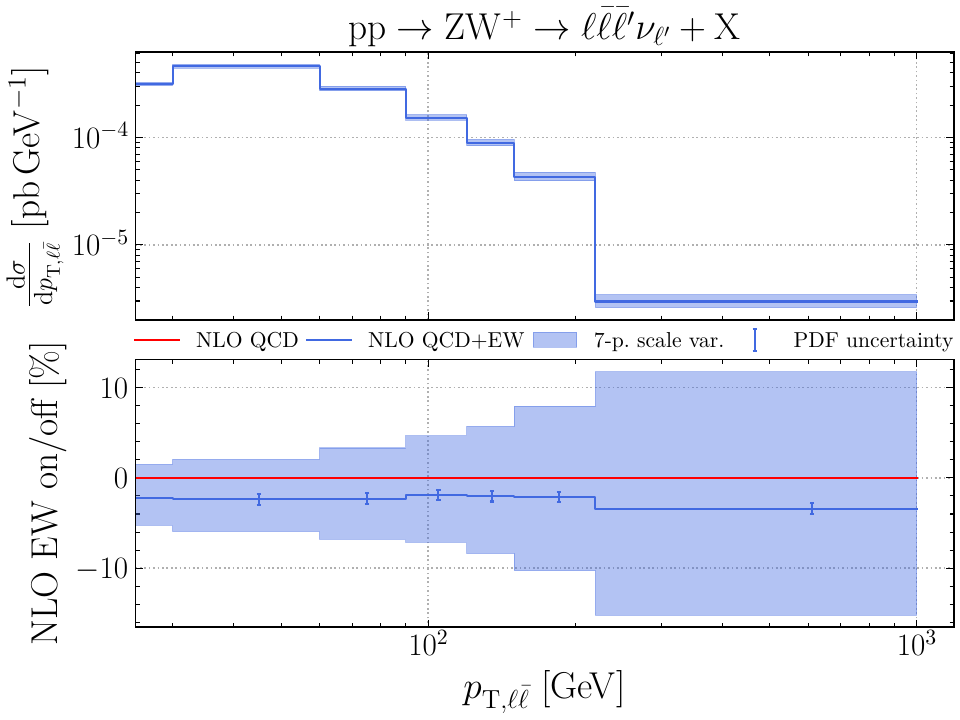}%
    \includegraphics[width=0.5\textwidth]{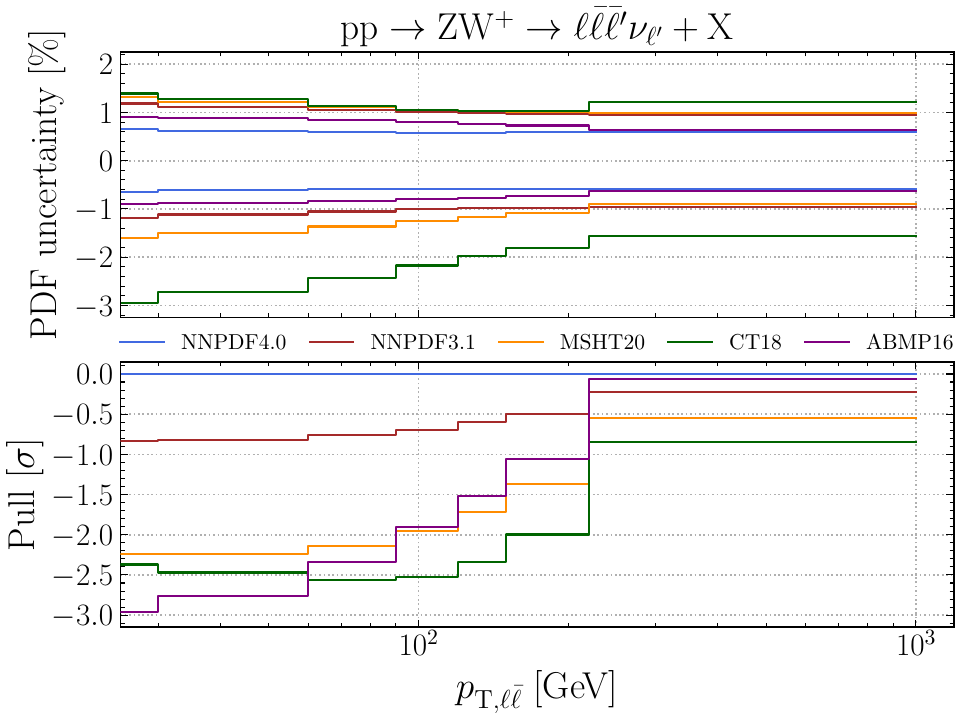}
    \caption{Same as Fig.~\ref{fig:pheno-wm} for $\mathrm{p}\mathrm{p} \to \ell \bar{\ell} \bar{\ell}' \nu_{\ell'} + \mathrm{X}$.}
    \label{fig:pheno-diboson-wpz}
\end{figure}

\begin{figure}[t]
    \centering
    \includegraphics[width=0.5\textwidth]{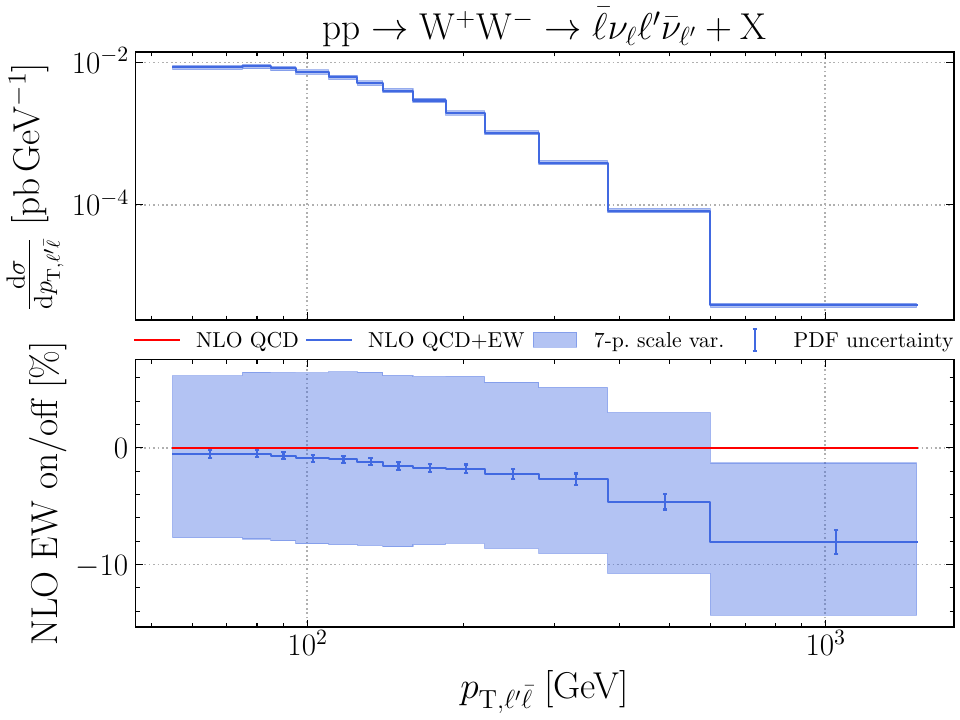}%
    \includegraphics[width=0.5\textwidth]{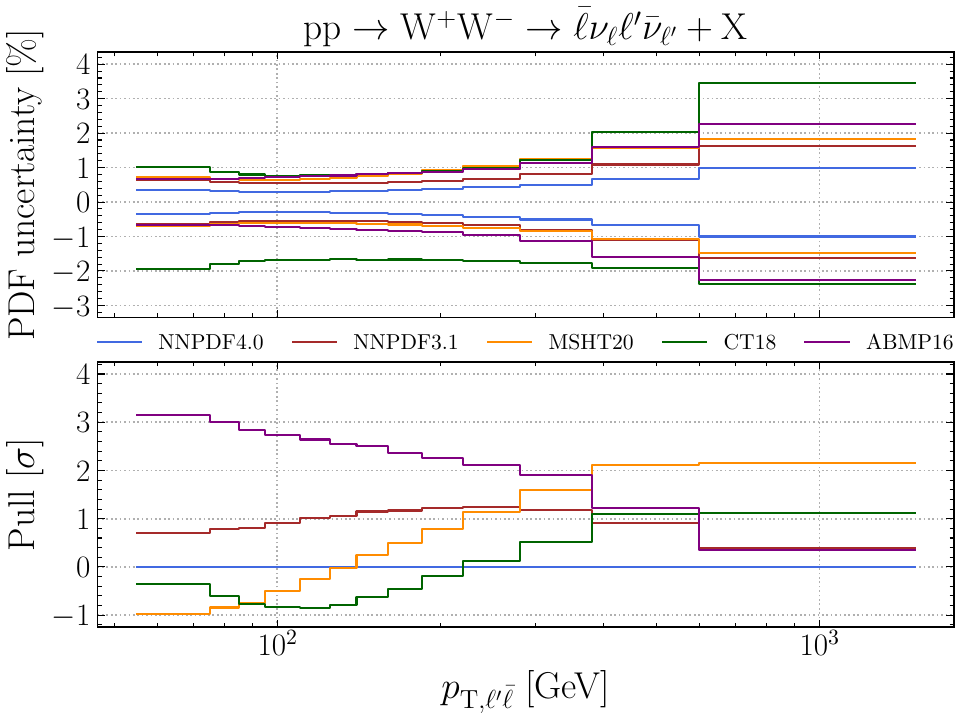}
    \caption{Same as Fig.~\ref{fig:pheno-wm} for $\mathrm{p}\mathrm{p} \to \bar{\ell} \nu_{\ell} \ell^\prime \bar{\nu}_{\ell^\prime} + \mathrm{X}$.}
    \label{fig:pheno-diboson-ww}
\end{figure}

\begin{figure}[t]
    \centering
    \includegraphics[width=0.5\textwidth]{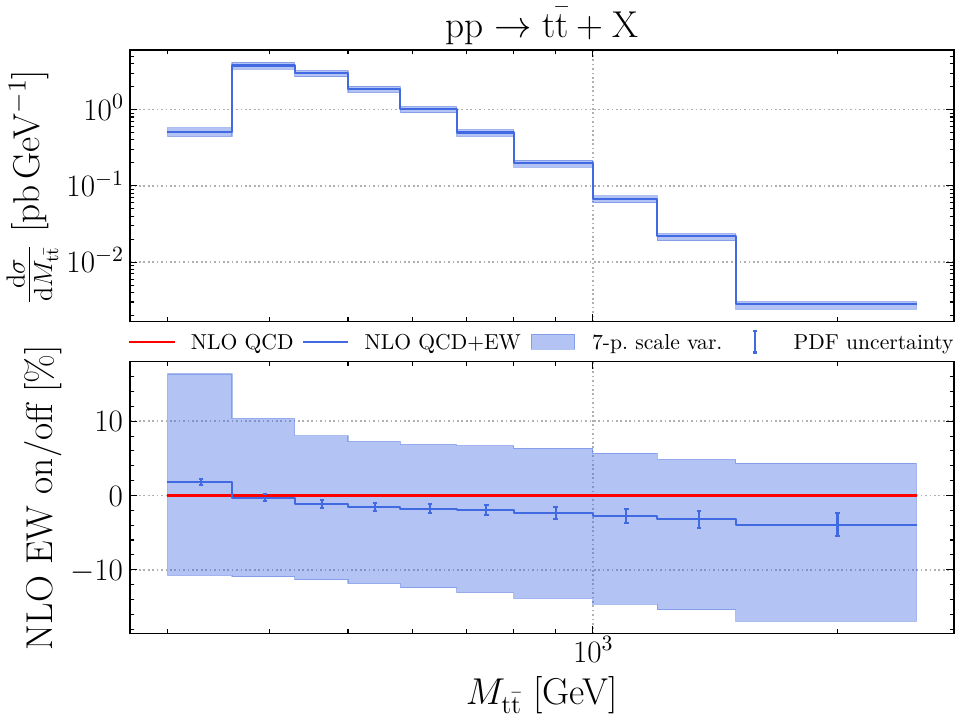}%
    \includegraphics[width=0.5\textwidth]{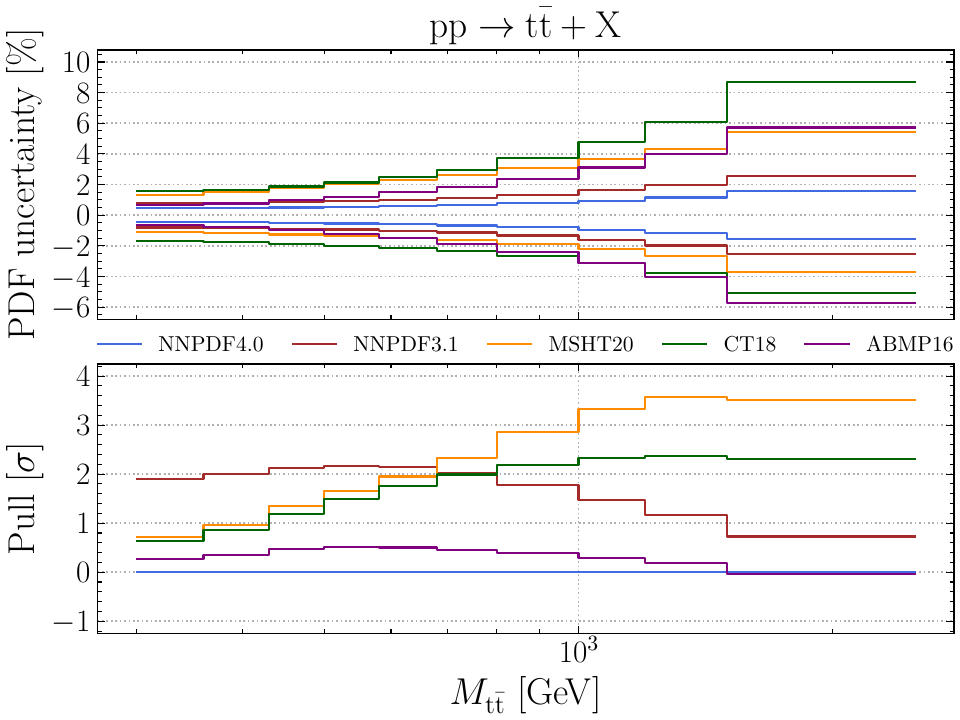}
    \caption{Same as Fig.~\ref{fig:pheno-wm} for $\mathrm{p}\mathrm{p} \to \mathrm{t}\bar{\mathrm{t}} + \mathrm{X}$.}
    \label{fig:pheno-ttbar}
\end{figure}

\begin{figure}[t]
    \centering
    \includegraphics[width=0.5\textwidth]{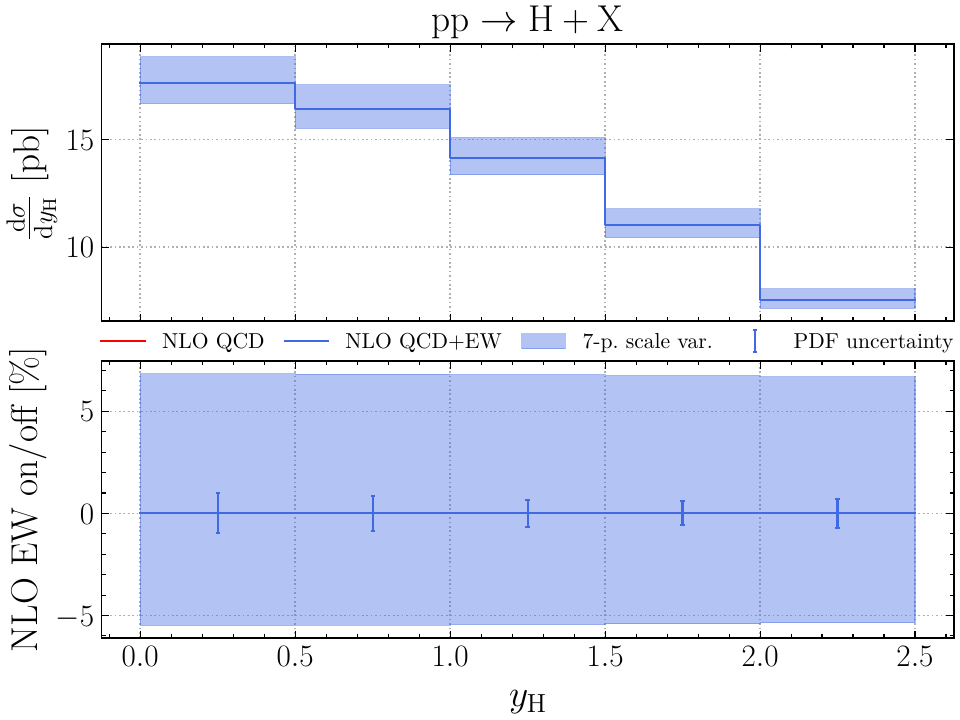}%
    \includegraphics[width=0.5\textwidth]{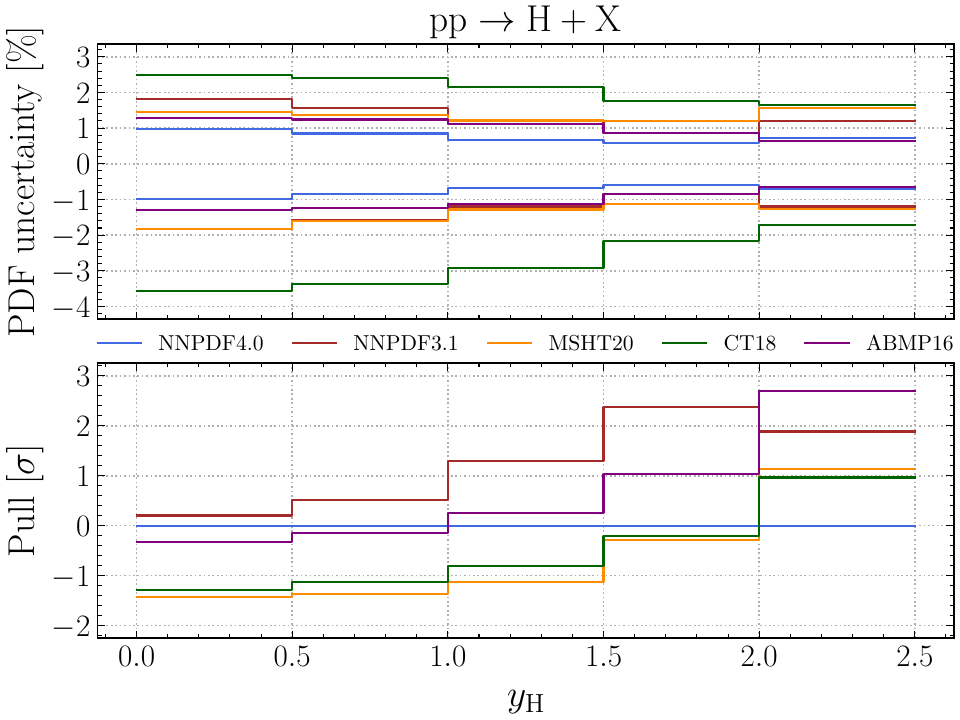}
    \caption{Same as Fig.~\ref{fig:pheno-wm} for $\mathrm{p}\mathrm{p} \to \mathrm{H} + \mathrm{X}$.}
    \label{fig:pheno-higgs}
\end{figure}

\begin{figure}[t]
    \centering
    \includegraphics[width=0.5\textwidth]{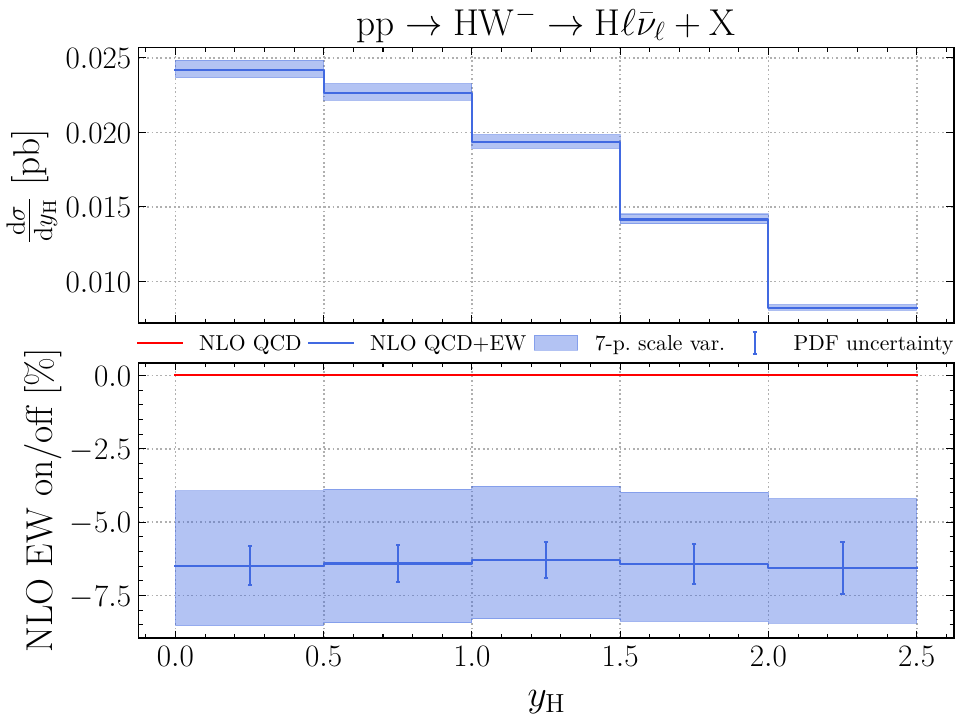}%
    \includegraphics[width=0.5\textwidth]{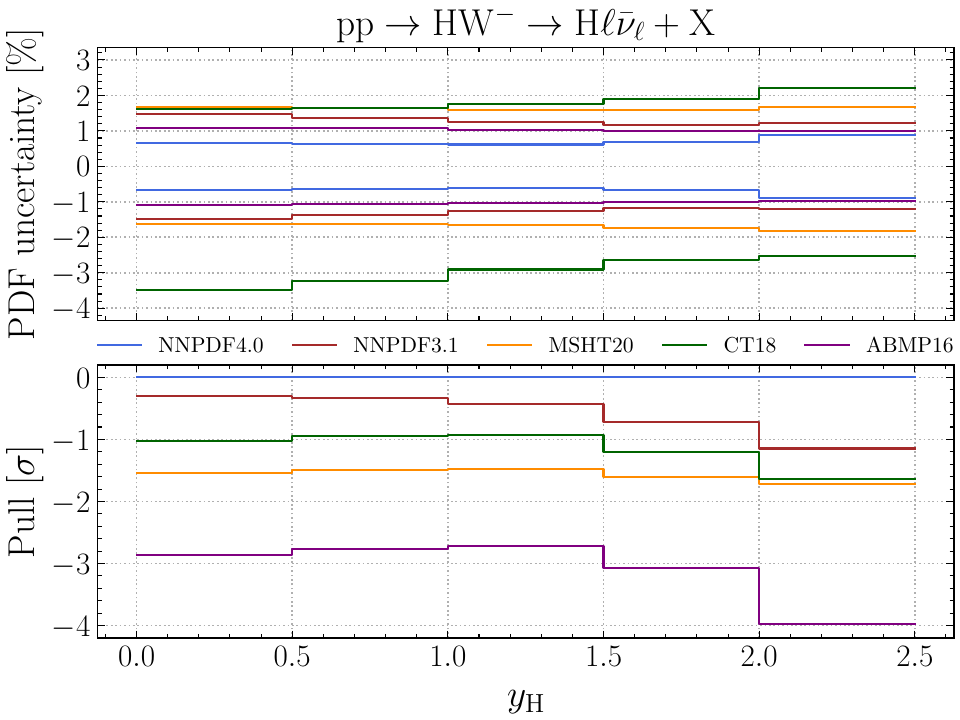}
    \caption{Same as Fig.~\ref{fig:pheno-wm} for $\mathrm{p}\mathrm{p} \to \mathrm{H} \ell \bar{\nu}_\ell + \mathrm{X}$.}
    \label{fig:pheno-higgs-wm}
\end{figure}

\begin{figure}[t]
    \centering
    \includegraphics[width=0.5\textwidth]{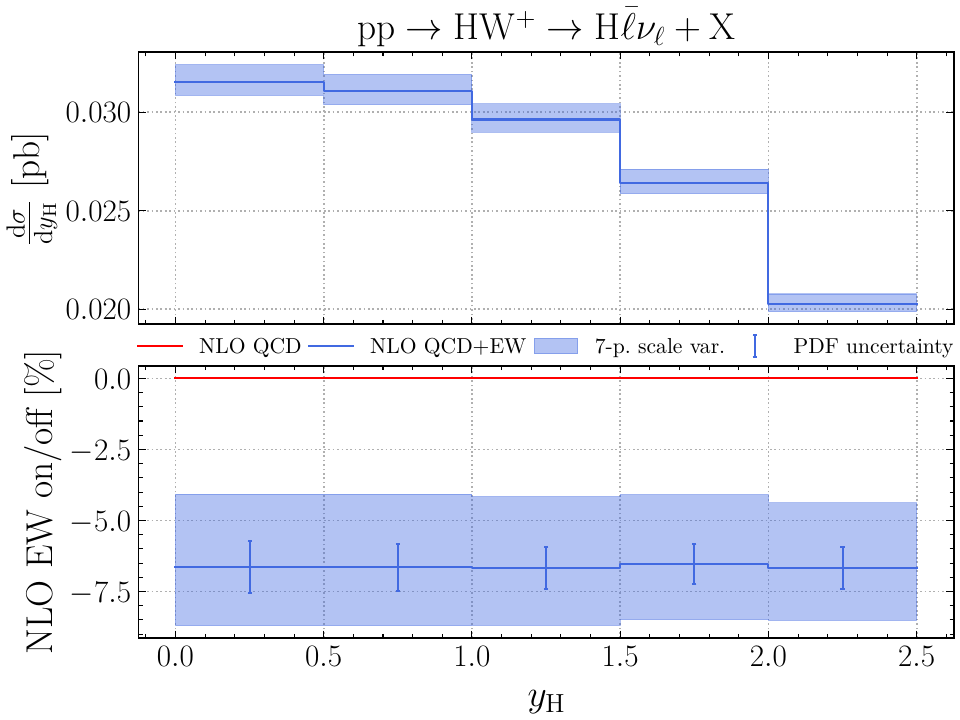}%
    \includegraphics[width=0.5\textwidth]{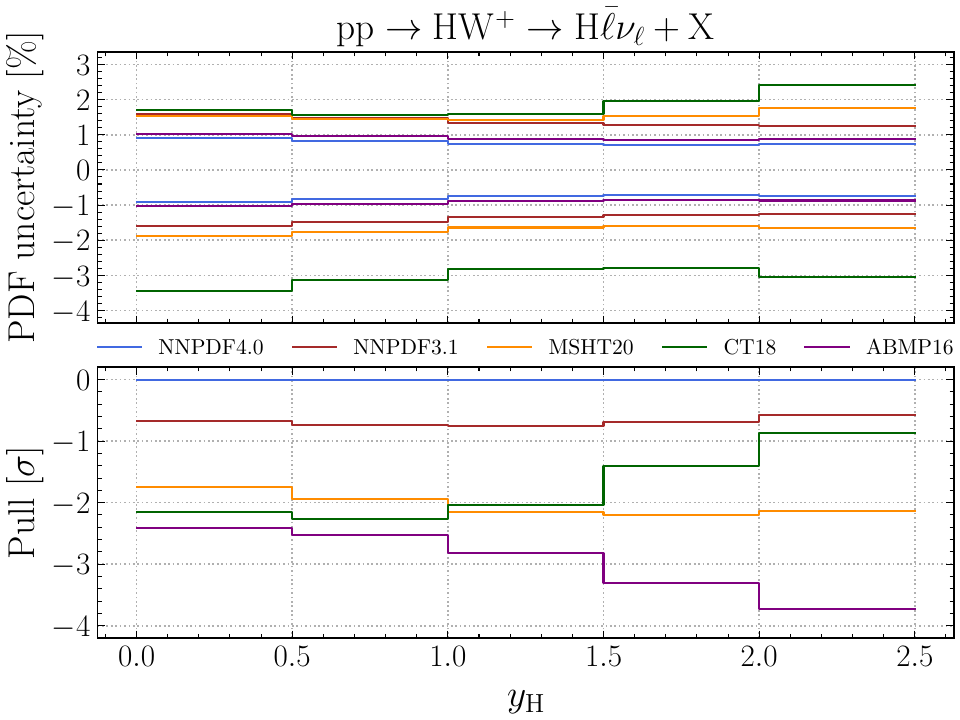}
    \caption{Same as Fig.~\ref{fig:pheno-wm} for $\mathrm{p}\mathrm{p} \to \mathrm{H} \bar{\ell} \nu_\ell + \mathrm{X}$.}
    \label{fig:pheno-higgs-wp}
\end{figure}

\begin{figure}[t]
    \centering
    \includegraphics[width=0.5\textwidth]{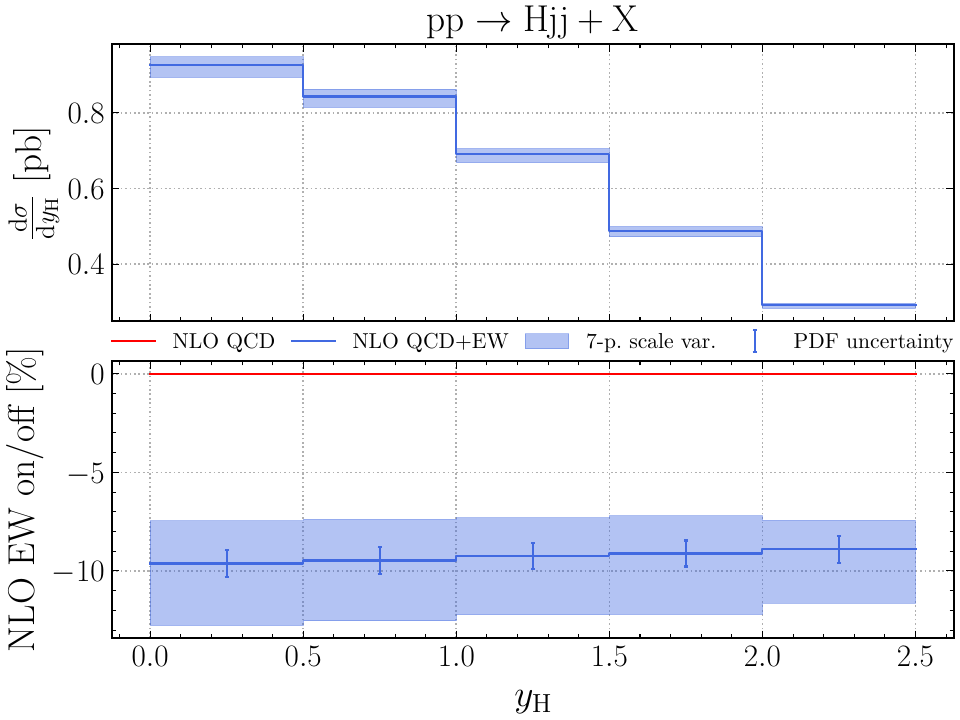}%
    \includegraphics[width=0.5\textwidth]{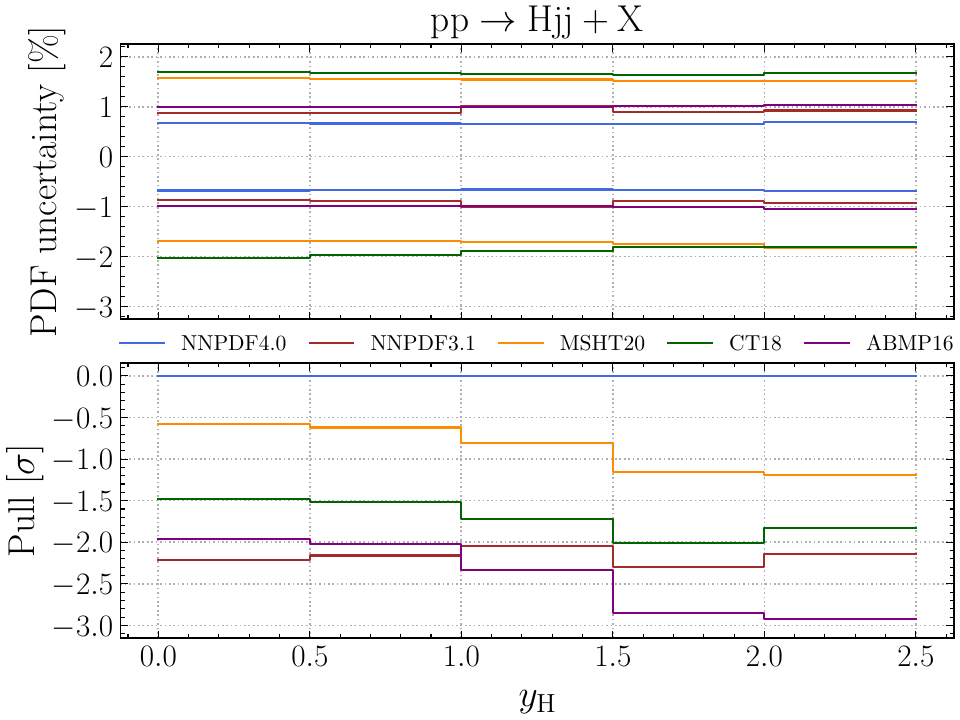}
    \caption{Same as Fig.~\ref{fig:pheno-wm} for $\mathrm{p}\mathrm{p} \to \mathrm{H} \mathrm{j}\mathrm{j} + \mathrm{X}$.}
    \label{fig:pheno-higgs-vbf}
\end{figure}

We consider differential distributions for the following processes:
charged current DY production (Figs.~\ref{fig:pheno-wm} and~\ref{fig:pheno-wp}),
neutral current DY (Fig.~\ref{fig:pheno-ttbar}),
gauge boson pair production (Figs.~\ref{fig:pheno-diboson-wmz},~\ref{fig:pheno-diboson-wpz},
and~\ref{fig:pheno-diboson-ww}),
top pair production (Fig.~\ref{fig:pheno-ttbar}), and then
Higgs production in the various channels (Figs.~\ref{fig:pheno-higgs},~\ref{fig:pheno-higgs-wm},~\ref{fig:pheno-higgs-wp},
and~\ref{fig:pheno-higgs-vbf}).
Recall that the fiducial cross-sections shown in the previous sections have been
obtained by integrating the differential distributions shown here.
Note also that in each case a fully off-shell calculation is presented,
including  nonfactorizable diagrams, and e.g.\ for diboson production  also
single and nonresonant contributions: so the heading in the
diagram merely indicates the dominant intermediate state.

While detailed conclusions can be reached by inspection of the plots,
we summarize here some generic features:

\begin{itemize}
  \item PDF uncertainties are uniformly smallest for NNPDF4.0, and
    largest for CT18, with ABMP16 uncertainties sometimes close to the
    NNPDF4.0 ones. However, when comparing uncertainties in different
    PDF sets recall the caveat discussed in
    Sect.~\ref{sec:pdflumis}.
 \item The pull is essentially always below one for NNPDF3.1, thus
   showing backward compatibility of NNPDF4.0 with its predecessor.   
\item The pull is generally largest for ABMP16, especially in
  regions sensitive to extrapolation where the uncertainties are very
  small for this PDF set, such as for instance highly boosted
  associate  Higgs
  production with $W^\pm$, where in the largest rapidity bins the pull
  can be as large as four.
\item The pulls of CT18 and MSHT20
  for the more inclusive observables, single gauge boson
  production and Higgs in gluon fusion, are generally below two and mostly
  below one. However, pulls for double gauge-boson production and associate
  Higgs production or Higgs in gauge fusion  are larger and
  sometimes exceed two.
\item Large pulls with CT18 and MSHT20
  are also seen for top production at large
    invariant mass, where the gluon at increasingly large $x$ is
    probed, in agreement with the comparison of gluon luminosities.
    \end{itemize}

\begin{figure}[t]
    \centering
    \includegraphics[width=0.5\textwidth]{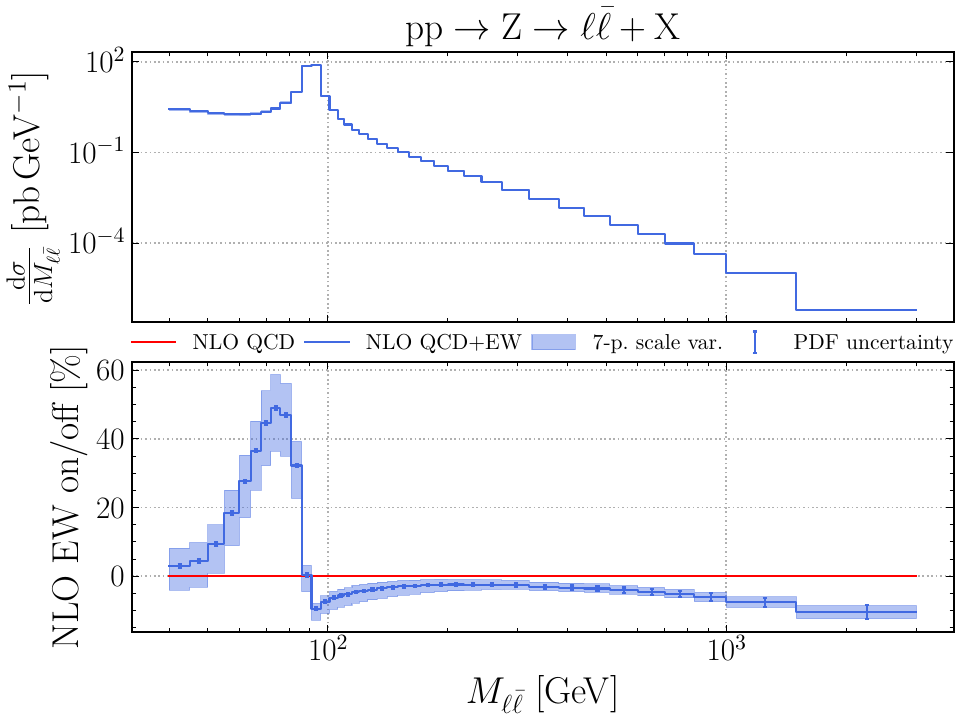}%
    \includegraphics[width=0.5\textwidth]{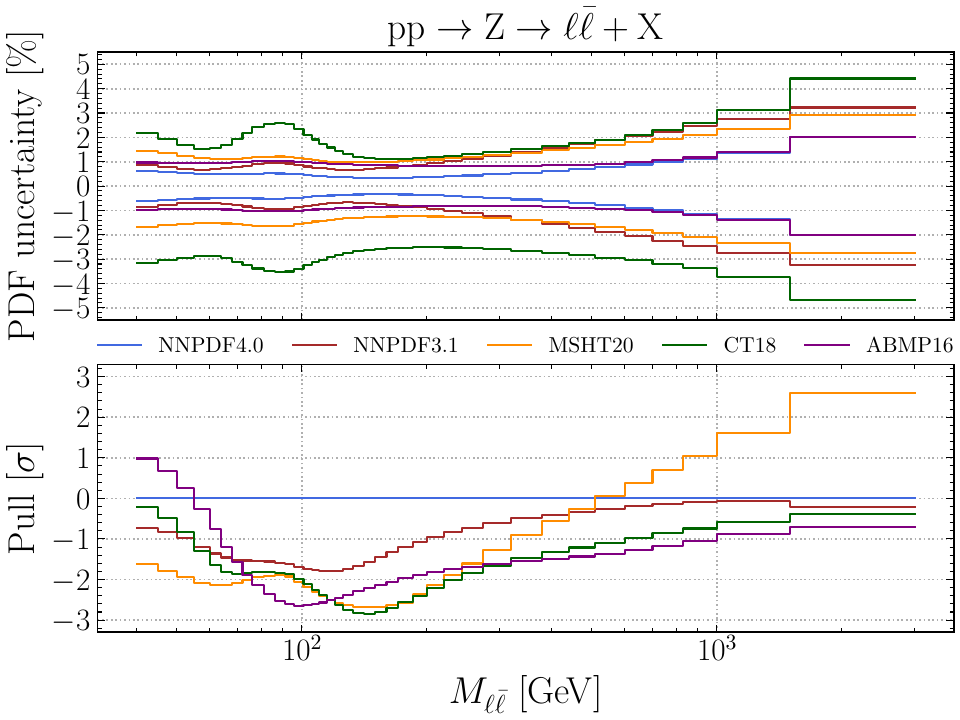}
    \caption{Same as Fig.~\ref{fig:pheno-wm} for $\mathrm{p}\mathrm{p} \to \ell\bar{\ell} + \mathrm{X}$.}
    \label{fig:pheno-dy}
\end{figure}

%% file: sec-summary.tex
\section{Deliverables, summary and outlook}
\label{sec:summary}

The NNPDF4.0 PDF set presented in this paper consists of two main classes
of deliverables.
The first class includes, as customary, the public release of various PDF sets, delivered in
standard {\sc\small LHAPDF6} interpolation grid format~\cite{Buckley:2014ana}.
The other class is, for the first time, the release of the
complete NNPDF fitting framework as an open-source code, including
extensive documentation and user-ready examples.
The availability of the NNPDF code as open source guarantees
the complete reproducibility of any aspect of the PDF determination presented in this work:
construction and hyperoptimization of the methodology, computation of
observables, PDF determination, statistical validation of results, and
visualization through suitable tools.
We believe that
the full open-source availability of our PDF fitting framework represents
a significant contribution to the LHC and QCD research communities,
as well as a major
step towards the achievement of the FAIR (findability, accessibility, interoperability, and reusability)
principles~\cite{FAIR}. As such, our code and data should be fully and
freely reusable by both humans and machines.

The publicly available open-source {\sc\small NNPDF}  code
is briefly described in Appendix~\ref{sec:nnpdffitter} and  more
extensively in a dedicated companion paper~\cite{NNPDF:2021uiq}. Below we
list the NNPDF4.0 PDF sets that we are making available. We then provide a
summary and outlook of this work.

\subsection{PDF grids}
\label{sec:delivery}

The NNPDF4.0 PDF sets are
made publicly available via the {\sc\small LHAPDF6}
interface, 
\begin{center}
{\bf \url{http://lhapdf.hepforge.org/}~} .
\end{center}
All sets are delivered as sets of  $N_{\rm rep}=100$ Monte Carlo replicas.
In the case of the baseline NNLO, the 100-replica set is obtained from
compression of  a larger  $N_{\rm rep}=1000$ replica
set, which is also made available, and a Hessian conversion with 
$N_{\rm eig}=50$ eigenvectors of the 1000-replica set is also made
available. We have checked that the 50 eigenvector set guarantees an
accuracy comparable to that of the PDF4LHC combined
set~\cite{Butterworth:2015oua,Ball:2022hsh}, namely at or better than  the 10-20\%
level on correlations and at the percent level on uncertainties. 

\begin{itemize}

\item {\it Baseline LO, NLO and NNLO NNPDF4.0 sets.}

The baseline LO, NLO, and NNLO NNPDF4.0 sets are based on the global
dataset, with $\alpha_s(m_Z)=0.118$ and a variable-flavor-number scheme with
up to five active flavors. These sets contain $N_{\rm rep}=100$  replicas each
and their file grid names are
\begin{flushleft}
\tt NNPDF40\_lo\_as\_01180 \\
\tt NNPDF40\_nlo\_as\_01180 \\
\tt NNPDF40\_nnlo\_as\_01180
\end{flushleft}

The NNLO set has been obtained from the optimized
compression~\cite{Carrazza:2015hva,Carrazza:2021hny} of
a dedicated $N_{\rm rep}=1000$ replica set, which is also made also available
\begin{flushleft}
\tt NNPDF40\_nnlo\_as\_01180\_1000
\end{flushleft}
and whose usage is recommended for applications that require a large replica sample,
such as Bayesian reweighting~\cite{Ball:2010gb,Ball:2011gg}.
This $N_{\rm rep}=1000$ replica set is also used as input
for the Hessian conversion~\cite{Carrazza:2015aoa,Carrazza:2016htc},
producing a set with $N_{\rm rep}=50$ eigenvectors with grid name
\begin{flushleft}
\tt NNPDF40\_nnlo\_as\_01180\_hessian 
\end{flushleft}

\item {\it PDF sets with $\alpha_s$ variations.}

NNLO PDF sets with baseline
  theory settings  are  made available for a variety of values of the
  strong coupling  spanning a range
  of  $\alpha_s(m_Z)$ from 0.116 to 0.120:
\begin{flushleft}
  \tt NNPDF40\_nnlo\_as\_01160 \\
  \tt NNPDF40\_nnlo\_as\_01170 \\
  \tt NNPDF40\_nnlo\_as\_01175 \\
  \tt NNPDF40\_nnlo\_as\_01185 \\
  \tt NNPDF40\_nnlo\_as\_01190 \\
  \tt NNPDF40\_nnlo\_as\_01200 \\
\end{flushleft}
Also, two NLO sets with $\alpha_s$ varied by $\pm0.001$ about
the central value are provided:
\begin{flushleft}
  \tt NNPDF40\_nlo\_as\_01170 \\
  \tt NNPDF40\_nlo\_as\_01190 \\
\end{flushleft}

In order to facilitate the computation of  combined PDF+$\alpha_s$
uncertainties, we provide  bundled  PDF+$\alpha_s$ variation
sets for $\alpha_s(m_Z)=0.118\pm0.001$ both for the NNLO Monte Carlo  and Hessian baseline sets:
\begin{flushleft}
  \tt
NNPDF40\_nnlo\_pdfas  \\	
NNPDF40\_nnlo\_hessian\_pdfas \\	
\end{flushleft}
These bundled PDF sets have been constructed as follows:
for the Monte Carlo sets
\begin{enumerate}
\item The central value (PDF member 0) is the central
  value of the corresponding $\alpha_s(m_Z)=0.118$ set;
\item PDF members 1 to 100 correspond to the
  $N_{\rep}=100$  Monte Carlo replicas;
\item the PDF members 101 and 102 are the central values
  of the sets with $\alpha_s(m_Z)=0.117$ and $\alpha_s(m_Z)=0.119$
  respectively;
\end{enumerate}
while for the Hessian sets
\begin{enumerate}
\item The central value (PDF member 0) is the central
  value of the corresponding $\alpha_s(m_Z)=0.118$ set;
\item members from 1 to 50 correspond to the $N_{\rm eig}=50$
  eigenvectors from the $\alpha_s(m_Z)=0.118$ set; 
\item members 51 and 52 are the central values
  of the sets with $\alpha_s(m_Z)=0.117$ and $\alpha_s(m_Z)=0.119$
  respectively.
\end{enumerate}

The usage of these bundled sets to evaluate the
combined PDF+$\alpha_s$ uncertainties for LHC cross-sections
is explained {\it e.g.} in~\cite{Butterworth:2015oua}.

\item {\it PDF sets with perturbative charm.}

In the  NNPDF4.0 baseline the charm PDF is independently parametrized,
along with light quark PDFs.
Variants  in which charm is not independently parametrized, but
rather obtained from perturbative  matching conditions and the FONLL
scheme is used, are also made available.
We release LO, NLO, and NNLO Monte Carlo sets with
$N_{\rm rep}=100$ PDF replicas each:
\begin{flushleft}
  \tt NNPDF40\_lo\_pch\_as\_01180 \\
\tt NNPDF40\_nlo\_pch\_as\_01180 \\
\tt NNPDF40\_nnlo\_pch\_as\_01180
\end{flushleft}

\item {\it PDF sets with flavor-number variations.}

  The baseline NNPDF4.0 PDFs are based on a variable-flavor-number
  scheme with a maximum of $n_f=5$ active flavors.
We have also produced sets, both at NLO and NNLO,
in which the maximum value of $n_f$ is either 4 or 6
\begin{flushleft}
   \tt NNPDF40\_nlo\_as\_01180\_nf\_4 \\
  \tt NNPDF40\_nlo\_as\_01180\_nf\_6 \\
  \tt NNPDF40\_nnlo\_as\_01180\_nf\_4 \\
  \tt NNPDF40\_nnlo\_as\_01180\_nf\_6 \\
\end{flushleft}
as well as variants of the perturbative charm fit in the $n_f=3$ scheme
\begin{flushleft}
   \tt NNPDF40\_nlo\_pch\_as\_01180\_nf\_3 \\
   \tt NNPDF40\_nnlo\_pch\_as\_01180\_nf\_3 \\
\end{flushleft}

Note that these grids are constructed by taking the baseline PDF sets
as a
fixed boundary condition
and then adjusting the settings
of perturbative
evolution and the running of $\alpha_s$ to the desired $n_f$ scheme.
For instance, the {\tt NNPDF40\_nnlo\_as\_0118\_nf\_4} is identical to the baseline
{\tt NNPDF40\_nnlo\_as\_0118} for $Q \le m_b$ but differs from it for $Q > m_b$ due
to the different number of active flavors in the evolution of
$\alpha_s(Q)$
and PDFs.

It is important to observe that, consequently, the value of the strong coupling in the
$n_f=3$ and $n_f=4$ schemes is modified and it is
$\alpha_s(m_Z)\ne 0.118$.
The naming convention adopted is that the value of $\alpha_s(m_Z)$ used is that
corresponding to $\alpha_s(m_Z)=0.118$ in the $n_f=5$ flavor scheme.

In the  $n_f=4$ case, bundled sets with $\alpha_s$ variations
are also constructed following the same strategy as in the baseline fits
\begin{flushleft}
   \tt NNPDF40\_nlo\_nf\_4\_pdfas \\
   \tt NNPDF40\_nnlo\_nf\_4\_pdfas \\
\end{flushleft}

\item {\it PDF sets with dataset variations.}

  The variants of NNPDF4.0 with different input datasets, in particular
  those discussed in Sect.~\ref{sec:dataset}, are made available
  in the {\sc\small LHAPDF6} format and have been linked to
  the  NNPDF website:
  
\begin{center}
\url{https://nnpdf.mi.infn.it/nnpdf4-0/}
\end{center}

Note that these consist of fits based both on subsets of the baseline dataset, such
as the collider-only PDFs, as well as fits where additional datasets have been
included, such as the those with the NOMAD or the HERA jet cross-sections.

\end{itemize}

In addition to the grid files explicitly listed here, the rest of the
PDF sets
 discussed in this paper  are also available upon request.
We also emphasize that since the fitting code is made public, see Appendix~\ref{sec:nnpdffitter},
arbitrary variants of the present NNPDF4.0 determination can be produced by interested users.

\subsection{Summary and outlook}
\label{sec:outlook}

The NNPDF4.0  set presented here is characterized by the feature of
exhibiting a remarkable precision, with PDF uncertainties of order of 1\% in a wide kinematic
region  for several PDF combinations. This is mostly a consequence of
having used in its determination a machine learned methodology, that
combines a significantly more general and flexible parametrization
with a very
efficient minimization.

The general
features of the underlying dataset support
the reliability of these small uncertainties. Specifically,
the determination is now dominated by collider data, which are
generally more reliable than older fixed-target data: indeed, DIS-only
and no-LHC PDFs now differ substantially from the global fit, and HERA
data are no longer needed in order to fix the small-$x$ behavior of PDFs.
Furthermore, there is generally good or excellent compatibility
between all the disparate pieces of information that enter the
global PDF determination, also thanks to the dataset selection procedure that has
been applied, as discussed in Sect.~\ref{sec:dataselection}, with
most of all data leading to mutually consistent constraints on PDF. This is
supported by the inspection of alternative PDF
determinations in which individual data or sets of data are removed, see
Sect.~\ref{sec:dataset}. Finally, the PDF fit includes many datasets
that provide mutually consistent constraints on the same PDF.
For instance, the $\bar{d}/\bar{u}$ ratio, that
is in principle constrained by
the SeaQuest data, is actually predicted with almost unchanged
precision by a fit in which this data are not used, and the same is
true for the charm PDF and the EMC structure function data, or for
strangeness and the NOMAD neutrino DIS data, or for the gluon and HERA DIS jet
data.

The excellent control on the individual PDF flavors achieved in the NNPDF4.0
determination
suggests that it would be interesting to carry out a detailed assessment of the
non-perturbative structure of the proton, specifically by comparing to 
models of proton structure and lattice QCD calculations
for quantities such as the $\bar{d}/\bar{u}$ and $d/u$
ratios in the large $x$ region, the strangeness content, and intrinsic charm. This analysis will be
presented in a dedicated  publication~\cite{ICpaper}.

In terms of methodology, the reliability of results for PDF
uncertainties is backed up by extensive closure testing and future
testing, see Sect.~\ref{sec:closure}, and by the stability 
upon the methodological variations considered in
Sect.~\ref{sec:tests}, specifically the lack of dependence on the
choice of fitting basis, which is a highly nontrivial check that we
performed here for the first time.

However, it is clear that percent-level PDF uncertainties must be
treated with caution, and in particular it is important to consider
carefully sources of uncertainty that might have been underestimated, or that
have not been included.

The first and most obvious one is  missing
higher order uncertainties, routinely estimated by scale
variation, that are not included in  PDF uncertainties. 
Their inclusion is possible using the methodology developed in 
Refs.~\cite{AbdulKhalek:2019bux,AbdulKhalek:2019ihb,Ball:2021icz}.
The inclusion of uncertainties related to missing higher perturbative
orders in QCD calculations will be crucial in ensuring full
reliability of central values and uncertainties to percent or
sub-percent accuracy.
A closely related aspect which deserves direct
investigation is the construction of PDF sets at N$^3$LO in QCD, which
is already possible by using suitable approximations, specifically for
anomalous dimensions~\cite{Vogt:2018miu}. These will be useful both
directly, for consistency with LHC calculations where matrix elements
are evaluated at the same perturbative order,
and as a means to accurately estimate uncertainties on NNLO
results.

Also, NNLO QCD corrections  at present are largely included through
$K$-factors.  Their full analytic inclusion should be soon possible, 
as fully differential Monte Carlo generators accurate to NNLO
QCD~\cite{Boughezal:2016wmq,Grazzini:2017mhc,Bothmann:2019yzt}
and fast-interpolation grids supporting NNLO QCD
corrections~\cite{Kluge:2006xs,Wobisch:2011ij,Britzger:2012bs,Carrazza:2020gss}
become more widely available.

Furthermore, at present the impact of
electroweak corrections is only verified a
posteriori. Rather, they should be included systematically in theory
predictions, alongside with a photon PDF. The construction of a
QED variant of NNPDF4.0 including coupled QED$\otimes$QCD  evolution, along the lines
of its NNPDF3.1QED predecessor~\cite{Bertone:2017bme}, will be an
immediate task, but a full PDF determination in which mixed QCD-EW
corrections are fully included up to NLO will be needed for full 
theoretical reliability. Such a determination should also include an estimate 
of the missing higher order electroweak corrections.

Finally, the machine-learning methodology that we have followed is
based on standard traditional $\chi^2$ minimization, and it has been
closure-tested to fully consistent pseudodata. This might miss
information contained in the full statistical features of the PDF
fit, of which the $\chi^2$  is only the simplest indicator, and specifically
it might lead to
uncertainty estimation in the presence
of incompatible data
and inaccuracies in the estimation of experimental
uncertainties. It should be improved through the
exploration and use of a more advanced methodology,
both for PDF determination and for validation (including closure
testing) in which the full statistical features of the dataset and the
ensuing PDFs are used.

All these developments are the focus of ongoing studies, with the goal
of achieving PDFs with fully reliable sub-percent accuracy.

\begin{center}
\rule{5cm}{.1pt}
\end{center}
\bigskip

\input{subsec-acknowledgements.tex}

%% file: subsec-acknowledgements.tex
\subsection*{Acknowledgements}

We thank Amanda Cooper-Sarkar, Thomas Cridge, Lucian Harland-Lang, Timothy
Hobbs, Joey Houston, Pavel Nadolsky, Gavin Salam and Robert Thorne for discussions.
We thank Andrew Pilkington and Reinhard Schwienhorst for help with the ATLAS
measurements; Maria Aldaya, Emanuela Barberis and Alberto
Orso Maria Iorio for assistance with the CMS data; and Stephen Farry and
Oliver Lupton for help with the LHCb data. We are grateful to Jun Gao for
providing us with the NNLO $K$-factors for the NuTeV, NOMAD, and ATLAS and CMS
single-top measurements, and for helping us with the benchmark of the
corresponding NLO calculations. We are grateful to Daniel Britzger for
providing us with NNLO fast interpolation grids for the HERA jet data and to
Thomas Gehrmann, Aude Gehrmann-De Ridder, Nigel Glover, Alexander Huss and Joao
Pires for providing us with NNLO $K$-factors for ATLAS and CMS jet measurements.
We thank Valerio Bertone for assistance with the correction of a bug in
{\tt APFEL}, Marco Zaro for support with {\tt MadGraph5\_aMC@NLO} and Luca
Rottoli for guidance in the computation of NNLO corrections for fixed-target
Drell--Yan measurements on deuteron targets. We acknowledge contributions from
Rabah Abdul Khalek, Alessandro Candido, Felix Hekhorn, Giacomo Magni,
Tanjona Rabemananjara and Giovanni Stagnitto at various stages of this work.

S.~C., S.~F., J.~C.-M., R.~S and C.~S. are supported by
the European Research Council under 
the European Union's Horizon 2020 research and innovation Programme
(grant agreement n.740006).
M. U. and Z. K. are supported by the European Research Council under the
European Union’s Horizon 2020 research and innovation Programme (grant agreement
n.950246). M. U. and S. I. are supported by the Royal Society grant RGF/EA/180148.
The work of M. U. is also funded by the Royal Society grant DH150088.
The work of M. U., S. I., C. V. and Z. K. is partially supported by the STFC consolidated grant ST/L000385/1.
The work of Z.K. was partly supported by supported by the European
Research Council Consolidator Grant “NNLOforLHC2”.
The work of ZK was previously supported by the European Commission grant (683211)
R.~D.~B, L.~D.~D. and E.~R.~N. are supported by the U.K.
Science and Technology Facility Council (STFC) grant ST/P000630/1. 
E.~R.~N. was also supported by the European Commission through the Marie
Sk\l odowska-Curie Action ParDHonS (grant number 752748).
J.~R. is partially supported by NWO, the Dutch Research Council.
T.~G. is supported by The Scottish Funding Council, grant H14027,
R.~L.~P.  and M.~W. by the STFC grant ST/R504737/1 and C.~V by the STFC grant ST/R504671/1.

%% file: app-codedoc.tex
\section{The open-source NNPDF code}
\label{sec:nnpdffitter}

The open-source NNPDF code is the subject of a separate dedicated
publication~\cite{NNPDF:2021uiq}, to which the reader is referred for
details, while  here we provide a brief summary.
The NNPDF fitting framework consists of an extensive array of tools for
global  analysis of non-perturbative QCD quantities.
While the current version of the code focuses
on unpolarized parton distributions, its flexible
infrastructure is extensible to many other related
objects such as polarized PDFs, nuclear PDFs, or
fragmentation functions.

The {\sc\small NNPDF} code is available from its
{\sc\small GitHub} repository
\begin{center}
  \url{https://github.com/NNPDF/nnpdf}
\end{center}
and is fully documented in
\begin{center}
  \url{https://docs.nnpdf.science/}
\end{center}
The main inputs of the {\sc\small NNPDF}  code
are the experimental data, stored in a common format,
and the theoretical calculations.
The latter are provided in the form of
fast grids such as
{\tt APPLgrid}~\cite{Carli:2010rw}, {\tt FastNLO}~\cite{Britzger:2012bs},
or {\tt PineAPPL}~\cite{Carrazza:2020gss}, which are then
combined~\cite{Bertone:2016lga}
with QCD evolution kernels to produce FK
interpolation tables~\cite{Ball:2008by,Ball:2010de}
and possibly supplemented with NNLO QCD and/or
NLO electroweak $K$-factors.
The user needs also to provide the settings
and parameters of the theory calculations,
such as the value of $\alpha_s(m_Z)$, the heavy quark masses
and the perturbative order.

All the settings that determine the outcome of a given PDF
determination, from input datasets to theory settings and the parameters
of the optimizer, can be adjusted in the corresponding run card
written in {\sc\small YAML}.
This runcard is the unique identifier of a determination and contains all required information to perform (and reproduce) a global PDF fit.
Instructions  on how to set up a runcard
and use it to launch a fit can be found in
\begin{center}
  \url{https://docs.nnpdf.science/tutorials/run-fit.html}
\end{center}
and examples of runcards used to produce the PDF sets  presented in this paper
are collected in
\begin{center}
  \url{https://github.com/NNPDF/nnpdf/tree/master/n3fit/runcards/reproduce_nnpdf40}
\end{center}
For instance, via the runcard the user can select
the neural network architecture or the specific variant
of the SGD optimizer.
They can also
define the ranges of hyperparameters
and determine automatically the optimal values
by means of the hyperopt method described in Sect.~\ref{sec:hyperparam}.

The {\sc\small NNPDF} code is composed of the following main packages:

\begin{description}

\item[\textbf{The {\tt buildmaster} experimental data formatter}]
  A {\tt C++} program that transforms the experimental
  data as originally provided by {\sc\small HepData}
  into a common standard format  tailored for PDF fitting,
  for  instance by allowing different treatments
  of the correlated systematic uncertainties.
  
\item[\textbf{The {\tt APFELcomb} interpolation table generator}]
  A code that takes as input fast NLO interpolation grids
  and combines them with the evolution kernels
  provided by {\tt APFEL} to produce FK tables, which
  map the PDFs at the input parametrization scale $Q_0$
  to physical observables.
  
\item[{\textbf{The {\tt n3fit} fitting code}}]
  A {\tt TensorFlow}-based code which carries out the training
  of neural networks by means of a variety of SGD algorithms.
  The hyperparameters that determine the outcome of the training
  can either be selected by hand or determined automatically
  by means of the hyperopt procedure.
  The flexibility and modularity of {\tt n3fit} ensures that,
  should that be required, the user can interface
  the code with new optimization algorithms of their choice.

 \item[\textbf{The {\tt validphys} analysis framework}]
    Implemented in {\tt reportengine}~\cite{zahari_kassabov_2019_2571601}, this
    code collects the outcome of a PDF fit and produces
    a complete HTML training report with all relevant
    information such as plots of the PDFs, parton luminosities
    and physical observables  and statistical estimators
    from the $\chi^2$ to the arc-length.
    The extensive functionalities of this framework
    are described in
\begin{center}
  \url{https://docs.nnpdf.science/vp/index.html}
  \end{center}
  All the plots and tables presented in this paper
  have been obtained using this {\tt validphys} framework,
  which also provides tools
  to interact with online resources such as the results of previous
  fits or precomputed sets of FK tables, which 
  are automatically downloaded when required by a runcard.

\item[\textbf{The {\tt libnnpdf} {\tt C++} fitting code}]
  This is the code underlying the NNPDF3 family of
  PDF determinations, and provides  backwards compatibility
  such that users can reproduce the outcome of NNPDF3.1-like
  global fits.
  
\end{description}

These various components of the
{\sc\small NNPDF} fitting code are available as binary packages
accessible through
{\tt conda}. This allows for a smooth installation
alongside with that of all relevant dependencies and
external packages.
Crucially, one can indicate specific versions
of both the fitting code and of its  dependencies,
ensuring the full reproducibility of the result of a given
PDF fit.

As discussed in~\cite{NNPDF:2021uiq}, as with any open-source
code we welcome suggestions and requests for improvements
and addition of new features.
For instance, a new feature can be requested by either
creating an issue in the {\sc\small GitHub} repository
or by means of a pull request.
The latter would undergo review, testing, and validation before being
eventually approved
and merged into the master code.

%% file: app-datacomp.tex
\section{Input datasets in PDF fits}
\label{app:datacomp}
In order to put the NNPDF4.0 PDF determination in context, we provide
here tables comparing the NNPDF4.0 dataset to that adopted for the
other PDF determinations considered in the main body of the paper,
namely NNPDF3.1~\cite{Ball:2017nwa}, ABMP16~\cite{Alekhin:2017kpj},
CT18~\cite{Hou:2019efy} and MSHT20~\cite{Bailey:2020ooq}.

Tables~\ref{tab:dataset_noLHC}--\ref{tab:dataset_LHCb}
collect all the data used for the construction of these PDF sets.
For each dataset, we provide the corresponding reference and indicate
with a blue tick or a red cross whether or not this dataset is part
of each PDF analysis. A parenthesized tick in a yellow box denotes that a
dataset was investigated, but for various reasons not included in the baseline fit:
\begin{itemize}
\item for NNPDF3.1 these are the datasets given in tables 2.1 and 2.3 of Ref.~\cite{Ball:2017nwa}, which are marked with square brackets,
\item similarly for NNPDF4.0, where the datasets are given in Tab.~\ref{tab:DIS_dataset}, \ref{tab:collider_dataset_1} and \ref{tab:collider_dataset_2}, also marked with square brackets,
\item for ABMP16 they are mentioned in the text in section II.A of Ref.~\cite{Alekhin:2017kpj},
\item for CT18 the datasets are discussed in section II.A.2 and II.B.6 of Ref.~\cite{Hou:2019efy}
\item and finally for MSHT20 the datasets are given in section 10 of Ref.~\cite{Bailey:2020ooq}.
\end{itemize}

\begin{table}[!t]
  \renewcommand*{\arraystretch}{1.60}
  \scriptsize
  \centering  
  \input{tables/tab-datacomp_noLHC.tex}\\
  \vspace{0.3cm}
  \caption{\small The fixed-target and collider DIS measurements
    used for PDF determination. For each PDF set, a blue tick indicates that
    the given dataset is included and a red cross that it is not included.
    A parenthesized tick denotes that a dataset was investigated but not
    included in the baseline fit.}
  \label{tab:dataset_noLHC}
\end{table}

\begin{table}[!t]
  \renewcommand*{\arraystretch}{1.60}
  \scriptsize
  \centering
  \input{tables/tab-datacomp_FTDY.tex}\\
   \vspace{0.3cm}
  \caption{Same as Table~\ref{tab:dataset_noLHC} for fixed-target Drell--Yan data sets.}
  \label{tab:dataset_FYDY}
\end{table}

\begin{table}[!t]
  \renewcommand*{\arraystretch}{1.60}
  \scriptsize
  \centering
  \input{tables/tab-datacomp_Tevatron.tex}\\
   \vspace{0.3cm}
  \caption{Same as Table~\ref{tab:dataset_noLHC} for Tevatron data sets.}
  \label{tab:dataset_Tevatron}
\end{table}

\begin{table}[!t]
  \renewcommand*{\arraystretch}{1.60}
  \scriptsize
  \centering  
  \input{tables/tab-datacomp_ATLAS.tex}\\
   \vspace{0.3cm}
  \caption{Same as Table~\ref{tab:dataset_noLHC} for ATLAS data sets.}
  \label{tab:dataset_ATLAS}
\end{table}

\begin{table}[!t]
  \renewcommand*{\arraystretch}{1.60}
  \scriptsize
  \centering  
  \input{tables/tab-datacomp_CMS.tex}\\
   \vspace{0.3cm}
  \caption{Same as Table~\ref{tab:dataset_noLHC} for CMS data sets.}
  \label{tab:dataset_CMS}
\end{table}

\begin{table}[!t]
  \renewcommand*{\arraystretch}{1.60}
  \scriptsize
  \centering  
  \input{tables/tab-datacomp_LHCb.tex}\\
   \vspace{0.3cm}
  \caption{Same as Table~\ref{tab:dataset_noLHC} for LHCb data sets.}
  \label{tab:dataset_LHCb}
\end{table}

This comparison allows one to quantify the differences in terms of the input
datasets between NNPDF4.0 and previous PDF fits.
For instance, considering the LHC measurements, one can read from this table
that the NNPDF4.0 determination considers 20, 18, and 5 datasets
from ATLAS, CMS, and LHCb respectively, to be compared
with 4, 6, 3 in the CT18 analysis.

%% file: tables/tab-datacomp_noLHC.tex
\begin{tabularx}{\textwidth}{Xcccccc}
  \toprule
  Data set
  & Ref.
  & NNPDF3.1
  & NNPDF4.0
  & ABMP16
  & CT18
  & MSHT20
  \\
  \midrule  
  NMC $F_2^d/F_2^p$
  & \cite{Arneodo:1996kd}
  & \cmark 
  & \cmark
  & \xmark
  & \xmark
  & \cmark
  \\
  NMC $\sigma^{{\rm NC},p}$
  & \cite{Arneodo:1996qe}
  & \cmark 
  & \cmark
  & \xmark
  & \cmark
  & \cmark
  \\
  SLAC $F_2^p, F_2^d$
  & \cite{Whitlow:1990gk,Whitlow:1991uw}
  & \cmark
  & \cmark
  & \cmark
  & \xmark
  & \cmark
  \\
  BCDMS $F_2^p$
  & \cite{Benvenuti:1989rh}
  & \cmark 
  & \cmark
  & \cmark
  & \cmark
  & \cmark
  \\
  BCDMS $F_2^d$
  & \cite{Benvenuti:1989fm}
  & \cmark
  & \cmark
  & \xmark
  & \cmark
  & \cmark 
  \\
  BCDMS, NMC, SLAC $F_\mathrm{L}$
  & \cite{Benvenuti:1989rh,Arneodo:1996qe,Whitlow:1990gk}
  & \xmark
  & \xmark
  & \xmark
  & \xmark
  & \cmark
  \\
  CHORUS $\sigma_{CC}^{\nu},\sigma_{CC}^{\bar{\nu}}$
  & \cite{Onengut:2005kv}
  & \cmark
  & \cmark
  & \xmark
  & \xmark
  & \cmark 
  \\
  CHORUS
  & \cite{Kayis-Topaksu:2011ols}
  & \xmark
  & \xmark
  & \cmark
  & \xmark
  & \xmark
  \\
  NuTeV $F_2, F_3$
  & \cite{Tzanov:2005kr}
  & \xmark
  & \xmark
  & \xmark
  & \xmark
  & \cmark
  \\
  NuTeV/CCFR $\sigma_{CC}^{\nu},\sigma_{CC}^{\bar{\nu}}$
  & \cite{Goncharov:2001qe}
  & \cmark
  & \cmark
  & \cmark
  & \cmark
  & \cmark
  \\
  EMC $F_2^c$
  & \cite{Aubert:1982tt}
  & \ymark
  & \ymark
  & \xmark
  & \xmark
  & \xmark
  \\
  NOMAD 
  & \cite{Samoylov:2013xoa}
  & \xmark
  & \ymark
  & \cmark
  & \xmark
  & \xmark
  \\
  CCFR $x F_3^p$
  & \cite{Seligman:1997mc}
  & \xmark
  & \xmark
  & \xmark
  & \cmark
  & \xmark
  \\
  CCFR $F_2^p$
  & \cite{CCFRNuTeV:2000qwc}
  & \xmark
  & \xmark
  & \xmark
  & \cmark
  & \xmark
  \\
  CDSHW $F_2^p, x F_3^p$
  & \cite{Berge:1989hr}
  & \xmark
  & \xmark
  & \xmark
  & \cmark
  & \xmark
  \\
  E665 $F_2^p, F_2^d$
  & \cite{E665:1996mob}
  & \xmark
  & \xmark
  & \xmark
  & \xmark
  & \cmark
  \\
  HERA NC, CC
  & \cite{Aaron:2009aa}
  & \xmark
  & \xmark
  & \xmark
  & \xmark
  & \cmark
  \\
  HERA I+II $\sigma^p_{\rm NC,CC}$
  & \cite{Abramowicz:2015mha}
  & \cmark 
  & \cmark
  & \cmark
  & \cmark
  & \xmark
  \\
  HERA I+II $\sigma_{c\bar c}^{\rm red}$
  & \cite{H1:2018flt}
  & \xmark
  & \cmark
  & \xmark
  & \ymark
  & \cmark
  \\
  HERA I+II $\sigma_{b\bar b}^{\rm red}$
  & \cite{H1:2018flt}
  & \xmark
  & \cmark
  & \xmark
  & \ymark
  & \xmark
  \\
  HERA I+II $\sigma_{c\bar c}^{\rm red}$
  & \cite{Abramowicz:1900rp}
  & \cmark
  & \xmark
  & \cmark
  & \cmark
  & \xmark
  \\
  H1 $F_2^{c\bar{c}}$
  & \cite{Aktas:2004az}
  & \xmark
  & \xmark
  & \xmark
  & \cmark
  & \xmark
  \\
  H1 $F_2^{b\bar{b}}$
  & \cite{Aaron:2009af}
  & \cmark
  & \xmark
  & \cmark
  & \xmark
  & \xmark
  \\
  ZEUS $\sigma_{b\bar{b}}^\mathrm{red}$
  & \cite{Abramowicz:2014zub}
  & \cmark
  & \xmark
  & \cmark
  & \xmark
  & \xmark
  \\
  H1 $F_\mathrm{L}$
  & \cite{Collaboration:2010ry}
  & \xmark
  & \xmark
  & \xmark
  & \cmark
  & \cmark
  \\
  H1 and ZEUS $F_\mathrm{L}$
  & \cite{h1fl,Chekanov:2009na}
  & \xmark
  & \xmark
  & \xmark
  & \xmark
  & \cmark
  \\
  ZEUS 820 (HQ) (1j)
  & \cite{ZEUS:2002nms}
  & \xmark
  & \ymark
  & \xmark
  & \xmark
  & \xmark
  \\
  ZEUS 920 (HQ) (1j)
  & \cite{ZEUS:2006xvn}
  & \xmark
  & \ymark
  & \xmark
  & \xmark
  & \xmark
  \\
  H1 (LQ) (1j-2j)
  & \cite{H1:2016goa}
  & \xmark
  & \ymark
  & \xmark
  & \xmark
  & \xmark
  \\
  H1 (HQ) (1j-2j)
  & \cite{H1:2014cbm}
  & \xmark
  & \ymark
  & \xmark
  & \xmark
  & \xmark
  \\
  ZEUS 920 (HQ) (2j)
  & \cite{ZEUS:2010vyw}
  & \xmark
  & \ymark
  & \xmark
  & \xmark
  & \xmark
  \\
  \bottomrule
\end{tabularx}

%% file: tables/tab-datacomp_FTDY.tex
\begin{tabularx}{\textwidth}{Xcccccc}
  \toprule
  Data set
  & Ref.
  & NNPDF3.1
  & NNPDF4.0
  & ABMP16
  & CT18
  & MSHT20
  \\
  \midrule
  DY E866 $\sigma^d_{\rm DY}/\sigma^p_{\rm DY}$ (NuSea)
  & \cite{Towell:2001nh}
  & \cmark
  & \cmark
  & \cmark
  & \cmark
  & \cmark
  \\
  DY E866 $\sigma^p_{\rm DY}$
  & \cite{Webb:2003ps}
  & \cmark
  & \cmark
  & \xmark
  & \cmark
  & \cmark
  \\
  DY E605 $\sigma^p_{\rm DY}$
  & \cite{Moreno:1990sf}
  & \cmark
  & \cmark
  & \cmark
  & \cmark
  & \xmark
  \\
  DY E906 $\sigma^d_{\rm DY}/\sigma^p_{\rm DY}$ (SeaQuest)
  & \cite{Dove:2021ejl}
  & \xmark
  & \cmark
  & \xmark
  & \xmark
  & \xmark
  \\
  \bottomrule
\end{tabularx}

%% file: tables/tab-datacomp_Tevatron.tex
\begin{tabularx}{\textwidth}{Xcccccc}
  \toprule
  Data set
  & Ref.
  & NNPDF3.1
  & NNPDF4.0
  & ABMP16
  & CT18
  & MSHT20
  \\
  \midrule
  CDF $Z$ rapidity
  & \cite{Aaltonen:2010zza}
  & \cmark
  & \cmark
  & \xmark
  & \cmark
  & \cmark
  \\
  CDF $W\to\ell\nu$ asymmetry ($1.8$~TeV)
  & \cite{CDF:1998uzn}
  & \xmark
  & \xmark
  & \xmark
  & \cmark
  & \xmark
  \\
  CDF $W\to e\nu$ asymmetry ($\mathcal{L} = 170$~pb${}^{-1}$)
  & \cite{Acosta:2005ud}
  & \xmark
  & \xmark
  & \xmark
  & \cmark
  & \xmark
  \\
  CDF $W\to e\nu$ asymmetry ($\mathcal{L} = 1$~fb${}^{-1}$)
  & \cite{Aaltonen:2009ta}
  & \xmark
  & \xmark
  & \xmark
  & \xmark
  & \cmark
  \\
  CDF $k_t$ inclusive jets
  & \cite{Abulencia:2007ez} 
  & \cmark 
  & \xmark
  & \xmark
  & \xmark
  & \cmark
  \\
  CDF cone-based inclusive jets
  & \cite{Aaltonen:2008eq}
  & \xmark
  & \xmark
  & \xmark
  & \cmark
  & \xmark
  \\
  D0 $Z$ rapidity
  & \cite{Abazov:2007jy}
  & \cmark
  & \cmark
  & \xmark
  & \cmark
  & \cmark
  \\
  D0 $W\to e\nu$ asymmetry ($\mathcal{L} = 0.75$~fb${}^{-1}$)
  & \cite{Abazov:2008qv}
  & \xmark
  & \xmark
  & \xmark
  & \xmark
  & \cmark
  \\
  D0 $W\to e\nu$ (prod.) asymmetry ($\mathcal{L} = 9.7$~fb${}^{-1}$)
  & \cite{D0:2013lql}
  & \xmark
  & \xmark
  & \ymark
  & \xmark
  & \cmark
  \\
  D0 $W\to e\nu$ (prod.\ and decay) asymmetry ($\mathcal{L} = 9.7$~fb${}^{-1}$)
  & \cite{D0:2014kma}
  & \cmark
  & \ymark
  & \cmark
  & \cmark
  & \xmark
  \\
  D0 $W\to \mu\nu$ asymmetry ($\mathcal{L} = 0.3$~fb${}^{-1}$)
  & \cite{D0:2007pcy}
  & \xmark
  & \xmark
  & \xmark
  & \cmark
  & \xmark
  \\
  D0 $W\to \mu\nu$ asymmetry ($\mathcal{L} = 7.3$~fb${}^{-1}$)
  & \cite{Abazov:2013rja}
  & \cmark
  & \cmark
  & \cmark
  & \xmark
  & \cmark
  \\
  D0 cone-based inclusive jets
  & \cite{D0:2008hua}
  & \xmark
  & \xmark
  & \xmark
  & \cmark
  & \cmark 
  \\
  CDF and D0 top-pair production
  & \cite{CDF:2013hmv}
  & \xmark
  & \xmark
  & \ymark
  & \xmark
  & \cmark
  \\
  CDF and D0 single-top production
  & \cite{CDF:2015gsg}
  & \xmark
  & \xmark
  & \cmark
  & \xmark
  & \xmark
  \\
  \bottomrule
\end{tabularx}

%% file: tables/tab-datacomp_ATLAS.tex
\begin{tabularx}{\textwidth}{Xcccccc}
  \toprule
  Data set
  & Ref.
  & NNPDF3.1
  & NNPDF4.0
  & ABMP16
  & CT18
  & MSHT20
  \\
  \midrule
  ATLAS $W,Z$ 7 TeV ($\mathcal{L}=35$~pb$^{-1}$)
  & \cite{Aad:2011dm}
  & \cmark
  & \cmark
  & \cmark
  & \cmark
  & \cmark
  \\
  ATLAS $W,Z$ 7 TeV ($\mathcal{L}=4.6$~fb$^{-1}$)
  & \cite{Aaboud:2016btc}
  & \cmark 
  & \cmark 
  & \xmark
  & \ymark
  & \cmark
  \\
  ATLAS low-mass DY 7 TeV
  & \cite{Aad:2014qja}
  & \cmark
  & \cmark
  & \xmark
  & \ymark
  & \xmark
  \\
  ATLAS high-mass DY 7 TeV
  & \cite{Aad:2013iua}
  & \cmark
  & \cmark
  & \xmark
  & \ymark
  & \cmark
  \\
  ATLAS $W$ 8 TeV
  & \cite{Aad:2019rou}
  & \xmark
  & \ymark
  & \xmark
  & \xmark
  & \cmark
  \\
  ATLAS DY 2D 8 TeV
  & \cite{Aaboud:2017ffb}
  & \xmark
  & \cmark
  & \xmark
  & \xmark
  & \cmark
  \\
  ATLAS high-mass DY 2D 8 TeV
  & \cite{Aad:2016zzw}
  & \xmark
  & \cmark
  & \xmark
  & \ymark
  & \cmark
  \\
  ATLAS $\sigma_{W,Z}$ 13 TeV
  & \cite{Aad:2016naf}
  & \xmark
  & \cmark
  & \cmark
  & \xmark
  & \xmark
  \\
  ATLAS $W$+jet 8 TeV
  & \cite{Aaboud:2017soa}
  & \xmark
  & \cmark
  & \xmark
  & \xmark
  & \cmark
  \\
  ATLAS $Z$ $p_T$ 7 TeV
  & \cite{Aad:2014xaa}
  & \ymark
  & \xmark
  & \xmark
  & \ymark
  & \xmark
  \\
  ATLAS $Z$ $p_T$ 8 TeV 
  & \cite{Aad:2015auj}
  & \cmark
  & \cmark
  & \xmark
  & \cmark 
  & \cmark
  \\
  ATLAS $W + c$ 7 TeV
  & \cite{Aad:2014xca}
  & \xmark
  & \cmark
  & \xmark
  & \ymark
  & \xmark
  \\
  ATLAS $\sigma_{tt}^{\rm tot}$ 7, 8 TeV
  & \cite{Aad:2014kva}
  & \cmark
  & \cmark
  & \cmark
  & \xmark
  & \xmark
  \\
  ATLAS $\sigma_{tt}^{\rm tot}$ 7, 8 TeV
  & \cite{Aad:2012vip,Aad:2014jra,Aad:2015pga,Aad:2015dya,ATLAS:2012gpa,ATLAS:2012uma}
  & \xmark
  & \xmark
  & \cmark
  & \xmark
  & \xmark
  \\
  ATLAS $\sigma_{tt}^{\rm tot}$ 13 TeV ($\mathcal{L}=3.2$~fb$^{-1}$)
  & \cite{Aaboud:2016pbd}
  & \cmark
  & \xmark
  & \cmark
  & \xmark
  & \xmark
  \\
  ATLAS $\sigma_{tt}^{\rm tot}$ 13 TeV ($\mathcal{L}=139$~fb$^{-1}$)
  & \cite{Aad:2020tmz}
  & \xmark
  & \cmark
  & \xmark
  & \xmark
  & \xmark
  \\
  ATLAS $\sigma_{tt}^{\rm tot}$ and $Z$ ratios
  & \cite{ATLAS:2019hau}
  & \xmark
  & \xmark
  & \xmark
  & \xmark
  & \ymark
  \\
  ATLAS $t\bar{t}$ lepton+jets 8 TeV
  & \cite{Aad:2015mbv}
  & \cmark
  & \cmark
  & \xmark
  & \cmark
  & \cmark
  \\
  ATLAS $t\bar{t}$ dilepton 8 TeV
  & \cite{Aaboud:2016iot}
  & \xmark
  & \cmark
  & \xmark
  & \xmark
  & \cmark
  \\
  ATLAS single-inclusive jets 7 TeV, R=0.6
  & \cite{Aad:2014vwa}
  & \cmark
  & \ymark
  & \xmark
  & \cmark
  & \cmark
  \\
  ATLAS single-inclusive jets 8 TeV, R=0.6
  & \cite{Aaboud:2017dvo}
  & \xmark
  & \cmark
  & \xmark
  & \xmark
  & \xmark
  \\
  ATLAS dijets 7 TeV, R=0.6
  & \cite{Aad:2013tea}
  & \xmark
  & \cmark
  & \xmark
  & \xmark
  & \xmark
  \\
  ATLAS direct photon production 8 TeV
  & \cite{Aad:2016xcr}
  & \xmark
  & \ymark
  & \xmark
  & \xmark
  & \xmark
  \\
  ATLAS direct photon production 13 TeV
  & \cite{ATLAS:2017nah}
  & \xmark
  & \cmark
  & \xmark
  & \xmark
  & \xmark
  \\
  ATLAS single top $R_{t}$ 7, 8, 13 TeV
  & \cite{Aad:2014fwa,Aaboud:2016ymp,Aaboud:2017pdi}
  & \xmark
  & \cmark
  & \cmark
  & \xmark
  & \xmark
  \\
  ATLAS single top diff. 7 TeV
  & \cite{Aad:2014fwa}
  & \xmark
  & \cmark
  & \xmark
  & \xmark
  & \xmark
  \\
  ATLAS single top diff. 8 TeV
  & \cite{Aaboud:2017pdi}
  & \xmark
  & \cmark
  & \xmark
  & \xmark
  & \xmark
  \\
  \bottomrule
\end{tabularx}

%% file: tables/tab-datacomp_CMS.tex
\begin{tabularx}{\textwidth}{Xcccccc}
  \toprule
  Data set
  & Ref.
  & NNPDF3.1
  & NNPDF4.0
  & ABMP16
  & CT18
  & MSHT20
  \\
  \midrule
  CMS $W$ asym.\ 7 TeV ($\mathcal{L}=36$~pb$^{-1}$)
  & \cite{Chatrchyan:2011jz}
  & \xmark
  & \xmark
  & \xmark
  & \xmark
  & \cmark
  \\
  CMS $Z$ 7 TeV ($\mathcal{L}=36$~pb$^{-1}$)
  & \cite{Chatrchyan:2011wt}
  & \xmark
  & \xmark
  & \xmark
  & \xmark
  & \cmark
  \\
  CMS $W$ electron asymmetry 7 TeV
  & \cite{Chatrchyan:2012xt}
  & \cmark
  & \cmark
  & \xmark
  & \cmark
  & \cmark
  \\
  CMS $W$ muon asymmetry 7 TeV
  & \cite{Chatrchyan:2013mza}
  & \cmark
  & \cmark
  & \cmark
  & \cmark
  & \xmark
  \\
  CMS Drell-Yan 2D 7 TeV
  & \cite{Chatrchyan:2013tia}
  & \cmark
  & \cmark
  & \xmark
  & \ymark
  & \cmark
  \\
  CMS Drell-Yan 2D 8 TeV
  & \cite{CMS:2014jea}
  & \ymark
  & \xmark
  & \xmark
  & \xmark
  & \xmark
  \\
  CMS $W$ rapidity 8 TeV
  & \cite{Khachatryan:2016pev}
  & \cmark
  & \cmark
  & \cmark
  & \cmark
  & \cmark
  \\
  CMS $W,Z$ $p_T$ 8 TeV ($\mathcal{L} = 18.4$~fb${}^{-1}$)
  & \cite{CMS:2016mwa}
  & \xmark
  & \xmark
  & \xmark
  & \ymark
  & \xmark
  \\
  CMS $Z$ $p_T$ 8 TeV
  & \cite{Khachatryan:2015oaa}
  & \cmark
  & \cmark
  & \xmark
  & \ymark
  & \xmark
  \\
  CMS $W+c$ 7 TeV
  & \cite{Chatrchyan:2013uja}
  & \cmark
  & \cmark
  & \xmark
  & \ymark
  & \cmark
  \\
  CMS $W+c$ 13 TeV
  & \cite{Sirunyan:2018hde}
  & \xmark
  & \cmark
  & \xmark
  & \xmark
  & \ymark
  \\
  CMS single-inclusive jets 2.76 TeV
  & \cite{Khachatryan:2015luy}
  & \cmark
  & \xmark
  & \xmark
  & \xmark
  & \cmark
  \\
  CMS single-inclusive jets 7 TeV
  & \cite{Chatrchyan:2014gia}
  & \cmark
  & \ymark
  & \xmark
  & \cmark
  & \cmark
  \\
  CMS dijets 7 TeV
  & \cite{Chatrchyan:2012bja}
  & \xmark
  & \cmark
  & \xmark
  & \xmark
  & \xmark
  \\
  CMS single-inclusive jets 8 TeV
  & \cite{Khachatryan:2016mlc}
  & \xmark
  & \cmark
  & \xmark
  & \cmark
  & \cmark
  \\
  CMS 3D dijets 8 TeV
  & \cite{Sirunyan:2017skj}
  & \xmark
  & \ymark
  & \xmark
  & \xmark
  & \xmark
  \\
  CMS $\sigma_{tt}^{\rm tot}$ 5 TeV
  & \cite{Sirunyan:2017ule}
  & \xmark
  & \cmark
  & \xmark
  & \xmark
  & \xmark
  \\
  CMS $\sigma_{tt}^{\rm tot}$ 7, 8 TeV
  & \cite{Spannagel:2016cqt}
  & \cmark
  & \cmark
  & \xmark
  & \xmark
  & \xmark
  \\
  CMS $\sigma_{tt}^{\rm tot}$ 8 TeV
  & \cite{Chatrchyan:2013faa}
  & \xmark
  & \xmark
  & \xmark
  & \xmark
  & \cmark
  \\
  CMS $\sigma_{tt}^{\rm tot}$ 5, 7, 8, 13 TeV
  & \cite{Khachatryan:2016mqs,Khachatryan:2016kzg,Khachatryan:2016yzq,CMS:2015toa,Chatrchyan:2012vs,Chatrchyan:2013kff,Chatrchyan:2013ual,CMS:2016rtp,Khachatryan:2014loa,CMS:2016pqu}
  & \xmark
  & \xmark
  & \cmark
  & \xmark
  & \xmark
  \\
  CMS $\sigma_{tt}^{\rm tot}$ 13 TeV
  & \cite{Khachatryan:2015uqb}
  & \cmark
  & \cmark
  & \cmark
  & \xmark
  & \xmark
  \\
  CMS $t\bar{t}$ lepton+jets 8 TeV
  & \cite{Khachatryan:2015oqa}
  & \cmark
  & \cmark
  & \xmark
  & \xmark
  & \cmark
  \\
  CMS $t\bar{t}$ 2D dilepton 8 TeV
  & \cite{Sirunyan:2017azo}
  & \xmark
  & \cmark
  & \xmark
  & \cmark
  & \cmark
  \\
  CMS $t\bar{t}$ lepton+jet 13 TeV
  & \cite{Sirunyan:2018wem}
  & \xmark
  & \cmark
  & \xmark
  & \xmark
  & \xmark
  \\
  CMS $t\bar{t}$ dilepton 13 TeV
  & \cite{Sirunyan:2018ucr}
  & \xmark
  & \cmark
  & \xmark
  & \xmark
  & \xmark
  \\
  CMS single top $\sigma_{t}+\sigma_{\bar{t}}$ 7 TeV
  & \cite{Chatrchyan:2012ep}
  & \xmark
  & \cmark
  & \cmark
  & \xmark
  & \xmark
  \\
  CMS single top $R_{t}$ 8, 13 TeV
  & \cite{Khachatryan:2014iya,Sirunyan:2016cdg}
  & \xmark
  & \cmark
  & \cmark
  & \xmark
  & \xmark
  \\
  CMS single top 13 TeV
  & \cite{CMS:2018lgn,CMS:2019jjp}
  & \xmark
  & \xmark
  & \xmark
  & \xmark
  & \ymark
  \\
  \bottomrule
\end{tabularx}

%% file: tables/tab-datacomp_LHCb.tex
\begin{tabularx}{\textwidth}{Xcccccc}
  \toprule
  Data set
  & Ref.
  & NNPDF3.1
  & NNPDF4.0
  & ABMP16
  & CT18
  & MSHT20
  \\
  \midrule
  LHCb $Z$ 7~TeV ($\mathcal{L}=940$~pb$^{-1}$)
  & \cite{Aaij:2012mda}
  & \cmark
  & \cmark
  & \xmark
  & \xmark
  & \cmark
  \\
  LHCb $Z\to ee$ 8~TeV ($\mathcal{L}=2$~fb$^{-1}$)
  & \cite{Aaij:2015vua}
  & \cmark
  & \cmark
  & \cmark
  & \cmark
  & \cmark
  \\
  LHCb $W$ 7~TeV ($\mathcal{L}=37$~pb$^{-1}$)
  & \cite{Aaij:2012vn}
  & \xmark
  & \xmark
  & \xmark
  & \xmark
  & \cmark
  \\
  LHCb $W,Z \to \mu$ 7 TeV
  & \cite{Aaij:2015gna}
  & \cmark
  & \cmark
  & \cmark
  & \cmark
  & \cmark
  \\
  LHCb $W,Z \to \mu$ 8 TeV
  & \cite{Aaij:2015zlq}
  & \cmark
  & \cmark
  & \cmark
  & \cmark
  & \cmark
  \\
  LHCb $W\to e$ 8 TeV
  & \cite{Aaij:2016qqz}
  & \xmark
  & \ymark
  & \xmark
  & \xmark
  & \xmark
  \\
  LHCb $Z\to \mu\mu, ee$ 13 TeV
  & \cite{Aaij:2016mgv}
  & \xmark
  & \cmark
  & \xmark
  & \xmark
  & \xmark
  \\
  \bottomrule
\end{tabularx}

%% file: nnpdf40.bbl
\providecommand{\href}[2]{#2}\begingroup\raggedright\begin{thebibliography}{100}

\bibitem{Salam:2018rwo}
G.~P. Salam, {\it {Theory vision}},  {\em PoS} {\bf LHCP2018} (2018) 304,
  [\href{http://arxiv.org/abs/1811.11282}{{\tt arXiv:1811.11282}}].

\bibitem{Heinrich:2020ybq}
G.~Heinrich, {\it {Collider Physics at the Precision Frontier}},
  \href{http://arxiv.org/abs/2009.00516}{{\tt arXiv:2009.00516}}.

\bibitem{Gao:2017yyd}
J.~Gao, L.~Harland-Lang, and J.~Rojo, {\it {The Structure of the Proton in the
  LHC Precision Era}},  {\em Phys. Rept.} {\bf 742} (2018) 1--121,
  [\href{http://arxiv.org/abs/1709.04922}{{\tt arXiv:1709.04922}}].

\bibitem{Ethier:2020way}
J.~J. Ethier and E.~R. Nocera, {\it {Parton Distributions in Nucleons and
  Nuclei}},  {\em Ann. Rev. Nucl. Part. Sci.} {\bf 70} (2020) 43--76,
  [\href{http://arxiv.org/abs/2001.07722}{{\tt arXiv:2001.07722}}].

\bibitem{Ball:2017nwa}
{\bf NNPDF} Collaboration, R.~D. Ball et~al., {\it {Parton distributions from
  high-precision collider data}},  {\em Eur. Phys. J.} {\bf C77} (2017), no.~10
  663, [\href{http://arxiv.org/abs/1706.00428}{{\tt arXiv:1706.00428}}].

\bibitem{Forte:2020yip}
S.~Forte and S.~Carrazza, {\it {Parton distribution functions}},
  \href{http://arxiv.org/abs/2008.12305}{{\tt arXiv:2008.12305}}.

\bibitem{Campbell:2018wfu}
J.~M. Campbell, J.~Rojo, E.~Slade, and C.~Williams, {\it {Direct photon
  production and PDF fits reloaded}},  {\em Eur. Phys. J. C} {\bf 78} (2018),
  no.~6 470, [\href{http://arxiv.org/abs/1802.03021}{{\tt arXiv:1802.03021}}].

\bibitem{Nocera:2019wyk}
E.~R. Nocera, M.~Ubiali, and C.~Voisey, {\it {Single Top Production in PDF
  fits}},  {\em JHEP} {\bf 05} (2020) 067,
  [\href{http://arxiv.org/abs/1912.09543}{{\tt arXiv:1912.09543}}].

\bibitem{AbdulKhalek:2020jut}
R.~Abdul~Khalek et~al., {\it {Phenomenology of NNLO jet production at the LHC
  and its impact on parton distributions}},  {\em Eur. Phys. J. C} {\bf 80}
  (2020), no.~8 797, [\href{http://arxiv.org/abs/2005.11327}{{\tt
  arXiv:2005.11327}}].

\bibitem{Faura:2020oom}
F.~Faura, S.~Iranipour, E.~R. Nocera, J.~Rojo, and M.~Ubiali, {\it {The
  Strangest Proton?}},  {\em Eur. Phys. J. C} {\bf 80} (2020), no.~12 1168,
  [\href{http://arxiv.org/abs/2009.00014}{{\tt arXiv:2009.00014}}].

\bibitem{Carrazza:2019mzf}
S.~Carrazza and J.~Cruz-Martinez, {\it {Towards a new generation of parton
  densities with deep learning models}},  {\em Eur. Phys. J. C} {\bf 79}
  (2019), no.~8 676, [\href{http://arxiv.org/abs/1907.05075}{{\tt
  arXiv:1907.05075}}].

\bibitem{Carrazza:2019agm}
S.~Carrazza, J.~Cruz-Martinez, J.~Urtasun-Elizari, and E.~Villa, {\it {Towards
  hardware acceleration for parton densities estimation}},  {\em Frascati Phys.
  Ser.} {\bf 69} (2019) 1--6, [\href{http://arxiv.org/abs/1909.10547}{{\tt
  arXiv:1909.10547}}].

\bibitem{Cruz-Martinez:2020tte}
J.~M. Cruz-Martinez, S.~Carrazza, and R.~Stegeman, {\it {Studying the parton
  content of the proton with deep learning models}},  {\em PoS} {\bf AISIS2019}
  (2020) 008, [\href{http://arxiv.org/abs/2002.06587}{{\tt arXiv:2002.06587}}].

\bibitem{Ball:2014uwa}
{\bf NNPDF} Collaboration, R.~D. Ball et~al., {\it {Parton distributions for
  the LHC Run II}},  {\em JHEP} {\bf 04} (2015) 040,
  [\href{http://arxiv.org/abs/1410.8849}{{\tt arXiv:1410.8849}}].

\bibitem{Cruz-Martinez:2021rgy}
J.~Cruz-Martinez, S.~Forte, and E.~R. Nocera, {\it {Future tests of parton
  distributions}},  {\em Acta Phys. Polon. B} {\bf 52} (2021) 243,
  [\href{http://arxiv.org/abs/2103.08606}{{\tt arXiv:2103.08606}}].

\bibitem{Carrazza:2020gss}
S.~Carrazza, E.~R. Nocera, C.~Schwan, and M.~Zaro, {\it {PineAPPL: combining EW
  and QCD corrections for fast evaluation of LHC processes}},  {\em JHEP} {\bf
  12} (2020) 108, [\href{http://arxiv.org/abs/2008.12789}{{\tt
  arXiv:2008.12789}}].

\bibitem{Ball:2018lag}
R.~D. Ball and A.~Deshpande, {\em {The proton spin, semi-inclusive processes,
  and measurements at a future Electron Ion Collider}}.
\newblock 2019.
\newblock \href{http://arxiv.org/abs/1801.04842}{{\tt arXiv:1801.04842}}.

\bibitem{Ball:2018twp}
{\bf NNPDF} Collaboration, R.~D. Ball, E.~R. Nocera, and R.~L. Pearson, {\it
  {Nuclear Uncertainties in the Determination of Proton PDFs}},  {\em Eur.
  Phys. J.} {\bf C79} (2019), no.~3 282,
  [\href{http://arxiv.org/abs/1812.09074}{{\tt arXiv:1812.09074}}].

\bibitem{Ball:2020xqw}
R.~D. Ball, E.~R. Nocera, and R.~L. Pearson, {\it {Deuteron Uncertainties in
  the Determination of Proton PDFs}},  {\em Eur. Phys. J. C} {\bf 81} (2021),
  no.~1 37, [\href{http://arxiv.org/abs/2011.00009}{{\tt arXiv:2011.00009}}].

\bibitem{AbdulKhalek:2020yuc}
R.~Abdul~Khalek, J.~J. Ethier, J.~Rojo, and G.~van Weelden, {\it {nNNPDF2.0:
  quark flavor separation in nuclei from LHC data}},  {\em JHEP} {\bf 09}
  (2020) 183, [\href{http://arxiv.org/abs/2006.14629}{{\tt arXiv:2006.14629}}].

\bibitem{Candido:2020yat}
A.~Candido, S.~Forte, and F.~Hekhorn, {\it {Can $ \overline{\mathrm{MS}} $
  parton distributions be negative?}},  {\em JHEP} {\bf 11} (2020) 129,
  [\href{http://arxiv.org/abs/2006.07377}{{\tt arXiv:2006.07377}}].

\bibitem{Forte:1992df}
S.~Forte, {\it {The Gottfried sum rule and the light flavor content of the
  nucleon}},  {\em Phys. Rev. D} {\bf 47} (1993) 1842--1853.

\bibitem{AbdulKhalek:2019bux}
{\bf NNPDF} Collaboration, R.~Abdul~Khalek et~al., {\it {A first determination
  of parton distributions with theoretical uncertainties}},  {\em Eur. Phys.
  J.} {\bf C} (2019) 79:838, [\href{http://arxiv.org/abs/1905.04311}{{\tt
  arXiv:1905.04311}}].

\bibitem{AbdulKhalek:2019ihb}
{\bf NNPDF} Collaboration, R.~Abdul~Khalek et~al., {\it {Parton Distributions
  with Theory Uncertainties: General Formalism and First Phenomenological
  Studies}},  {\em Eur. Phys. J. C} {\bf 79} (2019), no.~11 931,
  [\href{http://arxiv.org/abs/1906.10698}{{\tt arXiv:1906.10698}}].

\bibitem{Forte:2010ta}
S.~Forte, E.~Laenen, P.~Nason, and J.~Rojo, {\it {Heavy quarks in
  deep-inelastic scattering}},  {\em Nucl. Phys.} {\bf B834} (2010) 116--162,
  [\href{http://arxiv.org/abs/1001.2312}{{\tt arXiv:1001.2312}}].

\bibitem{Carrazza:2015aoa}
S.~Carrazza, S.~Forte, Z.~Kassabov, J.~I. Latorre, and J.~Rojo, {\it {An
  Unbiased Hessian Representation for Monte Carlo PDFs}},  {\em Eur. Phys. J.}
  {\bf C75} (2015), no.~8 369, [\href{http://arxiv.org/abs/1505.06736}{{\tt
  arXiv:1505.06736}}].

\bibitem{Carrazza:2016htc}
S.~Carrazza, S.~Forte, Z.~Kassabov, and J.~Rojo, {\it {Specialized minimal PDFs
  for optimized LHC calculations}},  {\em Eur. Phys. J.} {\bf C76} (2016),
  no.~4 205, [\href{http://arxiv.org/abs/1602.00005}{{\tt arXiv:1602.00005}}].

\bibitem{Carrazza:2015hva}
S.~Carrazza, J.~I. Latorre, J.~Rojo, and G.~Watt, {\it {A compression algorithm
  for the combination of PDF sets}},  {\em Eur. Phys. J.} {\bf C75} (2015) 474,
  [\href{http://arxiv.org/abs/1504.06469}{{\tt arXiv:1504.06469}}].

\bibitem{Carrazza:2021hny}
S.~Carrazza, J.~M. Cruz-Martinez, and T.~R. Rabemananjara, {\it {Compressing
  PDF sets using generative adversarial networks}},  {\em Eur. Phys. J. C} {\bf
  81} (2021), no.~6 530, [\href{http://arxiv.org/abs/2104.04535}{{\tt
  arXiv:2104.04535}}].

\bibitem{nnpdfcode}
R.~D. Ball, S.~Carrazza, J.~M. Cruz-Martinez, L.~Del~Debbio, S.~Forte,
  T.~Giani, S.~Iranipour, Z.~Kassabov, J.~I. Latorre, E.~R. Nocera, R.~L.
  Pearson, J.~Rojo, R.~Stegeman, C.~Schwan, M.~Ubiali, C.~Voisey, and
  M.~Wilson, {\it Nnpdf/nnpdf: nnpdf v4.0.3},  Sept., 2021.

\bibitem{NNPDF:2021uiq}
{\bf NNPDF} Collaboration, R.~D. Ball et~al., {\it {An open-source machine
  learning framework for global analyses of parton distributions}},
  \href{http://arxiv.org/abs/2109.02671}{{\tt arXiv:2109.02671}}.

\bibitem{Buckley:2014ana}
A.~Buckley, J.~Ferrando, S.~Lloyd, K.~Nordström, B.~Page, et~al., {\it
  {LHAPDF6: parton density access in the LHC precision era}},  {\em
  Eur.Phys.J.} {\bf C75} (2015) 132,
  [\href{http://arxiv.org/abs/1412.7420}{{\tt arXiv:1412.7420}}].

\bibitem{Arneodo:1996kd}
{\bf New Muon} Collaboration, M.~Arneodo et~al., {\it {Accurate measurement of
  $F_2^d/F_2^p$ and $R_d-R_p$}},  {\em Nucl. Phys.} {\bf B487} (1997) 3--26,
  [\href{http://arxiv.org/abs/hep-ex/9611022}{{\tt hep-ex/9611022}}].

\bibitem{Arneodo:1996qe}
{\bf New Muon} Collaboration, M.~Arneodo et~al., {\it {Measurement of the
  proton and deuteron structure functions, $F_2^p$ and $F_2^d$, and of the
  ratio $\sigma_L/\sigma_T$}},  {\em Nucl. Phys.} {\bf B483} (1997) 3--43,
  [\href{http://arxiv.org/abs/hep-ph/9610231}{{\tt hep-ph/9610231}}].

\bibitem{Whitlow:1991uw}
L.~W. Whitlow, E.~M. Riordan, S.~Dasu, S.~Rock, and A.~Bodek, {\it {Precise
  measurements of the proton and deuteron structure functions from a global
  analysis of the SLAC deep inelastic electron scattering cross-sections}},
  {\em Phys. Lett.} {\bf B282} (1992) 475--482.

\bibitem{Benvenuti:1989rh}
{\bf BCDMS} Collaboration, A.~C. Benvenuti et~al., {\it {A High Statistics
  Measurement of the Proton Structure Functions $F_2(x, Q^2)$ and $R$ from Deep
  Inelastic Muon Scattering at High $Q^2$}},  {\em Phys. Lett.} {\bf B223}
  (1989) 485.

\bibitem{Onengut:2005kv}
{\bf CHORUS} Collaboration, G.~Onengut et~al., {\it {Measurement of nucleon
  structure functions in neutrino scattering}},  {\em Phys. Lett.} {\bf B632}
  (2006) 65--75.

\bibitem{Goncharov:2001qe}
{\bf NuTeV} Collaboration, M.~Goncharov et~al., {\it {Precise measurement of
  dimuon production cross-sections in $\nu_{\mu}$Fe and $\bar{\nu}_{\mu}$Fe
  deep inelastic scattering at the Tevatron}},  {\em Phys. Rev.} {\bf D64}
  (2001) 112006, [\href{http://arxiv.org/abs/hep-ex/0102049}{{\tt
  hep-ex/0102049}}].

\bibitem{MasonPhD}
D.~A. Mason, {\it {Measurement of the strange - antistrange asymmetry at NLO in
  QCD from NuTeV dimuon data}}, . FERMILAB-THESIS-2006-01.

\bibitem{Abramowicz:2015mha}
{\bf ZEUS, H1} Collaboration, H.~Abramowicz et~al., {\it {Combination of
  measurements of inclusive deep inelastic ${e^{\pm }p}$ scattering cross
  sections and QCD analysis of HERA data}},  {\em Eur. Phys. J.} {\bf C75}
  (2015), no.~12 580, [\href{http://arxiv.org/abs/1506.06042}{{\tt
  arXiv:1506.06042}}].

\bibitem{Abramowicz:1900rp}
{\bf H1 , ZEUS} Collaboration, H.~Abramowicz et~al., {\it {Combination and QCD
  Analysis of Charm Production Cross Section Measurements in Deep-Inelastic ep
  Scattering at HERA}},  {\em Eur.Phys.J.} {\bf C73} (2013) 2311,
  [\href{http://arxiv.org/abs/1211.1182}{{\tt arXiv:1211.1182}}].

\bibitem{Aaron:2009af}
{\bf H1} Collaboration, F.~D. Aaron et~al., {\it {Measurement of the Charm and
  Beauty Structure Functions using the H1 Vertex Detector at HERA}},  {\em Eur.
  Phys. J.} {\bf C65} (2010) 89--109,
  [\href{http://arxiv.org/abs/0907.2643}{{\tt arXiv:0907.2643}}].

\bibitem{Abramowicz:2014zub}
{\bf ZEUS} Collaboration, H.~Abramowicz et~al., {\it {Measurement of beauty and
  charm production in deep inelastic scattering at HERA and measurement of the
  beauty-quark mass}},  {\em JHEP} {\bf 09} (2014) 127,
  [\href{http://arxiv.org/abs/1405.6915}{{\tt arXiv:1405.6915}}].

\bibitem{Aubert:1982tt}
{\bf European Muon} Collaboration, J.~J. Aubert et~al., {\it {Production of
  charmed particles in 250-GeV $\mu^+$ - iron interactions}},  {\em Nucl.
  Phys.} {\bf B213} (1983) 31--64.

\bibitem{Moreno:1990sf}
G.~Moreno et~al., {\it {Dimuon production in proton - copper collisions at
  $\sqrt{s}$ = 38.8-GeV}},  {\em Phys. Rev.} {\bf D43} (1991) 2815--2836.

\bibitem{Webb:2003ps}
{\bf NuSea} Collaboration, J.~C. Webb et~al., {\it {Absolute Drell-Yan dimuon
  cross sections in 800-GeV/c p p and p d collisions}},
  \href{http://arxiv.org/abs/hep-ex/0302019}{{\tt hep-ex/0302019}}.

\bibitem{Towell:2001nh}
{\bf FNAL E866/NuSea} Collaboration, R.~S. Towell et~al., {\it {Improved
  measurement of the anti-d/anti-u asymmetry in the nucleon sea}},  {\em Phys.
  Rev.} {\bf D64} (2001) 052002,
  [\href{http://arxiv.org/abs/hep-ex/0103030}{{\tt hep-ex/0103030}}].

\bibitem{Aaltonen:2010zza}
{\bf CDF} Collaboration, T.~A. Aaltonen et~al., {\it {Measurement of
  $d\sigma/dy$ of Drell-Yan $e^+e^-$ pairs in the $Z$ Mass Region from
  $p\bar{p}$ Collisions at $\sqrt{s}=1.96$ TeV}},  {\em Phys. Lett.} {\bf B692}
  (2010) 232--239, [\href{http://arxiv.org/abs/0908.3914}{{\tt
  arXiv:0908.3914}}].

\bibitem{Abazov:2007jy}
{\bf D0} Collaboration, V.~M. Abazov et~al., {\it {Measurement of the shape of
  the boson rapidity distribution for $p \bar{p} \to Z/\gamma^* \to e^{+}
  e^{-}$ + $X$ events produced at $\sqrt{s}$=1.96-TeV}},  {\em Phys. Rev.} {\bf
  D76} (2007) 012003, [\href{http://arxiv.org/abs/hep-ex/0702025}{{\tt
  hep-ex/0702025}}].

\bibitem{Abazov:2013rja}
{\bf D0} Collaboration, V.~M. Abazov et~al., {\it {Measurement of the muon
  charge asymmetry in $p\bar{p}$ $\to$ W+X $\to$ $\mu$$\nu$ + X events at
  $\sqrt{s}$=1.96 TeV}},  {\em Phys.Rev.} {\bf D88} (2013) 091102,
  [\href{http://arxiv.org/abs/1309.2591}{{\tt arXiv:1309.2591}}].

\bibitem{D0:2014kma}
{\bf D0} Collaboration, V.~M. Abazov et~al., {\it {Measurement of the electron
  charge asymmetry in $\boldsymbol{p\bar{p}\rightarrow W+X \rightarrow e\nu
  +X}$ decays in $\boldsymbol{p\bar{p}}$ collisions at
  $\boldsymbol{\sqrt{s}=1.96}$ TeV}},  {\em Phys. Rev.} {\bf D91} (2015), no.~3
  032007, [\href{http://arxiv.org/abs/1412.2862}{{\tt arXiv:1412.2862}}].
  [Erratum: Phys. Rev.D91,no.7,079901(2015)].

\bibitem{Abulencia:2007ez}
{\bf CDF - Run II} Collaboration, A.~Abulencia et~al., {\it {Measurement of the
  Inclusive Jet Cross Section using the $k_{\rm T}$ algorithm in
  $p\overline{p}$ Collisions at $\sqrt{s}$=1.96 TeV with the CDF II Detector}},
   {\em Phys. Rev.} {\bf D75} (2007) 092006,
  [\href{http://arxiv.org/abs/hep-ex/0701051}{{\tt hep-ex/0701051}}].

\bibitem{Aad:2011dm}
{\bf ATLAS} Collaboration, G.~Aad et~al., {\it {Measurement of the inclusive
  $W^{\pm}$ and $Z/\gamma^*$ cross sections in the electron and muon decay
  channels in pp collisions at $\sqrt{s}$= 7 TeV with the ATLAS detector}},
  {\em Phys.Rev.} {\bf D85} (2012) 072004,
  [\href{http://arxiv.org/abs/1109.5141}{{\tt arXiv:1109.5141}}].

\bibitem{Aaboud:2016btc}
{\bf ATLAS} Collaboration, M.~Aaboud et~al., {\it {Precision measurement and
  interpretation of inclusive $W^+$ , $W^-$ and $Z/\gamma ^*$ production cross
  sections with the ATLAS detector}},  {\em Eur. Phys. J.} {\bf C77} (2017),
  no.~6 367, [\href{http://arxiv.org/abs/1612.03016}{{\tt arXiv:1612.03016}}].

\bibitem{Aad:2014qja}
{\bf ATLAS} Collaboration, G.~Aad et~al., {\it {Measurement of the low-mass
  Drell-Yan differential cross section at $\sqrt{s}$ = 7 TeV using the ATLAS
  detector}},  {\em JHEP} {\bf 06} (2014) 112,
  [\href{http://arxiv.org/abs/1404.1212}{{\tt arXiv:1404.1212}}].

\bibitem{Aad:2013iua}
{\bf ATLAS} Collaboration, G.~Aad et~al., {\it {Measurement of the high-mass
  Drell--Yan differential cross-section in pp collisions at $\sqrt{s}$=7 TeV
  with the ATLAS detector}},  {\em Phys.Lett.} {\bf B725} (2013) 223,
  [\href{http://arxiv.org/abs/1305.4192}{{\tt arXiv:1305.4192}}].

\bibitem{Chatrchyan:2012xt}
{\bf CMS} Collaboration, S.~Chatrchyan et~al., {\it {Measurement of the
  electron charge asymmetry in inclusive W production in pp collisions at
  $\sqrt{s}$ = 7 TeV}},  {\em Phys.Rev.Lett.} {\bf 109} (2012) 111806,
  [\href{http://arxiv.org/abs/1206.2598}{{\tt arXiv:1206.2598}}].

\bibitem{Chatrchyan:2013mza}
{\bf CMS} Collaboration, S.~Chatrchyan et~al., {\it {Measurement of the muon
  charge asymmetry in inclusive pp to WX production at $\sqrt{s}$ = 7 TeV and
  an improved determination of light parton distribution functions}},  {\em
  Phys.Rev.} {\bf D90} (2014) 032004,
  [\href{http://arxiv.org/abs/1312.6283}{{\tt arXiv:1312.6283}}].

\bibitem{Chatrchyan:2013tia}
{\bf CMS} Collaboration, S.~Chatrchyan et~al., {\it {Measurement of the
  differential and double-differential Drell-Yan cross sections in
  proton-proton collisions at $\sqrt{s} =$ 7 TeV}},  {\em JHEP} {\bf 1312}
  (2013) 030, [\href{http://arxiv.org/abs/1310.7291}{{\tt arXiv:1310.7291}}].

\bibitem{Khachatryan:2016pev}
{\bf CMS} Collaboration, V.~Khachatryan et~al., {\it {Measurement of the
  differential cross section and charge asymmetry for inclusive $\mathrm
  {p}\mathrm {p}\rightarrow \mathrm {W}^{\pm }+X$ production at ${\sqrt{s}} =
  8$ TeV}},  {\em Eur. Phys. J.} {\bf C76} (2016), no.~8 469,
  [\href{http://arxiv.org/abs/1603.01803}{{\tt arXiv:1603.01803}}].

\bibitem{Aaij:2012mda}
{\bf LHCb} Collaboration, R.~Aaij et~al., {\it {Measurement of the
  cross-section for $Z \to e^+e^-$ production in $pp$ collisions at
  $\sqrt{s}=7$ TeV}},  {\em JHEP} {\bf 1302} (2013) 106,
  [\href{http://arxiv.org/abs/1212.4620}{{\tt arXiv:1212.4620}}].

\bibitem{Aaij:2015gna}
{\bf LHCb} Collaboration, R.~Aaij et~al., {\it {Measurement of the forward $Z$
  boson production cross-section in $pp$ collisions at $\sqrt{s}=7$ TeV}},
  {\em JHEP} {\bf 08} (2015) 039, [\href{http://arxiv.org/abs/1505.07024}{{\tt
  arXiv:1505.07024}}].

\bibitem{Aaij:2015vua}
{\bf LHCb} Collaboration, R.~Aaij et~al., {\it {Measurement of forward $\rm
  Z\rightarrow e^+e^-$ production at $\sqrt{s}=8$ TeV}},  {\em JHEP} {\bf 05}
  (2015) 109, [\href{http://arxiv.org/abs/1503.00963}{{\tt arXiv:1503.00963}}].

\bibitem{Aaij:2015zlq}
{\bf LHCb} Collaboration, R.~Aaij et~al., {\it {Measurement of forward W and Z
  boson production in $pp$ collisions at $ \sqrt{s}=8 $ TeV}},  {\em JHEP} {\bf
  01} (2016) 155, [\href{http://arxiv.org/abs/1511.08039}{{\tt
  arXiv:1511.08039}}].

\bibitem{Aad:2015auj}
{\bf ATLAS} Collaboration, G.~Aad et~al., {\it {Measurement of the transverse
  momentum and $\phi ^*_{\eta }$ distributions of Drell–Yan lepton pairs in
  proton–proton collisions at $\sqrt{s}=8$ TeV with the ATLAS detector}},
  {\em Eur. Phys. J.} {\bf C76} (2016), no.~5 291,
  [\href{http://arxiv.org/abs/1512.02192}{{\tt arXiv:1512.02192}}].

\bibitem{Khachatryan:2015oaa}
{\bf CMS} Collaboration, V.~Khachatryan et~al., {\it {Measurement of the Z
  boson differential cross section in transverse momentum and rapidity in
  proton–proton collisions at 8 TeV}},  {\em Phys. Lett.} {\bf B749} (2015)
  187--209, [\href{http://arxiv.org/abs/1504.03511}{{\tt arXiv:1504.03511}}].

\bibitem{Aad:2014kva}
{\bf ATLAS} Collaboration, G.~Aad et~al., {\it {Measurement of the $t\bar{t}$
  production cross-section using $e\mu $ events with b-tagged jets in pp
  collisions at $\sqrt{s}$ = 7 and 8 $\,\mathrm{TeV}$ with the ATLAS
  detector}},  {\em Eur. Phys. J.} {\bf C74} (2014), no.~10 3109,
  [\href{http://arxiv.org/abs/1406.5375}{{\tt arXiv:1406.5375}}]. [Addendum:
  Eur. Phys. J.C76,no.11,642(2016)].

\bibitem{Aaboud:2016pbd}
{\bf ATLAS} Collaboration, M.~Aaboud et~al., {\it {Measurement of the
  $t\bar{t}$ production cross-section using $e\mu$ events with b-tagged jets in
  pp collisions at $\sqrt{s}$=13 TeV with the ATLAS detector}},  {\em Phys.
  Lett.} {\bf B761} (2016) 136--157,
  [\href{http://arxiv.org/abs/1606.02699}{{\tt arXiv:1606.02699}}].

\bibitem{Aad:2015mbv}
{\bf ATLAS} Collaboration, G.~Aad et~al., {\it {Measurements of top-quark pair
  differential cross-sections in the lepton+jets channel in $pp$ collisions at
  $\sqrt{s}=8$ TeV using the ATLAS detector}},  {\em Eur. Phys. J.} {\bf C76}
  (2016), no.~10 538, [\href{http://arxiv.org/abs/1511.04716}{{\tt
  arXiv:1511.04716}}].

\bibitem{Khachatryan:2016mqs}
{\bf CMS} Collaboration, V.~Khachatryan et~al., {\it {Measurement of the t-tbar
  production cross section in the e-mu channel in proton-proton collisions at
  sqrt(s) = 7 and 8 TeV}},  {\em JHEP} {\bf 08} (2016) 029,
  [\href{http://arxiv.org/abs/1603.02303}{{\tt arXiv:1603.02303}}].

\bibitem{Khachatryan:2015uqb}
{\bf CMS} Collaboration, V.~Khachatryan et~al., {\it {Measurement of the top
  quark pair production cross section in proton-proton collisions at $\sqrt(s)
  =$ 13 TeV}},  {\em Phys. Rev. Lett.} {\bf 116} (2016), no.~5 052002,
  [\href{http://arxiv.org/abs/1510.05302}{{\tt arXiv:1510.05302}}].

\bibitem{Khachatryan:2015oqa}
{\bf CMS} Collaboration, V.~Khachatryan et~al., {\it {Measurement of the
  differential cross section for top quark pair production in pp collisions at
  $\sqrt{s} = 8\,\text {TeV} $}},  {\em Eur. Phys. J.} {\bf C75} (2015), no.~11
  542, [\href{http://arxiv.org/abs/1505.04480}{{\tt arXiv:1505.04480}}].

\bibitem{Aad:2011fc}
{\bf ATLAS} Collaboration, G.~Aad et~al., {\it {Measurement of inclusive jet
  and dijet production in pp collisions at $\sqrt{s}$ = 7 TeV using the ATLAS
  detector}},  {\em Phys. Rev.} {\bf D86} (2012) 014022,
  [\href{http://arxiv.org/abs/1112.6297}{{\tt arXiv:1112.6297}}].

\bibitem{Aad:2013lpa}
{\bf ATLAS} Collaboration, G.~Aad et~al., {\it {Measurement of the inclusive
  jet cross section in pp collisions at $\sqrt{s}$=2.76 TeV and comparison to
  the inclusive jet cross section at $\sqrt{s}$=7 TeV using the ATLAS
  detector}},  {\em Eur.Phys.J.} {\bf C73} (2013) 2509,
  [\href{http://arxiv.org/abs/1304.4739}{{\tt arXiv:1304.4739}}].

\bibitem{Aad:2014vwa}
{\bf ATLAS} Collaboration, G.~Aad et~al., {\it {Measurement of the inclusive
  jet cross-section in proton-proton collisions at $ \sqrt{s}=7 $ TeV using 4.5
  fb$^{-1}$ of data with the ATLAS detector}},  {\em JHEP} {\bf 02} (2015) 153,
  [\href{http://arxiv.org/abs/1410.8857}{{\tt arXiv:1410.8857}}]. [Erratum:
  JHEP09,141(2015)].

\bibitem{Chatrchyan:2012bja}
{\bf CMS} Collaboration, S.~Chatrchyan et~al., {\it {Measurements of
  differential jet cross sections in proton-proton collisions at $\sqrt{s}=7$
  TeV with the CMS detector}},  {\em Phys.Rev.} {\bf D87} (2013) 112002,
  [\href{http://arxiv.org/abs/1212.6660}{{\tt arXiv:1212.6660}}].

\bibitem{Khachatryan:2015luy}
{\bf CMS} Collaboration, V.~Khachatryan et~al., {\it {Measurement of the
  inclusive jet cross section in pp collisions at $\sqrt{s} = 2.76\,\text
  {TeV}$}},  {\em Eur. Phys. J.} {\bf C76} (2016), no.~5 265,
  [\href{http://arxiv.org/abs/1512.06212}{{\tt arXiv:1512.06212}}].

\bibitem{Chatrchyan:2013uja}
{\bf CMS} Collaboration, S.~Chatrchyan et~al., {\it {Measurement of associated
  W + charm production in pp collisions at $\sqrt{s}$ = 7 TeV}},  {\em JHEP}
  {\bf 02} (2014) 013, [\href{http://arxiv.org/abs/1310.1138}{{\tt
  arXiv:1310.1138}}].

\bibitem{Aad:2016zzw}
{\bf ATLAS} Collaboration, G.~Aad et~al., {\it {Measurement of the
  double-differential high-mass Drell-Yan cross section in pp collisions at $
  \sqrt{s}=8 $ TeV with the ATLAS detector}},  {\em JHEP} {\bf 08} (2016) 009,
  [\href{http://arxiv.org/abs/1606.01736}{{\tt arXiv:1606.01736}}].

\bibitem{Aaboud:2017ffb}
{\bf ATLAS} Collaboration, M.~Aaboud et~al., {\it {Measurement of the Drell-Yan
  triple-differential cross section in $pp$ collisions at $\sqrt{s} = 8$ TeV}},
   {\em JHEP} {\bf 12} (2017) 059, [\href{http://arxiv.org/abs/1710.05167}{{\tt
  arXiv:1710.05167}}].

\bibitem{Aad:2019rou}
{\bf ATLAS} Collaboration, G.~Aad et~al., {\it {Measurement of the
  cross-section and charge asymmetry of $W$ bosons produced in
  proton\textendash{}proton collisions at $\sqrt{s}=8~\text {TeV}$ with the
  ATLAS detector}},  {\em Eur. Phys. J. C} {\bf 79} (2019), no.~9 760,
  [\href{http://arxiv.org/abs/1904.05631}{{\tt arXiv:1904.05631}}].

\bibitem{Aaij:2016qqz}
{\bf LHCb} Collaboration, R.~Aaij et~al., {\it {Measurement of forward $W\to
  e\nu$ production in $pp$ collisions at $\sqrt{s}=8\,$TeV}},  {\em JHEP} {\bf
  10} (2016) 030, [\href{http://arxiv.org/abs/1608.01484}{{\tt
  arXiv:1608.01484}}].

\bibitem{Aad:2016naf}
{\bf ATLAS} Collaboration, G.~Aad et~al., {\it {Measurement of $W^{\pm}$ and
  $Z$-boson production cross sections in $pp$ collisions at $\sqrt{s}=13$ TeV
  with the ATLAS detector}},  {\em Phys. Lett.} {\bf B759} (2016) 601--621,
  [\href{http://arxiv.org/abs/1603.09222}{{\tt arXiv:1603.09222}}].

\bibitem{Aaij:2016mgv}
{\bf LHCb} Collaboration, R.~Aaij et~al., {\it {Measurement of the forward Z
  boson production cross-section in pp collisions at $\sqrt{s} = 13$ TeV}},
  {\em JHEP} {\bf 09} (2016) 136, [\href{http://arxiv.org/abs/1607.06495}{{\tt
  arXiv:1607.06495}}].

\bibitem{Aad:2014xca}
{\bf ATLAS} Collaboration, G.~Aad et~al., {\it {Measurement of the production
  of a $W$ boson in association with a charm quark in $pp$ collisions at
  $\sqrt{s} =$ 7 TeV with the ATLAS detector}},  {\em JHEP} {\bf 1405} (2014)
  068, [\href{http://arxiv.org/abs/1402.6263}{{\tt arXiv:1402.6263}}].

\bibitem{Sirunyan:2018hde}
{\bf CMS} Collaboration, A.~M. Sirunyan et~al., {\it {Measurement of associated
  production of a W boson and a charm quark in proton-proton collisions at
  $\sqrt{s} =$ 13 TeV}},  {\em Eur. Phys. J. C} {\bf 79} (2019), no.~3 269,
  [\href{http://arxiv.org/abs/1811.10021}{{\tt arXiv:1811.10021}}].

\bibitem{Czakon:2020coa}
M.~Czakon, A.~Mitov, M.~Pellen, and R.~Poncelet, {\it {NNLO QCD predictions for
  W+c-jet production at the LHC}},  {\em JHEP} {\bf 06} (2021) 100,
  [\href{http://arxiv.org/abs/2011.01011}{{\tt arXiv:2011.01011}}].

\bibitem{Aaboud:2017dvo}
{\bf ATLAS} Collaboration, M.~Aaboud et~al., {\it {Measurement of the inclusive
  jet cross-sections in proton-proton collisions at $ \sqrt{s}=8 $ TeV with the
  ATLAS detector}},  {\em JHEP} {\bf 09} (2017) 020,
  [\href{http://arxiv.org/abs/1706.03192}{{\tt arXiv:1706.03192}}].

\bibitem{Khachatryan:2016mlc}
{\bf CMS} Collaboration, V.~Khachatryan et~al., {\it {Measurement and QCD
  analysis of double-differential inclusive jet cross sections in pp collisions
  at $ \sqrt{s}=8 $ TeV and cross section ratios to 2.76 and 7 TeV}},  {\em
  JHEP} {\bf 03} (2017) 156, [\href{http://arxiv.org/abs/1609.05331}{{\tt
  arXiv:1609.05331}}].

\bibitem{Sirunyan:2017ule}
{\bf CMS} Collaboration, A.~M. Sirunyan et~al., {\it {Measurement of the
  inclusive $ \mathrm{t}\overline{\mathrm{t}} $ cross section in pp collisions
  at $ \sqrt{s}=5.02 $ TeV using final states with at least one charged
  lepton}},  {\em JHEP} {\bf 03} (2018) 115,
  [\href{http://arxiv.org/abs/1711.03143}{{\tt arXiv:1711.03143}}].

\bibitem{Aaboud:2016iot}
{\bf ATLAS} Collaboration, M.~Aaboud et~al., {\it {Measurement of top quark
  pair differential cross-sections in the dilepton channel in $pp$ collisions
  at $\sqrt{s}$ = 7 and 8 TeV with ATLAS}},  {\em Phys. Rev. D} {\bf 94}
  (2016), no.~9 092003, [\href{http://arxiv.org/abs/1607.07281}{{\tt
  arXiv:1607.07281}}]. [Addendum: Phys.Rev.D 101, 119901 (2020)].

\bibitem{Sirunyan:2017azo}
{\bf CMS} Collaboration, A.~M. Sirunyan et~al., {\it {Measurement of
  double-differential cross sections for top quark pair production in pp
  collisions at $\sqrt{s} = 8$ $\,\text {TeV}$ and impact on parton
  distribution functions}},  {\em Eur. Phys. J.} {\bf C77} (2017), no.~7 459,
  [\href{http://arxiv.org/abs/1703.01630}{{\tt arXiv:1703.01630}}].

\bibitem{Sirunyan:2018wem}
{\bf CMS} Collaboration, A.~M. Sirunyan et~al., {\it {Measurement of
  differential cross sections for the production of top quark pairs and of
  additional jets in lepton+jets events from pp collisions at $\sqrt{s} =$ 13
  TeV}},  {\em Phys. Rev.} {\bf D97} (2018), no.~11 112003,
  [\href{http://arxiv.org/abs/1803.08856}{{\tt arXiv:1803.08856}}].

\bibitem{Sirunyan:2018ucr}
{\bf CMS} Collaboration, A.~M. Sirunyan et~al., {\it {Measurements of
  $\mathrm{t\overline{t}}$ differential cross sections in proton-proton
  collisions at $\sqrt{s}=$ 13 TeV using events containing two leptons}},  {\em
  JHEP} {\bf 02} (2019) 149, [\href{http://arxiv.org/abs/1811.06625}{{\tt
  arXiv:1811.06625}}].

\bibitem{Aaboud:2017soa}
{\bf ATLAS} Collaboration, M.~Aaboud et~al., {\it {Measurement of differential
  cross sections and $W^+/W^-$ cross-section ratios for $W$ boson production in
  association with jets at $\sqrt{s}=8$ TeV with the ATLAS detector}},  {\em
  JHEP} {\bf 05} (2018) 077, [\href{http://arxiv.org/abs/1711.03296}{{\tt
  arXiv:1711.03296}}]. [Erratum: JHEP 10, 048 (2020)].

\bibitem{Aad:2014fwa}
{\bf ATLAS} Collaboration, G.~Aad et~al., {\it {Comprehensive measurements of
  $t$-channel single top-quark production cross sections at $\sqrt{s} = 7$ TeV
  with the ATLAS detector}},  {\em Phys. Rev. D} {\bf 90} (2014), no.~11
  112006, [\href{http://arxiv.org/abs/1406.7844}{{\tt arXiv:1406.7844}}].

\bibitem{Chatrchyan:2012ep}
{\bf CMS} Collaboration, S.~Chatrchyan et~al., {\it {Measurement of the
  Single-Top-Quark $t$-Channel Cross Section in $pp$ Collisions at $\sqrt{s}=7$
  TeV}},  {\em JHEP} {\bf 12} (2012) 035,
  [\href{http://arxiv.org/abs/1209.4533}{{\tt arXiv:1209.4533}}].

\bibitem{Aaboud:2017pdi}
{\bf ATLAS} Collaboration, M.~Aaboud et~al., {\it {Fiducial, total and
  differential cross-section measurements of $t$-channel single top-quark
  production in $pp$ collisions at 8 TeV using data collected by the ATLAS
  detector}},  {\em Eur. Phys. J. C} {\bf 77} (2017), no.~8 531,
  [\href{http://arxiv.org/abs/1702.02859}{{\tt arXiv:1702.02859}}].

\bibitem{Khachatryan:2014iya}
{\bf CMS} Collaboration, V.~Khachatryan et~al., {\it {Measurement of the
  t-channel single-top-quark production cross section and of the $\mid V_{tb}
  \mid$ CKM matrix element in pp collisions at $\sqrt{s}$= 8 TeV}},  {\em JHEP}
  {\bf 06} (2014) 090, [\href{http://arxiv.org/abs/1403.7366}{{\tt
  arXiv:1403.7366}}].

\bibitem{Aaboud:2016ymp}
{\bf ATLAS} Collaboration, M.~Aaboud et~al., {\it {Measurement of the inclusive
  cross-sections of single top-quark and top-antiquark $t$-channel production
  in $pp$ collisions at $\sqrt{s}$ = 13 TeV with the ATLAS detector}},  {\em
  JHEP} {\bf 04} (2017) 086, [\href{http://arxiv.org/abs/1609.03920}{{\tt
  arXiv:1609.03920}}].

\bibitem{Sirunyan:2016cdg}
{\bf CMS} Collaboration, A.~M. Sirunyan et~al., {\it {Cross section measurement
  of $t$-channel single top quark production in pp collisions at $\sqrt s =$ 13
  TeV}},  {\em Phys. Lett. B} {\bf 772} (2017) 752--776,
  [\href{http://arxiv.org/abs/1610.00678}{{\tt arXiv:1610.00678}}].

\bibitem{Aad:2016xcr}
{\bf ATLAS} Collaboration, G.~Aad et~al., {\it {Measurement of the inclusive
  isolated prompt photon cross section in pp collisions at $ \sqrt{s}=8 $ TeV
  with the ATLAS detector}},  {\em JHEP} {\bf 08} (2016) 005,
  [\href{http://arxiv.org/abs/1605.03495}{{\tt arXiv:1605.03495}}].

\bibitem{ATLAS:2017nah}
{\bf ATLAS} Collaboration, M.~Aaboud et~al., {\it {Measurement of the cross
  section for inclusive isolated-photon production in $pp$ collisions at $\sqrt
  s=13$ TeV using the ATLAS detector}},  {\em Phys. Lett. B} {\bf 770} (2017)
  473--493, [\href{http://arxiv.org/abs/1701.06882}{{\tt arXiv:1701.06882}}].

\bibitem{Aad:2019wmn}
{\bf ATLAS} Collaboration, G.~Aad et~al., {\it {Measurement of the transverse
  momentum distribution of Drell\textendash{}Yan lepton pairs in
  proton\textendash{}proton collisions at $\sqrt{s}=13$ TeV with the ATLAS
  detector}},  {\em Eur. Phys. J. C} {\bf 80} (2020), no.~7 616,
  [\href{http://arxiv.org/abs/1912.02844}{{\tt arXiv:1912.02844}}].

\bibitem{Sirunyan:2018owv}
{\bf CMS} Collaboration, A.~M. Sirunyan et~al., {\it {Measurement of the
  differential Drell-Yan cross section in proton-proton collisions at $
  \sqrt{\mathrm{s}} $ = 13 TeV}},  {\em JHEP} {\bf 12} (2019) 059,
  [\href{http://arxiv.org/abs/1812.10529}{{\tt arXiv:1812.10529}}].

\bibitem{Sirunyan:2017wgx}
{\bf CMS} Collaboration, A.~M. Sirunyan et~al., {\it {Measurement of the
  differential cross sections for the associated production of a $W$ boson and
  jets in proton-proton collisions at $\sqrt{s}=13$ TeV}},  {\em Phys. Rev. D}
  {\bf 96} (2017), no.~7 072005, [\href{http://arxiv.org/abs/1707.05979}{{\tt
  arXiv:1707.05979}}].

\bibitem{Aaboud:2017wsi}
{\bf ATLAS} Collaboration, M.~Aaboud et~al., {\it {Measurement of inclusive jet
  and dijet cross-sections in proton-proton collisions at $\sqrt{s}=13$ TeV
  with the ATLAS detector}},  {\em JHEP} {\bf 05} (2018) 195,
  [\href{http://arxiv.org/abs/1711.02692}{{\tt arXiv:1711.02692}}].

\bibitem{Khachatryan:2016wdh}
{\bf CMS} Collaboration, V.~Khachatryan et~al., {\it {Measurement of the
  double-differential inclusive jet cross section in proton–proton collisions
  at $\sqrt{s} = 13\,\text {TeV} $}},  {\em Eur. Phys. J.} {\bf C76} (2016),
  no.~8 451, [\href{http://arxiv.org/abs/1605.04436}{{\tt arXiv:1605.04436}}].

\bibitem{Aad:2019ntk}
{\bf ATLAS} Collaboration, G.~Aad et~al., {\it {Measurements of top-quark pair
  differential and double-differential cross-sections in the $\ell$+jets
  channel with $pp$ collisions at $\sqrt{s}=13$ TeV using the ATLAS detector}},
   {\em Eur. Phys. J. C} {\bf 79} (2019), no.~12 1028,
  [\href{http://arxiv.org/abs/1908.07305}{{\tt arXiv:1908.07305}}]. [Erratum:
  Eur.Phys.J.C 80, 1092 (2020)].

\bibitem{Aaij:2018imy}
{\bf LHCb} Collaboration, R.~Aaij et~al., {\it {Measurement of forward top pair
  production in the dilepton channel in $pp$ collisions at $\sqrt{s}=13$ TeV}},
   {\em JHEP} {\bf 08} (2018) 174, [\href{http://arxiv.org/abs/1803.05188}{{\tt
  arXiv:1803.05188}}].

\bibitem{Samoylov:2013xoa}
{\bf NOMAD} Collaboration, O.~Samoylov et~al., {\it {A Precision Measurement of
  Charm Dimuon Production in Neutrino Interactions from the NOMAD Experiment}},
   {\em Nucl.Phys.} {\bf B876} (2013) 339,
  [\href{http://arxiv.org/abs/1308.4750}{{\tt arXiv:1308.4750}}].

\bibitem{ZEUS:2002nms}
{\bf ZEUS} Collaboration, S.~Chekanov et~al., {\it {Inclusive jet
  cross-sections in the Breit frame in neutral current deep inelastic
  scattering at HERA and determination of alpha(s)}},  {\em Phys. Lett. B} {\bf
  547} (2002) 164--180, [\href{http://arxiv.org/abs/hep-ex/0208037}{{\tt
  hep-ex/0208037}}].

\bibitem{ZEUS:2006xvn}
{\bf ZEUS} Collaboration, S.~Chekanov et~al., {\it {Inclusive-jet and dijet
  cross-sections in deep inelastic scattering at HERA}},  {\em Nucl. Phys. B}
  {\bf 765} (2007) 1--30, [\href{http://arxiv.org/abs/hep-ex/0608048}{{\tt
  hep-ex/0608048}}].

\bibitem{ZEUS:2010vyw}
{\bf ZEUS} Collaboration, H.~Abramowicz et~al., {\it {Inclusive dijet cross
  sections in neutral current deep inelastic scattering at HERA}},  {\em Eur.
  Phys. J. C} {\bf 70} (2010) 965--982,
  [\href{http://arxiv.org/abs/1010.6167}{{\tt arXiv:1010.6167}}].

\bibitem{H1:2016goa}
{\bf H1} Collaboration, V.~Andreev et~al., {\it {Measurement of Jet Production
  Cross Sections in Deep-inelastic ep Scattering at HERA}},  {\em Eur. Phys. J.
  C} {\bf 77} (2017), no.~4 215, [\href{http://arxiv.org/abs/1611.03421}{{\tt
  arXiv:1611.03421}}].

\bibitem{H1:2014cbm}
{\bf H1} Collaboration, V.~Andreev et~al., {\it {Measurement of multijet
  production in $ep$ collisions at high $Q^2$ and determination of the strong
  coupling $\alpha _s$}},  {\em Eur. Phys. J. C} {\bf 75} (2015), no.~2 65,
  [\href{http://arxiv.org/abs/1406.4709}{{\tt arXiv:1406.4709}}].

\bibitem{Dove:2021ejl}
{\bf SeaQuest} Collaboration, J.~Dove et~al., {\it {The asymmetry of antimatter
  in the proton}},  {\em Nature} {\bf 590} (2021), no.~7847 561--565,
  [\href{http://arxiv.org/abs/2103.04024}{{\tt arXiv:2103.04024}}].

\bibitem{Bertone:2016lga}
V.~Bertone, S.~Carrazza, and N.~P. Hartland, {\it {APFELgrid: a high
  performance tool for parton density determinations}},  {\em Comput. Phys.
  Commun.} {\bf 212} (2017) 205--209,
  [\href{http://arxiv.org/abs/1605.02070}{{\tt arXiv:1605.02070}}].

\bibitem{Bertone:2013vaa}
V.~Bertone, S.~Carrazza, and J.~Rojo, {\it {APFEL: A PDF Evolution Library with
  QED corrections}},  {\em Comput.Phys.Commun.} {\bf 185} (2014) 1647,
  [\href{http://arxiv.org/abs/1310.1394}{{\tt arXiv:1310.1394}}].

\bibitem{Campbell:1999ah}
J.~M. Campbell and R.~K. Ellis, {\it {An Update on vector boson pair production
  at hadron colliders}},  {\em Phys. Rev. D} {\bf 60} (1999) 113006,
  [\href{http://arxiv.org/abs/hep-ph/9905386}{{\tt hep-ph/9905386}}].

\bibitem{Campbell:2011bn}
J.~M. Campbell, R.~K. Ellis, and C.~Williams, {\it {Vector boson pair
  production at the LHC}},  {\em JHEP} {\bf 1107} (2011) 018,
  [\href{http://arxiv.org/abs/1105.0020}{{\tt arXiv:1105.0020}}].

\bibitem{Campbell:2015qma}
J.~M. Campbell, R.~K. Ellis, and W.~T. Giele, {\it {A Multi-Threaded Version of
  MCFM}},  {\em Eur. Phys. J. C} {\bf 75} (2015), no.~6 246,
  [\href{http://arxiv.org/abs/1503.06182}{{\tt arXiv:1503.06182}}].

\bibitem{Bothmann:2019yzt}
{\bf Sherpa} Collaboration, E.~Bothmann et~al., {\it {Event Generation with
  Sherpa 2.2}},  {\em SciPost Phys.} {\bf 7} (2019), no.~3 034,
  [\href{http://arxiv.org/abs/1905.09127}{{\tt arXiv:1905.09127}}].

\bibitem{Frederix:2018nkq}
R.~Frederix, S.~Frixione, V.~Hirschi, D.~Pagani, H.~S. Shao, and M.~Zaro, {\it
  {The automation of next-to-leading order electroweak calculations}},  {\em
  JHEP} {\bf 07} (2018) 185, [\href{http://arxiv.org/abs/1804.10017}{{\tt
  arXiv:1804.10017}}].

\bibitem{Alwall:2014hca}
J.~Alwall, R.~Frederix, S.~Frixione, V.~Hirschi, F.~Maltoni, et~al., {\it {The
  automated computation of tree-level and next-to-leading order differential
  cross sections, and their matching to parton shower simulations}},  {\em
  JHEP} {\bf 1407} (2014) 079, [\href{http://arxiv.org/abs/1405.0301}{{\tt
  arXiv:1405.0301}}].

\bibitem{Nagy:2001fj}
Z.~Nagy, {\it {Three jet cross-sections in hadron hadron collisions at
  next-to-leading order}},  {\em Phys.Rev.Lett.} {\bf 88} (2002) 122003,
  [\href{http://arxiv.org/abs/hep-ph/0110315}{{\tt hep-ph/0110315}}].

\bibitem{Carli:2010rw}
T.~Carli et~al., {\it {A posteriori inclusion of parton density functions in
  NLO QCD final-state calculations at hadron colliders: The APPLGRID Project}},
   {\em Eur.Phys.J.} {\bf C66} (2010) 503,
  [\href{http://arxiv.org/abs/0911.2985}{{\tt arXiv:0911.2985}}].

\bibitem{Kluge:2006xs}
T.~Kluge, K.~Rabbertz, and M.~Wobisch, {\it {Fast pQCD calculations for PDF
  fits}},  \href{http://arxiv.org/abs/hep-ph/0609285}{{\tt hep-ph/0609285}}.

\bibitem{Wobisch:2011ij}
{\bf fastNLO} Collaboration, M.~Wobisch, D.~Britzger, T.~Kluge, K.~Rabbertz,
  and F.~Stober, {\it {Theory-Data Comparisons for Jet Measurements in
  Hadron-Induced Processes}},  \href{http://arxiv.org/abs/1109.1310}{{\tt
  arXiv:1109.1310}}.

\bibitem{Britzger:2012bs}
{\bf fastNLO} Collaboration, D.~Britzger, K.~Rabbertz, F.~Stober, and
  M.~Wobisch, {\it {New features in version 2 of the fastNLO project}},  in
  {\em {20th International Workshop on Deep-Inelastic Scattering and Related
  Subjects}}, 8, 2012.
\newblock \href{http://arxiv.org/abs/1208.3641}{{\tt arXiv:1208.3641}}.

\bibitem{DelDebbio:2013kxa}
L.~Del~Debbio, N.~P. Hartland, and S.~Schumann, {\it {MCgrid: projecting cross
  section calculations on grids}},  {\em Comput.Phys.Commun.} {\bf 185} (2014)
  2115--2126, [\href{http://arxiv.org/abs/1312.4460}{{\tt arXiv:1312.4460}}].

\bibitem{Bothmann:2015dba}
E.~Bothmann, N.~Hartland, and S.~Schumann, {\it {Introducing MCgrid 2.0:
  Projecting cross section calculations on grids}},  {\em Comput. Phys.
  Commun.} {\bf 196} (2015) 617--618.

\bibitem{Bertone:2014zva}
V.~Bertone, R.~Frederix, S.~Frixione, J.~Rojo, and M.~Sutton, {\it {aMCfast:
  automation of fast NLO computations for PDF fits}},  {\em JHEP} {\bf 08}
  (2014) 166, [\href{http://arxiv.org/abs/1406.7693}{{\tt arXiv:1406.7693}}].

\bibitem{Ball:2015tna}
R.~D. Ball, V.~Bertone, M.~Bonvini, S.~Forte, P.~Groth~Merrild, J.~Rojo, and
  L.~Rottoli, {\it {Intrinsic charm in a matched general-mass scheme}},  {\em
  Phys. Lett.} {\bf B754} (2016) 49--58,
  [\href{http://arxiv.org/abs/1510.00009}{{\tt arXiv:1510.00009}}].

\bibitem{Ball:2015dpa}
R.~D. Ball, M.~Bonvini, and L.~Rottoli, {\it {Charm in Deep-Inelastic
  Scattering}},  {\em JHEP} {\bf 11} (2015) 122,
  [\href{http://arxiv.org/abs/1510.02491}{{\tt arXiv:1510.02491}}].

\bibitem{Aad:2020tmz}
{\bf ATLAS} Collaboration, G.~Aad et~al., {\it {Measurement of the $t\bar{t}$
  production cross-section in the lepton+jets channel at $\sqrt{s}=13$ TeV with
  the ATLAS experiment}},  {\em Phys. Lett. B} {\bf 810} (2020) 135797,
  [\href{http://arxiv.org/abs/2006.13076}{{\tt arXiv:2006.13076}}].

\bibitem{Currie:2018xkj}
J.~Currie, A.~Gehrmann-De~Ridder, T.~Gehrmann, E.~W.~N. Glover, A.~Huss, and
  J.~Pires, {\it {Infrared sensitivity of single jet inclusive production at
  hadron colliders}},  {\em JHEP} {\bf 10} (2018) 155,
  [\href{http://arxiv.org/abs/1807.03692}{{\tt arXiv:1807.03692}}].

\bibitem{Ball:2018iqk}
{\bf NNPDF} Collaboration, R.~D. Ball, S.~Carrazza, L.~Del~Debbio, S.~Forte,
  Z.~Kassabov, J.~Rojo, E.~Slade, and M.~Ubiali, {\it {Precision determination
  of the strong coupling constant within a global PDF analysis}},  {\em Eur.
  Phys. J.} {\bf C78} (2018), no.~5 408,
  [\href{http://arxiv.org/abs/1802.03398}{{\tt arXiv:1802.03398}}].

\bibitem{Gao:2017kkx}
J.~Gao, {\it {Massive charged-current coefficient functions in deep-inelastic
  scattering at NNLO and impact on strange-quark distributions}},  {\em JHEP}
  {\bf 02} (2018) 026, [\href{http://arxiv.org/abs/1710.04258}{{\tt
  arXiv:1710.04258}}].

\bibitem{Berger:2016inr}
E.~L. Berger, J.~Gao, C.~S. Li, Z.~L. Liu, and H.~X. Zhu, {\it {Charm-Quark
  Production in Deep-Inelastic Neutrino Scattering at Next-to-Next-to-Leading
  Order in QCD}},  {\em Phys. Rev. Lett.} {\bf 116} (2016), no.~21 212002,
  [\href{http://arxiv.org/abs/1601.05430}{{\tt arXiv:1601.05430}}].

\bibitem{Zyla:2020zbs}
{\bf Particle Data Group} Collaboration, P.~A. Zyla et~al., {\it {Review of
  Particle Physics}},  {\em PTEP} {\bf 2020} (2020), no.~8 083C01.

\bibitem{Alekhin:2017kpj}
S.~Alekhin, J.~Blümlein, S.~Moch, and R.~Placakyte, {\it {Parton distribution
  functions, $\alpha_s$, and heavy-quark masses for LHC Run II}},  {\em Phys.
  Rev.} {\bf D96} (2017), no.~1 014011,
  [\href{http://arxiv.org/abs/1701.05838}{{\tt arXiv:1701.05838}}].

\bibitem{Hou:2019efy}
T.-J. Hou et~al., {\it {New CTEQ global analysis of quantum chromodynamics with
  high-precision data from the LHC}},  {\em Phys. Rev. D} {\bf 103} (2021),
  no.~1 014013, [\href{http://arxiv.org/abs/1912.10053}{{\tt
  arXiv:1912.10053}}].

\bibitem{Bailey:2020ooq}
S.~Bailey, T.~Cridge, L.~A. Harland-Lang, A.~D. Martin, and R.~S. Thorne, {\it
  {Parton distributions from LHC, HERA, Tevatron and fixed target data: MSHT20
  PDFs}},  {\em Eur. Phys. J. C} {\bf 81} (2021), no.~4 341,
  [\href{http://arxiv.org/abs/2012.04684}{{\tt arXiv:2012.04684}}].

\bibitem{H1:2018flt}
{\bf H1, ZEUS} Collaboration, H.~Abramowicz et~al., {\it {Combination and QCD
  analysis of charm and beauty production cross-section measurements in deep
  inelastic $ep$ scattering at HERA}},  {\em Eur. Phys. J.} {\bf C78} (2018),
  no.~6 473, [\href{http://arxiv.org/abs/1804.01019}{{\tt arXiv:1804.01019}}].

\bibitem{Spannagel:2016cqt}
S.~Spannagel, {\it {Top quark mass measurements with the CMS experiment at the
  LHC}},  {\em PoS} {\bf DIS2016} (2016) 150,
  [\href{http://arxiv.org/abs/1607.04972}{{\tt arXiv:1607.04972}}].

\bibitem{Chatrchyan:2014gia}
{\bf CMS} Collaboration, S.~Chatrchyan et~al., {\it {Measurement of the Ratio
  of Inclusive Jet Cross Sections using the Anti-$k_T$ Algorithm with Radius
  Parameters R=0.5 and 0.7 in pp Collisions at $\sqrt{s}=7$ TeV}},  {\em Phys.
  Rev. D} {\bf 90} (2014), no.~7 072006,
  [\href{http://arxiv.org/abs/1406.0324}{{\tt arXiv:1406.0324}}].

\bibitem{Aad:2013tea}
{\bf ATLAS Collaboration} Collaboration, G.~Aad et~al., {\it {Measurement of
  dijet cross sections in $pp$ collisions at 7 TeV centre-of-mass energy using
  the ATLAS detector}},  {\em JHEP} {\bf 1405} (2014) 059,
  [\href{http://arxiv.org/abs/1312.3524}{{\tt arXiv:1312.3524}}].

\bibitem{Sirunyan:2017skj}
{\bf CMS} Collaboration, A.~M. Sirunyan et~al., {\it {Measurement of the
  triple-differential dijet cross section in proton-proton collisions at
  $\sqrt{s}=8\,\text {TeV} $ and constraints on parton distribution
  functions}},  {\em Eur. Phys. J. C} {\bf 77} (2017), no.~11 746,
  [\href{http://arxiv.org/abs/1705.02628}{{\tt arXiv:1705.02628}}].

\bibitem{Maguire:2017ypu}
E.~Maguire, L.~Heinrich, and G.~Watt, {\it {HEPData: a repository for high
  energy physics data}},  {\em J. Phys. Conf. Ser.} {\bf 898} (2017), no.~10
  102006, [\href{http://arxiv.org/abs/1704.05473}{{\tt arXiv:1704.05473}}].

\bibitem{Harland-Lang:2017ytb}
L.~A. Harland-Lang, A.~D. Martin, and R.~S. Thorne, {\it {The Impact of LHC Jet
  Data on the MMHT PDF Fit at NNLO}},  {\em Eur. Phys. J. C} {\bf 78} (2018),
  no.~3 248, [\href{http://arxiv.org/abs/1711.05757}{{\tt arXiv:1711.05757}}].

\bibitem{Bailey:2019yze}
S.~Bailey and L.~Harland-Lang, {\it {Differential Top Quark Pair Production at
  the LHC: Challenges for PDF Fits}},  {\em Eur. Phys. J. C} {\bf 80} (2020),
  no.~1 60, [\href{http://arxiv.org/abs/1909.10541}{{\tt arXiv:1909.10541}}].

\bibitem{ATLAS:2018owm}
{\bf ATLAS} Collaboration, {\it {Determination of the parton distribution
  functions of the proton from ATLAS measurements of differential $W$ and
  $Z/\gamma^*$ and $t\bar{t}$ cross sections}},  tech. rep., 2018.
\newblock ATL-PHYS-PUB-2018-017.

\bibitem{Amoroso:2020lgh}
S.~Amoroso et~al., {\it {Les Houches 2019: Physics at TeV Colliders: Standard
  Model Working Group Report}},  in {\em {11th Les Houches Workshop on Physics
  at TeV Colliders}: {PhysTeV Les Houches}}, 3, 2020.
\newblock \href{http://arxiv.org/abs/2003.01700}{{\tt arXiv:2003.01700}}.

\bibitem{Ball:2010gb}
{\bf The NNPDF} Collaboration, R.~D. Ball et~al., {\it {Reweighting NNPDFs: the
  W lepton asymmetry}},  {\em Nucl. Phys.} {\bf B849} (2011) 112--143,
  [\href{http://arxiv.org/abs/1012.0836}{{\tt arXiv:1012.0836}}].

\bibitem{Ball:2011gg}
R.~D. Ball, V.~Bertone, F.~Cerutti, L.~Del~Debbio, S.~Forte, et~al., {\it
  {Reweighting and Unweighting of Parton Distributions and the LHC W lepton
  asymmetry data}},  {\em Nucl.Phys.} {\bf B855} (2012) 608--638,
  [\href{http://arxiv.org/abs/1108.1758}{{\tt arXiv:1108.1758}}].

\bibitem{H1:2007xjj}
{\bf H1} Collaboration, A.~Aktas et~al., {\it {Measurement of inclusive jet
  production in deep-inelastic scattering at high Q**2 and determination of the
  strong coupling}},  {\em Phys. Lett. B} {\bf 653} (2007) 134--144,
  [\href{http://arxiv.org/abs/0706.3722}{{\tt arXiv:0706.3722}}].

\bibitem{H1:2010mgp}
{\bf H1} Collaboration, F.~D. Aaron et~al., {\it {Jet Production in ep
  Collisions at Low Q**2 and Determination of alpha(s)}},  {\em Eur. Phys. J.
  C} {\bf 67} (2010) 1--24, [\href{http://arxiv.org/abs/0911.5678}{{\tt
  arXiv:0911.5678}}].

\bibitem{Britzger:2019kkb}
D.~Britzger et~al., {\it {Calculations for deep inelastic scattering using fast
  interpolation grid techniques at NNLO in QCD and the extraction of $\alpha_s$
  from HERA data}},  {\em Eur. Phys. J. C} {\bf 79} (2019), no.~10 845,
  [\href{http://arxiv.org/abs/1906.05303}{{\tt arXiv:1906.05303}}].

\bibitem{Anastasiou:2003ds}
C.~Anastasiou, L.~J. Dixon, K.~Melnikov, and F.~Petriello, {\it {High precision
  QCD at hadron colliders: Electroweak gauge boson rapidity distributions at
  NNLO}},  {\em Phys. Rev.} {\bf D69} (2004) 094008,
  [\href{http://arxiv.org/abs/hep-ph/0312266}{{\tt hep-ph/0312266}}].

\bibitem{Gavin:2010az}
R.~Gavin, Y.~Li, F.~Petriello, and S.~Quackenbush, {\it {FEWZ 2.0: A code for
  hadronic Z production at next-to-next-to-leading order}},  {\em Comput. Phys.
  Commun.} {\bf 182} (2011) 2388--2403,
  [\href{http://arxiv.org/abs/1011.3540}{{\tt arXiv:1011.3540}}].

\bibitem{Gavin:2012sy}
R.~Gavin, Y.~Li, F.~Petriello, and S.~Quackenbush, {\it {W Physics at the LHC
  with FEWZ 2.1}},  {\em Comput.Phys.Commun.} {\bf 184} (2013) 208--214,
  [\href{http://arxiv.org/abs/1201.5896}{{\tt arXiv:1201.5896}}].

\bibitem{Li:2012wna}
Y.~Li and F.~Petriello, {\it {Combining QCD and electroweak corrections to
  dilepton production in FEWZ}},  {\em Phys.Rev.} {\bf D86} (2012) 094034,
  [\href{http://arxiv.org/abs/1208.5967}{{\tt arXiv:1208.5967}}].

\bibitem{Catani:2007vq}
S.~Catani and M.~Grazzini, {\it {An NNLO subtraction formalism in hadron
  collisions and its application to Higgs boson production at the LHC}},  {\em
  Phys. Rev. Lett.} {\bf 98} (2007) 222002,
  [\href{http://arxiv.org/abs/hep-ph/0703012}{{\tt hep-ph/0703012}}].

\bibitem{Catani:2009sm}
S.~Catani, L.~Cieri, G.~Ferrera, D.~de~Florian, and M.~Grazzini, {\it {Vector
  boson production at hadron colliders: a fully exclusive QCD calculation at
  NNLO}},  {\em Phys. Rev. Lett.} {\bf 103} (2009) 082001,
  [\href{http://arxiv.org/abs/0903.2120}{{\tt arXiv:0903.2120}}].

\bibitem{Campbell:2019dru}
J.~Campbell and T.~Neumann, {\it {Precision Phenomenology with MCFM}},  {\em
  JHEP} {\bf 12} (2019) 034, [\href{http://arxiv.org/abs/1909.09117}{{\tt
  arXiv:1909.09117}}].

\bibitem{ATLAS:2021qnl}
{\bf ATLAS} Collaboration, G.~Aad et~al., {\it {Determination of the parton
  distribution functions of the proton from ATLAS measurements of differential
  $W^\pm$ and $Z$ boson production in association with jets}},  {\em JHEP} {\bf
  07} (2021) 223, [\href{http://arxiv.org/abs/2101.05095}{{\tt
  arXiv:2101.05095}}].

\bibitem{Boughezal:2015dva}
R.~Boughezal, C.~Focke, X.~Liu, and F.~Petriello, {\it {$W$-boson production in
  association with a jet at next-to-next-to-leading order in perturbative
  QCD}},  {\em Phys. Rev. Lett.} {\bf 115} (2015), no.~6 062002,
  [\href{http://arxiv.org/abs/1504.02131}{{\tt arXiv:1504.02131}}].

\bibitem{Ridder:2015dxa}
A.~Gehrmann-De~Ridder, T.~Gehrmann, E.~W.~N. Glover, A.~Huss, and T.~A. Morgan,
  {\it {Precise QCD predictions for the production of a Z boson in association
  with a hadronic jet}},  {\em Phys. Rev. Lett.} {\bf 117} (2016), no.~2
  022001, [\href{http://arxiv.org/abs/1507.02850}{{\tt arXiv:1507.02850}}].

\bibitem{Czakon:2016olj}
M.~Czakon, N.~P. Hartland, A.~Mitov, E.~R. Nocera, and J.~Rojo, {\it {Pinning
  down the large-x gluon with NNLO top-quark pair differential distributions}},
   {\em JHEP} {\bf 04} (2017) 044, [\href{http://arxiv.org/abs/1611.08609}{{\tt
  arXiv:1611.08609}}].

\bibitem{Czakon:2017dip}
M.~Czakon, D.~Heymes, and A.~Mitov, {\it {fastNLO tables for NNLO top-quark
  pair differential distributions}},
  \href{http://arxiv.org/abs/1704.08551}{{\tt arXiv:1704.08551}}.

\bibitem{Czakon:2019yrx}
M.~Czakon, S.~Dulat, T.-J. Hou, J.~Huston, A.~Mitov, A.~S. Papanastasiou,
  I.~Sitiwaldi, Z.~Yu, and C.~P. Yuan, {\it {An exploratory study of the impact
  of CMS double-differential top distributions on the gluon parton distribution
  function}},  {\em J. Phys. G} {\bf 48} (2020), no.~1 015003,
  [\href{http://arxiv.org/abs/1912.08801}{{\tt arXiv:1912.08801}}].

\bibitem{Czakon:2011xx}
M.~Czakon and A.~Mitov, {\it {Top++: A Program for the Calculation of the
  Top-Pair Cross-Section at Hadron Colliders}},  {\em Comput. Phys. Commun.}
  {\bf 185} (2014) 2930, [\href{http://arxiv.org/abs/1112.5675}{{\tt
  arXiv:1112.5675}}].

\bibitem{Gehrmann-DeRidder:2019ibf}
A.~Gehrmann-De~Ridder, T.~Gehrmann, E.~W.~N. Glover, A.~Huss, and J.~Pires,
  {\it {Triple Differential Dijet Cross Section at the LHC}},  {\em Phys. Rev.
  Lett.} {\bf 123} (2019), no.~10 102001,
  [\href{http://arxiv.org/abs/1905.09047}{{\tt arXiv:1905.09047}}].

\bibitem{Aaboud:2017cbm}
{\bf ATLAS} Collaboration, M.~Aaboud et~al., {\it {Measurement of the cross
  section for inclusive isolated-photon production in $pp$ collisions at $\sqrt
  s=13$ TeV using the ATLAS detector}},  {\em Phys. Lett. B} {\bf 770} (2017)
  473--493, [\href{http://arxiv.org/abs/1701.06882}{{\tt arXiv:1701.06882}}].

\bibitem{Campbell:2016lzl}
J.~M. Campbell, R.~K. Ellis, and C.~Williams, {\it {Direct Photon Production at
  Next-to\textendash{}Next-to-Leading Order}},  {\em Phys. Rev. Lett.} {\bf
  118} (2017), no.~22 222001, [\href{http://arxiv.org/abs/1612.04333}{{\tt
  arXiv:1612.04333}}]. [Erratum: Phys.Rev.Lett. 124, 259901 (2020)].

\bibitem{Frixione:1998jh}
S.~Frixione, {\it {Isolated photons in perturbative QCD}},  {\em Phys.Lett.}
  {\bf B429} (1998) 369, [\href{http://arxiv.org/abs/hep-ph/9801442}{{\tt
  hep-ph/9801442}}].

\bibitem{Campbell:2016yrh}
J.~M. Campbell, R.~K. Ellis, Y.~Li, and C.~Williams, {\it {Predictions for
  diphoton production at the LHC through NNLO in QCD}},  {\em JHEP} {\bf 07}
  (2016) 148, [\href{http://arxiv.org/abs/1603.02663}{{\tt arXiv:1603.02663}}].

\bibitem{Becher:2013zua}
T.~Becher and X.~Garcia~i Tormo, {\it {Electroweak Sudakov effects in $W, Z$
  and $\gamma$ production at large transverse momentum}},  {\em Phys. Rev.}
  {\bf D88} (2013), no.~1 013009, [\href{http://arxiv.org/abs/1305.4202}{{\tt
  arXiv:1305.4202}}].

\bibitem{Becher:2015yea}
T.~Becher and X.~Garcia~i Tormo, {\it {Addendum: Electroweak Sudakov effects in
  W, Z and gamma production at large transverse momentum}},
  \href{http://arxiv.org/abs/1509.01961}{{\tt arXiv:1509.01961}}. [Addendum:
  Phys.Rev.D 92, 073011 (2015)].

\bibitem{Berger:2016oht}
E.~L. Berger, J.~Gao, C.~P. Yuan, and H.~X. Zhu, {\it {NNLO QCD Corrections to
  t-channel Single Top-Quark Production and Decay}},  {\em Phys. Rev. D} {\bf
  94} (2016), no.~7 071501, [\href{http://arxiv.org/abs/1606.08463}{{\tt
  arXiv:1606.08463}}].

\bibitem{Berger:2017zof}
E.~L. Berger, J.~Gao, and H.~X. Zhu, {\it {Differential Distributions for
  t-channel Single Top-Quark Production and Decay at Next-to-Next-to-Leading
  Order in QCD}},  {\em JHEP} {\bf 11} (2017) 158,
  [\href{http://arxiv.org/abs/1708.09405}{{\tt arXiv:1708.09405}}].

\bibitem{Ball:2013gsa}
{\bf The NNPDF} Collaboration, R.~D. Ball et~al., {\it {Theoretical issues in
  PDF determination and associated uncertainties}},  {\em Phys.Lett.} {\bf
  B723} (2013) 330, [\href{http://arxiv.org/abs/1303.1189}{{\tt
  arXiv:1303.1189}}].

\bibitem{Carrazza:2021yrg}
S.~Carrazza, J.~M. Cruz-Martinez, and R.~Stegeman, {\it {A data-based
  parametrization of parton distribution functions}},
  \href{http://arxiv.org/abs/2111.02954}{{\tt arXiv:2111.02954}}.

\bibitem{Roberts:1990ww}
R.~G. Roberts, {\em {The Structure of the proton: Deep inelastic scattering}}.
\newblock {Cambridge University Press}, 1990.

\bibitem{Collins:2021vke}
J.~Collins, T.~C. Rogers, and N.~Sato, {\it {Positivity and renormalization of
  parton densities}},  \href{http://arxiv.org/abs/2111.01170}{{\tt
  arXiv:2111.01170}}.

\bibitem{Lin:2017snn}
H.-W. Lin et~al., {\it {Parton distributions and lattice QCD calculations: a
  community white paper}},  {\em Prog. Part. Nucl. Phys.} {\bf 100} (2018)
  107--160, [\href{http://arxiv.org/abs/1711.07916}{{\tt arXiv:1711.07916}}].

\bibitem{Constantinou:2020hdm}
M.~Constantinou et~al., {\it {Parton distributions and lattice-QCD
  calculations: Toward 3D structure}},  {\em Prog. Part. Nucl. Phys.} {\bf 121}
  (2021) 103908, [\href{http://arxiv.org/abs/2006.08636}{{\tt
  arXiv:2006.08636}}].

\bibitem{Ball:2010de}
{\bf {The NNPDF }} Collaboration, R.~D. Ball et~al., {\it {A first unbiased
  global NLO determination of parton distributions and their uncertainties}},
  {\em Nucl. Phys.} {\bf B838} (2010) 136,
  [\href{http://arxiv.org/abs/1002.4407}{{\tt arXiv:1002.4407}}].

\bibitem{Ball:2009qv}
{\bf The NNPDF} Collaboration, R.~D. Ball et~al., {\it {Fitting Parton
  Distribution Data with Multiplicative Normalization Uncertainties}},  {\em
  JHEP} {\bf 05} (2010) 075, [\href{http://arxiv.org/abs/0912.2276}{{\tt
  arXiv:0912.2276}}].

\bibitem{Perez-Salinas:2020nem}
A.~P\'erez-Salinas, J.~Cruz-Martinez, A.~A. Alhajri, and S.~Carrazza, {\it
  {Determining the proton content with a quantum computer}},  {\em Phys. Rev.
  D} {\bf 103} (2021), no.~3 034027,
  [\href{http://arxiv.org/abs/2011.13934}{{\tt arXiv:2011.13934}}].

\bibitem{efthymiou:2020qibo}
S.~Efthymiou, S.~Ramos-Calderer, C.~Bravo-Prieto, A.~P\'erez-Salinas,
  D.~Garc\'ia-Mart\'in, A.~Garcia-Saez, J.~I. Latorre, and S.~Carrazza, {\it
  Qibo: a framework for quantum simulation with hardware acceleration},  2020.

\bibitem{tensorflow2015:whitepaper}
M.~Abadi, A.~Agarwal, P.~Barham, E.~Brevdo, Z.~Chen, C.~Citro, G.~S. Corrado,
  A.~Davis, J.~Dean, M.~Devin, S.~Ghemawat, I.~Goodfellow, A.~Harp, G.~Irving,
  M.~Isard, Y.~Jia, R.~Jozefowicz, L.~Kaiser, M.~Kudlur, J.~Levenberg,
  D.~Man\'{e}, R.~Monga, S.~Moore, D.~Murray, C.~Olah, M.~Schuster, J.~Shlens,
  B.~Steiner, I.~Sutskever, K.~Talwar, P.~Tucker, V.~Vanhoucke, V.~Vasudevan,
  F.~Vi\'{e}gas, O.~Vinyals, P.~Warden, M.~Wattenberg, M.~Wicke, Y.~Yu, and
  X.~Zheng, {\it {TensorFlow}: Large-scale machine learning on heterogeneous
  systems},  2015.
\newblock Software available from tensorflow.org.

\bibitem{Bergstra:2013}
J.~Bergstra, D.~Yamins, and D.~D. Cox, {\it Making a science of model search:
  Hyperparameter optimization in hundreds of dimensions for vision
  architectures},  in {\em Proceedings of the 30th International Conference on
  International Conference on Machine Learning - Volume 28}, ICML'13,
  pp.~I--115--I--123, JMLR.org, 2013.

\bibitem{Bergstra:2011:AHO:2986459.2986743}
J.~Bergstra, R.~Bardenet, Y.~Bengio, and B.~K{\'e}gl, {\it Algorithms for
  hyper-parameter optimization},  in {\em Proceedings of the 24th International
  Conference on Neural Information Processing Systems}, NIPS'11, (USA),
  pp.~2546--2554, Curran Associates Inc., 2011.

\bibitem{Hawkins}
D.~M. Hawkins, {\it {The problem of overfitting}},  {\em J. Chem. Inf. Comput.
  Sci.} {\bf 44} (2004) 1.

\bibitem{Schaffer93selectinga}
C.~Schaffer, {\it Selecting a classification method by cross-validation},  in
  {\em Machine Learning}, pp.~135--143, 1993.

\bibitem{glorot:2010}
Y.~Bengio and X.~Glorot, {\it Understanding the difficulty of training deep
  feed forward neural networks},  {\em International Conference on Artificial
  Intelligence and Statistics} (01, 2010) 249--256.

\bibitem{Ball:2008by}
{\bf The NNPDF} Collaboration, R.~D. Ball et~al., {\it {A determination of
  parton distributions with faithful uncertainty estimation}},  {\em Nucl.
  Phys.} {\bf B809} (2009) 1--63, [\href{http://arxiv.org/abs/0808.1231}{{\tt
  arXiv:0808.1231}}].

\bibitem{Carrazza:2019sec}
S.~Carrazza, C.~Degrande, S.~Iranipour, J.~Rojo, and M.~Ubiali, {\it {Can New
  Physics hide inside the proton?}},  {\em Phys. Rev. Lett.} {\bf 123} (2019),
  no.~13 132001, [\href{http://arxiv.org/abs/1905.05215}{{\tt
  arXiv:1905.05215}}].

\bibitem{Greljo:2021kvv}
A.~Greljo, S.~Iranipour, Z.~Kassabov, M.~Madigan, J.~Moore, J.~Rojo, M.~Ubiali,
  and C.~Voisey, {\it {Parton distributions in the SMEFT from high-energy
  Drell-Yan tails}},  {\em JHEP} {\bf 07} (2021) 122,
  [\href{http://arxiv.org/abs/2104.02723}{{\tt arXiv:2104.02723}}].

\bibitem{Bertone:2017bme}
{\bf NNPDF} Collaboration, V.~Bertone, S.~Carrazza, N.~P. Hartland, and
  J.~Rojo, {\it {Illuminating the photon content of the proton within a global
  PDF analysis}},  {\em SciPost Phys.} {\bf 5} (2018), no.~1 008,
  [\href{http://arxiv.org/abs/1712.07053}{{\tt arXiv:1712.07053}}].

\bibitem{Bonvini:2015ira}
M.~Bonvini, S.~Marzani, J.~Rojo, L.~Rottoli, M.~Ubiali, R.~D. Ball, V.~Bertone,
  S.~Carrazza, and N.~P. Hartland, {\it {Parton distributions with threshold
  resummation}},  {\em JHEP} {\bf 09} (2015) 191,
  [\href{http://arxiv.org/abs/1507.01006}{{\tt arXiv:1507.01006}}].

\bibitem{Boughezal:2017nla}
R.~Boughezal, A.~Guffanti, F.~Petriello, and M.~Ubiali, {\it {The impact of the
  LHC Z-boson transverse momentum data on PDF determinations}},  {\em JHEP}
  {\bf 07} (2017) 130, [\href{http://arxiv.org/abs/1705.00343}{{\tt
  arXiv:1705.00343}}].

\bibitem{COV}
Z.~Kassabov, E.~R. Nocera, and M.~Wilson, ``On the necessity of robust
  estimates of experimental uncertainties in high energy physics.'' in
  preparation.

\bibitem{Forte:2020pyp}
S.~Forte and Z.~Kassabov, {\it {Why $\alpha _s$ cannot be determined from
  hadronic processes without simultaneously determining the parton
  distributions}},  {\em Eur. Phys. J. C} {\bf 80} (2020), no.~3 182,
  [\href{http://arxiv.org/abs/2001.04986}{{\tt arXiv:2001.04986}}].

\bibitem{Collins:2001es}
J.~C. Collins and J.~Pumplin, {\it {Tests of goodness of fit to multiple data
  sets}},  \href{http://arxiv.org/abs/hep-ph/0105207}{{\tt hep-ph/0105207}}.

\bibitem{Pumplin:2002vw}
J.~Pumplin et~al., {\it {New generation of parton distributions with
  uncertainties from global QCD analysis}},  {\em JHEP} {\bf 07} (2002) 012,
  [\href{http://arxiv.org/abs/hep-ph/0201195}{{\tt hep-ph/0201195}}].

\bibitem{Forte:2002fg}
S.~Forte, L.~Garrido, J.~I. Latorre, and A.~Piccione, {\it Neural network
  parametrization of deep-inelastic structure functions},  {\em JHEP} {\bf 05}
  (2002) 062, [\href{http://arxiv.org/abs/hep-ph/0204232}{{\tt
  hep-ph/0204232}}].

\bibitem{NNPDF:2011aa}
{\bf NNPDF} Collaboration, R.~D. Ball, V.~Bertone, F.~Cerutti, L.~Del~Debbio,
  S.~Forte, A.~Guffanti, J.~I. Latorre, J.~Rojo, and M.~Ubiali, {\it {On the
  Impact of NMC Data on NLO and NNLO Parton Distributions and Higgs Production
  at the Tevatron and the LHC}},  {\em Phys. Lett.} {\bf B704} (2011) 36--42,
  [\href{http://arxiv.org/abs/1102.3182}{{\tt arXiv:1102.3182}}].

\bibitem{Ball:2017otu}
R.~D. Ball, V.~Bertone, M.~Bonvini, S.~Marzani, J.~Rojo, and L.~Rottoli, {\it
  {Parton distributions with small-x resummation: evidence for BFKL dynamics in
  HERA data}},  {\em Eur. Phys. J.} {\bf C78} (2018), no.~4 321,
  [\href{http://arxiv.org/abs/1710.05935}{{\tt arXiv:1710.05935}}].

\bibitem{Guzzi:2021fre}
M.~Guzzi et~al., {\it {NNLO constraints on proton PDFs from the SeaQuest and
  STAR experiments and other developments in the CTEQ-TEA global analysis}},
  in {\em {28th International Workshop on Deep Inelastic Scattering and Related
  Subjects}}, 8, 2021.
\newblock \href{http://arxiv.org/abs/2108.06596}{{\tt arXiv:2108.06596}}.

\bibitem{Cacciari:2019qjx}
M.~Cacciari, S.~Forte, D.~Napoletano, G.~Soyez, and G.~Stagnitto, {\it
  {Single-jet inclusive cross section and its definition}},  {\em Phys. Rev. D}
  {\bf 100} (2019), no.~11 114015, [\href{http://arxiv.org/abs/1906.11850}{{\tt
  arXiv:1906.11850}}].

\bibitem{Cridge:2021qjj}
{\bf PDF4LHC21 combination group} Collaboration, T.~Cridge, {\it {PDF4LHC21:
  Update on the benchmarking of the CT, MSHT and NNPDF global PDF fits}},  in
  {\em {28th International Workshop on Deep Inelastic Scattering and Related
  Subjects}}, 8, 2021.
\newblock \href{http://arxiv.org/abs/2108.09099}{{\tt arXiv:2108.09099}}.

\bibitem{Ball:2022hsh}
R.~D. Ball et~al., {\it {The PDF4LHC21 combination of global PDF fits for the
  LHC Run III}},  \href{http://arxiv.org/abs/2203.05506}{{\tt
  arXiv:2203.05506}}.

\bibitem{Demartin:2010er}
F.~Demartin, S.~Forte, E.~Mariani, J.~Rojo, and A.~Vicini, {\it {The impact of
  PDF and $\alpha_s$ uncertainties on Higgs Production in gluon fusion at
  hadron colliders}},  {\em Phys. Rev.} {\bf D82} (2010) 014002,
  [\href{http://arxiv.org/abs/1004.0962}{{\tt arXiv:1004.0962}}].

\bibitem{Dulat:2015mca}
S.~Dulat, T.-J. Hou, J.~Gao, M.~Guzzi, J.~Huston, P.~Nadolsky, J.~Pumplin,
  C.~Schmidt, D.~Stump, and C.~P. Yuan, {\it {New parton distribution functions
  from a global analysis of quantum chromodynamics}},  {\em Phys. Rev.} {\bf
  D93} (2016), no.~3 033006, [\href{http://arxiv.org/abs/1506.07443}{{\tt
  arXiv:1506.07443}}].

\bibitem{Harland-Lang:2014zoa}
L.~A. Harland-Lang, A.~D. Martin, P.~Motylinski, and R.~S. Thorne, {\it {Parton
  distributions in the LHC era: MMHT 2014 PDFs}},  {\em Eur. Phys. J.} {\bf
  C75} (2015) 204, [\href{http://arxiv.org/abs/1412.3989}{{\tt
  arXiv:1412.3989}}].

\bibitem{NNPDF:datatheory}
{\bf NNPDF} Collaboration, R.~D. Ball et~al. Publicly available from
  \url{https://data.nnpdf.science/nnpdf40-reports/nnpdf40-vs-nnpdf31_like/#dataset-plots}.

\bibitem{demortier}
L.~Demortier, {\em {Proceedings, PHYSTAT 2011 Workshop on Statistical Issues
  Related to Discovery Claims in Search Experiments and Unfolding, CERN,Geneva,
  Switzerland 17-20 January 2011}}, ch.~Open Issues in the Wake of Banff 2011.
\newblock 2011.

\bibitem{Watt:2012tq}
G.~Watt and R.~S. Thorne, {\it {Study of Monte Carlo approach to experimental
  uncertainty propagation with MSTW 2008 PDFs}},  {\em JHEP} {\bf 1208} (2012)
  052, [\href{http://arxiv.org/abs/1205.4024}{{\tt arXiv:1205.4024}}].

\bibitem{DelDebbio:2021whr}
L.~Del~Debbio, T.~Giani, and M.~Wilson, {\it {Bayesian Approach to Inverse
  Problems: an Application to NNPDF Closure Testing}},
  \href{http://arxiv.org/abs/2111.05787}{{\tt arXiv:2111.05787}}.

\bibitem{Efron:1979bxm}
B.~Efron, {\it {Bootstrap Methods: Another Look at the Jackknife}},  {\em
  Annals Statist.} {\bf 7} (1979), no.~1 1--26.

\bibitem{Efron:1986hys}
B.~Efron and R.~Tibshirani, {\it {An introduction to the bootstrap}},  {\em
  Statist. Sci.} {\bf 57} (1986), no.~1 54--75.

\bibitem{DeRoeck:1995mt}
A.~De~Roeck, {\it {Proton structure function data and search for BFKL
  signatures at HERA}},  {\em Acta Phys. Polon. B} {\bf 27} (1995) 1175--1199.

\bibitem{Tung:2004ab}
W.~K. Tung, {\it {Progress of CTEQ global QCD analyses}},  in {\em {12th
  International Workshop on Deep Inelastic Scattering (DIS 2004)}},
  pp.~433--438, 4, 2004.

\bibitem{Butterworth:2014efa}
J.~Butterworth et~al., {\it {Les Houches 2013: Physics at TeV Colliders:
  Standard Model Working Group Report}},
  \href{http://arxiv.org/abs/1405.1067}{{\tt arXiv:1405.1067}}.

\bibitem{Ball:2016spl}
R.~D. Ball, E.~R. Nocera, and J.~Rojo, {\it {The asymptotic behaviour of parton
  distributions at small and large $x$}},  {\em Eur. Phys. J.} {\bf C76}
  (2016), no.~7 383, [\href{http://arxiv.org/abs/1604.00024}{{\tt
  arXiv:1604.00024}}].

\bibitem{Ball:2012cx}
R.~D. Ball et~al., {\it {Parton distributions with LHC data}},  {\em
  Nucl.Phys.} {\bf B867} (2013) 244,
  [\href{http://arxiv.org/abs/1207.1303}{{\tt arXiv:1207.1303}}].

\bibitem{Ball:2016neh}
{\bf NNPDF} Collaboration, R.~D. Ball, V.~Bertone, M.~Bonvini, S.~Carrazza,
  S.~Forte, A.~Guffanti, N.~P. Hartland, J.~Rojo, and L.~Rottoli, {\it {A
  Determination of the Charm Content of the Proton}},  {\em Eur. Phys. J.} {\bf
  C76} (2016), no.~11 647, [\href{http://arxiv.org/abs/1605.06515}{{\tt
  arXiv:1605.06515}}].

\bibitem{Brodsky:2015fna}
S.~J. Brodsky, A.~Kusina, F.~Lyonnet, I.~Schienbein, H.~Spiesberger, and
  R.~Vogt, {\it {A review of the intrinsic heavy quark content of the
  nucleon}},  {\em Adv. High Energy Phys.} {\bf 2015} (2015) 231547,
  [\href{http://arxiv.org/abs/1504.06287}{{\tt arXiv:1504.06287}}].

\bibitem{Rottoli:2016lsg}
L.~Rottoli, {\it {Fitting EMC structure functions with intrinsic charm}},  {\em
  PoS} {\bf DIS2016} (2016) 032, [\href{http://arxiv.org/abs/1606.09289}{{\tt
  arXiv:1606.09289}}].

\bibitem{ICpaper}
{\bf NNPDF} Collaboration , in preparation.

\bibitem{Nocera:2017zge}
E.~R. Nocera and M.~Ubiali, {\it {Constraining the gluon PDF at large x with
  LHC data}},  {\em PoS} {\bf DIS2017} (2018) 008,
  [\href{http://arxiv.org/abs/1709.09690}{{\tt arXiv:1709.09690}}].

\bibitem{Mangano:2016jyj}
M.~L. Mangano et~al., {\it {Physics at a 100 TeV pp collider: Standard Model
  processes}},  \href{http://arxiv.org/abs/1607.01831}{{\tt arXiv:1607.01831}}.

\bibitem{Martin:2003tt}
A.~D. Martin, R.~G. Roberts, W.~J. Stirling, and R.~S. Thorne, {\it {MRST
  partons and uncertainties}},  \href{http://arxiv.org/abs/hep-ph/0307262}{{\tt
  hep-ph/0307262}}.

\bibitem{Butterworth:2015oua}
J.~Butterworth et~al., {\it {PDF4LHC recommendations for LHC Run II}},  {\em J.
  Phys.} {\bf G43} (2016) 023001, [\href{http://arxiv.org/abs/1510.03865}{{\tt
  arXiv:1510.03865}}].

\bibitem{Pagani:2016caq}
D.~Pagani, I.~Tsinikos, and M.~Zaro, {\it {The impact of the photon PDF and
  electroweak corrections on $t \bar{t}$ distributions}},  {\em Eur. Phys. J.}
  {\bf C76} (2016), no.~9 479, [\href{http://arxiv.org/abs/1606.01915}{{\tt
  arXiv:1606.01915}}].

\bibitem{Denner:1999gp}
A.~Denner, S.~Dittmaier, M.~Roth, and D.~Wackeroth, {\it {Predictions for all
  processes e+ e- ---\ensuremath{>} 4 fermions + gamma}},  {\em Nucl. Phys. B}
  {\bf 560} (1999) 33--65, [\href{http://arxiv.org/abs/hep-ph/9904472}{{\tt
  hep-ph/9904472}}].

\bibitem{Denner:2005fg}
A.~Denner, S.~Dittmaier, M.~Roth, and L.~H. Wieders, {\it {Electroweak
  corrections to charged-current e+ e- ---\ensuremath{>} 4 fermion processes:
  Technical details and further results}},  {\em Nucl. Phys. B} {\bf 724}
  (2005) 247--294, [\href{http://arxiv.org/abs/hep-ph/0505042}{{\tt
  hep-ph/0505042}}]. [Erratum: Nucl.Phys.B 854, 504--507 (2012)].

\bibitem{Denner:2006ic}
A.~Denner and S.~Dittmaier, {\it {The Complex-mass scheme for perturbative
  calculations with unstable particles}},  {\em Nucl. Phys. B Proc. Suppl.}
  {\bf 160} (2006) 22--26, [\href{http://arxiv.org/abs/hep-ph/0605312}{{\tt
  hep-ph/0605312}}].

\bibitem{Artoisenet:2013puc}
P.~Artoisenet, P.~de~Aquino, F.~Demartin, R.~Frederix, S.~Frixione, et~al.,
  {\it {A framework for Higgs characterisation}},  {\em JHEP} {\bf 1311} (2013)
  043, [\href{http://arxiv.org/abs/1306.6464}{{\tt arXiv:1306.6464}}].

\bibitem{Demartin:2014fia}
F.~Demartin, F.~Maltoni, K.~Mawatari, B.~Page, and M.~Zaro, {\it {Higgs
  characterisation at NLO in QCD: CP properties of the top-quark Yukawa
  interaction}},  {\em Eur.Phys.J.} {\bf C74} (2014), no.~9 3065,
  [\href{http://arxiv.org/abs/1407.5089}{{\tt arXiv:1407.5089}}].

\bibitem{Bertone:2018dse}
V.~Bertone, R.~Gauld, and J.~Rojo, {\it {Neutrino Telescopes as QCD
  Microscopes}},  {\em JHEP} {\bf 01} (2019) 217,
  [\href{http://arxiv.org/abs/1808.02034}{{\tt arXiv:1808.02034}}].

\bibitem{FAIR}
M.~D. Wilkinson et~al., {\it {The FAIR Guiding Principles for scientific data
  management and stewardship}},  {\em Sci. data} {\bf 160018} (201).

\bibitem{Ball:2021icz}
R.~D. Ball and R.~L. Pearson, {\it {Correlation of Theoretical Uncertainties in
  PDF Fits and Theoretical Uncertainties in Predictions}},
  \href{http://arxiv.org/abs/2105.05114}{{\tt arXiv:2105.05114}}.

\bibitem{Vogt:2018miu}
A.~Vogt, F.~Herzog, S.~Moch, B.~Ruijl, T.~Ueda, and J.~A.~M. Vermaseren, {\it
  {Anomalous dimensions and splitting functions beyond the
  next-to-next-to-leading order}},  {\em PoS} {\bf LL2018} (2018) 050,
  [\href{http://arxiv.org/abs/1808.08981}{{\tt arXiv:1808.08981}}].

\bibitem{Boughezal:2016wmq}
R.~Boughezal, J.~M. Campbell, R.~K. Ellis, C.~Focke, W.~Giele, X.~Liu,
  F.~Petriello, and C.~Williams, {\it {Color singlet production at NNLO in
  MCFM}},  {\em Eur. Phys. J. C} {\bf 77} (2017), no.~1 7,
  [\href{http://arxiv.org/abs/1605.08011}{{\tt arXiv:1605.08011}}].

\bibitem{Grazzini:2017mhc}
M.~Grazzini, S.~Kallweit, and M.~Wiesemann, {\it {Fully differential NNLO
  computations with MATRIX}},  {\em Eur. Phys. J.} {\bf C78} (2018), no.~7 537,
  [\href{http://arxiv.org/abs/1711.06631}{{\tt arXiv:1711.06631}}].

\bibitem{zahari_kassabov_2019_2571601}
Z.~Kassabov, ``{Reportengine: A framework for declarative data analysis}.''
  \url{https://doi.org/10.5281/zenodo.2571601}, Feb., 2019.

\bibitem{Whitlow:1990gk}
L.~W. Whitlow, S.~Rock, A.~Bodek, E.~M. Riordan, and S.~Dasu, {\it {A Precise
  extraction of R = sigma-L / sigma-T from a global analysis of the SLAC deep
  inelastic e p and e d scattering cross-sections}},  {\em Phys. Lett. B} {\bf
  250} (1990) 193--198.

\bibitem{Benvenuti:1989fm}
{\bf BCDMS} Collaboration, A.~C. Benvenuti et~al., {\it {A High Statistics
  Measurement of the Deuteron Structure Functions $F_2(x, Q^2)$ and $R$ from
  Deep Inelastic Muon Scattering at High $Q^2$}},  {\em Phys. Lett.} {\bf B237}
  (1990) 592.

\bibitem{Kayis-Topaksu:2011ols}
A.~Kayis-Topaksu et~al., {\it {Measurement of charm production in neutrino
  charged-current interactions}},  {\em New J. Phys.} {\bf 13} (2011) 093002,
  [\href{http://arxiv.org/abs/1107.0613}{{\tt arXiv:1107.0613}}].

\bibitem{Tzanov:2005kr}
{\bf NuTeV} Collaboration, M.~Tzanov et~al., {\it {Precise measurement of
  neutrino and anti-neutrino differential cross sections}},  {\em Phys. Rev.}
  {\bf D74} (2006) 012008, [\href{http://arxiv.org/abs/hep-ex/0509010}{{\tt
  hep-ex/0509010}}].

\bibitem{Seligman:1997mc}
W.~G. Seligman et~al., {\it {Improved determination of alpha(s) from neutrino
  nucleon scattering}},  {\em Phys. Rev. Lett.} {\bf 79} (1997) 1213--1216,
  [\href{http://arxiv.org/abs/hep-ex/9701017}{{\tt hep-ex/9701017}}].

\bibitem{CCFRNuTeV:2000qwc}
{\bf CCFR/NuTeV} Collaboration, U.-K. Yang et~al., {\it {Measurements of $F_2$
  and $xF^{\nu}_3 - x F^{\bar{\nu}}_3$ from CCFR $\nu_\mu-$Fe and
  $\bar{\nu}_\mu-$Fe data in a physics model independent way}},  {\em Phys.
  Rev. Lett.} {\bf 86} (2001) 2742--2745,
  [\href{http://arxiv.org/abs/hep-ex/0009041}{{\tt hep-ex/0009041}}].

\bibitem{Berge:1989hr}
J.~P. Berge et~al., {\it {A Measurement of Differential Cross-Sections and
  Nucleon Structure Functions in Charged Current Neutrino Interactions on
  Iron}},  {\em Z. Phys. C} {\bf 49} (1991) 187--224.

\bibitem{E665:1996mob}
{\bf E665} Collaboration, M.~R. Adams et~al., {\it {Proton and deuteron
  structure functions in muon scattering at 470-GeV}},  {\em Phys. Rev. D} {\bf
  54} (1996) 3006--3056.

\bibitem{Aaron:2009aa}
{\bf H1 and ZEUS} Collaboration, F.~Aaron et~al., {\it {Combined Measurement
  and QCD Analysis of the Inclusive $e^{\pm}p$ Scattering Cross Sections at
  HERA}},  {\em JHEP} {\bf 1001} (2010) 109,
  [\href{http://arxiv.org/abs/0911.0884}{{\tt arXiv:0911.0884}}].

\bibitem{Aktas:2004az}
{\bf H1} Collaboration, A.~Aktas et~al., {\it {Measurement of F2($c \bar{c}$)
  and F2($b \bar{b}$) at high $Q^{2}$ using the H1 vertex detector at HERA}},
  {\em Eur. Phys. J.} {\bf C40} (2005) 349--359,
  [\href{http://arxiv.org/abs/hep-ex/0411046}{{\tt hep-ex/0411046}}].

\bibitem{Collaboration:2010ry}
{\bf H1} Collaboration, F.~Aaron et~al., {\it {Measurement of the Inclusive
  $e^{\pm}p$ Scattering Cross Section at High Inelasticity y and of the
  Structure Function $F_L$}},  {\em Eur.Phys.J.} {\bf C71} (2011) 1579,
  [\href{http://arxiv.org/abs/1012.4355}{{\tt arXiv:1012.4355}}].

\bibitem{h1fl}
{\bf H1} Collaboration, F.~D. Aaron et~al., {\it {Measurement of the Proton
  Structure Function $F_L$ at Low x}},  {\em Phys. Lett.} {\bf B665} (2008)
  139--146, [\href{http://arxiv.org/abs/0805.2809}{{\tt arXiv:0805.2809}}].

\bibitem{Chekanov:2009na}
{\bf ZEUS Collaboration} Collaboration, S.~Chekanov et~al., {\it {Measurement
  of the Longitudinal Proton Structure Function at HERA}},  {\em Phys.Lett.}
  {\bf B682} (2009) 8, [\href{http://arxiv.org/abs/0904.1092}{{\tt
  arXiv:0904.1092}}].

\bibitem{CDF:1998uzn}
{\bf CDF} Collaboration, F.~Abe et~al., {\it {Measurement of the Lepton Charge
  Asymmetry in $W$ Boson Decays Produced in $p \bar{p}$ Collisions}},  {\em
  Phys. Rev. Lett.} {\bf 81} (1998) 5754--5759,
  [\href{http://arxiv.org/abs/hep-ex/9809001}{{\tt hep-ex/9809001}}].

\bibitem{Acosta:2005ud}
{\bf CDF} Collaboration, D.~E. Acosta et~al., {\it {Measurement of the
  forward-backward charge asymmetry from $W \to e \nu$ production in $p\bar{p}$
  collisions at $\sqrt{s} = 1.96$ TeV}},  {\em Phys. Rev.} {\bf D71} (2005)
  051104, [\href{http://arxiv.org/abs/hep-ex/0501023}{{\tt hep-ex/0501023}}].

\bibitem{Aaltonen:2009ta}
{\bf CDF} Collaboration, T.~Aaltonen et~al., {\it {Direct Measurement of the
  $W$ Production Charge Asymmetry in $p\bar{p}$ Collisions at $\sqrt{s} = 1.96$
  TeV}},  {\em Phys. Rev. Lett.} {\bf 102} (2009) 181801,
  [\href{http://arxiv.org/abs/0901.2169}{{\tt arXiv:0901.2169}}].

\bibitem{Aaltonen:2008eq}
{\bf CDF} Collaboration, T.~Aaltonen et~al., {\it {Measurement of the Inclusive
  Jet Cross Section at the Fermilab Tevatron p-pbar Collider Using a Cone-Based
  Jet Algorithm}},  {\em Phys. Rev.} {\bf D78} (2008) 052006,
  [\href{http://arxiv.org/abs/0807.2204}{{\tt arXiv:0807.2204}}].

\bibitem{Abazov:2008qv}
{\bf D0} Collaboration, V.~M. Abazov et~al., {\it {Measurement of the electron
  charge asymmetry in $p \bar{p} \to W + X \to e \nu + X$ events at $\sqrt{s}$
  = 1.96-TeV}},  {\em Phys. Rev. Lett.} {\bf 101} (2008) 211801,
  [\href{http://arxiv.org/abs/0807.3367}{{\tt arXiv:0807.3367}}].

\bibitem{D0:2013lql}
{\bf D0} Collaboration, V.~M. Abazov et~al., {\it {Measurement of the W Boson
  Production Charge Asymmetry in $p\bar{p}\rightarrow W+X \rightarrow e\nu +X$
  Events at $\sqrt{s}=1.96$ TeV}},  {\em Phys. Rev. Lett.} {\bf 112} (2014),
  no.~15 151803, [\href{http://arxiv.org/abs/1312.2895}{{\tt
  arXiv:1312.2895}}]. [Erratum: Phys.Rev.Lett. 114, 049901 (2015)].

\bibitem{D0:2007pcy}
{\bf D0} Collaboration, V.~M. Abazov et~al., {\it {Measurement of the muon
  charge asymmetry from $W$ boson decays}},  {\em Phys. Rev. D} {\bf 77} (2008)
  011106, [\href{http://arxiv.org/abs/0709.4254}{{\tt arXiv:0709.4254}}].

\bibitem{D0:2008hua}
{\bf D0} Collaboration, V.~M. Abazov et~al., {\it {Measurement of the inclusive
  jet cross-section in $p \bar{p}$ collisions at $\sqrt{s}$=1.96-TeV}},  {\em
  Phys. Rev. Lett.} {\bf 101} (2008) 062001,
  [\href{http://arxiv.org/abs/0802.2400}{{\tt arXiv:0802.2400}}].

\bibitem{CDF:2013hmv}
{\bf CDF, D0} Collaboration, T.~A. Aaltonen et~al., {\it {Combination of
  Measurements of the Top-Quark Pair Production Cross Section from the Tevatron
  Collider}},  {\em Phys. Rev. D} {\bf 89} (2014), no.~7 072001,
  [\href{http://arxiv.org/abs/1309.7570}{{\tt arXiv:1309.7570}}].

\bibitem{CDF:2015gsg}
{\bf CDF, D0} Collaboration, T.~A. Aaltonen et~al., {\it {Tevatron Combination
  of Single-Top-Quark Cross Sections and Determination of the Magnitude of the
  Cabibbo-Kobayashi-Maskawa Matrix Element $\bf V_{tb}$}},  {\em Phys. Rev.
  Lett.} {\bf 115} (2015), no.~15 152003,
  [\href{http://arxiv.org/abs/1503.05027}{{\tt arXiv:1503.05027}}].

\bibitem{Aad:2014xaa}
{\bf ATLAS} Collaboration, G.~Aad et~al., {\it {Measurement of the $Z/\gamma^*$
  boson transverse momentum distribution in $pp$ collisions at $\sqrt{s}$ = 7
  TeV with the ATLAS detector}},  {\em JHEP} {\bf 09} (2014) 145,
  [\href{http://arxiv.org/abs/1406.3660}{{\tt arXiv:1406.3660}}].

\bibitem{Aad:2012vip}
{\bf ATLAS} Collaboration, G.~Aad et~al., {\it {Measurement of the ttbar
  production cross section in the tau+jets channel using the ATLAS detector}},
  {\em Eur. Phys. J. C} {\bf 73} (2013), no.~3 2328,
  [\href{http://arxiv.org/abs/1211.7205}{{\tt arXiv:1211.7205}}].

\bibitem{Aad:2014jra}
{\bf ATLAS} Collaboration, G.~Aad et~al., {\it {Simultaneous measurements of
  the $t\bar{t}$, $W^+W^-$, and $Z/\gamma^{*}\rightarrow\tau\tau$ production
  cross-sections in $pp$ collisions at $\sqrt{s} = 7$ TeV with the ATLAS
  detector}},  {\em Phys. Rev. D} {\bf 91} (2015), no.~5 052005,
  [\href{http://arxiv.org/abs/1407.0573}{{\tt arXiv:1407.0573}}].

\bibitem{Aad:2015pga}
{\bf ATLAS} Collaboration, G.~Aad et~al., {\it {Measurement of the top pair
  production cross section in 8 TeV proton-proton collisions using kinematic
  information in the lepton+jets final state with ATLAS}},  {\em Phys. Rev. D}
  {\bf 91} (2015), no.~11 112013, [\href{http://arxiv.org/abs/1504.04251}{{\tt
  arXiv:1504.04251}}].

\bibitem{Aad:2015dya}
{\bf ATLAS} Collaboration, G.~Aad et~al., {\it {Measurements of the top quark
  branching ratios into channels with leptons and quarks with the ATLAS
  detector}},  {\em Phys. Rev. D} {\bf 92} (2015), no.~7 072005,
  [\href{http://arxiv.org/abs/1506.05074}{{\tt arXiv:1506.05074}}].

\bibitem{ATLAS:2012gpa}
{\bf ATLAS} Collaboration, {\it {Measurement of the top quark pair production
  cross-section with ATLAS in $pp$ collisions at $\sqrt{s}=7$ TeV in the
  single-lepton channel using semileptonic $b$ decays}}, .

\bibitem{ATLAS:2012uma}
{\bf ATLAS} Collaboration, {\it {Measurement of the $t\bar{t}$ production cross
  section in the all-hadronic channel in ATLAS with $\sqrt s =7$ TeV data}}, .

\bibitem{ATLAS:2019hau}
{\bf ATLAS} Collaboration, G.~Aad et~al., {\it {Measurement of the $t\bar{t}$
  production cross-section and lepton differential distributions in $e\mu $
  dilepton events from $pp$ collisions at $\sqrt{s}=13\,\text {TeV}$ with the
  ATLAS detector}},  {\em Eur. Phys. J. C} {\bf 80} (2020), no.~6 528,
  [\href{http://arxiv.org/abs/1910.08819}{{\tt arXiv:1910.08819}}].

\bibitem{Chatrchyan:2011jz}
{\bf CMS} Collaboration, S.~Chatrchyan et~al., {\it {Measurement of the lepton
  charge asymmetry in inclusive $W$ production in pp collisions at $\sqrt{s} =
  7$ TeV}},  {\em JHEP} {\bf 04} (2011) 050,
  [\href{http://arxiv.org/abs/1103.3470}{{\tt arXiv:1103.3470}}].

\bibitem{Chatrchyan:2011wt}
{\bf CMS} Collaboration, S.~Chatrchyan et~al., {\it {Measurement of the
  Rapidity and Transverse Momentum Distributions of $Z$ Bosons in $pp$
  Collisions at $\sqrt{s}=7$ TeV}},  {\em Phys. Rev.} {\bf D85} (2012) 032002,
  [\href{http://arxiv.org/abs/1110.4973}{{\tt arXiv:1110.4973}}].

\bibitem{CMS:2014jea}
{\bf CMS} Collaboration, V.~Khachatryan et~al., {\it {Measurements of
  differential and double-differential Drell-Yan cross sections in
  proton-proton collisions at 8 TeV}},  {\em Eur. Phys. J.} {\bf C75} (2015),
  no.~4 147, [\href{http://arxiv.org/abs/1412.1115}{{\tt arXiv:1412.1115}}].

\bibitem{CMS:2016mwa}
{\bf CMS} Collaboration, V.~Khachatryan et~al., {\it {Measurement of the
  transverse momentum spectra of weak vector bosons produced in proton-proton
  collisions at $ \sqrt{s}=8 $ TeV}},  {\em JHEP} {\bf 02} (2017) 096,
  [\href{http://arxiv.org/abs/1606.05864}{{\tt arXiv:1606.05864}}].

\bibitem{Chatrchyan:2013faa}
{\bf CMS} Collaboration, S.~Chatrchyan et~al., {\it {Measurement of the $t
  \bar{t}$ production cross section in the dilepton channel in pp collisions at
  $\sqrt{s}$ = 8 TeV}},  {\em JHEP} {\bf 1402} (2014) 024,
  [\href{http://arxiv.org/abs/1312.7582}{{\tt arXiv:1312.7582}}].

\bibitem{Khachatryan:2016kzg}
{\bf CMS} Collaboration, V.~Khachatryan et~al., {\it {Measurement of the
  $t\bar{t}$ production cross section using events in the e$\mu$ final state in
  pp collisions at $\sqrt{s} =$ 13 TeV}},  {\em Eur. Phys. J. C} {\bf 77}
  (2017) 172, [\href{http://arxiv.org/abs/1611.04040}{{\tt arXiv:1611.04040}}].

\bibitem{Khachatryan:2016yzq}
{\bf CMS} Collaboration, V.~Khachatryan et~al., {\it {Measurements of the
  $\mathrm{t}\overline{\mathrm{t}}$ production cross section in lepton+jets
  final states in pp collisions at 8 $\,\text {TeV}$ and ratio of 8 to 7
  $\,\text {TeV}$ cross sections}},  {\em Eur. Phys. J. C} {\bf 77} (2017),
  no.~1 15, [\href{http://arxiv.org/abs/1602.09024}{{\tt arXiv:1602.09024}}].

\bibitem{CMS:2015toa}
{\bf CMS} Collaboration, {\it {Measurement of the inclusive and differential tt
  production cross sections in lepton + jets final states at 13 TeV}}, .

\bibitem{Chatrchyan:2012vs}
{\bf CMS} Collaboration, S.~Chatrchyan et~al., {\it {Measurement of the top
  quark pair production cross section in $pp$ collisions at $\sqrt{s} = 7$ TeV
  in dilepton final states containing a $\tau$}},  {\em Phys. Rev. D} {\bf 85}
  (2012) 112007, [\href{http://arxiv.org/abs/1203.6810}{{\tt
  arXiv:1203.6810}}].

\bibitem{Chatrchyan:2013kff}
{\bf CMS} Collaboration, S.~Chatrchyan et~al., {\it {Measurement of the
  $t\bar{t}$ Production Cross Section in the $\tau$ +Jets Channel in $pp$
  Collisions at $\sqrt{s} = 7$ TeV}},  {\em Eur. Phys. J. C} {\bf 73} (2013),
  no.~4 2386, [\href{http://arxiv.org/abs/1301.5755}{{\tt arXiv:1301.5755}}].

\bibitem{Chatrchyan:2013ual}
{\bf CMS} Collaboration, S.~Chatrchyan et~al., {\it {Measurement of the
  $t\bar{t}$ Production Cross Section in the All-Jet Final State in pp
  Collisions at $\sqrt{s}$ = 7 TeV}},  {\em JHEP} {\bf 05} (2013) 065,
  [\href{http://arxiv.org/abs/1302.0508}{{\tt arXiv:1302.0508}}].

\bibitem{CMS:2016rtp}
{\bf CMS} Collaboration, {\it {Measurement of the ${\rm t}{\rm \bar{t}}$
  production cross section at 13 TeV in the all-jets final state}},  tech.
  rep., 2016.
\newblock CMS-PAS-TOP-16-013.

\bibitem{Khachatryan:2014loa}
{\bf CMS} Collaboration, V.~Khachatryan et~al., {\it {Measurement of the $t
  \bar t$ Production Cross Section in $pp$ Collisions at $\sqrt s = 8$ TeV in
  Dilepton Final States Containing One $\tau$ Lepton}},  {\em Phys. Lett. B}
  {\bf 739} (2014) 23--43, [\href{http://arxiv.org/abs/1407.6643}{{\tt
  arXiv:1407.6643}}].

\bibitem{CMS:2016pqu}
{\bf CMS} Collaboration, {\it {First measurement of the top quark pair
  production cross section in proton-proton collisions at
  $\sqrt{s}=5.02~\mathrm{TeV}$}},  tech. rep., 2016.
\newblock CMS-PAS-TOP-16-015.

\bibitem{CMS:2018lgn}
{\bf CMS} Collaboration, A.~M. Sirunyan et~al., {\it {Measurement of the single
  top quark and antiquark production cross sections in the $t$ channel and
  their ratio in proton-proton collisions at $\sqrt{s}=$ 13 TeV}},  {\em Phys.
  Lett. B} {\bf 800} (2020) 135042,
  [\href{http://arxiv.org/abs/1812.10514}{{\tt arXiv:1812.10514}}].

\bibitem{CMS:2019jjp}
{\bf CMS} Collaboration, A.~M. Sirunyan et~al., {\it {Measurement of
  differential cross sections and charge ratios for t-channel single top quark
  production in proton\textendash{}proton collisions at $\sqrt{s}=13\,\text
  {Te}\text {V}$}},  {\em Eur. Phys. J. C} {\bf 80} (2020), no.~5 370,
  [\href{http://arxiv.org/abs/1907.08330}{{\tt arXiv:1907.08330}}].

\bibitem{Aaij:2012vn}
{\bf LHCb} Collaboration, R.~Aaij et~al., {\it {Inclusive $W$ and $Z$
  production in the forward region at $\sqrt{s} = 7$ TeV}},  {\em JHEP} {\bf
  1206} (2012) 058, [\href{http://arxiv.org/abs/1204.1620}{{\tt
  arXiv:1204.1620}}].

\end{thebibliography}\endgroup
